\documentclass[11pt,notitlepage,a4paper]{article}

\usepackage{graphicx}

\usepackage{booktabs}
\usepackage[round]{natbib}
\usepackage{algorithm}
\usepackage{algpseudocode}
\usepackage{amsfonts}
\usepackage{amsmath, amsthm, amssymb}
\usepackage[most]{tcolorbox}
\usepackage{soul}
\soulregister\ref7
\usepackage{geometry}
\usepackage{tabularx}
\usepackage{longtable}
\usepackage{pdflscape}
\usepackage[dvipsnames]{xcolor}
\definecolor{linkcolor}{rgb}{0, 0, 0.35}
\definecolor{linkbordercolor}{rgb}{.9, .9, .9}
\definecolor{block-gray}{gray}{0.85}
\newtcolorbox{blockquote}{colback=block-gray,grow to right by=-1mm,grow to left by=-1mm,boxrule=0pt,boxsep=0pt,breakable}

\usepackage{hyperref}
\usepackage{listings}

\usepackage{colortbl}
\hypersetup{
    colorlinks=true,
    linkcolor=linkcolor,
    anchorcolor=linkcolor,
    citecolor=linkcolor,
    filecolor=linkcolor,
    urlcolor=linkcolor,
    bookmarks=true,
    citebordercolor=linkbordercolor,
    filebordercolor=linkbordercolor,
    linkbordercolor=linkbordercolor,
}
\usepackage[singlespacing]{setspace}

\usepackage{subfig}

\usepackage{verbatim}
\usepackage{enumitem}

\definecolor{bg}{rgb}{0.95,0.95,0.95}
\definecolor{highlight}{rgb}{1,1,0.6}

\definecolor{rlblue}{HTML}{4878A8}
\definecolor{rlorange}{HTML}{D4915C}
\definecolor{rlgreen}{HTML}{6BA368}
\definecolor{rlred}{HTML}{C25B56}
\definecolor{rlpurple}{HTML}{8B7BAF}
\definecolor{rlgray}{HTML}{888888}
\definecolor{rlblack}{HTML}{2D2D2D}

\usepackage{tikz}
\usetikzlibrary{arrows.meta, positioning, calc, decorations.pathreplacing}

\usepackage{lmodern}

\usepackage{bbm}
\newcommand{\indicator}[1]{\mathbbm{1}\{#1\}}

\usepackage{thmtools, thm-restate}

\usepackage[normalem]{ulem}

\newtheorem{theorem}{Theorem}
\newtheorem{lemma}{Lemma}

\newtheorem{assumption}{Assumption}

\newtheorem{definition}{Definition}

\newtheorem{corollary}{Corollary}[theorem]

\def\argmin{\mathop{\it argmin}}
\def\argmax{\mathop{\it argmax}}

\usepackage{titlesec}
\usepackage[section]{placeins}

\setlist{noitemsep,topsep=0pt,parsep=0pt,partopsep=0pt}

\begin{document}

\begin{titlepage}
\title{\textsc{A Survey of Reinforcement Learning For Economics}}
\author{
\textsc{Pranjal Rawat, Georgetown University}\thanks{I am grateful to John Rust for encouraging me to undertake this survey. I thank Simon LeBastard for sharing his RL notes with me.}\\
}
\date{\textsc{March 2026}}

\maketitle
\thispagestyle{empty}

\begin{abstract}
\noindent
This survey (re)introduces reinforcement learning methods to researchers in the social sciences. The curse of dimensionality limits how far exact dynamic programming can be effectively applied, forcing us to rely on suitably ``small'' problems or our ability to convert ``big'' problems into smaller ones. While this reduction has been sufficient for many classical applications, a growing class of economic models resists such reduction. Reinforcement learning algorithms offer a natural, sample-based extension of dynamic programming, extending tractability to problems with high-dimensional states, continuous actions, and strategic interactions. I review the theory connecting classical planning to modern learning algorithms and demonstrate their mechanics through simulated examples in pricing, inventory control, strategic games, and preference elicitation. I also examine the practical vulnerabilities of these algorithms, noting their brittleness, sample inefficiency, sensitivity to hyperparameters, and the absence of global convergence guarantees outside of tabular settings. The successes of reinforcement learning remain strictly bounded by these constraints, as well as a reliance on accurate simulators. That said, when guided by economic structure, reinforcement learning provides a flexible and innovative framework. It stands as an imperfect, but promising, addition to the researcher's toolkit. A companion survey \citep{RustRawat2026} covers the inverse problem of inferring preferences from observed behavior. All simulation code is publicly available.\footnote{\url{https://github.com/rawatpranjal/survey-of-reinforcement-learning-in-economics}}

\bigskip


\end{abstract}

\end{titlepage}

\newpage
\setcounter{tocdepth}{2}  
\tableofcontents
\newpage

\section{Introduction}
\label{section:intro}
This survey (re)introduces reinforcement learning to researchers studying sequential decision problems. I review the theoretical connections between dynamic programming and reinforcement learning, demonstrating how value iteration, Q-learning, and policy gradient methods are common solution methods to the same class of optimization problems. I then examine applications across several domains, including structural economic models with high-dimensional state spaces; strategic games in which multi-agent algorithms compute equilibria under imperfect information; bandit problems in which economic structure yields tighter regret bounds; preference learning; and field deployments in recommendations, marketplaces, operations, physical systems, and post-training. The exposition combines formal theory, practical applications and computational illustration.

Both dynamic programming and reinforcement learning solve the Bellman equation; they differ in the information requirements and the way in which the solution is refined. First, dynamic programming requires knowledge of the transition in the environment and the reward function which allows the reduction of the average Bellman error, reinforcement learning estimates value functions only from sampled transitions (observed sets of state, action, reward, next-state) which only allows reduction of the sampled Bellman error \textit{at that state-action pair}. This allows us to improve policies in domains where it is easier to build a simulator than specify the model of the environment and rewards e.g. board games, physics simulators for robots. Second, dynamic programming makes a ``breadth-first'' (across all states and actions) update of the solution at each sweep, while reinforcement learning makes a ``incremental'' (only for the current state and action) update; this greatly reduces the computational burden and enhances scalability.

Reduction of average Bellman errors gives dynamic programming a geometric rate of convergence to the optimal solution, while the incremental updates and reduction of sampled Bellman errors, when combined with ``sufficient exploration'' of the state-action space, gives reinforcement learning only sublinear convergence guarantees. The trade is favourable where exact dynamic programming cannot be computed at all; where it can, the simulations in this survey generally find it matching or beating the sampled alternative. These approximation methods sacrifice theoretical guarantees. RL algorithms lack the convergence assurances of exact dynamic programming. They exhibit sensitivity to hyperparameters and initialization. They can converge to suboptimal policies without diagnostic indication. This survey presents reinforcement learning as a computationally flexible framework while acknowledging its methodological limitations, which Section~\ref{section:deeprl_practice} documents in detail.

Theory in reinforcement learning trails empirical success, often by years; convergence guarantees, sample complexity bounds, and approximation error characterizations typically arrive after practitioners have demonstrated that an algorithm works. The theoretical insights that eventually follow tend to be deep and structural, and the empirical frontier is itself a productive research frontier. Experiments are brittle, conducted on benchmark environments that are stylized approximations of deployment settings. These benchmarks nonetheless serve a critical coordination function, aligning research effort, enabling reproducible comparison, and exposing failure modes that motivate new theory. Details matter disproportionately in reinforcement learning; small implementation choices can determine whether an algorithm converges or diverges, and the practice of releasing code and documenting hyperparameters, seeds, and preprocessing has proven essential to progress.

Existing surveys of reinforcement learning for economists divide along scope. \citet{CharpentierElieRemlinger2021} present reinforcement learning techniques alongside applications in economics, game theory, operations research and finance, and treat inverse reinforcement learning and its dynamic discrete choice lineage. \citet{MosaviEtAl2020} review deep learning and deep reinforcement learning methods, with the reinforcement learning applications concentrated in stock trading. \citet{HamblyXuYang2023} work through the finance applications, namely optimal execution, portfolio optimization, option pricing and hedging, market making, robo-advising and order routing. \citet{AtashbarShi2022} cover deep reinforcement learning for macroeconomics, organized around solution methods and bounded rationality. \citet{iskhakov2021} contrast machine learning with structural econometrics, asking where machine learning can advance the goals of structural work, and set inverse reinforcement learning against structural estimation.

This survey differs in three respects. It organizes the material around the Bellman equation and the convergence of its solution methods rather than around application domains. It develops material that the earlier surveys do not, including offline reinforcement learning, learning from human feedback, and distributionally robust and constrained methods. Each chapter carries a working simulation benchmarked against a dynamic programming or analytical solution.

Several topics are treated elsewhere. Algorithmic collusion, in which independent pricing algorithms learn to sustain supra-competitive prices \citep{Calvano2020}, is treated in a companion thesis chapter \citep{Rawat2026collusion} and omitted here. Portfolio optimization, optimal execution, and asset pricing via reinforcement learning form a large body of work surveyed comprehensively elsewhere \citep{HamblyXuYang2023}. The inverse problem of inferring preferences from observed behavior using inverse reinforcement learning and structural estimation is treated in a companion survey \citep{RustRawat2026}.

The survey addresses the forward problem, that is, computing optimal policies given a known or simulated environment. Section~\ref{section:history} traces the parallel historical development of dynamic programming and reinforcement learning. Section~\ref{section:rl_algorithms} surveys reinforcement learning algorithms, and Section~\ref{sec:planning_learning} develops the unified theory connecting planning and learning, with Section~\ref{section:deeprl_practice} examining the empirics of deep RL. Section~\ref{section:rl_econ_models} applies reinforcement learning to structural estimation of econometric models. Section~\ref{section:rl_macro} treats macroeconomic models, and Section~\ref{section:rl_games} addresses strategic games. Section~\ref{section:bandits} surveys bandit problems with domain structure. Section~\ref{section:offline_rl} covers offline reinforcement learning and Section~\ref{section:rlhf} covers reinforcement learning from human feedback. Section~\ref{section:causal_rl} treats causal inference for RL, Section~\ref{section:rl_for_ci} treats off-policy evaluation and dynamic treatment effect estimation, and Section~\ref{section:adaptive_experiments} treats causal bandits and adaptive experimentation. Section~\ref{section:dist_robust_constrained} surveys quantile, robust, and constrained reinforcement learning, Section~\ref{section:world_models} treats world models and model-based RL, and Section~\ref{section:field_deployments} synthesizes the narrower evidence on real-world RL deployments. Section~\ref{section:conclusion} concludes.

\section{Two Cultures of Sequential Decision-Making}
\label{section:language}
Economics and reinforcement learning both study sequential decision-making under uncertainty, but they descend from different intellectual traditions. The first is fundamentally an \textit{inference culture}. Its central task is to understand the world. A ``model'' in this tradition is a specification of preferences, beliefs, constraints, and an equilibrium concept. The RL tradition is instead a \textit{control culture}. Its central task is to act in the world. An RL researcher's ``model'' is a transition kernel $P(s'|s,a)$, the probability law for the next state, and a reward function $r(s,a)$. These are different mathematical objects serving different scientific purposes.

The inference tradition concentrates its effort on specifying and estimating the objective function and the law of motion that governs the environment. The entire structural econometrics enterprise (demand estimation, dynamic discrete choice) is devoted to recovering these primitives from data, and the optimal policy is a byproduct that falls out once the primitives have been identified. Reduced-form econometrics and causal inference instead seek credible assumptions or research designs that identify causal effects without recovering the full structural model. The control tradition inverts this emphasis. Engineers typically take the objective and the dynamics as given and ask whether the optimal policy can be computed, approximated, and deployed under timing limits and modeling error. The controls enterprise, from proportional-integral-derivative (PID) feedback and linear-quadratic regulators (LQR) to model predictive control (MPC) and RL, develops methods for computing and implementing that policy under different information and computational constraints.\footnote{PID control reacts to current, accumulated, and changing errors. LQR chooses controls for linear dynamics under a quadratic loss, while MPC repeatedly solves a finite-horizon control problem using an explicit model.}

The two cultures maintain different relationships with data. Econometricians work primarily with observational data, where choices often correlate with unobserved determinants of outcomes. This endogeneity arises because agents sort and markets clear rather than receiving actions at random. Identification, the question of whether the data can distinguish the true model from observationally equivalent alternatives, is the defining challenge, and it disciplines functional forms, equilibrium assumptions, and research designs that isolate variation unrelated to unobserved outcome determinants.\footnote{An instrumental variable changes treatment without directly changing the outcome. A regression discontinuity compares units on either side of an assignment threshold, while a natural experiment uses an external event that approximates random assignment.} RL researchers have traditionally enjoyed what might be called \textit{simulator omnipotence}. They own the data-generating process, can inject arbitrary variation through exploration policies, and can generate millions of trajectories from the policy being studied at negligible marginal cost. Their binding constraint is computational (can the algorithm converge before the compute budget runs out?) rather than statistical (is the estimator consistent when actions and unobserved outcome determinants are related?). This asymmetry shapes everything downstream, from what counts as a valid result to what the word ``model'' means. To an econometrician, a model is a set of falsifiable restrictions on the joint distribution of observables and primitives. To an RL researcher, a model is a simulator you can call.

Offline RL is the clearest boundary case. The analyst receives a fixed log and cannot change the assignments that produced it. This data constraint resembles the inference culture, but the objective remains to choose a decision rule with high value.

The two fields also share a vocabulary (``agent,'' ``learning,'' ``model,'' ``policy'') whose meanings diverge in ways that create persistent confusion. These terms require explicit translation between the two traditions.

\subsection{Core Distinctions}
\label{subsec:core_distinctions}

In RL, \textit{prediction} means estimating the expected return of a fixed policy, either from state $s$, denoted $V^\pi(s)$, or after first taking action $a$, denoted $Q^\pi(s,a)$. This task is called policy evaluation and differs from forecasting observable variables. \textit{Control} means finding the policy $\pi^*$ that maximizes expected discounted return, a task called policy optimization rather than adding control variables to a regression. \textit{On-policy} methods evaluate and improve the same policy that generates the data. \textit{Off-policy} methods learn about a target policy $\pi$ from data generated by a different behavior or logging policy $\mu$.

When the RL state $s$ is the recorded history $h$, the logging probability $\mu(a\mid h)$ corresponds mathematically to the treatment-assignment propensity $e(a\mid h)$, the conditional probability of action $a$ given that history. The rule under study is the \textit{target policy} in RL and a dynamic treatment regime (DTR) in longitudinal causal inference. These labels describe the roles of the two rules; they do not provide the assumptions connecting them. \textit{Off-policy} means only that the data-generating and target rules differ. \textit{Counterfactual} means that the target concerns outcomes under $\pi$ rather than $\mu$.\footnote{``Counterfactual'' here is used in the interventional sense, asking what would happen under deployment of policy $\pi$ instead of $\mu$. This is distinct from the structural counterfactual in \citet{oberst2019counterfactual}, which conditions on the specific realized trajectory and asks what would have happened to \textit{this} individual under a different action, requiring a fully specified structural causal model rather than just the observational distribution under $\mu$.} A causal interpretation also requires consistency, meaning that the observed outcome equals the outcome under the actions actually taken, and sequential exchangeability, meaning that recorded history removes common causes of each action and later outcomes. Known logging probabilities do not establish exchangeability.

\textit{Coverage} and \textit{support} in RL correspond to \textit{positivity} and \textit{overlap} only after the target policy and the state representation are fixed. Coverage asks whether the logged state-action distribution contains the actions that the target policy would take. Positivity requires those actions to occur with positive probability at the histories the policy reaches, while practical overlap requires these probabilities not to be too small. Exploration can improve coverage, but it does not remove confounding.

\textit{Online} RL learns while interacting with the environment, collecting new data as a consequence of its own actions. \textit{Offline} RL (also called batch RL) learns exclusively from a fixed dataset of previously collected transitions, with no ability to gather additional samples. The offline setting is closer to standard empirical work, where the dataset is given and the analyst cannot run new experiments; the online setting corresponds more closely to adaptive experimental design or sequential decision problems. Note that ``online'' in RL carries no timing constraint; it means only that the agent generates fresh experience. \textit{Real-time} RL, by contrast, imposes hard deadlines on the perception-action loop, as in robotics and other physical control systems. Every real-time RL system is online, but most online RL (games, recommender systems) are not real-time.

In RL, the word \textit{model} refers strictly to the environment's dynamics, the transition kernel $P(s'|s,a)$ and reward function $r(s,a)$. A \textit{model-based} RL algorithm explicitly constructs or is given a mathematical representation of $P$ and $r$, then computes a policy by planning through that representation (for example, using a simulator or the known rules of a game, as in AlphaZero). A \textit{model-free} algorithm computes the value function or policy directly from experienced transitions without ever building an explicit representation of the transition probabilities. ``Model-free'' need not mean the algorithm lacks access to any model of the environment. A model-free algorithm could interact with a simulator that internally implements a complete computerized model of the environment; the distinction is that the algorithm never extracts or plans through the transition probabilities, treating the simulator as a black box that merely returns sample transitions. A model-free algorithm could also learn through direct interaction with the field environment when no simulator is available, although this is rarely done for reasons discussed later.

Neither ``model-free'' nor ``model-based'' maps onto the reduced-form versus structural distinction in econometrics. Both labels refer to whether the algorithm uses an explicit representation of $P$ and $r$, not to whether the analyst makes structural assumptions about preferences or equilibrium. A ``model-based'' RL algorithm needs only a representation of $P$ and $r$, regardless of whether those objects arise from structural assumptions about human behavior. Conversely, ``model-free'' does not mean ``assumption-free''; both variants operate within a Markov decision process (MDP), whose state must contain the information needed to predict the next state and reward given the action. This sufficiency requirement is the Markov property. In the inference tradition, ``model'' refers to agents, preferences, equilibrium concepts, and exclusion restrictions that rule out specified direct effects. It is therefore possible to specify a rich structural model but solve for its equilibrium using a model-free RL algorithm as a computational tool, as in Section~\ref{section:rl_econ_models}.

Every RL system passes through two phases. In the \textit{training phase}, the agent interacts with an environment (simulated or real) and updates its parameters. In the \textit{execution phase} (also called the deployment phase), the policy is frozen and used to make decisions without further updates. This distinction is critical for interpreting ``online.'' Online training in RL almost always takes place inside a simulator (and not in the ``real world''). AlphaGo Zero trained online through millions of self-play games; when it faced Lee Sedol, its parameters were frozen and it was purely executing its trained policy. Some deployed systems in Section~\ref{section:field_deployments} followed this pattern cleanly; DiDi's dispatch system trained value information from historical trip data and then embedded the learned signal in production dispatch. Others blur the boundary. The hotel revenue management system in Section~\ref{sec:hotel} updated Q-values from realized returns after each completed episode during live operation, making it an \textit{in-field} learner rather than a frozen executor.\footnote{``In-field'' is not standard RL vocabulary. The term is introduced here to distinguish live-market online learning, where exploration has real financial cost, from the far more common case of online learning inside a simulator.} Bandit algorithms (Section~\ref{section:bandits}) also learn in-field by design, updating demand estimates from real customer responses during deployment.

The word \textit{inference} is overloaded. In machine learning, ``inference'' refers to executing a frozen model, a forward pass producing outputs from inputs. In statistics, ``inference'' means statistical inference, the construction of standard errors, confidence intervals, and hypothesis tests. This survey uses ``inference'' exclusively in the statistical sense and ``execution'' or ``deployment'' when referring to applying a trained model (whether in a computer or in field). Established terms remain where appropriate; Bayesian inference updates probabilities with data, while variational inference approximates a difficult probability distribution by optimizing over a tractable family. Most RL convergence results refer to training, as do sample-complexity guarantees that bound the number of observations needed to learn; interpreting them as claims about deployed performance requires additional argument. Table~\ref{tab:lifecycle_grid} summarizes these distinctions.

\newcommand{\tabitem}{{\small$\cdot$}~}
\begin{table}[h]
\centering
\caption{The reinforcement learning lifecycle grid. Most RL research operates in the top-left cell. Readers from the inference tradition typically picture the bottom-left when they hear ``online.''}
\label{tab:lifecycle_grid}
\begin{tabular}{|p{2.8cm}|p{5.5cm}|p{5.5cm}|}
\hline
 & Training (parameters updated) & Execution (parameters frozen) \\
\hline
Simulator &
  \tabitem AlphaGo Zero self-play \newline
  \tabitem RL solver in structural estimation \newline
  \tabitem Simulation tuning for bandit algorithms &
  \tabitem Policy benchmarks on synthetic environments \\
\hline
Historical data &
  \tabitem Reward model from human comparisons \newline
  \tabitem Causal off-policy evaluation \newline
  \tabitem Policy learning from logged bandit data &
  \tabitem Off-policy evaluation of a target policy \\
\hline
Live market (``in-field'') &
  \tabitem Bandit pricing experiments \newline
  \tabitem Hotel revenue-management Q-learning &
  \tabitem DiDi dispatch with fixed parameters \newline
  \tabitem AlphaGo vs.\ Lee Sedol \\
\hline
\end{tabular}
\end{table}

A typical applied RL pipeline moves through the grid sequentially, from pre-training on historical logs (middle-left) to refinement in a simulator (top-left) to deployment with frozen weights (bottom-right). Bandits illustrate this fluidity; even a bandit algorithm that will ultimately learn in-field is typically calibrated in simulation and tuned on historical logs before any live deployment, because in-field exploration incurs real financial cost. The systems that do operate in-field arrive with exploration parameters, initial policies, and demand priors shaped by extensive offline preparation.

The Tmall e-commerce pricing project of \citet{Liu2019} (Section~\ref{sec:ecommerce_pricing}) illustrates this migration concretely. The team pre-trained a deep Q-network (DQN), a neural-network version of Q-learning that estimates action values, from logged specialist pricing decisions, then evaluated the candidate policy on held-out transactions that were not used for training. The evaluated policy was deployed for 15- to 30-day field experiments on live Tmall traffic, with the agent receiving reward and observation signals from the market environment. \citet{Liu2019} note that no accurate simulator exists for e-commerce pricing, so the project skipped the simulator row entirely and moved from historical pre-training to live deployment. Not every application traverses all six cells of Table~\ref{tab:lifecycle_grid}, but the grid clarifies which cells a given project could be working on.

\subsection{Overlapping Terminology}
\label{subsec:overlapping_terminology}

In the inference tradition, an \textit{agent} is always a human decision-maker (a consumer, worker, or firm) or a social planner whose choices are the object of study. In RL, ``the agent'' is the learning algorithm itself, or more precisely an algorithm deployed on behalf of a human decision-maker, and the human, if present at all, is part of the environment providing reward signals.

In RL the \textit{environment} is a formal object encompassing everything outside the agent, including the data-generating process, other agents, and market-clearing conditions, whereas in the inference tradition ``environment'' refers more loosely to market structure or institutional rules.\footnote{A source of confusion is \textit{generative model}. In machine learning broadly, this means a model of the joint distribution $P(X, Y)$ or a synthetic data generator such as a generative adversarial network or diffusion model. In RL theory, a generative model is a simulator oracle that, given any $(s,a)$, returns a sample $s' \sim P(\cdot|s,a)$ and reward $r(s,a)$; the usage implies random-access simulation, stronger than sequential online interaction but weaker than knowing $P$ analytically.} RL speaks of \textit{rewards} where the other tradition speaks of \textit{utility}. The mapping is not exact. A reward $r(s,a)$ is a known, externally specified function that the algorithm maximizes, whereas utility $u(x,d)$ in the inference tradition is a primitive of preferences that the analyst must recover or estimate from observed choices.

In RL, the return $G_t = \sum_{k=0}^{\infty} \gamma^k r_{t+k+1}$ is the discounted sum of future rewards from time $t$ onward, the random variable whose expectation defines the value function. The RL usage is closer to what is called the ``present discounted value'' of a stream of payoffs. A single complete sequence of interactions from an initial state to termination, $(s_0, a_0, r_1, s_1, \ldots, s_T)$, is called an \textit{episode} (synonymously, \textit{trajectory} or \textit{rollout}).\footnote{The closest statistical analogs are a single panel unit's time series, a realization of a stochastic process, or one ``history'' in a dynamic model.} Where a statistician speaks of the \textit{outcome}, meaning the dependent variable $Y$ in a regression, RL has no single analog: the reward $r_t$ is the per-period outcome, the return $G_t$ is the cumulative outcome, and the value function $V^\pi(s)$ is the expected cumulative outcome conditional on state.

\textit{Learning} has several distinct meanings. In decision theory (Bayesian learning, adaptive expectations), learning refers to agents forming and refining beliefs about unknown parameters of their environment. In supervised machine learning, learning means statistical estimation, fitting the weights of a parameterized model to minimize a loss function over data. In reinforcement learning, ``learning'' is primarily computation. When an RL agent ``learns'' a Q-function, it uses stochastic approximation, recursive updates from noisy samples, to find a fixed point of the Bellman operator, a value function left unchanged by one Bellman update. In many applications throughout this survey, particularly Section~\ref{section:rl_econ_models}, the RL algorithm is simply a numerical method for solving the Bellman equation; no actual human-like learning from experience is taking place. Finally, while RL draws inspiration from animal psychology (Section~\ref{sec:animal_psychology}), it is a drastic simplification of biological learning. Tabula rasa RL algorithms start without task-specific knowledge and require millions of iterations of trial and error to discover policies that animals acquire rapidly. Real-world animal learning relies on innate priors, basic physical knowledge, and parental nurturing and should be distinct from RL-style ``learning''.

Adaptive learning in macroeconomics has used ideas formally analogous to reinforcement learning for decades, though under different names and with different motivations. In the adaptive learning literature initiated by \citet{MarcetSargent1989}, boundedly rational agents update coefficient estimates as observations arrive using recursive least-squares and other Robbins-Monro-type stochastic approximation algorithms.\footnote{``Boundedly rational'' means that beliefs follow a specified updating rule rather than immediately satisfying model-consistent rational expectations. The ODE method studies a noisy recursive algorithm through a deterministic continuous-time approximation.} The convergence criterion in that literature, E-stability, asks whether the ordinary differential equation (ODE) mapping the perceived law of motion into the actual law converges locally to the rational expectations equilibrium. This fixed-point requirement resembles a contraction in RL, where each update shrinks differences between candidate values, and governs temporal-difference and Q-learning algorithms that update value estimates from observed transitions. \citet{borkar2000} make this mathematical connection explicit by proving convergence of Q-learning and actor-critic methods with the ODE method and citing \citet{Sargent1993} as a parallel application to boundedly rational agents.\footnote{An actor-critic method maintains both a policy model, the actor, and a value model, the critic, and updates them together.}

\textit{Active learning} appears in both fields but means different things. In the inference tradition, it denotes Bayesian experimentation in which an agent chooses what information to acquire by weighing its cost against the value of better future decisions. In machine learning, it is a supervised protocol in which an algorithm selects unlabeled examples for a labeling source. RL instead uses \textit{exploration} for actions chosen to improve future decisions, through methods ranging from $\varepsilon$-greedy randomization to an upper confidence bound (UCB), which selects the action with the highest plausible reward, or Thompson sampling, which chooses actions according to their data-updated probability of being best. The closest economic analogue is \textit{optimal experimentation} in the Bayesian bandit tradition (Rothschild 1974), where the agent's dynamic program determines the value of information. The adjective \textit{adaptive} adds another ambiguity because a DTR responds to an individual's recorded history, an adaptive experiment changes assignment probabilities as outcomes accumulate, and online RL updates its rule while collecting experience. A DTR can therefore be learned once from fixed data, while an adaptive experiment can change assignments without supporting valid post-experiment inference.

\textit{Bootstrapping} in statistics refers to Efron's resampling method \citep{Efron1979}; in RL, it means updating a value estimate using another value estimate rather than a complete realized return, as when a temporal-difference (TD) algorithm uses the target $r_{t+1} + \gamma V(s_{t+1})$ that depends on the current, uncertain estimate $V(s_{t+1})$ \citep{sutton1988}. \textit{Function approximation} in RL refers to representing value functions or policies with a parameterized class such as a linear basis or neural network. Statisticians will recognize the same idea as sieve or nonparametric series estimation, the approximation of an unknown function by projection onto a finite-dimensional basis. \textit{Calibration} carries unrelated meanings. In the inference tradition, calibration means choosing model parameters to match selected empirical summary statistics (Kydland and Prescott 1982). In machine learning, calibration refers to probability calibration, ensuring that predicted probabilities match observed frequencies, a statistical property of a classifier's outputs.

The term \textit{bandit} itself carries different mathematical content across disciplines. The classical multi-armed bandit in statistics \citep{Thompson1933, Rothschild1974, Gittins1979} allocates observations across actions while updating Bayesian beliefs about their rewards. The state is the agent's data-updated belief over the unknown reward distributions, and the Gittins index gives the optimal allocation by reducing it to separate decisions about when to stop sampling each action. In the RL and computer science literature, bandits are instead framed as regret-minimization problems that do not require Bayesian priors. Algorithms such as UCB bound cumulative regret $\sum_{t=1}^T (\mu^* - \mu_{A_t})$. The two traditions ask different questions, Bayesian optimality of the full sequential problem versus minimax regret rates over adversarial or stochastic environments, and their answers are not directly comparable.\footnote{A stochastic bandit assumes that each arm follows a fixed reward distribution. An adversarial bandit permits rewards to vary without that model, while a minimax guarantee gives the best worst-case rate over the stated environment class.} Section~\ref{section:bandits} adopts the regret framework because it connects more naturally to the number of observations required in field experiments.

The terms \textit{state}, \textit{history}, \textit{context}, and \textit{covariate} are not interchangeable. An MDP state is sufficient for the distribution of future rewards and transitions given future actions, whereas a causal history records observed variables and past treatments before the next decision. The full history can serve as the RL state, but any compression must preserve both the Markov property and the variables needed for sequential exchangeability. In a contextual bandit, ``context'' adds a further restriction because the action does not affect the next context, so that $P(x_{t+1} \mid x_t, a_t) = P(x_{t+1} \mid x_t)$. This separates exploration, learning which action is best for the current context, from planning, choosing actions that change future states. With exogenous contexts, the problem is repeated one-period optimization rather than long-horizon credit assignment.

The RL literature frames some recommender systems as contextual bandits, where user covariates are the context, the recommendation is the arm, and a click or rating is the reward. One might instead view movie recommendation as a two-sided learning problem in which the platform learns user preferences while users simultaneously explore the catalog and update their own tastes. The bandit formulation absorbs the user's utility maximization into the environment's reward signal and treats user arrivals as exogenous. This is a modeling choice, not a fact about the world. Also, the object called a ``bandit'' in this formulation, a one-step decision under exogenous context, is a slightly different mathematical object than the Bayesian sequential allocation problem of \citet{Gittins1979}, even though both carry the same name and are related. \footnote{\citet{Lattimore2016gittins} proves that the Gittins index with a flat Gaussian prior achieves finite-time regret of the same order as UCB, and that the index decomposes as posterior mean plus an exploration bonus that shrinks toward zero near the horizon, structurally resembling but differing from the UCB confidence bound.} RL abstracts away much of this complexity (human learning, strategic interaction) to fit the problem into the standard MDP framework ($\pi$, $V$, $P$, $r$).

The word \textit{policy} is overloaded. In the inference tradition, a ``policy'' is a rule set by a government, central bank, or regulator, such as a tax schedule, subsidy, interest rate rule, or licensing requirement. These rules form part of the \textit{environment} in which private agents optimize. ``Policy evaluation'' asks how those agents respond when a rule changes, as in an expansion of the earned income tax credit.

In RL, ``policy'' means the agent's decision rule $\pi(a\mid s)$, and ``policy evaluation'' means computing the expected return $V^\pi(s)$ from following that rule. In longitudinal causal inference, a treatment regime is the corresponding prescriptive rule when actions depend on recorded history. A conditional choice probability (CCP) instead describes behavior and may serve as a logging policy; it is not automatically the target policy. RL policy evaluation usually holds the environment fixed.\footnote{A notable exception is \citet{Tomasev2020}, where AlphaZero was used to evaluate alternative chess rule sets proposed by Kramnik, asking how optimal play changes under modified game rules. This is precisely an environment counterfactual.}

\textit{Identification} means different things in the two fields. In statistics, a parameter $\theta$ is identified if it is uniquely pinned down by the combination of the data-generating process and the model's maintained assumptions; formally, no two distinct parameter values $\theta \neq \tilde{\theta}$ can generate the same distribution of observable data \citep{Lewbel2019}. Identification is a logical property of the model, not a statistical one. If identification fails, no amount of additional data or more sophisticated estimation will recover the parameter because multiple values are observationally equivalent.

In RL, identification also has domain-specific meanings. In inverse reinforcement learning, reward identifiability asks whether observed optimal behavior uniquely determines the reward function. Strong identifiability means recovering every state-action reward up to a common additive constant. For deterministic maximum-entropy MDPs, whose objective rewards both return and randomized action choice, \citet{KimGarg2021} relate strong identifiability to reachability and cycle structure in the domain graph.\footnote{A strongly connected graph lets every state-action vertex reach every other, while coverability means that, for some step count, every vertex can be reached in exactly that many steps from a feasible initial vertex. In such a graph coverability is equivalent to aperiodicity, meaning that no integer greater than one divides every cycle length; the strong-identifiability result also requires a horizon long enough relative to the covering time.}\footnote{\citet{KimGarg2021} note that identifying dynamic discrete choice models \citep{rust1994structural} addresses ``an equivalent problem'' to reward identifiability. Inverse reinforcement learning is a longstanding RL subfield and is the subject of the companion survey \citep{RustRawat2026}.} In model-based RL, system identification asks whether learned transition dynamics $\hat{T}$ are adequate for control, a usage inherited from control theory rather than statistics \citep{RossBagnell2012}. Causal and offline RL use identification in the statistical sense given above, so \textit{identified}, \textit{estimated}, \textit{learned}, and \textit{computed} are not synonyms. Identification asks whether assumptions and the observed data distribution determine the target; estimation and statistical inference use finite data, while planning computes from a specified decision process. Bellman convergence establishes the last task, not identification of a counterfactual value from logged data.

\textit{Regret} diverges across fields. In decision theory, it is either an emotion that modifies preferences under uncertainty (Loomes and Sugden 1982) or a static minimax criterion for choosing among actions when probabilities are unknown (Savage 1951, Manski 2004). Online learning uses \textit{cumulative regret} for the payoff lost during data collection, pure exploration uses \textit{simple regret} for the value gap of the final recommendation, and offline policy learning often uses \textit{welfare regret}, $V(\pi^*)-V(\hat\pi)$, for the loss from the learned rule. None measures statistical uncertainty or supplies a confidence interval. \textit{Efficiency} is equally overloaded. Pareto efficiency means that no one can be made better off without making someone else worse off, allocative efficiency places resources in their highest-valued uses, and statistical efficiency asks whether an estimator attains the lowest variance permitted by the model. In RL, sample efficiency counts environment interactions and computational efficiency counts operations per timestep.

In the inference tradition, the \textit{discount factor} is a structural parameter encoding time preference, an agent's intrinsic willingness to trade present for future consumption. \citet{Koopmans1960} derived it from stationarity, preferences that depend on relative delays rather than calendar dates, and impatience, a preference for earlier rewards over otherwise equal later rewards \citep{Bleichrodt2008}. These axioms imply exponential discounting with a single factor strictly between zero and one. Economists treat this factor as a primitive of preferences and study departures such as quasi-hyperbolic present bias, which gives extra weight to immediate payoffs. In RL, discounting also keeps infinite-horizon returns finite and makes the Bellman operator a contraction, so each update shrinks differences between candidate value functions and converges to a unique fixed point. The factor is therefore often treated as a user-chosen tuning parameter rather than a structural claim about preferences, although \citet{Pitis2019} gives an axiomatic interpretation of this computational role.

\subsection{Structural Equivalences}
\label{subsec:structural_equivalences}

Beyond terminological differences, several formal objects in RL and the inference tradition are mathematically identical. The softmax (or Boltzmann) policy used throughout RL is the multinomial logit model of \citet{McFadden1974}. The RL softmax policy selects actions according to
\begin{equation}
\label{eq:softmax_logit}
\pi(a \mid s) = \frac{\exp(Q(s,a) / \tau)}{\sum_{a' \in \mathcal{A}} \exp(Q(s,a') / \tau)},
\end{equation}
where $\tau > 0$ is a temperature parameter controlling how diffuse the action probabilities are. In the discrete choice framework, $Q(s,a)$ plays the role of the deterministic component of utility $v(a \mid x)$, while $\varepsilon_a$ is an unobserved utility shock with a Type I extreme value (Gumbel) distribution whose scale is $\tau$. As $\tau \to 0$, the policy converges to the greedy (deterministic) policy, just as the logit choice probability concentrates on the utility-maximizing alternative as the variance of taste shocks vanishes.

Entropy regularization adds a bonus for policies that keep probability spread across actions rather than collapsing immediately onto one action. The resulting soft value function
\begin{equation}
\label{eq:soft_value}
V^{\text{soft}}(s) = \tau \log \sum_{a \in \mathcal{A}} \exp(Q(s,a) / \tau)
\end{equation}
is the inclusive value, or log-sum-exp, that summarizes the value of the full choice set in discrete choice. It is identical to the McFadden surplus function $W(x) = \tau \log \sum_a \exp(v(a|x)/\tau) + C$, where $C$ is Euler's constant. In structural estimation, the same object appears as the Emax function, the expected maximum continuation value, in dynamic discrete choice models following \citet{Rust1987}.

The action-value function $Q^\pi(s,a)$ is the choice-specific value function $v_\theta(x,d)$ of the dynamic discrete choice literature (Table~\ref{tab:notation}). The advantage function $A^\pi(s,a) = Q^\pi(s,a) - V^\pi(s)$ is therefore the choice-specific value net of the value before an action is chosen, a quantity that appears in the conditional choice probability estimator of \citet{HotzMiller1993}.\footnote{These equivalences share a convex-duality structure. Log-sum-exp and negative Shannon entropy are Fenchel conjugates, meaning that each can be recovered from the other through an optimization; \citet{ChiongGalichonShum2016} and \citet{FosgerauSorensen2022} use this duality for identification and estimation beyond the logit case.}

The two fields arrived at this shared mathematics from opposite directions. In RL, entropy regularization began as a pragmatic trick to prevent premature convergence and encourage exploration \citep{WilliamsPeng1991}, and only decades later did Ziebart (2010) and Levine (2018) give decision-theoretic foundations showing that the resulting policies tolerate errors in the assumed model. In the inference tradition, the softmax emerged not from any concern about exploration but from the random utility framework of \citet{McFadden1974}, where agents are fully informed and choose deterministically; the apparent randomness arises entirely from the econometrician's inability to observe all relevant taste variation. The RL agent randomizes because it is ignorant of the environment; the economic agent appears to randomize because the observer is ignorant of the agent's preferences.

\subsection{Notation}
\label{subsec:notation}

The survey uses one book-wide notation for sequential decisions. A state $s \in \mathcal{S}$ records the information available when an action $a \in \mathcal{A}$ is chosen. The policy $\pi(a\mid s)$ assigns actions to states, the reward $r(s,a)$ records the current payoff, and the transition kernel $P(s'\mid s,a)$ gives the distribution of the next state. The value $V^\pi(s)$ is expected discounted return before the action is chosen, while $Q^\pi(s,a)$ conditions on the current action. The realized return $G_t$ is the discounted sum of rewards from period $t$ onward.

When policy and data collection differ, $\pi$ denotes the target policy under study and $\mu$ denotes the behavior or logging policy that generated the observations. The distinction matters throughout the offline RL and causal chapters. Table~\ref{tab:notation} maps these symbols to their most common econometric equivalents.

\begin{table}[h]
\centering
\caption{Notation mapping between reinforcement learning and the inference tradition.}
\label{tab:notation}
\begin{tabular}{llll}
\toprule
RL Term & Symbol & Inference Tradition & Symbol \\
\midrule
State & $s \in \mathcal{S}$ & State or sufficient history & $x_t$, $\bar H_t$ \\
Action & $a \in \mathcal{A}$ & Choice, control variable & $d_t$, $u_t$\footnote{The letter $u$ appears twice in the inference-tradition column with different meanings: $u_t$ denotes the control variable (action), while $u(x,d)$ denotes per-period utility (reward). Context usually disambiguates, but readers should note the collision.} \\
Reward & $r(s,a)$ & Per-period utility, payoff & $u(x,d)$ \\
Discount factor & $\gamma$\footnote{RL permits $\gamma = 0$ (myopic, one-step) and $\gamma = 1$ (undiscounted, episodic tasks with guaranteed termination). The inference tradition requires $\beta \in (0,1)$ strictly; the contraction mapping argument for the Bellman operator relies on $\beta < 1$.} & Discount factor & $\beta$ \\
Target policy & $\pi(a\mid s)$ & Decision rule, treatment regime & $d(h)$ \\
Behavior policy & $\mu(a\mid s)$ & Assignment propensity, CCP & $e(a\mid h)$ \\
Value function & $V^\pi(s)$ & Ex-ante (pre-choice) value & $\bar{V}_\theta(x)$ \\
Policy value & $J(\pi)$ & Regime value, welfare & $\mathbb E[Y^\pi]$ \\
Q-function & $Q^\pi(s,a)$ & Choice-specific value function & $v_\theta(x,d)$ \\
Return & $G_t$ & Present discounted value & $\sum_{k=0}^\infty \beta^k u_{t+k}$ \\
Transition & $P(s'|s,a)$ & State transition law & $f(x_{t+1} | x_t, d_t)$ \\
Learning rate & $\alpha$ & Step size per update & $\alpha_n$ \\
TD error & $\delta_t$\footnote{The TD error $\delta_t = r_{t+1} + \gamma V(s_{t+1}) - V(s_t)$ is a stochastic, sample-based quantity. The ``Bellman residual'' in numerical methods and economics is the population object $\|V - \mathcal{T}V\|$, measuring how far a candidate $V$ is from satisfying the Bellman equation exactly. Bellman residual minimization (BRM), which directly minimizes $\mathbb{E}[(r + \gamma V(s') - V(s))^2]$, is a distinct estimation method from TD learning.} & Bellman residual at sample & -- \\
\bottomrule
\end{tabular}
\end{table}

\FloatBarrier
\subsection{The Engine Replacement MDP in Two Languages}
\label{engine:language}
\label{engine:model}

The algorithms in this survey, and much of the theory in the chapters that follow, are computed on one fixed model. A maintenance manager observes a bus engine at low or high mileage, so $\mathcal{S} = \{\text{low}, \text{high}\}$. Each period the manager keeps or replaces the engine, so $\mathcal{A} = \{\text{keep}, \text{replace}\}$. Keeping a low-mileage engine yields output $1$ and moves the engine to high mileage with probability $0.5$. Keeping a high-mileage engine yields output $0.2$ and leaves the engine there. Replacement costs $0.5$ from either state and returns the engine to low mileage with certainty. The discount factor is $\gamma = 0.9$. This two-state model is an instance of the bus engine replacement problem in \citet{Rust1987}. Per-period output corresponds to operating profit net of maintenance cost $\theta_1$, and the replacement charge corresponds to the fixed cost $RC$. Figure~\ref{fig:engine_digraph} shows the process.

\begin{figure}[h]
\centering
\begin{tikzpicture}[>=Stealth, semithick]
\node[circle, draw=rlblack, minimum size=1.15cm] (l) at (0,0) {low};
\node[circle, draw=rlblack, minimum size=1.15cm] (h) at (5,0) {high};
\draw[->, rlblue] (l) to[loop left, looseness=6] node[left, align=right, font=\footnotesize] {keep\\$0.5$, $r = 1$} (l);
\draw[->, rlblue] (l) to[bend left=22] node[above, font=\footnotesize] {keep, $0.5$} (h);
\draw[->, rlorange] (h) to[bend left=22] node[below, font=\footnotesize] {replace, $1$, $r = -0.5$} (l);
\draw[->, rlblue] (h) to[loop right, looseness=6] node[right, align=left, font=\footnotesize] {keep\\$1$, $r = 0.2$} (h);
\draw[->, rlorange] (l) to[loop above, looseness=6] node[above, font=\footnotesize] {replace, $1$, $r = -0.5$} (l);
\end{tikzpicture}
\caption{The Engine Replacement MDP as a labelled directed graph. Each arc carries its action and transition probability and, where the action is first named from that state, its reward. Blue arcs are the keep action and orange arcs the replace action.}
\label{fig:engine_digraph}
\end{figure}

The model is small enough to check by hand. The optimal policy keeps at low mileage and replaces at high mileage, with $V^\star = (5.3448,\ 4.3103)$. The action-value gap is the difference between the better and worse action value at a state. The gaps are $1.0345$ at low mileage and $0.2310$ at high mileage. Appendix~\ref{prelim:mdp} computes every step of the solution and describes the same states as good and worn condition. These primitives and benchmark values recur throughout the survey. Later Engine subsections hold the model fixed while changing the object under study, including value learning, structural parameters, reward equivalence, policy evaluation, robustness, and transition learning.

Table~\ref{tab:engine_dictionary} names each primitive in the vocabularies of reinforcement learning and dynamic economic models. The transition kernel is the controlled law of motion, the reward is the flow payoff, a policy is a replacement rule, and an action value is a choice-specific value.

\begin{table}[h]
\centering
\caption{The Engine Replacement MDP in the vocabularies of reinforcement learning and dynamic economic models.}
\label{tab:engine_dictionary}
\begin{tabular}{@{}lll@{}}
\hline
reinforcement learning & object & economic model \\
\hline
mileage grade $s$ & state & observed condition $x$ \\
keep or replace $a$ & action & discrete choice $d$ \\
$P(s'\mid s,a)$ & transition & controlled law of motion \\
$r(s,a)$ & reward & flow profit net of maintenance \\
$\pi(a\mid s)$ & policy & replacement rule \\
$Q^\pi(s,a)$ & action value & choice-specific value \\
$V^\pi(s)$ & value & pre-choice value \\
\hline
\end{tabular}
\end{table}

\FloatBarrier

\section{A Brief History of Reinforcement Learning}
\label{section:history}

Reinforcement learning draws on animal psychology, game-playing programs, and optimal control theory in roughly equal measure. Thorndike's law of effect and behaviourist trial-and-error learning provided the notion of ``reinforcement'' as formalized by Rescorla-Wagner.  Chess and checkers programs of Shannon and Samuel from the 1950s onward gave researchers concrete problems on which to test ideas about machine learning. The Bellman-Howard-Blackwell dynamic programming framework provided the recursive structure and language. 

\subsection{Animal Psychology}
\label{sec:animal_psychology}

Controlled experiments on rats, cats, and dogs inspired the ``gridworld" environments still used today, and shaped how the field conceptualizes ``training" agents through ``reward signals". \citet{Thorndike1898} placed cats in puzzle boxes with latched doors and food visible outside. Across 15 different box configurations, the cats initially engaged in undirected behavior such as clawing at the walls, pushing against the bars, reaching through openings. The first cat to escape the simplest box required 160 seconds of random activity before accidentally pressing the latch. By the 24th trial, the same cat pressed the latch directly within 6 seconds. The learning curves showed gradual, continuous improvement rather than sudden insight. From these experiments the Law of Effect was formulated: responses (actions) followed by satisfaction (positive rewards) are ``stamped in'' and more likely to recur, while those followed by discomfort (negative rewards) are ``stamped out.''


\citet{Pavlov1927} noted that dogs salivated not only at food itself but at the sight of an empty food bowl, the sound of footsteps, and other stimuli that preceded feeding (i.e. the state). To measure these ``psychic secretions'' precisely, he surgically implanted fistulas (tubes allowing external collection) to collect saliva. In the canonical experiment, a metronome sounded before food delivery. After 20--40 pairings, the metronome alone elicited salivation. The response followed from what the stimulus "predicted" rather than from the stimulus itself. In reinforcement learning terms, the conditioned stimulus is a state $s$, and the learned expectation of food is the value function $V(s)$.

\citet{Kamin1969} demonstrated that learning requires more than mere co-occurrence in time. In Phase I of his blocking experiment, rats learned that a noise predicted a shock and developed a conditioned fear response to the noise alone. In Phase II, a compound stimulus of noise plus light was paired with the same shock. When the light was subsequently presented alone, no fear response occurred. The light was ``blocked'' (ignored) because the noise already predicted the shock perfectly. There was no prediction error (or ``surprise'') to drive learning about the light. Once the noise was established as a predictor, the light added no new information and thus no new learning occurred.

\citet{RescorlaWagner1972} formalized the blocking phenomenon as a prediction-error learning rule. Translating to RL notation:\footnote{The original notation used $V_i$ for associative strength of stimulus $i$, $\alpha_i$ for stimulus salience (noticeability), $\beta_j$ for learning rate, $\lambda_j$ for maximum conditioning (1 if reward present, 0 otherwise), and $V_{\text{tot}} = \sum_k V_k$ for total prediction. The correspondence is: $V_i \to V(s)$, $\alpha_i \beta_j \to \alpha$, $\lambda_j \to r$, $V_{\text{tot}} \to V(s)$. See \citet{SuttonBarto1990} for details.}
\begin{equation}
V(s) \leftarrow V(s) + \alpha \, \delta, \quad \text{where } \delta = r - V(s)
\end{equation}
This is temporal difference learning with $\gamma = 0$, without discounting of future rewards. Each stimulus $s$ begins with $V(s) = 0$. On each trial, the organism observes the stimuli present, receives outcome $r \in \{0, 1\}$, and updates values according to $\delta$. The model is purely predictive; there are no actions, only learned expectations about reward.\footnote{The Rescorla-Wagner update is mathematically identical to the Widrow-Hoff least mean squares rule.} The model correctly predicted overexpectation, whereby two separately conditioned stimuli combined and reinforced together each lose value because their summed prediction exceeds $r$.


\subsection{Board Games}

Chess and checkers are sequential decision problems. The board position is a state $s \in \mathcal{S}$, a legal move is an action $a \in \mathcal{A}(s)$, and the resulting position is a successor state $s' = T(s,a)$ determined by the rules of the game. The game outcome provides a terminal reward $r \in \{+1, 0, -1\}$ for win, draw, or loss. An evaluation function $f(P)$ that scores a position corresponds to a value function $V(s)$ estimating expected outcome. These games are deterministic (no chance moves in chess), fully observable (both players see the entire board), and zero-sum (one player's gain is the other's loss). Being adversarial, the game adds a second player whose actions $a'$ must be anticipated.

\citet{Shannon1950} posed the fundamental question: can we program a general-purpose computer to play chess, and if so, what principles should guide the design? The challenge is computational. A chess game averages 40 moves per player with roughly 30 legal moves available at each position. Shannon estimated that exhaustive search through all possible games would require examining approximately $30^{80} \approx 10^{120}$ positions, a number exceeding the atoms in the observable universe. This is the curse of dimensionality applied to games, where state space size grows exponentially with the number of sequential decisions. Shannon calculated that brute-force enumeration would require $10^{90}$ years at any foreseeable computing speed. The curse intensifies with game complexity. Chess has approximately $10^{47}$ legal positions, shogi $10^{71}$, and Go $10^{171}$.\footnote{State space estimates from \citet{Igami2020}.}

Shannon distinguished two approaches. Type A strategies search all continuations to a fixed depth $H$, building a complete game tree and evaluating every leaf (``rote-learning''). Type B strategies search selectively, examining only variations deemed important by some criterion (``generalization''). Either approach requires an evaluation function $f(P)$ to score positions where search terminates. Shannon proposed linear evaluation:
\begin{equation}
f(P) = \sum_i w_i \phi_i(P)
\end{equation}
with features $\phi_i$ for material, mobility, pawn structure, and king safety. Weights $w_i$ were hand-tuned. The minimax principle governs adversarial search. In a two-player zero-sum game, the value of a position satisfies
\begin{equation}
V(s) = \max_{a \in \mathcal{A}(s)} \min_{a' \in \mathcal{A}'(s')} V(T(s, a, a'))
\end{equation}
where the maximizing player moves first and the minimizing opponent responds optimally. This is model-based planning, since the transition function $T$ is known exactly from the game rules. The computational problem is how to use limited search resources effectively given the exponential tree.

Both Type A and Type B strategies truncate the game tree at depth $H$ and substitute the evaluation function for exact continuation values. This is approximate dynamic programming. The true value $V^*(s)$ satisfies a recursive equation, but computing it exactly is infeasible, so the recursion is truncated and terminal values approximated. Deeper lookahead ($H$-step search) builds larger trees; rollout policies extend search by simulating play to the end using a fast base policy.\footnote{Monte Carlo tree search, developed later, samples rollouts rather than enumerating all branches, enabling deeper effective lookahead in games with large branching factors.} The evaluation function serves as a heuristic substitute for exact computation. Shannon did not implement a chess program; the 1950 paper is theoretical, outlining the architecture that shaped fifty years of game engines.

\citet{Samuel1959} built a checkers program for the IBM 704 that could improve through experience. The program played against itself, generating virtually unlimited training data at no cost. It parameterized the value function as a linear combination of hand-crafted features (piece advantage, mobility, king safety):
\begin{equation}
V(s; \mathbf{w}) = \sum_{i} w_i \phi_i(s)
\end{equation}
After each move from $s_t$ to $s_{t+1}$, weights were updated by temporal difference:
\begin{equation}
\mathbf{w} \leftarrow \mathbf{w} + \alpha \left[ V(s_{t+1}; \mathbf{w}) - V(s_t; \mathbf{w}) \right] \nabla_{\mathbf{w}} V(s_t; \mathbf{w})
\end{equation}

In a 1965 match, World Champion W.F. Hellman won all four games played by mail, but was played to a draw in one game. After learning from 173,989 book moves, the program agreed with the book-recommended move (or rated only 1 move higher) 64\% of the time without lookahead. With lookahead and minimaxing, it followed book moves "a much higher fraction of the time."

The linear parameterization compresses the value function from $10^{20}$ table entries to dozens of weights, the essential response to the curse of dimensionality. Samuel's architecture, minimax search to depth $H$ with the learned evaluation function scoring leaves, is an early instance of rollout, where tree search simulates forward, truncating at $H$ and substituting $V(s)$ for exact continuation values.\footnote{\citet{bertsekas2021lessons} interprets this as approximate dynamic programming, where offline training (learning $V$) combined with online planning (tree search) implements a Newton-like step for the Bellman equation.} The conceptual apparatus of modern game-playing AI (self-play, evaluation learning, tree search, and function approximation) was present in the 1950s.

\subsection{Optimal Control}

\citet{Bellman1957} considered multi-stage decision processes in which a system occupies state $s \in \mathcal{S}$, the decision-maker chooses action $a \in \mathcal{A}$, the system transitions to $s' \sim P(\cdot|s,a)$, and a reward $r(s,a,s')$ accrues. The objective is to maximize cumulative reward over a finite or infinite horizon. The classical approach treats an $N$-stage process as a single $N$-dimensional optimization. Bellman calculated the consequence. A 10-stage process with 10 grid points per variable requires $10^{10}$ function evaluations; at one evaluation per second, $10^{10}$ evaluations require 2.77 million hours. He called this exponential growth the curse of dimensionality. His solution was the principle of optimality, namely that an optimal policy has the property that, whatever the initial state and initial decision, the remaining decisions must constitute an optimal policy with regard to the state resulting from the first decision. This principle yields the Bellman equation:
\begin{equation}
V^*(s) = \max_{a \in \mathcal{A}} \left[ r(s,a) + \gamma \sum_{s'} P(s'|s,a) \, V^*(s') \right]
\label{eq:bellman}
\end{equation}
The equation reduces an $N$-dimensional problem to a sequence of $N$ one-dimensional problems. Value iteration computes $V^*$ by iterating $V_{k+1}(s) = \max_a [r(s,a) + \gamma \sum_{s'} P(s'|s,a) V_k(s')]$. The monograph applied the method to resource allocation, inventory control, bottleneck scheduling, gold mining under uncertainty, and multi-stage games.\footnote{Chapter IX shows how dynamic programming derives classical variational conditions from the functional equation. The continuous-time analogue is the Hamilton-Jacobi-Bellman equation.}

\citet{howard1960} observed that value iteration converges slowly for problems of indefinite duration. His alternative, policy iteration, solves for the value function of a fixed policy and then improves the policy directly. Given policy $\pi$, policy evaluation computes the gain $g$ (average reward per period) and relative values $v_i$ by solving the linear system $g + v_i = q_i + \sum_j p_{ij} v_j$ for $i = 1, \ldots, N$, where $q_i$ is the expected immediate reward in state $i$ and $p_{ij}$ is the transition probability under $\pi$. Policy improvement then selects, for each state $i$, the action $k$ maximizing $q_i^k + \sum_j p_{ij}^k v_j$. Howard proved that each iteration strictly increases the gain unless the policy is already optimal, and the algorithm terminates in finitely many steps. For a problem with 50 states and 50 actions per state, exhaustive enumeration must consider $50^{50} \approx 10^{85}$ policies; policy iteration finds the optimum in a handful of iterations. Howard demonstrated the method on a toymaker's production problem, taxicab dispatch in three city zones, and automobile replacement timing.

\citet{blackwell1965} established the measure-theoretic foundations for discounted dynamic programming with general state and action spaces. He proved that the Bellman operator $T$ defined by $Tu(s) = \sup_a [r(s,a) + \gamma \int u(s') P(ds'|s,a)]$ is a contraction with modulus $\gamma$: $\|Tu - Tv\|_\infty \leq \gamma \|u - v\|_\infty$. Banach's fixed-point theorem then guarantees a unique bounded solution $V^*$ to the Bellman equation, with $\|V_k - V^*\|_\infty \leq \gamma^k \|V_0 - V^*\|_\infty$ under value iteration. The central result concerns stationary policies, which use the same decision rule $f: \mathcal{S} \to \mathcal{A}$ at every period regardless of history. Blackwell proved that if the action space is finite, there exists an optimal stationary policy. For countable action spaces, $\epsilon$-optimal stationary policies exist for every $\epsilon > 0$. These results justify the focus on memoryless policies, since optimal behavior depends only on the current state, not on the history of past states and actions. The Bellman equation, Howard's policy iteration, and Blackwell's existence theorems constitute the planning framework. Given complete knowledge of $P$ and $r$, these methods compute optimal policies exactly. The challenge of learning without such knowledge is the central problem of reinforcement learning.\footnote{For comprehensive treatments of dynamic programming in economics, including numerical methods, computational complexity, and the curse of dimensionality, see \citet{rust2008dp} and \citet{rust1996numerical}.}

\section{Reinforcement Learning Algorithms}
\label{section:rl_algorithms}

\FloatBarrier
\subsection{The Classical Synthesis}
\label{sec:classical_synthesis}

\subsubsection{Monte Carlo Estimation}

When $P(s'|s,a)$ and $r(s,a)$ are unknown, the obvious approach is to use Monte Carlo (MC) to approximate them.\footnote{The Markov decision process itself, the tuple $(\mathcal{S}, \mathcal{A}, P, r, \gamma)$ together with policies, returns, and the value and action-value functions, is defined in Appendix~\ref{prelim:mdp}, which also fixes the notation used from here on.} These methods estimate value functions from sampled \textit{episodes} $(s_0, a_0, r_1, s_1, a_1, r_2, \ldots, s_T)$. The realized \textit{return} from state $s_t$ is
\begin{equation}
G_t = \sum_{k=0}^{T-t-1} \gamma^k r_{t+k+1}
\end{equation}
First-visit MC prediction averages $G_t$ over episodes for each state $s$, counting only its first occurrence per episode. Each first-visit return is an independent draw from the return distribution, so the sample mean converges almost surely to $V^\pi(s)$ by the strong law of large numbers \citep{sutton2018}. An incremental update is, $$V(s) \leftarrow V(s) + \alpha[G_t - V(s)]$$

For MC control, we can estimate action-values $Q(s,a)$ by averaging first-visit returns from each state-action pair, then improve the policy greedily: $\pi(s) = \argmax_a Q(s,a)$. Under exploring starts (every $(s,a)$ pair begins an episode infinitely often), this alternation converges to $Q^*$. \footnote{\citet{tsitsiklis2002} proved convergence even when the policy improves after every episode rather than waiting for complete evaluation. In practice, exploring starts is infeasible, so on-policy variants use $\varepsilon$-greedy exploration instead. These converge to $Q^*$ provided the exploration schedule satisfies the greedy-in-the-limit with infinite exploration (GLIE) condition: every state-action pair is visited infinitely often, and $\varepsilon_t \to 0$ so the policy converges to greedy in the limit.}

\subsubsection{Sutton (1988)}

Monte Carlo has two limitations. The agent must wait for episode termination to compute $G_t$, ruling out continuing tasks. And $G_t$ is unbiased ($\mathbb{E}[G_t \mid S_t = s] = V^\pi(s)$) but high-variance, because it sums random rewards over the entire \textit{trajectory}.

\citet{sutton1988} proposed temporal difference (TD) learning to fix both problems. TD(0) replaces the full return $G_t$ with a one-step target:
\begin{equation}
V(s_t) \leftarrow V(s_t) + \alpha \left[ r_{t+1} + \gamma V(s_{t+1}) - V(s_t) \right]
\end{equation}
The target $r_{t+1} + \gamma V(s_{t+1})$ depends on one random reward and one random transition, so its variance is low. The cost is bias. The bootstrap target $V(s_{t+1})$ is the agent's current estimate, not the true value. This is ``\textit{bootstrapping}''.\footnote{See Section~\ref{section:language} for the distinction between RL bootstrapping and Efron's resampling procedure.} As $V$ improves, the bias shrinks; the low variance persists regardless. Sutton demonstrated this tradeoff on a five-state random walk where TD(0) converged faster than Monte Carlo with less data.

The general TD($\lambda$) update interpolates between these extremes through an eligibility trace.\footnote{The eligibility trace records which states were recently visited, allowing credit assignment to propagate backward in time. States visited more recently receive stronger updates when the TD error is observed.}
\begin{equation}
V(s_t) \leftarrow V(s_t) + \alpha \, \delta_t \, e_t(s)
\end{equation}
where $\delta_t = r_{t+1} + \gamma V(s_{t+1}) - V(s_t)$ is the TD error and $e_t(s) = \gamma \lambda \, e_{t-1}(s) + \mathbbm{1}\{s = s_t\}$ is the eligibility trace for state $s$. Setting $\lambda = 0$ yields TD(0); setting $\lambda = 1$ recovers Monte Carlo returns. Intermediate $\lambda$ trades off variance against bias. \footnote{\citet{dayan1992} proved convergence of TD($\lambda$) for general $\lambda$ in the tabular case; \citet{jaakkola1994} gave a unified stochastic approximation proof covering both TD and Q-learning; \citet{tsitsiklis1997} extended the analysis to linear function approximation.}

\subsubsection{Watkins (1989)}

TD(0) learns value functions $V(s)$, but converting these to actions (the control problem) still requires knowing transition probabilities. Given $V(s')$ for all successor states, the agent needs to know which action leads to which successor. 

\citet{WatkinsDayan1992}, formalizing Watkins's 1989 PhD thesis, provided the solution. Instead of learning $V(s')$ learn $Q(s,a)$ (the action value function or the ``quality" of actions function) directly, the expected return from taking action $a$ in state $s$ and then behaving optimally. The optimal policy is then $\pi^*(s) = \argmax_a Q^*(s,a)$, requiring no model to act. The Bellman optimality equation provides the fixed-point condition:
\begin{equation}
Q^*(s,a) = \mathbb{E}\left[r + \gamma \max_{a'} Q^*(s',a') \mid s, a\right]
\end{equation}
Q-learning achieves this via the update
\begin{equation}
Q(s_t, a_t) \leftarrow Q(s_t, a_t) + \alpha \left[ r_{t+1} + \gamma \max_{a'} Q(s_{t+1}, a') - Q(s_t, a_t) \right]
\end{equation}
Q-learning converges to $Q^*$ under standard regularity conditions (Section~\ref{sec:stochastic_approx}). The maximization over $a'$ makes Q-learning \textit{off-policy}.\footnote{See Section~\ref{section:language} for the on-policy/off-policy distinction. Q-learning learns about the greedy policy $\pi^*(s) = \argmax_a Q(s,a)$ while collecting data with an exploratory policy. This exploratory policy needs only to be sufficiently ``exploratory'' and admits a wide range of policies; including a fully random policy.} The update target uses the greedy action at the next state regardless of the action actually taken. Therefore the agent can follow an $\varepsilon$-greedy exploration strategy or even a fully random policy, while learning about the optimal policy directly.

\subsubsection{Williams (1992)}

Value-based methods learn an action-value function and derive a policy from it. \citet{williams1992} derived an alternative, policy gradients, that optimize the policy directly by gradient ascent on expected return
\begin{equation}
\nabla_\theta J(\theta) = \mathbb{E}_{\pi_\theta} \left[ \sum_{t=0}^{T} \nabla_\theta \log \pi_\theta(a_t|s_t) \, G_t \right]
\end{equation}
where $G_t = \sum_{k=0}^{T-t} \gamma^k r_{t+k+1}$ is the discounted return from time $t$. The log-derivative trick allows the gradient to be estimated from sampled trajectories without differentiating through the environment dynamics.

Consider a robot arm that must apply a continuous torque $a \in \mathbb{R}$ to reach a target angle. Q-learning requires computing $\max_a Q(s,a)$ at every update, which becomes a nested optimization problem \footnote{In continuous action spaces, $\max_a Q(s,a)$ has no closed-form solution in general and must be solved numerically at every Bellman update. Discretizing a $d$-dimensional action space on a grid of $m$ points per dimension costs $O(m^d)$ evaluations per update step.} when the action space is continuous. A policy gradient method sidesteps the issue. Parameterize the policy as a Gaussian $\pi_\theta(a|s) = \mathcal{N}(\mu_\theta(s),\, \sigma_\theta^2(s))$,\footnote{The Gaussian distribution $\mathcal{N}(\mu, \sigma^2)$ has density $(2\pi\sigma^2)^{-1/2}\exp(-(a-\mu)^2/2\sigma^2)$. Here $\mu_\theta(s)$ and $\sigma_\theta(s)$ are neural network outputs parameterizing the policy mean and standard deviation.} sample an action, observe the return, and update $\theta$ by REINFORCE. Continuous-control methods in reinforcement learning are largely built on this policy gradient framework.


\subsubsection{Tesauro (1994)}

\citet{tesauro1994}'s TD-Gammon demonstrated that temporal difference learning with neural network function approximation could achieve expert-level play in a domain with approximately $10^{20}$ legal positions\footnote{The state $\mathbf{x}(s)$ was a 198-dimensional binary encoding of the raw board (four units per board point per player indicating checker counts, plus bar and borne-off counts). The output was a four-component vector estimating probabilities of each game outcome (White/Black $\times$ normal win/gammon), and the move maximizing expected outcome among all legal moves was selected at each step.}. 

Backgammon was far beyond tabular methods, yet TD-Gammon trained a feedforward neural network to estimate the probability of winning from any board position. A hidden layer\footnote{A feedforward neural network stacks an input layer, one or more hidden layers, and an output layer; each unit in a hidden layer computes a weighted sum of its inputs and applies a nonlinear activation, here the logistic sigmoid $\sigma(x) = 1/(1+e^{-x})$, which maps any real number to $(0,1)$ and is identical to the binary logit link function.} of 80 sigmoid units fed a single sigmoid output.
\begin{equation}
\hat{V}(s) = \sigma\!\bigl(\mathbf{w}^\top \sigma(W\mathbf{x}(s) + \mathbf{b}) + c\bigr)
\end{equation}
where $\sigma$ is the logistic sigmoid, $W$ is the input-to-hidden weight matrix, and $\mathbf{w}$ is the hidden-to-output weight vector. The network was trained by self-play using TD($\lambda$) with $\lambda = 0.7$. After each move from position $s_t$ to $s_{t+1}$, the weights $\boldsymbol{\theta}$\footnote{$\boldsymbol{\theta}$ denotes the full vector of network weights and biases, generalizing the scalar $\theta$ used for policy parameters in earlier sections.} were updated by
\begin{equation}
\boldsymbol{\theta} \leftarrow \boldsymbol{\theta} + \alpha \bigl[\hat{V}(s_{t+1}) - \hat{V}(s_t)\bigr] \, \mathbf{e}_t
\end{equation}
where $\alpha$\footnote{The \emph{learning rate} $\alpha$ controls the step size of each parameter update. TD learning uses a semi-gradient step rather than true gradient descent, but $\alpha$ plays the same role: too large and updates overshoot; too small and convergence is slow.} is the step size, and the eligibility trace $\mathbf{e}_t = \sum_{k=1}^{t} \lambda^{t-k} \nabla_{\boldsymbol{\theta}} \hat{V}(s_k)$ accumulates exponentially decayed gradients of past predictions.\footnote{In the neural network case, the eligibility trace $\mathbf{e}_t \in \mathbb{R}^{|\boldsymbol{\theta}|}$ is a vector accumulating exponentially-weighted gradients, extending the scalar state-based trace to parameter space.} At game's end, $\hat{V}(s_{t+1})$ is replaced by the outcome $z \in \{0, 1\}$.

A single neural network $\hat{V}$ serves as the evaluation function for both players. At each turn, the current player selects the legal move maximizing $\hat{V}(s')$ from its own perspective. As the network improves, it generates stronger play on both sides, producing harder training games that drive further improvement. The dice rolls ensure diverse board positions without requiring an explicit exploration mechanism.\footnote{Version 2.1, trained on 1,500,000 games with 2-ply search, achieved near-parity with former world champion Bill Robertie and discovered novel positional strategies subsequently adopted by the human backgammon community.} 

\subsubsection{SARSA (1994)}

Q-learning learns the optimal action-value function regardless of the policy generating experience. This off-policy property is useful but introduces complications when combined with function approximation. \citet{rummery1994} introduced SARSA as an \textit{on-policy}\footnote{See Section~\ref{section:language} for the on-policy/off-policy distinction.} alternative that learns the value of the policy actually being followed. The name derives from the quintuple $(s_t, a_t, r_{t+1}, s_{t+1}, a_{t+1})$ used in each update:
\begin{equation}
Q(s_t, a_t) \leftarrow Q(s_t, a_t) + \alpha \left[ r_{t+1} + \gamma Q(s_{t+1}, a_{t+1}) - Q(s_t, a_t) \right]
\end{equation}
The key difference from Q-learning is that SARSA bootstraps from the action $a_{t+1}$ actually taken at the next state, rather than the greedy action $\argmax_{a'} Q(s_{t+1}, a')$. This makes the algorithm on-policy, since the target depends on the behavior policy generating the data.

If the agent follows an $\varepsilon$-greedy policy, SARSA converges to $Q^{\varepsilon\text{-greedy}}$, not $Q^*$\footnote{SARSA converges to $Q^*$ under GLIE (greedy in the limit with infinite exploration). A GLIE schedule explores all state-action pairs infinitely often but converges to the greedy policy asymptotically. The standard $\varepsilon$-greedy policy with $\varepsilon_t \to 0$ is one such schedule.} policies and standard step-size conditions (Section~\ref{sec:stochastic_approx}). This distinction matters when exploration is costly. Consider the cliff-walking problem, where an agent must traverse a gridworld with a cliff along one edge. The optimal path runs along the cliff edge (shortest route), but the $\varepsilon$-greedy policy occasionally falls off. Q-learning learns the optimal path because it evaluates the greedy policy; the agent falls off during learning but the Q-values reflect the optimal route. SARSA learns a safer path further from the cliff because it evaluates the actual exploratory policy; it accounts for the fact that exploration sometimes leads to catastrophic states. 

\subsubsection{Baird (1995)}

\citet{Baird1995} constructed a six-state star MDP demonstrating divergence of Q-learning with linear function approximation. The MDP has five outer states that all transition to a single inner state under the target policy. The off-policy behavior samples states uniformly. With linear function approximation, the weights grow without bound under repeated Q-learning updates. The source of instability is the interaction of three components, namely bootstrapping (updating from estimated values rather than observed returns), off-policy learning (training on data from a different policy than the target), and function approximation (representing the value function with a parameterized model). \citet{sutton2018} later named this the deadly triad. Any two components can be combined safely; all three together permit divergence. The mechanism underlying this instability is analyzed in Section~\ref{sec:deadly_triad}.

This result explains the asymmetry between Tesauro's success and Baird's failure. TD-Gammon used bootstrapping and function approximation but was on-policy, with training data coming from the same self-play policy whose value was being estimated. Baird's counterexample used all three components and diverged. Baird also proposed a constructive solution, namely residual gradient algorithms that perform gradient descent on the mean-squared Bellman residual, guaranteeing convergence at the cost of a different fixed point.

\subsubsection{Actor-Critic Methods (2000)}

The idea of maintaining both a policy (\textit{actor}) and a value function (\textit{critic}) dates to \citet{barto1983neuronlike}, who used a two-component system to solve the pole-balancing task. Actor-critic methods address the high variance of REINFORCE by replacing Monte Carlo returns with bootstrapped TD targets as the learning signal. The critic learns a value function $V(s)$ by TD updates.
\begin{equation}
V(s_t) \leftarrow V(s_t) + \alpha_c \, \delta_t, \quad \delta_t = r_{t+1} + \gamma V(s_{t+1}) - V(s_t)
\end{equation}
The actor updates the policy using the TD error as a sample of the advantage.
\begin{equation}
\theta \leftarrow \theta + \alpha_a \nabla_\theta \log \pi_\theta(a_t|s_t) \, \delta_t
\end{equation}
The TD error $\delta_t$ estimates $A^\pi(s_t, a_t) = Q^\pi(s_t, a_t) - V^\pi(s_t)$, the advantage of action $a_t$ over the average action.\footnote{That $\delta_t$ is an unbiased estimate of the advantage follows from the policy gradient theorem \citep{SuttonMcAllester2000}.} Positive advantages indicate the action was better than expected; negative advantages indicate it was worse.

\citet{konda2000} provided the first convergence proof for actor-critic algorithms with function approximation, showing convergence to a stationary point under two-timescale learning rates ($\alpha_c \gg \alpha_a$) and a compatibility condition on the critic architecture (Section~\ref{sec:actor_critic}).


\subsubsection{Natural Policy Gradient (2001)}

Standard gradient descent treats all parameter directions equally, but policy parameters define probability distributions whose natural geometry is not Euclidean. \citet{Kakade2001} introduced the natural policy gradient, which measures progress in distribution space rather than parameter space. The update uses the Fisher information matrix $F(\theta)$.\footnote{Appendix~\ref{prelim:quadratic_forms} records its score outer-product form, local KL geometry, and conditioning.}
\begin{equation}
F(\theta) = \mathbb{E}_{s \sim d^\pi, a \sim \pi_\theta} \left[ \nabla_\theta \log \pi_\theta(a|s) \nabla_\theta \log \pi_\theta(a|s)^\top \right]
\end{equation}
The natural gradient is
\begin{equation}
\tilde{\nabla}_\theta J(\theta) = F(\theta)^{-1} \nabla_\theta J(\theta)
\end{equation}
This direction is invariant to reparameterization of the policy. In the tabular softmax\footnote{The softmax function maps a vector $\mathbf{z}$ to a probability distribution: $\text{softmax}(z_i) = \exp(z_i)/\sum_j \exp(z_j)$. A softmax policy parameterizes action probabilities as $\pi_\theta(a|s) = \text{softmax}(\theta_{s,a})$.} case, a single natural gradient step with unit step size recovers one step of exact policy iteration (Section~\ref{sec:policy_gradient}).

The computational bottleneck is inverting $F(\theta) \in \mathbb{R}^{d \times d}$. Practical implementations use conjugate gradient methods to solve $F(\theta) x = \nabla_\theta J(\theta)$ without forming $F$ explicitly. This approach was later scaled to deep neural networks by TRPO and PPO.

\subsubsection{Fitted Value Iteration and Fitted Q-Iteration (2005)}
\label{sec:fvi_fqi_algorithms}

Tabular Q-learning maintains a separate entry for every state-action pair. When $|\mathcal{S}|$ is large or the state space is continuous, as in most applied problems, this is infeasible. \emph{Fitted Q-Iteration} (FQI) \citep{Ernst2005} replaces the tabular update with a supervised regression step. Given a batch of transitions, it fits a function approximator to the Bellman targets.

Let $\mathcal{F}$ be a function class mapping $\mathcal{S} \times \mathcal{A} \to \mathbb{R}$.\label{def:fqi} Initialize $Q_0 \equiv 0$. At each iteration $k = 0, 1, \ldots, K-1$: (i) draw $N$ transitions $(s_i, a_i, r_i, s_i')$ from a generative model; (ii) construct regression targets $y_i^{(k)} = r_i + \gamma \max_{a'} Q_k(s_i', a')$; (iii) set $Q_{k+1} \leftarrow \arg\min_{f \in \mathcal{F}} \frac{1}{N} \sum_{i=1}^N \bigl( f(s_i, a_i) - y_i^{(k)} \bigr)^2$. The output is the greedy policy $\pi_K(s) = \arg\max_a Q_K(s,a)$.

The regression step replaces the exact Bellman application with a projection onto $\mathcal{F}$: $Q_{k+1} = \Pi_{\mathcal{F}} \mathcal{T} Q_k$, where $\mathcal{T}$ is the Bellman optimality operator and $\Pi_{\mathcal{F}}$ is the $L^2$-projection under the sample distribution. Fitted Value Iteration (FVI) \citep{MunosSzepesvari2008} applies the same idea to the value function directly: $V_{k+1} = \Pi_{\mathcal{F}} \mathcal{T}^* V_k$, where $(\mathcal{T}^* V)(s) = \max_a \{ r(s,a) + \gamma \sum_{s'} P(s'|s,a) V(s') \}$.

With feature matrix $\Phi \in \mathbb{R}^{|\mathcal{S}| \times d}$ (rows $\phi(s)^\top$) and per-action weight vectors $\theta_a \in \mathbb{R}^d$, each FQI regression step for action $a$ reduces to the normal equations.\footnote{Appendix~\ref{prelim:span_rank} interprets the columns of $\Phi$ as representable directions. Appendix~\ref{prelim:least_squares} derives the normal equations from orthogonality of the regression residual.}
\begin{equation}
\theta_a^{(k+1)} = \bigl(\Phi^\top \Phi\bigr)^{-1} \Phi^\top y_a^{(k)},
\label{eq:fqi_normal}
\end{equation}
where $y_a^{(k)}(s) = r(s,a) + \gamma \sum_{s'} P(s'|s,a) \max_{a'} \phi(s')^\top \theta_{a'}^{(k)}$. Computation is $O(d^2 |\mathcal{S}| + d^3)$ per action per iteration. The FVI update takes the same form with a single weight vector $\theta_V$:
\begin{equation}
\theta_V^{(k+1)} = \bigl(\Phi^\top \Phi\bigr)^{-1} \Phi^\top V_{\mathrm{target}}^{(k)},
\label{eq:fvi_normal}
\end{equation}
where $V_{\mathrm{target}}^{(k)}(s) = \max_a \bigl\{ r(s,a) + \gamma \sum_{s'} P(s'|s,a) \phi(s')^\top \theta_V^{(k)} \bigr\}$. The finite-sample error theory for these methods is developed in Section~\ref{sec:fvi_fqi_theory}.

\subsection{The Deep Learning Era}

\subsubsection{Deep Q-Networks (2015)}

\citet{mnih2015} trained a single convolutional neural network\footnote{A convolutional neural network applies learned spatial filters to detect local patterns in grid-structured data, commonly used for image inputs.} to play 49 Atari 2600 games directly from pixel inputs ($210 \times 160 \times 3$) and a scalar score, using no game-specific features.

The architecture processed four consecutive frames through three convolutional layers and a fully connected layer.\footnote{A fully connected layer computes $y = Wx + b$ where every input unit is connected to every output unit with learned weights $W$ and biases $b$; it is the affine transformation familiar from linear regression, followed by a nonlinear activation.} The network $Q(s, a; \theta)$ was trained to minimize the squared temporal difference error
\begin{equation}
L(\theta) = \mathbb{E}_{(s,a,r,s') \sim \mathcal{D}} \left[ \left( r + \gamma \max_{a'} Q(s', a'; \theta^-) - Q(s, a; \theta) \right)^2 \right]
\end{equation}

Two innovations stabilized learning. \textit{Experience replay} \citep{Lin1992} stored transitions $(s, a, r, s')$ in a buffer $\mathcal{D}$ and sampled uniformly for training, breaking the temporal correlation between consecutive updates. A \textit{target network} used a frozen copy of parameters $\theta^-$, updated periodically,\footnote{The buffer held $10^6$ transitions; the target network $\theta^-$ was synchronized to $\theta$ every $C = 10{,}000$ steps.} so the regression target does not shift with each gradient step.


DQN exceeded human-level performance on 29 of 49 games using a single architecture and hyperparameters.\footnote{\emph{Hyperparameters} are design choices fixed before training begins, such as network depth, learning rate, and replay buffer size; they are not estimated by gradient descent.} Games requiring long-horizon planning or sparse rewards, such as Montezuma's Revenge, remained difficult.

\subsubsection{TRPO and PPO (2015, 2017)}

Policy gradient methods suffer from a practical instability. A single large gradient step can move the policy into a region where performance collapses and recovery is slow.

\citet{Schulman2015} addressed this with Trust Region Policy Optimization (TRPO). TRPO solves the constrained optimization problem
\begin{equation}
\max_\theta \; L_{\theta_{\text{old}}}(\theta) \quad \text{subject to} \quad \bar{D}_{\mathrm{KL}}(\theta_{\text{old}}, \theta) \leq \delta
\end{equation}
where $L$ is a surrogate objective based on the advantage function $A^\pi(s,a) = Q^\pi(s,a) - V^\pi(s)$.\footnote{The Kullback-Leibler divergence $D_{\mathrm{KL}}(p \| q) = \sum_x p(x) \log(p(x)/q(x))$ measures statistical distance between distributions. The bar denotes expectation over states: $\bar{D}_{\mathrm{KL}} = \mathbb{E}_{s}[D_{\mathrm{KL}}(\pi_{\theta_{\text{old}}}(\cdot|s) \| \pi_\theta(\cdot|s))]$.}

\citet{Schulman2017} proposed Proximal Policy Optimization (PPO) as a simpler alternative. PPO replaces the KL constraint with a clipped surrogate objective:
\begin{equation}
L^{\text{CLIP}}(\theta) = \mathbb{E}_t \left[ \min\!\left( r_t(\theta) \hat{A}_t, \; \text{clip}(r_t(\theta), 1-\varepsilon, 1+\varepsilon) \hat{A}_t \right) \right]
\end{equation}
where $r_t(\theta) = \pi_\theta(a_t|s_t) / \pi_{\theta_{\text{old}}}(a_t|s_t)$ is the probability ratio. The clipping removes the incentive for $r_t(\theta)$ to move outside the interval $[1-\varepsilon, 1+\varepsilon]$, penalizing large policy updates without explicitly computing a divergence measure.

PPO outperformed A2C, TRPO, and the cross-entropy method on continuous control benchmarks and achieved the highest average reward on 30 of 49 Atari games among the methods tested. PPO became the default policy optimization algorithm for large-scale RL applications.

\subsubsection{Soft Actor-Critic (2018)}

\citet{Haarnoja2018} introduced Soft Actor-Critic (SAC), which adds entropy regularization to the actor-critic framework. The agent maximizes expected return plus an entropy bonus:
\begin{equation}
J(\theta) = \mathbb{E}_{\pi_\theta}\left[\sum_{t=0}^\infty \gamma^t \left(r_t + \tau \mathcal{H}(\pi_\theta(\cdot|s_t))\right)\right]
\end{equation}
where $\mathcal{H}(\pi) = -\sum_a \pi(a) \log \pi(a)$ is the entropy and $\tau > 0$ is a temperature parameter. The entropy bonus encourages exploration by penalizing deterministic policies. The optimal policy under this objective is softmax in the Q-values: $\pi^*(a|s) \propto \exp(Q^*(s,a)/\tau)$, connecting to discrete choice models in econometrics.

SAC maintains two Q-networks (to reduce overestimation bias) and a policy network. The soft Bellman operator for the critic is:
\begin{equation}
(T^\pi Q)(s,a) = r(s,a) + \gamma \mathbb{E}_{s'}\left[ V(s') \right], \quad V(s) = \mathbb{E}_{a \sim \pi}[Q(s,a) - \tau \log \pi(a|s)]
\end{equation}
SAC is off-policy (using experience replay), handles continuous actions naturally, and achieves state-of-the-art sample efficiency on continuous control benchmarks. The entropy regularization provides automatic exploration without $\varepsilon$-greedy schedules.

\begin{figure}[h]
\centering
\includegraphics[width=\textwidth]{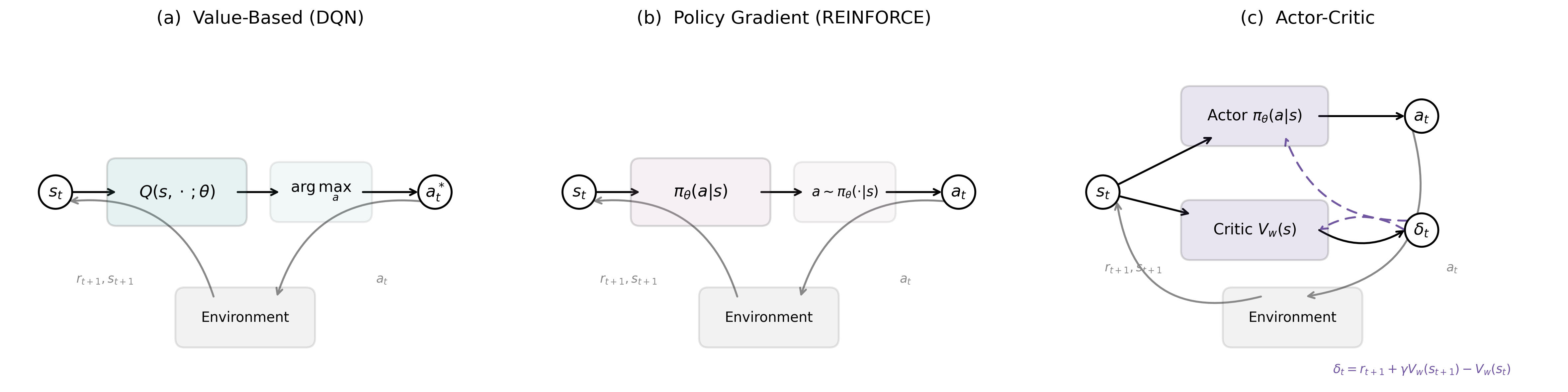}
\caption{Architecture comparison of three core algorithm families: value-based, policy gradient, and actor-critic. (a) DQN maps states to Q-values for all actions, selecting the argmax. (b) REINFORCE maps states to a probability distribution over actions, then samples. (c) Actor-Critic maintains separate policy and value networks; the critic's TD error $\delta_t$ feeds back into both during training, yielding lower-variance policy updates than REINFORCE's Monte-Carlo return. Each panel includes the environment feedback loop in which the agent's action produces a reward and next state.}
\label{fig:algorithm_architectures}
\end{figure}

\FloatBarrier
\subsubsection{Control as Probabilistic Inference}
\label{sec:control_inference}

The \emph{control-as-inference} framework \citep{Todorov2006, Kappen2011, Ziebart2010, Levine2018} recasts the preceding algorithms as instances of probabilistic inference (computation, not statistical inference) in a single graphical model.\footnote{The framework is a mathematical language, much like random utility in econometrics, that reveals structural connections rather than deriving algorithms from first principles.} The benefit is that every advance in approximate inference (variational methods, message passing, amortized inference) becomes a candidate RL algorithm, and the forward problem (finding the optimal policy given rewards) and the inverse problem (recovering rewards from observed behavior) become two queries in the same model. The construction introduces a binary ``optimality'' variable $\mathcal{O}_t \in \{0,1\}$ at each time step, appended to the standard state-action-dynamics chain (Figure~\ref{fig:control_inference_dag}). The distribution over this variable is defined as
\begin{equation}
P(\mathcal{O}_t = 1 \mid s_t, a_t) = \exp(r(s_t, a_t)/\tau)
\label{eq:optimality_variable}
\end{equation}
where $\tau > 0$ is a temperature parameter and rewards satisfy $r \leq 0$.\footnote{The non-positive reward assumption ensures $\exp(r/\tau) \in (0,1]$. Any bounded reward can be shifted to satisfy this without changing the optimal policy. Alternatively, the optimality variable can be replaced by an undirected potential $\Phi(s_t, a_t) = \exp(r(s_t,a_t)/\tau)$, yielding a conditional random field formulation that removes this restriction \citep{Ziebart2010}.} This is a modeling assumption, not a derivation from first principles; the exponential form is chosen so the resulting posterior matches entropy-regularized RL.

\begin{figure}[h]
\centering
\begin{tikzpicture}[
    state/.style={circle, draw, minimum size=8mm, inner sep=0pt, fill=rlblue!10, draw=rlblue!60, line width=0.5pt, font=\small},
    action/.style={circle, draw, minimum size=8mm, inner sep=0pt, fill=rlgray!15, line width=0.5pt, font=\small},
    opt/.style={circle, draw, minimum size=8mm, inner sep=0pt, fill=rlorange!20, draw=rlorange!70, line width=0.5pt, font=\small},
    arr/.style={-{Stealth[length=2mm]}, line width=0.5pt},
    dotsty/.style={font=\large, rlgray},
]
\node[state] (s1) at (-0.7, 0) {$s_1$};
\node[state] (s2) at (2.8, 0) {$s_2$};
\node[state] (s3) at (6.3, 0) {$s_3$};
\node[dotsty] at (8.8, 0) {$\cdots$};
\node[action] (a1) at (0.7, 1.5) {$a_1$};
\node[action] (a2) at (4.2, 1.5) {$a_2$};
\node[action] (a3) at (7.7, 1.5) {$a_3$};
\node[opt] (o1) at (0, 3) {$\mathcal{O}_1$};
\node[opt] (o2) at (3.5, 3) {$\mathcal{O}_2$};
\node[opt] (o3) at (7.0, 3) {$\mathcal{O}_3$};
\draw[arr] (s1) -- (s2);
\draw[arr] (a1) -- (s2);
\draw[arr] (s2) -- (s3);
\draw[arr] (a2) -- (s3);
\draw[arr] (s3) -- (8.3, 0);
\draw[arr] (a3) -- (8.3, 0);
\draw[arr] (s1) -- (o1);
\draw[arr] (a1) -- (o1);
\draw[arr] (s2) -- (o2);
\draw[arr] (a2) -- (o2);
\draw[arr] (s3) -- (o3);
\draw[arr] (a3) -- (o3);
\end{tikzpicture}
\caption{The control-as-inference graphical model. States $s_t$ and actions $a_t$ jointly determine the next state $s_{t+1}$ (dynamics) and the optimality variable $\mathcal{O}_t$ (reward signal). Optimality nodes $\mathcal{O}_t$ are observed as $\mathcal{O}_t = 1$; the posterior over actions $P(a_t \mid s_t, \mathcal{O}_{t:T} = 1)$ yields the optimal policy.}
\label{fig:control_inference_dag}
\end{figure}

The RL problem becomes a probabilistic inference query. Given that all future time steps are optimal ($\mathcal{O}_{t:T} = 1$), what is $P(a_t \mid s_t, \mathcal{O}_{t:T} = 1)$? Define backward messages $\beta_t(s, a) = P(\mathcal{O}_{t:T} = 1 \mid s_t = s,\, a_t = a)$, the probability that everything from $t$ onward is optimal given the agent is in state $s$ taking action $a$. These are computed by backward recursion:
\begin{align}
\beta_T(s, a) &= \exp(r(s, a)/\tau) \label{eq:beta_base} \\
\beta_t(s, a) &= \exp(r(s, a)/\tau) \sum_{s'} P(s'|s,a) \, \beta_{t+1}(s') \quad (t < T) \label{eq:beta_recursion}
\end{align}
where $\beta_{t+1}(s') \propto \sum_{a'} \beta_{t+1}(s', a')$ marginalizes over future actions under a uniform prior. By Bayes' rule, the posterior over actions is $P(a_t \mid s_t, \mathcal{O}_{t:T} = 1) = \beta_t(s_t, a_t) \,/\, \sum_{a'} \beta_t(s_t, a')$. Defining $Q(s,a) = \tau \log \beta_t(s,a)$, so that backward messages in log-space correspond to soft value functions, and $V(s) = \tau \log \sum_a \exp(Q(s,a)/\tau)$\footnote{Since $\beta_t(s) \propto \sum_a \beta_t(s,a) = \sum_a \exp(Q(s,a)/\tau)$, we have $V(s) = \tau \log \sum_a \exp(Q(s,a)/\tau)$.}, the posterior becomes $\pi(a \mid s) = \exp\bigl((Q(s,a) - V(s))/\tau\bigr)$, the softmax over Q-values.

Exact inference in this graphical model under stochastic dynamics produces risk-seeking policies. The backup becomes $Q(s,a) = r(s,a) + \tau \log \mathbb{E}_{s'}[\exp(V(s')/\tau)]$, which overweights unlikely favorable transitions because the agent implicitly assumes it can influence the dynamics.\footnote{Under exact inference, the posterior dynamics $P(s'|s,a,\mathcal{O}_{1:T}=1)$ differ from the true dynamics $P(s'|s,a)$, biasing the inferred transitions toward favorable outcomes. The log-expectation-of-exponentials exceeds the expectation by Jensen's inequality (Theorem~\ref{thm:prelim_jensen}). This optimism is an artifact of exact inference in the model, not a feature.} The framework corrects this by applying structured variational inference, restricting the variational family to distributions that match the true dynamics $P(s'|s,a)$. This yields the soft Bellman equations
\begin{align}
Q(s,a) &= r(s,a) + \gamma \, \mathbb{E}_{s' \sim P}[V(s')] \label{eq:soft_bellman_q} \\
V(s) &= \tau \log \sum_{a \in \mathcal{A}} \exp\!\left(\frac{Q(s,a)}{\tau}\right) \label{eq:soft_bellman_v}
\end{align}
with the optimal policy given by $\pi(a \mid s) = \exp\bigl((Q(s,a) - V(s))/\tau\bigr)$. The operator $\mathcal{T}^{\mathrm{soft}}$ defined by Equations~\eqref{eq:soft_bellman_q}--\eqref{eq:soft_bellman_v} is a $\gamma$-contraction in $\|\cdot\|_\infty$, with the same convergence rate as the standard Bellman operator.\footnote{The proof follows from the log-sum-exp being 1-Lipschitz in the sup norm; see \citet{Haarnoja2017}, Theorem~1.} As $\tau \to 0$, the log-sum-exp converges to the hard maximum and the soft Bellman equations reduce to the standard Bellman optimality equations. The value function in Equation~\eqref{eq:soft_bellman_v} is identical to McFadden's social surplus function from discrete choice theory \citep{McFadden1978}, completing the connection noted in the preceding subsection.

Classical RL already employs exploration strategies ($\varepsilon$-greedy, UCB), but these are designed separately from the optimization objective. In the probabilistic framework, the objective itself maximizes expected reward and policy entropy simultaneously.\footnote{The max-ent objective is $\max_\pi \sum_t \mathbb{E}[r(s_t,a_t) + \tau \mathcal{H}(\pi(\cdot|s_t))]$. This equals the evidence lower bound (ELBO) of the graphical model, a quantity from variational inference that lower-bounds the log-likelihood of the observed optimality evidence $\log P(\mathcal{O}_{1:T} = 1)$. Maximizing the ELBO with respect to the policy is equivalent to finding the best approximation to the true posterior, measured by KL divergence; see \citet{BleiKucukelbirMcAuliffe2017} for a review of variational inference.} The Q-function, the value function, and the exploration behavior are all derived from a single objective. Different approximation strategies applied to this single model recover familiar algorithms. When the backward messages in Equations~\eqref{eq:beta_base}--\eqref{eq:beta_recursion} are computed exactly in the tabular setting, the result is soft value iteration. Estimating the ELBO gradient via the likelihood ratio trick yields max-ent policy gradients, identical to REINFORCE with $-\tau \log \pi(a_t|s_t)$ added to the reward at each step. Fitting parameterized $Q_\phi$ and $V_\psi$ networks to approximate the backward messages gives Soft Actor-Critic \citep{Haarnoja2018}. Fitting $Q_\phi$ alone and extracting the policy implicitly via the softmax gives soft Q-learning \citep{Haarnoja2017}. Trust-region methods such as TRPO and PPO do not optimize the max-ent objective, but each policy update step solves a max-ent subproblem with the old policy as prior, so the per-step structure mirrors the framework even though the global target remains standard reward maximization \citep{Schulman2015, Schulman2017}. Hard-max algorithms (Q-learning, DQN, standard policy gradient) are not direct instances of the framework but are recovered in the zero-temperature limit $\tau \to 0$, where the softmax collapses to the argmax. Reading the graphical model in the reverse direction, treating the reward as the unknown and observed behavior as evidence of optimality, yields maximum entropy inverse reinforcement learning \citep{Ziebart2008}, connecting to the structural estimation methods discussed in Section~\ref{section:rl_econ_models}.

The framework has practical limitations. The converged fixed point is $Q^*_{\mathrm{soft}}$, not $Q^*$; at any $\tau > 0$, the policy is suboptimal for the original reward-maximization objective. The entropy bonus produces undirected exploration, spreading probability mass rather than targeting uncertain states. In safety-critical environments, the objective assigns nonzero probability to catastrophic actions. The practical benefits of the probabilistic perspective, including robustness to model perturbations and smooth optimization landscapes, are most apparent in continuous-control and robotics settings. In perfectly simulated environments with exact rewards (Atari, Go), the hard-max algorithms that dominate those domains do not require this probabilistic formulation, as the following subsection illustrates.

\subsubsection{AlphaGo Zero (2017)}
\label{subsubsec:alphago_zero}

The game of Go has approximately $10^{170}$ legal positions and a branching factor of roughly 250, far beyond the reach of brute-force search. Hand-crafted evaluation functions, which had succeeded in chess, failed here because positional concepts like influence, territory, and group viability are holistic and contextual. Monte Carlo tree search (MCTS) had achieved amateur-level play by using random simulations to estimate position values, but progress had stalled below professional strength. \citet{Silver2016} broke through by combining supervised learning from 30 million human expert positions, reinforcement learning via self-play, and MCTS with learned value and policy networks; the resulting system defeated Lee Sedol four games to one in March 2016. A year later, \citet{Silver2017} showed that none of the human data was necessary.

AlphaGo Zero uses a single convolutional neural network $f_\theta(s) = (\mathbf{p}, v)$ that takes a board position $s$ and outputs both a policy vector $\mathbf{p}$ over legal moves and a scalar value $v$ estimating the probability of winning. The input representation consists of 17 binary planes on the $19 \times 19$ board encoding the raw game state without hand-crafted features.\footnote{Eight planes for Black's stone positions over the last eight moves, eight for White's, and one indicating which color plays next. The history planes allow the network to detect ko situations and infer the trajectory of play.} The architecture uses residual blocks,\footnote{A residual block computes $\mathbf{x} + g(\mathbf{x})$ rather than just $g(\mathbf{x})$, where $g$ is a learned transformation. The skip connection allows gradients to flow through very deep networks without vanishing. AlphaGo Zero's 40-block architecture has 79 parameterized layers.} which allow training of very deep networks.

During play, each move is selected by running MCTS, which conducts 1,600 simulated games from the current position to estimate move quality. Each simulation proceeds in four phases, illustrated in Figure~\ref{fig:mcts_phases}. In the selection phase, the algorithm starts from the current position and traverses the partially built search tree by choosing at each node the action that maximizes $Q(s,a) + c_{\text{puct}} \cdot P(s,a) \cdot \sqrt{\sum_b N(s,b)} \,/\, (1 + N(s,a))$, where $Q(s,a)$ is the current average value of action $a$, $P(s,a)$ is the prior probability from the neural network, and $N(s,a)$ is the visit count.\footnote{The constant $c_{\text{puct}}$ controls exploration. Actions visited often have well-estimated $Q$ values but a shrinking exploration bonus; rarely visited actions have uncertain values but a large bonus. This is a continuous analogue of the upper confidence bound (UCB) strategy from bandit theory.} In the expansion and evaluation phase, when the traversal reaches a position not yet in the tree, the neural network evaluates it in a single forward pass, producing a policy vector $\mathbf{p}$ and a value estimate $v$; the policy initializes prior probabilities $P(s',a) = p_a$ for each child edge. In the backup phase, the value $v$ propagates back up the traversed path, incrementing each edge's visit count $N(s,a)$ and updating its mean value $Q(s,a)$. After all 1,600 simulations, the algorithm selects the move with the highest visit count at the root.

\begin{figure}[h]
\centering
\begin{tikzpicture}[
    nd/.style={circle, draw, minimum size=5mm, inner sep=0pt, fill=rlgray!12, line width=0.4pt},
    sel/.style={circle, draw, minimum size=5mm, inner sep=0pt, fill=rlblue!15, draw=rlblue!60, line width=0.7pt},
    newnd/.style={circle, draw, dashed, minimum size=5mm, inner sep=0pt, fill=rlgreen!10, draw=rlgreen!50!black, line width=0.7pt},
    edg/.style={-, rlgray!40, line width=0.3pt},
    sedg/.style={-, rlblue!60, line width=0.9pt},
    bedg/.style={-{Stealth[length=1.8mm]}, rlred!60, line width=0.7pt},
    phaselabel/.style={font=\footnotesize, anchor=north},
]
\begin{scope}[shift={(0,0)}]
    \node[sel] (r1) at (0,0) {};
    \node[nd] (a1) at (-0.9,-1.1) {};
    \node[sel] (b1) at (0.9,-1.1) {};
    \node[nd] (c1) at (-1.4,-2.2) {};
    \node[nd] (d1) at (-0.4,-2.2) {};
    \node[sel] (e1) at (0.5,-2.2) {};
    \node[nd] (f1) at (1.3,-2.2) {};
    \draw[edg] (r1) -- (a1);
    \draw[sedg] (r1) -- (b1);
    \draw[edg] (a1) -- (c1);
    \draw[edg] (a1) -- (d1);
    \draw[sedg] (b1) -- (e1);
    \draw[edg] (b1) -- (f1);
    \node[phaselabel] at (0,-2.9) {(a) Selection};
\end{scope}
\begin{scope}[shift={(3.8,0)}]
    \node[nd] (r2) at (0,0) {};
    \node[nd] (a2) at (-0.9,-1.1) {};
    \node[nd] (b2) at (0.9,-1.1) {};
    \node[nd] (c2) at (-1.4,-2.2) {};
    \node[nd] (d2) at (-0.4,-2.2) {};
    \node[nd] (e2) at (0.5,-2.2) {};
    \node[nd] (f2) at (1.3,-2.2) {};
    \node[newnd] (g2) at (0.5,-3.3) {};
    \draw[edg] (r2) -- (a2);
    \draw[edg] (r2) -- (b2);
    \draw[edg] (a2) -- (c2);
    \draw[edg] (a2) -- (d2);
    \draw[edg] (b2) -- (e2);
    \draw[edg] (b2) -- (f2);
    \draw[-, rlblue!60, dashed, line width=0.7pt] (e2) -- (g2);
    \node[phaselabel] at (0,-3.8) {(b) Expansion};
\end{scope}
\begin{scope}[shift={(7.6,0)}]
    \node[nd] (r3) at (0,0) {};
    \node[nd] (a3) at (-0.9,-1.1) {};
    \node[nd] (b3) at (0.9,-1.1) {};
    \node[nd] (c3) at (-1.4,-2.2) {};
    \node[nd] (d3) at (-0.4,-2.2) {};
    \node[nd] (e3) at (0.5,-2.2) {};
    \node[nd] (f3) at (1.3,-2.2) {};
    \node[newnd] (g3) at (0.5,-3.3) {};
    \draw[edg] (r3) -- (a3);
    \draw[edg] (r3) -- (b3);
    \draw[edg] (a3) -- (c3);
    \draw[edg] (a3) -- (d3);
    \draw[edg] (b3) -- (e3);
    \draw[edg] (b3) -- (f3);
    \draw[edg] (e3) -- (g3);
    \node[draw, rounded corners=2pt, fill=rlorange!12, font=\scriptsize, inner sep=3pt, draw=rlorange!60, line width=0.5pt] (nn3) at (-0.9,-3.3) {$f_\theta$};
    \draw[-{Stealth[length=1.8mm]}, rlorange!60, line width=0.6pt] (nn3) -- (g3) node[midway, above, font=\tiny, yshift=1pt] {$(\mathbf{p}, v)$};
    \node[phaselabel] at (0,-3.8) {(c) Evaluation};
\end{scope}
\begin{scope}[shift={(11.4,0)}]
    \node[sel] (r4) at (0,0) {};
    \node[nd] (a4) at (-0.9,-1.1) {};
    \node[sel] (b4) at (0.9,-1.1) {};
    \node[nd] (c4) at (-1.4,-2.2) {};
    \node[nd] (d4) at (-0.4,-2.2) {};
    \node[sel] (e4) at (0.5,-2.2) {};
    \node[nd] (f4) at (1.3,-2.2) {};
    \node[newnd] (g4) at (0.5,-3.3) {};
    \draw[edg] (r4) -- (a4);
    \draw[edg] (r4) -- (b4);
    \draw[edg] (a4) -- (c4);
    \draw[edg] (a4) -- (d4);
    \draw[edg] (b4) -- (e4);
    \draw[edg] (b4) -- (f4);
    \draw[edg] (e4) -- (g4);
    \draw[bedg] ([xshift=2pt]g4.north) -- ([xshift=2pt]e4.south);
    \draw[bedg] ([xshift=2pt]e4.north) -- ([xshift=-2pt]b4.south);
    \draw[bedg] ([xshift=2pt]b4.north) -- ([xshift=2pt]r4.south);
    \node[font=\tiny, rlred!60, anchor=west] at (0.85,-2.75) {$v$};
    \node[phaselabel] at (0,-3.8) {(d) Backup};
\end{scope}
\end{tikzpicture}
\caption{The four phases of a single MCTS simulation in AlphaGo Zero. (a) Selection traverses the tree from the root, choosing at each node the action maximizing a UCB-like score balancing exploitation ($Q$) and exploration ($P/N$). (b) Expansion adds a new leaf node when the traversal reaches an unexplored position. (c) The neural network $f_\theta$ evaluates the new position, producing move priors $\mathbf{p}$ and a value estimate $v$. (d) Backup propagates $v$ along the traversed path, updating mean values $Q(s,a)$ and visit counts $N(s,a)$ at each edge.}
\label{fig:mcts_phases}
\end{figure}

The training loop generates self-play games. At each board position $s_t$ during a game, the program runs MCTS to produce improved move probabilities $\boldsymbol{\pi}_t$, where $\pi_t(a) \propto N(s_t, a)^{1/\tau}$ and $\tau$ is a temperature parameter controlling exploration.\footnote{Early in the game ($t \leq 30$), $\tau = 1$ so moves are sampled proportionally to visit counts, encouraging diverse openings. Later, $\tau \to 0$ and the most-visited move is selected deterministically. Dirichlet noise is also added to root priors, $P(s,a) = (1 - \varepsilon)p_a + \varepsilon \eta_a$ with $\eta \sim \text{Dir}(0.03)$, ensuring all legal moves can be explored despite strong network priors.} A move $a_t$ is sampled from $\boldsymbol{\pi}_t$ and played. At game end, the outcome $z \in \{-1, +1\}$ is recorded. Each position becomes a training triple $(s_t, \boldsymbol{\pi}_t, z)$, and the network parameters are updated to minimize
\begin{equation}
\ell(\theta) = (z - v)^2 - \boldsymbol{\pi}^\top \log \mathbf{p} + c \|\theta\|^2
\end{equation}
where the first term is a value prediction loss, the second is a policy cross-entropy loss,\footnote{The cross-entropy loss $-\boldsymbol{\pi}^\top \log \mathbf{p}$ measures how well the predicted distribution $\mathbf{p}$ matches the target $\boldsymbol{\pi}$; it equals zero when the distributions are identical.} and the third is $L_2$ regularization.\footnote{$L_2$ regularization penalizes the squared magnitude of parameters, $c\|\theta\|^2$, analogous to ridge regression in econometrics.} The key mechanism is a self-reinforcing loop. MCTS serves as a policy improvement operator, since the search probabilities $\boldsymbol{\pi}$ are stronger than the raw network outputs $\mathbf{p}$. Training the network to match $\boldsymbol{\pi}$ distills the search improvements back into the network, and the improved network in turn produces better MCTS. After 72 hours of self-play on 4 TPUs, AlphaGo Zero surpassed all previous versions, including the one that defeated Lee Sedol, and discovered novel strategies not previously seen in human play.\footnote{The system that defeated Lee Sedol in March 2016 used fixed network weights throughout the match; no parameter updates occurred between or during games. This illustrates the training-execution distinction (Section~\ref{section:language}): the months of self-play constituted the training phase, while the five-game match was purely execution.}

Go was well-suited to this architecture. Its fixed $19 \times 19$ board maps naturally to convolutional networks, its perfect information and deterministic transitions make MCTS's tree structure exact, and the binary game outcome provides an unambiguous training signal. \citet{Igami2020} interprets the architecture in econometric terms, where the policy network is a conditional choice probability (CCP) estimator, the value network is a conditional value function (CVF) estimator, and the system performs CCP estimation and forward simulation jointly, connecting to the approach of \citet{HotzMiller1993} in dynamic discrete choice.

\subsubsection{Decision Transformers (2021)}
\label{subsubsec:decision_transformers}

\citet{Chen2021DT} proposed replacing Bellman backups with autoregressive
sequence modeling. The Decision Transformer conditions a causal GPT-style
Transformer on trajectories $\tau = (\hat{R}_1, s_1, a_1, \hat{R}_2, s_2,
a_2, \ldots)$, where $\hat{R}_t$ is the \emph{return-to-go} (desired future
cumulative reward). At test time, conditioning on a high target return
extracts a high-performing policy without any temporal-difference learning.
\citet{Janner2021TT} extended this to the Trajectory Transformer, modeling
entire trajectories as flat token sequences with continuous dimensions
discretized into bins, enabling planning via beam search over trajectories.

The approach has fundamental limitations that Bellman-based methods do not
share. \citet{Brandfonbrener2022} proved that return-conditioned supervised
learning (RCSL) recovers optimal policies only under assumptions strictly
stronger than those needed for dynamic programming: near-deterministic
dynamics, expert data coverage, and a unique mapping from returns to optimal
actions. In stochastic environments, high returns may reflect environmental
luck rather than good decisions, causing RCSL to imitate lucky-but-suboptimal
trajectories. \citet{Paster2022} demonstrated this concretely. In a simple
gambling MDP, the Decision Transformer conditioned on high returns selects
risky gambles over the optimal safe action, even with infinite
data.\footnote{\citet{Emmons2022} showed that a simple two-layer MLP matches
the Transformer architecture on D4RL benchmarks, suggesting the
autoregressive structure provides conditional density estimation rather than
temporal reasoning. The essential element is conditioning on the right outcome
variable, not the architecture.}

A second limitation is that RCSL cannot \emph{stitch} suboptimal trajectory
segments. If the dataset contains two trajectories that each visit a useful
intermediate state but from different starting points, Bellman-based methods
can compose the better segments by propagating values backward through the
shared state. Sequence models, which predict forward autoregressively, cannot
perform this backward composition.

\FloatBarrier
\subsection{Simulation Study: Algorithm Families on the Engine Replacement MDP}
\label{sec:engine_algorithms}

The classical and deep-learning sections differ in approximation architecture, update rule, and the object learned. The Engine Replacement MDP from Section~\ref{engine:model} holds the decision problem fixed so that the comparison measures those algorithmic differences. Prediction methods estimate $V^{\pi^\star}$, value-based control methods learn from exploratory behavior and report $\max_a \hat Q$, and policy-gradient methods report the exact value of their final learned policy.

\begin{table}[h]
\centering
\small
\caption{Eight algorithm families on the Engine Replacement MDP, 10 seeds each, ranked by mean sup-norm distance from $V^\star = (5.3448,\ 4.3103)$. Standard errors across seeds in parentheses. Rows carry different data budgets and three scoring modes. Prediction rows evaluate $\pi^\star$; control rows learn from exploratory behavior and report $\max_a \hat{Q}$; policy-gradient rows report the exact value of the final learned policy.}
\label{tab:engine_algorithms}
\begin{tabular}{@{}llrrr@{}}
\hline
algorithm & family & $\hat{V}(\text{low})$ & $\hat{V}(\text{high})$ & $\|\hat{V} - V^\star\|_\infty$ \\
\hline
MC & prediction, 2000 episodes & 5.3416 (0.0059) & 4.3078 (0.0039) & 0.0136 (0.0040) \\
Q-learning & off-policy control, 50000 steps & 5.3296 (0.0076) & 4.2928 (0.0064) & 0.0216 (0.0064) \\
FQI & batch VI, 20000 transitions & 5.3571 (0.0105) & 4.3214 (0.0095) & 0.0292 (0.0058) \\
SARSA & on-policy control, 50000 steps & 5.3122 (0.0087) & 4.2754 (0.0081) & 0.0389 (0.0065) \\
TD(0) & prediction, 20000 steps & 5.3172 (0.0124) & 4.2874 (0.0118) & 0.0413 (0.0073) \\
REINFORCE & policy gradient, 5000 episodes & 5.2967 (0.0038) & 4.2594 (0.0045) & 0.0510 (0.0045) \\
Actor-critic & policy gradient, 2000 episodes & 5.2909 (0.0008) & 4.2538 (0.0007) & 0.0566 (0.0007) \\
DQN & deep value learning, 40000 steps & 5.3852 (0.0248) & 4.3333 (0.0165) & 0.0821 (0.0123) \\
\hline
\end{tabular}
\end{table}

Table~\ref{tab:engine_algorithms} shows that every family recovers the optimal values to within a tenth of a unit. The rows are not direct efficiency comparisons because the families use different data budgets and three scoring modes. The two policy-gradient rows remain below $V^\star$ because their final policies are stochastic, while a sigmoid policy reaches the deterministic optimum only as its logits diverge. The table therefore connects the chapter's theory to the objects each algorithm estimates, rather than ranking the methods under a common computational budget.

\FloatBarrier


\section{The Theory of Reinforcement Learning}
\label{sec:planning_learning}

\subsection{The Geometry of Dynamic Programming}
\label{sec:newton_connection}

Value iteration (VI) and policy iteration (PI) are the workhorses of dynamic programming. VI applies the Bellman operator repeatedly until convergence; PI alternates between \emph{policy evaluation} (solving a linear system) and \emph{policy improvement} (taking the greedy action). PI converges faster. Why? The answer reveals a connection between dynamic programming and numerical optimization. Policy iteration is Newton's method applied to the Bellman equation.

\subsubsection{Value Iteration as Picard Iteration}

Consider the Bellman optimality operator $T$ acting on value functions.
\begin{equation}
(TV)(s) = \max_{a \in \mathcal{A}} \left\{ r(s,a) + \gamma \sum_{s' \in \mathcal{S}} P(s'|s,a) V(s') \right\}.
\label{eq:bellman_operator}
\end{equation}
This operator is nonlinear due to the $\max$. Its defining property is that it contracts distances in the supremum norm.

\begin{lemma}[Bellman contraction]
\label{lem:bellman_contraction}
The Bellman optimality operator $T$ of~\eqref{eq:bellman_operator} satisfies $\|TV_1 - TV_2\|_\infty \leq \gamma \|V_1 - V_2\|_\infty$ for all $V_1, V_2 \in \mathbb{R}^{\mathcal{S}}$.
\end{lemma}

\begin{proof}
The single idea is that swapping the action optimal for $V_1$ into the maximization for $V_2$ can only lower the second value, which removes both $\max$ operators and leaves a plain average.\footnote{The supremum norm is $\|f\|_\infty = \max_{s} |f(s)|$, the largest entry of $f$ across the (finite) state space. Calling $T$ a $\gamma$-contraction in this norm, a Lipschitz map of constant $\gamma$ (Theorem~\ref{thm:prelim_lipschitz}), means it shrinks the worst-case gap between any two value functions by at least the factor $\gamma$ at each application.} Fix a state $s$ and let $a^* \in \argmax_{a} \{ r(s,a) + \gamma \sum_{s'} P(s'|s,a) V_1(s') \}$ attain the maximum that defines $(TV_1)(s)$. The same $a^*$ is a feasible, though not necessarily optimal, choice in the maximization defining $(TV_2)(s)$, so $(TV_2)(s) \geq r(s,a^*) + \gamma \sum_{s'} P(s'|s,a^*) V_2(s')$. Subtracting this from $(TV_1)(s) = r(s,a^*) + \gamma \sum_{s'} P(s'|s,a^*) V_1(s')$,
\begin{align*}
(TV_1)(s) - (TV_2)(s)
  &\leq \gamma \sum_{s'} P(s'|s,a^*)\big[ V_1(s') - V_2(s') \big]
    && \text{(the reward terms $r(s,a^*)$ cancel)} \\
  &\leq \gamma \underbrace{\sum_{s'} P(s'|s,a^*)}_{=\,1}\, \|V_1 - V_2\|_\infty
    && \text{(each bracket $\leq \|V_1 - V_2\|_\infty$)} \\
  &= \gamma \|V_1 - V_2\|_\infty .
\end{align*}
The last two steps use that each row of $P$ is a probability distribution.\footnote{Each row of the transition matrix is a distribution, $\sum_{s'} P(s'|s,a) = 1$ with $P \geq 0$, so $\sum_{s'} P(s'|s,a^*)\, x(s')$ is a weighted average of the entries of $x$ and cannot exceed $\max_{s'} |x(s')| = \|x\|_\infty$.} Exchanging the roles of $V_1$ and $V_2$ bounds $(TV_2)(s) - (TV_1)(s)$ by the same quantity, so $|(TV_1)(s) - (TV_2)(s)| \leq \gamma \|V_1 - V_2\|_\infty$. Taking the supremum over $s$ yields the result.
\end{proof}

\begin{corollary}[Value iteration]
\label{cor:vi_convergence}
The operator $T$ has a unique fixed point $V^*$, and the iterates $V_{k+1} = TV_k$ satisfy $\|V_k - V^*\|_\infty \leq \gamma^k \|V_0 - V^*\|_\infty$.
\end{corollary}

\begin{proof}
Since $\gamma < 1$, Lemma~\ref{lem:bellman_contraction} makes $T$ a contraction on the complete metric space $(\mathbb{R}^{\mathcal{S}}, \|\cdot\|_\infty)$, so the Banach fixed-point theorem (Theorem~\ref{thm:prelim_banach}) applies and supplies a unique fixed point $V^*$ with $V_k \to V^*$.\footnote{Banach fixed-point theorem: a map $T$ on a complete metric space satisfying $\|Tx - Ty\| \leq \gamma \|x - y\|$ with $\gamma < 1$ has exactly one fixed point $x^* = Tx^*$, and the iterates $x_{k+1} = Tx_k$ from any starting point converge to it, with $\|x_k - x^*\| \leq \gamma^k \|x_0 - x^*\|$. The space $\mathbb{R}^{\mathcal{S}}$ under the supremum norm is complete because $\mathcal{S}$ is finite.} Applying the contraction once per iteration to the identity $V^* = TV^*$,
\begin{align*}
\|V_k - V^*\|_\infty
  &= \|TV_{k-1} - TV^*\|_\infty
    && \text{(since $V_k = TV_{k-1}$ and $V^* = TV^*$)} \\
  &\leq \gamma \|V_{k-1} - V^*\|_\infty
    && \text{(Lemma~\ref{lem:bellman_contraction})} \\
  &\leq \cdots \leq \gamma^k \|V_0 - V^*\|_\infty
    && \text{(unfold the bound $k$ times).}
\end{align*}
\end{proof}

Value iteration applies $T$ repeatedly, $V_{k+1} = TV_k$, and Corollary~\ref{cor:vi_convergence} delivers the geometric bound $\|V_k - V^*\|_\infty \leq \gamma^k \|V_0 - V^*\|_\infty$ \citep{denardo1967}.\footnote{This is the Contraction Mapping Theorem. The identical mathematical structure governs convergence of value function iteration in consumption-savings models, competitive equilibrium computation, and Bellman equation solution.} This is Picard iteration, with linear convergence at rate $\gamma$.\footnote{Picard iteration is $x_{k+1} = f(x_k)$ for finding roots of $x = f(x)$. When $f$ is a contraction, convergence is geometric. The rate $\gamma$ means each iteration reduces the error by a fixed proportion; more patient agents (higher $\gamma$) face slower convergence because the operator contracts less per step.} The iteration count to reduce error by a factor of $\delta$ is $k = \log(\delta) / \log(1/\gamma)$ \citep{bertsekas1996}.

Corollary~\ref{cor:vi_convergence} shows the iteration converges, but not yet that its limit is the object of interest. The fixed point $V^*$ was defined by the equation $V^* = TV^*$, not as the best value achievable across policies, and the \emph{principle of optimality} identifies the two.

\begin{theorem}[Principle of optimality]
\label{thm:verification}
Let $\Pi$ denote the class of all policies, including those that randomize and condition on the entire past history. The fixed point $V^*$ of Corollary~\ref{cor:vi_convergence} is the optimal value function, $V^*(s) = \sup_{\pi \in \Pi} V^\pi(s)$ for every $s \in \mathcal{S}$, and the stationary deterministic policy $\pi^*(s) \in \argmax_{a} \{ r(s,a) + \gamma \sum_{s'} P(s'|s,a) V^*(s') \}$ attains the supremum, $V^{\pi^*} = V^*$.\footnote{This is the fundamental theorem of Markov decision processes \citep[Ch.~6]{puterman1994}. On a finite state and action space the greedy $\argmax$ is always attained; on the continuous, unbounded problems of Section~\ref{sec:continuous_unbounded} a measurable selection theorem supplies a measurable $\pi^*$ that attains the maximum.}
\end{theorem}

\begin{proof}
The argument has two halves, that no policy exceeds $V^*$ and that the greedy policy meets it. For the first, fix any $\pi \in \Pi$ and let $a_t$ be its (possibly randomized, history-conditioned) action at time $t$. Because $V^* = TV^*$ takes the maximum over actions state by state, an average over the first action drawn by $\pi$ cannot exceed that maximum,
\[
V^*(s) \;\geq\; \mathbb{E}_{a_0 \sim \pi}\Big[ r(s,a_0) + \gamma \sum_{s'} P(s'|s,a_0)\, V^*(s') \Big] = \mathbb{E}_\pi\big[ r_0 + \gamma V^*(s_1) \,\big|\, s_0 = s \big].
\]
Applying the same one-step inequality at $s_1$ under $\pi$'s continuation and taking expectations forward,
\begin{align*}
V^*(s)
  &\geq \mathbb{E}_\pi\Big[ \textstyle\sum_{t=0}^{n-1} \gamma^t r_t \,\Big|\, s_0 = s \Big] + \gamma^n\, \mathbb{E}_\pi\big[ V^*(s_n) \,\big|\, s_0 = s \big]
    && \text{(unroll $n$ steps)} \\
  &\longrightarrow V^\pi(s)
    && \text{(let $n \to \infty$),}
\end{align*}
since the tail $\gamma^n \mathbb{E}_\pi[V^*(s_n)]$ vanishes as $n \to \infty$ with $V^*$ bounded and $\gamma < 1$. Thus $V^*(s) \geq V^\pi(s)$, and as $\pi$ was arbitrary, $V^* \geq \sup_{\pi \in \Pi} V^\pi$. For the second half, the greedy policy $\pi^*$ satisfies $T^{\pi^*} V^* = TV^* = V^*$ by construction, so $V^*$ is a fixed point of the affine operator $T^{\pi^*} V = r^{\pi^*} + \gamma P^{\pi^*} V$. That operator is a $\gamma$-contraction with unique fixed point $V^{\pi^*}$, the value of following $\pi^*$ forever, by the Banach argument of Corollary~\ref{cor:vi_convergence} applied to $T^{\pi^*}$. Hence $V^{\pi^*} = V^*$, the supremum is attained, and $V^* = \sup_{\pi \in \Pi} V^\pi = V^{\pi^*}$.
\end{proof}

\subsubsection{Policy Iteration as Newton's Method}

Policy iteration takes a different approach. At the current value estimate $\tilde{V}$, define the greedy policy $\tilde{\pi}(s) = \argmax_a \{ r(s,a) + \gamma \sum_{s'} P(s'|s,a) \tilde{V}(s') \}$. The \emph{policy evaluation} step solves the linear fixed-point equation $V = T^{\tilde{\pi}} V$ exactly, where $T^{\tilde{\pi}}$ is the policy-specific Bellman operator
\begin{equation}
(T^{\tilde{\pi}} V)(s) = r(s, \tilde{\pi}(s)) + \gamma \sum_{s'} P(s'|s,\tilde{\pi}(s)) V(s').
\end{equation}

The geometric structure, formalized by \citet{puterman1979}, is that the linear operator $T^{\tilde{\pi}}$ is a supporting hyperplane to the nonlinear operator $T$ at the current iterate.\footnote{A supporting hyperplane to a convex function at $x_0$ is a linear function $\ell$ with $\ell(x_0) = f(x_0)$ and $\ell(x) \leq f(x)$ everywhere. In plainer terms: $T^{\tilde{\pi}}$ is the tangent-line approximation to $T$ from elementary calculus, extended to function spaces. The policy operator $T^{\tilde{\pi}}$ plays this role for the Bellman operator.} Specifically, the operators satisfy tangency: $T^{\tilde{\pi}} \tilde{V} = T\tilde{V}$, so the linearization agrees with the nonlinear operator at the current iterate. They also satisfy support: $T^{\tilde{\pi}} V \leq TV$ for all $V$, meaning the linear operator lies weakly below the nonlinear one everywhere, just as a tangent line lies below a convex function. Policy evaluation solves for the fixed point of this linearization exactly. This is precisely the structure of Newton's method; linearize the nonlinear equation at the current point, solve the linearized system, and iterate.\footnote{The Newton interpretation of policy iteration has precursors in \citet{kleinman1968} for Riccati equations in linear-quadratic control and \citet{pollatschek1969} for stochastic games.}\footnote{Algebraically: consider finding the root of $G(V) = V - TV = 0$. The Bellman operator $T$ is piecewise affine, not smooth: $T$ is affine on each region where the greedy policy is constant, with kinks at boundaries where the optimal action switches. This makes $G$ a semismooth function in the sense of \citet{qi1993}. At any iterate $V_k$ where the greedy policy $\tilde{\pi}$ is unique (a generic condition), $T$ is locally affine: $TV = r^{\tilde{\pi}} + \gamma P^{\tilde{\pi}} V$, so $G'(V_k) = I - \gamma P^{\tilde{\pi}}$. The Newton step $V_{k+1} = V_k - [G'(V_k)]^{-1} G(V_k) = (I - \gamma P^{\tilde{\pi}})^{-1} r^{\tilde{\pi}}$ is exactly the policy evaluation solution. At the non-smooth boundary points where two actions tie, any element of the B-subdifferential yields the same iterate because the two candidate linearizations produce the same fixed point.}

\begin{theorem}[Policy Improvement, \citet{howard1960}]
\label{thm:policy_improvement}
Let $\pi_k$ be the current policy with value $V^{\pi_k}$, and let $\pi_{k+1}$ be the greedy policy with respect to $V^{\pi_k}$:
\begin{equation}
\pi_{k+1}(s) = \argmax_{a \in \mathcal{A}} \left\{ r(s,a) + \gamma \sum_{s'} P(s'|s,a)\, V^{\pi_k}(s') \right\}.
\end{equation}
Then $V^{\pi_{k+1}}(s) \geq V^{\pi_k}(s)$ for all $s \in \mathcal{S}$, with strict inequality at some state unless $\pi_k$ is already optimal.
\end{theorem}

\begin{proof}
The strategy is to show the greedy operator improves $V^{\pi_k}$ in a single step, then let its own iteration carry that gain all the way to the greedy value. Write $\pi' = \pi_{k+1}$. By the greedy construction,
\begin{align*}
(T^{\pi'} V^{\pi_k})(s)
  &= \max_{a} \Big\{ r(s,a) + \gamma \sum_{s'} P(s'|s,a) V^{\pi_k}(s') \Big\}
    && \text{($\pi'$ is greedy for $V^{\pi_k}$)} \\
  &\geq r(s,\pi_k(s)) + \gamma \sum_{s'} P(s'|s,\pi_k(s)) V^{\pi_k}(s')
    && \text{($\pi_k(s)$ is one feasible action)} \\
  &= (T^{\pi_k} V^{\pi_k})(s) = V^{\pi_k}(s)
    && \text{($V^{\pi_k}$ is the fixed point of $T^{\pi_k}$),}
\end{align*}
so $T^{\pi'} V^{\pi_k} \geq V^{\pi_k}$ pointwise. The policy operator $T^{\pi'}$ is monotone,\footnote{An operator is monotone (order-preserving) when $U \leq W$ pointwise implies $T^{\pi'} U \leq T^{\pi'} W$. For $T^{\pi'} V = r^{\pi'} + \gamma P^{\pi'} V$ this holds because $P^{\pi'} \geq 0$ entrywise, so it sends a nonnegative difference to a nonnegative difference.} so applying it repeatedly preserves the inequality:
\[
V^{\pi_k} \leq T^{\pi'} V^{\pi_k} \leq (T^{\pi'})^2 V^{\pi_k} \leq \cdots \longrightarrow V^{\pi'} .
\]
The increasing sequence converges to $V^{\pi'}$ because $T^{\pi'}$ is itself a $\gamma$-contraction with unique fixed point $V^{\pi'}$, by the Banach argument of Corollary~\ref{cor:vi_convergence} applied to the linear operator $T^{\pi'}$. Therefore $V^{\pi'} \geq V^{\pi_k}$. If equality held at every state, then
\[
V^{\pi_k} = T^{\pi'} V^{\pi_k} = T V^{\pi_k},
\]
where the second equality is the greedy identity,\footnote{Applying the policy greedy for $V^{\pi_k}$ reproduces exactly the pointwise maximum that defines the Bellman optimality operator, so $T^{\pi'} V^{\pi_k} = T V^{\pi_k}$.} making $V^{\pi_k}$ a fixed point of $T$. By the uniqueness in Corollary~\ref{cor:vi_convergence}, $V^{\pi_k} = V^*$ and $\pi_k$ is already optimal. Otherwise the improvement is strict at some state.
\end{proof}

The consequence is finite termination. Since there are at most $|\mathcal{A}|^{|\mathcal{S}|}$ deterministic policies and each PI step strictly improves the value function (Theorem~\ref{thm:policy_improvement}), PI reaches the exact optimum in finitely many iterations. While VI requires $k = \log(100) / \log(1/\gamma)$ iterations to reduce error by a factor of 100 \citep{bertsekas1996}, PI typically converges in 5--10 iterations regardless of $\gamma$.\footnote{\citet{ye2011} proves PI is strongly polynomial with iteration count $O\!\left(\frac{|\mathcal{S}||\mathcal{A}|}{1-\gamma}\log\frac{|\mathcal{S}|}{1-\gamma}\right)$, resolving a long-standing conjecture. This bound is for fixed $\gamma$; \citet{fearnley2010} constructs examples requiring exponentially many iterations when $\gamma$ is allowed to vary with $|\mathcal{S}|$.}\footnote{For continuous-state problems discretized on a grid (the norm in economics), the finite-termination argument still applies to the discretized problem, but the number of grid points $n$ enters the bound. \citet{santos2004} establish a three-tier convergence result for PI applied to discretized dynamic programs: order $\approx 1.5$ globally for general interpolation schemes, quadratic convergence locally when the value function approximation is concave and piecewise linear, and superlinear convergence for general smooth interpolation. The formal error constants $C(h)$ in their quadratic bound degrade as the grid mesh $h \to 0$, but the iteration count is empirically independent of grid size.} At $\gamma = 0.90$, VI needs 44 iterations while PI needs only 5--8; at $\gamma = 0.95$, VI requires 90 versus 5--8; at $\gamma = 0.99$, VI needs 459 iterations while PI still converges in 5--10.

\citet{bertsekas2022newton} extends this interpretation to a broad class of dynamic programming problems. The Newton structure applies whenever the Bellman operator can be written as a pointwise maximum over linear operators: $T = \max_\pi T^\pi$. This includes not only infinite horizon problems with discounting but also optimal stopping problems (job search, option exercise), average cost optimization (inventory, queueing), and minimax formulations for adversarial settings.\footnote{\citet{rust1996numerical} surveys successive approximation and policy iteration methods for economic models, comparing their performance on the bus engine replacement problem. \citet{zhang2023distributed} extends randomized policy iteration to multi-agent problems where the control is $m$-dimensional, reducing per-iteration complexity from exponential to linear in $m$.}\footnote{These problems appear under different names such as ``stochastic shortest path'' for optimal stopping, ``average-cost MDP'' for long-run average optimization, and ``model predictive control'' (or receding-horizon control) for finite-horizon replanning.} The practical implication is that algorithms with policy-improvement structure (evaluate a policy exactly or approximately, then improve) inherit Newton-like convergence behavior, while pure value-iteration methods (apply $T$ directly) are limited to linear convergence.\footnote{\citet{blackwell1965} proves that for discounted MDPs with finite state and action spaces, a stationary deterministic policy $\pi: \mathcal{S} \to \mathcal{A}$ exists that is optimal for all initial states simultaneously. This ``uniform optimality'' has three implications: (1) the search space reduces from history-dependent or stochastic policies to static maps, justifying neural networks that condition only on current state; (2) the optimal policy is independent of the initial distribution $d_0$, so changing where episodes start does not require retraining; (3) PI and VI are guaranteed to converge to the same globally optimal policy regardless of initialization.}

\subsubsection{Dynamic Programming as a Linear Program}
\label{sec:dp_as_lp}

Value iteration and policy iteration both solve the Bellman equation by successive approximation. A third classical route replaces the fixed-point equation with a \emph{linear program}, which exposes the convex geometry that the discounted occupancy $d^{\pi_\theta}$ of the policy gradient theorem (Section~\ref{sec:policy_gradient}) later makes central.

The primal program searches over value functions. Any $V$ with $V \geq TV$ pointwise dominates $V^*$, since iterating the monotone operator from such a $V$ drives it down toward the fixed point, $V \geq TV \geq T^2 V \geq \cdots \to V^*$, and $V^*$ itself is feasible with equality. Writing the maximum inside $TV$ as one linear constraint per action turns the search for the smallest feasible $V$ into
\begin{equation}
\min_{V} \; \sum_s \nu_0(s)\, V(s) \quad \text{s.t.} \quad V(s) \geq r(s,a) + \gamma \sum_{s'} P(s'|s,a)\, V(s') \ \ \forall (s,a),
\label{eq:primal_lp}
\end{equation}
where $\nu_0$ is any strictly positive state weighting, whose choice leaves the minimizer $V^*$ unchanged \citep{manne1960,depenoux1963}. The objective is linear and the feasible region is an intersection of half-spaces, so \eqref{eq:primal_lp} is a linear program with $|\mathcal{S}|$ variables and $|\mathcal{S}||\mathcal{A}|$ constraints.

The dual searches over \emph{state-action occupancy measures}. Assigning a multiplier $d(s,a) \geq 0$ to each primal constraint gives
\begin{equation}
\max_{d \geq 0} \; \sum_{s,a} d(s,a)\, r(s,a) \quad \text{s.t.} \quad \sum_a d(s',a) = \nu_0(s') + \gamma \sum_{s,a} P(s'|s,a)\, d(s,a) \ \ \forall s'.
\label{eq:dual_lp}
\end{equation}
The dual constraint is a flow-conservation identity, the discounted rate at which the process enters $s'$ equals the rate at which it leaves, and $d(s,a)$ counts the discounted visits to $(s,a)$. Every stationary policy induces a feasible occupancy $d^\pi(s,a) = \sum_{t \geq 0} \gamma^t\, \mathbb{P}(s_t = s, a_t = a)$, whose state marginal, normalized to a probability distribution, is the discounted state occupancy $d^{\pi_\theta}$ of the policy gradient theorem. Conversely every feasible $d$ recovers a stationary policy through $\pi(a|s) = d(s,a) / \sum_{a'} d(s,a')$, a bijection between stationary policies and the feasible set \citep[Ch.~6.9]{puterman1994}. That set is a convex polytope whose vertices are the deterministic policies, and maximizing the linear objective over it lands the optimum at a vertex. Policy optimization, nonconvex in the policy parameters, is thus a linear program in occupancy space.

For an economist the reading is immediate. The primal minimizes a cost over shadow values $V(s)$, the dual allocates discounted visitation subject to flow conservation, and strong duality equates the two optima. The multipliers on the flow constraints are the shadow prices of an extra unit of occupancy, the same marginal-value reading carried by $V^*$. This is the native formulation of constrained Markov decision processes \citep{Altman1999}, where a budget constraint $\sum_{s,a} d(s,a)\, c(s,a) \leq b$ appends one linear inequality to \eqref{eq:dual_lp} and its multiplier prices the constraint, the structure the robust and constrained methods later in this survey build on.

\subsubsection{Continuous States and Unbounded Returns}
\label{sec:continuous_unbounded}

The contraction theory above assumes a finite state space and bounded rewards. The Brock--Mirman economy just solved satisfies neither. Its capital state is continuous and its felicity $\log(c_t)$ is unbounded below as consumption approaches zero, so the value function has infinite supremum norm and Lemma~\ref{lem:bellman_contraction} does not apply as stated. Discretizing to a finite grid restores the finite-state theory for the computed problem, but the underlying model, like the recursive models of \citet{stokey1989} and most of macroeconomics, lives on a continuous state with returns unbounded above or below.

The remedy is to measure distance in a \emph{weighted supremum norm}. Fix a weight $w: \mathcal{S} \to [1, \infty)$ that grows at least as fast as the value function and define
\begin{equation}
\|V\|_w = \sup_{s \in \mathcal{S}} \frac{|V(s)|}{w(s)}.
\label{eq:weighted_norm}
\end{equation}
A value function unbounded in the ordinary supremum norm can be finite in $\|\cdot\|_w$ once $w$ absorbs its growth; for the log-utility model $w(k) = 1 + |\log k|$ suffices. The Bellman operator contracts in this norm under a drift condition on the transition kernel.

\begin{theorem}[Weighted-norm contraction]
\label{thm:weighted_contraction}
Suppose there exist a weight $w \geq 1$ and constants $\bar{r} < \infty$ and $\lambda < 1/\gamma$ such that $|r(s,a)| \leq \bar{r}\, w(s)$ and $\sum_{s'} P(s'|s,a)\, w(s') \leq \lambda\, w(s)$ for all $(s,a)$. Then $T$ maps $\|\cdot\|_w$-bounded functions to $\|\cdot\|_w$-bounded functions and is a contraction of modulus $\gamma\lambda < 1$ in $\|\cdot\|_w$, so it has a unique fixed point $V^*$ with $\|V^*\|_w < \infty$, and value iteration converges to it geometrically \citep{wessels1977,bertsekas2018abstract}.
\end{theorem}

The drift condition $\sum_{s'} P(s'|s,a)\, w(s') \leq \lambda\, w(s)$ is the discrete-time analogue of a Lyapunov condition, bounding the expected weight after one transition by a factor $\lambda$ that discounting must outpace, $\gamma\lambda < 1$. The weight may itself grow in expectation, $\lambda > 1$ is allowed, and it is discounting that keeps future returns summable even when per-period returns grow with the state. The proof repeats the swap argument of Lemma~\ref{lem:bellman_contraction} with every quantity divided by $w(s)$, and the reward bound $|r| \leq \bar{r}\, w$ makes $T$ carry $\|\cdot\|_w$-bounded functions to $\|\cdot\|_w$-bounded functions.\footnote{The economics literature reaches the same conclusion through order-theoretic rather than metric arguments. \citet{boyd1990} solves the Ramsey problem in a weighted norm of exactly this form, and \citet{stokey1989} treat bounded returns directly and unbounded returns through monotone convergence. On a continuous state the greedy policy of Theorem~\ref{thm:verification} exists only if the maximization admits a measurable selection, guaranteed under continuity of $r$ and $P$ with a compact feasible action set.} With the contraction in hand, the fixed-point, verification, and policy-iteration results of this section carry over to the continuous, unbounded setting unchanged, now measured in $\|\cdot\|_w$.

\subsubsection{Simulation Study: The Brock--Mirman Economy}
\label{sec:brock_mirman_sim}

The \citet{brockmirman1972} optimal growth model tests the contraction and policy-iteration results above. A planner chooses capital $k'$ to maximize $\sum_{t=0}^\infty \beta^t \log(c_t)$ subject to the resource constraint $c_t + k_{t+1} = z_t k_t^\alpha$, where productivity $z_t \in \{0.9, 1.1\}$ follows a Markov chain with persistence 0.8. The simulation sets $\alpha = 0.36$ and $\beta = 0.96$, then discretizes capital on a 500-point grid covering two productivity states (1,000 states). The closed-form policy $k'(k,z) = \alpha\beta z k^\alpha$ provides an exact benchmark.

The discretized model makes the PI--Newton equivalence concrete. Define the Bellman residual $G(V) = V - TV$ on $\mathbb{R}^n$ where $n = 1{,}000$ (500 capital grid points $\times$ 2 productivity states). At iterate $V_k$, let $\pi_k$ denote the greedy policy. Since $\pi_k$ is unique at $V_k$ (generically), $T$ is locally affine: $TV = r^{\pi_k} + \gamma P^{\pi_k} V$, so the residual becomes $G(V) = (I - \gamma P^{\pi_k})V - r^{\pi_k}$ with Jacobian $G'(V_k) = I - \gamma P^{\pi_k}$. The Newton update is
\begin{equation}
V_{k+1} = V_k - [G'(V_k)]^{-1}\, G(V_k) = (I - \gamma P^{\pi_k})^{-1}\, r^{\pi_k},
\label{eq:pi_newton}
\end{equation}
which is exactly the policy evaluation step, the Neumann-series resolvent of Theorem~\ref{thm:prelim_neumann}: solving $V = r^{\pi_k} + \gamma P^{\pi_k} V$ for $V$. Each PI iteration is one Newton step on the Bellman residual, explaining the 11-iteration convergence on a 1,000-state problem.\footnote{\citet{santos2004} is the definitive reference on PI convergence for discretized economic models of precisely this type. Their analysis of the Brock--Mirman growth model establishes that PI iteration counts are empirically independent of grid resolution (7--11 iterations across all grid sizes in Table~\ref{tab:brock_mirman}), consistent with the Newton interpretation: Newton's method converges in a number of steps determined by the nonlinearity of the operator, not the dimension of the discretization.}

The VI iteration count follows from the contraction bound: each iteration reduces the Bellman residual by the factor $\beta = 0.96$, requiring $k = \lceil \log(\epsilon / \|TV_0 - V_0\|_\infty) / \log \beta \rceil = 567$ iterations for tolerance $\epsilon = 10^{-10}$.\footnote{At $\beta = 0.99$, the weaker contraction yields approximately four times as many iterations, since $\log(0.96)/\log(0.99) \approx 4$.}

\begin{figure}[h]
\centering
\includegraphics[width=\textwidth]{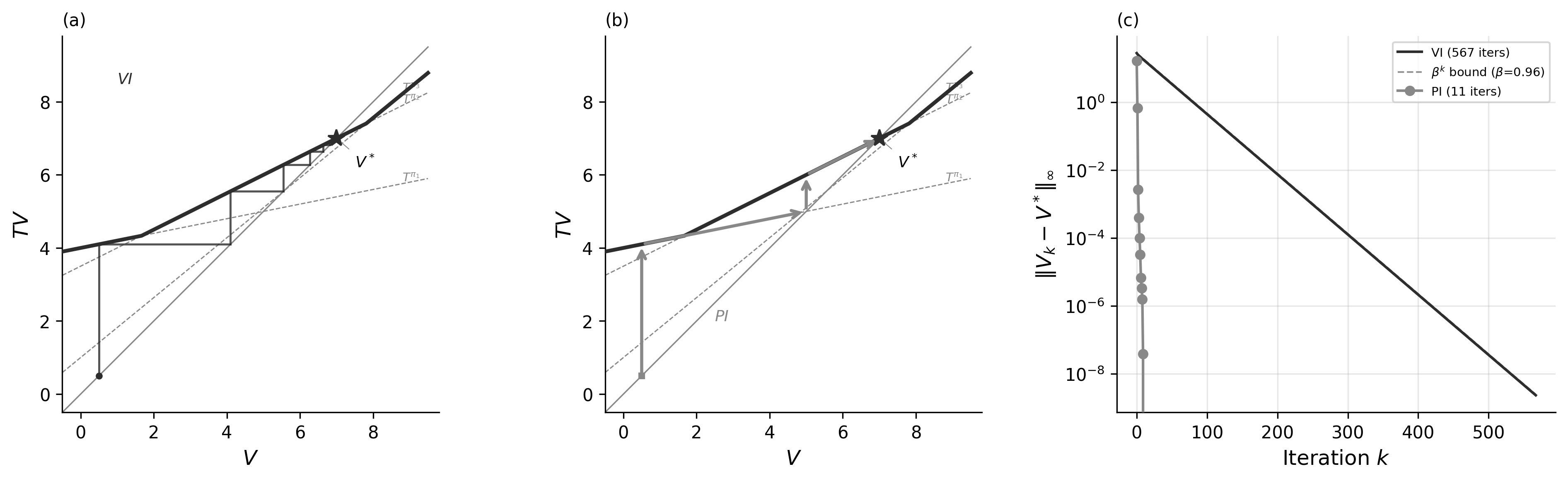}
\caption{The Brock--Mirman economy ($\alpha=0.36$, $\beta=0.96$, 1{,}000 states).
(a)~Value iteration on a scalar Bellman equation. The staircase iterates $V_{k+1} = TV_k$, converging at the linear rate $\gamma$.
(b)~Policy iteration as Newton's method. Each step solves for the fixed point of the active policy operator $T^{\pi_k}$, jumping to the tangent line's intersection with the diagonal.
(c)~Sup-norm error $\|V_k - V^*\|_\infty$ for the discretized model;
VI requires 567 iterations, PI converges in 11.}
\label{fig:brock_mirman_convergence}
\end{figure}

Figure~\ref{fig:brock_mirman_convergence}(a)--(b) provide a geometric interpretation for a scalar Bellman equation with three policies. Each policy operator $T^{\pi_i} V = r^{\pi_i} + \gamma_i V$ is affine; the Bellman operator $T = \max_i T^{\pi_i}$ is their upper envelope, a convex piecewise-linear function.\footnote{An affine function is a straight line, $f(x) = a + bx$, and the pointwise maximum of finitely many such lines is a convex piecewise-linear curve with a kink wherever the maximizing line switches. A two-line worked example, $f_1 = 1 + \tfrac{1}{2}x$ against $f_2 = 2 - x$ with the kink at $x = \tfrac{2}{3}$, is in Appendix~\ref{prelim:affine_envelope}, one of the elementary objects collected in Appendix~\ref{prelim:elementary}.} Panel~(a) shows VI: the staircase iterates $V_{k+1} = TV_k$ by alternating between the $T$ curve and the $45^\circ$ line, converging at the linear rate $\gamma$. Panel~(b) shows PI: at each iterate, the algorithm identifies the active policy operator and solves for its fixed point on the $45^\circ$ line, jumping directly to the intersection. This is a Newton step, where each $T^{\pi_k}$ is a supporting hyperplane to $T$ at $V_k$, and the fixed point of the linearization is the Newton iterate. The scalar picture extends to $\mathbb{R}^n$: the affine operator $T^{\pi_k} V = r^{\pi_k} + \gamma P^{\pi_k} V$ supports $T$ at $V_k$, and its fixed point $(I - \gamma P^{\pi_k})^{-1} r^{\pi_k}$ is the Newton iterate from equation~\eqref{eq:pi_newton}. Finite termination follows because $T$ has finitely many affine pieces; the iteration count depends on the number of policy switches, not the state-space dimension.

Table~\ref{tab:brock_mirman} and Figure~\ref{fig:brock_mirman_convergence} match the contraction and Newton predictions. VI requires 567 iterations at rate $\beta^n = 0.96^n$, while PI converges in 11. The \citet{manne1960} LP recovers the same value function to solver precision ($\|V_{\text{LP}} - V_{\text{VI}}\|_\infty < 10^{-8}$).\footnote{PI wall-clock time scales favorably: at $n_k = 200$, PI is roughly $50\times$ faster than VI (Table~\ref{tab:brock_mirman}), because the $O(n^3)$ per-iteration cost of policy evaluation is offset by 7--10 total iterations versus 567.}

\begin{table}[t]
\centering
\caption{Brock--Mirman Economy: VI vs PI vs LP. Wall-clock times are medians of 5 independent runs; iteration counts are deterministic.}
\label{tab:brock_mirman}
\begin{tabular}{llrrr}
\hline
Regime & Method & Iterations & Time (s) & Final error \\
\hline
1. Contraction & VI & 567 & 30.55 & 9.7e-11 \\
 & PI & 11 & 0.76 & --- \\
2. LP dual & VI & --- & 0.007 & --- \\
 & LP & --- & 0.017 & 2.3e-09 \\
3. Rate ($n_k$=10) & VI & 567 & 0.003 & --- \\
 & PI & 7 & 0.000 & --- \\
3. Rate ($n_k$=200) & VI & 567 & 2.467 & --- \\
 & PI & 10 & 0.051 & --- \\
\hline
\end{tabular}
\end{table}

\FloatBarrier
\subsubsection{Engine Replacement MDP: Value Space, Iteration, and Improvement}
\label{engine:value_space}

The Bellman operator $T$ from Lemma~\ref{lem:bellman_contraction} and the greedy-policy fixed point from Theorem~\ref{thm:verification} determine the value- and policy-iteration paths in this subsection.

\begin{figure}[h]
  \centering
  \includegraphics[width=\textwidth]{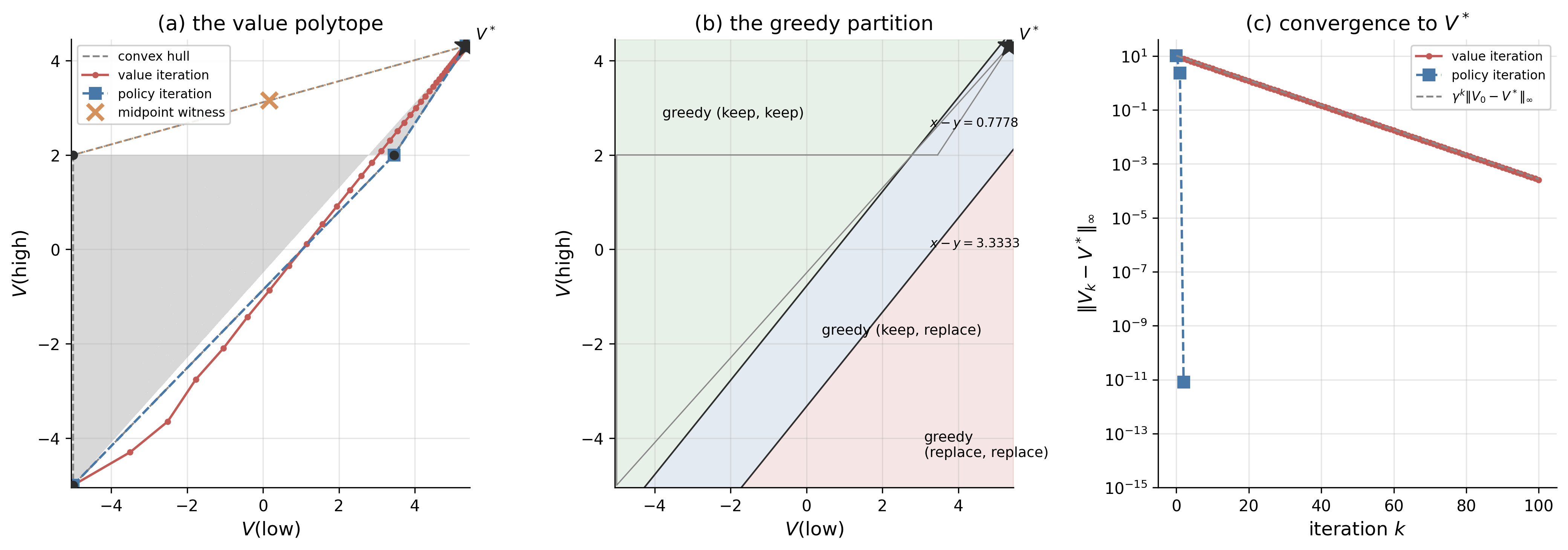}
  \caption{Value space of the Engine Replacement MDP, $x = V(\text{low})$, $y = V(\text{high})$. (a)~The value polytope (shaded), its convex hull (dashed), the four deterministic-policy values (dots), the midpoint witness (cross) on the chord between $V^{(\text{keep}, \text{replace})}$ and $V^{(\text{replace}, \text{keep})}$ (dotted), and the paths of value iteration (circles) and policy iteration (squares) from $V^{(\text{replace}, \text{replace})}$. (b)~The greedy partition, three parallel cells separated by the lines $x - y = 0.7778$ and $x - y = 3.3333$, with the polytope boundary and $V^*$ overlaid. (c)~The sup-norm error $\|V_k - V^*\|_\infty$ against iteration on a log scale, with the contraction bound $\gamma^k \|V_0 - V^*\|_\infty$ (dashed); policy iteration's last point sits at machine precision and the axis floor is clipped at $10^{-15}$.}
  \label{fig:engine_value_polytope}
\end{figure}

On the Engine Replacement MDP (Section~\ref{engine:model}) the set of all value functions, $\mathcal{V} = \{V^\pi\} \subset \mathbb{R}^2$, can be plotted in the plane. \citet{dadashi2019polytope} characterize this set for finite MDPs as the \emph{value function polytope}, and their line theorem states that the policies agreeing at every state but one trace a straight segment in value space. Sweeping the two replacement probabilities $(\pi(\text{replace} \mid \text{low}), \pi(\text{replace} \mid \text{high}))$ over $[0,1]^2$ and solving the resolvent at each point produces the region in Figure~\ref{fig:engine_value_polytope}(a). The four deterministic policies have computed values $V^{(\text{keep}, \text{replace})} = (5.3448, 4.3103)$, $V^{(\text{keep}, \text{keep})} = (3.4545, 2.0000)$, $V^{(\text{replace}, \text{keep})} = (-5.0000, 2.0000)$, and $V^{(\text{replace}, \text{replace})} = (-5.0000, -5.0000)$. In this MDP, these four values are exactly the vertices of the convex hull of $\mathcal{V}$. This is a property of this MDP, not a general fact, since deterministic values and polytope vertices can fail to coincide in either direction \citep{dadashi2019polytope}. The polytope has area $28.00$, strictly less than the hull's $39.36$, so it is non-convex. The midpoint of $V^{(\text{keep}, \text{replace})}$ and $V^{(\text{replace}, \text{keep})}$ is $(0.172, 3.155)$. The midpoint lies in the hull but sits $1.155$ away from the nearest value function. Table~\ref{tab:engine_value_polytope} lists the four deterministic-policy values.

\begin{table}[h]
\centering
\caption{The four deterministic-policy values of the Engine Replacement MDP, exact resolvent solves listed by $V(\text{low})$. Both iteration paths start at $V^{(\text{replace}, \text{replace})}$.}
\label{tab:engine_value_polytope}
\begin{tabular}{lrr}
\hline
policy & $V(\text{low})$ & $V(\text{high})$ \\
\hline
$(\text{keep}, \text{replace})$ & 5.3448 & 4.3103 \\
$(\text{keep}, \text{keep})$ & 3.4545 & 2.0000 \\
$(\text{replace}, \text{keep})$ & -5.0000 & 2.0000 \\
$(\text{replace}, \text{replace})$ & -5.0000 & -5.0000 \\
\hline
\end{tabular}
\end{table}

Value iteration and policy iteration move through this geometry differently. Value iteration applies $T$ directly from $V^{(\text{replace}, \text{replace})} = (-5, -5)$, and its first thirteen iterates leave the polytope, by as much as $0.662$, because value iteration generates vectors that are no policy's value function \citep{dadashi2019polytope}. Policy iteration from the same start stays on deterministic-policy values throughout, jumping $(\text{replace}, \text{replace}) \to (\text{keep}, \text{keep}) \to (\text{keep}, \text{replace})$ in two improvement steps. Figure~\ref{fig:engine_value_polytope}(b) overlays the greedy map on the same plane. Because replacement returns the engine to the low grade from either state, the greedy action at each state depends on $V$ only through the difference $V(\text{low}) - V(\text{high})$. Keep is greedy at high mileage when the difference is at most $\tfrac{7}{9}$ and at low mileage when it is at most $\tfrac{10}{3}$, so the plane splits into three parallel cells and the action pair $(\text{replace}, \text{keep})$ is greedy nowhere. $V^*$ has $V^*(\text{low}) - V^*(\text{high}) = 1.034 \in (\tfrac{7}{9}, \tfrac{10}{3})$, so it lies in the cell of its own greedy policy, an instance of the fixed-point property of Theorem~\ref{thm:verification}. Figure~\ref{fig:engine_value_polytope}(c) tracks the sup-norm error of both algorithms against the contraction bound $\gamma^k \|V_0 - V^*\|_\infty$. Value iteration's per-step contraction settles to exactly $\gamma = 0.9$ once a faster-decaying transient dies out. Policy iteration reaches $8 \times 10^{-12}$, machine precision at these magnitudes, after two improvement steps.

The computed paths verify the contraction bound for value iteration and the greedy fixed-point condition for policy iteration.

\FloatBarrier
\subsection{Value Learning Methods}
\label{sec:stochastic_approx}

\subsubsection{Stochastic Approximation Foundations}

When $P$ is unknown, a single sampled transition $(s, a, r, s')$ can replace the expectation $\mathbb{E}_{s' \sim P(\cdot|s,a)}[V(s')]$. The mathematical foundation is stochastic approximation, developed by \citet{robbinsmonro1951} (Theorem~\ref{thm:prelim_robbins_monro}).\footnote{The modern theory of stochastic approximation, including convergence rates and the ODE (ordinary differential equation) method for analyzing iterates, is developed in \citet{kushner1978} and \citet{borkar2000}. The ODE method shows that the expected trajectory of the stochastic iterates tracks the solution of a deterministic differential equation $\dot{x} = -g(x)$, providing stability conditions via Lyapunov theory.} Consider the problem of finding $x^*$ such that $g(x^*) = 0$, where $g$ cannot be evaluated directly but one can observe noisy samples $g(x) + \epsilon$. The Robbins-Monro iteration is:
\begin{equation}
x_{t+1} = x_t - \alpha_t [g(x_t) + \epsilon_t],
\label{eq:robbins_monro}
\end{equation}
where $\epsilon_t$ is zero-mean noise. Under two conditions on the step sizes, this converges to $x^*$ with probability one. The conditions are $\sum_{t=0}^\infty \alpha_t = \infty$ (sufficient exploration, ensuring that learning never ceases) and $\sum_{t=0}^\infty \alpha_t^2 < \infty$ (diminishing noise, ensuring the variance of cumulative updates is finite). The canonical choice $\alpha_t = 1/(t+1)$ satisfies both conditions.

\subsubsection{Q-Learning and SARSA}

Q-learning \citep{WatkinsDayan1992} is Robbins-Monro applied to the Bellman equation for action-value functions.\footnote{The ``Q'' in Q-learning stands for ``quality,'' following \citet{WatkinsDayan1992}, who used $Q(s,a)$ to denote the quality (expected return) of taking action $a$ in state $s$. The term ``Q-factor'' is used interchangeably with ``action-value function'' throughout the RL literature.} Define the Q-factor Bellman operator:
\begin{equation}
(FQ)(s,a) = r(s,a) + \gamma \sum_{s'} P(s'|s,a) \max_{a'} Q(s',a').
\end{equation}

\begin{theorem}[Q-factor contraction]
\label{thm:q_factor_contraction}
The operator $F$ is a $\gamma$-contraction in the supremum norm, $\|FQ_1 - FQ_2\|_\infty \leq \gamma \|Q_1 - Q_2\|_\infty$, and hence has a unique fixed point $Q^*$ to which $Q_{k+1} = FQ_k$ converges at rate $\gamma^k$.
\end{theorem}

\begin{proof}
The difference of the two operators is $\gamma$ times an average of differences of maxima, which the elementary fact that a difference of maxima is dominated by the largest pointwise difference collapses to a sup-norm bound.\footnote{For any two functions $f, g$ on a finite set, $|\max_a f(a) - \max_a g(a)| \leq \max_a |f(a) - g(a)|$. To see it, let $a_1$ attain $\max_a f$; then $\max_a f - \max_a g \leq f(a_1) - g(a_1) \leq \max_a |f(a) - g(a)|$, and the same argument with $f, g$ swapped handles the other sign. A numerical example is in Appendix~\ref{prelim:maxdiff}.} For any $(s,a)$, the reward terms cancel and
\[
(FQ_1)(s,a) - (FQ_2)(s,a) = \gamma \sum_{s'} P(s'|s,a) \big[ \max_{a'} Q_1(s',a') - \max_{a'} Q_2(s',a') \big].
\]
Taking absolute values and bounding term by term,
\begin{align*}
|(FQ_1)(s,a) - (FQ_2)(s,a)|
  &\leq \gamma \sum_{s'} P(s'|s,a) \big| \max_{a'} Q_1(s',a') - \max_{a'} Q_2(s',a') \big|
    && \text{(triangle ineq.)} \\
  &\leq \gamma \sum_{s'} P(s'|s,a) \max_{a'} |Q_1(s',a') - Q_2(s',a')|
    && \text{(max lemma)} \\
  &\leq \gamma \underbrace{\sum_{s'} P(s'|s,a)}_{=\,1}\, \|Q_1 - Q_2\|_\infty
    && \text{($\textstyle\sum P = 1$).}
\end{align*}
Taking the supremum over $(s,a)$ gives the contraction, and the fixed-point claim follows from Banach's theorem exactly as in Corollary~\ref{cor:vi_convergence}.
\end{proof}

The Q-learning update, upon observing transition $(s_t, a_t, r_t, s_{t+1})$, is
\begin{equation}
Q(s_t, a_t) \leftarrow Q(s_t, a_t) + \alpha_t \left[ r_t + \gamma \max_{a'} Q(s_{t+1}, a') - Q(s_t, a_t) \right].
\label{eq:qlearning}
\end{equation}

The logic of Q-learning is best understood as a Monte Carlo approximation of the Bellman contraction. The true Bellman operator involves an integral over the transition distribution, $(FQ)(s,a) = r + \gamma \int \max_{a'} Q(s',a') \, dP(s'|s,a)$. Since $P$ is unknown, this integral cannot be computed analytically. However, a sample transition $(s,a,r,s')$ acts as a single-point Monte Carlo estimate of this integral, which repeated sampling averages to the true value by the law of large numbers (Theorem~\ref{thm:prelim_lln_clt}). The Q-learning update is simply an exponential moving average (with weight $\alpha_t$) between the current estimate and this noisy Monte Carlo target. Because $F$ is a $\gamma$-contraction in the supremum norm, the expected update drives the estimate toward the fixed point $Q^*$, provided the noise in the Monte Carlo sample averages out over time (which the Robbins-Monro conditions ensure) \citep{tsitsiklis1994}.\footnote{Q-learning can be viewed as root-finding for the expected Bellman residual: the goal is to find parameters such that $\mathbb{E}_{(s,a,r,s')}[\delta_t] = 0$, where $\delta_t = r + \gamma \max_{a'} Q(s',a') - Q(s,a)$ is the temporal difference error. The update is a stochastic gradient step on the mean-squared Bellman error, but with a critical distinction: the gradient is ``semi-gradient'' because the target $\max_{a'} Q(s',a')$ is treated as a fixed constant rather than a function of the parameters being updated. This simplifies computation but disconnects the update from true gradient descent, requiring the specific stability conditions of \citet{tsitsiklis1994}.}

Convergence requires two conditions. Exploration (visiting all state-action pairs infinitely often) ensures identification. The Robbins-Monro step-size conditions ($\sum_t \alpha_t = \infty$, $\sum_t \alpha_t^2 < \infty$) balance tracking versus noise suppression. These conditions turn the contraction of Theorem~\ref{thm:q_factor_contraction} into almost-sure convergence of the sampled iteration.

\begin{theorem}[Q-learning convergence]
\label{thm:qlearning_convergence}
Suppose the state and action spaces are finite, every state-action pair is visited infinitely often, rewards are bounded, and for each $(s,a)$ the step sizes satisfy $\sum_t \alpha_t(s,a) = \infty$ and $\sum_t \alpha_t^2(s,a) < \infty$. Then the Q-learning iterates~\eqref{eq:qlearning} converge to $Q^*$ with probability one.
\end{theorem}

\begin{proof}
The plan is to recast the error $\Delta_t = Q_t - Q^*$ as a stochastic-approximation recursion, one whose conditional mean contracts by $\gamma$ and whose noise is mean-zero with controlled variance, then invoke a standard convergence lemma for such recursions. Write the update~\eqref{eq:qlearning} in incremental form. At the visited pair $(s,a) = (s_t, a_t)$,
\[
Q_{t+1}(s,a) = (1 - \alpha_t(s,a))\, Q_t(s,a) + \alpha_t(s,a)\big[\, \underbrace{(FQ_t)(s,a)}_{\text{target}} + \underbrace{w_t}_{\text{noise}} \big],
\]
where $\alpha_t(s,a) = 0$ for pairs not visited at step $t$ and the noise is the gap between the one-sample target and its expectation,
\[
w_t = \underbrace{r_t + \gamma \max_{a'} Q_t(s_{t+1}, a')}_{\text{sampled target}} - (FQ_t)(s,a).
\]
Given the history $\mathcal{F}_t$, the successor $s_{t+1}$ is drawn from $P(\cdot \mid s,a)$, so the sampled target is an unbiased draw of $(FQ_t)(s,a)$ and $\mathbb{E}[w_t \mid \mathcal{F}_t] = 0$.\footnote{$\mathbb{E}[\,r_t + \gamma \max_{a'} Q_t(s_{t+1},a') \mid \mathcal{F}_t] = r(s,a) + \gamma \sum_{s'} P(s'|s,a) \max_{a'} Q_t(s',a') = (FQ_t)(s,a)$ by the definition of $F$. So $w_t$ is a martingale difference, mean-zero given the past, and the almost-sure convergence of the resulting recursion is a martingale convergence argument (Theorem~\ref{thm:prelim_martingale}).} Because rewards are bounded and $\max_{a'} Q_t$ grows at most linearly in $\|Q_t - Q^*\|_\infty$ (since $Q^*$ is fixed and $|\max_{a'} Q_t(s',a') - \max_{a'} Q^*(s',a')| \leq \|\Delta_t\|_\infty$), the conditional variance obeys $\mathbb{E}[w_t^2 \mid \mathcal{F}_t] \leq C(1 + \|Q_t - Q^*\|_\infty^2)$ for a constant $C$. Subtracting $Q^*$ and writing $\Delta_t = Q_t - Q^*$,
\[
\Delta_{t+1}(s,a) = (1 - \alpha_t(s,a))\, \Delta_t(s,a) + \alpha_t(s,a)\big[\, \underbrace{(FQ_t)(s,a) - Q^*(s,a)}_{\text{contracts } \Delta_t} + w_t \big].
\]
Because $Q^* = FQ^*$, the Q-factor contraction (Theorem~\ref{thm:q_factor_contraction}) controls the conditional mean of the bracket,
\[
\big\|\mathbb{E}[(FQ_t) - Q^* \mid \mathcal{F}_t]\big\|_\infty = \|FQ_t - FQ^*\|_\infty \leq \gamma \|\Delta_t\|_\infty, \qquad \gamma < 1 .
\]
The process $\Delta_t$ therefore satisfies all the hypotheses of the stochastic-approximation lemma for contractive updates,\footnote{The lemma \citep[Theorem~1]{jaakkola1994}, a Q-learning-adapted form of the Dvoretzky stochastic-approximation theorem, states that a process $\Delta_{t+1}(x) = (1 - \alpha_t(x))\Delta_t(x) + \alpha_t(x) F_t(x)$ converges to $0$ with probability one provided (i) $\sum_t \alpha_t(x) = \infty$ and $\sum_t \alpha_t^2(x) < \infty$ at each $x$; (ii) $\|\mathbb{E}[F_t \mid \mathcal{F}_t]\|_\infty \leq \gamma \|\Delta_t\|_\infty$ with $\gamma < 1$; and (iii) $\mathrm{Var}[F_t \mid \mathcal{F}_t] \leq C(1 + \|\Delta_t\|_\infty^2)$. Here $F_t = (FQ_t - Q^*) + w_t$.} namely componentwise Robbins-Monro step sizes, a conditional mean that contracts $\Delta_t$ by $\gamma$ in the supremum norm, and conditional noise variance bounded by $C(1 + \|\Delta_t\|_\infty^2)$. The lemma yields $\Delta_t \to 0$ with probability one, so $Q_t \to Q^*$.
\end{proof}

\citet{WatkinsDayan1992} and \citet{jaakkola1994} establish this; \citet{tsitsiklis1994} gives the general asynchronous stochastic-approximation framework.

The choice of step-size schedule has quantitative consequences. \citet{evendar2003} show that polynomial schedules $\alpha_t = 1/t^\omega$ with $\omega \in (1/2, 1)$ achieve convergence rate $O(1/t^{1-\omega})$, creating an explicit tradeoff between speed and stability. Recent work by \citet{li2024minimax} establishes that Q-learning with variance-reduced updates achieves minimax-optimal sample complexity $\tilde{O}(|\mathcal{S}||\mathcal{A}|/(1-\gamma)^3\epsilon^2)$, matching information-theoretic lower bounds.\footnote{Vanilla Q-learning (without variance reduction) has tight complexity $\tilde{\Theta}(|\mathcal{S}||\mathcal{A}|/(1-\gamma)^4\epsilon^2)$, worse by a factor of $1/(1-\gamma)$ due to maximization bias inflating variance. The optimal cubic rate requires variance-reduced updates \citep{wainwright2019,sidford2018}. By contrast, naive model-based RL (estimate $\hat{P}$ from samples, then solve by planning) achieves the optimal $(1-\gamma)^{-3}$ rate with no special tricks \citep{agarwalKakadeYang2020}, illustrating the statistical cost of discarding transition structure.} Model-free learning is possible. The optimal value function can be found without ever estimating the transition probabilities.\footnote{Model-free methods are essential when (a) the environment is a physical system with dynamics too complex to write down, or (b) the agent learns directly from interaction. Note that ``model-free'' does not mean ``atheoretical.'' The agent does not store $P(s'|s,a)$ explicitly, but the Q-function serves as an implicit model encoding long-run consequences.}

SARSA provides an on-policy variant.\footnote{SARSA is named for the quintuple $(S_t, A_t, R_t, S_{t+1}, A_{t+1})$ used in each update \citep{rummery1994}. Convergence requires the GLIE (Greedy in the Limit with Infinite Exploration) condition: the behavior policy must explore all actions infinitely often while converging to a greedy policy. $\varepsilon$-greedy with $\varepsilon_t \to 0$ satisfies this.} Instead of taking the maximum over next actions, SARSA uses the action actually taken:
\begin{equation}
Q(s_t, a_t) \leftarrow Q(s_t, a_t) + \alpha_t \left[ r_t + \gamma Q(s_{t+1}, a_{t+1}) - Q(s_t, a_t) \right].
\end{equation}
This solves for the value function of the behavior policy $\pi$ rather than the optimal policy. \citet{singh2000} prove convergence under the same step-size conditions, provided the behavior policy converges to a stationary distribution (Theorem~\ref{thm:prelim_markov}). Q-learning is noisy value iteration on Q-factors; SARSA is noisy \emph{policy evaluation}.\footnote{\citet{bhandari2021finite} provide finite-time analysis of TD learning, showing that the convergence rate depends on the mixing time of the Markov chain under the behavior policy. Faster mixing (less serial correlation in the state sequence) yields faster convergence.} The Robbins-Monro conditions ensure that the noise averages out faster than the signal decays, allowing asymptotic convergence despite using only single-sample estimates.

\subsubsection{Multi-Step Returns and TD($\lambda$)}
\label{sec:td_lambda}

The updates in~\eqref{eq:qlearning} bootstrap from a single successor state. More generally, one can bootstrap from $n$ steps ahead. The $n$-step return is
\begin{equation}
G_t^{(n)} = \sum_{k=0}^{n-1} \gamma^k R_{t+k+1} + \gamma^n V(S_{t+n}).
\label{eq:n_step_return}
\end{equation}
Setting $n=1$ recovers the TD(0) target; letting $n \to \infty$ (or reaching a terminal state) gives the Monte Carlo return.

The \emph{$\lambda$-return} \citep{sutton1988} averages all $n$-step returns with geometrically decaying weights:
\begin{equation}
G_t^\lambda = (1-\lambda) \sum_{n=1}^{\infty} \lambda^{n-1} G_t^{(n)}, \qquad \lambda \in [0,1].
\label{eq:lambda_return}
\end{equation}
This is the \emph{forward view}: at time $t$, look forward at all possible truncation horizons and take a weighted average. The parameter $\lambda$ controls a bias-variance tradeoff: lower $\lambda$ gives higher bias (more bootstrapping) but lower variance; higher $\lambda$ gives lower bias but higher variance, since more of the update relies on stochastic returns rather than value estimates.

The forward view requires waiting until the end of the episode to compute $G_t^\lambda$. The \emph{backward view} computes the same total update incrementally. At each step, compute one TD error $\delta_t = R_{t+1} + \gamma V(S_{t+1}) - V(S_t)$ and distribute it to all states via an \emph{eligibility trace}:
\begin{equation}
e_t(s) = \gamma\lambda\, e_{t-1}(s) + \mathbbm{1}\{s = S_t\}, \qquad V(s) \leftarrow V(s) + \alpha\,\delta_t\,e_t(s).
\label{eq:eligibility_trace}
\end{equation}
The trace $e_t(s)$ acts as a fading memory of recently visited states: it spikes when $s$ is visited and decays by $\gamma\lambda$ per step. Each TD error $\delta_t$ updates every state in proportion to its current trace, enabling $O(|\mathcal{S}|)$ per-step credit assignment without storing trajectories.\footnote{The forward and backward views produce identical total weight changes over a complete episode \citep{sutton1988}. The backward view is preferred in practice because it operates online (updating after each transition) rather than requiring the full episode to be stored. Practical variants include \emph{replacing traces} ($e_t(s) = 1$ when $s = S_t$, capping the trace at 1 instead of accumulating), \emph{Dutch traces} \citep{vanseijen2016}, and off-policy extensions such as Retrace($\lambda$) \citep{munos2016retrace} and V-trace \citep{espeholt2018impala}.} The historical development of eligibility traces is discussed in Section~\ref{section:history}.

Under linear function approximation, TD($\lambda$) converges to a unique fixed point with approximation error bounded by $\frac{1-\lambda\gamma}{1-\gamma}$ times the best-in-class error (Theorem~\ref{thm:proj_bellman}; \citealp{tsitsiklis1997}). Higher $\lambda$ tightens this bound, approaching the projection of $V^\pi$ as $\lambda \to 1$.

\subsubsection{Finite-Sample Theory of Fitted Methods}
\label{sec:fvi_fqi_theory}

Fitted Q-Iteration and Fitted Value Iteration (Definition~\ref{def:fqi}, Section~\ref{sec:fvi_fqi_algorithms}) replace exact Bellman applications with projected regression steps. The projection introduces approximation error that compounds across iterations.

Define the inherent Bellman approximation error
\begin{equation}
\varepsilon_{\mathrm{approx}} = \inf_{f \in \mathcal{F}} \| \mathcal{T} f - f \|_{p,\mu},
\end{equation}
the smallest residual achievable when the Bellman operator maps any element of $\mathcal{F}$ back to itself. \citet{MunosSzepesvari2008} show that after $K$ iterations with $N$ i.i.d.\ samples per iteration,
\begin{equation}
\| V_{K} - V^* \|_{p,\rho} \;\leq\; C_{\rho,\mu} \left[ \gamma^K \|V_0 - V^*\|_{p,\rho} + \frac{\varepsilon_{\mathrm{approx}}}{(1-\gamma)^2} + O\!\left(\frac{1}{\sqrt{N}}\right) \right],
\label{eq:fvi_error_bound}
\end{equation}
where $C_{\rho,\mu}$ is a \emph{concentrability coefficient} bounding the ratio of future-state distributions under the evaluation distribution $\rho$ relative to the data distribution $\mu$.\footnote{The concentrability coefficient $C_{\rho,\mu}$ measures how well the data distribution $\mu$ covers future states reachable under optimal policies from $\rho$. When $\mu$ is the state-action distribution of the optimal policy itself, $C_{\rho,\mu} = 1$. Distribution mismatch, common when batch data comes from a sub-optimal behavior policy, inflates $C_{\rho,\mu}$ and worsens the bound. \citet{Antos2008} extend these results to continuous action spaces and single-trajectory data.} The error is driven by three terms, the geometric decay $\gamma^K$ (initialization bias), the approximation error $(1-\gamma)^{-2} \varepsilon_{\mathrm{approx}}$ (bias from function class), and the estimation error $O(1/\sqrt{N})$ (variance from finite samples). When $\mathcal{F}$ contains $V^*$ exactly, $\varepsilon_{\mathrm{approx}} = 0$ and the bound recovers exact convergence as $K \to \infty$. The $(1-\gamma)^{-2}$ amplification, one factor of $(1-\gamma)^{-1}$ more than in tabular Q-learning, reflects error accumulation across approximate DP steps: each regression step introduces bias, and this bias compounds over $K$ iterations.\footnote{The simulation in Section~\ref{sec:lqc_fvi_fqi} exercises only the bias / projection-error term of the Munos--Szepesvari bound; the variance / concentrability term is not stressed because we use a full $301 \times 201$ deterministic grid with the true transition kernel rather than a Monte Carlo sample, so there is no $N$ to vary and the $O(1/\sqrt{N})$ contribution is absent.}

Both algorithms solve projected Bellman equations. When $V^* \in \mathrm{span}(\Phi)$ exactly, the projection step is exact and FVI reduces to iterating a $\gamma$-contraction in coefficient space: from $V_0 = V^*$ it would terminate after a single projected iteration via the normal equations \eqref{eq:fvi_normal}, but from $V_0 = 0$ (the initial condition used in Section~\ref{sec:lqc_fvi_fqi}) the contraction drives $\theta_V \to \theta_V^*$ geometrically, so the algorithm reaches the tolerance $\|\theta_{k+1} - \theta_k\|_\infty < 10^{-9}$ in nine fitted iterations. For FQI, the per-action Q-functions satisfy $Q^*(s, a^*(s)) = V^*(s)$ at the optimal action $a^*(s)$, so when $Q^*(\cdot, a)$ is also representable in $\mathrm{span}(\Phi)$ for each $a$, FQI recovers consistent value estimates $\phi(s)^\top \theta_{a^*(s)}^* = \phi(s)^\top \theta_V^*$ at convergence. Whether FQI succeeds therefore depends on the geometry of the problem. When $Q^*(\cdot, a) \notin \mathrm{span}(\Phi)$, as on the Brock--Mirman economy, where the per-action Q-function requires fractional-power terms $k^{-n\alpha}$ outside the log-polynomial span, FQI stalls at error 1.65 while FVI converges to 0.001; when $Q^*(x,u)$ is exactly quadratic in $(x,u)$, as on the linear-quadratic control problem (Section~\ref{sec:lqc_fvi_fqi}), both FVI and FQI converge to near-zero error. Section~\ref{sec:bm_fvi_fqi} shows that replacing the linear basis with a nonlinear parametric model that matches the log-Cobb--Douglas structure of $Q^*$ restores FQI convergence on the Brock--Mirman economy.

\subsubsection{Rollout, Lookahead, and AlphaZero}

Two constructions bridge the Newton interpretation of Section~\ref{sec:newton_connection} to practical algorithms. Given a base policy $\mu$ with value function $V^\mu$, the \emph{rollout} policy selects
\begin{equation}
\mu_R(s) = \argmax_{a \in \mathcal{A}} \left\{ r(s,a) + \gamma \sum_{s'} P(s'|s,a)\, V^\mu(s') \right\}.
\end{equation}
This is one step of policy iteration starting from $\mu$: the Policy Improvement Theorem (Theorem~\ref{thm:policy_improvement}) guarantees $V^{\mu_R}(s) \geq V^\mu(s)$ for all $s$, with strict inequality unless $\mu$ is already optimal \citep{bertsekas2021lessons}.\footnote{Bertsekas uses cost-minimization notation throughout his work, writing $\min$ where standard RL uses $\max$ and defining value as accumulated cost rather than reward. I translate to the reward-maximization convention used elsewhere in this chapter; the mathematics are equivalent with reversed inequalities.} Given an arbitrary approximate value function $\tilde{V}$ (not necessarily the value of any policy), the \emph{one-step lookahead} policy selects
\begin{equation}
\tilde{\pi}(s) = \argmax_{a \in \mathcal{A}} \left\{ r(s,a) + \gamma \sum_{s'} P(s'|s,a)\, \tilde{V}(s') \right\}.
\end{equation}
When $\tilde{V} = V^\mu$, lookahead and rollout coincide. The distinction matters because rollout inherits the monotone improvement guarantee (it starts from the value of a policy), while lookahead from an arbitrary $\tilde{V}$ has no such monotonicity. The Newton interpretation from Section~\ref{sec:newton_connection} explains why lookahead nevertheless helps: both constructions solve the linearized Bellman equation at the current iterate.\footnote{Rollout requires a simulator (generative model) that can be queried from arbitrary states, a stronger assumption than the trajectory-based access of Q-learning and SARSA.}

An $\ell$-step lookahead extends this by applying $(\ell - 1)$ steps of value iteration before the final greedy selection. The first $(\ell - 1)$ steps are ordinary Bellman contractions, each shrinking the approximation error by a factor of $\gamma$. Only the final step, the greedy policy improvement, constitutes the Newton step.\footnote{\citet{bertsekas2021lessons} states this explicitly: ``whatever follows the first step of the lookahead is preparation for the Newton step.'' The preceding value iteration steps have linear convergence at rate $\gamma$; only the terminal improvement step has superlinear character.} The resulting error bound is
\begin{equation}
\|V^{\tilde{\pi}} - V^*\|_\infty \leq \gamma^\ell \, \|\tilde{V} - V^*\|_\infty,
\end{equation}
where the $\gamma^\ell$ factor reflects $\ell$ total contractions \citep[Prop.~2.3.1]{bertsekas2022}. Deep lookahead compensates for poor approximation through repeated contraction, not through repeated Newton steps.

Recall the AlphaGo Zero system from Section~\ref{subsubsec:alphago_zero}, where a neural network $f_\theta(s) = (\mathbf{p}, v)$ outputs a prior policy $\mathbf{p}$ and a value estimate $v$ for any board position $s$. During play, the network does not act alone: Monte Carlo Tree Search runs simulated games from the current position, using $v$ to evaluate leaf nodes and $\mathbf{p}$ to guide which branches to explore.\footnote{AlphaGo Zero uses 1,600 MCTS simulations per move for Go; the generalized AlphaZero algorithm uses 800 simulations per move across chess, shogi, and Go \citep[Table~S3]{Silver2018}.} The network provides $\tilde{V}$; MCTS applies multi-step lookahead through selective tree expansion. Table~\ref{tab:pi_alphazero} makes the correspondence explicit.

\begin{table}[h]
\centering
\small
\begin{tabular}{lll}
\hline
Step & Policy Iteration & AlphaZero \\
\hline
Initialize & Arbitrary $\pi_0$ & Random network $f_\theta$ \\
Evaluate & Solve $V = T^{\pi_k} V$ exactly & Network value head $v \approx V^{\pi_k}$ \\
Improve & $\pi_{k+1} = \argmax_a \{r + \gamma P V^{\pi_k}\}$ & MCTS simulations from $v$ \\
Iterate & Repeat until $\pi$ is stationary & Retrain $f_\theta$ on self-play outcomes \\
\hline
\end{tabular}
\caption{Policy iteration and AlphaZero follow the same evaluate-improve loop. The network provides approximate policy evaluation; MCTS provides approximate policy improvement via selective tree search.}
\label{tab:pi_alphazero}
\end{table}

The gap between network-only play and network-plus-search is the contraction factor $\gamma^H$, where $H$ is the effective search depth. The network provides a rough starting point $\tilde{V}$; MCTS applies the Bellman operator through deep lookahead, shrinking the approximation error by $\gamma$ per level of search.\footnote{MCTS adds UCB exploration and selective tree expansion beyond the literal lookahead framework. The Newton interpretation explains why lookahead helps at all, namely, the final greedy selection over the search tree is a policy improvement step. It does not explain the specific mechanisms (upper confidence bounds, progressive widening) that make MCTS computationally efficient.}

\subsubsection{Simulation Study: Credit Assignment in a Corridor}
\label{sec:sim_td_lambda_corridor}

A 20-state deterministic corridor ($s \in \{0, \ldots, 19\}$, action: move right, reward $+1$ on the transition into the terminal state $s = 19$, $\gamma = 0.99$) provides a clean setting to study the credit-assignment mechanism. The true value function is $V^*(s) = \gamma^{18-s}$ for $s \le 18$ (and $V^*(19) = 0$). TD($\lambda$) performs policy evaluation for four values of $\lambda$ across 20 seeds and 200 episodes.

\begin{figure}[h]
\centering
\includegraphics[width=0.7\textwidth]{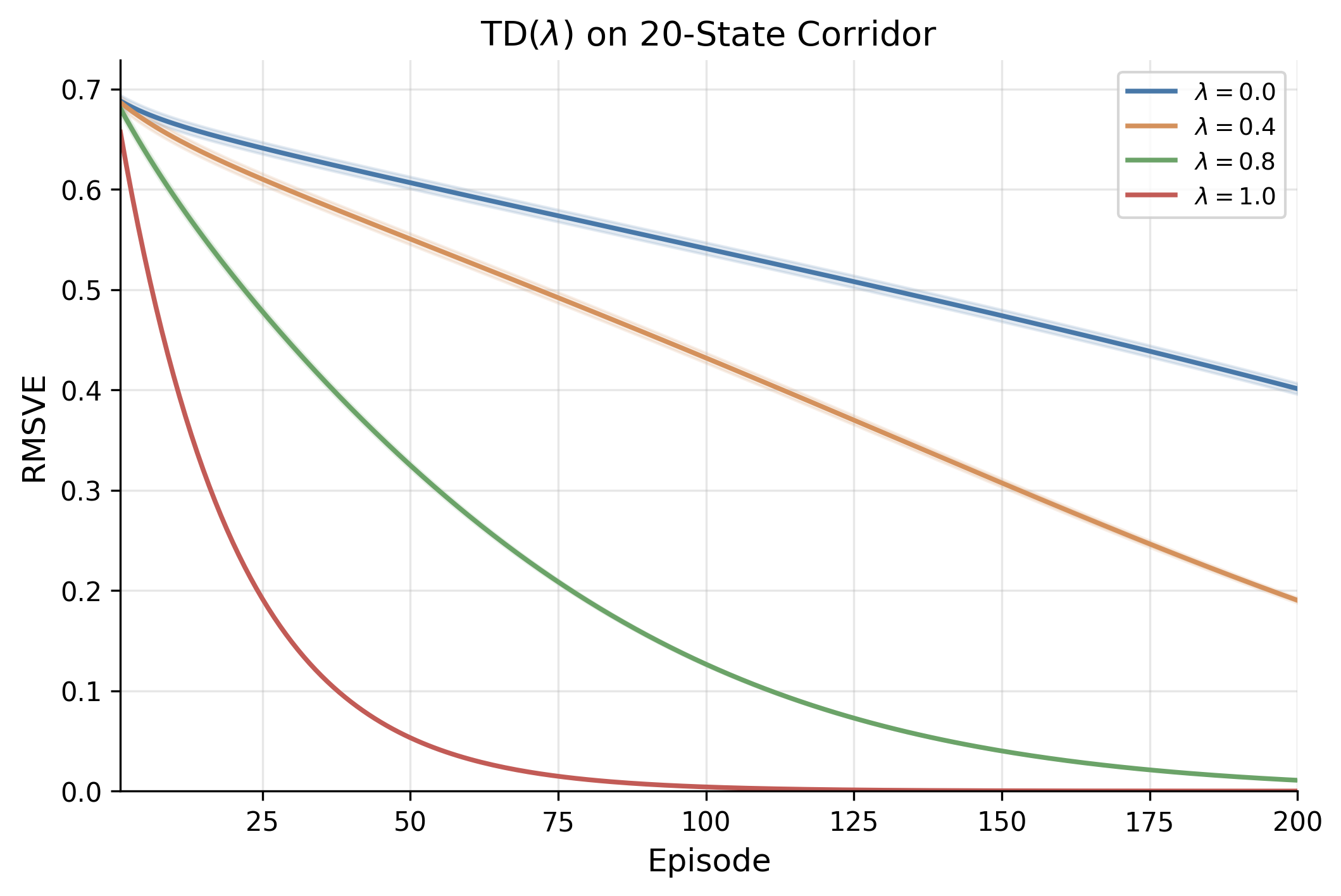}
\caption{RMSVE vs.\ episodes for TD($\lambda$) on the 20-state corridor. Shaded regions show $\pm 1$ SE over 20 seeds.}
\label{fig:td_lambda_corridor}
\end{figure}

Table~\ref{tab:td_lambda_corridor} and Figure~\ref{fig:td_lambda_corridor} show that higher $\lambda$ propagates the sparse terminal reward signal backward through the corridor faster: TD($\lambda = 1$) crosses RMSVE $< 0.05$ at episode 52 and converges to $V^*$ to numerical precision, while TD(0) does not cross the threshold within 200 episodes because the reward must diffuse back through one-step bootstrapping alone.

\begin{table}[h]
\centering
\caption{TD($\lambda$) on 20-state corridor. Mean $\pm$ SE over 20 seeds, 200 episodes, $\gamma = 0.99$, $\alpha = 0.05$.}
\label{tab:td_lambda_corridor}
\begin{tabular}{ccc}
\hline
$\lambda$ & Final RMSVE & Episodes to RMSVE $< 0.05$ \\
\hline
0.0 & 0.4012 $\pm$ 0.0056 & $> 200$ \\
0.4 & 0.1902 $\pm$ 0.0028 & $> 200$ \\
0.8 & 0.0108 $\pm$ 0.0002 & 141 $\pm$ 1 \\
1.0 & 0.0000 $\pm$ 0.0000 & 52 $\pm$ 0 \\
\hline
\end{tabular}
\end{table}

\subsubsection{Simulation Study: Fitted Methods on Linear-Quadratic Control}
\label{sec:lqc_fvi_fqi}

Linear-quadratic control (LQC) provides exact analytical benchmarks for the fitted methods because both $V^*$ and $Q^*$ are quadratic polynomials. The model has scalar state $x \in [-4, 4]$ and action $u \in [-2, 2]$, deterministic dynamics $x' = ax + bu$ with $a = 0.5$, $b = 1.0$, reward $r(x,u) = -(x^2 + u^2)$, and discount $\gamma = 0.95$. The parameters ensure that $x' = 0.5x + u \in [-4, 4]$ whenever $(x, u) \in [-4,4] \times [-2,2]$, so the grid is strictly invariant. The optimal value function satisfies $V^*(x) = -Px^2$, where $P$ solves $\gamma b^2 P^2 + P(1 - \gamma(a^2 + b^2)) - 1 = 0$, yielding $P \approx 1.129$. The optimal Q-function is $Q^*(x,u) = -(1+\gamma Pa^2)x^2 - 2\gamma Pab\,xu - (1+\gamma Pb^2)u^2 \approx -1.268x^2 - 1.073xu - 2.073u^2$, which lies exactly in $\mathrm{span}\{x^2, xu, u^2\} \subset \mathrm{span}\{x, x^2, u, u^2, xu\}$. FVI uses state features $\phi_V(x) = [x, x^2]^\top \in \mathbb{R}^2$ with no intercept because $V^*(0) = 0$. FQI uses state-action features $\phi_Q(x,u) = [x, x^2, u, u^2, xu]^\top \in \mathbb{R}^5$. Both use a 301-point state grid and 201-point action grid. DQN uses a two-layer network of 64 units per layer with ReLU activations, an experience replay buffer of 50,000 transitions, a hard target-network update every 500 gradient steps, and rewards scaled by a factor of $1/20$ to stabilize training.

\begin{figure}[h]
\centering
\includegraphics[width=\textwidth]{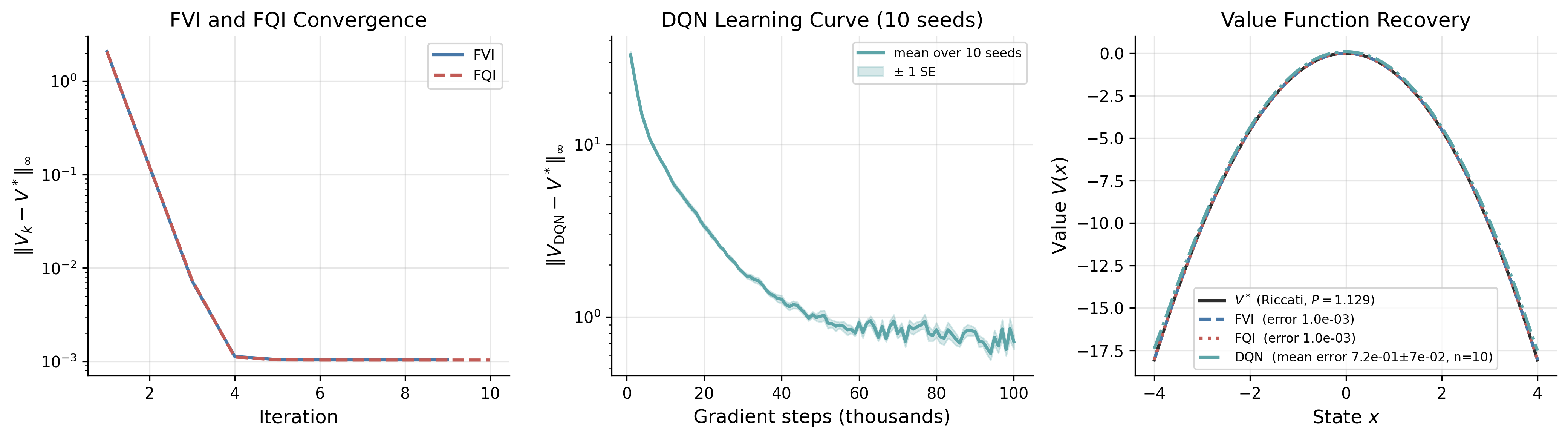}
\caption{LQC convergence of FVI and FQI (left), DQN learning curve (middle, mean and $\pm 1$ standard-error band over ten seeds), and value function recovery for all three methods (right; the DQN trace is the cross-seed mean). FVI and FQI reduce $\|V_k - V^*\|_\infty$ to near-zero in under 10 iterations, exploiting the known polynomial structure of $Q^*$. DQN's learning curve over 100,000 gradient steps is shown with no feature basis specified.}
\label{fig:lqc_fvi_fqi}
\end{figure}

Table~\ref{tab:lqc_fvi_fqi} and Figure~\ref{fig:lqc_fvi_fqi} report recovery of the analytical solution. FVI and FQI converge in under 10 iterations with errors below $10^{-3}$ because the FVI feature space contains $V^*$ and the FQI feature space contains $Q^*$. Across ten independent seeds, DQN reaches a mean error of order $10^{-1}$ against $V^*$ after 100,000 gradient steps without prior knowledge of the feature basis; Table~\ref{tab:lqc_fvi_fqi} reports the exact mean and standard error. This is not a direct performance contest. FVI and FQI use the full deterministic $301 \times 201$ grid, the true transition map, and feature bases containing their respective analytical solutions, while DQN learns a generic two-layer network from sampled transitions without a prior on the polynomial structure. The Brock--Mirman comparison isolates the role of the function class. FQI succeeds here because $Q^*(x,u) \in \mathrm{span}(\Phi_Q)$, while it fails on Brock--Mirman because $Q^*(\cdot,a) \notin \mathrm{span}(\Phi)$.

\begin{table}[h]
\centering
\small
\begin{tabular}{lrrrr}
\hline
Method & Iterations & Error vs $V^*$ & $P$ (recovered) & Key coefficient \\
\hline
Exact VI (discrete) & 25 & 1.12e-03 & --- & --- \\
FVI & 9 & 3.23e-04 & 1.1294 & $\hat\theta_V^{x^2} = -1.1294$ \\
FQI & 10 & 9.37e-05 & 1.2682 & $\hat\theta_Q^{xu} = -1.0729$ \\
DQN ($2 \times 64$ ReLU, 10 seeds) & 100000 & 7.16e-01 $\pm$ 6.80e-02 & --- & --- \\
Analytical ($V^* = -Px^2$) & --- & 0 & 1.1294 & $c_{xu} = -1.0729$ \\
\hline
\end{tabular}

\caption{Fitted weights and convergence metrics for FVI, FQI, and DQN on linear-quadratic control. The FVI $x^2$ coefficient recovers the Riccati solution $P \approx 1.129$. The FQI quadratic coefficients match the analytical $Q^*$ to four decimal places. FVI and FQI achieve max error below $10^{-3}$ against the analytical $V^*$; the DQN row reports mean $\pm$ standard error across ten independent seeds after 100,000 gradient steps with no feature basis specified. FVI and FQI exercise the full deterministic grid with the true transition kernel, so seed-to-seed variation is exactly zero and no standard error is reported for them.}
\label{tab:lqc_fvi_fqi}
\end{table}

\subsubsection{Simulation Study: Basis Representability on the Brock--Mirman Economy}
\label{sec:bm_fvi_fqi}

The Brock--Mirman stochastic growth model \citep{brockmirman1972} provides the negative case. The economy has $|\mathcal{S}| = 100$ states ($N_K = 50$ capital grid points, $N_Z = 2$ productivity levels) and $|\mathcal{A}| = 50$ actions. We use the same log-polynomial basis $\phi(k,z) = [1,\, \log k,\, k/\bar{k},\, (k/\bar{k})^2,\, (k/\bar{k})^3] \otimes [\mathbbm{1}_{z = z_\ell},\, \mathbbm{1}_{z = z_h}]$ from the theory discussion above, which contains $V^*$ up to residual $\|\Pi_\Phi V^* - V^*\|_\infty = 0.0002$. To test whether the failure is inherent to FQI or to the basis, we add two methods that use the structurally correct per-action feature $\log(z k^\alpha - k')$, the log-consumption implied by the Cobb--Douglas technology. Oracle-FQI treats $\alpha = 0.36$ as known and runs standard OLS per action with three parameters $[\mathbbm{1}_{z = z_\ell},\, \mathbbm{1}_{z = z_h},\, \log(z k^\alpha - k')]$. NLLS-FQI estimates $\alpha$ jointly via concentrated least squares: for each candidate $\alpha$, it solves conditional OLS for intercepts and slope, then optimizes $\alpha$ to minimize total residual sum of squares, initialized at the deliberately wrong value $\alpha_0 = 0.5$.\footnote{Observations where $z k^\alpha - k' \leq 0$ (infeasible consumption) contribute a penalty equal to the mean squared target, preventing the optimizer from improving RSS by shrinking the feasible set.}

\begin{figure}[h]
\centering
\includegraphics[width=\textwidth]{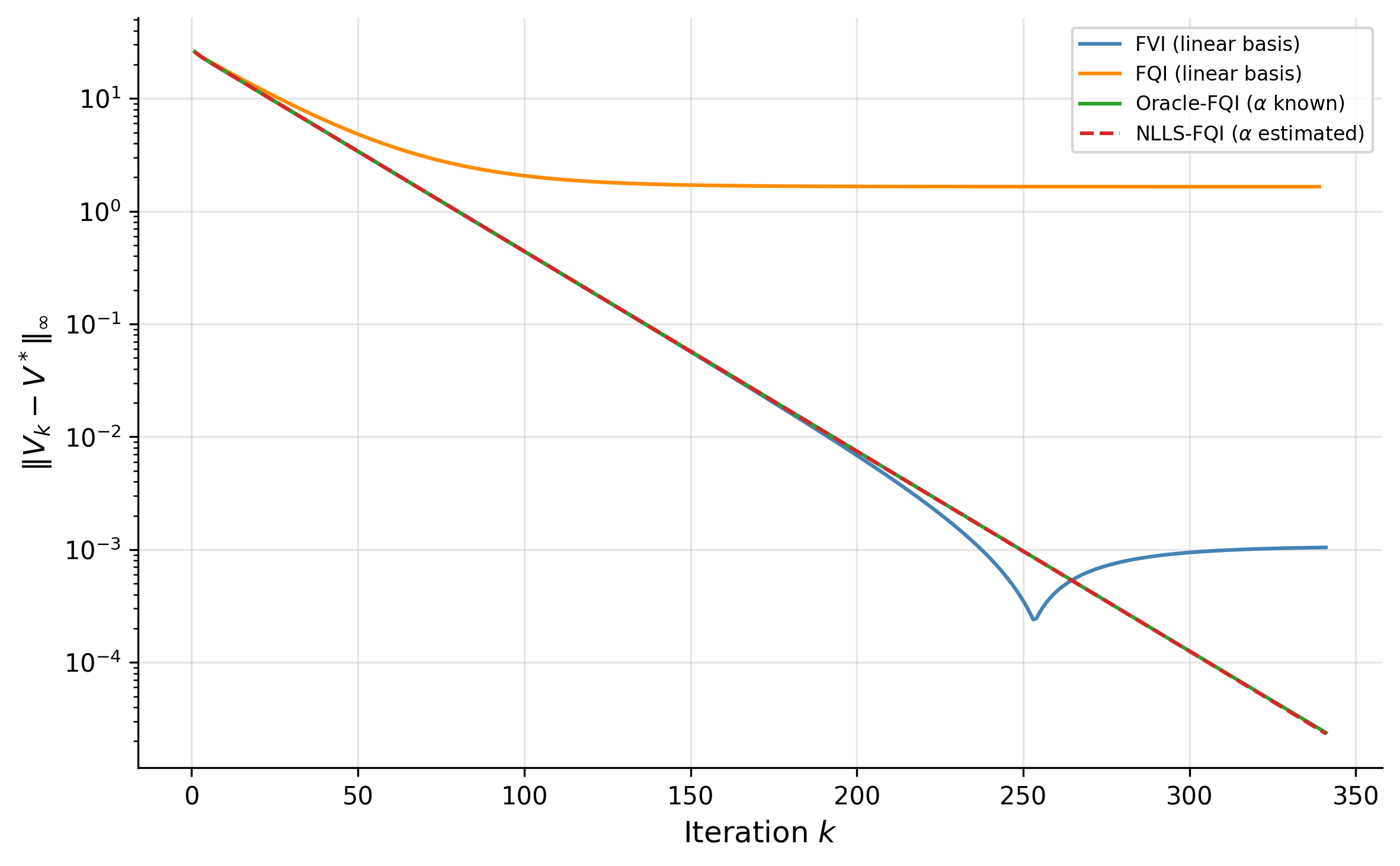}
\caption{Convergence of $\|V_k - V^*\|_\infty$ for FVI, linear FQI, Oracle-FQI, and NLLS-FQI on the Brock--Mirman economy.}
\label{fig:bm_fvi_fqi}
\end{figure}

Table~\ref{tab:bm_fvi_fqi} and Figure~\ref{fig:bm_fvi_fqi} confirm that the failure is one of basis representability rather than algorithmic failure. FVI converges near the projection floor of the log-polynomial basis; linear FQI stalls at error 1.65, confirming $Q^*(\cdot, a) \notin \mathrm{span}(\Phi)$. Oracle-FQI and NLLS-FQI, using the structurally correct log-consumption feature, match exact VI with error below $10^{-4}$. NLLS-FQI recovers $\hat{\alpha} = 0.3600$ in a single iteration, demonstrating that the same FQI algorithm succeeds when the function class contains $Q^*$.

\begin{table}[h]
\centering
\small
\begin{tabular}{lrrrr}
\hline
Method & $\|V - V^*\|_\infty$ & $\|V - V^*\|_\text{rms}$ & Policy agreement (\%) & Iterations \\
\hline
Exact VI & 0.0000 & 0.0000 & 98.0 & --- \\
FVI (linear) & 0.0010 & 0.0008 & 99.0 & 341 \\
FQI (linear) & 1.6521 & 1.4805 & 4.0 & 339 \\
Oracle-FQI & 0.0000 & 0.0000 & 98.0 & 341 \\
NLLS-FQI ($\hat{\alpha} = 0.3600$) & 0.0000 & 0.0000 & 98.0 & 341 \\
\hline
\end{tabular}
\caption{Convergence metrics for five methods on the Brock--Mirman economy ($N_K = 50$, $N_Z = 2$, $\gamma = 0.96$). Policy agreement is measured against the closed-form optimal policy $k' = \alpha \beta z k^\alpha$.}
\label{tab:bm_fvi_fqi}
\end{table}

\subsubsection{Simulation Study: Bus Engine Replacement as DP versus DQN}
\label{sec:bus_engine}

The fleet replacement benchmark extends the single-engine decision in \citet{Rust1987} to compare exact dynamic programming with approximate value learning. A fleet manager decides each month whether to replace engines based on accumulated mileage. Replacement incurs a fixed cost but resets mileage to zero; continued operation incurs maintenance costs increasing in mileage.

The simulation extends the single-engine problem to a fleet of $N$ engines with a capacity constraint on replacements per period. The state $s = (m_1, \ldots, m_N)$ records discretized mileage for each engine. Actions are subsets of engines to replace, subject to the capacity constraint. The per-period cost is $c(s, a) = \alpha \sum_{i} m_i + \beta |a|$, where $\alpha$ is the operating cost per unit mileage and $\beta$ is the replacement cost.\footnote{The mileage-dependent operating cost $\alpha \sum_i m_i$ follows Rust's original specification $c(x, \theta_1) = \theta_{11} x$, creating a non-trivial threshold replacement policy. The fleet extension uses deterministic mileage increments to isolate the combinatorial scaling challenge that arises from the joint state of multiple engines.} Mileage evolves deterministically. Replaced engines reset to $m = 0$, while the others increment by one bin.\footnote{With $M = 6$ mileage bins, the state space is $6^N$: 1,296 states at $N = 4$, 7,776 at $N = 5$, 46,656 at $N = 6$.} The comparison uses $\gamma = 0.95$ rather than Rust's monthly $\beta = 0.9999$. This choice shortens the effective discount horizon from roughly $10{,}000$ periods to roughly $20$ and stabilises DQN training without changing the qualitative DQN-vs-DP comparison.\footnote{Rust's $\beta = 0.9999$ produces Q-values of order $10^4$, which destabilise gradient updates on a small replay buffer. The substantive scaling story (combinatorial blow-up of $|\mathcal{S}|$ with $N$) is invariant to the discount choice.}

\begin{figure}[htbp]
\centering
\includegraphics[width=\textwidth]{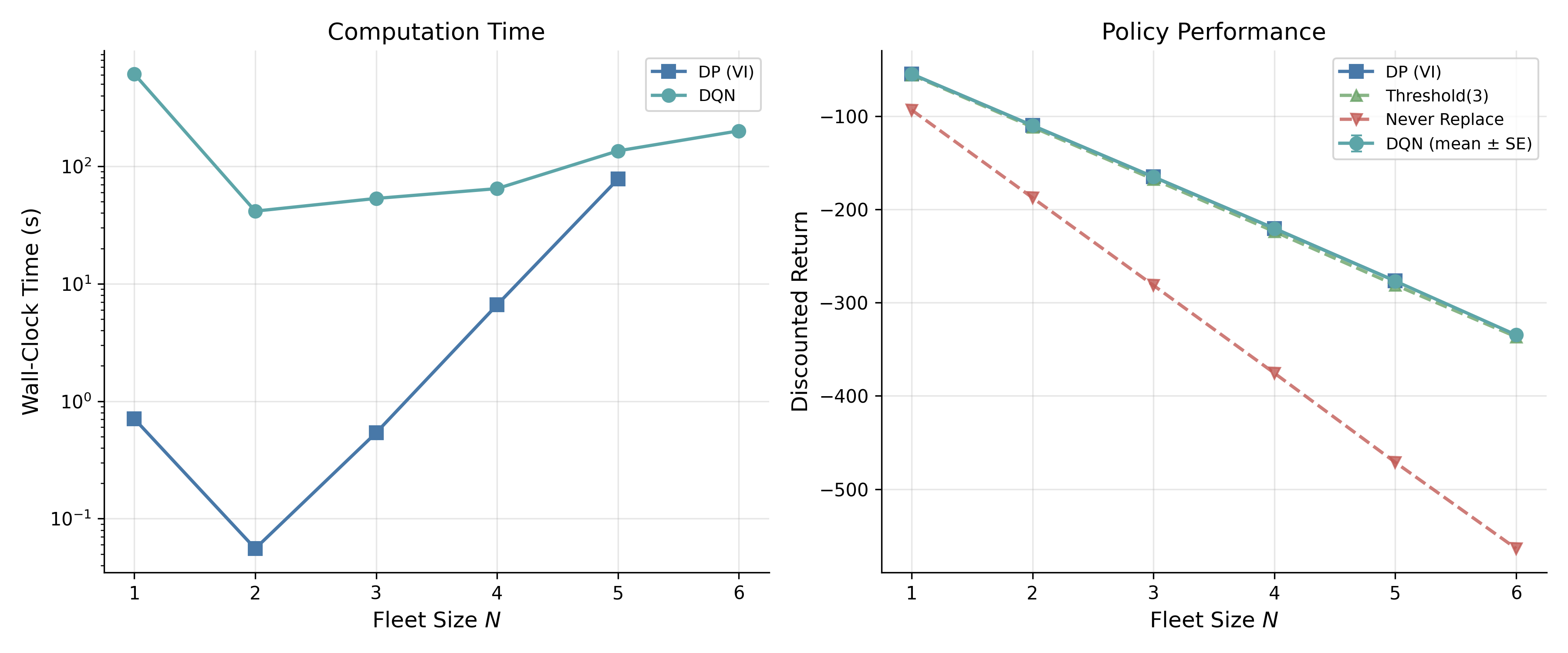}
\caption{Bus engine replacement benchmark. Left: computation time vs.\ fleet size (log scale). Right: discounted return vs.\ fleet size for DP, DQN, and heuristic baselines; DQN error bars are mean $\pm$ standard error over ten seeds. At $N = 6$ (46,656 states), DP is omitted because plain-Python value iteration becomes wall-clock prohibitive at this scale.}
\label{fig:bus_engine_scaling}
\end{figure}

Figure~\ref{fig:bus_engine_scaling} compares dynamic programming, DQN, and heuristic baselines across fleet sizes. All policies are evaluated on the same pre-sampled set of initial states so that the comparison is paired across methods.

\begin{table}[htbp]
\centering
\caption{Bus engine replacement benchmark across fleet sizes $N$. DP time is wall-clock for value iteration; returns are discounted returns on the shared evaluation states, with DQN reported as mean $\pm$ standard error over ten seeds.}
\label{tab:bus_engine}
\begin{tabular}{rrrrrrrr}
\toprule
$N$ & $|\mathcal{S}|$ & DP Time & DP Return & DQN Return & Thresh(3) & Never Replace \\
\midrule
1 & 6 & $0.71$s & $-54.8$ & $-54.8 \pm 0.00$ & $-55.6$ & $-93.3$ \\
2 & 36 & $0.06$s & $-110.0$ & $-110.0 \pm 0.00$ & $-111.8$ & $-187.5$ \\
3 & 216 & $0.54$s & $-165.1$ & $-165.1 \pm 0.01$ & $-167.7$ & $-281.4$ \\
4 & 1,296 & $6.61$s & $-220.4$ & $-220.4 \pm 0.00$ & $-223.8$ & $-375.6$ \\
5 & 7,776 & $77.82$s & $-276.7$ & $-276.9 \pm 0.02$ & $-281.1$ & $-471.3$ \\
6 & 46,656 & --- & --- & $-334.7 \pm 1.86$ & $-336.7$ & $-563.8$ \\
\bottomrule
\end{tabular}
\end{table}

Table~\ref{tab:bus_engine} reports the results. For $N = 1$ through $5$ where both methods are computed, DQN matches DP within 1\% of the optimal discounted return. At $N = 6$ (46,656 states), the experiment omits DP rather than running it. Plain-Python value iteration is not numerically infeasible at this size, but a single backup sweep costs $|\mathcal{S}|^2 |\mathcal{A}|$ in Python-loop time and exceeds the wall-clock budget by an order of magnitude.\footnote{The DP wall-clock curve in the left panel reflects this Python-loop implementation; a vectorised backup using \texttt{numpy} broadcasts would shrink the absolute timings by one to two orders of magnitude. The scaling exponent (slope of the log-time curve in $N$) is the right asymptotic shape, but the absolute timings are implementation-dependent.} The threshold heuristic replaces engines above a mileage cutoff, so it cannot account for capacity constraints or the joint state of multiple engines. The never-replace heuristic accumulates mileage costs and produces lower returns. These two baselines establish that the cost structure creates a non-trivial replacement decision.

\FloatBarrier
\subsubsection{Engine Replacement MDP: Sampling the Same Operator}
\label{engine:value_learning}

\begin{figure}[h]
\centering
\includegraphics[width=0.7\textwidth]{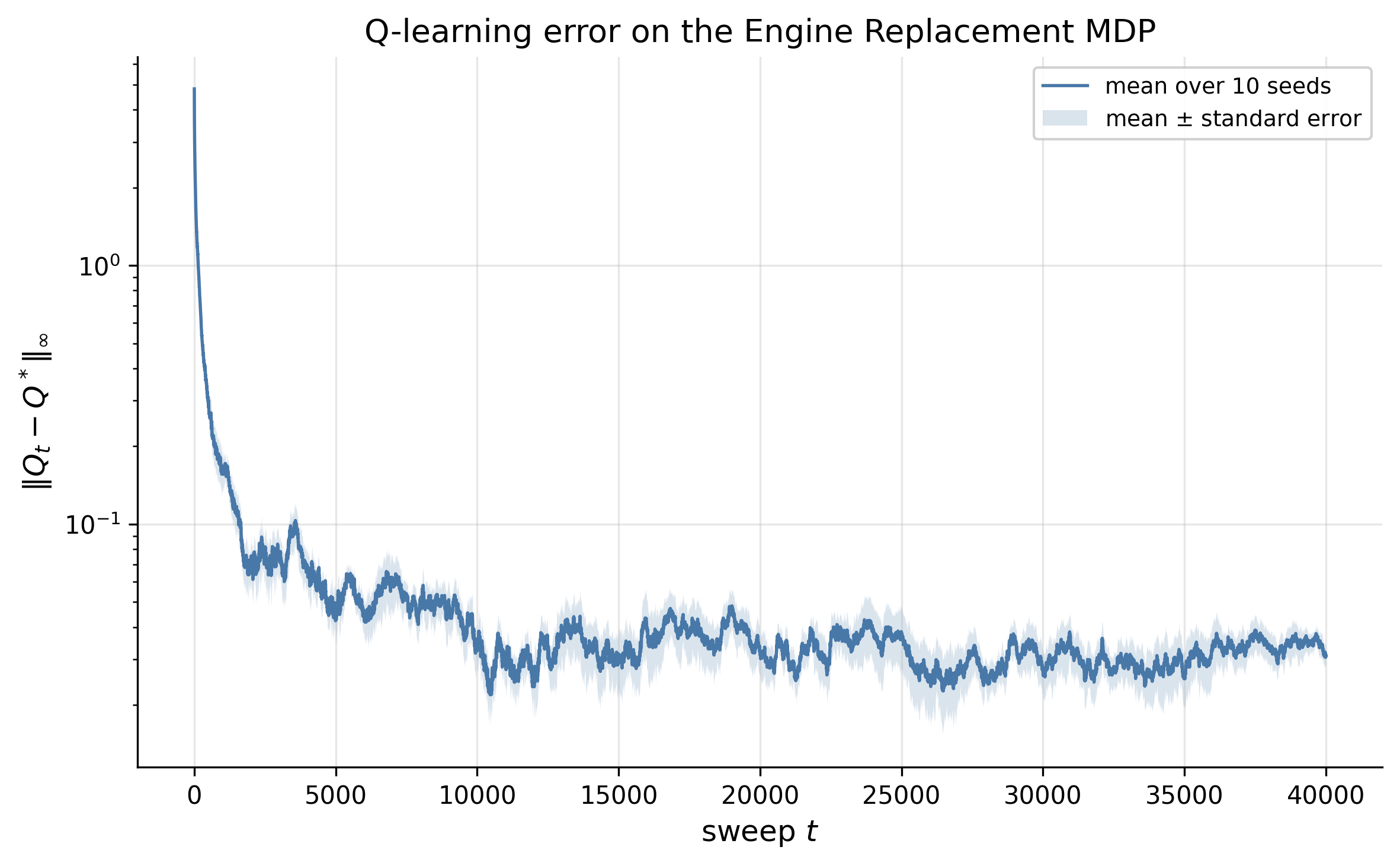}
\caption{Tabular Q-learning on the Engine Replacement MDP over ten fixed seeds, with step size $\alpha_n=n^{-0.6}$ per state-action pair and $40{,}000$ synchronous sweeps. The line is the mean sup-norm error and the band is one standard error.}
\label{fig:engine_value_learning}
\end{figure}

In the Engine Replacement MDP of Section~\ref{engine:model}, the Q-factor operator $F$ from Theorem~\ref{thm:q_factor_contraction} and the Q-learning update~\eqref{eq:qlearning} differ only in whether they compute or sample the conditional expectation. Five exact Q-iteration backups from $Q_0=0$ give the intermediate table $Q_5$. Its next backup at the engine's stochastic state-action pair is $(FQ_5)(\text{low},\text{keep})=2.8268$. Ten Monte Carlo replications with one million draws each give $2.8268$ with standard error $0.0001$ and mean absolute error $0.0004$. Tabular Q-learning over the same ten fixed seeds reduces the mean sup-norm error from $4.8103$ after the first synchronous sweep to $0.0312$ after $40{,}000$ sweeps. Figure~\ref{fig:engine_value_learning} and Table~\ref{tab:engine_value_learning} report the trace and computed quantities.

\begin{table}[h]
\centering
\caption{Value-learning quantities computed on the Engine Replacement MDP. The Bellman-operator identity and Monte Carlo estimate use the intermediate table $Q_{5}$ at $(s,a) = (\text{low}, \text{keep})$, the engine's one stochastic transition. The Q-learning trace uses 10 fixed seeds and 40,000 synchronous sweeps with step size $\alpha_n = n^{-0.6}$. Projected TD($\lambda$) moduli use $\phi=(1,-1)$ and the stationary distribution of $P^{\pi^\star}$. Concentrability uses the discounted occupancy from the low-mileage start.}
\label{tab:engine_value_learning}
\begin{tabular}{p{0.68\textwidth}r}
\hline
quantity & value \\
\hline
exact target $(FQ_5)(\text{low},\text{keep})$ & 2.8268 \\
Monte Carlo mean, $N = 1,000,000$ & 2.8268 $\pm$ 0.0001 \\
mean absolute sampling error, $N = 1,000,000$ & 0.0004 \\
Q-learning final error, mean $\pm$ SE & 0.0312 $\pm$ 0.0027 \\
projected TD($\lambda=0$) norm, bound 0.9000 & 0.5196 \\
projected TD($\lambda=0.5$) norm, bound 0.8182 & 0.3231 \\
projected TD($\lambda=0.9$) norm, bound 0.4737 & 0.1608 \\
projected TD($\lambda=1$) norm, bound 0.0000 & 0.0000 \\
state-action concentrability $C_\infty$ & 7.3180 \\
\hline
\end{tabular}
\end{table}

The projected TD($\lambda$) calculation uses the one-dimensional feature $\phi=(1,-1)$ and the $d^{\pi^\star}$-orthogonal projection from Theorem~\ref{thm:proj_bellman}. Its realized operator norm lies below $\kappa_\lambda=\gamma(1-\lambda)/(1-\lambda\gamma)$ at every reported value of $\lambda$. The state-action concentrability calculation uses the low-mileage start, the optimal target policy, and a logging policy that keeps with probability $0.1$ in each state. The largest discounted occupancy ratio is $C_\infty=7.3180$.

The computed backups verify that sampling targets the same fixed-point operator $F$, while the projected calculation remains within the bound of Theorem~\ref{thm:proj_bellman}.

\FloatBarrier
\subsection{The Central Challenge: The Deadly Triad}
\label{sec:deadly_triad}

State spaces are too large for lookup tables (Go has $10^{170}$ states; most economic models have continuous state variables). Practitioners must combine three ingredients: \textit{function approximation} (to handle large state spaces), bootstrapping (to learn from single transitions rather than complete episodes), and off-policy learning (to learn about the optimal policy while exploring, or to reuse old data). Each is desirable in isolation. Their interaction, known as the \emph{deadly triad} \citep[Ch.~11]{sutton2018}, is the central open problem in reinforcement learning theory.

Off-policy learning is preferred for three reasons. First, sample efficiency: transitions collected under any behavioral policy can be reused to evaluate or improve a different target policy, amortizing the cost of data collection. Discarding data because it was generated by a superseded policy is wasteful. Second, exploration and exploitation separate cleanly: the agent can follow an exploratory policy (e.g., $\varepsilon$-greedy) to ensure adequate state-space coverage while simultaneously learning the optimal deterministic policy. On-policy methods such as SARSA entangle the two, learning the value of the exploratory policy rather than the optimal one. Third, off-policy evaluation answers ``what would have happened under policy $\pi$?'' from data generated by policy $\mu$, the counterfactual question central to policy comparison.

\subsubsection{The Projected Bellman Operator}

With function approximation $V(s) \approx \phi(s)^\top \theta$, the parameter vector $\theta$ is shared across states. Updating $\theta$ to improve the value estimate at one state simultaneously changes the estimate at every other state. The algorithm can no longer apply the Bellman operator $T^\pi$ to each state independently; instead, it applies $T^\pi$ to compute a target, then \emph{projects} the result back onto the function space (the span of the features $\phi$).\footnote{Appendix~\ref{prelim:span_rank} defines the feature span, and Appendix~\ref{prelim:least_squares} connects regression to projection onto that span.} The composed operator is $\Pi T^\pi$, the \textit{projected Bellman operator}, where $\Pi$ denotes this projection \citep{tsitsiklis1997}. Convergence of the approximate iteration $\theta_{k+1} = \Pi T^\pi \theta_k$ requires $\Pi T^\pi$ to be a contraction.

In the on-policy setting, $\Pi$ minimizes squared error weighted by $d^\pi$, the stationary distribution of the policy being evaluated, so $\Pi V = \arg\min_{\hat{V} \in \text{span}(\Phi)} \|V - \hat{V}\|_{d^\pi}$, where $\|V\|_{d^\pi}^2 = \sum_s d^\pi(s) V(s)^2$. The Bellman operator $T^\pi$ is a $\gamma$-contraction in the same $d^\pi$-norm. Because both operators use the same norm, the projection $\Pi$ is \emph{orthogonal}, meaning the residual $V - \Pi V$ is perpendicular to the approximation subspace. The Pythagorean theorem then gives $\|\Pi V\|_{d^\pi}^2 + \|V - \Pi V\|_{d^\pi}^2 = \|V\|_{d^\pi}^2$, so $\|\Pi V\|_{d^\pi} \leq \|V\|_{d^\pi}$.\footnote{The same argument holds in $\mathbb{R}^n$. Projecting a vector onto a subspace never makes it longer. This is the geometric content of the Cauchy-Schwarz inequality. In the function-approximation setting, the ``vector'' is a value function, the ``subspace'' is the span of features, and ``length'' is the $d^\pi$-weighted $L^2$ norm.} The projection cannot expand distances (Theorem~\ref{thm:prelim_hilbert}). The composition therefore contracts:
\begin{equation}
\|\Pi T^\pi V_1 - \Pi T^\pi V_2\|_{d^\pi} \leq \underbrace{\|\Pi\|}_{\leq\, 1} \cdot \underbrace{\|T^\pi V_1 - T^\pi V_2\|_{d^\pi}}_{\leq\, \gamma \|V_1 - V_2\|_{d^\pi}} < \|V_1 - V_2\|_{d^\pi}.
\label{eq:on_policy_contraction}
\end{equation}
The same argument extends from TD(0) to TD($\lambda$) through the $\lambda$-averaged operator $T^{(\lambda)}$, whose $d^\pi$-contraction modulus is $\kappa_\lambda = \gamma(1-\lambda)/(1-\lambda\gamma) \leq \gamma$.

\begin{theorem}[On-policy TD convergence]
\label{thm:proj_bellman}
Sample states from the stationary distribution $d^\pi$ and let $\Pi$ be the $d^\pi$-orthogonal projection onto $\mathrm{span}(\Phi)$. For each $\lambda \in [0,1]$ the projected operator $\Pi T^{(\lambda)}$ is a contraction in the $d^\pi$-norm with modulus $\kappa_\lambda = \gamma(1-\lambda)/(1-\lambda\gamma)$. It has a unique fixed point $\Phi\theta^*$, to which TD($\lambda$) converges with probability one, and
\[
\|\Phi\theta^* - V^\pi\|_{d^\pi} \leq \frac{1-\lambda\gamma}{1-\gamma}\, \|\Pi V^\pi - V^\pi\|_{d^\pi}.
\]
\end{theorem}

The $\lambda = 0$ contraction is exactly~\eqref{eq:on_policy_contraction}. The bound equals the best achievable approximation error at $\lambda = 1$ and loosens as $\lambda$ falls. The general case and the error bound follow \citep[Theorem~1]{tsitsiklis1997}.

\begin{proof}
Two facts drive the proof. The projection never lengthens a vector, and the $\lambda$-averaged Bellman operator contracts in the $d^\pi$-norm. Composing them contracts, and one triangle inequality converts that contraction into the error bound. The projection $\Pi$ is orthogonal in the $d^\pi$-weighted inner product,\footnote{The $d^\pi$-weighted norm is $\|v\|_{d^\pi}^2 = \sum_s d^\pi(s)\, v(s)^2$, weighting each state by how often the policy visits it. Orthogonal projection onto a subspace returns the closest point in that subspace under this norm. Appendix~\ref{prelim:norms} places this weighted norm alongside the $L^p$ family and the supremum norm.} hence nonexpansive,
\[
\|\Pi V\|_{d^\pi}^2 + \|V - \Pi V\|_{d^\pi}^2 = \|V\|_{d^\pi}^2 \;\;\Longrightarrow\;\; \|\Pi V\|_{d^\pi} \leq \|V\|_{d^\pi},
\]
by the Pythagorean identity. The $\lambda$-averaged Bellman operator\footnote{$T^{(\lambda)} = (1-\lambda)\sum_{m\geq0}\lambda^m (T^\pi)^{m+1}$ is the geometric average of the $m$-step Bellman operators TD($\lambda$) implicitly targets. Setting $\lambda = 0$ recovers the one-step $T^\pi$ and $\lambda \to 1$ the full-return Monte Carlo operator. Its fixed point is $V^\pi$ for every $\lambda$.} $T^{(\lambda)} = (1-\lambda)\sum_{m\geq0}\lambda^m (T^\pi)^{m+1}$ has linear part $(1-\lambda)\sum_{m\geq0}\lambda^m (\gamma P^\pi)^{m+1}$, and because $d^\pi$ is the stationary distribution of $P^\pi$, the transition operator is itself nonexpansive in this norm.\footnote{With $d^\pi P^\pi = d^\pi$, Jensen's inequality gives $\|P^\pi v\|_{d^\pi}^2 = \sum_s d^\pi(s)\big(\sum_{s'} P^\pi(s'|s) v(s')\big)^2 \leq \sum_s d^\pi(s) \sum_{s'} P^\pi(s'|s) v(s')^2 = \sum_{s'} d^\pi(s') v(s')^2 = \|v\|_{d^\pi}^2$, the last equality by stationarity.} Hence
\begin{align*}
\|T^{(\lambda)} V_1 - T^{(\lambda)} V_2\|_{d^\pi}
  &\leq (1-\lambda)\sum_{m\geq0} \lambda^m \gamma^{m+1} \|V_1 - V_2\|_{d^\pi}
    && \text{($\|P^\pi v\|_{d^\pi} \leq \|v\|_{d^\pi}$, term by term)} \\
  &= \underbrace{\frac{\gamma(1-\lambda)}{1-\lambda\gamma}}_{=\,\kappa_\lambda}\, \|V_1 - V_2\|_{d^\pi}
    && \text{(sum the geometric series $\sum_m (\lambda\gamma)^m$).}
\end{align*}
Composing with the nonexpansive $\Pi$ makes $\Pi T^{(\lambda)}$ a $\kappa_\lambda$-contraction ($\kappa_\lambda \leq \gamma < 1$), so Banach's theorem gives the unique fixed point $\Phi\theta^*$, and TD($\lambda$), which realizes $\theta_{k+1} = \Pi T^{(\lambda)} \theta_k$ as stochastic approximation, converges to it almost surely by the argument of Theorem~\ref{thm:qlearning_convergence}. For the error bound, the true value $V^\pi$ is the fixed point of $T^{(\lambda)}$ for every $\lambda$ (each $T^{(\lambda)}$ is built from operators that fix $V^\pi$), so $\Pi T^{(\lambda)} V^\pi = \Pi V^\pi$, and
\begin{align*}
\|\Phi\theta^* - \Pi V^\pi\|_{d^\pi}
  &= \|\Pi T^{(\lambda)} \Phi\theta^* - \Pi T^{(\lambda)} V^\pi\|_{d^\pi} \\
  &\leq \kappa_\lambda \|\Phi\theta^* - V^\pi\|_{d^\pi}
    && \text{($\Pi T^{(\lambda)}$ is a $\kappa_\lambda$-contraction).}
\end{align*}
Combining with the triangle inequality $\|\Phi\theta^* - V^\pi\|_{d^\pi} \leq \|\Phi\theta^* - \Pi V^\pi\|_{d^\pi} + \|\Pi V^\pi - V^\pi\|_{d^\pi}$ gives $(1 - \kappa_\lambda)\|\Phi\theta^* - V^\pi\|_{d^\pi} \leq \|\Pi V^\pi - V^\pi\|_{d^\pi}$, and $1/(1-\kappa_\lambda) = (1-\lambda\gamma)/(1-\gamma)$ delivers the stated bound.
\end{proof}

\subsubsection{Why Off-Policy Learning Diverges}

In the off-policy setting, samples come from a behavior distribution $\mu \neq d^\pi$. The projection now minimizes error under $\mu$, but the Bellman operator $T^\pi$ still contracts in the $d^\pi$-norm. The two operators measure distance in different norms. The projection becomes \emph{oblique} in the $d^\pi$-norm rather than orthogonal. Unlike orthogonal projections, oblique projections can expand distances, with $\|\Pi_\mu\|_{d^\pi}$ exceeding 1 in the worst case. If $\|\Pi_\mu\|_{d^\pi} > 1/\gamma$, the expansion from projection overwhelms the $\gamma$-contraction from the Bellman operator, and the fixed-point iteration diverges.

This divergence differs from overfitting. Overfitting occurs when the approximator memorizes training data at the expense of generalization; collecting more data helps. Divergence means the parameter vector $\theta$ grows without bound, producing arbitrarily large value estimates that bear no relation to the true values. More data does not help; the algorithm itself is unstable. The distinction matters because the remedies are entirely different. Regularization and early stopping address overfitting, while the deadly triad requires structural changes to the algorithm.

\citet{Baird1995} constructed a six-state star MDP that makes this failure concrete. All rewards are zero, so the true value is $V^*(s) = 0$ for every state. A lookup table learns this immediately. The MDP has a star topology: states 1 through 5 each transition to state 6, and state 6 transitions to itself. Linear function approximation gives each state's value as a combination of a shared weight, common to all states, and a state-specific weight, so that adjusting one state's estimate shifts its neighbors through the shared component. Training samples all transitions equally often (uniform distribution, not $d^\pi$). The dynamics are as follows.\footnote{\citet{Baird1995} also presents an MDP variant with two actions per state, demonstrating that Q-learning diverges under the same mechanism. The mechanism is identical: shared parameters create cross-state coupling that uniform sampling cannot counterbalance.} When $V(6)$ is large and positive, the TD target $\gamma V(6)$ exceeds $V(s)$ for states 1 through 5, producing positive TD errors that push the shared weight upward. At state 6, the TD target is $\gamma V(6) < V(6)$, producing a negative TD error that pushes it downward. But states 1 through 5 are each visited as often as state 6, so the shared weight receives five upward pushes for every one downward push. The shared weight diverges to $+\infty$. The on-policy distribution would concentrate mass on state 6 (the absorbing state), counterbalancing the upward pressure; uniform sampling destroys this balance.\footnote{The gradient of $\|Q - TQ\|^2$ requires two independent next-state samples from the same $(s,a)$, since $\nabla\mathbb{E}[(Q - \mathbb{E}[r + \gamma V(s')])^2]$ involves $\mathbb{E}[\cdot] \cdot \nabla\mathbb{E}[\cdot]$ and $\mathbb{E}[XY] \neq \mathbb{E}[X]\mathbb{E}[Y]$. This ``double-sampling'' requirement is impractical \citep{Baird1995}, so practitioners use semi-gradient TD, treating the bootstrap target as a constant. This semi-gradient structure makes off-policy TD vulnerable to projection mismatch.}

\begin{theorem}[Off-policy divergence, \citealp{Baird1995}]
\label{thm:baird}
On the star MDP with zero rewards, whose linear features share a common weight across states and represent the true value $V^* \equiv 0$ exactly, semi-gradient TD(0) with all transitions sampled equally often diverges. There is an initialization, one in which the absorbing state's value is sufficiently larger than the others, from which the shared weight grows without bound, so the iterate leaves every bounded set even though $V^*$ lies in the representable class.
\end{theorem}

\begin{proof}
The mechanism is a counting imbalance. The shared weight collects a positive push from each of the five spoke states and a single negative push from the absorbing state, so it drifts up without bound. Rewards are zero, so the semi-gradient TD(0) increment for a transition $s \to s'$ is $\alpha\, \delta(s)\, \nabla_\theta V(s)$ with temporal-difference error $\delta(s) = \gamma V(s') - V(s)$ and the bootstrap target held fixed under differentiation.\footnote{Semi-gradient TD treats the target $\gamma V(s')$ as a constant when differentiating, rather than as a function of the shared parameters it actually depends on. This is what breaks the descent interpretation and, off-policy, admits divergence. A true gradient step on the squared TD error would require two independent successor samples and does not diverge here.} States $1$ through $5$ each transition to the absorbing state $6$, and state $6$ transitions to itself. Let $g_s = \nabla_{w} V(s) > 0$ be the positive gradient component of the shared weight $w$ in each state's value. Summing one increment per transition over an epoch, the change in $w$ is proportional to
\[
\Delta w \;\propto\; \underbrace{\sum_{s=1}^{5} \big[\gamma V(6) - V(s)\big]\, g_s}_{>\,0 \text{ once } \gamma V(6) > V(s)} \;+\; \underbrace{(\gamma - 1) V(6)\, g_6}_{<\,0} .
\]
At an initialization where $V(6)$ is large enough that $\gamma V(6) > V(s)$ for $s \in \{1, \dots, 5\}$, the five spoke terms are each positive and the lone absorbing-state term is negative. The five upward contributions outweigh the one downward contribution,\footnote{The imbalance is sharp as $\gamma \to 1$. The single downward term $(\gamma-1) V(6)\, g_6$ is $O(1-\gamma)$ and vanishes in the limit, while each of the five upward terms tends to the strictly positive $[V(6) - V(s)]\, g_s$. So for $\gamma$ close enough to $1$ the positive terms dominate regardless of the exact feature magnitudes $g_s, g_6$; \citet{Baird1995} tracks the full weight-dynamics with the specific feature values.} so the epoch raises $w$, which raises $V(6)$ further and preserves $\gamma V(6) > V(s)$ into the next epoch. The configuration is self-sustaining, $w$ grows without bound, and the iterate diverges. Since $V^* \equiv 0$ lies in the span of the features, this is a failure of the off-policy semi-gradient iteration rather than of representability. The full weight-dynamics and the two-action MDP variant appear in \citet{Baird1995}.
\end{proof}

Each element of the triad is individually necessary for divergence. Without function approximation (tabular), the projection is the identity and Q-learning's contraction applies directly. Without bootstrapping (Monte Carlo returns), targets are independent of current value estimates and the problem reduces to supervised regression. Without off-policy learning, samples come from $d^\pi$, the projection is orthogonal, and the Tsitsiklis-Van Roy convergence guarantee holds.

\begin{figure}[t]
\centering
\includegraphics[width=\textwidth]{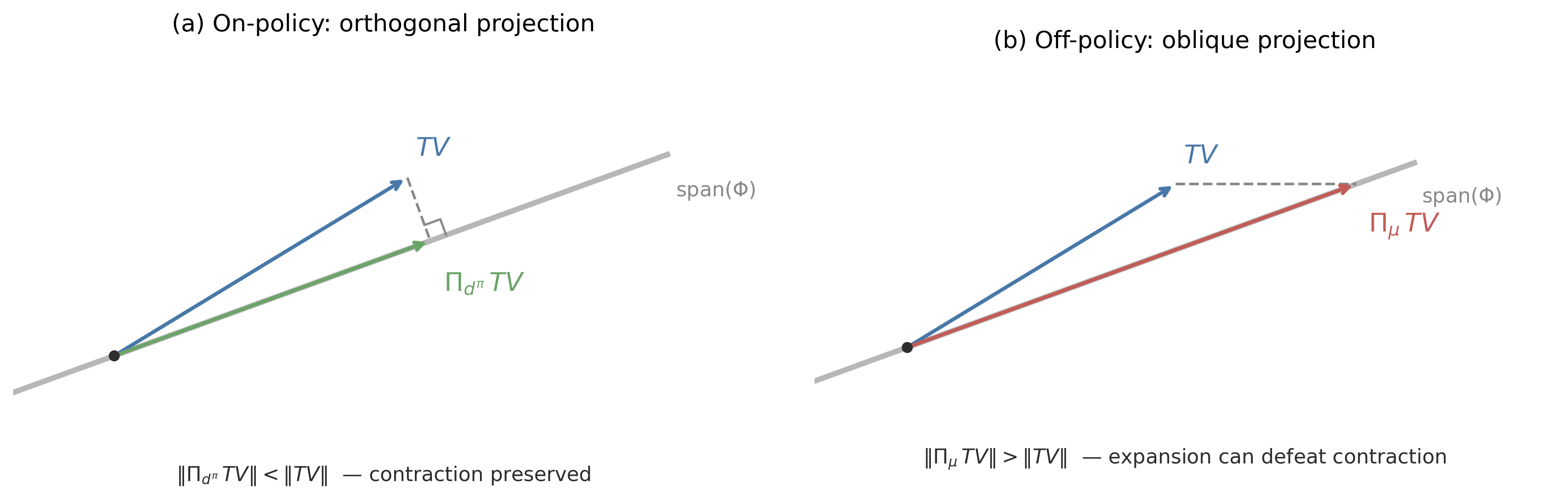}
\caption{Geometric intuition for the projection-norm leg of the deadly triad in $\mathbb{R}^2$. The gray line is the function approximation subspace span($\Phi$); the blue arrow is $TV$, the Bellman update. (a) On-policy: the orthogonal projection $\Pi_{d^\pi}$ (under the target-policy stationary distribution) drops $TV$ perpendicularly onto the subspace, preserving the contraction. (b) Off-policy: the oblique projection $\Pi_\mu$ (under the behavior distribution) reaches a point further from the origin than $TV$ itself, so the projection can expand distances. Bootstrapping and off-policy sampling, the other two legs of the triad, are not depicted; their joint effect on divergence is exhibited numerically in Baird's six-state star counterexample \citep{Baird1995}. The displayed expansion ratio is illustrative; the subspace and projection-direction angles are chosen so contraction in (a) and expansion in (b) are visible at figure scale rather than derived from a specific MDP.}
\label{fig:deadly_triad_geometry}
\end{figure}

\subsubsection{Resolutions}
\label{sec:deadly_triad_resolutions}

Three classes of algorithms restore convergence, each neutralizing a different component of the triad.

\emph{Target networks} weaken bootstrapping. Instead of updating toward $r + \gamma Q(s'; \theta)$, where the target moves with each parameter update, DQN \citep{mnih2015} updates toward $r + \gamma Q(s'; \theta^-)$, where $\theta^-$ is a slowly-updated copy of the parameters.\footnote{Experience replay \citep{Lin1992} complements target networks by breaking temporal correlation in the training data. The two mechanisms address different sources of instability: target networks stabilize the bootstrap target, while replay stabilizes the sampling distribution.} The regression target becomes quasi-static, converting the coupled fixed-point problem into a sequence of supervised learning problems. \citet{zhang2021target} prove that this two-timescale scheme converges to a regularized TD fixed point with linear function approximation. \citet{Fellows2023} show that target networks recondition the Jacobian of the TD update: the spectral radius of the composed update operator (Theorem~\ref{thm:prelim_spectral}) depends on the target network update frequency $k$, and for sufficiently large $k$ the spectral radius drops below 1 even in off-policy settings with nonlinear function approximation.

\emph{Gradient TD methods} fix the projection mismatch. \citet{sutton2009gtd} reformulate the projected Bellman error as a saddle-point problem $\min_\theta \max_y L(\theta, y)$, yielding algorithms (GTD, GTD2, TDC) that perform true stochastic gradient descent on the mean-squared projected Bellman error.\footnote{The saddle-point formulation introduces auxiliary variables $y$ of the same dimension as $\theta$, doubling the parameter count and requiring a second learning rate. \citet{bhandari2021finite} provide finite-time convergence rates for these two-timescale algorithms.} These methods converge off-policy with linear function approximation because they eliminate the semi-gradient approximation that causes the norm mismatch.

\emph{Regularization} shrinks the projection operator. \citet{lim2024regularized} add an $\ell_2$ penalty $-\eta\theta$ to the Q-learning update. This changes the projection from $\Pi = X(X^\top D X)^{-1} X^\top D$ to $\Pi_\eta = X(X^\top D X + \eta I)^{-1} X^\top D$. As the regularization strength $\eta$ increases, the projection ``shrinks'' toward the origin. For sufficiently large $\eta$, $\gamma\|\Pi_\eta\| < 1$, restoring the contraction property. The algorithm converges to a biased but stable fixed point, with the bias controlled by $\eta$.

\subsection{When Value-Based RL Works}
\label{sec:when_value_works}

The deadly triad gives conditions for divergence. The converse question, when value-based methods with function approximation provably converge, has a clean answer organized by one distinction. \emph{Realizability}, the assumption that the function class $\mathcal{F}$ contains $V^*$, is not by itself enough. The class must also satisfy \emph{Bellman completeness}, meaning it is closed under a single Bellman backup so that the regression target $T\hat{V}$ stays inside $\mathcal{F}$ for every $\hat{V} \in \mathcal{F}$, or else the data must cover the states a good policy visits.\footnote{Realizability asks only that the answer $V^*$ be representable; Bellman completeness asks that the intermediate targets produced along the way never leave the class. Realizability without completeness is provably insufficient for value-based methods, which is why fitted iteration is analyzed under completeness and offline methods under coverage instead.}

The coverage route is quantified by the concentrability coefficient already met in the finite-sample bound for fitted value iteration (Section~\ref{sec:fvi_fqi_theory}, the Munos--Szepesvari term $C_{\rho,\mu}$). Its worst-case form ranges over all policies and can be exponentially large in the horizon, but the pessimistic methods central to offline reinforcement learning need only \emph{single-policy concentrability}, coverage of a single comparator policy rather than of every policy \citep{Rashidinejad2021}. This weaker condition is the statistical price of learning from a fixed dataset, and it is the spine of the offline and constrained methods treated later in this survey.

The structural route imposes closure on the dynamics rather than coverage on the data. When the transition kernel and reward are linear in a known feature map, the Bellman operator preserves linearity, and that closure sidesteps the projection mismatch driving the triad; the same idea generalizes to low Bellman rank and its relatives. Section~\ref{sec:breaking_curse} develops these structural classes and their sample-complexity guarantees, together with the exponential lower bounds that show why realizability without closure or coverage does not suffice. Policy gradient methods make the opposite tradeoff. They converge globally with no concentrability assumption \citep[Theorem~5.3]{agarwal2021theory} but need on-policy data and carry high variance, while value-based methods reuse off-policy data efficiently, only under coverage or structure.

\subsection{Policy Learning Methods}
\label{sec:policy_gradient}

Value-based methods find fixed points of the Bellman operator. Policy-based methods parameterize the policy directly as $\pi_\theta(a|s)$ and maximize expected return $J(\theta) = \mathbb{E}_{\pi_\theta}[\sum_{t=0}^\infty \gamma^t R_t]$ by gradient ascent. This formulation sidesteps the Bellman equation entirely and frames reinforcement learning as constrained optimization.

\subsubsection{The Policy Gradient Theorem}

The policy gradient theorem, proved independently by \citet{williams1992} for the episodic case and \citet{SuttonMcAllester2000} for the general discounted setting, provides a tractable expression for the gradient.

\begin{theorem}[Policy gradient]
\label{thm:policy_gradient}
For the discounted objective $J(\theta) = \mathbb{E}_{\pi_\theta}\big[\sum_{t \geq 0} \gamma^t R_t\big]$ with normalized discounted state distribution $d^{\pi_\theta}(s) = (1-\gamma) \sum_{t=0}^\infty \gamma^t \mathbb{P}(s_t = s \mid \pi_\theta)$,
\begin{equation}
\nabla_\theta J(\theta) = \frac{1}{1-\gamma}\, \mathbb{E}_{s \sim d^{\pi_\theta},\, a \sim \pi_\theta} \left[ \nabla_\theta \log \pi_\theta(a|s) \, Q^{\pi_\theta}(s,a) \right].
\label{eq:policy_gradient}
\end{equation}
\end{theorem}

\begin{proof}
The derivation differentiates the value recursively, notices that the reward and transition terms carry no $\theta$-dependence, and unrolls what remains into a discounted sum over visited states. Write the value and action-value recursions,
\[
V^{\pi}(s) = \sum_a \pi_\theta(a|s)\, Q^\pi(s,a), \qquad Q^\pi(s,a) = r(s,a) + \gamma \sum_{s'} P(s'|s,a)\, V^\pi(s').
\]
Differentiating the first by the product rule,
\[
\nabla_\theta V^\pi(s) = \underbrace{\sum_a \nabla_\theta \pi_\theta(a|s)\, Q^\pi(s,a)}_{\text{policy term}} + \underbrace{\sum_a \pi_\theta(a|s)\, \nabla_\theta Q^\pi(s,a)}_{\text{value term}} ,
\]
and since $r(s,a)$ and $P(s'|s,a)$ carry no $\theta$-dependence, the second recursion gives $\nabla_\theta Q^\pi(s,a) = \gamma \sum_{s'} P(s'|s,a)\, \nabla_\theta V^\pi(s')$. Substituting turns the value term into $\gamma$ times the same gradient one step ahead, a recursion whose unrolling from $s_0$ collects the discounted visitation weights:
\begin{align*}
\nabla_\theta V^\pi(s_0)
  &= \sum_s \underbrace{\Big(\sum_{t \geq 0} \gamma^t \mathbb{P}(s_t = s \mid s_0, \pi)\Big)}_{=\; d^{\pi_\theta}(s)/(1-\gamma)} \sum_a \nabla_\theta \pi_\theta(a|s)\, Q^\pi(s,a) \\
  &= \frac{1}{1-\gamma} \sum_s d^{\pi_\theta}(s) \sum_a \nabla_\theta \pi_\theta(a|s)\, Q^\pi(s,a)
    && \text{($d^{\pi_\theta}$ is the normalized occupancy).}
\end{align*}
Taking the expectation over the initial state gives $\nabla_\theta J(\theta)$. Finally the score-function (log-derivative) identity converts the inner sum into an expectation over $a \sim \pi_\theta$,\footnote{The log-derivative identity $\nabla_\theta \pi_\theta(a|s) = \pi_\theta(a|s)\, \nabla_\theta \log \pi_\theta(a|s)$ turns a sum weighted by $\nabla\pi$ into an expectation under $\pi$. It is the same score that appears in maximum likelihood, and it implies $\mathbb{E}_{a\sim\pi_\theta}[\nabla_\theta \log \pi_\theta(a|s)] = 0$, which is why subtracting any baseline $b(s)$ from $Q^\pi$ leaves the gradient unbiased.} giving~\eqref{eq:policy_gradient}. The transition gradients $\nabla_\theta P$ never appear because $P$ is independent of $\theta$.
\end{proof}

The distribution $d^{\pi_\theta}$ depends on $\theta$ through the environment's dynamics, so a naive derivative would seem to require $\nabla_\theta d^{\pi_\theta}(s)$ and hence differentiation of the unknown transition matrix $P(s'|s,a)$.

The policy gradient theorem sidesteps this entirely via a likelihood ratio (or score function) trick.\footnote{The score function $\nabla_\theta \log \pi_\theta(a|s)$ is the same mathematical object that appears in the Cram\'er-Rao bound and the score test in maximum likelihood estimation. The policy gradient is proportional to a covariance, $\nabla J = \frac{1}{1-\gamma}\text{Cov}_{d^{\pi_\theta} \times \pi_\theta}(\nabla \log \pi_\theta, Q^{\pi_\theta})$, measuring how sensitive the log-likelihood of the policy is to parameter changes, weighted by action quality.} The theorem transforms a sensitivity analysis problem (how does the system evolve?) into a simpler expectation problem (what is the correlation between the score $\nabla \log \pi$ and the value $Q$?). The gradient can be written as an expectation under the current policy, weighted by action-values, without requiring $\nabla_\theta d^{\pi_\theta}$. The transition dynamics $P(s'|s,a)$ do not appear; the gradient is estimable via sample averages from trajectories alone.

\subsubsection{REINFORCE and Variance Reduction}

REINFORCE \citep{williams1992} is the simplest policy gradient algorithm. Sample a trajectory $(s_0, a_0, r_0, s_1, \ldots)$, compute the return $G_t = \sum_{k=0}^\infty \gamma^k r_{t+k}$ from each time step, and update:
\begin{equation}
\theta \leftarrow \theta + \alpha \sum_t \nabla_\theta \log \pi_\theta(a_t|s_t) \, G_t.
\end{equation}
This is an unbiased estimator of $\nabla_\theta J(\theta)$, but its variance is high because a single trajectory provides a noisy estimate of $Q^{\pi_\theta}$. Despite high variance, REINFORCE converges to a globally optimal policy in the tabular setting.\footnote{The baseline $b(s)$ subtracted from $G_t$ reduces variance while preserving unbiasedness, since $\mathbb{E}[\nabla_\theta \log \pi_\theta(a|s) b(s)] = 0$ for any baseline independent of $a$.}

\subsubsection{Natural Policy Gradient and Gradient Domination}

Standard gradient descent treats all parameter directions equally. But small changes in $\theta$ can cause large changes in the policy distribution $\pi_\theta$. The natural policy gradient \citep{Kakade2001}, building on the natural gradient framework of \citet{amari1998natural}, accounts for this curvature by preconditioning with the Fisher information matrix.\footnote{The Fisher information $F(\theta) = \mathbb{E}[\nabla \log p \nabla \log p^\top]$ measures curvature of the log-likelihood and appears in the Cram\'er-Rao bound. Here it measures how policy distributions change with parameters.} The relationship between NPG and standard PG parallels that between Fisher scoring and gradient ascent in MLE. Both precondition with the inverse Fisher information matrix $F(\theta)^{-1}$, achieving parameterization invariance and quadratic convergence near the optimum.
\begin{equation}
\tilde{\nabla}_\theta J(\theta) = F(\theta)^{-1} \nabla_\theta J(\theta), \quad F(\theta) = \mathbb{E}_{s,a} \left[ \nabla_\theta \log \pi_\theta(a|s) \nabla_\theta \log \pi_\theta(a|s)^\top \right].
\label{eq:npg_fisher}
\end{equation}
Why does NPG recover policy iteration? Standard gradient ascent is sensitive to parameterization: it takes the steepest step in Euclidean parameter space, where units depend on how the policy is parameterized. NPG takes the steepest step in distribution space (measured by KL-divergence), which is invariant to reparameterization. In the tabular case, \citet[Theorem~2]{Kakade2001} proves that this geometric correction aligns the gradient exactly with the greedy policy $\tilde{\pi}$ from policy iteration. With step size 1 (and exact estimation), NPG performs one full Newton step; with smaller step sizes, it performs damped Newton updates. This explains its rapid convergence: NPG approximates the quadratic convergence of finding a fixed point rather than the linear convergence of hill-climbing.

The RL objective $J(\theta)$ is non-convex in $\theta$. For researchers trained to distrust gradient methods on non-convex objectives, the natural concern is convergence to spurious local optima. For tabular softmax policies (one free parameter per state-action pair), this concern is unfounded. The landscape is ``benign'' in a precise sense. \citet{agarwal2021theory} prove that $J(\theta)$ satisfies a \emph{gradient domination} condition (also called Polyak-\L{}ojasiewicz, or PL). The PL condition has the same functional form as the strong convexity condition for guaranteeing linear convergence of gradient descent (Theorem~\ref{thm:prelim_gd}), but it applies to non-convex functions: whenever $\|\nabla J(\theta)\|$ is small, the policy must be near-optimal. Formally, the sub-optimality $J(\pi^*) - J(\pi_\theta)$ is bounded by a constant times $\|\nabla J(\theta)\|^2$. The implication is immediate: any point where the gradient vanishes is globally optimal, so gradient ascent cannot get trapped at a spurious local maximum.

\citet{mei2020} sharpen this result for softmax parameterization, proving explicit convergence rates. These guarantees are specific to the tabular parameterization. With function approximation, the PL condition does not hold. \citet[Theorem~6.2]{agarwal2021theory} show that NPG with log-linear or smooth policy classes (including neural networks) converges to a neighborhood of the optimum whose radius depends on the approximation error of the policy class, not to the global optimum itself.\footnote{However, ``no spurious local optima'' does not imply ``easy optimization.'' The landscape is dominated by vast plateaus (saddle points) where gradients vanish. Without sufficient exploration, the probability of visiting relevant states decays exponentially with the horizon, rendering the gradient exponentially small. Global convergence requires the starting distribution to have adequate coverage relative to the optimal policy's visitation distribution, formalized as the ``distribution mismatch coefficient'' by \citet{agarwal2021theory}. The Natural Policy Gradient addresses this by preconditioning with the Fisher Information Matrix, making the update direction covariant: invariant to invertible linear transformations of the parameter space. This standardizes units across parameters, preventing stalling on plateaus caused by poor parameter scaling. \citet{li2022softmax} make this quantitatively precise: vanilla softmax policy gradient requires iterations doubly exponential in the effective horizon $1/(1-\gamma)$ because score functions are exponentially small in directions corresponding to suboptimal actions.}

In the tabular setting, NPG achieves more: \emph{dimension-free} convergence. Standard gradient ascent on $J(\theta)$ has a convergence rate that depends on the smoothness constant, which scales with $|\mathcal{S}|$. NPG circumvents this by preconditioning with the Fisher information matrix $F(\theta)^{-1}$. The mechanism is that the state-visitation distribution $d^{\pi_\theta}(s)$ appears in both $\nabla J(\theta)$ and $F(\theta)$; when computing $F^{-1} \nabla J$, these terms cancel analytically. The resulting update rule is equivalent to soft policy iteration and converges at rate $O(1/(1-\gamma)^2 \epsilon)$, independent of $|\mathcal{S}|$ and $|\mathcal{A}|$ \citep{xiao2022convergence}.

\subsubsection{Trust Region Methods}

Trust region methods rest on an exact accounting of how the objective changes between two policies.

\begin{lemma}[Performance difference, \citealp{Kakade2002}]
\label{lem:pdl}
For any policies $\pi$ and $\pi'$, with advantage $A^\pi(s,a) = Q^\pi(s,a) - V^\pi(s)$,
\[
J(\pi') - J(\pi) = \frac{1}{1-\gamma}\, \mathbb{E}_{s \sim d^{\pi'},\, a \sim \pi'}\big[ A^\pi(s,a) \big].
\]
\end{lemma}

\begin{proof}
The trick is to insert a telescoping sum of $V^\pi$ along a $\pi'$-trajectory, which reconstructs $-J(\pi)$ for free and turns each summand into an advantage. Along any trajectory the discounted increments of $V^\pi$ telescope,\footnote{A telescoping sum collapses because consecutive terms cancel: $\sum_{t=0}^{T} \big(\gamma^{t+1} V^\pi(s_{t+1}) - \gamma^t V^\pi(s_t)\big) = \gamma^{T+1} V^\pi(s_{T+1}) - V^\pi(s_0)$, and the leading term vanishes as $T \to \infty$ since $\gamma < 1$ and $V^\pi$ is bounded. Appendix~\ref{prelim:telescoping} pairs this identity with the geometric series it is usually used alongside.}
\[
\sum_{t \geq 0} \gamma^t \big(\gamma V^\pi(s_{t+1}) - V^\pi(s_t)\big) = -V^\pi(s_0).
\]
Taking the expectation over trajectories generated by $\pi'$ and using $J(\pi) = \mathbb{E}_{s_0}[V^\pi(s_0)]$ to identify the telescoped term with $-J(\pi)$,
\[
J(\pi') - J(\pi) = \mathbb{E}_{\pi'}\Big[ \sum_{t \geq 0} \gamma^t \big( \underbrace{r(s_t,a_t) + \gamma V^\pi(s_{t+1}) - V^\pi(s_t)}_{\text{conditional mean }=\, A^\pi(s_t,a_t)} \big) \Big].
\]
The bracketed term has conditional mean
\[
r(s_t,a_t) + \gamma \mathbb{E}[V^\pi(s_{t+1}) \mid s_t, a_t] - V^\pi(s_t) = Q^\pi(s_t,a_t) - V^\pi(s_t) = A^\pi(s_t,a_t),
\]
by the Bellman consistency $Q^\pi(s,a) = r(s,a) + \gamma \mathbb{E}[V^\pi(s') \mid s,a]$. Collecting the discounted state-visitation weights of $\pi'$ into the normalized $d^{\pi'}$ produces the $1/(1-\gamma)$ factor and the stated identity.
\end{proof}

\begin{figure}[t]
  \centering
  \includegraphics[width=\textwidth]{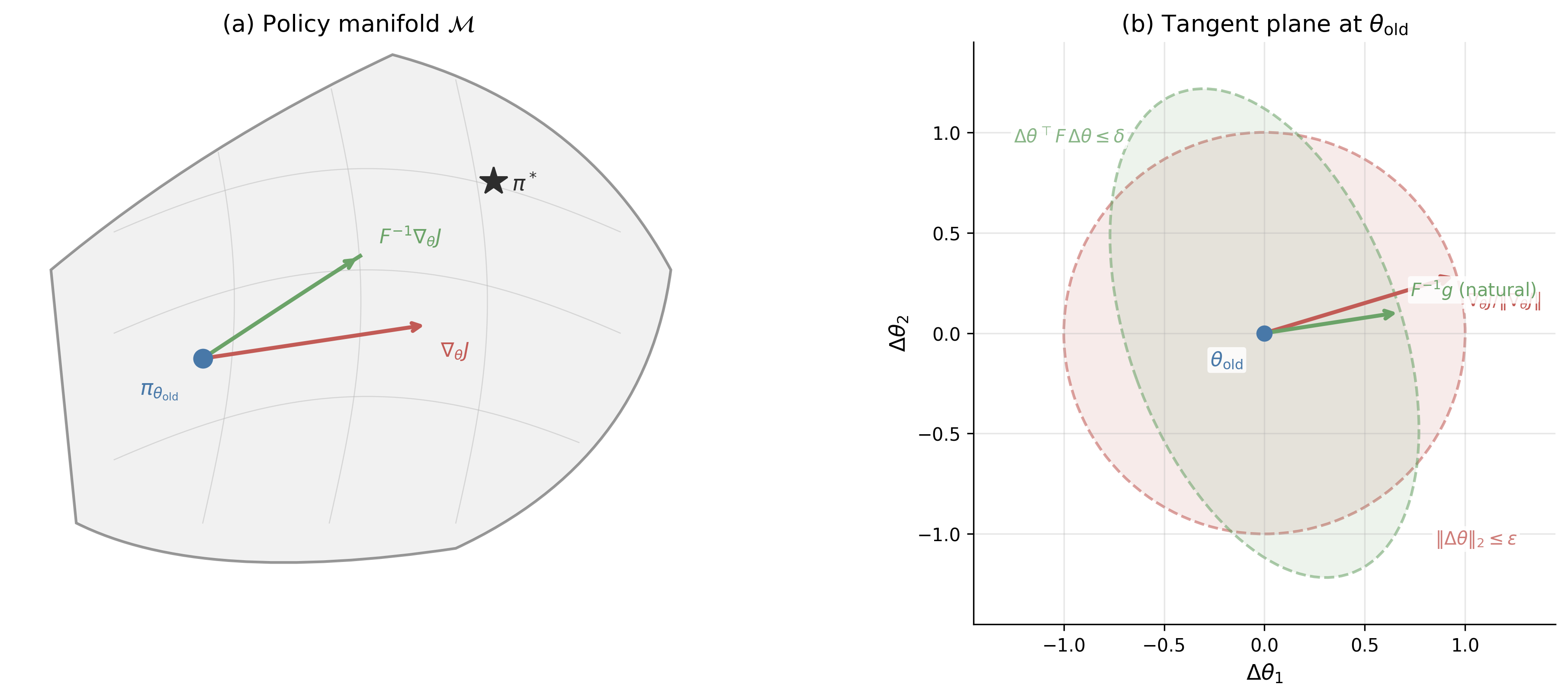}
  \caption{Information geometry of the natural policy gradient.
  \emph{Left}: schematic of the policy manifold $\mathcal{M}$ with Euclidean
  gradient $\nabla_\theta J$ (red) and natural gradient
  $F^{-1}\nabla_\theta J$ (green) from the current iterate
  $\pi_{\theta_{\mathrm{old}}}$ toward the optimal policy $\pi^*$. The two
  arrows are placed schematically to convey the convergence advantage of NPG;
  quantitative behavior depends on the loss landscape and is shown in the
  right panel.
  \emph{Right}: tangent plane at $\theta_{\mathrm{old}}$ showing the Euclidean
  unit ball $\|\Delta\theta\|_2 \leq \varepsilon$ (red) and the KL unit ball
  $\Delta\theta^\top F \Delta\theta \leq \delta$ (green), with the
  respective steepest-ascent directions. The Euclidean step (red arrow) leaves
  the KL ball, violating the trust-region constraint; the natural step (green
  arrow) saturates the KL budget by construction.}
  \label{fig:info_geometry}
\end{figure}

NPG requires computing and inverting the Fisher information matrix $F(\theta)$, which scales as $O(d^2)$ in parameters and is impractical for neural networks. TRPO \citep{Schulman2015} approximates the natural gradient using conjugate gradient methods\footnote{Conjugate gradient is an iterative method for solving linear systems $Ax = b$ without forming $A$ explicitly, requiring only matrix-vector products $Av$. With $k$ iterations it costs $O(kd)$ versus $O(d^3)$ for direct inversion, making it feasible for neural networks with millions of parameters.} without forming $F$ explicitly, and enforces trust regions via line search.\footnote{The performance difference lemma (Lemma~\ref{lem:pdl}) bounds how much policy improvement is possible and motivates constraining updates to regions where advantage estimates remain accurate \citep{Kakade2002}.} \citet{shani2020} prove convergence for adaptive trust region methods that adjust the constraint radius dynamically. PPO \citep{Schulman2017} simplifies further by replacing the hard KL constraint with a clipped surrogate objective, trading theoretical guarantees for computational simplicity.
The geometric foundation of trust region methods lies in information geometry. The space of policies $\{\pi_\theta : \theta \in \mathbb{R}^d\}$ forms a statistical manifold, and the natural distance between two nearby policies is the KL divergence, not the Euclidean distance between their parameters \citep{amari1998natural}. To second order, $\mathrm{KL}(\pi_\theta \| \pi_{\theta + \Delta\theta}) \approx \frac{1}{2} \Delta\theta^\top F(\theta) \Delta\theta$, where $F(\theta)$ is the Fisher information matrix. Two parameter vectors $\theta$ and $\theta'$ that are far apart in Euclidean distance may correspond to nearly identical distributions, while nearby parameters may produce radically different policies. The natural gradient corrects for this by measuring steepest ascent in KL-divergence rather than Euclidean norm. Figure~\ref{fig:info_geometry} illustrates the distinction. On the policy manifold, the Euclidean gradient $\nabla_\theta J$ points in a direction that ignores curvature, while the natural gradient $F^{-1} \nabla_\theta J$ follows the manifold's intrinsic geometry toward the optimum.

TRPO formalizes this insight as a constrained optimization problem. At each iteration, TRPO maximizes the importance-weighted surrogate
\begin{equation}
\label{eq:trpo}
\max_\theta \; L(\theta) = \mathbb{E}_{s \sim d^{\pi_{\theta_{\mathrm{old}}}}} \left[ \sum_a \frac{\pi_\theta(a|s)}{\pi_{\theta_{\mathrm{old}}}(a|s)} A^{\pi_{\theta_{\mathrm{old}}}}(s,a) \right] \quad \text{s.t.} \quad \mathrm{KL}(\pi_{\theta_{\mathrm{old}}} \| \pi_\theta) \leq \delta,
\end{equation}
where $A^{\pi}(s,a) = Q^{\pi}(s,a) - V^{\pi}(s)$ is the advantage function. Linearizing $L(\theta)$ around $\theta_{\mathrm{old}}$ and applying the quadratic KL approximation yields a Lagrangian whose closed-form solution is
\begin{equation}
\label{eq:trpo_step}
\theta_{\mathrm{new}} = \theta_{\mathrm{old}} + \sqrt{\frac{\delta}{g^\top F^{-1} g}} \, F^{-1} g, \qquad g = \nabla_\theta L(\theta)\big|_{\theta_{\mathrm{old}}}.
\end{equation}
This is precisely the natural gradient direction, scaled so that the step saturates the KL budget $\delta$. The step size is determined entirely by the trust region geometry, not by a learning rate hyperparameter. In practice, TRPO solves the linear system $F v = g$ via conjugate gradient and performs a backtracking line search to enforce the KL constraint exactly.\footnote{Appendix~\ref{prelim:least_squares} distinguishes solving a linear system from explicitly forming an inverse. Conjugate gradient solves $Fv = g$ iteratively using only matrix-vector products $Fv$, which can be computed via automatic differentiation without forming $F$ explicitly. With $k$ iterations it costs $O(kd)$ versus $O(d^3)$ for direct inversion, making it feasible for neural networks with millions of parameters.}

\begin{figure}[t]
  \centering
  \includegraphics[width=\textwidth]{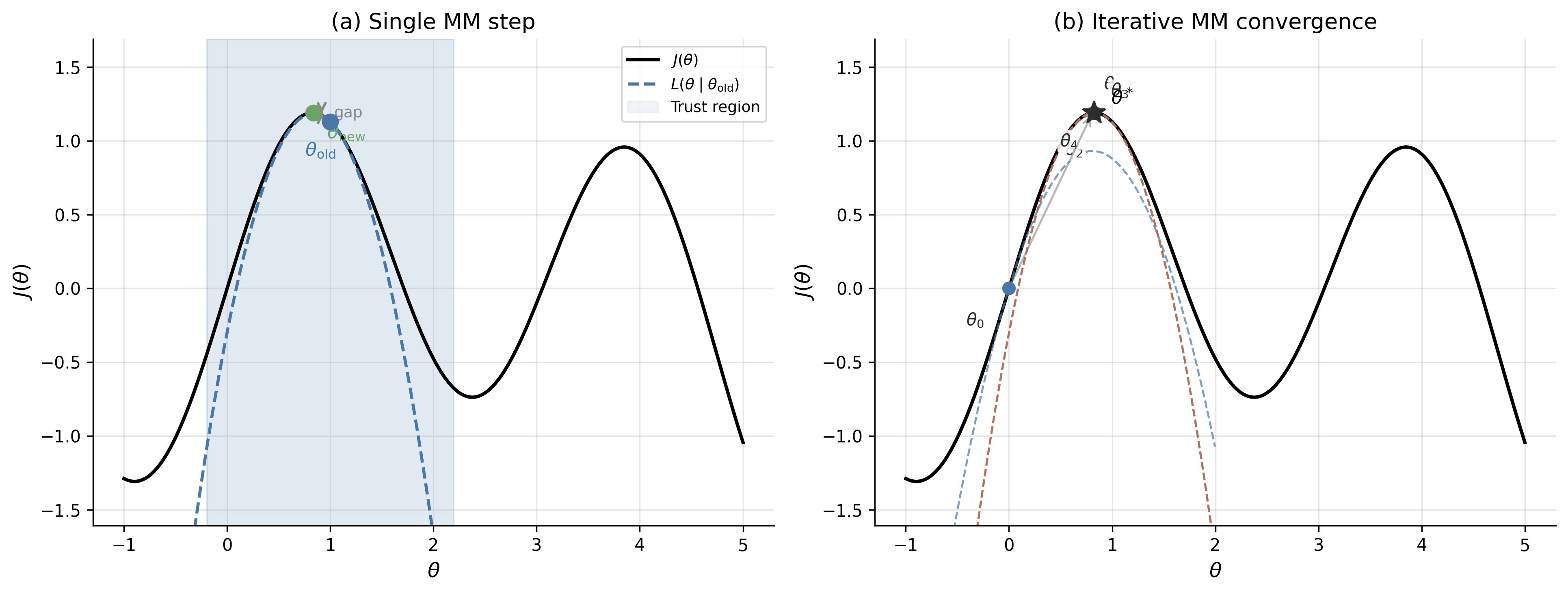}
  \caption[Schematic of the MM step underlying TRPO]{Schematic of the majorize-minimize step underlying TRPO: a surrogate
  $L(\theta)$ lower-bounds the true objective $J(\theta)$ and is monotonically
  improved by maximizing $L$.
  \emph{Left}: the surrogate (dashed) is tight at $\theta_{\mathrm{old}}$;
  the trust region (shaded) constrains the step. The gap between
  $L(\theta_{\mathrm{new}})$ and $J(\theta_{\mathrm{new}})$ is the
  guaranteed improvement.
  \emph{Right}: iterative MM convergence from $\theta_0$ through four
  surrogates (dashed, colored by iteration) to $\theta^*$.\protect\footnotemark}
  \label{fig:mm_surrogate}
\end{figure}

The theoretical guarantee underlying TRPO is a majorization-minimization (MM) argument. The surrogate $L(\theta)$ is a local lower bound on $J(\theta)$ that is tight at $\theta_{\mathrm{old}}$: $L(\theta_{\mathrm{old}}) = J(\theta_{\mathrm{old}})$ and $L(\theta) \leq J(\theta)$ within the trust region.\footnote{The bound follows from the performance difference lemma: $J(\pi') - J(\pi) = \frac{1}{1-\gamma} \mathbb{E}_{s \sim d^{\pi'}}[\sum_a \pi'(a|s) A^\pi(s,a)]$. Replacing $d^{\pi'}$ with $d^{\pi}$ introduces error controlled by the KL divergence between the two policies \citep{Kakade2002}.} Maximizing $L$ within the trust region therefore guarantees monotonic improvement: $J(\theta_{\mathrm{new}}) \geq L(\theta_{\mathrm{new}}) \geq L(\theta_{\mathrm{old}}) = J(\theta_{\mathrm{old}})$. This is the same pattern as the EM algorithm in statistics, where the E-step constructs a surrogate (the ELBO) and the M-step maximizes it.\footnote{In EM, $Q(\theta|\theta^{(t)})$ lower-bounds the log-likelihood and is tight at $\theta^{(t)}$. Each M-step guarantees $\ell(\theta^{(t+1)}) \geq Q(\theta^{(t+1)}|\theta^{(t)}) \geq Q(\theta^{(t)}|\theta^{(t)}) = \ell(\theta^{(t)})$. The TRPO bound has the same structure with $L$ playing the role of $Q$ and $J$ playing the role of $\ell$.} \citet{shani2020} prove convergence for adaptive trust region methods that adjust $\delta$ dynamically. Figure~\ref{fig:mm_surrogate} illustrates this mechanism. Each surrogate is a lower bound that touches $J$ at the current iterate, and sequential maximization produces monotonically improving iterates converging to $\theta^*$.

\footnotetext{The quadratic majorizer here uses a fitted constant $c$ chosen so the surrogate touches the true objective at $\theta_{\mathrm{old}}$ and remains below it on the visible grid. \citet{Schulman2015}'s theoretical constant $C = 4\varepsilon\gamma/(1-\gamma)^2$ is so conservative the resulting step is unusably small in practice; modern TRPO implementations replace the penalty with a hard KL constraint. This illustrative cartoon visualizes the \emph{form} of the MM bound, not its sharpness.}

\begin{figure}[t]
  \centering
  \includegraphics[width=\textwidth]{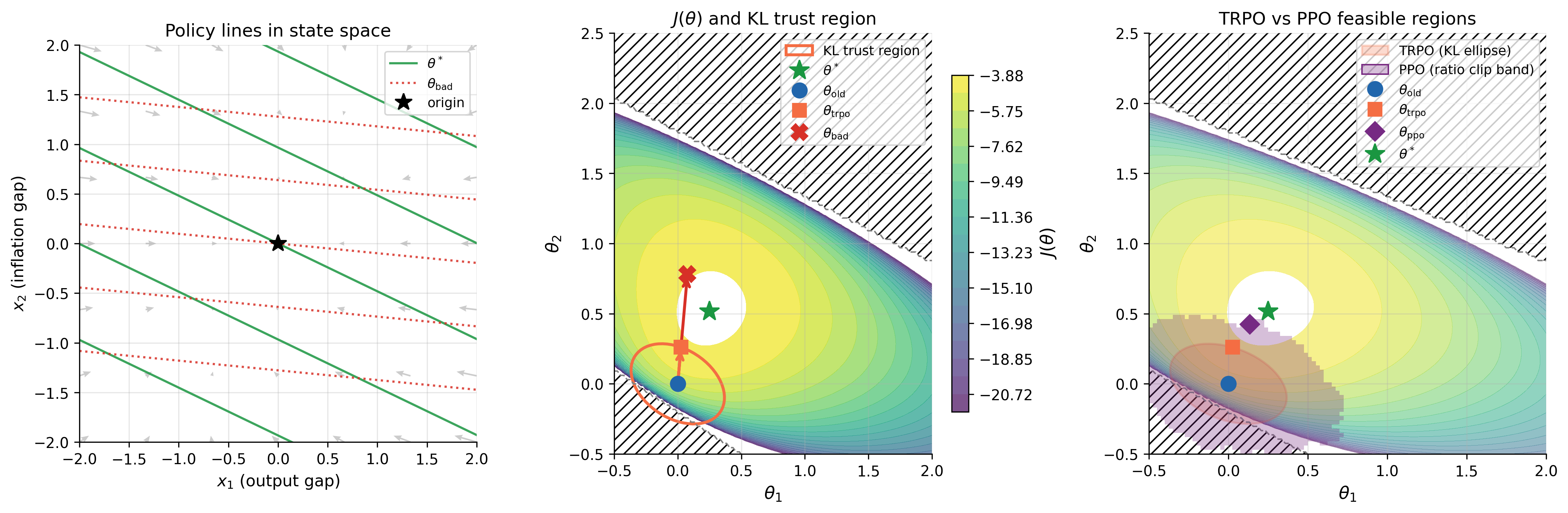}
  \caption{Trust region methods in the LQC monetary policy setting. A central bank
  learns a Taylor rule $u_t = -(\theta_1 x_{1t} + \theta_2 x_{2t})$ mapping output
  gap $x_1$ and inflation gap $x_2$ to an interest rate instrument.
  \emph{Left}: policy contour lines in state space for the current iterate
  $\theta_{\mathrm{old}}$, optimal weights $\theta^*$, and an illustrative
  KL-violating iterate $\theta_{\mathrm{bad}}$ (constructed as a $3\times$ scaling
  of the TRPO step, not a literal unconstrained gradient step), with phase arrows
  showing closed-loop dynamics under $\theta_{\mathrm{old}}$.
  \emph{Center}: expected return $J(\theta_1, \theta_2)$ in parameter space;
  hatching marks the unstable region. The KL trust region ellipse bounds the TRPO
  step; $\theta_{\mathrm{bad}}$ lands in the unstable region, illustrating the
  consequence of violating the KL constraint.
  \emph{Right}: TRPO feasible region (KL ellipse) and PPO feasible region (50\%
  ratio-clip band over 200 sampled state-action pairs) overlaid on $J(\theta)$,
  with the respective constrained steps marked.}
  \label{fig:trust_region_lqc}
\end{figure}

PPO \citep{Schulman2017} replaces the hard KL constraint with a clipped surrogate objective. Let $r_t(\theta) = \pi_\theta(a_t|s_t) / \pi_{\theta_{\mathrm{old}}}(a_t|s_t)$ denote the importance ratio. PPO maximizes
\begin{equation}
\label{eq:ppo}
L^{\mathrm{clip}}(\theta) = \mathbb{E}_t \left[ \min\!\big( r_t(\theta) A_t, \; \mathrm{clip}(r_t(\theta), 1-\varepsilon, 1+\varepsilon) A_t \big) \right],
\end{equation}
where $\varepsilon$ (typically 0.1--0.2) bounds the ratio. When $A_t > 0$, clipping prevents $r_t$ from exceeding $1+\varepsilon$; when $A_t < 0$, it prevents $r_t$ from falling below $1-\varepsilon$. The resulting feasible region is not an ellipsoid in parameter space but rather a non-convex set determined by the ratio constraint at each sampled state-action pair. PPO requires no Fisher information computation and uses only first-order gradients, making it the dominant method in large-scale applications including RLHF (Section~\ref{section:rlhf}). Figure~\ref{fig:trust_region_lqc} illustrates all three mechanisms in the LQC monetary policy setting, where the non-ellipsoidal PPO feasible region is visible in contrast to TRPO's KL ellipse.

The trust region framework connects naturally to econometric optimization. The Levenberg-Marquardt algorithm for nonlinear least squares uses a similar trust region mechanism, interpolating between gradient descent and Gauss-Newton steps.\footnote{Levenberg-Marquardt solves $\min_\theta \|r(\theta)\|^2$ by adding a damping term $\lambda I$ to the Gauss-Newton Hessian approximation $J^\top J$, which is equivalent to constraining the step to a trust region whose radius decreases with $\lambda$. TRPO replaces $J^\top J$ with the Fisher information matrix $F(\theta)$.} More broadly, the Fisher information matrix that defines TRPO's trust region is the same object that appears in the Cram\'er-Rao bound: it measures the statistical precision of the policy parameterization. The natural gradient adapts step sizes to this precision, taking large steps in well-identified directions and small steps where the data provide little information about the policy.

\FloatBarrier
\subsubsection{Engine Replacement MDP: The Policy Square}
\label{engine:policy_space}

In the Engine Replacement MDP of Section~\ref{engine:model}, the Fisher-preconditioned operator $F^{-1}\nabla J$ from equation~\eqref{eq:npg_fisher} determines the natural-gradient path. Under the parameterization with one logit per state, $\pi_\theta(\text{replace} \mid s) = \sigma(\theta_s)$, the policy class fills the open unit square with coordinates $(\pi(\text{replace} \mid \text{low}),\ \pi(\text{replace} \mid \text{high}))$, reaching the four deterministic vertices only in the limit of infinite logits. The objective is $J(\theta) = V^{\pi_\theta}(\text{low})$ for the start distribution $\nu = \delta_{\text{low}}$. At the uniform policy the unnormalized occupancy is $\rho = (7.0968,\ 2.9032)$, and the Fisher matrix of~\eqref{eq:npg_fisher} specializes to the diagonal $F = \mathrm{diag}\big(\rho(s)\, \pi_k(s)\, \pi_r(s)\big) = \mathrm{diag}(1.7742,\ 0.7258)$ with determinant $1.2877$, so preconditioning is an entrywise division rather than a pseudo-inverse.\footnote{Under the usual overparameterized softmax with one logit per state-action pair, the per-state Fisher block is $\rho(s)\,\pi_k \pi_r \left[\begin{smallmatrix} 1 & -1 \\ -1 & 1 \end{smallmatrix}\right]$, which has rank one with null direction $(1,1)$, because adding a constant to both logits at a state leaves the policy unchanged. Any ``invert the Fisher'' step there is silently a pseudo-inverse. Dropping the redundant coordinate removes the direction along which the objective is exactly flat. The Fisher here is weighted by the same unnormalized occupancy $\rho$ that appears in the gradient; weighting by the normalized $d^{\pi}$ instead rescales $F$ by $1-\gamma$ and the natural gradient by $1/(1-\gamma)$, leaving its direction unchanged.} The natural gradient $F^{-1} \nabla J$ then equals the action-value gap $Q(s, \text{replace}) - Q(s, \text{keep})$ exactly, here $(-1.2677,\ -0.2355)$. Both gaps are negative at the uniform start. Under uniform continuation the engine is worth keeping in both states, so the first pull of both gradient methods is toward the $(\text{keep}, \text{keep})$ vertex, away from the optimal action at high mileage. Table~\ref{tab:engine_policy_square} collects these quantities.

\begin{table}[h]
\centering
\caption{Policy-gradient objects on the Engine Replacement MDP at the uniform policy $\theta = (0, 0)$, start distribution $\nu = \delta_{\text{low}}$. The Fisher matrix is diagonal under the one-logit-per-state parameterization; its inverse is taken entrywise, never a pseudo-inverse. Convergence counts iterations until $J^\star - J < 10^{-3}$ at step size 0.5.}
\label{tab:engine_policy_square}
\begin{tabular}{lrr}
\hline
 & low & high \\
\hline
occupancy $\rho(s)$ & 7.0968 & 2.9032 \\
Fisher diagonal $\rho(s)\pi_k(s)\pi_r(s)$ & 1.7742 & 0.7258 \\
gradient $\partial J / \partial \theta_s$ & -2.2492 & -0.1709 \\
action-value gap $Q(s, \text{r}) - Q(s, \text{k})$ & -1.2677 & -0.2355 \\
natural gradient $F^{-1} \nabla J$ & -1.2677 & -0.2355 \\
policy-gradient iterations to $J^\star-J<10^{-3}$ & \multicolumn{2}{c}{not reached in 300} \\
natural-gradient iterations to $J^\star-J<10^{-3}$ & \multicolumn{2}{c}{58} \\
\hline
\end{tabular}
\end{table}

\begin{figure}[t]
  \centering
  \includegraphics[width=0.72\textwidth]{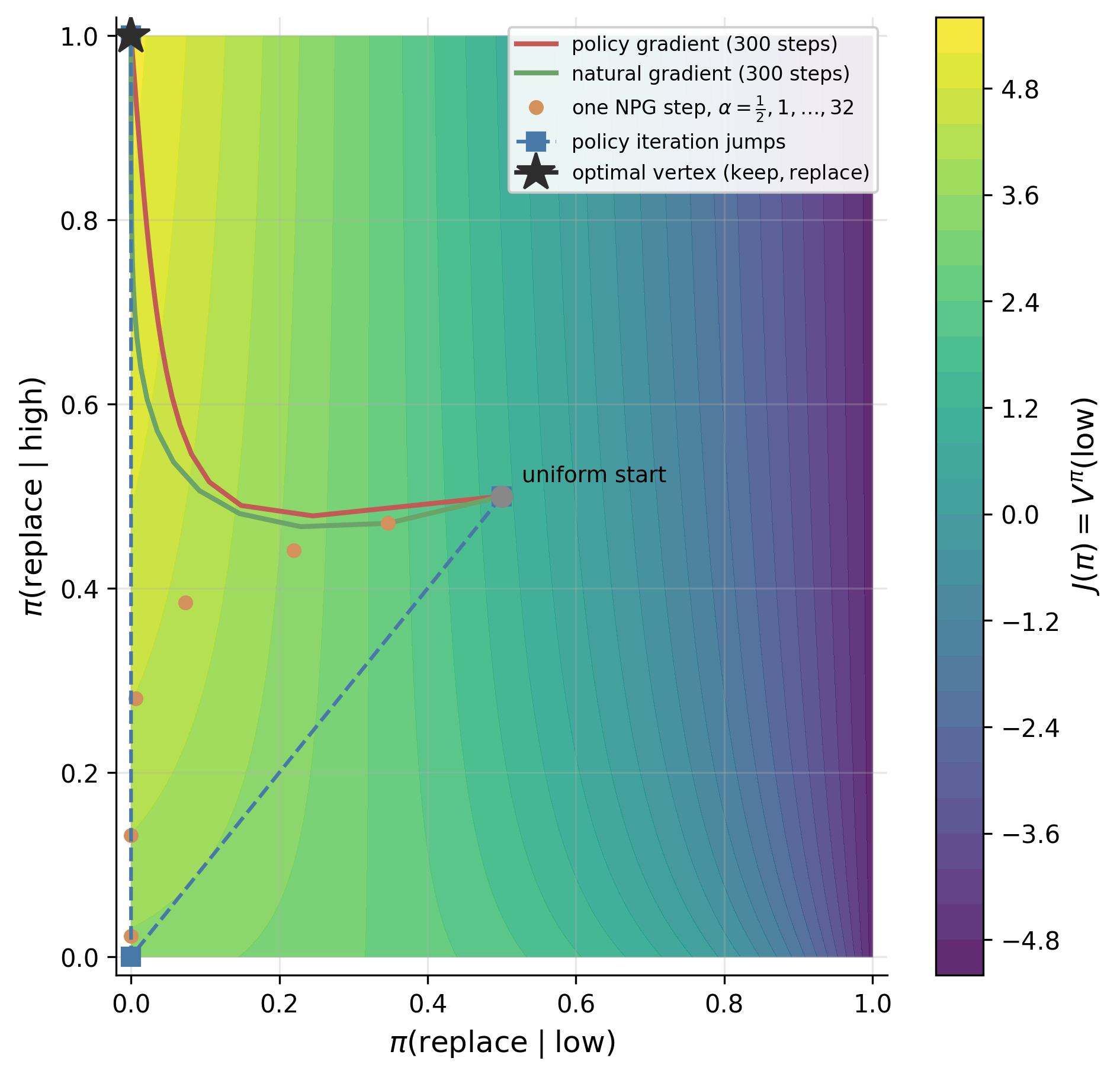}
  \caption{The policy square of the Engine Replacement MDP. Contours show $J(\pi) = V^\pi(\text{low})$ over $(\pi(\text{replace} \mid \text{low}),\ \pi(\text{replace} \mid \text{high}))$. Overlaid are vanilla and natural gradient ascent from the uniform policy (300 steps each, step size $0.5$), a single natural-gradient step at $\alpha = \tfrac{1}{2}, 1, 2, \ldots, 32$ (dots), the two policy-iteration jumps (dashed), and the optimal vertex (star).}
  \label{fig:engine_policy_square}
\end{figure}

Figure~\ref{fig:engine_policy_square} overlays the methods on the square. Vanilla and natural gradient ascent both dip toward the $(\text{keep}, \text{keep})$ corner before curving up the left edge. The natural path brings $J^\star - J$ below $10^{-3}$ in 58 iterations, while the vanilla path has not done so after 300 (Table~\ref{tab:engine_policy_square}). A single natural-gradient step from the uniform policy with growing step size $\alpha$ traces the orange dots, and as $\alpha \to \infty$ it lands on the greedy vertex of the \emph{current} policy, the $(\text{keep}, \text{keep})$ corner, not on the optimum. The large-step natural gradient therefore supplies the improvement half-step of policy iteration \citep{Kakade2001}. Policy iteration itself jumps vertex to vertex, from the uniform start to $(\text{keep}, \text{keep})$ and then to the optimal $(\text{keep}, \text{replace})$. Policy iteration takes two improvement steps, while the natural-gradient path takes fifty-eight interior steps.

The computed paths verify that Fisher preconditioning follows the action-value gaps and that a large natural-gradient step reproduces the policy-improvement half-step.

\FloatBarrier
\subsection{Hybrid Methods}
\label{sec:actor_critic}

REINFORCE estimates policy gradients from sample returns (unbiased, high variance); TD methods use bootstrapped targets $r + \gamma V(s')$ (lower variance, biased when $V$ is approximate). Actor-critic methods combine both. The critic estimates the value function, the actor updates the policy using the critic's estimates.

\subsubsection{Actor-Critic Architecture and Two-Timescale Convergence}

The theoretical foundation is two-timescale stochastic approximation \citep{konda2000}, building on the two-timescale ODE convergence theory of \citet{borkar1997}.\footnote{The original analysis of \citet{konda2000} uses the average-cost formulation with TD error $\delta_t = c(X_t, U_t) - \Lambda + V(X_{t+1}) - V(X_t)$, where $\Lambda$ is the average cost. The discounted variant presented here follows by replacing the average-cost baseline with $\gamma V(s')$. A related but distinct ODE stability framework for single-timescale stochastic approximation, including Q-learning and TD learning, appears in \citet{borkar2000}.} Run two concurrent learning processes:
\begin{align}
\text{Critic (fast):} \quad & \theta_{t+1} = \theta_t + \alpha_t^{(c)} \delta_t \nabla_\theta \hat{V}(s_t; \theta_t), \\
\text{Actor (slow):} \quad & \omega_{t+1} = \omega_t + \alpha_t^{(a)} \delta_t \nabla_\omega \log \pi_\omega(a_t|s_t),
\end{align}
where $\delta_t = r_t + \gamma \hat{V}(s_{t+1}; \theta_t) - \hat{V}(s_t; \theta_t)$ is the TD error. The critic updates the value function parameters $\theta$; the actor updates the policy parameters $\omega$.

Convergence requires the critic to learn faster than the actor.
\begin{equation}
\lim_{t \to \infty} \frac{\alpha_t^{(a)}}{\alpha_t^{(c)}} = 0, \quad \text{with both satisfying Robbins-Monro conditions.}
\end{equation}
Under this separation, the actor sees a quasi-stationary critic: from the actor's perspective, the critic provides approximately correct value estimates at each step.\footnote{The two-timescale structure is analogous to nested optimization in structural estimation, where an inner loop solves for equilibrium given parameters and an outer loop searches over parameters. The critic's inner loop (policy evaluation) must converge before the actor's outer loop (policy improvement) takes a step. \citet{wu2020twotimescale} provide finite-time convergence rates ($\tilde{O}(\epsilon^{-2.5})$ sample complexity) for two-timescale actor-critic with linear approximation, and \citet{tian2023} establish analogous rates for single-timescale actor-critic with multi-layer neural networks.} The actor's updates are then approximately unbiased policy gradient steps. \citet{konda2000} prove convergence to a stationary point of $J(\omega)$ (i.e., $\nabla J(\omega) \to 0$).

Convergence requires a structural condition on the critic. The critic's feature vectors must span the actor's score functions $\nabla_\omega \log \pi_\omega(a|s)$, so that the critic's approximation error lies orthogonal to the policy gradient direction. Under this compatibility condition, the critic's projection error does not bias the actor's gradient estimates.\footnote{The compatible function approximation theorem first appears in \citet{SuttonMcAllester2000} and is the key structural requirement in \citet{konda2000}. It constrains the critic architecture to be ``compatible'' with the actor parameterization, the same condition that makes the natural policy gradient equal to the critic's weight vector in \citet{Kakade2002}.}

A2C (Advantage Actor-Critic) is the synchronous variant: collect a batch of transitions, compute TD errors, and update both networks. A3C \citep{mnih2016a3c} parallelizes this across multiple workers updating a shared parameter server asynchronously.\footnote{Parallel workers decorrelate gradient estimates by exploring different parts of the state space simultaneously, removing the need for an experience replay buffer. Rigorous convergence theory for A3C's lock-free asynchronous parameter updates remains an open problem; existing analyses of asynchronous stochastic approximation \citep{qu2020async} address classical asynchrony (different state-action pairs updated at different times on a single trajectory), not the parallel-worker gradient setting.}

\subsubsection{Entropy Regularization and Soft Actor-Critic}

SAC (Soft Actor-Critic) \citep{Haarnoja2018} extends the actor-critic framework with entropy regularization, building on the soft Q-learning algorithm of \citet{Haarnoja2017}. The agent maximizes the entropy-augmented objective:
\begin{equation}
J_\tau(\theta) = \mathbb{E}_{\pi_\theta}\left[\sum_{t=0}^\infty \gamma^t \left(R_t + \tau \mathcal{H}(\pi_\theta(\cdot|s_t))\right)\right],
\end{equation}
where $\mathcal{H}(\pi) = -\sum_a \pi(a) \log \pi(a)$ is the entropy and $\tau > 0$ is the temperature parameter. \citet{geist2019regularized} later provided the unifying theoretical framework, showing that entropy regularization converts the Bellman optimality operator's non-smooth hard max into a smooth log-sum-exp, and that the resulting soft Bellman operator remains a $\gamma$-contraction, preserving the convergence guarantees of standard dynamic programming.\footnote{The optimal policy under entropy regularization is $\pi^*(a|s) \propto \exp(Q^*(s,a)/\tau)$, which is precisely the \citet{McFadden1974} logit choice probability with systematic utility $Q^*(s,a)$ and scale parameter $\tau$. The entropy-regularized value function is the log-sum-exp of Q-values, corresponding to the inclusive value (log-sum) operator in nested logit models. This equivalence between the soft-control framework and dynamic discrete choice models with EV1 taste shocks is developed formally in \citet{RustRawat2026}, Appendix~A.}

Entropy regularization also addresses the deadly triad directly. By maintaining policy stochasticity, the behavior policy used for data collection remains close to the target policy being optimized. This reduces the distribution mismatch between the stationary distribution under the behavior policy and the update targets, mitigating the off-policy instability leg of the triad. \citet{cen2022fast} formalize a second benefit: entropy regularization accelerates convergence of the natural policy gradient from $O(1/\epsilon)$ to $O(\log(1/\epsilon))$, providing a precise sense in which smoothing the policy landscape aids optimization. The actor-critic structure separates identification from optimization: the critic solves a regression problem (estimate $V^\pi$ from data), while the actor solves an optimization problem (improve $\pi$ using the estimated values).

\subsubsection{Error Amplification Under Approximate Value Functions}

Two questions remain important: how does approximation error propagate to policy quality, and how does computational complexity scale with problem size? \citet{singh1994} bound the policy degradation from value-function error (an independent derivation is \citealt{bertsekastsitsiklis1996}, Proposition~6.1).

\begin{theorem}[Greedy error amplification, \citealp{singh1994}]
\label{thm:singh_yee}
If $\|\hat{V} - V^*\|_\infty \leq \epsilon$ and $\hat{\pi}$ is greedy with respect to $\hat{V}$, then
\begin{equation}
\|V^* - V^{\hat{\pi}}\|_\infty \leq \frac{2\gamma}{1-\gamma}\, \epsilon.
\label{eq:singh_yee}
\end{equation}
\end{theorem}

\begin{proof}
Split the loss into a policy-mismatch term and a value-propagation term, bound each by $\gamma\epsilon$ or by $\gamma$ times the loss itself, then solve the resulting self-referential inequality. Fix $s$ and write $T^{\pi^*}$ and $T^{\hat\pi}$ for the operators of the optimal and greedy policies. Using $T^{\pi^*} V^* = V^*$ and $T^{\hat\pi} V^{\hat\pi} = V^{\hat\pi}$, add and subtract $(T^{\hat\pi} V^*)(s)$:
\[
V^*(s) - V^{\hat\pi}(s) = \underbrace{(T^{\pi^*} V^*)(s) - (T^{\hat\pi} V^*)(s)}_{\text{policy mismatch}} + \underbrace{(T^{\hat\pi} V^*)(s) - (T^{\hat\pi} V^{\hat\pi})(s)}_{\text{value propagation}}.
\]
The value-propagation term equals $\gamma \sum_{s'} P(s'|s,\hat\pi)\,(V^* - V^{\hat\pi})(s') \leq \gamma \|V^* - V^{\hat\pi}\|_\infty$. For the policy-mismatch term, insert $\hat V$ into both operators and use that $\hat\pi$ is greedy for $\hat V$:\footnote{Greedy means $\hat\pi$ maximizes the one-step lookahead under $\hat V$, so $(T^{\hat\pi} \hat V)(s) \geq (T^{\pi} \hat V)(s)$ for every policy $\pi$, in particular $\pi^*$. This lets the middle term $(T^{\pi^*}\hat V)(s) - (T^{\hat\pi}\hat V)(s) \leq 0$ be dropped in the chain below.}
\begin{align*}
&(T^{\pi^*} V^*)(s) - (T^{\hat\pi} V^*)(s) \\
&\qquad \leq \underbrace{(T^{\pi^*} V^* - T^{\pi^*} \hat V)(s)}_{\leq\, \gamma\epsilon} + \underbrace{(T^{\hat\pi} \hat V - T^{\hat\pi} V^*)(s)}_{\leq\, \gamma\epsilon} \;\leq\; 2\gamma\epsilon,
\end{align*}
each summand being $\gamma$ times a $P$-average of the gap $|V^* - \hat V| \leq \epsilon$.
Adding the two terms, $V^*(s) - V^{\hat\pi}(s) \leq 2\gamma\epsilon + \gamma \|V^* - V^{\hat\pi}\|_\infty$. Because $V^{\hat\pi} \leq V^*$ pointwise,\footnote{$V^*$ is by definition the largest value function achievable, the pointwise supremum over all policies, so $V^{\hat\pi} \leq V^*$ and $V^* - V^{\hat\pi} \geq 0$. The supremum of the difference is then exactly $\|V^* - V^{\hat\pi}\|_\infty$.} taking the supremum over $s$ gives $\|V^* - V^{\hat\pi}\|_\infty \leq 2\gamma\epsilon + \gamma \|V^* - V^{\hat\pi}\|_\infty$, and solving for the norm yields the bound.
\end{proof}

At $\gamma = 0.99$, the amplification factor is $2 \cdot 0.99 / 0.01 = 198$. A 1\% error in value function approximation yields at most 198\% error in policy value.\footnote{The Singh-Yee bound is worst-case and not tight in general; \citet{farahmand2010} derive tighter bounds under smoothness assumptions on the MDP. The amplification factor $2\gamma/(1-\gamma)$ is a sensitivity analysis: it quantifies how errors in the ``inputs'' (value estimates) propagate to ``outputs'' (policy quality), analogous to errors-in-variables bias in regression. The discount factor $\gamma$ controls sensitivity; more patient agents face larger amplification.} This bound is pessimistic but finite: approximate value functions do not cause unbounded policy degradation.

\subsubsection{Sample Complexity of Planning}

Classical dynamic programming complexity scales with the state space size $|\mathcal{S}|$. For problems like Go, where $|\mathcal{S}| \approx 10^{170}$, exact computation is impossible. \citet{kearns2002} prove that with access to a generative model\footnote{A ``generative model'' in RL is a simulator that, given any state-action pair $(s,a)$, returns a sampled next state $s' \sim P(\cdot|s,a)$ and reward $r$. This is unrelated to ``generative models'' in machine learning (GANs, diffusion models) or ``generative processes'' in Bayesian econometrics. The distinction matters: planning with a generative model is strictly easier than learning from a single trajectory, because the agent can query arbitrary states rather than following a sequential path.} (a simulator that samples transitions from any state-action pair), near-optimal planning is possible with \emph{no dependence on $|\mathcal{S}|$}. The cost is exponential dependence on the effective horizon $H = \log(R_{\max}/(\epsilon(1-\gamma))) / \log(1/\gamma)$: the sparse sampling algorithm requires $O((|\mathcal{A}|/\epsilon)^{H})$ simulator calls.\footnote{The sparse sampling algorithm builds a random tree of depth $H$ from the current state, sampling $C$ successor states per action at each node, then estimates values by averaging leaf rewards back up the tree. Its running time is $O((C|\mathcal{A}|)^H)$, which is exponential in $H$ but entirely independent of $|\mathcal{S}|$. \citet{kearns2002} also prove a lower bound of $\Omega(2^H)$ generative model calls for any planning algorithm, so exponential horizon dependence is unavoidable in the worst case.} For $\gamma$ near 1, $H \approx (1-\gamma)^{-1} \log(R_{\max}/\epsilon)$, so the method is practical only for short effective horizons or moderate discount factors. Classical DP scales linearly in $|\mathcal{S}|$ but polynomially in $H$; sparse sampling eliminates state-space dependence at the cost of exponential horizon dependence. This explains why MCTS succeeds in large state spaces with bounded lookahead.

The minimax-optimal sample complexity for planning with a generative model, when queries to arbitrary state-action pairs are permitted, is $\Theta(|\mathcal{S}||\mathcal{A}|/((1-\gamma)^3\epsilon^2))$ \citep{azar2013}. This bound scales linearly in $|\mathcal{S}|$ but polynomially in $1/(1-\gamma)$, the opposite regime from sparse sampling. \citet{agarwalKakadeYang2020} show that the plug-in model-based approach (learn $\hat{P}$ from samples, then plan with $\hat{P}$) achieves this minimax rate, establishing that model-based RL is statistically optimal. \citet{li2024breaking} further tighten this result by breaking the $|\mathcal{S}||\mathcal{A}|/(1-\gamma)^2$ sample-size barrier, showing that the minimax rate is achievable with total sample size as low as $|\mathcal{S}||\mathcal{A}|/(1-\gamma)$.\footnote{Recent extensions push these results beyond standard MDPs. \citet{clavier2024robust} study the robust MDP setting where the agent must plan under model uncertainty with sa-rectangular or s-rectangular uncertainty sets, establishing minimax rates for robust policy optimization. \citet{wang2025cvar} extend the analysis to risk-sensitive objectives under the iterated CVaR criterion.}

%
%

\subsection{The Curse of Dimensionality}
\label{sec:curse_of_dimensionality}

Exact solution methods must visit every state, and the number of states grows exponentially in the problem's natural dimension. With $d$ state variables each taking $n$ discrete values, the state space has $n^d$ elements and value iteration requires $O(n^{2d} \cdot |\mathcal{A}|)$ time per sweep.\footnote{The term ``curse of dimensionality'' is due to \citet{Bellman1957}.} Table~\ref{tab:curse_grid_dp} evaluates this for models that appear elsewhere in this survey. The barrier is not reached only by problems a reader would call large. The bus engine of \citet{Rust1987}, on its published 175-point mileage grid, sweeps in microseconds. Three such engines with a joint replacement decision, a model whose transition law still fits on one line, take days. Five put a single sweep into geological time and need more than a terabyte to hold the value function alone.

\begin{table}[htbp]
\centering
\caption{Cost of one exact value-iteration sweep on a uniform grid, for models used elsewhere in this survey. States are $n^d$ over $d$ state variables at $n$ points each, one sweep costs $n^{2d}|\mathcal{A}|$ operations, and wall-clock assumes $10^{9}$ operations per second. Memory is one float64 value per state. Rows ordered by state count.}
\label{tab:curse_grid_dp}
\begin{tabular}{lrrrrrr}
\toprule
Model & $d$ & $n$ & $|\mathcal{A}|$ & $|\mathcal{S}|$ & Value table & One sweep \\
\midrule
Bus engine, one bus \citep{Rust1987} & 1 & 175 & 2 & 175 & 1.4 kB & 61 $\mu$s \\
Wind farm, base state & 3 & 7 & 11 & 343 & 2.7 kB & 1 ms \\
Brock--Mirman growth & 1 & 500 & 500 & 1,000 & 8.0 kB & 500 ms \\
Wind farm, six state variables & 6 & 7 & 11 & $1.2 \times 10^{5}$ & 941 kB & 3 min \\
Inventory, five products & 5 & 20 & 7,776 & $3.2 \times 10^{6}$ & 26 MB & 922 days \\
Bus engine, fleet of 3 & 3 & 175 & 8 & $5.4 \times 10^{6}$ & 43 MB & 3 days \\
Macro model, ten state variables & 10 & 10 & 10 & $10^{10}$ & 80 GB & $3.2 \times 10^{4}$ yr \\
Bus engine, fleet of 5 & 5 & 175 & 32 & $1.6 \times 10^{11}$ & 1.3 TB & $2.7 \times 10^{7}$ yr \\
\bottomrule
\end{tabular}
\end{table}

The theoretical foundations of this barrier were established through a sequence of results in computational complexity. \citet{papadimitriou1987} proved that computing an optimal policy for finite-horizon MDPs is P-complete, placing it among the hardest problems solvable in polynomial time. For infinite-horizon discounted MDPs, they showed that policy iteration terminates in polynomial time, but the polynomial depends on the number of states and actions, which may themselves be exponential in the natural problem dimension.

The decisive complexity result for continuous-state MDPs comes from \citet{chow1989complexity}. The class of interest is discounted MDPs with $d_s$-dimensional continuous state spaces and $d_a$-dimensional continuous action spaces, where the transition density and reward function are Lipschitz continuous. The computational problem is to compute an $\epsilon$-accurate approximation of the optimal value function in the supremum norm, from oracle evaluations of the reward function and transition density in a real-number model of computation. \citeauthor{chow1989complexity} characterize the worst-case complexity of this problem, and the answer turns on one extra condition on the transition density. Say that the problem \emph{mixes} at rate $\rho > 0$ if $\int_{\mathcal{S}} \min_{s, a} P(s'|s,a) \, ds' \geq \rho$, so that from every state and every action some fixed mass of the next-state distribution lands on a common set and the influence of the initial condition washes out at a fixed rate.

\begin{theorem}[Chow-Tsitsiklis, 1989]
\label{thm:chow_tsitsiklis}
Consider Lipschitz-continuous discounted MDPs with $d_s$-dimensional states, $d_a$-dimensional actions, and discount factor $\gamma$, and let $C(\gamma, \epsilon)$ be the minimum, over correct algorithms, of the worst-case number of oracle queries needed to compute an $\epsilon$-approximation of the optimal value function. On the subclass that mixes at some rate $\rho > 0$,
\begin{equation}
C(\gamma, \epsilon) = \Theta\left( \left( \frac{1}{(1-\gamma)\,\epsilon} \right)^{2d_s + d_a} \right).
\label{eq:ct_mixing}
\end{equation}
Without the mixing condition, and likewise when the kernel is only a subprobability measure,
\begin{equation}
C(\gamma, \epsilon) = \Theta\left( \left( \frac{1}{(1-\gamma)^2\,\epsilon} \right)^{2d_s + d_a} \right).
\label{eq:ct_general}
\end{equation}
\end{theorem}

Both are tight, the upper bounds coming from discretization procedures and the lower bounds from an adversary argument that constructs two instances an algorithm making too few queries cannot distinguish.\footnote{The lower bound without mixing is proved for some Lipschitz constant rather than for every one. \citeauthor{chow1989complexity} note that a small enough constant satisfies the mixing condition automatically, in which case the tighter bound \eqref{eq:ct_mixing} is the best available.} Counting queries rather than arithmetic is the natural measure here, since the oracle supplies the only information the algorithm has about the reward and the transition density. Translating the bounds into arithmetic operations is the subject of a companion multigrid algorithm \citep{chow1989multigrid}, which attains \eqref{eq:ct_mixing} exactly under mixing and comes within a factor $(1-\gamma)^{-1}$ of \eqref{eq:ct_general} without it.

Three things follow. Backward induction on uniform discretization grids attains the exponent $2d_s + d_a$, so no cleverer deterministic scheme improves on simple grids by more than constant and logarithmic factors.\footnote{The exponent falls if more smoothness is available. \citeauthor{chow1989complexity} expect that bounding derivatives of order $r$ replaces $2d_s + d_a$ by a smaller exponent depending on $r$, which is the same mechanism Section~\ref{sec:smoothness_pathway} develops for neural approximation; they state this as an extension rather than prove it. When the action space is a single point the problem is a linear Fredholm equation of the second kind, and there they do show the exponent drops to $2d_s$.} The lower bound proves that no algorithm can do better without additional structural assumptions. And the discount factor is exponentiated alongside the dimensions, so a horizon that a reader would call long is as costly as several extra state variables; mixing is worth exactly one power of $(1-\gamma)$ inside that exponentiated base. Table~\ref{tab:curse_chow_tsitsiklis} evaluates both regimes.

\begin{table}[htbp]
\centering
\caption{Oracle queries required by Theorem~\ref{thm:chow_tsitsiklis} at $\epsilon = 0.01$. The mixing column is \eqref{eq:ct_mixing} and the general column is \eqref{eq:ct_general}; both are tight, so the gap between them is a real difference in problem difficulty and not slack in the analysis. The last column is their ratio, $(1-\gamma)^{-(2d_s+d_a)}$, the price of losing the mixing condition.}
\label{tab:curse_chow_tsitsiklis}
\begin{tabular}{rrrrrrr}
\toprule
$d_s$ & $d_a$ & $\gamma$ & $2d_s + d_a$ & Mixing & General & Ratio \\
\midrule
1 & 1 & 0.9 & 3 & $10^{9}$ & $10^{12}$ & 1,000 \\
1 & 1 & 0.95 & 3 & $8.0 \times 10^{9}$ & $6.4 \times 10^{13}$ & 8,000 \\
1 & 1 & 0.99 & 3 & $10^{12}$ & $10^{18}$ & $10^{6}$ \\
2 & 1 & 0.9 & 5 & $10^{15}$ & $10^{20}$ & $10^{5}$ \\
2 & 1 & 0.95 & 5 & $3.2 \times 10^{16}$ & $10^{23}$ & $3.2 \times 10^{6}$ \\
2 & 1 & 0.99 & 5 & $10^{20}$ & $10^{30}$ & $10^{10}$ \\
3 & 2 & 0.9 & 8 & $10^{24}$ & $10^{32}$ & $10^{8}$ \\
3 & 2 & 0.95 & 8 & $2.6 \times 10^{26}$ & $6.6 \times 10^{36}$ & $2.6 \times 10^{10}$ \\
3 & 2 & 0.99 & 8 & $10^{32}$ & $10^{48}$ & $10^{16}$ \\
\bottomrule
\end{tabular}
\end{table}

The exponent $2d_s + d_a$ admits a heuristic decomposition into the three numerical subproblems nested in the Bellman operator \citep{rust1996numerical}. Maximizing over the action space contributes $\Theta(\epsilon^{-d_a})$, integrating over the transition density contributes $\Theta(\epsilon^{-d_s})$, and approximating the value function over the state space contributes another $\Theta(\epsilon^{-d_s})$; the complexity bound is the product of the three. The result holds for continuous-action problems, deterministic algorithms, and worst-case complexity, and relaxing each of these three qualifications is a way to escape the barrier, most notably the random Bellman operators discussed below.

This complexity result extends the work of \citet{traub1988} on information-based complexity, which established similar exponential lower bounds for multivariate integration and function approximation. The connection is direct. Computing expected values under the Bellman operator involves integrating over the transition density, inheriting the curse from numerical integration. \citeauthor{traub1988} showed that for functions in standard Sobolev spaces, no integration method can achieve error $\epsilon$ with fewer than $\Theta(\epsilon^{-d})$ function evaluations. The MDP problem compounds this with the need to solve a fixed-point equation.

\citet{rust1996numerical} surveys these computational barriers in the context of structural econometric estimation of dynamic discrete choice models. The nested fixed-point algorithm (NFXP) requires solving the Bellman equation to compute choice probabilities at each parameter guess, making the curse of dimensionality a binding constraint on the complexity of estimable models.\footnote{This limitation motivated the development of conditional choice probability methods \citep{HotzMiller1993} and simulation-based estimators that avoid full solution of the dynamic program. \citet{rust1997randomization} showed that a random Bellman operator breaks the curse of dimensionality for discrete-choice MDPs, exploiting the fact that randomization breaks the curse of multivariate integration. \citet{bray2022comment} shows that this class of dynamic programs is limited, in that it requires all but a vanishingly small fraction of the state variables to behave arbitrarily similarly to i.i.d.\ uniform random variables.}

\subsection{Breaking the Curse: Three Pathways}
\label{sec:breaking_curse}

\begin{figure}[htbp]
\centering
\includegraphics[width=0.8\textwidth]{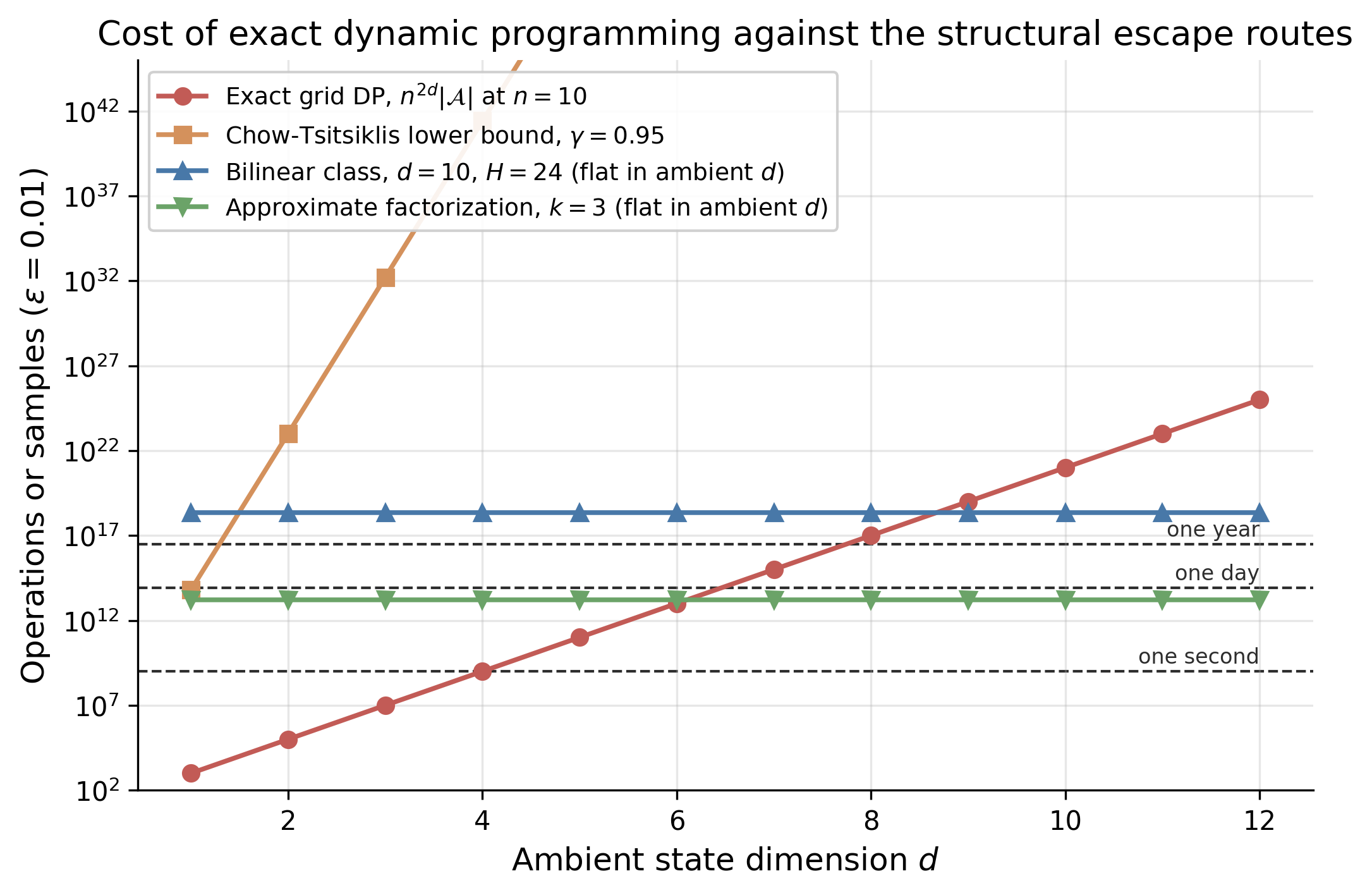}
\caption{Operation and sample counts against ambient state dimension, on a log scale, at $\epsilon = 0.01$. Exact grid dynamic programming and the Chow-Tsitsiklis lower bound of Theorem~\ref{thm:chow_tsitsiklis} grow exponentially in $d$; the bilinear and factored bounds of Theorems~\ref{thm:bilinear_sample} and~\ref{thm:factored_complexity} are constant in $d$ at fixed structural parameters. Dashed lines mark one second, one day and one year at $10^9$ operations per second. The Chow-Tsitsiklis curve leaves the panel because it bounds a harder class, continuous states and actions to a fixed supremum-norm accuracy.}
\label{fig:curse_arithmetic}
\end{figure}

Recent theoretical work has identified three distinct structural conditions under which the curse of dimensionality can be circumvented: bilinear structure in the Bellman error, smoothness exploitable by neural networks, and factorization arising from weak coupling. What the three share is visible in Figure~\ref{fig:curse_arithmetic}. Each replaces the ambient state dimension with a structural quantity, an inherent rank or a parent-set size, so its cost is flat in the dimension that makes exact dynamic programming explode. Each also carries a large constant and a condition that has to be checked, which is what the rest of this section quantifies.

\subsubsection{Bilinear Structure}
\label{sec:bilinear_pathway}

\citet{du2021} introduce the Bilinear Class framework, which unifies and extends prior structural conditions for sample-efficient reinforcement learning. The unifying observation is that many tractable RL settings share a common algebraic property, namely that the Bellman error can be written as a bilinear form.

\begin{definition}[Bilinear Class, Du et al. 2021]
\label{def:bilinear_class}
Consider an MDP $M$, a hypothesis class $\mathcal{H}$, a discrepancy function $\ell_f$, and a set of estimation policies $\Pi_{\text{est}}$. The tuple $(\mathcal{H}, \ell_f, \Pi_{\text{est}}, M)$ is a Bilinear Class if $\mathcal{H}$ is realizable in $M$ and there exist functions $W_h: \mathcal{H} \to \mathcal{V}$ and $X_h: \mathcal{H} \to \mathcal{V}$ for some Hilbert space $\mathcal{V}$ such that for all $f \in \mathcal{H}$ and $h \in [H]$:
\begin{enumerate}
\item The Bellman error decomposes as $\mathbb{E}_{a_{0:h} \sim \pi_f}[Q_{h,f}(s_h, a_h) - r_h - V_{h+1,f}(s_{h+1})] = \langle W_h(f) - W_h(f^*), X_h(f) \rangle$.
\item The discrepancy measure permits estimation. For any $g \in \mathcal{H}$, $|\mathbb{E}_{a_{0:h-1} \sim \pi_f, a_h \sim \pi_{\text{est}}}[\ell_f(o_h, g)]| = |\langle W_h(g) - W_h(f^*), X_h(f) \rangle|$.
\end{enumerate}
\end{definition}

The bilinear structure enables a key computational advantage. Data collected under one hypothesis can be reused to estimate the Bellman error for all hypotheses in the class simultaneously, analogous to uniform convergence in supervised learning. The resulting algorithm, BiLin-UCB, has a sample complexity stated as a reduction to the generalization error of that supervised problem. Writing $\varepsilon_{\text{gen}}(m, \mathcal{H})$ for the uniform-convergence rate of the discrepancy functions over $\mathcal{H}$ at batch size $m$, the suboptimality of the returned hypothesis is bounded by $3H \varepsilon_{\text{gen}}(m, \mathcal{H})$ times $1 + \sqrt{\tilde{d}_m}\,\mathrm{conf}(\cdot)$, where $\tilde{d}_m = H \lceil 3d \ln(1 + 3B_X^2 B_W^2 / \varepsilon_{\text{gen}}^2) \rceil$ is an effective dimension. The amplification is therefore of order $H^{3/2}\sqrt{d}$ up to logarithmic and confidence factors, not $H$ alone, and it is what produces the horizon powers in the explicit count below. Any hypothesis class whose generalization error decays polynomially in $m$ is learnable. Specializing to a finite hypothesis class makes the count explicit:

\begin{theorem}[Du et al. 2021, Corollary 5.1]
\label{thm:bilinear_sample}
Let $(\mathcal{H}, \ell, \Pi_{\text{est}}, M)$ be a Bilinear Class of inherent dimension $d$, with a finite hypothesis class $\mathcal{H}$ and bounded discrepancy. For any $\epsilon \in (0,1)$ and $\delta \in (0, 1/3)$, BiLin-UCB returns an $\epsilon$-optimal policy with probability at least $1 - \delta$, using
\begin{equation}
O\left( \frac{d^2 H^7 \ln(d H^2)\, \ln(|\mathcal{H}|)\, \ln(1/\delta)}{\epsilon^2} \cdot \ln^2\!\left( \frac{d H B_X B_W \ln(|\mathcal{H}|) \ln(1/\delta)}{\epsilon} \right) \right)
\label{eq:bilin_samples}
\end{equation}
trajectories, where $B_X$ and $B_W$ bound the norms of the two bilinear factors.
\end{theorem}

The dependence on the ambient state dimension is gone, replaced by the inherent dimension $d$ of the bilinear factorization. What replaces it is a seventh power of the horizon, and Table~\ref{tab:curse_sample_complexity} shows what that costs at horizons an applied problem would treat as short. Two qualifications limit what the theorem delivers. It bounds trajectories, not arithmetic, and BiLin-UCB searches over $\mathcal{H}$ at each iteration, so a class large enough to be realizable need not be one that can be optimized over. And realizability is assumed, not tested.

\begin{table}[htbp]
\centering
\caption{Trajectories required by \eqref{eq:bilin_samples}, with the absolute constant set to one, at hypothesis-class cardinality $|\mathcal{H}| = 10^{6}$, failure probability $\delta = 0.05$ and $B_X = B_W = 1$. The horizon column isolates the factor $H^7$. Environment steps are trajectories times $H$. The inherent dimension $d$ is the rank of the bilinear factorization, not the ambient state dimension.}
\label{tab:curse_sample_complexity}
\begin{tabular}{rrrrrr}
\toprule
$d$ & $H$ & $\epsilon$ & $H^7$ & Trajectories & Environment steps \\
\midrule
5 & 10 & 0.1 & $10^{7}$ & $6.4 \times 10^{14}$ & $6.4 \times 10^{15}$ \\
10 & 24 & 0.1 & $4.6 \times 10^{9}$ & $2.2 \times 10^{18}$ & $5.2 \times 10^{19}$ \\
10 & 24 & 0.05 & $4.6 \times 10^{9}$ & $9.8 \times 10^{18}$ & $2.3 \times 10^{20}$ \\
20 & 50 & 0.1 & $7.8 \times 10^{11}$ & $2.3 \times 10^{21}$ & $1.2 \times 10^{23}$ \\
\bottomrule
\end{tabular}
\end{table}

The Bilinear Class framework subsumes several previously studied settings, including tabular MDPs, linear MDPs \citep{jin2020}, linear mixture MDPs \citep{AyoubVTR2020}, linear Bellman complete models \citep{zanette2020}, and factored MDPs \citep{kearns1999}. It also introduces new tractable settings, including the Linear $Q^*/V^*$ model where both the optimal Q-function and V-function are linear in known features. The unifying observation is that low Bellman rank (the dimension of the bilinear factorization) implies polynomial learnability.\footnote{The Bilinear Class is related to but distinct from the Bellman Eluder dimension of \citet{jin2021}. Neither framework strictly contains the other; they capture different structural properties that enable efficient learning.}

\subsubsection{Smoothness and Neural Function Approximation}
\label{sec:smoothness_pathway}

The second pathway exploits smoothness of the value function, which neural networks can approximate efficiently. \citet{liu2022deep} analyze deep Q-learning with $\epsilon$-greedy exploration, characterizing when neural function approximation provably works beyond the linear regime.

The key assumption is that the optimal Q-function lies in a Besov space $\mathcal{B}^{\alpha}_{p,q}(\mathcal{X})$, where $\alpha > 0$ is the smoothness parameter. Besov spaces generalize Sobolev and H\"{o}lder spaces, allowing for spatially inhomogeneous smoothness with spikes and jumps. The smoothness parameter $\alpha$ controls the approximation rate, with smoother functions (larger $\alpha$) admitting more efficient approximation.

\begin{assumption}[Bellman Smoothness, Liu et al. 2022]
\label{ass:bellman_smoothness}
Let $\tilde{R}$ be a fixed constant. Define $\mathcal{B}_{\tilde{R}} = \{f \in \mathcal{B}^{\alpha}_{p,q}(\mathcal{X}): \|f\|_{\mathcal{B}} \leq \tilde{R}\}$. For any $h \in [H]$ and $Q: \mathcal{S} \times \mathcal{A} \to [0, H]$, the Bellman operator satisfies $T^*_h Q \in \mathcal{B}_{\tilde{R}}$.
\end{assumption}

Since $Q$ takes values in $[0, H]$, the radius $\tilde{R}$ grows with the horizon, $\tilde{R} \asymp H$. Under this assumption, \citeauthor{liu2022deep} prove that value iteration with deep ReLU networks achieves sublinear regret:

\begin{theorem}[Liu et al. 2022, Theorem 1]
\label{thm:deep_rl_regret}
Under Assumption~\ref{ass:bellman_smoothness} with smoothness $\alpha > d(1/p - 1/4)_+$, value iteration via deep ReLU networks with depth and width
\begin{equation}
L \asymp \frac{d}{2\alpha + d} \log T, \qquad m \asymp \frac{d}{2\alpha + d}\, T^{\frac{d}{2\alpha + d}} \log T
\label{eq:liu_architecture}
\end{equation}
achieves, for an MDP-dependent constant $K \in [1, H]$ and an action set of size $A$, regret
\begin{equation}
\text{Regret}(T) = \tilde{O}\left( H^{\frac{H+4}{H+2}} K^{\frac{2}{K+2}} A^{\frac{K}{K+2}}\, T^{\frac{\alpha K + (\alpha + d)(K+2)}{(2\alpha + d)(K+2)}} \right).
\label{eq:liu_regret}
\end{equation}
\end{theorem}

The constant $K$ measures how myopic the problem is under $\epsilon$-greedy exploration, $K = 1$ under dense informative rewards and $K = H$ in the sparse-reward worst case. The regret exponent improves with smoothness. As $\alpha \to \infty$ it approaches $(K+1)/(K+2)$, matching the $\epsilon$-greedy contextual bandit rate; at $K = 1$ that is $T^{2/3}$. For the Barron space of two-layer networks, width $m = \Omega(\sqrt{T})$ suffices and the exponent becomes $(2K+3)/(2K+4)$, free of $d$ altogether.

The mechanism is that smooth functions can be approximated by neural networks with polynomially many parameters in the accuracy $\epsilon$, rather than the exponential scaling that generic Lipschitz functions require. The curse shifts from state-space size to the smoothness-to-dimension ratio $\alpha/d$. That is a shift, not a removal. Table~\ref{tab:curse_smoothness} evaluates the exponent and the width \eqref{eq:liu_architecture} the theorem demands: at a ratio below roughly one the exponent sits close to unity, the guarantee is sublinear only in the formal sense, and the network the theorem prescribes is wider than any that gets trained. The condition $\alpha > d(1/p - 1/4)_+$ binds before that: at ten state variables a twice-differentiable $Q$-function is already inadmissible, and at fifty even an eight-times-differentiable one is.

\begin{table}[htbp]
\centering
\caption{What Theorem~\ref{thm:deep_rl_regret} asks of the problem, at Besov index $p = 2$, benign exploration $K = 1$ and $T = 10^{6}$ episodes. The admissibility condition $\alpha > d(1/p - 1/4)_+$ demands a minimum smoothness that grows linearly in the ambient dimension, reported in the second column. Each smoothness level then gives the regret exponent and the width \eqref{eq:liu_architecture} the theorem prescribes; n/a marks a level that fails the condition, where the theorem gives no guarantee at all. Regret is sublinear in every admissible cell, but the exponent approaches one as the smoothness-to-dimension ratio falls.}
\label{tab:curse_smoothness}
\begin{tabular}{rrrrrr}
\toprule
 & & \multicolumn{2}{c}{$\alpha = 2$} & \multicolumn{2}{c}{$\alpha = 8$} \\
\cmidrule(lr){3-4} \cmidrule(lr){5-6}
$d$ & Required $\alpha$ & Exponent & Width $m$ & Exponent & Width $m$ \\
\midrule
1 & 0.25 & 0.733 & 44 & 0.686 & 2 \\
2 & 0.50 & 0.778 & 461 & 0.704 & 7 \\
3 & 0.75 & 0.810 & 2,207 & 0.719 & 19 \\
5 & 1.25 & 0.852 & $1.7 \times 10^{4}$ & 0.746 & 88 \\
10 & 2.50 & n/a & n/a & 0.795 & 1,079 \\
20 & 5.00 & n/a & n/a & 0.852 & $1.7 \times 10^{4}$ \\
50 & 12.50 & n/a & n/a & n/a & n/a \\
\midrule
\multicolumn{6}{l}{As $\alpha \to \infty$ the exponent falls to 0.667 for every $d$, and the width requirement vanishes.} \\
\bottomrule
\end{tabular}
\end{table}

\subsubsection{Factorization and Weak Coupling}
\label{sec:factored_pathway}

The third pathway exploits structural sparsity in the transition dynamics. \citet{lu2025} study factored MDPs where the state vector $s = (s_1, \ldots, s_n)$ consists of $n$ components, and transitions factorize across components:
\begin{equation}
P(s'|s, a) = \prod_{i=1}^{n} P_i(s'_i | s_{\text{pa}(i)}, a),
\end{equation}
where $\text{pa}(i) \subseteq [n]$ denotes the parent set of component $i$. In an exactly factored MDP, each component's next value depends only on a small subset of current components, encoded in a directed graph.

The challenge is that exact factorization rarely holds in practice. \citeauthor{lu2025} extend the theory to approximate factorization, where the true transition kernel is close to a product of component kernels but not equal to it. Writing $x = (s,a)$ for a state-action pair, $Z^S_k$ and $Z^P_k$ for the scopes of component $k$ on the state and on the state-action space, and $x[Z]$ for the restriction of $x$ to a scope $Z$, the approximation error of a scheme $\omega$ is
\begin{equation}
\Delta_P^\omega = \sup_{P_1, \ldots, P_{K_\omega}} \; \max_{s' \in \mathcal{S},\, x \in \mathcal{X}} \left| P(s' \mid x) - \prod_{k=1}^{K_\omega} P_k\!\left( s'[Z^S_k] \mid x[Z^P_k] \right) \right|,
\label{eq:lu_coupling}
\end{equation}
the largest pointwise gap between the true kernel and the factored one, over every admissible choice of component marginals.\footnote{This is a pointwise maximum, not a total-variation distance, so it carries neither a sum over successor states nor a factor of one half. The supremum over component marginals makes the guarantee hold for whichever choice the algorithm makes.} A matching quantity $\Delta_R^\omega$ measures the reward decomposition. Weak coupling, a common property of networked systems, is what makes $\Delta_P^\omega$ small.

\begin{definition}[Weak Coupling]
\label{def:weak_coupling}
An MDP exhibits weak coupling if the transition of each state component $s_i$ depends strongly on a small set of parents $\text{pa}(i)$ and only weakly on the remaining components, so that the approximation error $\Delta_P^\omega$ of a scheme that drops the remaining dependencies is small.
\end{definition}

Approximate factorization buys a smaller sample requirement but does not deliver an exact solution. The error it introduces never vanishes with more data:

\begin{theorem}[Lu et al. 2025, Theorem 5.1]
\label{thm:factored_complexity}
Fix a factorization scheme $\omega$ with scope sets $Z^P_k$ for the transition components and $Z^R_i$ for the reward components, and write $|\mathcal{X}[Z]|$ for the number of state-action configurations a scope $Z$ can take. Define the misspecification bias
\begin{equation}
\mathcal{E}_\omega = \frac{\gamma}{(1-\gamma)^2}\, \Delta_P^\omega + \frac{1}{1-\gamma}\, \Delta_R^\omega .
\label{eq:lu_bias}
\end{equation}
For any $\epsilon \in (0,1)$ and $\delta > 0$, with probability at least $1 - \delta$, model-based Q-value iteration on the factored estimate returns $\hat{Q}^*_\omega$ satisfying
\begin{equation}
\|\hat{Q}^*_\omega - Q^*\|_\infty \leq \epsilon + \mathcal{E}_\omega
\label{eq:lu_accuracy}
\end{equation}
once the sample count satisfies
\begin{equation}
D_\omega \gtrsim \frac{\left( \sum_{k \in [\kappa_p]} |\mathcal{X}[Z^P_k]| \right) \log\left( |\mathcal{X}[\cup_k Z^P_k]| / \delta \right)}{\epsilon^2 (1-\gamma)^3} + \sum_{i \in [\kappa_r]} |\mathcal{X}[Z^R_i]| .
\label{eq:lu_samples}
\end{equation}
\end{theorem}

The comparison is against the minimax rate $\tilde{O}(|\mathcal{S}||\mathcal{A}| / (\epsilon^2 (1-\gamma)^3))$ for an unstructured MDP \citep{azar2013}. The dependence on the discount factor and the accuracy is identical; what changes is that the full state-action count $|\mathcal{S}||\mathcal{A}|$ is replaced by a sum of scope sizes, which is exponentially smaller because each scope covers a few components rather than all of them.

Equation~\eqref{eq:lu_accuracy} is where the restriction sits. The guarantee is $\epsilon + \mathcal{E}_\omega$, not $\epsilon$, and $\mathcal{E}_\omega$ is a floor that no amount of additional data lowers. Chasing an accuracy below the floor is wasted sampling. The $(1-\gamma)^{-2}$ multiplying $\Delta_P$ makes the floor sharply sensitive to the discount factor, so a coupling error that is negligible at $\gamma = 0.9$ can dominate the target accuracy at $\gamma = 0.99$. Table~\ref{tab:curse_factored} reports both blocks, the scope sizes that make the escape work and the coupling error at which it stops working. Panel (b) puts a number on the word small. At $\gamma = 0.9$ a coupling error of one part in ten thousand leaves the floor just inside a target accuracy of $0.01$; at $\gamma = 0.99$ the same coupling error puts the floor two orders of magnitude above that target, and the coupling would have to fall to roughly one part in a million to recover it.

\begin{table}[htbp]
\centering
\caption{Theorem~\ref{thm:factored_complexity} in two parts, for a system of 100 components with 10 states each and $|\mathcal{A}| = 2$, so the unstructured problem has $|\mathcal{S}||\mathcal{A}| = 2.0 \times 10^{100}$. Panel (a) is what the factorization buys: the scope sum that enters \eqref{eq:lu_samples} in place of $|\mathcal{S}||\mathcal{A}|$, as the maximum parent-set size $k$ grows. Panel (b) is what it costs: the bias floor $\mathcal{E}_\omega$ of \eqref{eq:lu_bias} at $\Delta_R = 0$, which bounds accuracy from below however many samples are drawn. Cells exceeding a target $\epsilon = 0.01$ are the settings where the target is unreachable.}
\label{tab:curse_factored}
\begin{tabular}{rrr}
\multicolumn{3}{l}{(a) Scope sizes against the full state-action count} \\
\toprule
Parent-set size $k$ & One scope & Scope sum \\
\midrule
1 & 20 & 2,000 \\
2 & 200 & $2.0 \times 10^{4}$ \\
3 & 2,000 & $2.0 \times 10^{5}$ \\
4 & $2.0 \times 10^{4}$ & $2.0 \times 10^{6}$ \\
6 & $2.0 \times 10^{6}$ & $2.0 \times 10^{8}$ \\
\bottomrule
\end{tabular}

\vspace{1em}

\begin{tabular}{rrrrrr}
\multicolumn{6}{l}{(b) Misspecification bias floor $\mathcal{E}_\omega$ by coupling error} \\
\toprule
$\gamma$ & $\Delta_P = 10^{-4}$ & $\Delta_P = 10^{-3}$ & $\Delta_P = 10^{-2}$ & $\Delta_P = 10^{-1}$ & Break-even $\Delta_P$ \\
\midrule
0.9 & 0.009 & 0.09 & 0.9 & 9 & $1.1 \times 10^{-4}$ \\
0.95 & 0.038 & 0.38 & 3.8 & 38 & $2.6 \times 10^{-5}$ \\
0.99 & 0.99 & 9.9 & 99 & 990 & $10^{-6}$ \\
\bottomrule
\end{tabular}
\end{table}

The practical implication is that networked systems, such as supply chains, power grids, and multi-agent coordination problems, are amenable to methods that exploit the factored structure even when exact independence fails. The scheme $\omega$ is a choice, and it trades the two terms against each other. Coarser scopes shrink $\sum_k |\mathcal{X}[Z^P_k]|$ and so the sample requirement \eqref{eq:lu_samples}, while raising $\Delta_P^\omega$ and so the floor \eqref{eq:lu_bias}. Consistency holds only in the exactly factored case, where $\mathcal{E}_\omega = 0$.

\subsubsection{Simulation Study: Wind Farm Storage Control}
\label{sec:wind_farm_simulation}

The barrier of Theorem~\ref{thm:chow_tsitsiklis} and the structural escape routes of Section~\ref{sec:breaking_curse} can both be exhibited in a single control problem, a wind farm with battery storage adapted from the scaling experiment in \citet{lu2025}. Each hour the operator chooses a charge or discharge rate $a_t \in [-20, 20]$ kW, facing stochastic wind generation, spot prices, and demand, over a finite horizon of $H = 24$ hours with discount factor $\gamma = 0.95$. The base state ($d = 3$) is $s_t = (w_t, p_t, c_t)$, where $w_t \in [0, 100]$ is wind generation (kW), $p_t \in [0, 1]$ is the spot price (\$/kWh), and $c_t \in [0, 50]$ is the battery state of charge (kWh); scaling experiments append auxiliary state variables $x_t^{(4)}, \ldots, x_t^{(d)} \in [0, 1]$ that follow AR(1) processes weakly coupled to wind. Transitions are
\begin{align}
w_{t+1} &= 0.7 w_t + \epsilon_t^w, \quad \epsilon_t^w \sim \mathcal{N}(30, 25), \\
p_{t+1} &= 0.6 p_t + 0.05 (w_t / 100) + \epsilon_t^p, \quad \epsilon_t^p \sim \mathcal{N}(0.4, 0.01), \\
c_{t+1} &= c_t + 0.9 a_t, \\
x_{t+1}^{(i)} &= 0.8 x_t^{(i)} + 0.01(w_t/100 - 0.5) + 0.1 + \epsilon_t^{(i)}, \quad \epsilon_t^{(i)} \sim \mathcal{N}(0, 0.0025),
\end{align}
with all variables clipped to their bounds, and the reward is
\begin{equation}
r(s_t, a_t) = p_t \cdot \min(w_t + a_t, D_t) - 0.01 c_t - 5 \cdot \max(0, D_t - w_t - a_t),
\end{equation}
where demand $D_t \sim \text{Poisson}(50 + 10 \sin(2\pi t / 24))$ follows a diurnal pattern. The reward depends only on the base state, so the auxiliary dimensions are payoff-irrelevant yet inflate the state space that any tabular method must enumerate. Four solution methods run at each $d \in \{3, 4, 5, 6\}$. Tabular DP discretizes each dimension into 7 bins and solves by backward induction with Monte Carlo integration over transitions, a benchmark under coarse discretization rather than an exact optimum, granted a 10 minute budget per dimension. Factored RL runs Q-learning with one Q-table per state dimension, aggregated additively; it is given the identity of the three reward-relevant dimensions in advance, so it instantiates the known-structure premise of factored MDPs rather than the algorithms of \citet{lu2025}. DQN is a standard deep Q-network with experience replay and a target network, illustrating the neural function-approximation pathway rather than the specific algorithm analyzed by \citet{liu2022deep}. Linear AC is an actor-critic with linear value and policy functions in polynomial features (intercept, linear, squared, and pairwise cross terms), illustrating the linear function-approximation pathway rather than the BiLin-UCB algorithm of Theorem~\ref{thm:bilinear_sample}. The three RL implementations are illustrative analogies to the three pathways of Section~\ref{sec:breaking_curse}, not implementations of the cited algorithms.\footnote{All methods share the same 11-point action grid. RL methods train for 3{,}000 episodes per seed across 10 seeds; DP uses 20 Monte Carlo samples per state-action pair with a fixed integration seed. Policies are evaluated greedily over 50 episodes on a common evaluation environment (seed 99), so all methods face the same shock sequences. Factored RL uses 30 bins per dimension, learning rate 0.2, and exploration rate 0.2. DQN uses two hidden layers of 128 units, learning rate $5 \times 10^{-4}$, a replay buffer of 10{,}000 transitions, batch size 64, target updates every 100 steps, and $\epsilon$-greedy exploration decaying from 0.3 to 0.01. Linear AC uses actor and critic learning rates of 0.02 and 0.08.}

\begin{figure}[htbp]
\centering
\includegraphics[width=0.7\textwidth]{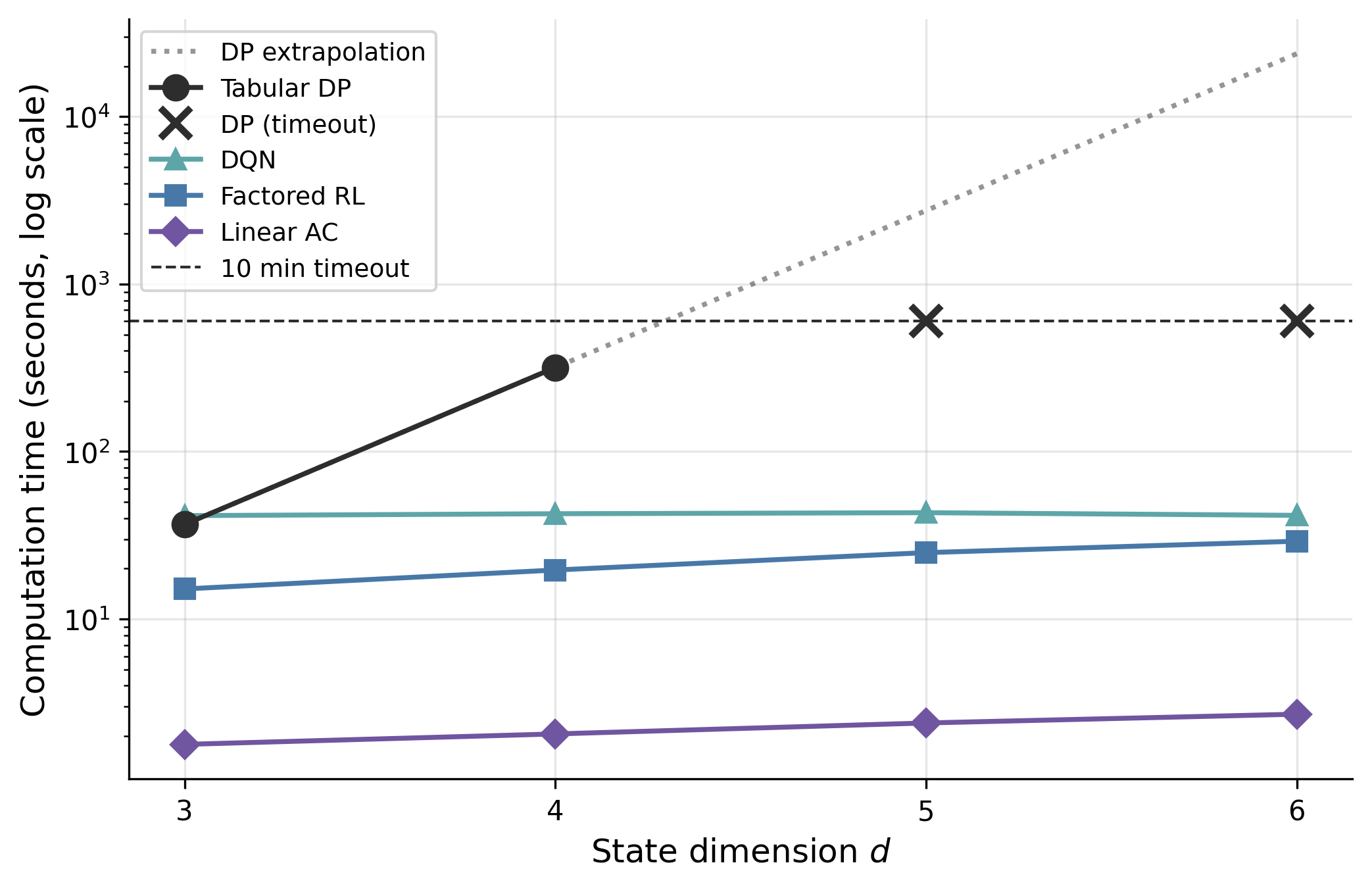}
\caption{Computation time versus state dimension on a log scale. Circles trace completed tabular DP runs and crosses mark runs stopped at the 10 minute budget (dashed reference line); the dotted line extrapolates the exponential fit through the completed DP runs. RL curves show mean training time per seed over 10 seeds.}
\label{fig:curse_scaling}
\end{figure}

Table~\ref{tab:curse_results} reports average returns and Figure~\ref{fig:curse_scaling} reports computation times. Tabular DP completes at $d = 3$ (36.7 seconds) and $d = 4$ (317.5 seconds) and exhausts its budget at $d \geq 5$; the exponential fit through its two completed runs grows by a factor of 8.6 per added dimension, implying roughly 45.7 minutes at $d = 5$ and 6.6 hours at $d = 6$. That measured factor is well below the $n^2 = 49$ that Table~\ref{tab:curse_grid_dp} charges an exact sweep at seven bins, because this solver integrates each transition with a fixed 20-sample Monte Carlo draw rather than summing over all $n^d$ successor cells, so it pays one factor of $n^d$ rather than two. The barrier is the same; the constant is smaller because the integral is approximated. The three RL methods train in under a minute per seed at every dimension. Where DP completes, DQN reaches within one percent of the DP return, Factored RL sits roughly two to three percent below it, and Linear AC attains DP-level returns on some seeds while collapsing to substantially worse policies on others, visible in its standard errors. The ranking is unchanged at $d = 5$ and $d = 6$ where DP is intractable. The cost profiles are consistent with the polynomial-complexity scaling predicted by the structural results of Section~\ref{sec:breaking_curse}, with the caveat that the cheapest method is also the least reliable.

\begin{table}[htbp]
\centering
\caption{Average return per episode (\$/day) by method and state dimension. RL cells are mean $\pm$ standard error over 10 training seeds, each evaluated over 50 episodes with common random shocks; the DP row is a single deterministic solve, with TIMEOUT marking dimensions where backward induction exceeded its 10 minute budget. Rows ordered by mean return at $d = 3$.}
\label{tab:curse_results}
\begin{tabular}{lcccc}
\toprule
Method & $d=3$ & $d=4$ & $d=5$ & $d=6$ \\
\midrule
Tabular DP & 1110 & 1113 & TIMEOUT & TIMEOUT \\
DQN & 1108 $\pm$ 0.4 & 1107 $\pm$ 0.7 & 1107 $\pm$ 0.2 & 1100 $\pm$ 0.4 \\
Factored RL & 1091 $\pm$ 4.2 & 1083 $\pm$ 4.1 & 1094 $\pm$ 1.8 & 1084 $\pm$ 4.9 \\
Linear AC & 1002 $\pm$ 29.4 & 1016 $\pm$ 33.9 & 1023 $\pm$ 25.4 & 1009 $\pm$ 30.7 \\
\bottomrule
\end{tabular}

\end{table}

\FloatBarrier
\subsubsection{Engine Replacement MDP: Search as a Partial Improvement Step}
\label{engine:hybrid}

In the Engine Replacement MDP of Section~\ref{engine:model}, one greedy lookahead applies the approximate-value policy improvement bounded by Theorem~\ref{thm:singh_yee}. The calculation is a one-step proxy for the evaluate-improve pattern in Table~\ref{tab:pi_alphazero}, not an AlphaZero or tree-search simulation. The approximate evaluation is the uniform policy's exact value, $v=V^{(\mathrm{unif})}=(1.3387,0.8226)$. One greedy lookahead lands on the $(\text{keep},\text{keep})$ vertex in Figure~\ref{fig:engine_policy_square}. The same vertex is the $\alpha\to\infty$ limit of one natural-gradient step. The improved policy has value $V^{\hat\pi}=(3.4545,2.0000)$ rather than the optimum $V^\star=(5.3448,4.3103)$.

\begin{table}[h]
\centering
\caption{A one-step evaluate-improve proxy on the Engine Replacement MDP. The approximate value $v$ is the uniform policy's exact value; one greedy lookahead lands on the $(\text{keep},\text{keep})$ vertex of Figure~\ref{fig:engine_policy_square}. The bound is Theorem~\ref{thm:singh_yee} at $\gamma = 0.9$.}
\label{tab:engine_hybrid_search}
\begin{tabular}{lrr}
\hline
 & low & high \\
\hline
network value $v = V^{(\mathrm{unif})}$ & 1.3387 & 0.8226 \\
searched policy value $V^{\hat\pi}$ & 3.4545 & 2.0000 \\
optimal value $V^\star$ & 5.3448 & 4.3103 \\
\hline
 & \multicolumn{2}{c}{value} \\
\hline
$\varepsilon = \|v - V^\star\|_\infty$ & \multicolumn{2}{c}{4.0061} \\
actual loss $\|V^\star - V^{\hat\pi}\|_\infty$ & \multicolumn{2}{c}{2.3103} \\
Singh--Yee constant $2\gamma/(1-\gamma)$ & \multicolumn{2}{c}{18.0000} \\
Singh--Yee bound $2\gamma\varepsilon/(1-\gamma)$ & \multicolumn{2}{c}{72.1101} \\
actual loss / bound & \multicolumn{2}{c}{0.0320} \\
\hline
\end{tabular}
\end{table}

For $\varepsilon=4.0061$ and $\gamma=0.9$, Theorem~\ref{thm:singh_yee} gives the bound $72.1101$. The computed loss is $2.3103$, or $0.032$ of the bound (Table~\ref{tab:engine_hybrid_search}). The computed loss therefore verifies the Singh--Yee bound; repeated evaluation and search reaches the optimal vertex in the two policy-improvement steps reported in Section~\ref{engine:policy_space}.

\FloatBarrier
\subsection{Fundamental Tradeoffs}
\label{sec:tradeoffs}

The choice between methods involves distinct tradeoffs rooted in DP structure. Value-based methods target $Q^*$ via the Bellman contraction \citep{szepesvari2010}. In the tabular case convergence is guaranteed, but function approximation introduces the deadly triad. Policy-based methods optimize $\pi_\theta$ directly. Modern theory \citep{agarwal2021theory} establishes global convergence for softmax policies, with high variance as the practical weakness rather than local traps. Actor-critic methods combine both \citep{konda2000}, using the critic for low-variance value estimates while the actor inherits policy gradient's global convergence. Each family traces to DP foundations.

Four additional trade-offs pervade reinforcement learning. First, \emph{exploration versus exploitation}: should the agent act on its current best estimate or gather information to improve future decisions? \citet{LaiRobbins1985} establish the fundamental lower bound: any consistent policy must incur regret at least logarithmic in the number of periods. Naive exploration ($\varepsilon$-greedy) requires samples exponential in the horizon; strategic exploration (UCB, optimism in the face of uncertainty) reduces this to polynomial \citep{auer2002}, formalizing the value of targeted experimentation.\footnote{The exploration-exploitation tradeoff is the subject of the bandits chapter, where the multi-armed bandit framework provides the sharpest analysis. In full MDPs, exploration is harder because the agent must learn not just reward distributions but also transition dynamics, compounding the information requirement.}

Second, \emph{model-based versus model-free}: model-based methods learn a transition model $\hat{P}(s'|s,a)$ and plan with it \citep{sutton1990}; model-free methods learn value functions or policies directly from transitions. The Dyna architecture \citep{sutton1990} bridges these by generating simulated experience from the learned model to supplement real transitions. Model-based methods are sample-efficient (each transition updates the entire model, which improves value estimates for all states) but suffer asymptotic bias if the model class is misspecified; model-free methods are asymptotically unbiased but sample-inefficient, using each transition for a single gradient step.\footnote{Model misspecification in RL is the analog of omitted variable bias in econometrics: if the learned model omits relevant state variables or misspecifies the functional form of transitions, the resulting policy is biased regardless of sample size. \citet{moerland2023} provide a comprehensive survey of model-based RL, analyzing the model-bias versus sample-efficiency tradeoff across method families.}

Third, \emph{on-policy versus off-policy} \citep[Ch.~5--7]{sutton2018}: on-policy methods (SARSA, REINFORCE) learn from data generated by the current policy, ensuring stability but discarding past experience; off-policy methods (Q-learning, DQN) reuse stored experience via replay buffers, gaining sample efficiency but risking the instabilities of the deadly triad.

Fourth, \emph{bias versus variance in advantage estimation}: REINFORCE uses the full Monte Carlo return (unbiased but high variance); actor-critic methods \citep{konda2000} use bootstrapped TD targets (low variance but biased by the critic's approximation error). Generalized Advantage Estimation in PPO \citep{Schulman2015} interpolates between these extremes via a parameter $\lambda \in [0,1]$, where $\lambda = 1$ recovers Monte Carlo returns and $\lambda = 0$ recovers one-step TD. The value function baseline is the variance-minimizing choice, motivating the actor-critic architecture as a bias-variance compromise.

\subsection{Conclusion}
\label{sec:ch2_conclusion}

RL algorithms are asymptotic approximations to classical dynamic programming operators, justified by the mathematics of contractions, stochastic approximation, and gradient domination. Value iteration becomes Q-learning \citep{WatkinsDayan1992, tsitsiklis1994} when expectations are replaced by single samples. Policy iteration becomes the natural policy gradient \citep{Kakade2001, agarwal2021theory} when the greedy improvement step is approximated by gradient ascent, and NPG recovers PI exactly in the tabular case. The stochastic approximation framework, from the foundational work of \citet{robbinsmonro1951} through the ODE method of \citet{borkar2000}, guarantees that under appropriate step-size conditions, noisy iterates converge to the same fixed points as their deterministic counterparts. Reinforcement learning extends dynamic programming. Tabular RL and RL with linear function approximation rest on solid theoretical foundations. Deep RL lacks comparable guarantees: convergence remains an open problem, and empirical successes remain case-specific.

\section{The Empirics of Deep Reinforcement Learning}
\label{section:deeprl_practice}

I review the empirical pathologies of deep reinforcement learning, their causes, and the diagnostic tools that expose them.

\subsection{The Moving Target Problem}
\label{sec:moving_target}

In supervised learning, the loss function is a \textit{fixed function} of the training data and model parameters. Deep reinforcement learning does not enjoy this property. Each gradient step moves both the current value estimates and the targets simultaneously, creating a ``nonstationary" optimization landscape. The target network heuristic introduced by \citet{mnih2015} slows target drift by periodically freezing a copy of the network, but does not eliminate it. Therefore Bellman residual is a poor proxy for the accuracy of the value function.

\citet{Fujimoto2022} formalize this observation. Let $Q^\pi$ denote the true value function for policy $\pi$, let $\Delta(s,a) = Q(s,a) - Q^\pi(s,a)$ denote the value error, and let $\varepsilon(s,a) = Q(s,a) - (r + \gamma \mathbb{E}_{s',a'}[Q(s',a')])$ denote the Bellman error. Substituting the definition of $Q^\pi$ yields the key identity: $\varepsilon(s,a) = \Delta(s,a) - \gamma \mathbb{E}_{s',a'}[\Delta(s',a')]$. The Bellman error is a \emph{difference} of value errors at consecutive states, not the value error itself. If the errors $\Delta(s,a)$ and $\Delta(s',a')$ are correlated across time, they can partly cancel even when the individual errors are large.\footnote{A constant shift $C$ does not make the residual zero. It adds $(1-\gamma)C$ to the Bellman error, so discounting attenuates the shift but does not eliminate it.}

The second failure mode is specific to finite datasets: \citet[Corollary~1]{Fujimoto2022} show that over an incomplete dataset, zero Bellman error is consistent with arbitrarily large value error, because the network can fit unobserved successor pairs to whatever values make the observed residuals vanish.\footnote{The Bellman equation uniquely identifies $Q^\pi$ when enforced over the entire MDP, but over an incomplete dataset it admits infinitely many solutions. Whenever a transition $(s',a')$ that is reachable from the dataset is not itself in the dataset, the network is free to set $Q(s',a')$ at unobserved pairs to whatever value makes the residual vanish on the observed ones, unconstrained by any loss. A network can thus reach near-zero training loss while the value function remains arbitrarily inaccurate over the full state space.}

The dual failure mode appears in policy gradient methods: \citet{IlyasFang2020} find that even when PPO's surrogate objective improves monotonically, episode return can plateau or decline, because the surrogate gradient is poorly aligned with the gradient of the true return.\footnote{\citet{IlyasFang2020} examine the surrogate objective used in Proximal Policy Optimization \citep{Schulman2017}. The PPO clipping mechanism is designed to keep policy updates within a trust region by bounding the probability ratio $\pi_\theta(a|s)/\pi_{\theta_{\text{old}}}(a|s)$. The gradient of the surrogate is poorly aligned with the gradient of the true return, particularly in later training phases where the policy has diverged from the behavior policy used to collect the replay data. The loss metric that practitioners monitor throughout training is measuring something other than what they care about.}

\subsection{The Reproducibility Crisis and Sensitivity to Random Seeds}
\label{sec:reproducibility}

\citet{henderson2018deep} trained five leading policy gradient algorithms (PPO, TRPO, DDPG, TD3, SAC) on six MuJoCo benchmark environments, holding all hyperparameters fixed and varying only the random seed. The resulting learning curves from different seeds were non-overlapping: a seed that performed well under one algorithm performed comparably to a different algorithm's best seeds, making cross-algorithm comparison unreliable.\footnote{Differences as large as 2,000 points in final episode return arose from seed variation alone.} \citet{Agarwal2021} quantify the problem: comparing point estimates from 5 runs per task on Atari 100k yields Type I error exceeding 50\%, meaning a random noise injection appears beneficial in half of all comparisons.

\citet{Agarwal2021} propose the \emph{interquartile mean} (IQM) as a replacement for mean and median when comparing algorithms. The IQM discards the top and bottom 25\% of runs before averaging, reducing sensitivity to outlier seeds. Using these tools and stratified bootstrap confidence intervals, \citet{Agarwal2021} find that several widely-cited algorithmic improvements on Atari 100k vanish or reverse when statistical uncertainty is accounted for.\footnote{They also introduce performance profiles, which plot the fraction of tasks and seeds where an algorithm achieves performance above a threshold $\tau$, as $\tau$ varies from 0 to the maximum. Performance profiles reveal the full shape of the score distribution rather than collapsing it to a single statistic.}\footnote{The fragility extends to hyperparameters. \citet{Eimer2023} conduct a systematic study of hyperparameter sensitivity across 6 algorithms and 17 environments, finding that default hyperparameters from published papers perform competitively in the specific environments used in those papers but generalize poorly across environments. \citet{Patterson2024} synthesize these findings into an empirical design handbook, recommending at minimum 10 seeds per configuration, IQM-based comparisons, and preregistration of hyperparameter search protocols.}

\subsection{Value Overestimation and Spikes}
\label{sec:overestimation}

Q-learning uses the Bellman optimality update
\begin{equation}
    Q(s,a) \leftarrow r + \gamma \max_{a'} Q(s', a'),
    \label{eq:q_update}
\end{equation}
where the maximum is taken over the estimated Q-values of all actions at the successor state. \citet{ThrunSchwartz1993} identify a positive bias intrinsic to this update: if the Q-value estimates contain noise with mean zero, the maximum over noisy estimates is biased upward by Jensen's inequality. An agent that uses a single network for both action selection ($\arg\max$) and value estimation ($\max Q$) systematically overestimates the values of every state it visits, biasing the Bellman target upward at every update step. The bias compounds through bootstrapping: overestimated targets produce overestimated updates, which produce further overestimated targets.

\citet{vanHasselt2010} propose double Q-learning as a remedy: maintain two independent Q-networks $Q_A$ and $Q_B$. Use $Q_A$ to select the greedy action at $s'$, but use $Q_B$ to evaluate that action. Because the two networks are trained on different data, their errors are approximately independent, and the positive bias largely cancels. \citet{vanHasselt2016ddqn} implement this as Double DQN, using the online network for action selection and the periodically-frozen target network for evaluation. On 49 Atari games, Double DQN reduces overestimation by a factor of 3--5 and improves median performance by 20\% relative to DQN.

\citet{fujimoto2018td3} observe that Double DQN's correction is incomplete in continuous-action settings, where the target network and online network remain correlated through shared updates. They propose Clipped Double Q-learning: compute two Q-value estimates $Q_1, Q_2$ with separate networks trained on the same data, and use $y = r + \gamma \min(Q_1(s',a'), Q_2(s',a'))$ as the Bellman target. The minimum operator introduces pessimistic underestimation, which is conservative but avoids the explosive positive bias.\footnote{\citet{Ciosek2019} note that clipped double Q can cause excessive pessimism under high uncertainty, proposing optimistic actor-critic as a counterweight.}

DQN prevents outright divergence, but \citet{vanHasselt2018} find that \emph{soft divergence}, defined as a temporary spike in value estimates by more than 10\% followed by recovery, occurs in the majority of DQN training runs across 57 Atari games.\footnote{Larger networks diverge more frequently than smaller ones, counter to the usual intuition that more expressive models should generalize better.} The deadly triad produces no obvious training failure: the value estimates may spike and recover, leaving no visible trace in the loss curve while corrupting the policy.

\citet{KumarImplicit2021} describe a continuous manifestation of the deadly triad: \emph{implicit under-parameterization}.\footnote{Neural networks trained with bootstrapped TD objectives progressively lose their effective rank, with fewer and fewer neurons contributing distinct directions in the representation. This rank collapse is silent (training loss continues to decrease) but the network's ability to represent new information degrades over time, a form of capacity loss distinct from but related to plasticity loss.}

\FloatBarrier
\subsection{Engine Replacement MDP: Zero Bellman Error with Large Value Error}
\label{engine:ch03b}

The Engine Replacement MDP of Section~\ref{engine:model} separates two failure modes. Fixed-policy Bellman equations expose residual cancellation on observed state-action pairs. A separate noisy maximum calculation exposes maximization bias.

\begin{table}[H]
\centering
\caption{Bellman residuals and Gaussian maximum bias in the Engine Replacement MDP. The target policy keeps at low mileage and replaces at high mileage. The data omit the high-mileage replacement pair.}
\label{tab:engine_bellman_error}
\small
\begin{tabular}{llcrrr}
\hline
state & action & in data & $Q^\pi$ & $\widehat Q$ & $|\widehat Q-Q^\pi|$ \\
\hline
low & keep & yes & 5.3448 & 13.5266 & 8.1818 \\
low & replace & yes & 4.3103 & 11.6740 & 7.3636 \\
high & keep & yes & 4.0793 & 13.0793 & 9.0000 \\
high & replace & no & 4.3103 & 14.3103 & 10.0000 \\
\hline
\multicolumn{3}{l}{maximum observed-pair residual} & \multicolumn{3}{c}{1.8e-15} \\
\hline
$\sigma$ & \multicolumn{2}{c}{Gaussian maximum bias} & \multicolumn{3}{c}{bias divided by the $0.2310$ gap} \\
\hline
0.05 & \multicolumn{2}{c}{0.000010} & \multicolumn{3}{c}{0.0000} \\
0.10 & \multicolumn{2}{c}{0.003035} & \multicolumn{3}{c}{0.0131} \\
0.20 & \multicolumn{2}{c}{0.033003} & \multicolumn{3}{c}{0.1428} \\
0.30 & \multicolumn{2}{c}{0.078233} & \multicolumn{3}{c}{0.3386} \\
0.40 & \multicolumn{2}{c}{0.128723} & \multicolumn{3}{c}{0.5572} \\
0.50 & \multicolumn{2}{c}{0.181502} & \multicolumn{3}{c}{0.7856} \\
\hline
\end{tabular}
\end{table}

The target policy keeps the engine at low mileage and replaces it at high mileage. The data contain three of the four state-action pairs and omit high-mileage replacement, even though low-mileage keeping reaches that pair with positive probability. Setting the missing estimate ten units above $Q^\pi$ and solving the three observed Bellman equations gives zero residual on every observed pair while their absolute value errors range from $7.36$ to $9.00$.

The same table evaluates Gaussian maximum bias against the model's true high-mileage action gap of $0.2310$. With independent mean-zero errors of standard deviation $\sigma$, the bias is $0.000010$ at $\sigma=0.05$ and $0.181502$ at $\sigma=0.50$. At $\sigma=0.50$, the bias equals $78.6$ percent of the gap. Maximization can therefore change the Bellman target even when each action-value estimate is unbiased.

The computed residuals verify the finite-data construction in Corollary~1 of \citet{Fujimoto2022}, while the bias calculation quantifies the effect of the maximum in update~\eqref{eq:q_update}.
\FloatBarrier

\subsection{Plasticity Loss and Primacy Bias}
\label{sec:plasticity}

A network that trains well at step $t = 100$ may be incapable of learning at step $t = 100{,}000$, even if the data quality at the later step is higher. \citet{Lyle2022} call this \emph{capacity loss}: the network's ability to update its own weights degrades progressively during training, measured by the fraction of effective parameters and the network's ability to fit new random labels. \citet{Lyle2023} extend this to \emph{plasticity loss}, identifying dead ReLU neurons, weight norm growth, and feature rank collapse as three distinct mechanisms, not all of which co-occur.

\citet{Nikishin2022} identify \emph{primacy bias} as a specific cause. Because the replay buffer is filled incrementally, early transitions are oversampled relative to later ones, and the network over-fits to early environment transitions, corrupting representations throughout the remainder of training.\footnote{\citet{Nikishin2022} show that 100 initial ``priming'' steps of excessive gradient updates degrade a SAC agent's performance for hundreds of thousands of subsequent steps. \citet{Sokar2023} measure the fraction of dormant neurons (units with near-zero activation across the replay buffer) accumulating monotonically during training. \citet{DohareEtAl2024} find that standard deep networks lose all plasticity within a few million gradient steps in continual learning tasks.}

The proposed remedies fall into three categories: periodic resets, continual backpropagation, and architectural interventions.\footnote{Periodic resets reinitialize the last layers of the network while retaining the replay buffer, allowing the agent to forget overfit representations without discarding experience \citep{Nikishin2022, Doro2023}. Continual backpropagation replaces neurons with near-zero utility at each gradient step rather than waiting for a global reset \citep{DohareEtAl2024}. Architectural interventions, layer normalization \citep{Lyle2025}, orthogonal initialization, spectral normalization, reduce the rate at which plasticity is lost by stabilizing gradient magnitudes and preventing weight norm growth.}

\subsection{Implementation Dominates Algorithmic Innovation}
\label{sec:implementation}

\citet{Engstrom2020} find that PPO with the clipping mechanism disabled performs indistinguishably from the full algorithm; TRPO \citep{Schulman2015} with the same code-level additions matches PPO.\footnote{The non-clipping components that suffice are: observation normalization, reward normalization, value function clipping, global gradient norm clipping, orthogonal weight initialization, and the Adam optimizer.} \citet{andrychowicz2021matters} identify observation normalization, orthogonal initialization, and learning rate annealing as the three choices accounting for most variance across 250,000 agents and 250 hyperparameter configurations. \citet{huang2022ppo} catalog 37 implementation details required to reproduce PPO on Atari;\footnote{These include reward clipping to $[-1, 1]$, frame stacking to 4 frames, a specific episode termination convention, and a numerically stable normalization of the advantage estimate.} omitting any produces materially different results.

The most consequential implementation detail is the distinction between termination and truncation \citep{pardo2018time}. In reinforcement learning, an episode can end for two distinct reasons: \emph{termination}, where the environment reaches a natural absorbing state (the pole falls in CartPole, the robot falls in locomotion tasks), and \emph{truncation}, where the episode is cut short by an external time limit. At a termination, the value of the successor state is zero: $V(s_\text{term}) = 0$. At a truncation, the episode is merely paused and the successor state has non-zero value: $V(s_\text{trunc}) \neq 0$. Treating truncated transitions as terminated substitutes zero for a non-zero bootstrap value at every time limit boundary, corrupting every Bellman update in the vicinity of episode boundaries. \citet{pardo2018time} show that this conflation degrades performance by 20--40\% on standard MuJoCo benchmarks.\footnote{The Gymnasium API \citep{Towers2024} enforces the distinction by returning separate \texttt{terminated} and \texttt{truncated} flags, but most pre-2022 codebases conflate them in the \texttt{done} flag.}

\subsection{Replay Buffer Pathologies and Reward Scaling}
\label{sec:replay_reward}

Experience replay \citep{Lin1992} decouples the data collection and learning processes, allowing a single transition to be used for multiple gradient updates. The replay ratio (the number of gradient updates per environment step) governs the trade-off between sample efficiency and data staleness. \citet{Zhang2017replay} show that increasing the replay ratio beyond a modest threshold degrades performance.\footnote{At high replay ratios, the distribution shift between the current policy and the behavior policy that generated the stored data grows large enough to violate the off-policy assumptions of Q-learning. The relationship is non-monotone and environment-dependent, making replay buffer size a sensitive hyperparameter with no universal default.}

\citet{schaul2016prioritized} propose prioritized experience replay (PER).\footnote{PER samples transitions with probability proportional to the magnitude of their TD error, on the argument that high-error transitions are the most informative.} \citet{fedus2020revisiting} revisit PER on large-scale Atari experiments and find that uniform sampling from a large enough buffer matches or outperforms PER, while being simpler to implement and tune.

Reward scaling introduces a separate class of failure modes. Standard DQN \citep{mnih2015} clips rewards to $[-1, +1]$ across all environments to stabilize training. \citet{vanHasselt2016popart} observe that reward clipping changes the objective; clipped rewards make all positive events equivalent regardless of magnitude, so the agent learns to maximize the frequency of positive events rather than their cumulative value. This substitution can produce policies that are locally rational under the clipped reward but qualitatively suboptimal under the true reward. \citet{vanHasselt2016popart} propose PopArt as a remedy.\footnote{PopArt normalizes targets to have unit variance while adjusting the output layer so that the policy remains invariant to the normalization. PopArt allows consistent learning across reward scales spanning several orders of magnitude without reward clipping.}

\citet{Skalse2022} formalize reward hacking and show that any non-constant proxy reward can in principle be exploited by a sufficiently capable optimizer.\footnote{\citet{Skalse2022} define a proxy as \emph{unhackable} if increasing expected proxy return cannot decrease expected true return. Their main result states that for the set of all stochastic policies, two reward functions are unhackable only if one of them is constant. For deterministic policies and finite policy sets, non-trivial unhackable pairs exist, but the conditions are stringent.}

\subsection{Simulation Study: Bellman Error and Value Error in Offline Policy Evaluation}
\label{sec:bellman_sim}

The MDP uses $s = (k, z)$ with $k$ on a 50-point log-spaced capital grid and $z \in \{0.9, 1.1\}$ following a Markov chain with persistence 0.8. Actions are next-period capital choices on the same grid. Reward is $\log(zk^\alpha - k')$ with $\alpha = 0.36$, $\beta = 0.96$. Rewards are shifted by $-\bar{r}$ (mean reward over feasible pairs) to center $Q^*$ near zero, which avoids initialization issues without altering the optimal policy. The offline dataset $\mathcal{D}$ consists of $T = 2{,}000$ transitions simulated from the value-iteration optimal policy on the discretized grid, which agrees with the closed-form rule $k^*(k,z) = \alpha\beta z k^\alpha$ on 94\% of states (the residual disagreement is single-grid-step rounding at the bottom of the capital grid). Since the optimal policy concentrates capital near its steady state, $\mathcal{D}$ covers only 11 of the 4,795 feasible $(s,a)$ pairs (0.2\% coverage), the distribution mismatch condition in \citet[Corollary~1]{Fujimoto2022}.

The experiment trains two algorithms on $\mathcal{D}$.\footnote{50,000 gradient steps per seed, 10 seeds; two-layer MLP with 64 hidden units, Adam at $5 \times 10^{-4}$; target network updated every 500 steps.} BRM minimizes $\mathbb{E}_\mathcal{D}[(Q(s,a) - (r + \gamma \max_{a'} Q(s',a')))^2]$ with the current network on both $Q(s,a)$ and $Q(s',a')$, so gradients flow through both sides simultaneously.\footnote{The implementation uses the Bellman-optimality target $\max_{a'} Q(s',a')$ rather than the policy-evaluation target $Q(s', a' \sim \pi)$ of \citet[Section~2]{Fujimoto2022}; both algorithms are therefore FQI-flavored variants of the Fujimoto formulations. Because $\mathcal{D}$ is generated by the optimal policy $\pi^*$, the two targets coincide on the data manifold and diverge off-manifold (where value error is evaluated). The qualitative BE-vs-VE pattern reported below still holds under the substitution; the comparison concerns value estimation, not control. Feasibility is enforced at the target step by a $-10^9$ mask on infeasible $(s',a')$ pairs, so the network is told the constraint set but not the optimal action within it.} The network can therefore zero the residual at an observed pair $(s,a)$ by moving $Q(s,a)$ and $Q(s',a')$ together rather than moving either toward $Q^*$. Supervised-learning labels, by contrast, are fixed external targets that do not move with the model weights. FQE uses a frozen target network for $Q(s',a')$, so the network must reduce $Q(s,a)$ toward a target that does not respond to its own gradient steps. Every 500 steps we record the Bellman error on $\mathcal{D}$ with the current network on both sides and the value error $\frac{1}{|\mathcal{F}|}\sum_{(s,a)\in\mathcal{F}}|Q_\theta(s,a) - Q^*(s,a)|$ over all 4,795 feasible pairs. An OLS regression of $\log c$ on $\log k$ and $\log z$, estimated on expanding windows, supplies a definitional reference panel.\footnote{Noise $\sigma = 0.30$; windows from $n = 10$ to $2{,}000$; held-out test set of 500 transitions.}

The OLS panel records Pearson $r(\mathrm{MSE}, R^2) = -1.000$ (Table~\ref{tab:bellman_sim}). On a fixed test set with fixed $SS_\text{tot}$, held-out $R^2 = 1 - n_\text{test} \cdot \mathrm{MSE}/SS_\text{tot}$ is an exact affine transform of held-out MSE. Their correlation is therefore $-1$ by construction rather than an empirical finding. The panel is an algebraic reference. Bellman error and full-MDP value error have no corresponding identity.

Across 10 seeds, BRM drives the Bellman residual on $\mathcal{D}$ to a mean of $1.4 \times 10^{-4}$, roughly $1{,}500\times$ lower than FQE's mean of $0.209$ (median ratio $1{,}846\times$; three of ten BRM seeds drove BE below $5 \times 10^{-5}$, i.e.\ to numerical noise). The full-MDP value errors are statistically indistinguishable. BRM has mean VE $7.380$, FQE has mean VE $7.310$, and the Welch test gives $t = 0.57$ on $n=10$ seeds each ($p = 0.58$), a difference under one percent of either mean. This comparison concerns value estimation rather than control. Both methods produce $Q$-functions whose greedy policies agree with $\pi^*$ on roughly $6$--$8\%$ of states, so neither delivers a usable decision rule under $0.2\%$ coverage. These results reproduce the value-estimation critique of \citet{Fujimoto2022}, since minimizing Bellman residual on $\mathcal{D}$ does not approximate $Q^*$ off-support. The 11 observed pairs permit error cancellation, while the 4,784 unobserved pairs leave $Q(s',a')$ unconstrained. These are the two mechanisms from Section~\ref{sec:moving_target}.

\begin{table}[h]
\centering
\small
\begin{tabular}{@{}lrrrr@{}}
\toprule
Method & BE on $\mathcal{D}$ & VE & VE/BE & Policy agree. \\
\midrule
BRM (mean $\pm$ SE)   & $(1.39 \pm 0.42) \times 10^{-4}$ & $7.380 \pm 0.064$ & $(1.24 \pm 0.68) \times 10^{6}$ & $7.2\% \pm 0.3\%$ \\
BRM (median)          & $1.04 \times 10^{-4}$            & $7.364$           & $72{,}306$                 & --- \\
FQE (mean $\pm$ SE)   & $0.2087 \pm 0.0394$              & $7.310 \pm 0.107$ & $44.4 \pm 6.8$             & $6.0\% \pm 0.0\%$ \\
FQE (median)          & $0.1911$                          & $7.237$           & $37.2$                     & --- \\
\midrule
OLS & OOS MSE $= 0.0959$ & OOS $R^2 = 0.112$ & $r(\mathrm{MSE}, R^2) = -1.000$ & --- \\
\bottomrule
\end{tabular}
\caption{Bellman error on dataset $\mathcal{D}$ and value error on the full MDP for BRM and FQE trained on offline Brock--Mirman data, with $n = 10$ seeds per method. BE is mean squared Bellman error evaluated with the current network on both sides; VE is mean absolute deviation from $Q^*$ over all 4,795 feasible state-action pairs; SE is the standard error of the mean across seeds. Policy agreement is the fraction of states where $\arg\max_a Q_\theta(s,a) = \pi^*(s)$. The OLS row reports a single fixed-test-set fit with $n = 2{,}000$; on that test set $r(\mathrm{MSE}, R^2) = -1$ is an algebraic identity (see prose).}
\label{tab:bellman_sim}
\end{table}

\begin{figure}[h]
\centering
\includegraphics[width=0.9\textwidth]{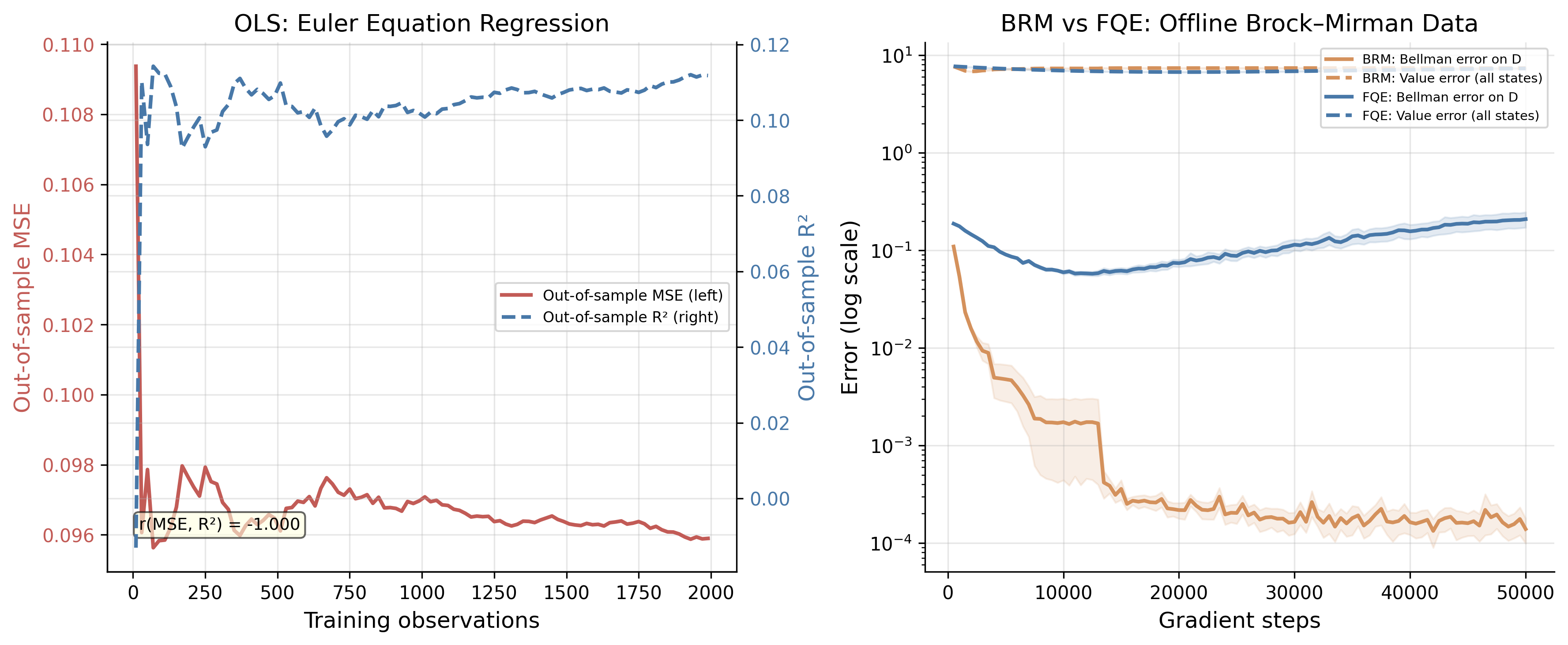}
\caption{Left: OLS regression of log consumption on log capital and log productivity, estimated on expanding windows from the Brock--Mirman optimal policy. Out-of-sample MSE (left axis, red) and out-of-sample $R^2$ (right axis, blue) track each other with Pearson $r = -1.000$. Right: BRM (orange) and FQE (blue) trained on offline Brock--Mirman data $\mathcal{D}$ ($T = 2{,}000$ transitions, 0.2\% state-space coverage); note the log scale on the $y$-axis. Solid lines show Bellman error on $\mathcal{D}$ (current network both sides); dashed lines show mean absolute value error against $Q^*$ over all 4,795 feasible state-action pairs; shaded bands are $\pm 1$ SE over 10 seeds.}
\label{fig:bellman_vs_return}
\end{figure}

\subsection{Discussion and Recommendations}
\label{sec:checklist}

Track episode return and policy entropy alongside training loss; entropy collapse and stagnating return are early warning signs of plasticity loss (Section~\ref{sec:plasticity}). Use PPO or SAC as default baselines before implementing custom algorithms. Report at least 10 seeds per configuration with IQM-based comparisons (Section~\ref{sec:reproducibility}).

\section{Structural Estimation with Reinforcement Learning}
\label{section:rl_econ_models}
Several recent papers have used RL training loops,\footnote{Throughout this chapter, the RL training loop runs entirely inside the econometrician's computational model; the agent never interacts with real economic agents or markets but serves as a numerical method for solving the Bellman equation within a structural estimation procedure (Section~\ref{section:language}). There is no execution phase in the usual sense.} namely Q-learning, temporal-difference learning, policy gradient, and actor-critic methods, to solve structural economic models at scales where conventional dynamic programming fails.\footnote{I exclude papers that use neural network function approximation without an RL training mechanism. Inverse reinforcement learning is treated in the sister survey \citep{RustRawat2026}.} Throughout this chapter I adopt a unified notation. An MDP is a tuple $(\mathcal{S}, \mathcal{A}, P, r, \gamma)$ where $\mathcal{S}$ is the state space, $\mathcal{A}$ is the action space, $P(s' | s, a)$ is the transition kernel, $r(s,a)$ is the per-period reward, and $\gamma \in [0,1)$ is the discount factor.\footnote{Several of the papers reviewed here use $\beta$ for the discount factor, following economics convention. I translate all results to $\gamma$ for consistency with the RL literature and the rest of this survey.} A policy $\pi: \mathcal{S} \to \Delta(\mathcal{A})$ maps states to distributions over actions. The value function under $\pi$ is $V^\pi(s) = \mathbb{E}_\pi\left[\sum_{t=0}^\infty \gamma^t r(s_t, a_t) \mid s_0 = s\right]$, and the action-value function is $Q^\pi(s,a) = r(s,a) + \gamma \mathbb{E}_{s' \sim P(\cdot|s,a)}[V^\pi(s')]$. The optimal value function satisfies $V^*(s) = \max_{a \in \mathcal{A}} Q^*(s,a)$.

The canonical structural estimation framework for MDPs was formulated by \citet{rust1994structural}, whose nested fixed-point (NFXP) algorithm embeds the Bellman equation inside a maximum likelihood estimator. The methods reviewed in this chapter replace the inner fixed-point computation with RL-based approximations.

\begin{figure}[t]
\centering
\includegraphics[width=0.85\textwidth]{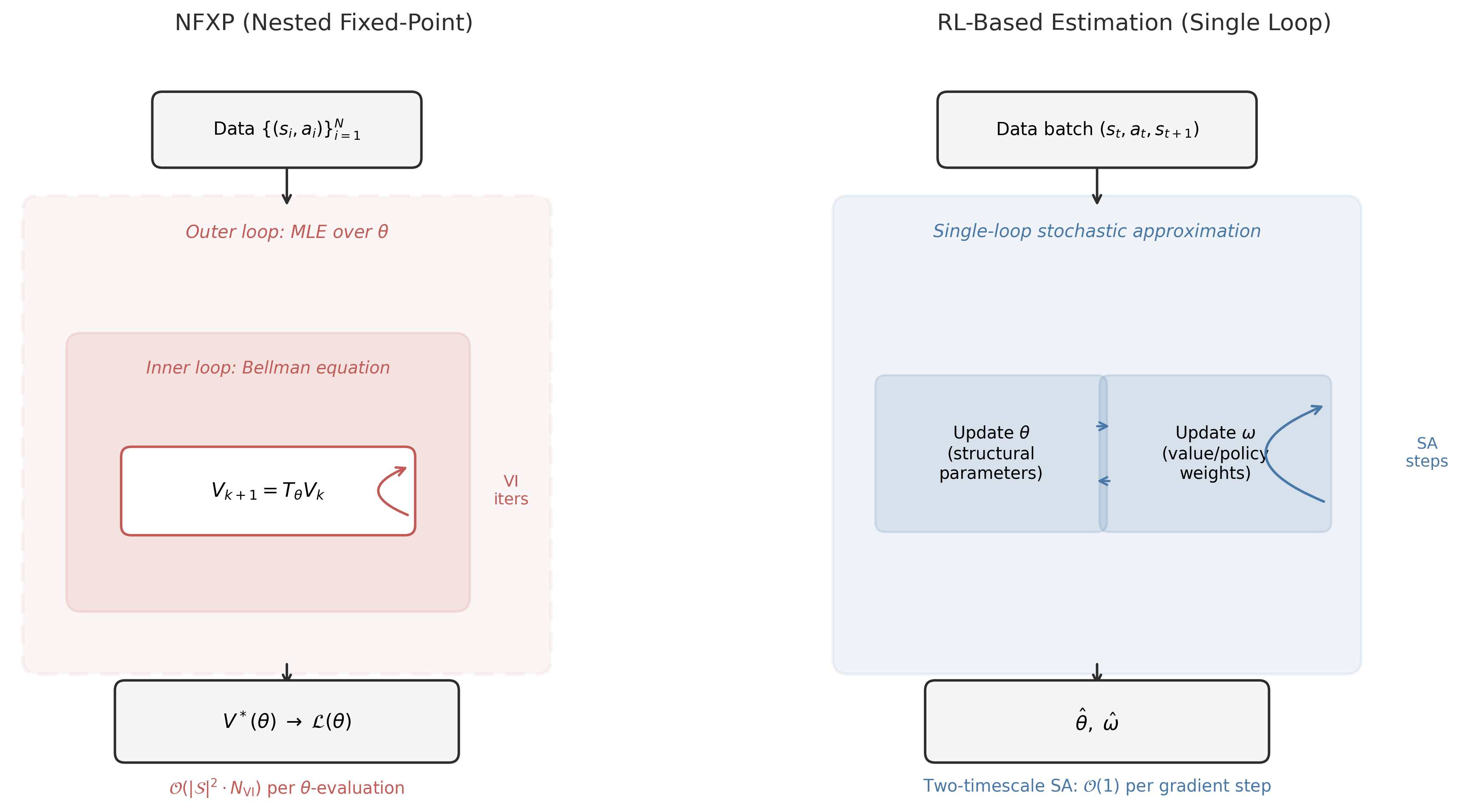}
\caption{NFXP versus RL-based structural estimation. Top: the nested fixed-point algorithm evaluates the likelihood by solving the Bellman equation to convergence inside each optimizer step. Bottom: single-loop stochastic approximation updates structural parameters $\theta$ and value/policy weights $\omega$ simultaneously from data batches. The NFXP algorithm is due to \citet{Rust1987}.}
\label{fig:estimation_flowcharts}
\end{figure}

\subsection{Single-Agent Structural Estimation}

\subsubsection{TD Learning for CCP Estimation}

\citet{AdusumilliEckardt2022} adapt temporal-difference (TD) learning to estimate the recursive terms that arise in CCP-based estimation, entirely avoiding specification or estimation of transition densities.

The CCP approach requires computing two functions $h: \mathcal{A} \times \mathcal{S} \to \mathbb{R}^d$ and $g: \mathcal{A} \times \mathcal{S} \to \mathbb{R}$ that solve the recursive equations
\begin{align}
h(a,s) &= z(s,a) + \gamma \, \mathbb{E}[h(a', s') \mid a, s], \label{eq:ae_h}\\
g(a,s) &= e(a,s) + \gamma \, \mathbb{E}[g(a', s') \mid a, s], \label{eq:ae_g}
\end{align}
where $e(a,s) = \gamma_{\text{E}} - \ln P(a|s)$ under logit errors ($\gamma_{\text{E}}$ denoting the Euler constant), and the expectation is over the next-period state-action pair $(s', a')$ given the transition kernel $P(s'|s,a)$ and the observed policy $P(a|s)$. Both $h$ and $g$ satisfy a Bellman-like recursion under the observed (data-generating) policy, not the optimal policy, so standard TD learning applies directly. \citet{AdusumilliEckardt2022} propose two methods.

The first is the linear semi-gradient method.\footnote{The method is called ``semi-gradient'' because it computes the gradient of only the prediction $\phi(a,s)^\top w$ with respect to $w$, not the full TD error including the bootstrap target $\phi(a',s')^\top w$. This avoids differentiating through the target but sacrifices guaranteed convergence in some settings.} Approximate $h(a,s) \approx \phi(a,s)^\top w$ where $\phi: \mathcal{A} \times \mathcal{S} \to \mathbb{R}^p$ is a vector of basis functions (e.g., polynomials in state variables) and $w \in \mathbb{R}^p$ are the weights to be estimated. The TD(0) fixed-point equation is
\begin{equation}
\mathbb{E}\left[\phi(a,s)\left(\phi(a,s) - \gamma \phi(a', s')\right)^\top\right] w = \mathbb{E}\left[\phi(a,s) \, z(s,a)\right]. \label{eq:ae_semigradient}
\end{equation}
The sample analog replaces population expectations with averages over the observed panel.
\begin{equation}
\hat{w} = \left(\frac{1}{n(T-1)} \sum_{i=1}^n \sum_{t=1}^{T-1} \phi_{it}\left(\phi_{it} - \gamma \phi_{i,t+1}\right)^\top\right)^{-1} \left(\frac{1}{n(T-1)} \sum_{i=1}^n \sum_{t=1}^{T-1} \phi_{it} \, z_{it}\right),
\end{equation}
where $\phi_{it} = \phi(a_{it}, s_{it})$ and $z_{it} = z(s_{it}, a_{it})$. This requires inverting a $p \times p$ matrix, where $p$ is the number of basis functions, making computation trivial in most settings. No transition density estimation is needed, since the method uses only observed sequences of current and next-period state-action pairs.

The second method is approximate value iteration (AVI), which iterates the Bellman-like operator using nonparametric regression. At iteration $k$, one constructs pseudo-outcomes
\begin{equation}
Y_{it}^{(k)} = z_{it} + \gamma \, \hat{h}^{(k-1)}(a_{i,t+1}, s_{i,t+1})
\end{equation}
and then fits $\hat{h}^{(k)}$ by regressing $Y_{it}^{(k)}$ on $(a_{it}, s_{it})$ using any machine learning method, including LASSO, random forests, or neural networks.\footnote{\emph{LASSO} \citep{Tibshirani1996} adds an $\ell_1$ penalty to a regression loss, shrinking many coefficients to zero for sparse solutions; \emph{random forests} average many decision trees fit to random subsamples and feature subsets for nonparametric estimation; \emph{neural networks} compose affine transformations with elementwise nonlinearities across multiple layers.} This is the first DDC estimator compatible with arbitrary ML prediction methods, enabling application to very high-dimensional state spaces.

With $\hat{h}$ and $\hat{g}$ in hand, structural parameters are recovered by pseudo-maximum likelihood estimation (PMLE). For continuous state spaces, \citet{AdusumilliEckardt2022} derive a locally robust correction to the PMLE criterion that accounts for the nonparametric first-stage estimation of value terms, restoring $\sqrt{n}$-convergence of $\hat{\theta}$.\footnote{An estimator has $\sqrt{n}$-convergence if its error shrinks at rate $n^{-1/2}$ where $n$ is the sample size. This is the standard parametric rate; slower rates (e.g., $n^{-1/4}$) indicate efficiency loss from nonparametric first stages.} The PMLE score is $m(a,s;\theta,h,g) = \partial_\theta \ln \pi(a,s;\theta,h,g)$, where $\pi$ is the logit choice probability with continuation value $V(a,s) = h(a,s)^\top \theta + g(a,s)$. The naive estimator solves $\mathbb{E}_n[m(a,s;\theta,\hat{h},\hat{g})] = 0$, but with continuous states this moment condition is not orthogonal to the first-stage estimates and converges slower than $\sqrt{n}$. The locally robust moment adds a debiasing correction:
\begin{equation}
\zeta = m(a,s;\theta,h,g) - \lambda(a,s;\theta)\left\{z(s,a)^\top \theta + \gamma \, e(a',s') + \gamma \, V(a',s') - V(a,s)\right\},
\label{eq:ae_robust}
\end{equation}
where $\lambda(a,s;\theta)$ solves a backward recursion that propagates the influence of estimation error through the dynamic structure.\footnote{The term in braces is the temporal-difference error of the continuation value $V$. The adjoint $\lambda$ weights this TD error by its marginal impact on the PMLE score. See Online Appendix B.3 of \citet{AdusumilliEckardt2022} for the derivation.} The corrected estimator $\hat{\theta}_{LR}$ solves $\mathbb{E}_n[\zeta_n] = 0$ and is computationally no harder than the naive PMLE, since the correction is constant in $\theta$.

\begin{theorem}[\citet{AdusumilliEckardt2022}, Theorem 1]
Under regularity conditions, the linear semi-gradient estimator $\hat{h}$ satisfies $\|\hat{h} - h\|_2 = O_P(n^{-1/2}(T-1)^{-1/2})$, where $\|\cdot\|_2$ denotes the $L^2(P)$ norm.\footnote{The $L^2(P)$ norm is $\|f\|_2 = (\int f(x)^2 \, dP(x))^{1/2}$, measuring average squared deviation under the probability measure $P$. This is the natural norm for mean-squared-error analysis.}
\end{theorem}

\begin{theorem}[\citet{AdusumilliEckardt2022}, Theorem 5]
Under regularity conditions on the ML method used in AVI, the locally robust PMLE estimator $\hat{\theta}_{LR}$ satisfies $\sqrt{n}(\hat{\theta}_{LR} - \theta^*) \xrightarrow{d} \mathcal{N}(0, \Sigma)$ for an explicit variance $\Sigma$, even when the state space is continuous.
\end{theorem}

Monte Carlo experiments on a dynamic firm entry model with seven structural parameters and five continuous state variables show that the TD-based estimators achieve a 4- to 11-fold reduction in mean squared error compared to CCP estimators using state-space discretization.

For dynamic discrete games, the method extends naturally. Standard CCP-based estimation of games requires integrating out other players' actions, which becomes intractable with many players or continuous states. TD learning avoids this entirely, since it works directly with the joint empirical distribution of states and their successors. The ``integrating out'' is done implicitly within sample expectations.

\FloatBarrier
\subsection{Engine Replacement MDP: Estimating the Engine's Parameters}
\label{engine:ch05}

\begin{table}[h]
\centering
\caption{Structural parameter recovery in the $+\mathrm{EV}$ Engine Replacement MDP. Population columns use exact choice and transition probabilities. Finite-sample columns report means and standard errors across 20 fixed seeds, with 9,000 transitions per seed.}
\label{tab:engine_estimation}
\small
\begin{tabular}{lrrrr}
\hline
 & \multicolumn{2}{c}{high-mileage keep reward} & \multicolumn{2}{c}{replacement cost} \\
method & population & finite sample & population & finite sample \\
\hline
NFXP & 0.200000 & 0.1993 (0.0038) & 0.500000 & 0.4943 (0.0060) \\
CCP & 0.200000 & 0.1997 (0.0038) & 0.500000 & 0.4934 (0.0059) \\
TD-CCP & 0.200000 & 0.2024 (0.0039) & 0.500000 & 0.4908 (0.0062) \\
\hline
generating value & \multicolumn{2}{c}{0.200000} & \multicolumn{2}{c}{0.500000} \\
\hline
\end{tabular}
\end{table}

The Engine Replacement experiment compares NFXP, CCP, and TD-CCP on the same two structural parameters. Its $+\mathrm{EV}$ variant adds independent Type~I extreme-value shocks with scale $0.4$ to the Engine Replacement MDP. The low-mileage keep reward remains normalized to one, while the high-mileage keep reward and replacement cost have generating values $0.2$ and $0.5$. NFXP recomputes the smoothed Bellman fixed point inside the likelihood \citep{Rust1987}. CCP uses the binary log-odds inversion of \citet{HotzMiller1993}. TD-CCP uses a complete two-state basis to solve sample state-value TD equations, estimates the transition kernel from the same sample, and then applies the same inversion. This finite-state construction is an analogue of the TD approach in \citet{AdusumilliEckardt2022}, not its action-state full-basis estimator.

All three population estimators recover both parameters with maximum absolute error below $10^{-6}$. The finite-sample experiment uses 20 fixed seeds and 9,000 transitions per seed. The mean high-mileage reward estimates range from $0.1993$ to $0.2024$, and the mean replacement-cost estimates range from $0.4908$ to $0.4943$. The reported standard errors are computed across seeds.
\FloatBarrier

\subsubsection{Policy Gradient for DDC Estimation}

\citet{HuYang2025} combine policy gradient methods with the Simulated Method of Moments (SMM) to estimate DDCs, with particular focus on models with unobserved state variables.

The outer loop is SMM.\footnote{SMM \citep{GourierouxMonfort1993} estimates structural parameters by matching moments from simulated data to moments from observed data, avoiding direct evaluation of the likelihood function.} Define a vector of data moments $\mathbf{M}_d$ computed from the observed panel. For candidate structural parameters $\theta$ and transition parameters $\xi$, simulate the model to produce simulated moments $\mathbf{M}_s(\theta, \xi)$. The estimator minimizes
\begin{equation}
(\hat{\theta}, \hat{\xi}) = \arg\min_{\theta, \xi} \left(\mathbf{M}_d - \mathbf{M}_s(\theta, \xi)\right)^\top \mathbf{W} \left(\mathbf{M}_d - \mathbf{M}_s(\theta, \xi)\right), \label{eq:hy_outer}
\end{equation}
where $\mathbf{W}$ is a positive definite weight matrix. Computing $\mathbf{M}_s(\theta, \xi)$ requires solving for the optimal policy under $(\theta, \xi)$, which is the inner-loop problem.

The inner loop parametrizes the choice probability directly as a logistic function of state variables. For a binary choice $J_t \in \{0, 1\}$, the general form is $\Pr(J_t = 1 \mid \boldsymbol{X}_t; \boldsymbol{\gamma}(\theta)) = \text{logistic}(\boldsymbol{X}_t \boldsymbol{\gamma}(\theta))$, where $\boldsymbol{\gamma}(\theta)$ are policy parameters that depend on the structural parameters.\footnote{The linear index can be replaced by higher-order terms of $\boldsymbol{X}_t$ or deep neural networks; the method requires only that the gradient $\nabla_{\boldsymbol{\gamma}} \log \pi_{\boldsymbol{\gamma}}$ has a closed form.} In their application to a Rust bus engine model with unobserved bus condition $S_t^*$, this takes the form
\begin{equation}
\Pr(J_t = 1 \mid X_t, S_t^*, t; \boldsymbol{\gamma}) = \frac{\exp(\gamma_0 + \gamma_1 t + \gamma_2 X_t + \gamma_3 S_t^*)}{1 + \exp(\gamma_0 + \gamma_1 t + \gamma_2 X_t + \gamma_3 S_t^*)}, \label{eq:hy_policy}
\end{equation}
where $t$ enters the index directly to capture time-varying replacement incentives. The policy parameters are updated by REINFORCE-style gradient ascent, applying the policy gradient theorem \citep{Sutton1999}:
\begin{equation}
\nabla_{\boldsymbol{\gamma}} V(\boldsymbol{\gamma}) = \mathbb{E}\left[\sum_{t=0}^{T} \nabla_{\boldsymbol{\gamma}} \log \pi_{\boldsymbol{\gamma}}(J_t \mid X_t, S_t^*, t) \, Q^{\pi_{\boldsymbol{\gamma}}}(X_t, S_t^*, J_t)\right], \label{eq:hy_pg}
\end{equation}
where $Q^{\pi_{\boldsymbol{\gamma}}}$ is the action-value function under the current policy, estimated by Monte Carlo returns from forward-simulated trajectories.

The main contribution is handling unobserved state variables. When state variables are only partially observed, the policy in (\ref{eq:hy_policy}) is parametrized as a function of both $X_t$ and $S_t^*$, and the algorithm forward-simulates trajectories of both observed and unobserved variables.

Building on the nonparametric identification results of \citet{HuShum2012}, the outer-loop SMM targets moments from five consecutive periods of observed data, which suffice to separately identify the structural parameters $\theta$ and transition parameters $\xi$ without requiring the econometrician to observe $S_t^*$. No discretization of continuous unobserved states is needed; the same algorithm handles both discrete and continuous unobserved heterogeneity.

For each candidate $(\theta, \xi)$ in the outer minimization (\ref{eq:hy_outer}), the inner loop runs policy gradient until convergence, producing optimal policy parameters $\boldsymbol{\gamma}^*(\theta, \xi)$. These are used to simulate data and compute $\mathbf{M}_s(\theta, \xi)$.

On an extended Rust bus engine model with a continuous unobserved bus condition following an AR(1) process, estimates of seven structural parameters are centered around their true values across 400 simulations. On a discrete-unobservable variant, the method matches the precision of \citet{ArcidiaconoMiller2011}'s two-step EM algorithm at comparable computation times, though the advantage diminishes as more inner-loop iterations are used for precision.

\subsection{Dynamic Oligopoly and Strategic Interaction}

Dynamic oligopoly models combine game theory and dynamic programming. Firms choose actions strategically while anticipating competitors' strategies, and the state space grows combinatorially in the number of firms.

\subsubsection{Q-Learning in Dynamic Procurement Auctions}

\citet{AskerEtAl2020} develop a computational framework for analyzing dynamic procurement auctions with serially correlated asymmetric information. Their approach builds on the Experience-Based Equilibrium (EBE) concept of \citet{FershtmanPakes2012}, which computes equilibria by simulating industry trajectories and updating strategies toward best responses. EBE evaluates values only on recurrently visited states rather than the full state space, making it feasible for large state spaces. While \citet{FershtmanPakes2012} use a stochastic approximation algorithm to update continuation values from simulated industry trajectories, \citet{AskerEtAl2020} add explicit value-function updates via stochastic approximation.

The model is a repeated first-price sealed-bid auction with two firms. Each firm $i$ maintains a private inventory state $\omega_{i,t}$ (stock of unharvested timber, in their application). The state evolves endogenously, as winning an auction increases inventory while harvesting depletes it. Each firm's private state is not observed by its competitor except at periodic revelation events.

The key computational innovation is the use of Q-learning to compute equilibrium strategies. Each firm $i$ maintains a Q-function $Q_i: \mathcal{S}_i \times \mathcal{A}_i \to \mathbb{R}$, where $\mathcal{S}_i$ encodes firm $i$'s information set (its own inventory, beliefs about the competitor's inventory, public history) and $\mathcal{A}_i$ is its action set (participation decision and bid level). The Q-function satisfies
\begin{equation}
Q_i(s, a) = \mathbb{E}\left[r_i(s, a, a_{-i}) + \gamma \max_{a' \in \mathcal{A}_i} Q_i(s', a') \;\middle|\; s, a\right], \label{eq:asker_bellman}
\end{equation}
where $r_i(s, a, a_{-i})$ is firm $i$'s per-period profit given the state, its own action $a$, and the competitor's action $a_{-i}$, and the expectation is taken over the competitor's strategy and the stochastic transitions.

Firms update Q-values using sample averaging.
\begin{equation}
Q_i(s, a) \leftarrow Q_i(s, a) + \frac{1}{h_k(s,a)}\left[r_i + \gamma \max_{a' \in \mathcal{A}_i} Q_i(s', a') - Q_i(s, a)\right], \label{eq:asker_qupdate}
\end{equation}
where $h_k(s,a)$ is the number of times state-action pair $(s,a)$ has been visited.\footnote{This sample-averaging rule ($\alpha_k = 1/h_k$) is equivalent to maintaining the running mean of observed returns, as distinct from fixed-$\alpha$ Q-learning which gives exponentially decaying weight to older observations. The update is applied for all actions, including counterfactual actions not taken, using the observed state transition.} The equilibrium computation proceeds iteratively, with firms simultaneously updating their Q-functions based on simulated play against each other's current strategies. Strategies are derived from Q-values using an $\varepsilon$-greedy rule or Boltzmann exploration.\footnote{$\varepsilon$-greedy selects the greedy action $\argmax_a Q(s,a)$ with probability $1-\varepsilon$ and a uniformly random action otherwise. Boltzmann exploration selects action $a$ with probability $\propto \exp(Q(s,a)/\tau)$ where $\tau > 0$ is a temperature parameter.}

\citet{AskerEtAl2020} add a boundary consistency condition to the EBE concept that restricts behavior at the boundary of the recurrent state class, reducing the multiplicity of equilibria. Their numerical analysis reveals that information sharing between firms can, through increased precision of beliefs about competitor states, induce firms to spend more time in states where competition is less intense. The dynamic RL-computed equilibrium yields qualitatively different predictions from both static analysis and myopic ($\gamma = 0$) benchmarks. With dynamics, information sharing decreases average bids and increases average profits, while the myopic benchmark shows negligible effects.

The limitation of this approach is the tabular representation; the Q-function is stored as a lookup table over discretized states and actions, restricting applicability to models with moderate state-space dimension.

\subsubsection{TD Learning for Merger Analysis with Innovation}

\citet{Hollenbeck2019} uses RL to solve a dynamic oligopoly model with endogenous mergers, entry, exit, and quality investment. The model extends the Ericson-Pakes framework to study how horizontal mergers affect innovation incentives.

The industry state is $\Omega = (\omega_1, \ldots, \omega_n)$ where $\omega_i \in \{1, \ldots, \omega_{\max}\}$ is firm $i$'s product quality. Firms produce differentiated goods and compete in prices (Bertrand competition with logit demand). In each period, firms simultaneously choose investment levels, entry/exit decisions, and potentially initiate merger negotiations.

Each firm $i$ computes its continuation value $V_i(\Omega)$ from the industry state. Because the state space is the product of all firms' quality levels plus industry structure (number of active firms, recent mergers), exact dynamic programming is infeasible for industries with more than two or three firms. \citet{Hollenbeck2019} instead uses temporal-difference learning to estimate values from simulated industry trajectories.

The value function update for firm $i$ is\footnote{This stochastic approximation update, introduced by \citet{PakesMcGuire1994} for dynamic oligopoly computation, is closely related to temporal-difference learning in the RL literature. The original algorithm uses visit-count averaging ($\alpha_k = 1/k$ where $k$ counts visits to each state) rather than a fixed learning rate.}
\begin{equation}
V_i(\Omega) \leftarrow V_i(\Omega) + \alpha\left[\Pi_i(\Omega, \mathbf{a}) + \gamma V_i(\Omega') - V_i(\Omega)\right], \label{eq:hollenbeck_td}
\end{equation}
where $\Pi_i(\Omega, \mathbf{a})$ denotes firm $i$'s per-period profit given industry state $\Omega$ and the joint action profile $\mathbf{a}$ (investment, entry/exit, merger decisions),\footnote{I use $\Pi_i$ for firm $i$'s per-period profit to avoid confusion with policy $\pi$.} $\gamma$ is the discount factor, and $\Omega'$ is the realized next-period state. The algorithm uses $\varepsilon$-decreasing exploration to prevent convergence to locally suboptimal equilibria.

The equilibrium computation follows the Pakes-McGuire iterative scheme.\footnote{The Pakes-McGuire algorithm \citep{PakesMcGuire1994} computes Markov perfect equilibria by iterating: (1) given current value functions, compute best-response strategies; (2) given strategies, update value functions via simulation. Convergence is not guaranteed but works well in practice.} The algorithm simulates long industry histories, updates each firm's value function via (\ref{eq:hollenbeck_td}), re-derives best-response strategies from updated values, and repeats.

The central finding is that horizontal mergers, while reducing static consumer surplus in the short run, create a strong incentive for entry and investment. Firms enter with negative static profits because the prospect of a lucrative buyout justifies the initial investment. The result is substantially higher long-run innovation and consumer welfare with mergers than without. This finding reverses the standard static antitrust prediction and can only emerge in a dynamic model where firms are forward-looking.

Several recent papers bring RL machinery to dynamic discrete choice and inverse reinforcement learning estimation from other directions. \citet{Kang2025} recast offline inverse RL and DDC estimation as a single empirical risk minimization problem, training the reward function by gradient descent rather than by nested fixed-point iteration. \citet{Khwaja2025} use reinforcement learning to speed up conditional-choice-simulation estimation, \citet{Nguyen2025} estimate high-dimensional DDC models with neural networks, and \citet{Lee2026} apply adversarial IRL to serialized dynamic discrete choice. \citet{DearingBlevins2025} give a convergent sequential pseudo-likelihood estimator for dynamic discrete games.

Two related papers merit brief mention. \citet{LomysMagnolfi2024} develop structural estimation methods for strategic settings where agents use learning algorithms (specifically regret-minimizing rules) rather than playing a fixed equilibrium. They impose an ``asymptotic no-regret'' condition as a minimal rationality requirement and derive identification results for payoff parameters. \citet{Covarrubias2022} uses deep RL to study oligopolistic pricing in a New Keynesian framework, representing firms' pricing policies as neural networks $\pi_\phi(a|s)$. The method uncovers multiple equilibria ranging from competitive to collusive pricing.\footnote{The most prominent example of algorithmic collusion is \citet{Calvano2020}, who showed that independent Q-learning agents in a repeated Bertrand pricing game learn to sustain supra-competitive prices and punish deviators without explicit communication. This result and its implications for competition policy are treated in the companion thesis chapter \citep{Rawat2026collusion}.}

\subsection{Auction Equilibria and Mechanism Design}

Closed-form equilibrium bidding strategies exist only for narrow families of valuation distributions and auction formats, and numerical methods scale poorly with the number of bidders and items.

\subsubsection{RL for Sequential Price Mechanisms}

\citet{BreroEtAl2021} use RL to design optimal sequential price mechanisms (SPMs), a class of indirect auction mechanisms where agents are approached in sequence and offered menus of items at posted prices. SPMs generalize both serial dictatorship and posted-price mechanisms and essentially characterize all strongly obviously strategyproof (SOSP) mechanisms \citep{PyciaAndTroyan2019}.\footnote{A mechanism is strategyproof if truthful reporting is a dominant strategy. It is obviously strategyproof if this dominance is apparent even to boundedly rational agents; strongly obviously strategyproof (SOSP) adds that dominance holds at every information set \citep{PyciaAndTroyan2019}.}

The mechanism design problem is formulated as a partially observable Markov decision process (POMDP).\footnote{In a POMDP, the agent cannot directly observe the full state. It maintains a belief distribution over possible states, updated via Bayes' rule as new observations arrive. This captures the mechanism designer's uncertainty about bidder valuations.} The state at round $t$ includes the set of remaining items $\rho_t^{\text{items}} \subseteq [m]$, remaining agents $\rho_t^{\text{agents}} \subseteq [n]$, the partial allocation $\mathbf{x}_t$, and the agents' (unobserved) valuation functions. The action $a_t = (i_t, \{p_j^t\}_{j \in \rho_{t-1}^{\text{items}}})$ specifies which agent to visit next and what prices to offer. The observation is the agent's purchase decision, from which the mechanism can update beliefs about valuations. The reward is the objective function evaluated at the final allocation:
\begin{equation}
r = g(\mathbf{x}_T, \boldsymbol{\tau}_T; \mathbf{v}), \label{eq:brero_reward}
\end{equation}
where $g$ can be social welfare $\sum_{i} v_i(x_i)$, revenue $\sum_i \tau_i$, or max-min fairness $\min_i v_i(x_i)$.

The theorem below states when adaptive mechanisms outperform static ones.

\begin{theorem}[\citet{BreroEtAl2021}, Propositions 1--4, informal]
Each feature of adaptive mechanisms is necessary for welfare optimality, even in simple settings. Personalized prices are needed with one item and two i.i.d.\ agents (Proposition 1); adaptive prices are needed with two identical items and three i.i.d.\ agents (Proposition 2); adaptive ordering is needed with six agents whose valuations are correlated (Proposition 3); both adaptive prices and ordering are needed with four agents whose valuations are independently but non-identically distributed (Proposition 4).
\end{theorem}

The policy maps from a sufficient statistic of the observation history to actions. \citet{BreroEtAl2021} show that this statistic can be represented compactly. The set of remaining items and agents suffices for independent valuations, while the full allocation matrix is needed for correlated valuations. They train the mechanism policy using Proximal Policy Optimization (PPO),\footnote{Proximal Policy Optimization \citep{Schulman2017} is a policy gradient algorithm that constrains each update to a trust region, preventing large destabilizing policy changes. It is described in Chapter 2.} which handles the discrete action space (agent selection) and continuous action space (price setting) of the POMDP.

Experimental results show that the learned SPMs achieve near-optimal welfare across settings with up to 20 agents and 5 items (with similar results noted for up to 30 of each), significantly outperforming static pricing benchmarks. The improvement is largest when agent valuations are correlated, since adaptive prices allow the mechanism to infer information about remaining agents from earlier purchases.

The limitation is that the POMDP formulation requires knowledge of the prior distribution over valuations, which in practice must be estimated from data. The method also does not scale easily to very large numbers of items due to the combinatorial action space.

\subsubsection{Fitted Policy Iteration for Combinatorial Auctions}

\citet{RavindranathEtAl2024} address revenue-maximizing mechanism design for combinatorial auctions with multiple items and strategic bidders. Their innovation is integrating differentiable auction structure into a fitted policy iteration framework, enabling analytical gradient computation where standard RL methods struggle with high variance.

The mechanism visits agents one at a time in sequence. Each agent $i$, upon being visited, selects a bundle of items from those still available, given posted prices. Valuations are drawn once from distributions $V_i$ and remain fixed throughout the mechanism. Complementarities arise from the structure of the valuation function over bundles, not from dynamic evolution. The MDP state at step $t$ is $s_t = (i_t, S_t)$, where $i_t$ is the current bidder and $S_t$ is the set of remaining items.

The mechanism's policy maps from state to a price vector over available items. The key technical contribution is making the auction clearing differentiable. In a standard auction, the allocation is an argmax over bids (non-differentiable), and payments depend discontinuously on the allocation. \citet{RavindranathEtAl2024} replace the hard bundle selection with a softmax relaxation,\footnote{The softmax relaxation replaces the hard allocation $\argmax_i b_i$ with a soft allocation $\exp(b_i/\tau)/\sum_j \exp(b_j/\tau)$, which is differentiable and approaches the hard allocation as $\tau \to 0$.} enabling analytical gradient computation through the mechanism. The actor loss is the negative expected revenue, and gradients flow directly through the softmax-relaxed allocation and payment rules. The paper explicitly avoids REINFORCE-style estimators, noting that analytical gradients overcome the sample inefficiency and high variance of score-function methods.

The method follows fitted policy iteration \citep{bertsekastsitsiklis1996}, alternating between evaluating the current policy (computing expected revenue) and improving it via gradient ascent on the actor loss. This differs from model-free RL approaches (PPO, SAC) that the paper uses as baselines.

Experiments on settings with additive and combinatorial valuations show that the learned mechanisms achieve up to 13\% higher revenue than item-wise Myerson optimal auctions, with the largest gains in combinatorial settings where bundle complementarities make item-wise pricing suboptimal. Standard RL baselines (PPO, SAC) are also outperformed, confirming the advantage of exploiting differentiable structure.\footnote{DDPG was also tested but found to be unstable. DQN is not applicable to continuous price-setting.}

\subsection{Simulation Study: DDC Estimation at Scale}
\label{sec:ddc_estimation_sim}

The test bed extends the \citet{Rust1987} bus engine replacement model to multiple wear components. In the original formulation a maintenance superintendent observes discretized mileage $m\in\{0,\ldots,M{-}1\}$ and makes a binary keep-or-replace decision, facing running cost $c(s;\theta)=\theta_1 x+\theta_2 x^2$ with $x=m/M$, replacement cost $RC$, and Type~I extreme value additive errors that yield logit conditional choice probabilities. The simulation uses $K$ independent wear components, each evolving in $\{0,\ldots,M{-}1\}$, with aggregate normalized wear $x(s)=\sum_k m_k/M$ entering the same cost function. The state space is $|\mathcal{S}|=M^K$. Increasing $K$ therefore holds the data-generating process fixed while the computational burden grows exponentially.

\begin{figure}[t]
\centering
\includegraphics[width=0.85\textwidth]{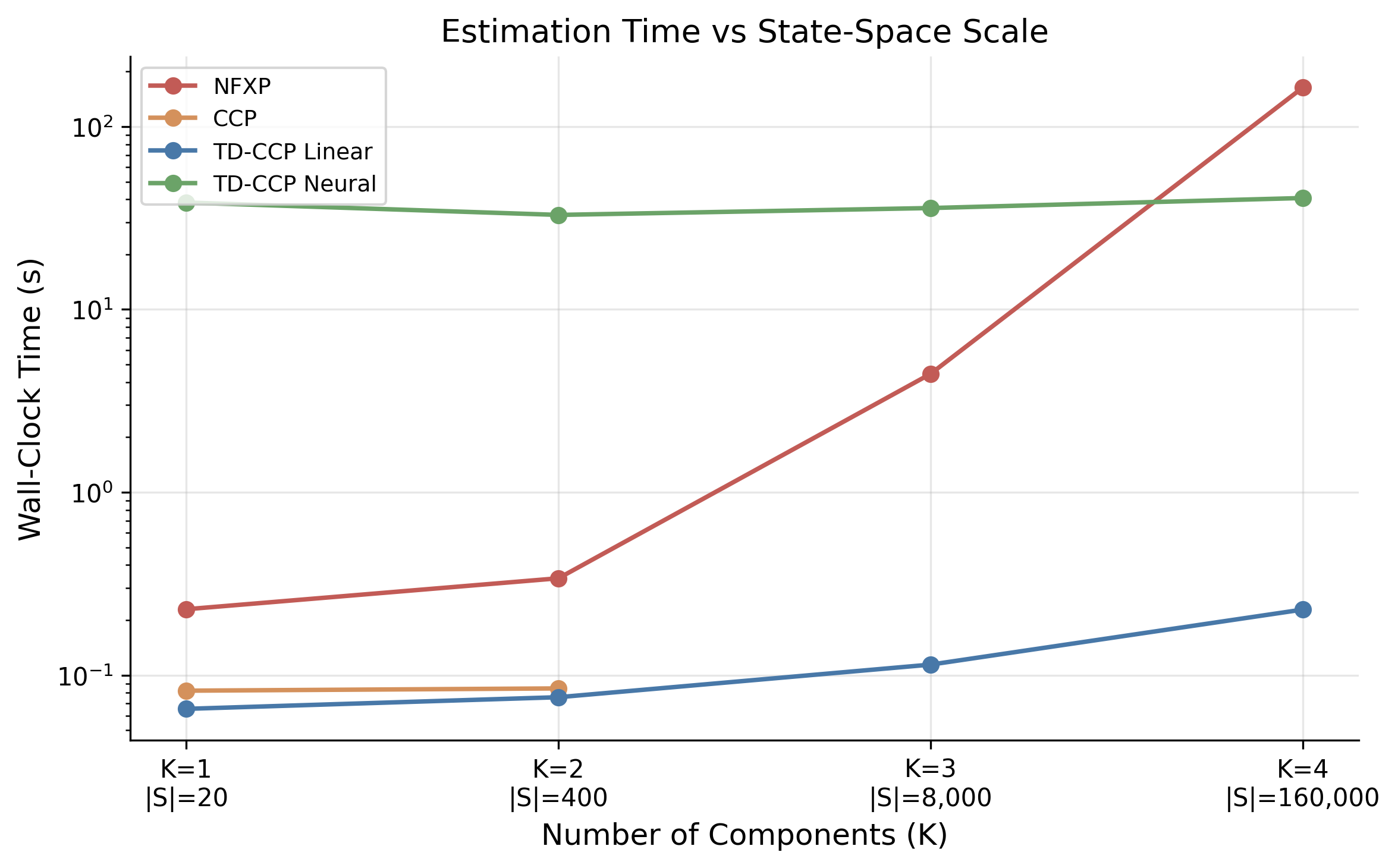}
\caption{Wall-clock estimation time versus state-space scale for the four DDC estimators. Vertical axis is log-scaled. Each point is the mean over 10 seeds.}
\label{fig:ddc_scaling_time}
\end{figure}

The experiment compares four estimation methods for $K \in \{1,2,3,4\}$, producing state spaces from 20 to 160,000 states. Panel data consist of $N=500$ agents observed for $T=100$ periods. The methods are NFXP (nested fixed-point MLE), CCP (Hotz-Miller inversion), TD-CCP Linear (semi-gradient TD with polynomial basis), and TD-CCP Neural (approximate value iteration with a two-layer MLP). Both TD-CCP variants follow \citet{AdusumilliEckardt2022}.\footnote{The TD-CCP variants implemented here are simplified relative to \citet{AdusumilliEckardt2022}. We omit the locally robust PMLE correction (Theorem~5 of that paper, the orthogonal moment construction in Section~3.3 and Online Appendix~B.3 that yields $\sqrt{n}$-consistency); the reported point estimates are for the simpler plug-in PMLE and do not inherit the paper's $\sqrt{n}$-convergence guarantee. Empirically, the correction would absorb the first-stage bias from the value-function estimate, narrow the seed-to-seed RMSE, and deliver nominal coverage for confidence intervals built from $\Sigma$ in Theorem~5; the directional scaling story reported below would not change. We also reformulate the bootstrap target by decomposing $EV(s;\theta)$ into $\theta$-linear pieces and learning each piece separately, rather than fitting $h(a,s)$ directly as in the paper; the two are algebraically related under our binary action with one-period reset, but the change is undocumented in the original. Finally, the marketed advantage of TD-CCP that ``no transition density is needed'' is qualified in our implementation: the PMLE step still evaluates $v_0 = -c + \gamma P_{\text{keep}} \widehat{EV}$ using the empirical $P_{\text{keep}}$ assumed known under the Rust convention. A fully density-free variant that replaces $P_{\text{keep}} \widehat{EV}$ by an out-of-sample target average is left for future work.} Each configuration is replicated across 10 seeds with the PyTorch optimizer also seeded. Table~\ref{tab:ddc_estimation} and Figure~\ref{fig:ddc_scaling_time} report means with seed-level standard errors.

\begin{table}[t]
\centering
\caption{DDC estimation results across state-space scales. Rows show method, number of components $K$, state-space size $|\mathcal{S}|$, mean wall-clock time (seconds), RC bias, RC RMSE, and root mean squared error for $\theta_1$ and $\theta_2$. Standard errors in adjacent columns; means over 10 seeds, dashes indicate method failure (sparse state coverage).}
\label{tab:ddc_estimation}
\footnotesize
\begin{tabular}{@{}ll rr cc cc cc cc@{}}
\toprule
 & & & & \multicolumn{2}{c}{RC Bias} & \multicolumn{2}{c}{RC RMSE} & \multicolumn{2}{c}{$\theta_1$ RMSE} & \multicolumn{2}{c}{$\theta_2$ RMSE} \\
\cmidrule(lr){5-6} \cmidrule(lr){7-8} \cmidrule(lr){9-10} \cmidrule(lr){11-12}
Method & K & $|S|$ & Time (s) & Mean & SE & Value & SE & Value & SE & Value & SE \\
\midrule
NFXP & 1 & 20 & 0.2 & +0.053 & 0.027 & 0.097 & 0.015 & 0.307 & 0.055 & 0.495 & 0.111 \\
CCP & 1 & 20 & 0.1 & +0.055 & 0.027 & 0.099 & 0.015 & 0.316 & 0.055 & 0.510 & 0.112 \\
TD-CCP Linear & 1 & 20 & 0.1 & -0.087 & 0.017 & 0.101 & 0.016 & 0.255 & 0.041 & 0.296 & 0.044 \\
TD-CCP Neural & 1 & 20 & 38.4 & +0.054 & 0.026 & 0.094 & 0.013 & 0.260 & 0.054 & 0.423 & 0.082 \\
\midrule
NFXP & 2 & 400 & 0.3 & +0.027 & 0.028 & 0.087 & 0.018 & 0.270 & 0.061 & 0.344 & 0.077 \\
CCP & 2 & 400 & 0.1 & +0.089 & 0.027 & 0.120 & 0.021 & 0.527 & 0.068 & 0.797 & 0.089 \\
TD-CCP Linear & 2 & 400 & 0.1 & -0.171 & 0.021 & 0.182 & 0.021 & 0.131 & 0.027 & 0.888 & 0.060 \\
TD-CCP Neural & 2 & 400 & 32.9 & +0.012 & 0.030 & 0.090 & 0.021 & 0.247 & 0.065 & 0.334 & 0.067 \\
\midrule
NFXP & 3 & 8,000 & 4.4 & +0.011 & 0.014 & 0.044 & 0.009 & 0.236 & 0.046 & 0.364 & 0.081 \\
CCP & 3 & 8,000 & --- & --- & --- & --- & --- & --- & --- & --- & --- \\
TD-CCP Linear & 3 & 8,000 & 0.1 & -0.246 & 0.011 & 0.248 & 0.010 & 0.125 & 0.022 & 1.178 & 0.071 \\
TD-CCP Neural & 3 & 8,000 & 35.9 & +0.007 & 0.017 & 0.051 & 0.009 & 0.215 & 0.050 & 0.318 & 0.081 \\
\midrule
NFXP & 4 & 160,000 & 163.5 & -0.034 & 0.017 & 0.061 & 0.012 & 0.153 & 0.027 & 0.143 & 0.024 \\
CCP & 4 & 160,000 & --- & --- & --- & --- & --- & --- & --- & --- & --- \\
TD-CCP Linear & 4 & 160,000 & 0.2 & -0.328 & 0.014 & 0.331 & 0.014 & 0.226 & 0.028 & 1.103 & 0.027 \\
TD-CCP Neural & 4 & 160,000 & 40.7 & -0.031 & 0.019 & 0.066 & 0.011 & 0.258 & 0.042 & 0.236 & 0.039 \\
\bottomrule
\end{tabular}
\end{table}

Table~\ref{tab:ddc_estimation} reports the numerical comparison. NFXP wall-clock grows by roughly three orders of magnitude across the four scales because each likelihood evaluation repeats value iteration. Sparse state coverage makes CCP infeasible beyond $K{=}2$. TD-CCP Neural maintains near-constant runtime across all $K$ levels with RC RMSE comparable to NFXP, consistent with the AVI approach of \citet{AdusumilliEckardt2022}. At $K{=}4$, TD-CCP Neural has RC RMSE $0.066$ versus $0.061$ for NFXP and wall time $40.7$ seconds versus $163.5$ seconds. This small RMSE difference occurs in the Zurcher-style setup before applying the locally robust correction described above. TD-CCP Linear runs in well under a second at every scale. Its polynomial basis is misspecified at higher $K$, and its RC RMSE rises monotonically with $K$. Figure~\ref{fig:ddc_scaling_time} shows the corresponding runtime curves.

\section{Reinforcement Learning for Macroeconomic Models}
\label{section:rl_macro}

Reinforcement learning is a recent arrival in macroeconomics. Most of
the algorithms surveyed in this chapter were published in the past
few years, and the literature is small enough that few results have
been independently replicated. The arrival is nevertheless of some
interest, because RL admits two distinct interpretations in a macro
context. As
a solver, RL is a computational tool that finds rational-expectations
or Nash equilibria for models whose state space breaks dynamic
programming. As a behavioural model, RL describes a boundedly-rational
agent who learns a policy from realised utility under no a priori
knowledge of the economy. The two interpretations share the same
training code; the distinction is in what the analyst uses the output
to claim.

The Engine Replacement MDP does not represent the continuous states,
continuous actions, mean-field coupling, or partially observed population
statistics that define the macroeconomic problems in this chapter.

This chapter is organised around four roles RL plays for
macroeconomics. Section~\ref{sec:macro:setup} fixes notation and
defines the four problem classes the chapter uses.
Section~\ref{sec:macro:solver_single_agent} and
Section~\ref{sec:macro:solver} treat RL as a global solver, first for
representative-agent macro models that classical dynamic programming
already handles, and then for heterogeneous-agent models where DP
breaks down. The simulation in
Section~\ref{sec:macro:simulation} makes the comparison concrete on a
textbook RBC. Section~\ref{sec:macro:mfg} takes the continuum limit
to mean field games with common noise, in two variants by game
structure: symmetric mean-field and hierarchical Stackelberg.
Section~\ref{sec:macro:behavioral} reframes RL as a model of the
household itself, using learning dynamics to reproduce empirical
anomalies that rational expectations cannot match jointly.
Section~\ref{sec:macro:policy} surveys RL for policy design under
two methodologically distinct settings, multi-agent ABM in which
the planner and the citizens are all RL learners co-adapting, and
single-planner RL in which only the planner uses RL and the rest of
the macro model is fixed.
Section~\ref{sec:macro:discussion} closes with a short synthesis.

\subsection{Setup and notation}
\label{sec:macro:setup}

We collect the definitions here so that subsequent sections can map a
paper's notation to a single chapter-wide convention without
restating the machinery.

\subsubsection{The agent's MDP}

A Markov decision process (MDP) is a tuple $(\mathcal S, \mathcal A,
P, r, \beta)$ collecting state space, action space, transition rule,
reward, and discount. Concretely, the state space is $\mathcal S$,
the action space is $\mathcal A$, the transition kernel
$P(s' \mid s, a)$ gives the probability of moving to state $s'$ after
taking action $a$ in state $s$, the reward function is
$r(s, a) \in \mathbb R$, and the discount factor is
$\beta \in (0, 1)$. A
policy $\pi(a \mid s)$ maps states to action distributions. The
return is $G_t = \sum_{k \ge 0} \beta^k r_{t+k}$, the value is
$V^\pi(s) = \mathbb E^\pi[G_t \mid s_t = s]$, and the Q-function is
$Q^\pi(s, a) = \mathbb E^\pi[G_t \mid s_t = s, a_t = a]$. The optimal
value satisfies the Bellman equation
$V^\star(s) = \max_a \{ r(s, a) + \beta\, \mathbb E[V^\star(s') \mid s,
a]\}$. Reinforcement learning finds an approximately optimal policy
from interaction with the environment rather than from a fully
specified model. We refer to RL algorithms (value iteration,
Q-learning, policy gradient, actor-critic, soft actor-critic, proximal
policy optimisation, deep deterministic policy gradient) by name
without re-deriving them; the algorithmic taxonomy is covered in
Sections~\ref{section:rl_algorithms} and~\ref{sec:planning_learning} of
the survey.

\subsubsection{Generalisations used in this chapter}

Four generalisations of the MDP appear in the macro literature
surveyed below.

A partially observed MDP (POMDP) supplements the MDP with an
observation space $\mathcal O$ and an emission distribution
$U(o \mid s)$. The agent does not see the state $s$ directly; it
conditions its policy on the observation history $o_{0:t}$, typically
through a recurrent network or a sufficient statistic (a function of
the history that preserves all decision-relevant information).
POMDPs arise in macro whenever a household sees only prices or
summary statistics rather than the full cross-sectional distribution,
that is, the population-wide distribution of individual states such
as wealth and productivity. The cross-sectional distribution matters
to the household because equilibrium prices, and therefore its own
budget constraint, depend on what the rest of the population is
doing.

A Markov game extends the MDP to $n$ agents who share the state but
choose actions independently. Each agent $i \in \{1, \ldots, n\}$ has
its own action space $\mathcal A_i$ and reward $r_i(s, a)$, where
$a = (a_1, \ldots, a_n)$ is the joint action. Each agent treats the
other agents' policies as part of a non-stationary environment and
maximises its own return. Two-level games (a planner and a population
of agents) are a special case.

A mean field game with common noise takes the continuum limit
$n \to \infty$. A representative agent interacts with the
cross-sectional distribution $\mu \in \Delta(\mathcal S)$ of
individual states and with a common noise $z \in \mathcal Z$ that
captures aggregate shocks. The individual transition is
$T(s' \mid s, a, \mu, z)$; the mean field evolves under
$\mu_{t+1} = \Phi^\pi(\mu_t, z_t)$ for an operator $\Phi^\pi$ induced
by the policy. The continuum limit makes the otherwise exponential
per-agent state space tractable at the cost of losing finite-population
effects.

A multi-objective MDP replaces the scalar reward $r$ with a vector
$r \in \mathbb R^d$. Without a scalar preference over its components,
the solution is a set of non-dominated policies rather than a single
optimum.

\subsubsection{Equilibrium concepts}

A policy $\pi^\star_i$ is a best response to the other agents'
policies $\pi_{-i}$ when $V_i^{(\pi^\star_i, \pi_{-i})}(s) \ge
V_i^{(\pi'_i, \pi_{-i})}(s)$ for every alternative $\pi'_i$. A Nash
equilibrium is a profile $(\pi^\star_1, \ldots, \pi^\star_n)$ in which
every component is a best response to the others. A mean-field Nash
equilibrium is the continuum analogue, in which the representative
agent's policy is a best response to the cross-sectional distribution
it induces. A restricted perceptions equilibrium is a best response
within a restricted policy class, for example policies that condition
on prices but not on the full distribution; it is weaker than
rational expectations and is the equilibrium concept SRL targets in
Section~\ref{sec:macro:solver}. An
$\varepsilon$-equilibrium weakens Nash by an $\varepsilon$ slack on
each player's incentive. Exploitability, $\sup_{\pi'} V^{\pi'} -
V^\pi$ evaluated at the current mean field (the gap between the best
response and the current policy), is the convergence metric used in
MFG-RL; it upper-bounds the gap to a mean-field Nash equilibrium.

\subsubsection{Notation summary}

Throughout the chapter we use $s$ for state, $a$ for action, $r$ for
reward, $\pi$ for policy, $V$ and $Q$ for values, $\beta$ for the
discount factor, $P$ or $T$ for the transition kernel,
$\mu \in \Delta(\mathcal S)$ for the cross-sectional distribution,
$z \in \mathcal Z$ for the aggregate shock or common noise, $p$ for
the equilibrium price vector when explicit, $\theta$ for policy
parameters, $\phi$ for auxiliary parameters when a second network is
needed (a critic in single-agent RL, a planner in two-level RL), $n$
for the number of agents in a Markov game, and $i$ for an agent
index. Algorithm-specific scalars (training-step counts $N$, $T$,
trajectory counts $E$) are introduced inside the relevant algorithm
boxes.

The chapter applies RL to macro models under two interpretations,
introduced above and used without further comment hereafter. Under
the solver interpretation, the policy network is a numerical device
for finding an equilibrium that closed-form or DP-based methods
cannot reach. Under the behavioural interpretation, the policy
network is the household's actual cognition, learning a policy from
realised utility under no prior model knowledge.
Section~\ref{sec:macro:behavioral} leans on the behavioural
interpretation; Sections~\ref{sec:macro:solver_single_agent},
\ref{sec:macro:solver}, and~\ref{sec:macro:mfg} lean on the solver
interpretation; the policy-design section uses one or the other
depending on whether the planner is the focus.

A growing parallel literature uses deep learning rather than
reinforcement learning to solve heterogeneous-agent and DSGE models
by training networks to satisfy equilibrium equations as nonlinear
regression objectives \citep{Maliar2021dl, Azinovic2022den,
Azinovic2026sequence, Kase2025hank, Kase2025gem,
HallHoffarth2023constraints, Carmona2021mfg, Fang2026dpi,
Thoeni2025nmfg}; \citet{Fernandez2024dl} survey this line. The
distinction is that those methods compute equation residuals offline,
while RL trains the network from a reward signal under the agent's
own policy. The chapter restricts attention to the RL approach and
notes the equation-residual literature only for comparison
where relevant.

\subsection{RL for single-agent macro models}
\label{sec:macro:solver_single_agent}

The simplest macroeconomic environment in which RL can be benchmarked
against dynamic programming is the representative-agent stochastic
real business cycle. The state space is two-dimensional, the
analytical and numerical benchmarks are well understood, and the
training dynamics expose how RL behaves on a problem for which DP
remains the right tool. The interest is in the comparison rather
than in any new claim about the RBC.

\citet{Atashbar2023imf} train a representative household with deep
deterministic policy gradient on a canonical stochastic RBC with
log utility, Cobb-Douglas production, and AR(1) total factor
productivity. The state is $(K, A)$ with $K$ capital and $A$
productivity, the action is consumption $C$, and the transition is
the capital accumulation equation together with the AR(1) for
productivity. In both a deterministic variant without TFP shocks and
a stochastic variant with shocks, the trained agent's consumption and
investment policy lies close to the deterministic steady-state
benchmark. The paper also documents training runs that fail to
converge under high discount factors and suboptimal critic
hyperparameters; the same instability reappears in the heterogeneous-agent
setting in
Section~\ref{sec:macro:solver}, where long-horizon credit assignment
across an idiosyncratic-shock distribution amplifies the same problem.
Section~\ref{sec:macro:simulation} returns to the same RBC with a
side-by-side comparison across PPO, DDPG, value-function iteration,
and a Blanchard-Kahn log-linearisation.

\citet{Shi2022consumption} trains an actor-critic agent on a
stochastic optimal-growth environment with no a priori knowledge of
preferences or the productivity process. The agent recovers the
rational-expectations optimal saving rule asymptotically. Higher
exploration accelerates learning at the cost of lower realised welfare
during the exploration phase, which gives the explore-vs-exploit
trade-off a macroeconomic interpretation that
Section~\ref{sec:macro:behavioral} returns to under the behavioural
lens.

\subsection{RL as a global solver for heterogeneous-agent macro models}
\label{sec:macro:solver}

Heterogeneous-agent (HA) macro models replace the representative
agent of Section~\ref{sec:macro:simulation} with a continuum of
households facing uninsurable idiosyncratic shocks on top of an
aggregate shock. Canonical examples are \citet{Aiyagari1994},
\citet{KrusellSmith1998}, and \citet{KaplanMollViolante2018}. The
state of the agent's problem now includes the cross-sectional
distribution $\mu_t \in \Delta(\mathcal S)$ of individual states,
because equilibrium prices depend on $\mu_t$ and the distribution
itself depends on the policy. Dynamic programming runs on an
infinite-dimensional state and the value function lives on
$\mathcal S \times \Delta(\mathcal S) \times \mathcal Z$. The
traditional macro response summarises $\mu_t$ by a handful of moments
\citep{KrusellSmith1998} or linearises around a stationary
equilibrium \citep{Reiter2009, Auclert2021}. Both reductions impose
ex ante which features of the distribution matter.

\citet{Yang2025srl} propose Structural Reinforcement Learning (SRL),
which sidesteps the distribution entirely. The structural assumption
is that the household knows its own budget constraint, preferences,
and idiosyncratic shock process exactly. The only unknown part of
the environment from the household's point of view is the joint
process for equilibrium prices and aggregate shocks. SRL trains the
policy by simulating the full economy and treating prices as
exogenous to the agent through a stop-gradient. The sequence-space
view of \citet{Auclert2021} justifies replacing $\mu$ in the agent
state with the equilibrium price sequence: under rational
expectations the price path is itself a sufficient statistic for the
part of $\mu$ that affects forward-looking household decisions,
because the household's budget constraint depends on $\mu$ only
through equilibrium prices. \citet{Han2022deepham}, the immediate
predecessor of SRL, kept the full cross-sectional distribution $\mu$
in the agent state via a deep-network policy that conditions on
$\mu$ directly; SRL replaces $\mu$ with prices.

The SRL household's state is
$s_t = (b_t, y_t, z_t, p_t)$, with $b_t$ wealth, $y_t$ the
idiosyncratic productivity shock, $z_t$ the aggregate shock, and
$p_t$ the equilibrium price vector. The action is $a_t = c_t$,
consumption, with savings $b_{t+1}$ determined residually by the
budget. The next state is generated by three structural pieces: the
budget constraint for $b_{t+1}$, the exogenous Markov chain for
$(y_{t+1}, z_{t+1})$, and the market-clearing update $p_{t+1} =
P^\star(\mu_{t+1}, z_{t+1})$ that the agent treats as exogenous. The
reward is $r_t = u(c_t)$. The policy class is restricted to
$\pi(s, z, p)$, which means the agent never conditions on $\mu$ at
any step. \citet{Yang2025srl} call the resulting fixed point a
sequential restricted perceptions equilibrium: given the price
process induced by $P^\star$, the policy $\pi^\star$ maximises
expected discounted utility within the restricted class, and market
clearing holds at $p_t = P^\star(\mu_t, z_t)$ for every $t$. Agents'
price beliefs are statistically consistent with realised prices on
the equilibrium path, and the equilibrium is a fixed point in policy
space rather than in the perceived-law-of-motion regression of
Krusell-Smith.

The policy is parameterised as a tabular vector
$\theta \in \mathbb R^{J \times K \times L}$ on the discretised
$(s, z, p)$ grid, and training uses the structural policy gradient
algorithm of Algorithm~\ref{alg:macro:spg}. The cross-sectional
distribution rolls forward to compute prices while a separate
value-computation distribution rolls forward to score the policy.
The Monte Carlo value is differentiated analytically by
backpropagating through the known one-step transition matrix
$A_\pi(z, p)$, the grid-restricted version of the transition kernel.
Convergence is declared when the $\ell_\infty$ distance between
successive policy iterates falls below a tolerance $\varepsilon$.

\begin{algorithm}[t]
\caption{Structural Policy Gradient (SPG) for HA macro models}
\label{alg:macro:spg}
\begin{algorithmic}[1]
\Require initial tabular policy $\theta_0 \in \mathbb{R}^{J \times K \times L}$ on the $(s, z, p)$ grid; step sizes $\{\eta_k\}$; trajectory count $N$; horizon $T$; initial distributions $\psi_g, \psi_z$; convergence tolerance $\varepsilon$
\For{$k = 0, 1, 2, \ldots$}
  \State Sample $N$ initial conditions $(\mu^n_0, z^n_0) \sim (\psi_g, \psi_z)$ for $n = 1, \ldots, N$
  \State Initialise the value-computation distribution $d^n_0 = d_0$, uniform over the individual-state grid
  \For{$n = 1, \ldots, N$}
    \For{$t = 0, 1, \ldots, T-1$}
      \State Compute the equilibrium price $p^n_t = P^*(\mu^n_t, z^n_t)$ from market clearing, with a stop-gradient applied to $p^n_t$
      \State Roll the cross-sectional distribution: $\mu^n_{t+1} = A_{\pi_\theta}(z^n_t, p^n_t)^\top \mu^n_t$ (used only to update prices)
      \State Roll the value-computation distribution: $d^n_{t+1} = A_{\pi_\theta}(z^n_t, p^n_t)^\top d^n_t$ (used only for the value)
      \State Sample the next aggregate shock $z^n_{t+1} \sim \Xi(\cdot \mid z^n_t)$
    \EndFor
  \EndFor
  \State Form the Monte Carlo value $\widehat{v}^{\pi}(\theta_k) = \dfrac{1}{N}\sum_{n=1}^N \sum_{t=0}^{T-1} \beta^t \,\bigl\langle d^n_t,\, u^{\pi_\theta}(z^n_t, p^n_t)\bigr\rangle$
  \State Compute the analytic gradient $\nabla_\theta \widehat{v}^{\pi}(\theta_k)$ by backpropagating through $A_{\pi_\theta}$ with prices held fixed by the stop-gradient
  \State Update $\theta_{k+1} \leftarrow \theta_k + \eta_k \nabla_\theta \widehat{v}^{\pi}(\theta_k)$
  \If{$\|\theta_{k+1} - \theta_k\|_\infty < \varepsilon$}
    \State \Return $\theta_{k+1}$
  \EndIf
\EndFor
\end{algorithmic}
\end{algorithm}

\citet{Yang2025srl} report runtimes on a single A100 GPU. Krusell-Smith
converges in roughly $55$ seconds and the authors report that its
dynamics match the deep-learning rational-expectations benchmark.
The Huggett model with aggregate risk, which has a nontrivial
market-clearing condition for bonds, converges in roughly $75$
seconds. A one-account HANK model with a forward-looking Phillips
curve and sticky prices converges in roughly three
minutes. Allowing agents to condition on a short price history rather
than only the current price does not change the solution materially,
which suggests that current prices encode most of the information
relevant for behaviour in these calibrations.

\subsection{Simulation: representative-agent RBC under DP and DRL}
\label{sec:macro:simulation}

The comparison tests whether RL matches dynamic programming on a
problem for which DP is the appropriate benchmark. The environment is
a standard representative-agent stochastic RBC. The
household chooses consumption $C_t$ and saves the residual, with
capital evolving as $K_{t+1} = (1-\delta) K_t + A_t K_t^\alpha - C_t$
and total factor productivity following $\log A_{t+1} = \rho \log A_t
+ \varepsilon_{t+1}$ for $\varepsilon_{t+1} \sim \mathcal{N}(0,
\sigma^2)$. Utility is $u(C_t) = \log C_t$. Parameters are calibrated
as $\beta = 0.96$, $\alpha = 0.36$, $\delta = 0.10$, $\rho = 0.95$,
$\sigma = 0.007$, with horizon $T = 200$. The deterministic steady
state is $K^\star = 4.29$, $C^\star = 1.26$.

The state is $s = (K, A)$, the action
is $a = C$, the reward is $r = \log C$, and the transition is the
capital accumulation equation together with the AR(1) for TFP. Four
methods are compared: a Blanchard-Kahn log-linearisation around the
deterministic steady state (KPR), value-function iteration on a
$400 \times 41$ tensor grid with a Tauchen-discretised TFP process
(VFI), proximal policy optimisation with a Gaussian actor (PPO), and
deep deterministic policy gradient (DDPG).\footnote{Both RL methods
use a 64-64 multilayer perceptron for actor and critic. Each is
trained from scratch with $10$ independent seeds. Every method is
evaluated on a fixed set of $30$ initial conditions $(K_0, A_0) \sim
\mathcal{U}[0.5, 8] \times \mathcal{U}[0.95, 1.05]$ and shock paths
$\{\varepsilon_t\}_{t=1}^T$, identical across methods, so cross-method
differences are not driven by evaluation noise. PPO is trained for
roughly $102{,}400$ environment steps per seed ($100$ updates of
$1024$ steps), DDPG for $60{,}000$, reflecting per-algorithm tuning
rather than equalised compute; the qualitative ordering (DRL within
the VFI standard-error band on welfare) is robust to this asymmetry.
The PPO entropy coefficient is set to zero, which reduces actor
variance in this one-dimensional control problem but may need to be
positive in higher-dimensional MDPs to avoid premature collapse.
Wall-clock training time per seed is reported in
Table~\ref{tab:macro:rbc-results}. The training budgets and
convergence-monitoring schedules are recorded in the simulation
script.}

\begin{table}[ht]
\centering
\begin{tabular}{lrrrr}
\toprule
Method & Mean return & SE & Policy MSE vs VFI & Wall clock \\
\midrule
KPR & 45.88 & 0.587 & 0.0004 & $-$ \\
VFI & 45.86 & 0.589 & 0.0000 & 13.8 s \\
PPO & 45.82 & 0.843 & 0.0087 & 32.9 s \\
DDPG & 45.17 & 1.476 & 0.0365 & 579.4 s \\
\bottomrule
\end{tabular}

\caption{Representative-agent RBC under four methods. Mean episode return averaged over 30 evaluation episodes with shared initial conditions and shock paths. The SE column reflects cross-seed standard error of per-seed mean returns for the stochastic methods (PPO and DDPG, $10$ seeds) and cross-episode standard error of returns within the single deterministic solution for KPR and VFI ($30$ episodes); the two are not strict comparables but are reported in the same column for compactness. Policy mean-squared error reported against VFI consumption on 1000 random states drawn uniformly from the evaluation support. Wall-clock time is per seed for PPO and DDPG, and total solver time for VFI.}
\label{tab:macro:rbc-results}
\end{table}

\begin{figure}[ht]
\centering
\includegraphics[width=0.85\linewidth]{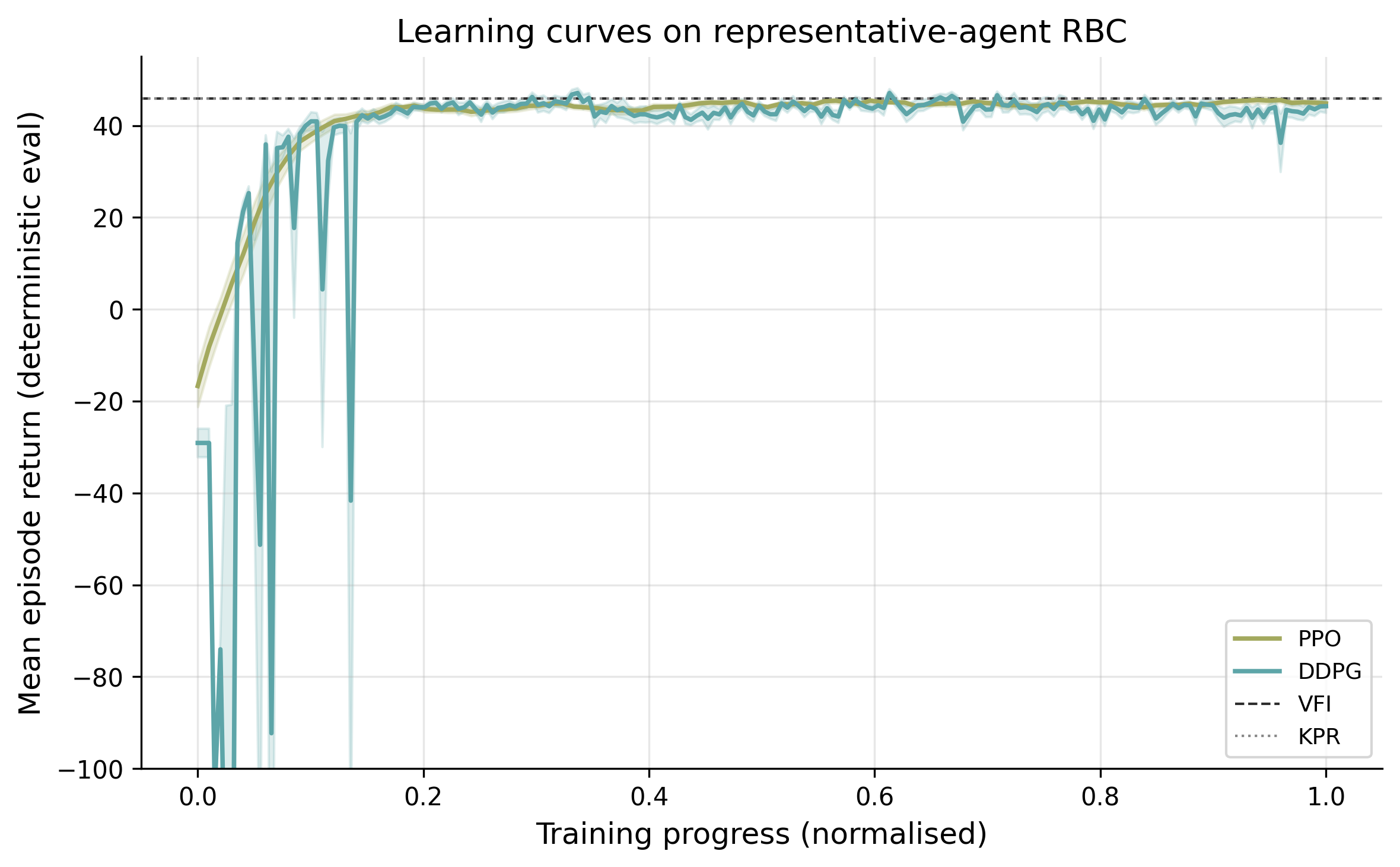}
\caption{Mean deterministic-evaluation return as a function of normalised training progress for PPO (olive) and DDPG (cyan), shaded by cross-seed standard error across ten training seeds. Horizontal lines mark the VFI benchmark mean return (dashed black) and the KPR log-linear benchmark (dotted gray).}
\label{fig:macro:rbc-curves}
\end{figure}

Table~\ref{tab:macro:rbc-results} reports mean episode return,
standard error, and policy mean-squared error against VFI. KPR
matches VFI to within $0.001$ in policy MSE and within statistical
noise on mean return. PPO converges to within the standard error of
VFI with a policy MSE of $0.009$. DDPG attains a mean return within
$2\%$ of VFI but has roughly double the cross-seed standard error
and a policy MSE of $0.037$. PPO and DDPG approach the VFI and KPR
benchmarks within their respective training budgets
(Figure~\ref{fig:macro:rbc-curves}). The comparison therefore links
the model-free policies to the Bellman benchmark defined in
Section~\ref{sec:macro:setup}; it does not imply that RL replaces
dynamic programming on this textbook case.

\subsection{Mean Field Games as a large-population RL problem}
\label{sec:macro:mfg}

The mean-field game generalisation of Section~\ref{sec:macro:setup}
applies when large populations of identical, anonymous agents
interact through aggregate quantities. Direct multi-agent RL does
not scale, since the joint state and action spaces grow exponentially
in $n$. Taking the continuum limit reduces the problem to a single
representative agent whose environment includes the population
distribution $\mu \in \Delta(\mathcal S)$ and the common noise $z$.
The RL problem is to find a policy that is a best response to the
distribution it induces under the noise process. Convergence to a
mean-field Nash equilibrium is monitored by exploitability.

The Krusell-Smith environment appears here in a different form
than in Section~\ref{sec:macro:solver}. SRL conditions agents on the
current price vector and treats it as a Markovian sufficient
statistic; RSPG instead lets agents carry a recurrent memory of past
observations, relaxing the Markov assumption and replacing restricted
perceptions with a partially observable mean-field Nash equilibrium.
The two are alternative solution concepts for the same underlying
household problem.

\subsubsection{Recurrent Structural Policy Gradient}
\label{sec:macro:rspg}

\citet{Wibault2026rspg} propose the Recurrent Structural Policy
Gradient (RSPG) algorithm for partially observable mean field games
with common noise. In macroeconomic applications agents typically
observe equilibrium prices but not the cross-sectional distribution,
so the natural formulation is a POMFG. RSPG exploits the fact that
individual transitions $T(s' \mid s, a, \mu, z)$ are known to the
agent (the household knows its own budget constraint and
idiosyncratic shock process), while the aggregate process for
$(\mu, z)$ must be learned. The setting therefore sits between
dynamic programming, which would integrate over both, and model-free
RL, which would sample both.

In chapter notation the POMFG with common noise is a tuple
$(p_{\mu_0}, p_{z_0}, \mathcal S, \mathcal Z, \mathcal A, T, \Xi, R,
\beta, U, \mathcal O)$, collecting initial distributions for the
individual state and common noise, individual and aggregate state
and action spaces, individual and common-noise transitions, reward,
discount, and observation map. It combines the MFG primitives of
Section~\ref{sec:macro:setup} with the POMDP observation pair
$(U, \mathcal O)$ and a common-noise transition $\Xi(\cdot \mid z)$.
The mean field evolves under the analytic operator
$\mu_{t+1} = \Phi^\pi(\mu_t, z_t)$ with $(s, s')$ entry
$\mathbb{E}_{a \sim \pi}[T(s' \mid s, a, \mu, z)]$. The RSPG
architectural choice is to restrict the policy memory to a history
of shared aggregate observations $o_{0:t}$, parameterised as the
hidden state of a recurrent neural network. The reduced policy
$\pi(a \mid s, h(o_{0:t}))$ retains history-dependence while keeping
the asymptotic cost of the analytic mean-field update independent of
the individual state.

For an initial common-noise distribution $p_{z_0}$, RSPG samples $E$
trajectories in parallel, rolls out the endogenous mean field
$\mu^\pi_{0:T}$ via the analytic operator $\Phi^\pi$, and computes
the discounted return
\begin{equation}
J^\pi \;=\; \mathbb{E}_{z_{0:T} \sim \Xi}\,\Bigl[\sum_{t=0}^{T-1} \beta^t \, \langle \mu^\pi_t,\; R(\cdot,\, \pi,\, \mu^\pi_t,\, z_t)\rangle\Bigr].
\label{eq:macro:rspg-return}
\end{equation}
Gradients flow through the individual transitions $T$ and the
expected-reward over actions. Gradients do not flow through the
mean-field transitions; the analytic mean-field update is treated as
a fixed environment, mirroring SRL's treatment of equilibrium
prices.\footnote{The training loop is parameterised by the
common-noise-trajectory batch size $E$, the rollout horizon $T$, and
the RNN hidden dimension. Convergence is monitored by exploitability,
the gap between the value of a best-response policy and the value of
the agent's current policy at the current mean field. The
exploitability metric requires a separate best-response oracle that
\citet{Wibault2026rspg} train as a single-agent RL agent against the
frozen mean-field path; reported exploitabilities are therefore upper
bounds.}

\begin{algorithm}[t]
\caption{Recurrent Structural Policy Gradient (RSPG)}
\label{alg:macro:rspg}
\begin{algorithmic}[1]
\Require initial recurrent policy $\theta_0$; trajectories $E$; horizon $T$
\For{$k = 0, 1, 2, \ldots$ until convergence}
  \State Sample $E$ common-noise trajectories $z^e_{0:T}$ with $z^e_0 \sim p_{z_0}$ and $z^e_{t+1} \sim \Xi(\cdot \mid z^e_t)$
  \State Roll forward $\mu^e_{t+1} = \Phi^{\pi_\theta}(\mu^e_t, z^e_t)$ analytically, with policy memory $h^e_{t+1} = \mathrm{RNN}(h^e_t, o^e_t)$
  \State Form the sample return $\widehat{J}(\theta_k)$ from \eqref{eq:macro:rspg-return} averaged over $e = 1, \ldots, E$
  \State Update $\theta_{k+1} \leftarrow \theta_k + \eta_k \nabla_\theta \widehat{J}(\theta_k)$ with gradients flowing through $T$ and $R$ only
\EndFor
\State \Return $\theta^*$
\end{algorithmic}
\end{algorithm}

\citet{Wibault2026rspg} evaluate RSPG on partially observable
Linear-Quadratic, Beach-Bar, and Krusell-Smith environments. Across
the three, RSPG attains the lowest or second-lowest exploitability
among the algorithms compared and converges roughly an order of
magnitude faster in wall-clock time than independent and recurrent
PPO baselines. Finite-horizon agents under RSPG spend wealth just
before episode end, pushing interest rates up and wages down.
Memoryless variants of SPG and IPPO do not show this pattern. The
paper positions RSPG as the first method to solve a partially
observable Krusell-Smith model in which agents observe prices but
not the cross-sectional distribution.

\subsubsection{Hierarchical Stackelberg mean-field RL}
\label{sec:macro:dsmfg}

\citet{Mi2025dsmfg} extend the mean-field formulation of
Section~\ref{sec:macro:rspg} from a symmetric population to a
hierarchical setting in which a government (leader) sets policy
first and a continuum of heterogeneous individual followers
optimises against the announced policy. The framework formalises a
Dynamic Stackelberg Mean Field Game (DSMFG) as the tuple
$(\mathcal S^l, \mathcal S^f, \mathcal A^l, \mathcal A^f, P, r^l,
r^f, T)$ collecting leader state space, follower state space, leader
action space, follower action space, joint transition, leader
reward, follower reward, and horizon. A mean field
$L_t \in \Delta(\mathcal S^f \times \mathcal A^f)$ summarises the
population state-action distribution.

The follower's best response to a leader policy $\pi^l$ is the
representative-agent policy
\begin{equation}
\pi^{f,*}(\pi^l) \;\in\; \arg\max_{\pi^f}\;
\mathbb E\!\left[\sum_{t=0}^{T} \beta^t\, r^f(s^f_t, a^f_t, L_t,
\pi^l_t)\right],
\label{eq:macro:dsmfg-follower}
\end{equation}
under the McKean-Vlasov fixed point that requires $L_t$ to be the
distribution induced by $\pi^{f,*}(\pi^l)$ at $t$. The leader then
chooses
\begin{equation}
\pi^{l,*} \;\in\; \arg\max_{\pi^l}\;
\mathbb E\!\left[\sum_{t=0}^{T} \beta^t\, r^l(s^l_t, a^l_t, L_t)\right]
\label{eq:macro:dsmfg-leader}
\end{equation}
anticipating the follower best response. The dynamic feedback is
critical, as a static Stackelberg formulation cannot capture the
temporal coupling that arises when followers observe and respond to
the leader's policy over $T$ periods.

The Stackelberg Mean Field Reinforcement Learning (SMFRL) algorithm
trains both levels jointly under centralised training with
decentralised execution. A Stackelberg mean-field Q-function
$\widetilde Q^l(s^l, a^l, L)$ scores leader actions given the
population state-action distribution; a follower-shared Q-network
conditions on the same mean field. Best-response updates alternate.
One step fixes the leader, trains followers to a $\varepsilon$-best-response and
updates $L$; the next fixes followers and optimises the leader. The mean-field
approximation reduces the pairwise complexity of agent-agent and
agent-government interactions from $O(N^2)$ to $O(N)$.

\citet{Mi2025dsmfg} report that DSMFG scales to $1{,}000$ followers
where the prior multi-agent benchmark in the same TaxAI environment
\citep{Mi2024taxai} ran at one hundred agents. On the optimal-tax
benchmark the learned leader policy achieves a four-fold per-capita
GDP gain over the analytical Saez optimum \citep{Saez2001} and a
nineteen-fold improvement over the static 2022 US federal income tax
schedule, both within the calibrated TaxAI environment. Ablations
confirm that removing either the Stackelberg hierarchy or the
mean-field approximation loses welfare or training stability.
Exploitability decreases monotonically over training, indicating
convergence to an approximate Stackelberg mean-field equilibrium.
DSMFG bridges the symmetric mean-field methods of
Section~\ref{sec:macro:rspg} and the single-planner setting of
Section~\ref{sec:macro:policy}, in that the leader-follower
structure makes the planner an explicit second decision-maker
without dropping the continuum-population approximation that keeps
the problem tractable.

\subsubsection{Offline, asynchronous, and non-stationary MFG-RL}
\label{sec:macro:other-mfg-rl}

Three recent algorithms extend MFG-RL along orthogonal axes.
\citet{Brunnbauer2024offmmd} learn an equilibrium policy from a
static dataset of past trajectories without online interaction, by
adapting Munchausen online mirror descent to offline data with an
overestimation-mitigating regulariser; their experiments are on
crowd-exploration and navigation, not macro.
\citet{YangShan2026tmf} drop the synchronous-move assumption by
replacing the mean-action statistic $\bar a$ with the population
distribution $\mu \in \Delta(\mathcal O)$, derive a TMF Bellman
equation, and prove an $O(1/\sqrt{N})$ finite-population
exploitability bound that holds for any batch size of acting agents.
\citet{Magnino2025nonstationary} consider non-stationary
continuous-space MFGs and report roughly an order-of-magnitude
reduction in sampling cost on their benchmarks via a fictitious-play
loop with a time-conditional normalising flow on the equilibrium
density.\footnote{\citet{Peng2025mfcg} apply PPO with generalised
advantage estimation and target networks to mean-field control
games, a hybrid that combines internal coordination as in mean-field
control with external competition as in MFGs.}

\subsection{Simulation: a partially observable linear-quadratic mean field game}
\label{sec:macro:mfg_sim}

The simulation uses a compact version of the partially observable
Linear-Quadratic environment in
\citet{Wibault2026rspg}. The individual state is
$s \in \{0,\ldots,98\}$ in the public MFAX implementation, the action
is $a \in \{-3,\ldots,3\}$, the common noise is
$z \in \{-1,1\}$, and the horizon is $T = 30$. The common noise is
persistent within an episode. It shifts the whole population down
early in the episode, is neutral in the middle, and shifts it in the
opposite direction late in the episode. Let
$\bar{s}_t = \sum_{s'} \mu_t(s')s'$ denote the population mean. The
agent observes its own state and $o_t = \bar{s}_t$, but not the full
distribution, the common-noise realisation, or the calendar time. The
mean field evolves by the analytic population transition operator
$\mu_{t+1} = \Phi^{\pi}(\mu_t,z_t)$.

The step reward follows the paper's quadratic form,
\begin{equation}
R_t(s,a,\mu,z)
= -c_a a^2 + q a(\bar{s}_{t+1} - s)
  - \frac{\kappa}{2}(\bar{s}_{t+1} - s)^2,
\label{eq:macro:lqmfg-reward}
\end{equation}
with terminal reward
$R_T(s,\mu,z) = -c_{\mathrm{term}}(\bar{s}_T - s)^2/2$.
The MFAX defaults that we adopt unchanged are $c_a = 0.5$, $q = 0.1$,
$\kappa = 0.5$, and $c_{\mathrm{term}} = 1.0$, with idiosyncratic-noise
scale $\sigma = 1.0$ and common-noise sensitivity $\rho = 0.5$. The
horizon $T = 30$ is the MFAX truncation; the discount factor
$\gamma = 0.99$ used during training is finite-horizon-compatible and
matches the paper's configuration. The implementation follows MFAX in
evaluating the reward after the population transition, using
$\bar{s}_{t+1}$, and using the terminal reward on the final
transition. The reward encourages agents to coordinate with the moving
population while paying an action cost.

The reported results come from the public MFAX implementation of the
paper's two structural policy-gradient algorithms on this environment.%
\footnote{Specifically MFAX commit \texttt{9acc1eb}
(\url{https://github.com/CWibault/mfax.git}). We apply only mechanical
Python~3.11 dataclass \texttt{default\_factory} fixes and a stdout
addition that prints the per-iteration mean policy return; the diff is
checked in at \texttt{patches/mfax/py311\_compat.patch} with a README.
None of the patched lines touches the SPG, RSPG, or environment update
rules.}
SPG uses the analytic mean-field update and a memoryless categorical
policy that conditions on current state and current mean observation.
RSPG uses the same analytic update, but encodes the sequence of shared
mean-state observations with a GRU; the recurrent hidden state is
independent of the individual state, as in the paper. Both methods are
evaluated with the analytic mean-field path induced by the final
policy. Approximate exploitability is computed by backward induction
against that path, so lower values mean a smaller gain from deviating
to a best response. We do not include the paper's sampled recurrent PPO
grid in the table, because that baseline is a much heavier finite-agent
RL experiment rather than the structural mean-field update studied here.

\begin{table}[ht]
\centering
\begin{tabular}{lrrrr}
\toprule
Method & LR & Exploitability & Expected return & Train time (s) \\
\midrule
SPG & $10^{-2}$ & $86.64 \pm 16.29$ & $-1021.3 \pm 22.3$ & $24.39 \pm 0.06$ \\
RSPG & $10^{-3}$ & $60.37 \pm 4.11$ & $-1186.9 \pm 10.1$ & $28.79 \pm 0.58$ \\
\bottomrule
\end{tabular}

\caption{Partially observable Linear-Quadratic mean field game. Approximate
exploitability is computed by a best-response dynamic program on the
analytic mean-field path induced by the final policy. Expected return and
training time are averaged over 10 training seeds, with standard errors
across seeds. Learning rates are selected from the official MFAX grid
$\{10^{-4},10^{-3},10^{-2}\}$ by mean final exploitability for each
method.}
\label{tab:macro:lqmfg-results}
\end{table}

\begin{figure}[ht]
\centering
\includegraphics[width=0.80\linewidth]{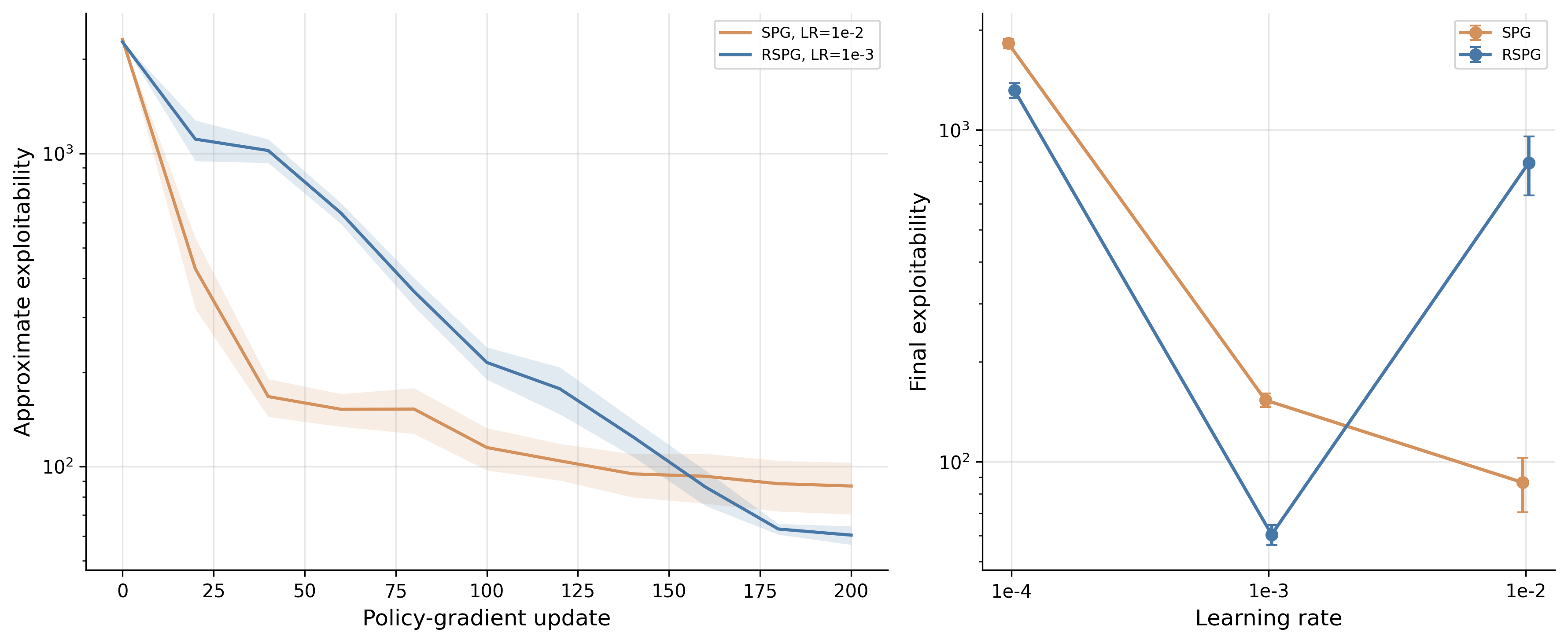}
\caption{Official MFAX Linear-Quadratic grid. The left panel shows
approximate exploitability over policy-gradient updates for the selected
SPG and RSPG learning rates, shaded by cross-seed standard error across
ten training seeds. The right panel shows final exploitability across
the swept learning rates.}
\label{fig:macro:lqmfg-curves}
\end{figure}

Table~\ref{tab:macro:lqmfg-results} reports exploitability, expected
return, and training time across ten training seeds after sweeping the
official learning-rate grid. RSPG has lower exploitability than SPG.
The ranking is consistent with the partial-observation mechanism in
Section~\ref{sec:macro:rspg}, since the recurrent policy can use
observation history to infer the common-noise path. The experiment does
not isolate recurrence as the cause.%
\footnote{RSPG's mean final exploitability is non-monotone in the
learning rate (1316, 60, 797 at $10^{-4}, 10^{-3}, 10^{-2}$), an
interior optimum at $10^{-3}$. SPG's minimum sits at the grid edge
$10^{-2}$ (1824, 153, 87). A wider sweep is required before comparing
the best attainable values of the two methods.}
Expected returns do not provide a welfare ranking because SPG and RSPG
induce different mean-field paths.
Figure~\ref{fig:macro:lqmfg-curves} plots exploitability over training
and final exploitability across learning rates.

\subsection{RL as a model of bounded rationality}
\label{sec:macro:behavioral}

The preceding sections used RL as a numerical device for finding
equilibrium policies that closed-form or DP-based methods cannot
reach. The interpretation switches here. The household itself
learns from realised utility under no a priori knowledge of the
income process, and the question is whether the resulting consumption
behaviour matches empirical patterns that rational expectations does
not.

\citet{Kaplowitz2025} models a household solving a Markovian
consumption-savings problem with cash-on-hand $x = y + Ra$,
consumption share $\psi \in [0, 1]$, and savings residual
$a' = (1 - \psi) x$, where $y \in \{y_e, y_u\}$ denotes employed and
unemployed income drawn from a Bernoulli transition matrix
$P_{yy'}$. The household does not know $P_{yy'}$ and learns an
expected-continuation-value estimator
$\widehat{EV}(y, a'; \phi)$ as a neural network with weights $\phi$
updated by stochastic gradient descent under a Robbins-Monro
$1/\sqrt{t}$ learning-rate schedule. The Q-function is
\begin{equation}
Q(a, y, a') \;=\; u(\psi x) + \beta\, \widehat{EV}(y, a'; \phi),
\label{eq:macro:kap-q}
\end{equation}
and the consumption policy is a polynomial fit over $Q$. The agent
is initialised at the rational-expectations solution to the same
problem, so any deviation from rational behaviour reflects the
learning dynamics rather than initial misspecification. The agent
runs for fifty quarters with parameters drawn from the standard
household-finance calibration (Krueger et al., 2016; Ganong et al.,
2024 as cited in the source).

The same RL mechanism reproduces two empirical patterns that the
combination of full-information rational expectations and
borrowing constraints cannot generate jointly. First, the
Ganong-et-al. marginal propensity to consume out of stimulus
transfers is around $0.50$ for previously-low-asset unemployed
households and around $0.34$ for previously-high-asset unemployed
households, even when neither group is at a borrowing constraint.
The simulated agents reproduce a gap of roughly the same magnitude
(approximately $0.20$). Second, the Malmendier-Shen scarring effect,
in which past unemployment experience persistently lowers consumption
controlling for current state, appears with the correct sign in the
simulated paths. Both patterns trace to a single
mechanism, value-function approximation errors that evolve with
experience and generate ex-post heterogeneity from ex-ante identical
agents. Belief-updating explanations, in which the household revises
beliefs about $P_{yy'}$ after each unemployment spell, can produce
the scarring effect but force consumption levels and MPCs to co-move
in a way the data do not show.

\citet{Shi2022consumption} (Section~\ref{sec:macro:solver_single_agent})
shows that a single-agent actor-critic on stochastic optimal growth
converges to the rational-expectations optimal saving rule
asymptotically; KAP's contribution is the non-asymptotic regime in
which the agent has not yet converged. \citet{Kuriksha2021} embeds heterogeneous
neural-network learners in an Aiyagari-class economy. Each household
updates its saving rule from its own realised consumption history,
and the resulting wealth distribution is fat-tailed and exhibits the
high average MPCs and excess sensitivity documented in the empirical
literature, even without any single agent matching specific anomalies.
The two papers together support the reading that RL households fail
to satisfy rational expectations during learning in ways that
reproduce stylised micro and macro facts, and that the mechanism is
generic to value-function approximation rather than a peculiarity of
any one architecture.

\subsection{Multi-agent reinforcement learning for policy design}
\label{sec:macro:policy_marl}
\label{sec:macro:policy}

The first variant of RL for policy design treats the planner and the
citizens as RL learners co-adapting in a shared environment. The
planner's tax schedule, monetary rule, or climate-cooperation
protocol is the action of one RL agent; the citizens' consumption,
labour, or mitigation responses are the actions of other RL agents.
Both sides learn from realised payoffs, so the planner sees the
emergent strategic responses of the citizens to its policy choices.
This setting is a Markov game in the sense of
Section~\ref{sec:macro:setup}, with $n$ heterogeneous agents whose
actions are coordinated either by market clearing or by explicit
search-and-match protocols. The contribution of RL is the discovery
of mechanism designs and the surfacing of emergent behaviours that
classical optimisation cannot reach. The single-planner variant in
Section~\ref{sec:macro:policy_single} drops the multi-agent training
and treats the rest of the macroeconomy as a fixed environment.

\subsubsection{MARL on canonical macro environments}
\label{sec:macro:marl-rbc}

\citet{Gabriele2026marlbc} build MARL-BC, an $n$-household RBC in
which the representative-agent RBC and the Krusell-Smith mean-field
economy appear as limit cases. Each household chooses a consumption
fraction and labour supply at every period. Production is
Cobb-Douglas in productivity-weighted aggregate capital and labour.
In the chapter's Markov-game notation, the state is shared across
agents and contains aggregate capital, labour, and the per-agent
productivity draws; each agent's action is its consumption fraction
and labour choice; each agent's reward is the log-plus-leisure
utility
\begin{equation}
r^i_t \;=\; \log c^i_t + b\,\log(1 - \ell^i_t).
\label{eq:macro:marl-bc-reward}
\end{equation}
The authors benchmark PPO, SAC, DDPG, and TD3 and report that the
off-policy methods are more sample-efficient than PPO on the
configurations they
test.\footnote{Sample efficiency is reported in terms of environment
steps to reach a fixed return threshold. The exact threshold and the
number of seeds vary across the three population configurations,
$n=1$, $n=20$ ex-ante identical, $n=20$ heterogeneous. The paper
does not provide a single hyperparameter table for the four
algorithms.} With $n = 1$ the framework reproduces the textbook RBC
solution. With a large population of ex-ante identical agents (in
their largest run, on a $23 \times 23$ productivity grid yielding
$n = 529$) it reproduces the Krusell-Smith mean-field wealth
distribution and aggregate dynamics. With a population of
heterogeneous agents it produces wealth and consumption distributions
that mean-field methods cannot represent.

\citet{Curry2022marl} consider a richer micro-founded RBC with $100$
worker-consumers, $10$ firms, and a tax-setting government,
formulated as a partially observable Markov game in which firms set
prices and wages and goods are rationed proportionally if demand
exceeds supply. Each agent type is trained by independent PPO with
parameter sharing within type, accelerated by the WarpDrive GPU
framework.\footnote{Without a curriculum, independent learners fail
to recover non-trivial behaviour. A multi-phase curriculum that
stages consumer, firm, and government training recovers non-trivial
$\varepsilon$-meta-equilibria. The $\varepsilon$ metric is evaluated
by training single-agent best-response oracles against frozen others,
so reported values are upper bounds.} In the open-economy variant
with an external export market, the model exhibits multiple
qualitatively distinct $\varepsilon$-meta-equilibria.

\subsubsection{RL agents inside macro ABMs}
\label{sec:macro:rmabm}

\citet{Brusatin2024rmabm} extend an existing macroeconomic ABM with
capital and credit, in which $1{,}000$ worker-households, $20$
capital-goods firms, $100$ consumption-goods firms, and one bank
interact through search-and-match consumption markets, a labour
market, and a credit market. The consumption-goods firms in the
original ABM follow a hand-coded trend-following rule for setting
prices and target quantities. The authors replace $N$ of these $100$
firms with tabular Q-learners.

Each RL firm $i$ has state $s_{i,t}$
consisting of its log price-delta and log stock-delta relative to
market averages, both binned on a finite grid; the action $a_{i,t}$
is a discrete multiplicative adjustment to next-period price and
quantity; the reward $r_{i,t}$ is profit normalised by assets, with
a bankruptcy penalty. The Q-table update is the standard tabular
Q-learning rule
\begin{equation}
Q_{i,t+1}(s_{i,t}, a_{i,t}) \;=\; (1 - \alpha)\, Q_{i,t}(s_{i,t}, a_{i,t}) + \alpha\,\bigl[\, r_{i,t} + \beta \max_{a'} Q_{i,t}(s_{i,t+1}, a')\,\bigr]
\label{eq:macro:rmabm-q}
\end{equation}
with learning rate $\alpha$ and $\varepsilon$-greedy
exploration.\footnote{The grid discretisation, the exploration-rate
schedule, and the training-step budget per firm are left as
configurable inputs in the published code. The shared-policy variant
uses a single Q-table across all RL firms; the independent-policy
variant maintains a separate Q-table per firm. The market-competition
level is controlled by the number of firms a consumer compares
prices across.}

\citet{Brusatin2024rmabm} report that RL firms learn three distinct
profit-maximising strategies depending on competition level and the
share of rational firms. Independent learners with separate Q-tables
segregate into distinct strategic groups, raising aggregate market
power and profits. Raising the share of RL firms always raises total
output in the configurations tested, sometimes at the cost of higher
output volatility.

\subsubsection{Two-level deep RL for optimal taxation}
\label{sec:macro:aie}

\citet{Zheng2022aie} propose the AI Economist, a two-level deep RL
framework for optimal taxation. The inner level is a population of
agents who maximise discounted post-tax utility; the outer level is
a social planner who sets a tax schedule to maximise a social-welfare
function. Both levels are parameterised by deep neural networks and
trained with proximal policy optimisation.

In the chapter's Markov-game notation, agent $i$ has policy
$\pi_\theta$ and solves
\begin{equation}
\max_{\pi_\theta}\; \mathbb{E}\sum_{t=0}^{H} \beta^t\, u_i(C_{i,t}, L_{i,t}),
\label{eq:macro:aie-agent}
\end{equation}
where $u_i$ is isoelastic in post-tax income $C_{i,t}$ with curvature
$\eta > 0$ and linear in cumulative labour $L_{i,t}$; post-tax income
depends on a bracketed tax schedule $\mathcal T(z; \tau)$. The
planner has policy $\pi_\phi$ and solves
\begin{equation}
\max_{\pi_\phi}\; \mathbb{E}\sum_{t=0}^{H} \beta^t\, \text{swf}_t,
\label{eq:macro:aie-planner}
\end{equation}
with $\text{swf}_t$ either inverse-income-weighted utilitarian welfare
or the product of equality and productivity.

The two-level formulation is unstable because each side's effective
MDP is non-stationary in the other side's policy. \citet{Zheng2022aie}
stabilise training with two mechanisms. A curriculum gradually
introduces labour costs and then taxes, so that early in training
the agents experience low costs and explore freely. Entropy
regularisation on $\pi_\phi$ keeps the tax schedule diffuse, so that
agents experience a wide range of tax rates and learn to condition
on them. These mechanisms stage training in two phases: agents adapt
to random taxes first; the planner then enters and the two levels
co-adapt.\footnote{Both networks are LSTM-based and trained with PPO;
the agent and planner have separate replay buffers $\mathcal{D}$ and
$\mathcal{D}_p$. The planner acts every $T$ environment steps at a
tax-period boundary; agents act every step. Curriculum and entropy
schedules are advanced between training iterations and reported in
the paper's experimental appendix. The reported wall-clock cost is
several thousand GPU-hours for the Gather-Trade-Build runs, the
largest training budget of the anchor papers in this chapter.}

\begin{algorithm}[t]
\caption{Two-level Reinforcement Learning, after \citet{Zheng2022aie}}
\label{alg:macro:aie}
\begin{algorithmic}[1]
\Require initial agent parameters $\theta$, planner parameters $\phi$; sampling horizon $H$; tax-period length $T$; inner learner $\mathcal{A}$ (PPO in this work)
\State Initialise curriculum and planner-entropy schedules
\While{not converged}
  \State Reset world state $s_0$ and replay buffers $\mathcal{D}$ (agents), $\mathcal{D}_p$ (planner)
  \For{$t = 0, \ldots, H$}
    \If{$t \bmod T = 0$}
      \State Planner samples tax rates $\tau \sim \pi_\phi(\cdot \mid o_{p,t}, h_{p,t})$
    \EndIf
    \State Each agent samples action $a_{i,t} \sim \pi_\theta(\cdot \mid o_{i,t}, h_{i,t})$ in parallel
    \State Step the environment; compute post-tax rewards; deliver the planner reward at tax-period boundaries
    \State Append transitions to $\mathcal{D}$ and $\mathcal{D}_p$
  \EndFor
  \State Update $\theta$ via $\mathcal{A}$ on $\mathcal{D}$; update $\phi$ via $\mathcal{A}$ on $\mathcal{D}_p$
  \State Advance curriculum and entropy schedules
\EndWhile
\State \Return $(\theta, \phi)$
\end{algorithmic}
\end{algorithm}

\citet{Zheng2022aie} test the framework in two environments. In a
one-step economy with closed-form Saez optimum \citep{Saez2001}, the
AI Economist's learned tax schedule matches the Saez optimum to
within statistical noise. In the four-agent Open-Quadrant variant of
the Gather-Trade-Build spatial environment, the learned schedule
reportedly outperforms a Saez baseline calibrated to the same
environment by $8\%$ on utilitarian welfare and $12\%$ on the product
of equality and productivity. The schedule is non-monotonic, with
highest marginal rates on middle income brackets. Emergent behaviour
includes specialisation by skill into gatherers and builders and
intertemporal tax avoidance by high-skill agents. The result is
suggestive rather than decisive. The two-level RL planner discovers
tax schedules that the analytical Saez framework does not, but the
welfare ranking depends on the choice of SWF and on the absence of
behavioural responses outside the modelled action
space.\footnote{\citet{Tacchetti2025dmd} survey the deep mechanism
design landscape, recommending bilevel optimisation as the central
stabilisation strategy and noting that human-response models built
from imitation data are not robust to strategic adversaries who
anticipate being modelled.}

\subsubsection{Scaling MARL-based optimal taxation}
\label{sec:macro:taxai}

\citet{Mi2024taxai} extend the AI Economist agenda to a Bewley-Aiyagari
environment calibrated to US data, scaling from the four-to-ten-agent
Gather-Trade-Build sandbox of \citet{Zheng2022aie} to ten thousand
heterogeneous households. Households face idiosyncratic labour-income
shocks and choose consumption, saving, and labour supply; firms
produce with Cobb-Douglas technology; a financial intermediary
intermediates household savings to firm investment; the government
sets income-tax brackets and lump-sum transfers. The framework
benchmarks seven multi-agent RL algorithms (DDQN, MAAC, MADDPG,
BiCNet, and centralised-training/decentralised-execution variants)
against a genetic-algorithm baseline and a dynamic-programming
single-planner baseline. The seven MARL methods uniformly outperform
the classical baselines on welfare metrics in the configurations
tested. Households trained under MARL exhibit emergent tax-avoidance
strategies, a strategic-response phenomenon that the smaller-scale
spatial environment of \citet{Zheng2022aie} cannot surface. The
finding is a scale-up rather than a methodological alternative; AIE
remains the conceptual anchor for two-level RL policy design.

\subsubsection{Multi-region RL on RICE-N}
\label{sec:macro:ricen}

Climate-economy policy combines multiple regions interacting through
global temperature and trade, no central authority enforcing
cooperation, and a multi-decade horizon. The benchmark family of
models is the integrated assessment model, which couples a climate
block to an economic block and informs IPCC reports. The canonical
IAMs are DICE \citep{Nordhaus1992dice} and its regional extension
RICE. RICE-N replaces the optimal-control solver inside RICE with a
multi-agent system in which each region is a strategic learner.

In chapter notation RICE-N is a Markov game with $n$ regions. The
state $s_t = (s^{\mathrm{nat}}_t, s^{\mathrm{soc}}_t)$ has a natural
component, the atmospheric carbon mass and global mean temperature,
and a social component collecting each region's capital, technology,
population, trade balance, and the negotiation state. Region $i$'s
action $a_{i,t}$ is a tuple of a savings rate, mitigation rate,
import-demand and import-tariff vectors, and a vector of negotiation
actions. The transition factorises into a natural block (carbon mass
and temperature evolve by the carbon-cycle and energy-balance
equations inherited from DICE, with aggregate emissions
$\sum_i E_{i,t}$ as the only social-to-natural channel) and a social
block (Cobb-Douglas output scaled by a damage factor
$1 - \Omega(T_t)$, with capital accumulating in the standard way and
the negotiation state updating under the protocol in force). Region
$i$'s reward is the isoelastic utility of per-capita consumption,
where consumption is an Armington aggregate of domestic and imported
goods net of foreign tariffs.

RICE-N uses RL rather than DP because the game has no central
planner, is non-stationary from any one region's perspective, has a
combinatorial negotiation action space, and is only partially
observed; classical RICE sidesteps all four by computing a
single-planner Nash equilibrium. The agents themselves remain
utility-maximisers; RL is the method that finds their strategic
policies and does not imply that regions are boundedly rational. The cost
of the RL solution is the loss of the optimality certificate and the
seed-independent reproducibility that the classical RICE optimum
provides.

The contribution of RICE-N is to formalise the negotiation stage
that precedes economic activity at each timestep, with one timestep
representing five years. \citet{Zhang2025ricen} compare two
negotiation protocols. Bilateral Negotiation has each region propose
a pair $(\hat{m}_i, \hat{m}_j)$ to every other region $j$, stating
its own promised mitigation rate and the rate it requests; regions
accept or reject; each commits to the maximum mitigation rate across
accepted proposals. The Basic Club protocol \citep{Nordhaus2015climateclubs}
proposes a club rate $\hat{m}_i$; regions accept or reject; a club
forms with a common minimum rate, and non-members face a tariff
proportional to the gap. Each region's policy is a neural network
with weights shared across regions and region-specific inputs,
trained with advantage actor-critic.\footnote{Both negotiation
protocols reduce temperature growth at the cost of a small drop in
production, while distributing emission-reduction costs more
equitably than the no-negotiation baseline in the published runs.
\citet{Zhang2025ricen} are deliberately non-committal on what
equilibrium concept the learners reach; the framework produces a
family of outcomes indexed by protocol design rather than converging
to a single solution. The AAAI workshop predecessor
\citep{Zhang2022ricen} introduced the modelling framework without
the Basic Club protocol; \citet{Heitzig2023ccf} proposes a Conditional
Commitments mechanism intended to enhance participation stability.}

Three findings carry policy interpretation that the classical DICE
and RICE optimisations cannot produce. A climate club enforced by
border tariffs is self-sustaining in the configurations tested. The
negotiation institution is itself a policy lever, in that Bilateral
Negotiation and Basic Club reach materially different climate and
economic outcomes from the same underlying economy. Both protocols
lower the cross-region inequality of abatement cost and mitigation
rate relative to the no-negotiation baseline, and the authors
caution that a uniform border tariff can still operate as a de
facto carbon tax on developing regions absent redistribution or
technology transfer.

\subsubsection{Multi-objective MARL for equitable policy}
\label{sec:macro:justice}

\citet{Biswas2025justice} keep the multi-region structure of RICE-N
but change the reward. JUSTICE incorporates the economy, damage, and
abatement modules from the $57$-region RICE50+ integrated assessment
model and couples them with the FAIR climate emulator. It is a
multi-objective multi-agent Markov decision process in which the
regions share a team reward that is a two-component vector, the
negative of global mean temperature and global net economic output,
both to be maximised.
Training uses a multi-objective multi-agent actor-critic that
decomposes the vector-reward problem into scalarised single-objective
problems via weighted-sum scalarisation.\footnote{\citet{Biswas2025justice}
sample one hundred weight vectors, train one policy per weight, and
collect the non-dominated policies into the solution set. The output
is a frontier of policies rather than a single policy. A Gini-based
equity index across regions is reported alongside the climate-output
frontier.} JUSTICE reports the non-dominated policies found across the
one hundred weighted-sum runs. The reported set is an empirical
approximation to the climate-output frontier, not an exhaustive
enumeration of every efficient policy.

\subsection{Single-planner reinforcement learning for policy design}
\label{sec:macro:policy_single}

The second variant of RL for policy design treats the planner as the
sole RL agent. The rest of the macroeconomic environment is a fixed
DSGE, a fitted VAR, or an empirically calibrated structural model
populated by rational households whose responses are taken as given.
The contribution of RL is the ability to learn nonlinear and
state-dependent reaction functions in environments where closed-form
optimal control is intractable. The setting is closer to optimal
control theory than to mechanism design, with RL in the role that
nonlinear model predictive control plays in classical engineering
applications.

\subsubsection{Single-planner regime learning in a monetary DSGE}
\label{sec:macro:cjk}

\citet{Chen2021drl} apply deep RL to the \citet{Benhabib2001} model
of monetary-fiscal interaction. A representative household solves
\begin{equation}
\max\, \mathbb{E}_0 \sum_{t=0}^{\infty} \beta^t \bigl[\,u(c_t) + v(m_t) - w(n_t)\,\bigr]
\quad \text{s.t.} \quad c_t + m_t + b_t = w_t n_t - \tau_t + \frac{R_{t-1} b_{t-1} + m_{t-1}}{\pi_t},
\label{eq:macro:cjk-household}
\end{equation}
where $m_t = M_t/P_t$ is real money, $b_t = B_t/P_t$ is real bonds,
$\tau_t$ is a lump-sum tax, $w_t$ is the real wage, $\pi_t =
P_t/P_{t-1}$ is gross inflation, and output is linear in labour,
$y_t = \varepsilon^y_t n_t$, with $\varepsilon^y_t$ an i.i.d.\
technology shock. Fiscal policy follows a Leeper-style rule
$\tau_t = \gamma_0 + \gamma\, b_{t-1} + \varepsilon^\tau_t$ and
monetary policy follows a global nonlinear interest-rate rule
$R_t = f(\pi_t)\, \varepsilon^R_t$ with $f$ strictly positive and
strictly increasing. The deterministic Fisher equation
$R = \pi/\beta$ intersects $f(\pi)$ at the inflation target
$\pi^\star$ and at a liquidity trap $\pi_L$.\footnote{Each policy
rule is locally active or passive depending on the sign of a
coefficient. Fiscal policy is active when $\gamma < \beta^{-1} - 1$.
Monetary policy is locally active at inflation $\pi$ when
$f'(\pi) > \beta^{-1}$. The cross-product of the two stances
generates four regimes, of which the doubly-passive case is locally
indeterminate under rational expectations and the doubly-active case
is locally explosive. \citet{EvansHonkapohja2005} show that
recursive-least-squares adaptive learning reaches only the
determinate regimes.}

The household's state is
$s_t = (m_{t-1}, b_{t-1}, \pi_{t-1}, c_{t-1}, n_{t-1},
\varepsilon^\tau_t, \varepsilon^R_t, \varepsilon^y_t)$, the action
$a_t = (c_t, b_t, n_t)$ is a continuous tuple of consumption, bond
holdings, and labour, the reward
$r_t = u(c_t) + v(m_t) - w(n_t)$ is the instantaneous utility, and
the transition $P(s_{t+1} \mid s_t, a_t)$ is induced by the
household's intertemporal Euler equation
\begin{equation}
u'(c_t) \;=\; \beta\, R_t\, \mathbb{E}_t\!\left[\frac{u'(c_{t+1})}{\pi_{t+1}}\right]
\label{eq:macro:cjk-euler}
\end{equation}
together with the equilibrium fiscal-monetary rules and the
exogenous shock processes. The policy is learned by soft actor-critic
\citep{Haarnoja2018}.\footnote{SAC is an off-policy maximum-entropy
actor-critic. The Gaussian actor $\pi_\theta$ and the Q-function
$Q_\phi$ are deep neural networks trained by stochastic gradient
descent on minibatches drawn from a replay buffer. The critic update
follows the temporal-difference loss with a target network and the
actor update is the reparameterised maximum-entropy gradient.
\citet{Chen2021drl} report that the trained policy converges to a
stationary outcome in all four monetary-fiscal regimes, including
the doubly-passive case (locally indeterminate under rational
expectations) and the doubly-active case (locally explosive). The
interpretation is that the deep policy is constrained by the global
utility-maximisation problem rather than by the linearised dynamics
around any one steady state.}

The substantive claim is that the set of monetary-fiscal regimes in
which a household can converge to a stationary policy is strictly
larger under deep RL than under the recursive-least-squares
adaptive-learning benchmark of \citet{EvansHonkapohja2005}. RL extends
learnability to the indeterminate and explosive regimes that adaptive
learning cannot reach.

\subsubsection{Benchmarking RL on monetary policy}
\label{sec:macro:mpu}

\citet{Wang2025mpu} compare nine RL algorithms on a monetary-policy
MDP fitted to US Federal Reserve quarterly data (1955--2025). In
chapter notation the state is $s_t = (\pi_t, U_t, \mathrm{gap}_t,
R_t)$ comprising inflation, unemployment, the output gap, and the
policy rate; the action $a_t$ is a discrete adjustment to the policy
rate; the reward encodes the Fed's dual mandate as a quadratic loss
in deviations from a $2\%$ inflation target and a $4.5\%$ natural
unemployment rate. The transition is a linear-Gaussian VAR fitted to
historical transitions.\footnote{The nine algorithms span tabular
Q-learning variants, SARSA, actor-critic, deep Q-networks, Bayesian
Q-learning, and a POMDP particle-filter variant. The training and
evaluation protocols are aligned across algorithms, with each agent
trained on the same fitted VAR dynamics and evaluated on a held-out
window.} Standard tabular Q-learning achieves the best mean
discounted return, outperforming both the enhanced RL methods and a
Taylor-rule baseline. The dynamics are linear-Gaussian rather than a
structural DSGE and the paper has not been peer-reviewed, so the
finding is preliminary. It suggests, no more strongly than that, that
sophisticated RL machinery does not dominate the combination of
simple rules and tabular RL in low-dimensional monetary-policy MDPs.

\subsubsection{Single-planner RL on data-grounded monetary models}
\label{sec:macro:ht-ks}

\citet{HinterlangTaenzer2024} train a central-bank RL agent on a
neural-network surrogate of the US economy estimated from quarterly
data 1987Q3--2007Q2. The pipeline first fits transition equations
from data, in linear (SVAR) and nonlinear (feedforward ANN) variants,
and then trains an RL reaction function inside the fitted environment
with the zero lower bound, asymmetric central-bank preferences, and
the dual mandate baked into the reward. The RL-optimised linear rule
reduces the central-bank loss by more than thirty-five per cent
relative to common Taylor-style rules and the realised federal funds
rate over the sample window. The nonlinear ANN reaction function
extends this to a forty-three per cent reduction. A
cross-validation exercise feeds the learned rules into eleven
alternative DSGE models from the Macroeconomic Model Database and
finds smaller unconditional inflation and output variances under the
RL rules than under the Taylor benchmarks, a generalisation result
absent from the MPU comparison above.

\citet{Khundadze2025} extend the single-planner setup to the Eurozone,
using soft actor-critic to control inflation, the output gap, and
debt in a two-bloc (North-South) macro environment with asymmetric
shocks. The RL policy is benchmarked against a nonlinear Model
Predictive Control (NMPC) baseline. SAC achieves comparable or better
stabilisation on average and is more robust under asymmetric shocks
because the learned feedback policy generalises where NMPC must
re-solve an open-loop trajectory each period. Cross-bloc transfers
emerge as a learned component of the joint fiscal-monetary policy.

\subsection{Discussion}
\label{sec:macro:discussion}

\subsubsection{Four roles, one toolbox}

Reinforcement learning enters macroeconomics in four distinct roles
covered above. As a global solver, structural policy gradient methods
solve heterogeneous-agent models on timescales that classical
projection and perturbation cannot reach. As a game-theoretic
formalism, mean-field RL handles partial observability and
hierarchical Stackelberg structure. As a model of the household,
RL learning dynamics reproduce empirical anomalies that rational
expectations with borrowing constraints cannot generate jointly.
As a planner, RL discovers tax schedules, climate-cooperation
mechanisms, and monetary-fiscal reaction functions that closed-form
optimisation does not. The four roles share the same algorithms but answer different questions, and the empirical and theoretical
literatures around each are at different stages of maturity.

\subsubsection{Common methodological issues}
\label{sec:macro:common_issues}

Reading across the four roles reveals issues that no single section
makes explicit. The first concerns convergence theory. Of the
algorithms surveyed in this chapter, only the temporal mean-field
construction of \citet{YangShan2026tmf} carries a classical
guarantee, with existence and uniqueness of the equilibrium under a
discrete monotonicity condition, finite-population
$O(1/\sqrt{N})$ Nash approximation regardless of the per-step batch
size, and provable convergence of the policy-gradient iterate (the
sequence of policy updates produced by the algorithm). The
remaining methods, including \citet{Yang2025srl},
\citet{Wibault2026rspg}, \citet{Kaplowitz2025},
\citet{Zheng2022aie}, \citet{Chen2021drl}, \citet{Mi2025dsmfg},
\citet{Han2022deepham}, \citet{Brunnbauer2024offmmd}, and
\citet{Magnino2025nonstationary}, monitor convergence empirically by
exploitability, $L^{\infty}$ norms of policy updates, or Bellman
residuals against application-specific tolerances. This is not a
failure of rigour. The macro equilibrium concepts these algorithms
target, including restricted perceptions equilibrium, mean-field
Nash, Stackelberg mean-field, and co-adaptation (joint training of
planner and agents against each other's current policy), live in
spaces where classical fixed-point theorems (existence results such
as Brouwer or Kakutani) either do not apply or do not
yield computable bounds. The honest reading of the literature is
that convergence here means training stopped changing the policy,
not that the iterate provably reaches a model equilibrium.

The four roles also target four different equilibrium concepts,
and papers within a role diverge further
(Section~\ref{sec:macro:setup}). The restricted perceptions
equilibrium of \citet{Yang2025srl} assumes agents condition on the
current price vector as if it were Markov, which is not satisfied by
true Krusell-Smith equilibrium prices but is what makes structural
policy gradient tractable. \citet{Wibault2026rspg} extends this to a
partially observable mean-field Nash with a shared observation and
recurrent memory. \citet{Mi2025dsmfg} targets a Stackelberg
mean-field with a single leader and a continuum of homogeneous
followers. \citet{Zheng2022aie} trains planner and citizens by PPO
against each other's current policy without naming the equilibrium
the joint iterate is supposed to reach. \citet{Chen2021drl} accept
whichever DSGE steady state the SAC iterate stabilises at, including
the indeterminate and explosive regimes that adaptive learning
cannot reach. \citet{Kaplowitz2025} abandons equilibrium altogether
in favour of single-agent Q-learning calibrated to micro-panel
moments. A claim that RL solves Krusell-Smith therefore means
different things under \citet{Yang2025srl} than under
\citet{Han2022deepham}, where the same model is solved as a
recursive equilibrium in which $\mu$ is summarised by a finite set
of moments (moment closure), with neural-network value and policy
functions.

Several recurring pitfalls follow from these design choices.
Multiplicity is acknowledged but resolved silently by initialisation,
through the warm-up phase of \citet{Yang2025srl}, the two-phase
entropy schedule of \citet{Zheng2022aie}, and the regime-by-regime
training of \citet{Chen2021drl}; none of these papers tests
initialisation robustness. Curriculum learning is empirically
necessary but theoretically unmotivated, with \citet{Curry2022marl}
showing independent multi-agent RL fails to recover non-trivial
behaviour without one and \citet{Zheng2022aie} confirming the same
pattern in two-level policy design. Hyperparameter sensitivity is
severe and unevenly characterised, with \citet{Wibault2026rspg}
reporting that off-the-shelf IPPO and recurrent-IPPO require 81 to
243 configurations on partially observable Krusell-Smith while
structural policy gradient methods converge with three; no paper
offers transferable tuning guidance. Discount-factor sensitivity is
essentially unexamined, with \citet{Atashbar2023imf} flagging
instability at high $\beta$ but no paper sweeping $\beta$ over the
macro-realistic range. The single-agent to equilibrium gap is open,
since the household learning rule of \citet{Kaplowitz2025}
reproduces marginal propensity to consume heterogeneity and
scarring at the household level but has not been tested for
consistency once embedded in an Aiyagari-style equilibrium where
its policy would shift the cross-sectional distribution.

Three cross-cutting findings are worth stating separately from
their parent sections. The first is that tabular and grid-based RL
competes with or beats deep RL on macro problems.
\citet{Wang2025mpu} benchmarks nine algorithms on US Federal
Reserve quarterly data and finds tabular Q-learning beats DQN, soft
actor-critic, and Bayesian variants on a quadratic-loss monetary
objective, and \citet{Yang2025srl} reports grid policies training
faster than deep networks on Krusell-Smith. The ordering familiar
from robotics and games does not transfer mechanically to macro,
whose state spaces are low-dimensional once correctly compressed.
The second finding is that RL reaches equilibrium regions adaptive
learning cannot. \citet{Chen2021drl} shows deep RL stabilises in
indeterminate and explosive DSGE regimes that least-squares
learning rules out by E-stability (the expectational-stability
condition under which a perceived law of motion is learnable by
least-squares updating), suggesting the same set of algorithms
extends the equilibrium set under bounded rationality rather than
merely reproducing it. The third is sociological. The three
methodological clusters covered in this chapter, structural solvers
\citep{Yang2025srl,Wibault2026rspg,Han2022deepham,Azinovic2026sequence},
behavioural single-agent learners
\citep{Kaplowitz2025,Kuriksha2021,Shi2022consumption}, and policy
designers
\citep{Zheng2022aie,Mi2024taxai,Chen2021drl,HinterlangTaenzer2024,Khundadze2025,Wang2025mpu},
have evolved in parallel with limited cross-citation. The solver
papers do not engage the behavioural ones, the behavioural paper of
\citet{Kaplowitz2025} cites no RL-macro work, and the two monetary
policy papers \citep{HinterlangTaenzer2024,Wang2025mpu} appear
unaware of each other. The risk is that solver-driven equilibrium
claims, behavioural calibration evidence, and policy-design
exercises drift apart without methodological reconciliation.

\section{Reinforcement Learning in Games}
\label{section:rl_games}

With multiple agents adapting simultaneously, each agent's environment includes the others' changing policies, so the stationary-transition assumption behind single-agent convergence results fails. \citet{shoham2007multiagent} enumerate five desiderata for multi-agent learning: (1) convergence to a stationary strategy in self-play; (2) rationality (best-responding against stationary opponents); (3) equilibrium attainment; (4) safety (guaranteeing at least the Nash-value payoff); (5) social welfare. No existing algorithm satisfies all five in general games.

Two paradigms emerged. Value-based methods generalize the Bellman operator to games, replacing the $\max$ with game-theoretic solution concepts (minimax, Nash), targeting stochastic games with simultaneous moves and observable payoffs. Regret-based methods (CFR) accumulate counterfactual regrets and let the time-averaged strategy converge, targeting extensive-form games with sequential moves and private information. Computing Nash equilibria is PPAD-complete \citep{Daskalakis2009}, so neither approach escapes the fundamental hardness, but both achieve convergence in the game classes they target.

The Engine Replacement MDP has one decision-maker, so it cannot represent strategic opponents or the information sets required for counterfactual regret minimization.

\subsection{Stochastic Games and Equilibrium Learning}

\subsubsection{The Stochastic Game Framework}

An $n$-player stochastic game $\Gamma = (n, \mathcal{S}, \mathcal{A}_1, \ldots, \mathcal{A}_n, P, r_1, \ldots, r_n, \gamma)$ consists of a finite state space $\mathcal{S}$; finite action sets $\mathcal{A}_i$ for each player $i$; a transition function $P: \mathcal{S} \times \mathcal{A}_1 \times \cdots \times \mathcal{A}_n \to \Delta(\mathcal{S})$; reward functions $r_i: \mathcal{S} \times \mathcal{A}_1 \times \cdots \times \mathcal{A}_n \to \mathbb{R}$; and a common discount factor $\gamma \in [0,1)$. At each stage, all players simultaneously choose actions, receive individual rewards, and the game transitions to a new state.\footnote{\citet{shapley1953stochastic} introduced stochastic games in 1953 for the two-player zero-sum case, proving existence of the value via a contraction argument on the Bellman operator. The general-sum extension to $n$ players is due to Fink (1964) and Takahashi (1964).}

A Markov decision process is a stochastic game with $n = 1$; a matrix game is a stochastic game with $|\mathcal{S}| = 1$. Each player $i$ seeks a policy $\pi_i: \mathcal{S} \to \Delta(\mathcal{A}_i)$ maximizing discounted return $\mathbb{E}[\sum_{t=0}^\infty \gamma^t r_i(s_t, a_{1,t}, \ldots, a_{n,t})]$.

Standard Q-learning convergence \citep{WatkinsDayan1992} requires stationary transition and reward dynamics; with multiple learners, this assumption fails. \citet{BowlingVeloso2002} formalized two properties a learning algorithm should satisfy. It is \emph{rational} if, when all other players converge to stationary policies, it converges to a best response; it is \emph{convergent} if, in self-play, it converges to a stationary policy.

If all players use rational, convergent algorithms, the resulting profile is a Nash equilibrium by construction. The challenge is achieving both properties simultaneously.

\subsubsection{Minimax-Q Learning}

\citet{littman1994markov} proposed the first Q-learning algorithm for stochastic games, targeting two-player zero-sum games. The key modification replaces the $\max$ operator in the standard Q-learning backup with a minimax operator. Each agent maintains $Q_i(s, a_i, a_{-i})$ over the joint action space. The update rule is
\begin{equation}
Q_i(s, a_i, a_{-i}) \leftarrow (1-\alpha)\, Q_i(s, a_i, a_{-i}) + \alpha\left[r_i + \gamma\, V_i(s')\right],
\end{equation}
where the value backup solves a linear program:
\begin{equation}
V_i(s) = \max_{\pi_i \in \Delta(\mathcal{A}_i)} \min_{a_{-i} \in \mathcal{A}_{-i}} \sum_{a_i \in \mathcal{A}_i} \pi_i(a_i)\, Q_i(s, a_i, a_{-i}).
\label{eq:minimaxv}
\end{equation}
This is the RL analogue of Shapley's value iteration for zero-sum stochastic games \citep{Shapley1964}. The resulting policy $\pi_i(s)$ is generally a mixed strategy, since deterministic policies are exploitable in adversarial settings.

Minimax-Q converges to the minimax Q-values under Robbins-Monro learning rates ($\sum_t \alpha_t = \infty$, $\sum_t \alpha_t^2 < \infty$) and infinite exploration of all state-action tuples.\footnote{The convergence proof extends the contraction argument for standard Q-learning. Because the zero-sum minimax operator is a contraction with modulus $\gamma$ under the $\ell^\infty$ norm, the stochastic approximation converges to the fixed point. See \citet{littman1994markov} and the general treatment in Szepesvári and Littman (1999).} However, the algorithm sacrifices rationality: it plays the equilibrium strategy even against exploitable opponents.\footnote{In the soccer game of \citet{littman1994markov}, minimax-Q won 53.7\% against a hand-built opponent versus 26.1\% for Q-learning. Against an adversarial challenger, Q-learning won 0\% (its deterministic policy was fully predictable); minimax-Q won 37.5\% through mixed strategies.}

\subsubsection{Nash-Q Learning}

\citet{hu2003nash} extended the framework to general-sum stochastic games, where players may have aligned, opposed, or mixed incentives. Each agent $i$ maintains a Q-function over the joint action space $Q_i(s, a_1, \ldots, a_n)$ and updates via
\begin{equation}
Q_i(s, \mathbf{a}) \leftarrow (1-\alpha)\, Q_i(s, \mathbf{a}) + \alpha\left[r_i + \gamma\, \text{Nash}_i\bigl(Q_1(s'), \ldots, Q_n(s')\bigr)\right],
\label{eq:nash_q_update}
\end{equation}
where $\text{Nash}_i(\cdot)$ denotes player $i$'s payoff under a Nash equilibrium of the stage game defined by the current Q-values $(Q_1(s'), \ldots, Q_n(s'))$. At each backup, the algorithm treats the Q-values as payoff matrices, computes a Nash equilibrium of this matrix game, and uses the equilibrium payoffs for the value estimate.

\begin{theorem}[\citet{hu2003nash}]
\label{thm:nash_q}
Nash-Q converges to Nash Q-values under Robbins-Monro learning rates and infinite exploration, provided either (a) every stage game encountered during learning has a global optimal point (all agents receive their highest payoff at the same joint action), or (b) every such stage game has a saddle-point Nash equilibrium, with agents consistently updating at the corresponding equilibrium.
\end{theorem}

\begin{proof}
The argument recasts the joint tuple of Q-factors as a stochastic-approximation recursion driven by the Nash backup, shows that this backup fixes the equilibrium values and contracts toward them under the stage-game condition, and then applies the same pseudo-contraction lemma that delivers single-agent Q-learning convergence (Theorem~\ref{thm:qlearning_convergence}). Collect the agents' tables into a tuple $Q = (Q_1, \ldots, Q_n)$ and write the Nash-backup operator
\[
(P_t Q)_i(s, \mathbf{a}) = r_i(s, \mathbf{a}) + \gamma\, \mathrm{Nash}_i\bigl(Q_1(s'), \ldots, Q_n(s')\bigr), \qquad s' \sim P(\cdot \mid s, \mathbf{a}),
\]
so the update~\eqref{eq:nash_q_update} is the exponential moving average $Q_i^{t+1}(s,\mathbf{a}) = (1-\alpha_t)\, Q_i^t(s,\mathbf{a}) + \alpha_t\, (P_t Q^t)_i(s,\mathbf{a})$, applied only at the visited tuple. Measure distance between tuples in the supremum norm $\|Q - \hat{Q}\| = \max_i \max_s \max_{\mathbf{a}} |Q_i(s,\mathbf{a}) - \hat{Q}_i(s,\mathbf{a})|$.

The backup fixes the equilibrium values. A general-sum discounted stochastic game admits a stationary Nash equilibrium, and its equilibrium action values $Q^* = (Q_1^*, \ldots, Q_n^*)$ satisfy $Q_i^*(s,\mathbf{a}) = r_i(s,\mathbf{a}) + \gamma \sum_{s'} P(s'|s,\mathbf{a})\, \mathrm{Nash}_i\bigl(Q^*(s')\bigr)$ \citep{hu2003nash}. Taking the conditional expectation of $P_t$ over the sampled successor therefore returns $\mathbb{E}[P_t Q^*]_i(s,\mathbf{a}) = Q_i^*(s,\mathbf{a})$, so $Q^*$ is the fixed point of the mean backup and the one-sample noise $P_t Q^* - \mathbb{E}[P_t Q^*]$ is mean-zero given the history.

The backup contracts, provided every stage game encountered has a global optimum or a saddle and agents update at that equilibrium. The only nonlinearity in $P_t$ is the map from a stage game's payoff matrices to its Nash value; rewards are fixed and the discount is $\gamma$, so it suffices that this map be nonexpansive. Let $\sigma$ and $\hat{\sigma}$ be the selected equilibria of the stage games $Q(s')$ and $\hat{Q}(s')$, and write $\sigma Q_i$ for the expected payoff of the joint profile $\sigma$ under $Q_i$. When both are global optima, $\hat{\sigma}$ gives every agent its highest payoff under $\hat{Q}$, so $\hat{\sigma} \hat{Q}_i \geq \sigma \hat{Q}_i$, and
\[
\sigma Q_i - \hat{\sigma} \hat{Q}_i \;\leq\; \sigma Q_i - \sigma \hat{Q}_i \;=\; \sum_{\mathbf{a}} \sigma(\mathbf{a})\bigl[Q_i(s',\mathbf{a}) - \hat{Q}_i(s',\mathbf{a})\bigr] \;\leq\; \|Q_i(s') - \hat{Q}_i(s')\|;
\]
the symmetric bound gives $|\mathrm{Nash}_i(Q(s')) - \mathrm{Nash}_i(\hat{Q}(s'))| \leq \|Q(s') - \hat{Q}(s')\|$.\footnote{The saddle case is identical after replacing the global-optimum inequalities with the two saddle inequalities, using that all saddle points of a stage game carry the same value \citep[Lemma~14]{hu2003nash}. Both arguments require the selected equilibria to be value-unique; where a stage game has multiple non-equivalent equilibria the Nash-value selection is neither single-valued nor nonexpansive, the contraction fails, and different agents backing up different equilibria can drive the iterates apart. This is the divergence documented above.} Hence $\|P_t Q - P_t \hat{Q}\| \leq \gamma \|Q - \hat{Q}\|$, so $P_t$ is a $\gamma$-contraction in the supremum norm.

With $\mathbb{E}[P_t Q^*] = Q^*$ and the $\gamma$-contraction established, the stochastic-approximation lemma for pseudo-contractive updates \citep[Lemma~8]{hu2003nash}, the joint-space counterpart of the single-agent result behind Theorem~\ref{thm:qlearning_convergence}, yields $Q^t \to Q^*$ with probability one under infinite visitation of every $(s,\mathbf{a})$ and Robbins-Monro step sizes.
\end{proof}

These conditions are restrictive. When stage games have multiple Nash equilibria, agents may select different equilibria for their backups, causing divergence.\footnote{In the experiments of \citet{hu2003nash}, a grid game with a unique equilibrium Q-function converged in 100\% of trials under Nash-Q versus 20\% under independent Q-learning; a game with three equilibrium Q-functions converged in only 68--90\% of trials. Nash-Q requires each agent to observe all other agents' rewards and to maintain Q-values over the joint action space $\mathcal{A}_1 \times \cdots \times \mathcal{A}_n$, with storage $O(n |\mathcal{S}| \prod_i |\mathcal{A}_i|)$, exponential in the number of players.} Nash-Q reduces to standard Q-learning in the single-agent case.

\subsubsection{The Convergence Problem}

Table~\ref{tab:marl_comparison} summarizes the trade-offs. No single algorithm achieves both rationality and convergence in general games.

\begin{table}[htbp]
\centering
\caption{Multi-agent Q-learning algorithms: convergence and information requirements}
\label{tab:marl_comparison}
\begin{tabular}{l cc ll}
\hline
Algorithm & Rational & Convergent & Game class & Information \\
\hline
Indep.\ Q-learning & Yes & No & Any & Own reward \\
Minimax-Q & No & Yes & Zero-sum & Joint actions \\
Nash-Q & Cond. & Cond. & General-sum & All rewards \\
WoLF-PHC & Yes & Yes ($2{\times}2$) & General-sum & Own reward \\
\hline
\end{tabular}
\end{table}

WoLF-PHC (Win or Learn Fast, Policy Hill-Climbing) of \citet{BowlingVeloso2002} maintains a policy $\pi_i(a|s)$ and a running average policy $\bar{\pi}_i(a|s)$, updating Q-values as in standard Q-learning. The policy moves toward the greedy action at a variable rate:
\begin{equation}
\delta = \begin{cases}
\delta_l & \text{if } \sum_a \pi_i(a|s)\, Q_i(s,a) < \sum_a \bar{\pi}_i(a|s)\, Q_i(s,a) \quad \text{(losing)} \\
\delta_w & \text{otherwise} \quad \text{(winning)}
\end{cases}
\end{equation}
with $\delta_l > \delta_w$. When the current policy underperforms the historical average (losing), the agent adapts quickly; when outperforming (winning), it adapts slowly to avoid destabilizing the opponent. \citet{BowlingVeloso2002} proved that WoLF-IGA (the infinitesimal gradient ascent variant) converges to Nash equilibrium in all two-player, two-action games. The trajectory traces piecewise ellipses around the equilibrium, shrinking by a factor of $\ell^4 < 1$ per orbit, where $\ell = \sqrt{\delta_w / \delta_l}$. WoLF-PHC requires only own-reward observations, the same information as independent Q-learning.

Two further algorithms deserve mention. Friend-or-Foe Q-learning \citep{littman2001friend} decomposes agents as cooperative (friend) or adversarial (foe), using $\max$ for friends and minimax for foes; it always converges but requires knowing the relationship type a priori.\footnote{Correlated-Q learning \citep{greenwald2003correlated} generalizes both Nash-Q and Minimax-Q by using correlated equilibrium, a probability distribution over joint actions enforced by a correlating device. The set of correlated equilibria contains all Nash equilibria. Correlated-Q converges under conditions analogous to Nash-Q.} The evolutionary perspective of \citet{borgers1997learning} shows that reinforcement learning dynamics in matrix games converge to the replicator equation from evolutionary game theory. The cycling of Q-learning in games such as matching pennies is structurally identical to the cycling of replicator dynamics in Rock-Paper-Scissors games.

\subsubsection{Simulation Study: Cournot and Bertrand Duopoly}
\label{sec:cournot_bertrand}

Two canonical games from industrial organization serve as benchmarks.
In Cournot duopoly, two firms choose quantities
$q_i \in \{0, 1, \ldots, 9\}$ with inverse demand $P(Q) = 10 - Q$ and
marginal cost $c = 1$. The continuous Nash equilibrium is $q^* = 3$
with profit $\pi^* = 9$. In Bertrand duopoly with differentiated
products, two firms choose prices $p_i \in \{0, 1, \ldots, 9\}$ with
demand $d_i = 10 - 2 p_i + p_j$ and marginal cost $c = 1$. The
symmetric first-order condition gives
$p^* = (a + b c)/(2 b - e) = 4$ with profit $\pi^* = 18$.

Bertrand admits a single pure-strategy Nash equilibrium on the integer
grid at $(p_0, p_1) = (4, 4)$. Cournot admits three pure-strategy Nash
equilibria on the integer grid, $(2, 4)$, $(3, 3)$, and $(4, 2)$.
Only the symmetric Cournot equilibrium coincides with the continuous
solution.\footnote{Each configuration runs for 50{,}000 iterations
across 20 seeds. The Nash-Q implementation here selects the
joint-payoff-maximizing equilibrium when multiple pure Nash equilibria
exist. This deviates from the canonical \citet{hu2003nash} backup by
using an exogenous selection rule to pin down a single value function.
In Cournot it selects $(3, 3)$.} The comparison includes independent
Q-learning (IQL), Nash-Q, and WoLF-PHC. The stateless design has no
memory of past actions, so it cannot represent the trigger strategies
studied in the Q-learning collusion literature
\citep{Calvano2020,AskerEtAl2020}.

\begin{table}[htbp]
\centering
\caption{Learned actions and profits in Cournot and Bertrand duopoly}
\label{tab:cournot_bertrand}
\begin{tabular}{ll rr r}
\hline
Game & Algorithm & Action & Profit & $|a - a^*|$ \\
\hline
Cournot & IQL & $2.95 \pm 0.05$ & $9.1 \pm 0.0$ & $0.05$ \\
 & Nash-Q & $2.89 \pm 0.33$ & $8.8 \pm 1.3$ & $0.17$ \\
 & WoLF-PHC & $3.00 \pm 0.00$ & $9.0 \pm 0.0$ & $0.00$ \\
\hline
Bertrand & IQL & $4.00 \pm 0.00$ & $18.0 \pm 0.0$ & $0.00$ \\
 & Nash-Q & $4.00 \pm 0.00$ & $18.0 \pm 0.0$ & $0.00$ \\
 & WoLF-PHC & $3.95 \pm 0.05$ & $17.8 \pm 0.2$ & $0.05$ \\
\hline
\end{tabular}
\begin{minipage}{0.9\textwidth}
\footnotesize
Notes: Action and Profit report mean $\pm$ standard error across 20 seeds over the final 5,000 iterations. $|a - a^*|$ is the mean distance from the symmetric continuous Nash action ($q^* = 3$ for Cournot, $p^* = 4$ for Bertrand).
\end{minipage}
\end{table}

\begin{figure}[htbp]
\centering
\includegraphics[width=\textwidth]{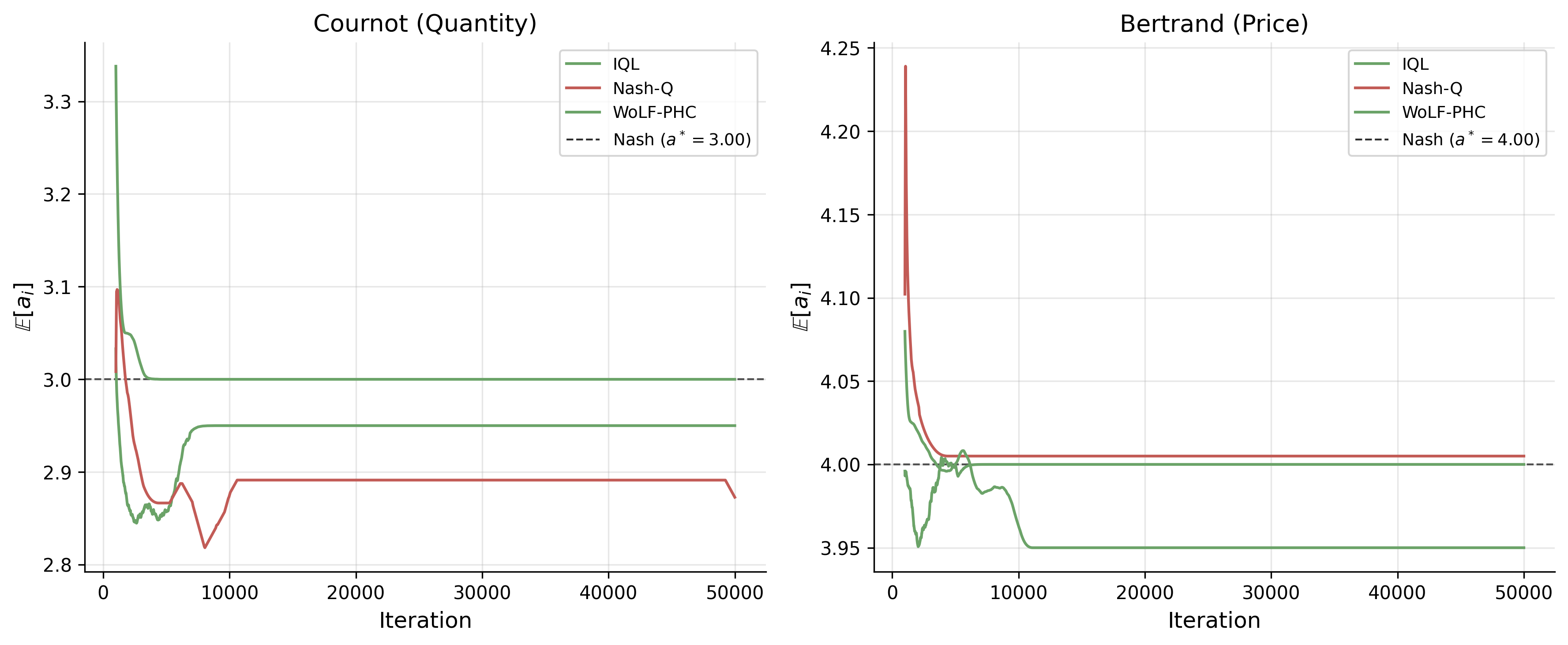}
\caption{Expected actions over training in Cournot and Bertrand
duopoly. The left panel reports Cournot and the right panel reports
Bertrand. Curves use 1{,}000-iteration smoothing windows and averages
across 20 seeds.}
\label{fig:cournot_bertrand}
\end{figure}

Table~\ref{tab:cournot_bertrand} reports that all three algorithms
settle near the symmetric Nash action in both games. IQL has no
game-theoretic backup but matches the game-aware methods on these two
pure-action benchmarks. This comparison does not test games whose Nash
equilibria require randomized strategies. Figure~\ref{fig:cournot_bertrand}
shows the corresponding training paths.

\subsection{Counterfactual Regret Minimization}

Extensive-form games require a different approach. CFR bypasses equilibrium selection. Instead of computing Nash equilibria at each step, it minimizes cumulative regret and lets the time-averaged strategy converge to equilibrium.

An extensive-form game consists of a game tree with information sets $\mathcal{I}_i$ partitioning player $i$'s decision nodes. An information set groups nodes where the player cannot distinguish between them due to hidden opponent actions or private information. A behavioral strategy $\sigma_i: \mathcal{I}_i \to \Delta(\mathcal{A})$ assigns action probabilities at each information set.

The \emph{counterfactual value} of action $a$ at information set $I$ is
\begin{equation}
v^\sigma_i(I, a) = \sum_{h \in I} \pi^\sigma_{-i}(h) \sum_{z \sqsupseteq ha} \pi^\sigma(ha, z) u_i(z)
\end{equation}
where $\pi^\sigma_{-i}(h)$ is the probability opponents reach $h$, and $\pi^\sigma(ha, z)$ is the probability of reaching terminal $z$ from $ha$. The counterfactual formulation weights by opponent reach $\pi^\sigma_{-i}(h)$ rather than joint reach $\pi^\sigma(h)$, ensuring non-zero updates even at rarely-visited information sets. Cumulative regret for action $a$ is $R^T(I,a) = \sum_{t=1}^T [v^{\sigma^t}(I,a) - v^{\sigma^t}(I)]$. CFR updates via \emph{regret matching}:\footnote{Regret matching is an action selection rule where the probability of choosing action $a$ is proportional to the cumulative regret for not having chosen $a$ in the past (truncated at zero). Actions with high regret receive higher probability; actions with negative regret (having performed worse than average) receive zero probability.}
\begin{equation}
\sigma^{T+1}(I, a) = \frac{[R^T(I, a)]^+}{\sum_{a'} [R^T(I, a')]^+}, \quad [x]^+ = \max(x,0).
\end{equation}
The average strategy $\bar{\sigma}^T$ converges to $\varepsilon$-Nash with $\varepsilon = O(|\mathcal{I}|\sqrt{|\mathcal{A}|}\Delta / \sqrt{T})$, where $\Delta$ is the range of payoffs \citep{zinkevich2008}.\footnote{An $\varepsilon$-Nash equilibrium is a strategy profile where no player can improve her expected payoff by more than $\varepsilon$ through unilateral deviation. As $\varepsilon \to 0$, this converges to an exact Nash equilibrium.}

CFR+ \citep{tammelin2014solving} replaces standard regret matching with Regret Matching+, which truncates cumulative regrets to zero after each update rather than only at action selection. The update becomes $R^{T+1}(I, a) = \max(R^T(I, a) + r^{T+1}(I, a),\, 0)$, where $r^{T+1}(I,a) = v^{\sigma^{T+1}}(I,a) - v^{\sigma^{T+1}}(I)$ is the instantaneous counterfactual regret. CFR+ also weights iteration $t$ by $t$ when computing the average strategy. While vanilla CFR converges at $O(1/\sqrt{T})$, CFR+ empirically converges at $O(1/T)$.\footnote{The $O(1/T)$ rate for CFR+ is empirically observed but not yet proven in full generality. \citet{tammelin2014solving} conjectured this rate based on extensive experiments across poker variants.} CFR+ enabled \citet{bowling2015heads} to essentially solve heads-up limit Texas hold'em, a game with $3.16 \times 10^{17}$ states, the first non-trivial imperfect-information game played competitively by humans to be essentially solved. Their program Cepheus achieved exploitability below 1 mbb/g (milli-big-blind per game).

\subsubsection{Simulation Study: Equilibrium Computation in Kuhn Poker}
\label{sec:kuhn_poker}

Kuhn poker is a three-card simplified poker with twelve information sets, six per player, and a known one-parameter family of Nash equilibria; player 0's equilibrium strategies are indexed by a bluffing probability $\alpha \in [0, 1/3]$ while player 1's equilibrium strategy is unique. The simulation compares vanilla CFR, CFR+, and fictitious play \citep{Brown1951}, a classical learning rule in which each player best responds to the opponent's empirical average strategy. Each method runs for 2{,}000 iterations, and exploitability is computed exactly as the sum of both players' best-response values, obtained by enumerating each player's $2^6$ pure strategies.\footnote{The fictitious play variant operates at the information-set level and computes each best response with uniform weights over the opponent's cards rather than reach-weighted posteriors. All three algorithms are deterministic given the game, so exploitability trajectories are exact; wall-clock times are averaged over ten runs.}

\begin{table}[htbp]
\centering
\caption{Exploitability in Kuhn poker at selected iterations, first iteration below $\varepsilon = 0.01$, and wall-clock time (mean $\pm$ standard deviation over ten runs).}
\label{tab:kuhn_poker}
\begin{tabular}{lcccccc}
\toprule
Method & \multicolumn{4}{c}{Exploitability at Iteration} & Iter to $\varepsilon < 0.01$ & Time (s) \\
\cmidrule(lr){2-5}
 & 100 & 500 & 1000 & 2000 & & \\
\midrule
CFR+ & 0.0216 & 0.0098 & 0.0099 & 0.0076 & 340 & 0.87 $\pm$ 0.03 \\
Vanilla CFR & 0.0466 & 0.0244 & 0.0130 & 0.0090 & 930 & 0.86 $\pm$ 0.03 \\
Fictitious Play & 0.3235 & 0.3313 & 0.3323 & 0.3328 & $>$2000 & 0.68 $\pm$ 0.03 \\
\bottomrule
\end{tabular}
\end{table}

\begin{figure}[htbp]
\centering
\includegraphics[width=0.75\textwidth]{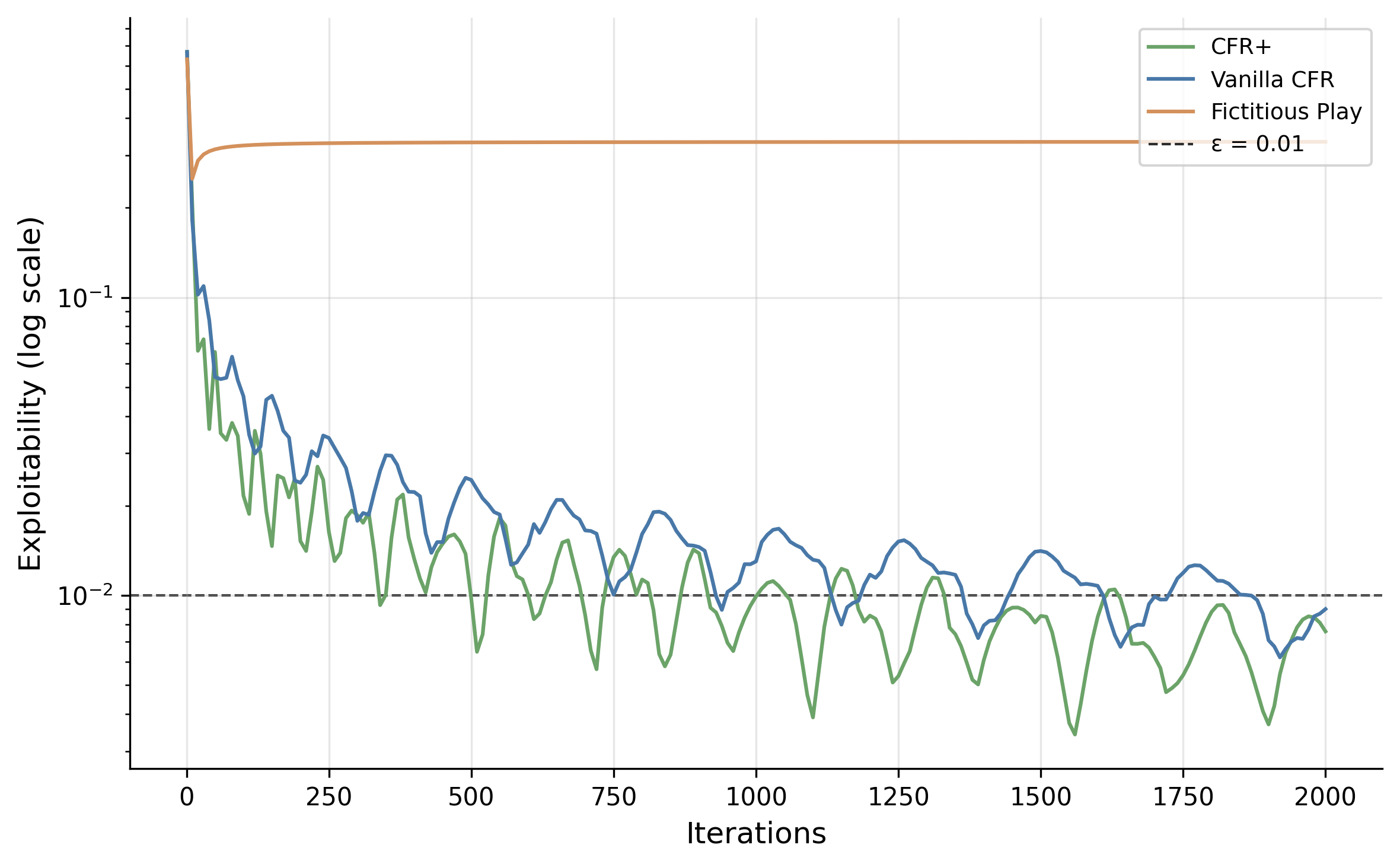}
\caption{Exploitability versus iteration (log scale) for CFR+, vanilla CFR, and fictitious play in Kuhn poker. The dashed reference line marks $\varepsilon = 0.01$.}
\label{fig:kuhn_poker}
\end{figure}

Table~\ref{tab:kuhn_poker} reports that CFR+ reaches exploitability
below $0.01$ within 340 iterations and vanilla CFR within 930. Their
exploitabilities after 2{,}000 iterations are $0.0076$ and $0.0090$.
Both trajectories oscillate as the average strategy moves along the
equilibrium family. Fictitious play plateaus at exploitability $0.33$.
The classical convergence result for fictitious play in two-player
zero-sum games applies to the normal form. The information-set
implementation here weights opponent cards uniformly when computing
best responses and stalls away from equilibrium. This difference links
the result to the counterfactual reach weighting in the CFR update
above. Figure~\ref{fig:kuhn_poker} shows the exploitability paths.

\subsection{Neural Extensions}

Tabular CFR stores regrets at every information set, infeasible when $|\mathcal{I}| > 10^{14}$. Two neural approaches scale to large games.

\subsubsection{Deep CFR}

\citet{Brown2019} approximate cumulative regrets with a neural network $V_\theta: \mathcal{I} \times \mathcal{A} \to \mathbb{R}$ trained on sampled (information set, iteration, regret) tuples:
\begin{equation}
L_V(\theta) = \mathbb{E}_{(I, t', r) \sim \mathcal{M}} \left[ t' \cdot (V_\theta(I, a) - r)^2 \right].
\end{equation}
Weighting by iteration index $t'$ gives more recent regret estimates higher importance.
A separate network $\Pi_\phi$ approximates the average strategy, trained via weighted MSE on strategy samples with the same iteration weighting. The current strategy derives from regret matching, with $\sigma(I, a) \propto [V_\theta(I, a)]^+$.

\subsubsection{Neural Fictitious Self-Play}

NFSP \citep{heinrich2016deep} combines fictitious play (Section~\ref{sec:kuhn_poker}) with deep Q-learning.\footnote{Deep Q-Network (DQN) \citep{mnih2015} approximates the Q-function with a neural network, using experience replay and a target network to stabilize training. The target network $\theta^-$ in the loss function is a delayed copy of the main network, updated periodically.} Each player maintains two networks: a best-response network $Q_\theta$ trained via DQN, and an average strategy network $\bar{\pi}_\phi$ trained via supervised learning on the best-response actions:
\begin{equation}
L_{\text{RL}}(\theta) = \mathbb{E}\left[ \left( r + \gamma \max_{a'} Q_{\theta^-}(I', a') - Q_\theta(I, a) \right)^2 \right], \quad
L_{\text{SL}}(\phi) = \mathbb{E}\left[ -\log \bar{\pi}_\phi(a | I) \right].
\end{equation}
During play, agents follow $\bar{\pi}_\phi$ with probability $1-\eta$ and $Q_\theta$ with probability $\eta$.

\subsubsection{Poker Results}

These methods achieved superhuman performance in poker. Deep CFR attained \emph{exploitability} of 37 mbb/g\footnote{Exploitability measures how far a strategy is from Nash equilibrium, defined as the maximum expected gain an adversary could achieve by best-responding. Zero exploitability means the strategy is unexploitable (Nash). In poker, exploitability is measured in milli-big-blinds per game (mbb/g).} in heads-up flop hold'em \citep{Brown2019}. Libratus defeated top human professionals in heads-up no-limit hold'em using nested subgame solving with CFR \citep{brown2017superhuman}. Pluribus extended to six-player no-limit hold'em \citep{brown2019superhuman}. The game tree of heads-up no-limit hold'em contains $\sim 10^{161}$ states; these results demonstrate that CFR-based methods scale to practically relevant game sizes.

\subsection{The Coase Conjecture in a Durable Goods Monopoly}
\label{subsec:coase}

The durable goods monopoly is a canonical extensive-form bargaining problem with private information: the seller does not know the buyer's valuation.

\citet{coase1972durability} conjectured that a durable goods monopolist\footnote{A durable good provides utility over multiple periods (e.g., a car, appliance, or software license) rather than being consumed immediately. Unlike non-durable goods, durable goods create intertemporal competition, as the seller at time $t$ competes with her own future self at $t+1$.} loses market power when buyers are patient. Unable to commit to future prices, the seller competes with her future self, eroding rents. As the inter-offer interval shrinks ($\delta \to 1$), price collapses to marginal cost and all surplus is extracted by buyers. \citet{stokey1981rational} showed that a monopolist facing rational consumers cannot price discriminate intertemporally; \citet{bulow1982durable} proved that in the no-gap case the monopolist prefers renting to selling. \citet{gul1986foundations} formalized the Coase conjecture for the gap case, where the seller's cost is strictly below all buyer valuations, establishing that the unique stationary Markov-perfect equilibrium price collapses to marginal cost as the period length shrinks. \citet{ausubel1989reputation} and the survey of \citet{ausubel2002bargaining} extend the analysis to the no-gap case and to bargaining with two-sided incomplete information.

The Coase conjecture is asymptotic in $T$ and $\delta$. A two-period game cannot exhibit it; the surplus erosion only operates when the seller can re-optimize many times and buyers can credibly wait. The simulation below uses backward induction on the seller's recursive problem with a continuum of buyers, varying $T$ and $\delta$ on a grid, and compares the Markov-perfect no-commitment value against the commitment benchmark. Section~\ref{subsec:durable_goods_screening} below presents a companion two-period exercise that uses CFR on the extensive-form game tree to recover the screening-versus-pooling equilibrium structure underlying the durable-goods game.

\subsubsection{Model and Dynamic Programming}

A monopolist with marginal cost $c = 0$ faces a continuum of buyers with valuations $v$ drawn from the uniform distribution on $[0, 1]$. The mass of buyers is normalized to one. In each period $t = 1, \ldots, T$, the seller posts a price $p_t$; each remaining buyer of type $v$ accepts iff $v - p_t \ge \delta \, \mathbb{E}_t[v - p_{t+1}^*]$, where $p_{t+1}^*$ is the equilibrium price the seller will post at $t+1$. The seller's per-period revenue is $p_t$ times the mass of buyers who accept.

Define the state at the start of period $t$ as the cutoff $v_t \in [0, 1]$: remaining buyers have valuations uniformly distributed on $[0, v_t]$, with mass $v_t$. The seller's action is the next-period cutoff $w \in [0, v_t]$; buyers with type in $[w, v_t]$ accept at price $p_t$ and exit, while those in $[0, w]$ remain. Marginal-buyer indifference at $w$ pins down the posted price,
\begin{equation}
p_t(w) = (1 - \delta) w + \delta \, p_{t+1}^*(w),
\label{eq:coase_marginal_price}
\end{equation}
where $p_{t+1}^*(w)$ is the seller's equilibrium price at state $w$ in period $t+1$. The seller's Bellman equation is
\begin{equation}
V_t(v) = \max_{w \in [0, v]} \left\{ p_t(w) \, (v - w) + \delta \, V_{t+1}(w) \right\},
\label{eq:coase_bellman}
\end{equation}
with terminal condition $V_T(v) = \max_{w \in [0, v]} w (v - w) = v^2 / 4$ at $w = v/2$.

With uniform $F$ on $[0, v]$, the value, equilibrium price, and equilibrium cutoff functions are homogeneous in $v$: $V_t(v) = \beta_t v^2$, $p_t^*(v) = \mu_t v$, $w_t^*(v) = \lambda_t v$. Substituting into \eqref{eq:coase_marginal_price}--\eqref{eq:coase_bellman} reduces the dynamic program to a scalar recursion in $(\mu_t, \lambda_t, \beta_t)$ with terminal values $\mu_T = \lambda_T = 1/2$ and $\beta_T = 1/4$.\footnote{Define $A_{t+1} = 1 - \delta + \delta \mu_{t+1}$. The first-order condition for the seller's choice of $w$ in \eqref{eq:coase_bellman} gives $\lambda_t = A_{t+1} / (2(A_{t+1} - \delta \beta_{t+1}))$, after which $\mu_t = \lambda_t A_{t+1}$ and $\beta_t = \lambda_t (1 - \lambda_t) A_{t+1} + \delta \beta_{t+1} \lambda_t^2$. The recursion is closed-form per period; the entire backward induction over $T = 200$ periods runs in under a millisecond.}

The commitment benchmark is the optimal pre-committed price path. With forward-looking buyers, uniform $F$, and $c = 0$, the optimal commitment is the static-monopoly price $p^* = 1/2$ in every period. Buyers with $v \ge 1/2$ buy in period 1; nobody else buys (any later discount would induce strategic waiting and lower expected revenue). The commitment value is $V^{\mathrm{com}} = p^* (1 - p^*) = 1/4$ independent of $(T, \delta)$.

\subsubsection{Results}

\begin{figure}[htbp]
\centering
\includegraphics[width=\textwidth]{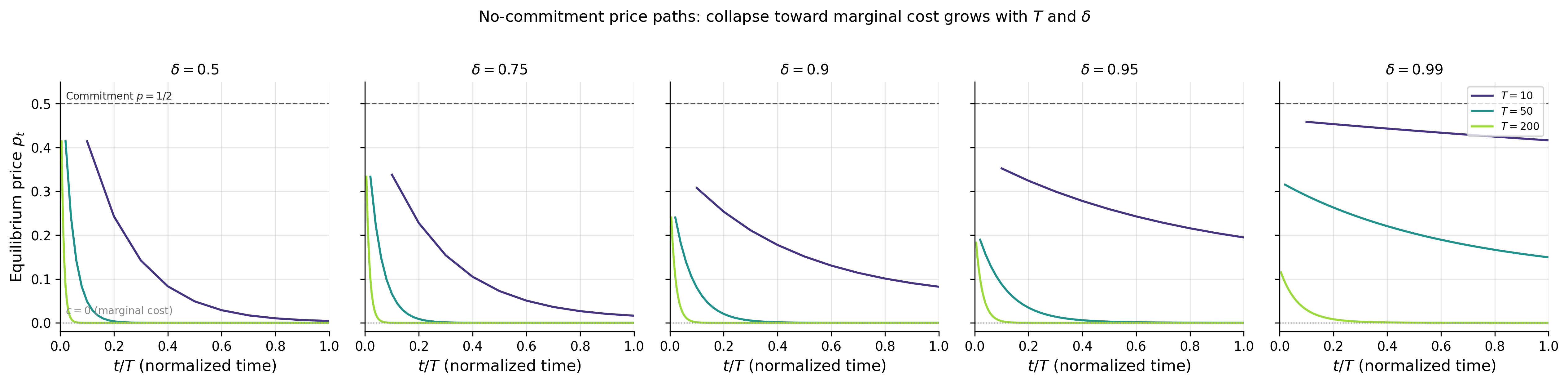}
\caption{No-commitment price paths $p_t$ versus normalized period $t/T$ for representative horizons $T \in \{10, 50, 200\}$ at $\delta \in \{0.5, 0.75, 0.9, 0.95, 0.99\}$. The dashed line marks the commitment price $p = 1/2$. The dotted line marks marginal cost $c = 0$.}
\label{fig:coase_paths}
\end{figure}

\begin{figure}[htbp]
\centering
\includegraphics[width=\textwidth]{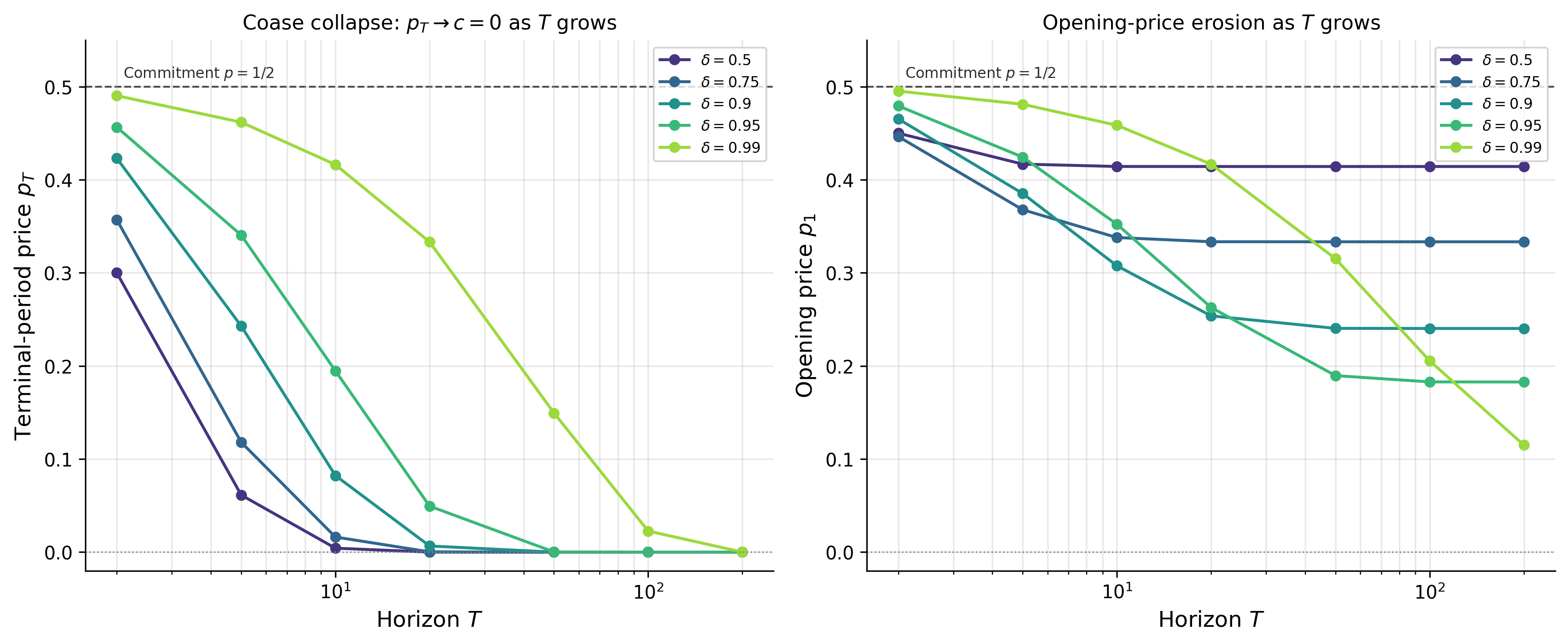}
\caption{Coase collapse. The left panel reports terminal-period price
$p_T$ against $T$ on a log-$T$ axis. The right panel reports opening
price $p_1$ on the same axis. Each curve fixes $\delta$.}
\label{fig:coase_collapse}
\end{figure}

\begin{table}[htbp]
\centering
\caption{Commitment value $V^{\mathrm{com}}$, no-commitment Markov-perfect value $V^{\mathrm{nc}}$, ratio $V^{\mathrm{nc}}/V^{\mathrm{com}}$, opening price $p_1$, and terminal price $p_T$ across the $(T, \delta)$ grid. Uniform valuations on $[0, 1]$, $c = 0$.}
\label{tab:coase_dp}
\begin{tabular}{ccccccc}
\toprule
$T$ & $\delta$ & $V^{\mathrm{com}}$ & $V^{\mathrm{nc}}$ & $V^{\mathrm{nc}}/V^{\mathrm{com}}$ & $p_1$ & $p_T$ \\
\midrule
2 & 0.50 & 0.2500 & 0.2250 & 0.900 & 0.4500 & 0.3000 \\
2 & 0.75 & 0.2500 & 0.2232 & 0.893 & 0.4464 & 0.3571 \\
2 & 0.90 & 0.2500 & 0.2327 & 0.931 & 0.4654 & 0.4231 \\
2 & 0.95 & 0.2500 & 0.2397 & 0.959 & 0.4793 & 0.4565 \\
2 & 0.99 & 0.2500 & 0.2476 & 0.990 & 0.4952 & 0.4903 \\
\midrule
5 & 0.50 & 0.2500 & 0.2084 & 0.834 & 0.4168 & 0.0613 \\
5 & 0.75 & 0.2500 & 0.1839 & 0.736 & 0.3678 & 0.1180 \\
5 & 0.90 & 0.2500 & 0.1926 & 0.770 & 0.3852 & 0.2428 \\
5 & 0.95 & 0.2500 & 0.2121 & 0.849 & 0.4243 & 0.3405 \\
5 & 0.99 & 0.2500 & 0.2405 & 0.962 & 0.4811 & 0.4618 \\
\midrule
10 & 0.50 & 0.2500 & 0.2071 & 0.828 & 0.4142 & 0.0042 \\
10 & 0.75 & 0.2500 & 0.1690 & 0.676 & 0.3379 & 0.0162 \\
10 & 0.90 & 0.2500 & 0.1538 & 0.615 & 0.3076 & 0.0823 \\
10 & 0.95 & 0.2500 & 0.1761 & 0.705 & 0.3523 & 0.1948 \\
10 & 0.99 & 0.2500 & 0.2293 & 0.917 & 0.4585 & 0.4163 \\
\midrule
20 & 0.50 & 0.2500 & 0.2071 & 0.828 & 0.4142 & 0.0000 \\
20 & 0.75 & 0.2500 & 0.1667 & 0.667 & 0.3334 & 0.0003 \\
20 & 0.90 & 0.2500 & 0.1269 & 0.508 & 0.2538 & 0.0066 \\
20 & 0.95 & 0.2500 & 0.1314 & 0.526 & 0.2628 & 0.0494 \\
20 & 0.99 & 0.2500 & 0.2084 & 0.834 & 0.4168 & 0.3333 \\
\midrule
50 & 0.50 & 0.2500 & 0.2071 & 0.828 & 0.4142 & 0.0000 \\
50 & 0.75 & 0.2500 & 0.1667 & 0.667 & 0.3333 & 0.0000 \\
50 & 0.90 & 0.2500 & 0.1202 & 0.481 & 0.2404 & 0.0000 \\
50 & 0.95 & 0.2500 & 0.0948 & 0.379 & 0.1896 & 0.0002 \\
50 & 0.99 & 0.2500 & 0.1576 & 0.630 & 0.3152 & 0.1494 \\
\midrule
100 & 0.50 & 0.2500 & 0.2071 & 0.828 & 0.4142 & 0.0000 \\
100 & 0.75 & 0.2500 & 0.1667 & 0.667 & 0.3333 & 0.0000 \\
100 & 0.90 & 0.2500 & 0.1201 & 0.481 & 0.2403 & 0.0000 \\
100 & 0.95 & 0.2500 & 0.0914 & 0.366 & 0.1828 & 0.0000 \\
100 & 0.99 & 0.2500 & 0.1028 & 0.411 & 0.2056 & 0.0227 \\
\midrule
200 & 0.50 & 0.2500 & 0.2071 & 0.828 & 0.4142 & 0.0000 \\
200 & 0.75 & 0.2500 & 0.1667 & 0.667 & 0.3333 & 0.0000 \\
200 & 0.90 & 0.2500 & 0.1201 & 0.481 & 0.2403 & 0.0000 \\
200 & 0.95 & 0.2500 & 0.0914 & 0.365 & 0.1827 & 0.0000 \\
200 & 0.99 & 0.2500 & 0.0576 & 0.230 & 0.1151 & 0.0000 \\
\bottomrule
\end{tabular}
\end{table}

Table~\ref{tab:coase_dp} reports the sweep. At $T = 200$, $\delta = 0.99$ the no-commitment monopolist captures $V^{\mathrm{nc}} = 0.058$ against the commitment benchmark $V^{\mathrm{com}} = 0.25$, a ratio of $0.23$; the terminal price $p_T$ rounds to zero to four decimals, and the opening price $p_1 = 0.115$ is well below the static monopoly price $1/2$. Holding $\delta$ at $0.95$ and varying $T$, the value ratio drops monotonically from $0.96$ at $T = 2$ to $0.37$ at $T = 200$, and $p_T$ from $0.46$ to numerical zero. Holding $T$ at $200$ and varying $\delta$, the ratio drops monotonically from $0.83$ at $\delta = 0.5$ to $0.23$ at $\delta = 0.99$. As a cross-check, the finite-horizon values at $T = 200$ for $\delta \le 0.95$ match the analytical stationary Markov-perfect equilibrium values $\beta^*$ and $\mu^*$ to five decimals.\footnote{The stationary fixed point of the recursion in the footnote above satisfies $\mu = \lambda (1 - \delta) / (1 - \delta \lambda)$, $\beta = \lambda(1-\lambda) A / (1 - \delta \lambda^2)$, and $\lambda = A / (2 (A - \delta \beta))$. At $\delta = 0.9$ this gives $\lambda \approx 0.760$, $\mu \approx 0.240$, $\beta \approx 0.120$, matching the table's $T = 200$ entries. At $\delta = 0.99$ the stationary $\mu \approx 0.091$, $\beta \approx 0.045$, so $T = 200$ is still slightly above the asymptotic limit; the residual gap closes as $T$ grows further.}

Figure~\ref{fig:coase_paths} traces the equilibrium price path, where long horizons and patient buyers force the seller to start lower and end near marginal cost. Figure~\ref{fig:coase_collapse} summarizes the collapse rate, where $p_T \to 0$ at all $\delta < 1$ once $T$ is large enough, and $p_1$ falls toward the stationary level $\mu^*(\delta)$ that itself tends to zero as $\delta \to 1$.

The numerical sweep illustrates the finite-horizon mechanism behind the
Coase conjecture. Without commitment power, the durable-goods
monopolist captures only a fraction of the commitment surplus, and that
fraction shrinks toward zero as $T$ and $\delta$ jointly grow.
Section~\ref{subsec:durable_goods_screening} restricts the same model
to two periods and a two-point type distribution, where it reduces to
a screening-versus-pooling problem.

\subsection{Screening versus Pooling in the Durable Goods Monopoly}
\label{subsec:durable_goods_screening}

The two-period analogue of the model in Section~\ref{subsec:coase} reduces to an extensive-form game with a discrete buyer type and a discrete seller action set, small enough to admit equilibrium computation via counterfactual regret minimization (CFR). Replacing the continuum of valuations with a two-point distribution $v \in \{v_L, v_H\}$ and restricting the seller to two prices $\{v_L, P^*(\delta)\}$ recovers the screening-versus-pooling threshold $\pi^* = 1/2$ in closed form; CFR finds this threshold without being told the analytical solution. The horizon is fixed at $T = 2$ throughout this subsection; the asymptotic price-collapse mechanism documented in Section~\ref{subsec:coase} does not operate here.

\subsubsection{Model}

A seller with zero cost faces a buyer with private valuation $v \in \{v_L, v_H\}$, where $\Pr(v = v_H) = \pi$. In each period $t = 1, 2, \ldots$, the seller posts price $p_t$; the buyer accepts or rejects. Upon acceptance at $t$, the seller receives $\delta^{t-1} p_t$ and the buyer receives $\delta^{t-1}(v - p_t)$, where $\delta \in (0,1)$ is the common discount factor. The game ends upon acceptance or after $T$ periods. Parameters: $v_L = 100$, $v_H = 200$, $T = 2$ periods.

\subsubsection{Equilibrium Analysis}

The screening price $P^*(\delta)$ makes the high-type buyer indifferent between accepting now and waiting.
\begin{equation}
v_H - P^* = \delta(v_H - v_L) \implies P^*(\delta) = v_H - \delta(v_H - v_L).
\end{equation}
The seller's optimal strategy depends on $\pi$. Let $\Pi_{\text{screen}} = \pi P^* + (1-\pi)\delta v_L$ and $\Pi_{\text{pool}} = v_L$. The seller screens if $\Pi_{\text{screen}} > \Pi_{\text{pool}}$, yielding threshold
\begin{equation}
\pi^* = \frac{v_L(1-\delta)}{P^* - \delta v_L} = \frac{1}{2} \quad \text{when } v_H = 2v_L.
\end{equation}
For $\pi < \pi^*$, the seller pools (offers $v_L$ immediately). For $\pi > \pi^*$, the seller screens (offers $P^*$, then $v_L$ if rejected).\footnote{In a screening equilibrium, the seller uses price to separate buyer types: high-value buyers accept immediately at a high price, while low-value buyers reject and receive a lower offer. In a pooling equilibrium, the seller offers a single price that all types accept. The gap case ($0 = c < v_L$) guarantees trade in equilibrium.} As $\delta \to 1$, the screening price $P^*(\delta) \to v_L$ and the seller cannot extract a screening premium from high types, but in a two-period game with action set $\{v_L, P^*(\delta)\}$ the asymptotic price-collapse mechanism of the Coase conjecture does not operate.

\subsubsection{Computational Results}

The simulation represents the bargaining problem as an extensive-form
game and applies CFR.\footnote{The game tree has two information sets
for the seller, indexed by rejection history, and two for the buyer,
indexed by private type. CFR runs for 5{,}000 iterations per parameter
configuration. Vanilla CFR is deterministic given zero initial
regrets, so perturbing the initial regret table with Gaussian noise of
standard deviation $0.05$ supplies seed-level variability. Ten seeds
are run per $(\pi, \delta)$ cell. The reported P(Screen) and NashConv
are means across seeds. The SE columns report the standard error of the
mean. NashConv is the sum of both players' exploitabilities on the
utility scale where the maximum single-player payoff is $v_H = 200$.}

\begin{table}[htbp]
\centering
\caption{CFR strategy versus the analytical equilibrium in the
$\pi$-sweep at $\delta = 0.5$, with $n = 10$ seeds per cell.}
\label{tab:coase}
\begin{tabular}{lcccccccc}
\toprule
$\pi$ & $P^*$ & P(Screen) & SE & Theory & $|\Delta|$ & NashConv & SE & Eq.\ Type \\
\midrule
0.10 & 150 & 0.000 & 0.000 & 0.0 & 0.000 & 4.04 & 0.59 & Pooling \\
0.15 & 150 & 0.000 & 0.000 & 0.0 & 0.000 & 6.04 & 0.88 & Pooling \\
0.20 & 150 & 0.000 & 0.000 & 0.0 & 0.000 & 8.04 & 1.17 & Pooling \\
0.25 & 150 & 0.000 & 0.000 & 0.0 & 0.000 & 10.05 & 1.46 & Pooling \\
0.30 & 150 & 0.001 & 0.000 & 0.0 & 0.001 & 12.05 & 1.75 & Pooling \\
0.35 & 150 & 0.001 & 0.000 & 0.0 & 0.001 & 14.06 & 2.04 & Pooling \\
0.40 & 150 & 0.001 & 0.001 & 0.0 & 0.001 & 16.58 & 2.37 & Pooling \\
0.45 & 150 & 0.202 & 0.133 & 0.0 & 0.202 & 23.36 & 4.80 & Pooling \\
0.50 & 150 & 0.351 & 0.150 & 0.5 & 0.149 & 23.85 & 4.32 & Indifferent \\
0.55 & 150 & 0.401 & 0.163 & 1.0 & 0.599 & 23.59 & 3.97 & Screening \\
0.60 & 150 & 0.598 & 0.144 & 1.0 & 0.402 & 23.20 & 3.78 & Screening \\
0.65 & 150 & 0.801 & 0.132 & 1.0 & 0.199 & 22.41 & 3.11 & Screening \\
0.70 & 150 & 0.900 & 0.100 & 1.0 & 0.100 & 19.41 & 2.75 & Screening \\
0.75 & 150 & 0.900 & 0.100 & 1.0 & 0.100 & 16.16 & 2.29 & Screening \\
0.80 & 150 & 1.000 & 0.000 & 1.0 & 0.000 & 14.92 & 1.27 & Screening \\
0.85 & 150 & 1.000 & 0.000 & 1.0 & 0.000 & 11.17 & 0.96 & Screening \\
0.90 & 150 & 1.000 & 0.000 & 1.0 & 0.000 & 7.42 & 0.65 & Screening \\
\bottomrule
\end{tabular}
\begin{minipage}{0.9\textwidth}
\footnotesize
Notes. P(Screen) is the probability the seller offers $P^* = 150$ in period 1, averaged across seeds. SE columns report the across-seed standard error. Theory gives the analytical pure-strategy prediction (0 for pooling, 1 for screening, 0.5 at the indifference point $\pi^* = 1/2$). $|\Delta| = |\text{P(Screen)} - \text{Theory}|$. NashConv is the sum of both players' exploitabilities at iteration 5{,}000, on the same utility scale as the payoffs.
\end{minipage}
\end{table}

Table~\ref{tab:coase} reports results from varying $\pi$ at fixed
$\delta = 0.5$. Away from the indifference threshold
($\pi \le 0.40$ and $\pi \ge 0.80$), CFR's average strategy matches the
analytical step function. The error satisfies $|\Delta| \le 0.001$ in
the pooling region and equals $0.000$ in the deep screening region.
Across $\pi \in [0.45, 0.65]$, the seed-averaged screening
probabilities lie between the two pure predictions, with across-seed
standard errors of $0.13$--$0.16$. Only $\pi = 0.50$ is the analytical
indifference point. CFR gives $\text{P(Screen)} \approx 0.60$ at
$\pi = 0.60$ and reaches $0.90$ only at $\pi = 0.70$.
Residual NashConv after 5{,}000 iterations ranges from $4$ to $24$
across $\pi$, equivalent to up to $12\%$ of the maximum single-player
payoff. CFR is therefore not at $\varepsilon$-Nash for small
$\varepsilon$ in these cells.

A $\delta$-sweep at $\pi = 0.7$ reproduces the screening price formula
$P^*(\delta) = 200 - 100\delta$ with zero error across
$\delta \in [0.1, 0.9]$. CFR's average P(Screen) drops from $1.0$ at
$\delta = 0.1$ to $0.50 \pm 0.17$ at $\delta \ge 0.75$. The analytical
column predicts pure screening (P(Screen) $= 1$) throughout this range
because $\pi = 0.7 > \pi^* = 1/2$, so the decline is not a Coase regime
switch. As $\delta$ grows, the screening price $P^*(\delta)$ shrinks
toward $v_L$ and the profit gap between screening and pooling narrows.
The seed-perturbed estimates become more variable as that gap
narrows.\footnote{Four stress tests give the following results.
(1) Awkward primes $(v_L, v_H) = (37, 83)$ yield $P^* = 55$, while
theory gives 55.4. (2) The information-leak test uses a single seller
information set at the root. (3) Removing 150 from the grid recovers
145 with action probability $0.994$. (4) The three-period game gives
$P_1 = 120$ against the theoretical value 136, missing the test
tolerance by one unit. Three of the four stress tests pass.}

\subsection{Discussion}

Stochastic-game Q-learning and CFR target complementary game classes,
simultaneous-move games with observable payoffs and extensive-form
games with private information, respectively. The duopoly simulations
produce actions near the symmetric pure Nash benchmarks. CFR and CFR+
reach low exploitability in Kuhn poker, while the two-period screening
exercise retains material NashConv in several parameter cells.

\section{Bandits and Dynamic Pricing}
\label{section:bandits}
Dynamic pricing is the canonical in-field reinforcement-learning problem for economics. Unlike simulator-based training, the agent learns from real customers, so \textit{exploration} is not an abstract sampling cost: a bad price loses revenue today and may affect the customer relationship. A seller faces $T$ customers in sequence, sets one price for each customer, and observes only whether the customer bought.\footnote{\citet{Rothschild1974} posed pricing under demand uncertainty as a two-armed bandit problem. His insight was that a \textit{myopic} seller can get stuck at a suboptimal price forever, because exploiting the currently best-looking price generates no information about alternatives.} The seller does not observe the customer's willingness to pay.\footnote{\citet{denBoer2015} surveys dynamic pricing with demand uncertainty across operations research, marketing, computer science and econometrics, recovering much of the older literature on the problem. This chapter is narrower, asking which economic structures sharpen the regret rate.}

\textit{Regret}, the total revenue gap between the seller's policy and the best benchmark price or pricing rule, is the central measure. If the benchmark earns $r^*$ per customer, a policy with regret $R(T)$ earns $T r^* - R(T)$ in total. The central question is what economic structure turns pricing from a slow arm-learning problem into a fast demand-learning problem. With no useful structure, regret is polynomial in $T$. With enough structure, such as well-separated parametric demand, sparse contextual demand, revealed-preference bounds, or smooth monotone demand curves, learning can be much faster. But the assumptions matter, since an unknown noise distribution, strategic buyers, or fairness constraints can restore high regret even when the demand model looks rich.

The Engine Replacement MDP does not appear because a bandit has no action-dependent state transition. Removing that transition is the defining reduction from an MDP to the model studied here.

\subsection{Foundations}
\label{sec:dp_foundations}

\subsubsection{No Structure on Demand}
\label{sec:kleinberg}

\citet{Kleinberg2003} study the benchmark case where the seller knows almost nothing. Customers arrive with valuations drawn i.i.d.\ from an unknown distribution on $[0,1]$, and the seller posts a price from a continuous set. The demand curve $D(p) = \Pr(v \geq p)$ is unknown; the only feedback is whether each customer bought. The seller's goal is to minimize regret against the single price $p^*$ that maximizes $p \cdot D(p)$. Kleinberg and Leighton prove that the minimax regret is $\Theta(\sqrt{T})$.\footnote{The upper bound, $O(\sqrt{T \log T})$, discretizes $[0,1]$ into $K = \lceil (T/\log T)^{1/4} \rceil$ prices and runs the UCB1 algorithm of \citet{auer2002}. The lower bound, $\Omega(\sqrt{T})$, constructs a family of demand curves parameterized by the location of the optimal price $p^* \in [0.3, 0.4]$. The key tension: posting prices far from $p^*$ is informative about demand but costly in revenue; posting prices near $p^*$ is cheap but uninformative. Resolving this tension costs at least $\Omega(\sqrt{T})$ in cumulative revenue. UCB1 \citep{auer2002} selects the price maximizing $\hat{\mu}_{p_k}(t) + \sqrt{2\ln t / N_{p_k}(t)}$, where $\hat{\mu}_{p_k}(t)$ is the empirical mean profit and $N_{p_k}(t)$ the number of trials; the second term is an exploration bonus that shrinks as a price is tried more, implementing the principle of optimism in the face of uncertainty.}

The rate is a useful baseline. After 10,000 customers, $\sqrt{T}$ regret is roughly 100 customers' worth of revenue lost to learning. No algorithm can improve on this worst-case rate without adding structure to $D$. For adversarial valuations, where a worst-case opponent rather than a fixed distribution chooses customer values, the minimax regret rises to $\Theta(T^{2/3})$, achieved by the Exp3 algorithm\footnote{\emph{Exp3} \citep{auer2002nonstochastic} maintains a weight $w_k(t)$ for each price, selecting $p_k$ with probability proportional to $w_k(t)$ and updating the chosen price's weight by $\exp(\eta \hat{r}_{k,t})$ where $\hat{r}_{k,t}$ is the revenue importance-weighted by the selection probability; because no model of demand is assumed, the guarantee holds against an adversary who chooses valuations after observing the algorithm.} on a discretized price grid. I focus on the stochastic setting throughout this chapter.

\subsubsection{Parametric Demand}
\label{sec:broder}

\citet{Broder2012} show that parametric demand is not automatically enough. Let $d(p; z) = \Pr(V \geq p)$, where $z \in \mathbb{R}^n$ is an unknown parameter governing the demand curve. The seller observes binary purchase decisions and updates a maximum likelihood estimate of $z$. Under standard regularity conditions (bounded demand, unique optimal price, smooth revenue function), the minimax regret is again $\Theta(\sqrt{T})$.\footnote{The lower bound (Theorem 3.1 of \citet{Broder2012}) constructs a linear demand family where all demand curves pass through the same point at the optimal price $p^*(z_0)$. Observing purchases at this price provides no information about $z$. An MLE-Cycle policy that interleaves dedicated exploration rounds with greedy pricing achieves the matching upper bound $O(\sqrt{T})$ (Theorem 3.6).} What matters is not whether the model has a finite-dimensional parameter, but whether profitable prices are also informative about that parameter.

The picture changes under a ``well-separated'' condition requiring that every price in the feasible set is informative about the demand parameter. Formally, the Fisher information $I(p, z)$ is bounded below by $c_f > 0$ for all prices $p$ and parameters $z$.\footnote{This means no two demand curves $d(p; z)$ and $d(p; z')$ cross on the pricing interval, so every purchase observation distinguishes the two hypotheses. In the lower-bound family of Theorem 3.1, the curves all cross at $p = 1$, which is why the $\sqrt{T}$ floor emerges.} Under well-separation, a pure greedy policy (MLE-Greedy, pricing at $p^*(\hat{z}_t)$ each period) achieves $O(\log T)$ regret (Theorem 4.8), because every price is informative and dedicated exploration rounds become unnecessary.

\begin{figure}[t]
\centering
\includegraphics[width=0.7\textwidth]{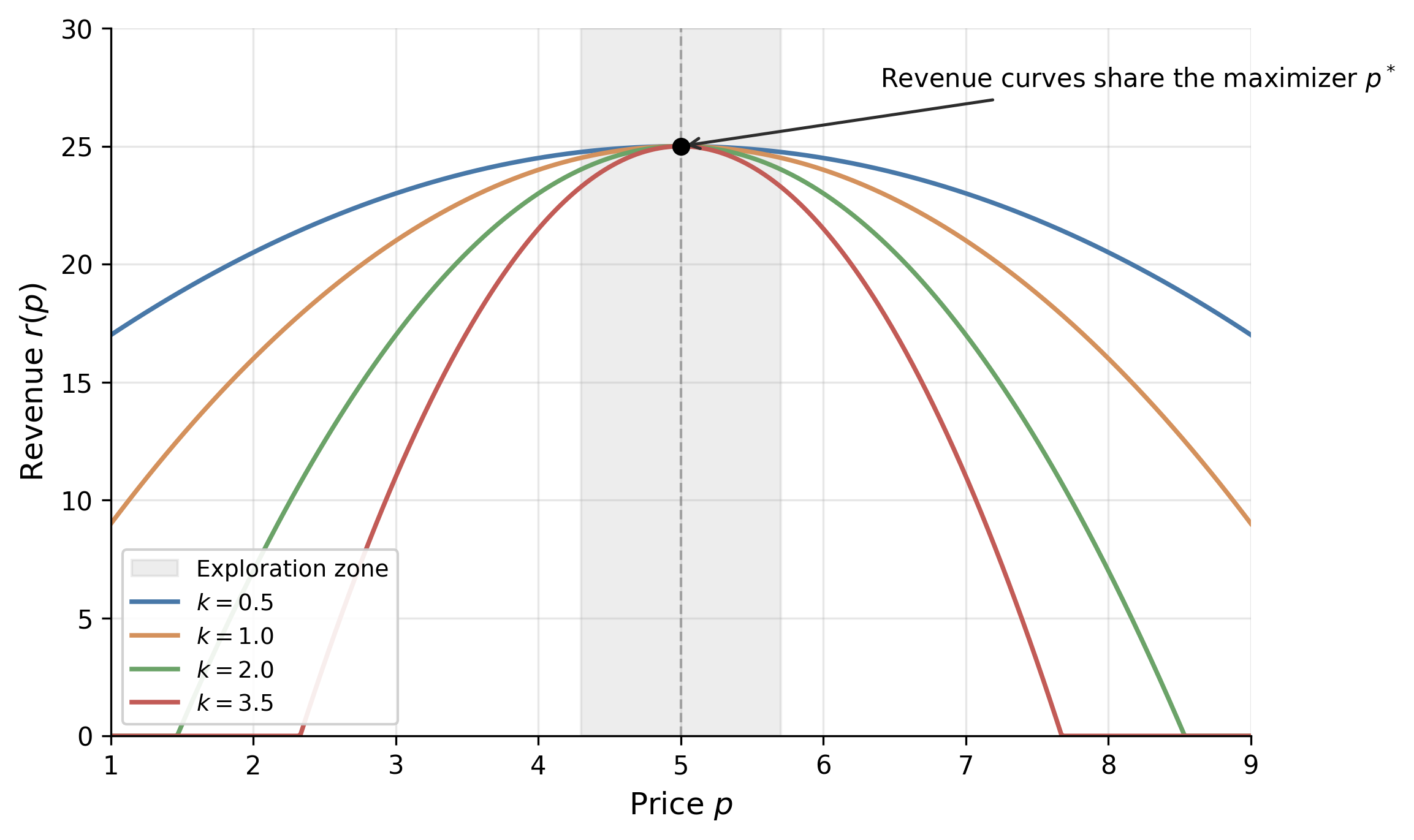}
\caption{Revenue curves $r(p) = r^* - k(p - p^*)^2$ for four demand curvatures $k \in \{0.5, 1.0, 2.0, 3.5\}$. All revenue curves share the maximizer $p^* = 5$; the underlying demand curves (not shown here, see \citet{Broder2012} Theorem 3.1) are linearly distinct but cross at $p^*$. Within the illustrative exploration window $|p - p^*| \leq 0.7$ (shaded; half-width is pedagogical, not derived), revenue separation across the four curves is at most $1.72$ units, about 7\% of $r^*$, which is small relative to typical purchase-noise variance at this scale.}
\label{fig:uninformative_price}
\end{figure}

\subsubsection{High-Dimensional Features with Sparsity}
\label{sec:javanmard}

\citet{Javanmard2019} extend the setting to products described by $d$-dimensional feature vectors $x_t$. The market value is $v_t = x_t^\top \theta_0 + z_t$, where $\theta_0 \in \mathbb{R}^d$ is unknown and $z_t$ is i.i.d.\ noise from a known log-concave distribution $F$.\footnote{Log-concavity of $F$ and $1 - F$ is satisfied by the normal, logistic, uniform, Laplace, and exponential distributions. It ensures that expected revenue $p \cdot [1 - F(p - x_t^\top \theta_0)]$ is strictly quasi-concave in $p$, giving a unique optimal price.} Only $s_0$ of the $d$ coordinates of $\theta_0$ are nonzero, but the seller does not know which ones.

Their RMLP algorithm (Regularized Maximum Likelihood Pricing) operates in episodes whose lengths double (1, 2, 4, 8, \ldots). At each episode boundary, the seller fits a LASSO-penalized maximum likelihood estimate of $\theta_0$ using data from the previous episode, then prices greedily throughout the current episode. The regret is $O(s_0 \log d \cdot \log T)$ (Theorem 4.1), with a matching lower bound of $\Omega(s_0 (\log d + \log T))$ (Theorem 5.1). The reason this beats $\sqrt{T}$ connects back to the Broder lower bound. In the parametric model of Section~\ref{sec:broder}, all demand curves can cross at the optimal price, making that price uninformative. Here, customer features vary across periods, so the aggregate demand function at any fixed price changes in proportion to the estimation error in $\theta_0$. Every price is informative, and dedicated exploration rounds become unnecessary.\footnote{If some feature directions are rarely observed, the seller cannot learn all coordinates of $\theta_0$ quickly, and regret degrades to $O(\sqrt{\log(d) \cdot T})$ (Theorem 4.2). If the noise distribution belongs to a known parametric family but its scale parameter is unknown, regret reverts to $\Omega(\sqrt{T})$ (Theorem 7.1), foreshadowing the result of \citet{Xu2021} discussed in Section~\ref{sec:xu}.} Even with hundreds of features, if only a handful matter, learning is fast.

\subsection{Revealed Preference and Partial Identification}
\label{sec:misra}

\subsubsection{WARP-Based Elimination}

\citet{Misra2019} bring economic theory into the bandit pricing framework. Their model has $K$ discrete prices, $S$ consumer segments (segment membership observed, but valuations unknown), and within-segment heterogeneity $\delta$. A consumer in segment $s$ has valuation $v_i = v_s + n_i$, where $v_s$ is the segment midpoint and $n_i \in [-\delta, \delta]$ is idiosyncratic noise. The key structural assumption is WARP (Weak Axiom of Revealed Preference): if a consumer buys at price $p$, she would buy at any lower price $p' < p$.

WARP turns a binary purchase into more information than a single arm reward. If a customer in segment $s$ buys at \$80, then prices below \$80 are also feasible for that customer; if she refuses \$120, prices above \$120 are also ruled out. Aggregating these inequalities gives partial identification of each segment's valuation. Suppose the seller has offered several prices to segment $s$. Define $p_s^{\min} = \max\{p_k : \text{all observed segment-$s$ customers bought at } p_k\}$ and $p_s^{\max} = \min\{p_k : \text{no observed segment-$s$ customer bought at } p_k\}$. Then the segment midpoint lies in $[p_s^{\min}, p_s^{\max}]$, and within-segment heterogeneity satisfies $\hat{\delta}_s \leq (p_s^{\max} - p_s^{\min})/2$. These bounds propagate to aggregate demand: for each price $p_k$, the seller constructs upper and lower bounds on total profit $\pi(p_k) = p_k \cdot D(p_k)$. When the profit upper bound at some price falls below the best profit lower bound across all prices, that price is dominated and permanently eliminated from consideration.

The UCB-PI algorithm combines dominance elimination with a price-scaled exploration bonus:
\begin{equation}
\label{eq:ucbpi}
I_{p_k}(t) = \hat{\mu}_{p_k}(t) + p_k \sqrt{\frac{2 \ln t}{N_{p_k}(t)}}
\end{equation}
if $p_k$ is not dominated, and $I_{p_k}(t) = 0$ otherwise, where $\hat{\mu}_{p_k}(t)$ is the average profit observed at price $p_k$ and $N_{p_k}(t)$ is the number of trials.\footnote{Standard UCB1 uses an exploration bonus of $\sqrt{2\ln t / N_{p_k}(t)}$, which assumes rewards in $[0,1]$. Since profit at price $p_k$ is bounded by $p_k$, scaling the bonus by $p_k$ tightens exploration for cheap prices that cannot contribute much profit regardless.} The improvement over standard UCB1 comes from two design choices. First, dominance elimination reduces the effective number of arms the algorithm must explore. Second, the $p_k$ scaling focuses exploration on prices where uncertainty actually matters for profit. These gains come from pruning dominated arms and calibrating arm-level exploration; they do not fully exploit smoothness or monotone dependence across nearby prices. Together, within the partial-identification model, these yield $O(\log T)$ regret.
\begin{equation}
\mathbb{E}[R_T(\text{UCB-PI})] \leq \sum_{k \neq k^*} \frac{8 p_k \log T}{\Delta_k} + \left(1 + \frac{\pi^2}{3}\right) \sum_{k=1}^K \Delta_k
\end{equation}
where $\Delta_k = \mu_{k^*} - \mu_k$ is the gap between arm $k$ and the optimal arm.\footnote{The first sum is the leading term, scaling as $O(\log T)$. The second sum is a constant that does not grow with $T$. Replacing $p_k$ with 1 recovers the standard UCB1 bound, which is looser.}

\citet{Misra2019} calibrate the model to an in-field deployment at ZipRecruiter, a B2B online recruiting platform. With 7,870 customers per month, 1,000 segments, and 10 price points from \$19 to \$399, UCB-PI achieves 98\% of oracle profit and produces 43\% higher profits during the first month of testing compared to a learn-then-earn alternative. The algorithm has both higher mean profit and lower variance, reflecting the value of eliminating dominated prices early.\footnote{For multi-product settings, \citet{Mueller2019} impose low-rank structure on the price-sensitivity matrix, achieving regret $O(T^{3/4}\sqrt{d})$ that scales with the latent demand dimension $d$ rather than the number of products. \citet{Badanidiyuru2013} extend the bandit framework to handle inventory constraints (``bandits with knapsacks''), relevant when the seller faces limited stock alongside the pricing decision.}

\subsubsection{Demand-Curve Learning}

\citet{Weaver2025} clarify what UCB-PI still fails to exploit. After dominated prices have been removed, the remaining prices are still learned mostly as separate arms. But pricing arms are not independent pulls; they lie on one demand curve. \citet{Misra2019} use revealed-preference bounds to eliminate dominated prices; \citet{Weaver2025} use smoothness and monotonicity to create informational externalities across prices.\footnote{For example, if demand at \$40 is estimated around 60\%, then a curve-learning bandit treats demand at \$41 as nearby rather than unknown from scratch; a monotone variant also rules out demand at \$41 being systematically higher than demand at \$40. Price scaling matters because the same five-percentage-point demand uncertainty creates \$0.50 of profit uncertainty at a \$10 price but \$5.00 at a \$100 price, so whether the optimal price is low, middle, or high on the grid changes the exploration problem.}

GP-UCB and GP-TS place a nonparametric prior on demand $D(p)$ rather than on each price's reward separately. An observation at one price therefore updates beliefs about nearby prices. The monotone variants add the economic restriction that demand weakly decreases with price by modeling both $D(p)$ and its derivative $D'(p)$ and constraining the derivative to be nonpositive. This is a stronger form of economic structure than arm elimination. It says not only which prices can be discarded but how evidence at one price should move beliefs about the rest of the price grid. That is why the incremental gains from partial identification can be small once a curve-learning bandit is already transferring information efficiently across prices.

\subsection{The Value of Knowing the Noise Distribution}
\label{sec:xu}

\citet{Xu2021} ask how much it helps to know the shape of demand uncertainty. In their model, a feature vector $x_t \in \mathbb{R}^d$ describes each sales session, the customer's valuation is $w_t = x_t^\top \theta^* + N_t$ where $\theta^*$ is unknown and $N_t$ is zero-mean i.i.d.\ noise with CDF $F$, and the seller observes only whether the customer bought at the posted price. The question is whether the seller knows $F$. The error distribution determines how purchase probabilities translate into valuations.\footnote{The regret benchmark here differs from Section~\ref{sec:kleinberg}. \citet{Kleinberg2003} and \citet{Broder2012} measure regret against the best fixed price; \citet{Xu2021} and \citet{Liu2024strategic} measure regret against the clairvoyant contextual policy that sets the optimal price $p_t^*$ for each customer's features $x_t$. The contextual benchmark is harder.}

If $F$ is known (the seller knows that demand shocks are, say, normally distributed with known variance), the EMLP algorithm (Epoch-based Maximum Likelihood Pricing) achieves regret $O(d \log T)$ (Theorem 3).\footnote{EMLP runs in doubling epochs of length $\tau_k = 2^{k-1}$. At each epoch boundary, the seller fits a maximum likelihood estimate $\hat{\theta}_k$ using data from the previous epoch, then prices greedily at $p_t = J(x_t^\top \hat{\theta}_k)$ throughout the epoch, where $J(u) = \arg\max_v \, v[1 - F(v - u)]$ is the revenue-maximizing price function. The key technical insight is that the negative log-likelihood is strongly convex (Lemma 7 of \citet{Xu2021}), so MLE concentrates at rate $O(d / \tau_k)$. Since regret is quadratic in the parameter estimation error (Lemma 5) and there are $O(\log T)$ epochs, the total regret is $O(d \log T)$.} After 10,000 customers with $d = 5$ features, the revenue loss is roughly 50 customers' worth, an improvement over the $\sqrt{T} \approx 100$ customers' worth that Kleinberg's lower bound imposes without structural knowledge.

If $F$ is unknown (even if only the variance of a Gaussian is unknown, with everything else known), the regret is at least $\Omega(\sqrt{T})$ (Theorem 12).\footnote{The lower bound constructs two noise variances $\sigma_1 = 1$ and $\sigma_2 = 1 - T^{-1/4}$. Any algorithm that performs well under both must spend $\Omega(\sqrt{T})$ revenue distinguishing the two cases. This extends the ``uninformative price'' phenomenon of \citet{Broder2012}: when the seller does not know $F$, there exist prices at which observed purchase behavior is nearly identical under different demand parameters.} The seller is back to the Kleinberg baseline: additional features alone do not buy logarithmic regret.

For fixed $d$, the gap between $O(d \log T)$ and $\Omega(\sqrt{T})$ grows like $\sqrt{T}/\log T$. This is not a constant-factor tuning issue; it is a change in the learning rate. When a modeler specifies a logit or probit demand model, the assumed noise distribution can buy logarithmic regret. Semiparametric approaches that leave the error distribution unspecified pay a concrete cost, reverting from logarithmic to polynomial regret.\footnote{\citet{Tullii2024} establish the tightest known bound under minimal distributional assumptions: if the noise distribution (c.d.f.) is merely Lipschitz continuous, the minimax regret is $\Theta(T^{2/3})$, strictly between the $\log T$ rate with known $F$ and the $\sqrt{T}$ rate with unknown $F$. \citet{Fan2024} consider a semiparametric setting where the noise density is smooth and connect the pricing problem to the econometrics of semiparametric estimation.}

\subsection{Strategic Buyers}
\label{sec:liu}

\citet{Liu2024strategic} introduce a different economic friction: buyers may respond strategically to the learning rule itself. At time $t$, a buyer arrives with true covariates $x_t^0 \in \mathbb{R}^d$ and valuation $v_t = \theta_0^\top x_t^0 + z_t$. The seller announces a pricing rule $p_t = g(\hat{\theta}_k^\top x_t)$, where $\hat{\theta}_k$ is the current parameter estimate. Crucially, the buyer observes this rule and can distort her reported features. She solves a cost-minimization problem:
\begin{equation}
\label{eq:manipulation}
\min_{\tilde{x}} \; (p - v_t) + \frac{1}{2}(\tilde{x} - x_t^0)^\top A (\tilde{x} - x_t^0)
\end{equation}
where $A$ is a positive definite matrix governing manipulation costs.\footnote{The matrix $A$ captures how costly it is for the buyer to distort each feature dimension. High eigenvalues of $A$ mean manipulation is expensive. This is the standard model of strategic classification \citep{Hardt2016}, adapted to pricing.} The first-order condition yields a systematic bias: the buyer shifts her features to make the seller's model predict a lower valuation, securing a lower price. The seller observes only the distorted features $\tilde{x}_t$, not the true $x_t^0$.

\begin{theorem}[Theorem 1 of \citet{Liu2024strategic}]
\label{thm:liu_impossibility}
Under standard regularity conditions, any pricing policy that treats reported features as truthful accumulates regret $\Omega(T)$.
\end{theorem}

Regret here is linear in $T$. Every standard dynamic pricing algorithm, including EMLP and RMLP, systematically underprices because it bases decisions on manipulated features, and the bias does not shrink with more data because the manipulation is endogenous to the pricing rule.

The fix is to jointly estimate demand parameters and manipulation behavior. \citet{Liu2024strategic} propose an episodic algorithm with two phases per episode. During the exploration phase, the seller posts uniform random prices that do not depend on features. Since the price is independent of $\tilde{x}_t$, buyers have no incentive to manipulate, and the seller observes true features $x_t^0$. During the exploitation phase, the seller uses the corrected pricing rule that anticipates the manipulation:
\begin{equation}
\label{eq:strategic_price}
p_t = g\left(\hat{\theta}_k^\top x_t + \hat{\beta}_k^\top A^{-1} \hat{\beta}_k \cdot g'(\hat{\theta}_k^\top x_t)\right)
\end{equation}
where $\hat{\beta}_k$ is the estimated coefficient on the manipulable features and $g$ is the optimal pricing function. This correction adds a markup that offsets the anticipated feature distortion.

\begin{theorem}[Theorem 2 of \citet{Liu2024strategic}]
\label{thm:liu_fix}
With known manipulation cost matrix $A$, the strategic pricing algorithm achieves regret $O(d\sqrt{T})$.
\end{theorem}

When $A$ is unknown, the seller can still recover $O(d\sqrt{T/\tau})$ regret by tracking repeat buyers across exploration and exploitation phases, where $\tau$ is the fraction of buyers who appear in both phases (Theorem 3). Higher repeat rates mean more matched pairs for estimating manipulation behavior.

A pricing rule changes buyer incentives, and those incentives change the data the seller sees. Incentive compatibility matters even in settings where the seller is ``just'' learning demand.\footnote{\citet{Agrawal2024ref} document a related phenomenon in pricing with reference effects. If consumers anchor on past prices, a static pricing policy that ignores reference dependence accumulates linear regret $\Omega(T)$.}

\subsection{Comparison of Regret Rates}
\label{sec:comparison}

\begin{figure}[!htbp]
\centering
\includegraphics[width=0.75\textwidth]{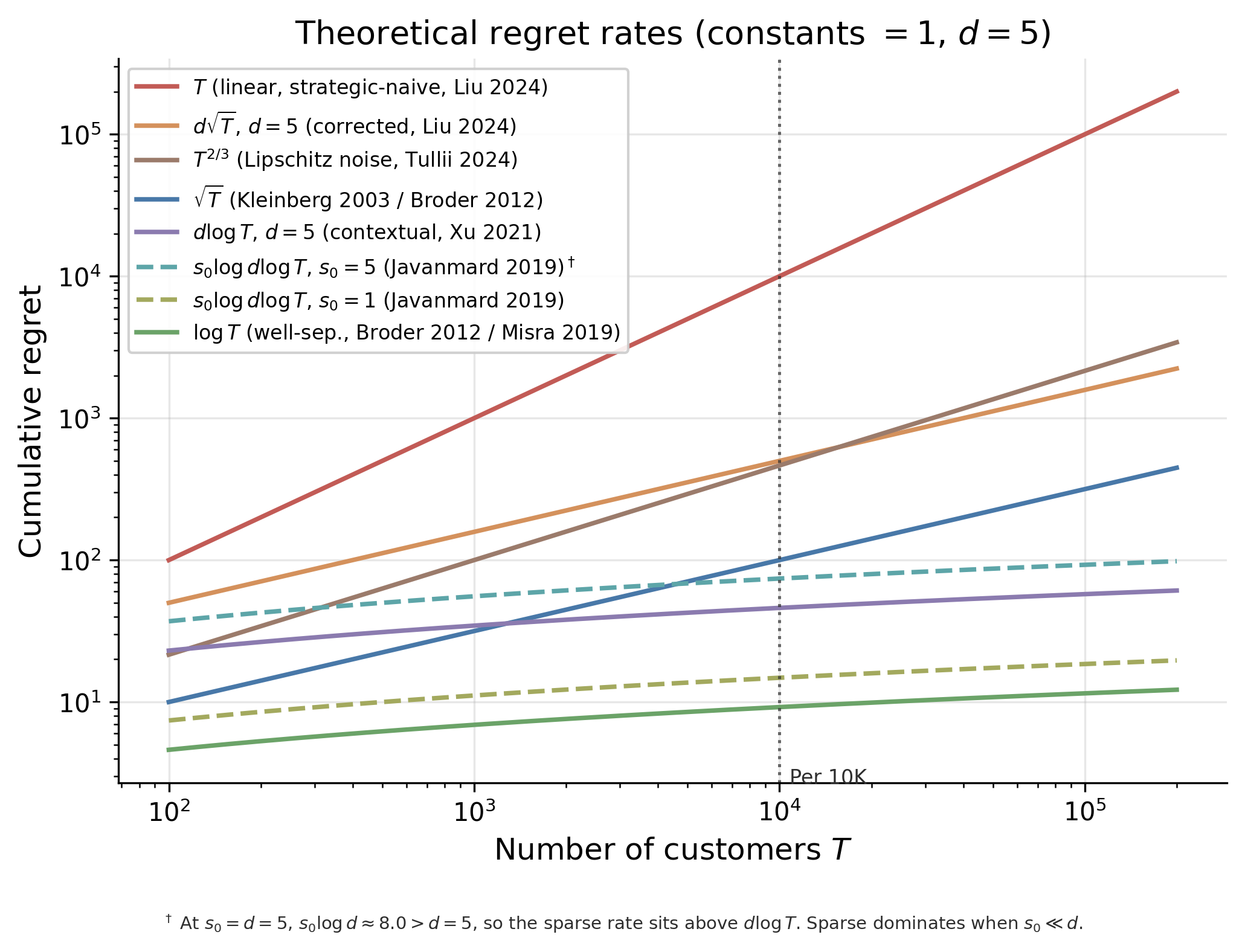}
\caption{Theoretical regret rate functions at constants equal to 1, $d = 5$. Legend names the source paper for each rate. Two cases for the $s_0 \log d \log T$ rate are shown: $s_0 = 1$ (very sparse, $\approx 15$ lost at $T = 10{,}000$) and $s_0 = 5$ (here $s_0 = d$, $\approx 74$ lost), which at this constants-equal-to-1 calibration sits above $d \log T$ because $s_0 \log d \approx 8.0 > d = 5$; sparse dominates when $s_0 \ll d$. The vertical line marks $T = 10{,}000$, matching the ``Per 10K'' column of Table~\ref{tab:regret_comparison}.}
\label{fig:regret_rates}
\end{figure}

Table~\ref{tab:regret_comparison} collects the regret rates discussed above. The dominant pattern is that stronger assumptions yield faster learning, but only when the assumptions make profitable actions informative about demand. For fixed dimension, the gap between $\log T$ and $\sqrt{T}$ grows without bound as $T \to \infty$, so the distinction is more than a constant factor. Strategic behavior is the outlier: it produces linear regret that no amount of data can overcome without explicit correction. Figure~\ref{fig:regret_rates} plots each rate on a log-log scale with all constants set to 1. The figure makes a qualitative asymptotic point; in practice the implicit constants in front of $\log T$ can dwarf those in front of $\sqrt{T}$ at modest horizons, so the visual ordering near $T = 10{,}000$ is illustrative rather than predictive of finite-sample behavior.

\begin{table}[!htbp]
\centering
\small
\caption{Regret rates in dynamic pricing under progressively stronger assumptions. $T$ is the number of customers, $d$ the feature dimension, $s_0$ the sparsity level. The last column translates asymptotic rates into concrete terms for $T = 10{,}000$ with $d = 5$, setting constants to 1.}
\label{tab:regret_comparison}
\resizebox{\textwidth}{!}{%
\begin{tabular}{@{}llllp{2.8cm}@{}}
\toprule
Paper & Demand & Noise & Regret & Per 10K ($d{=}5$) \\
\midrule
\citet{Kleinberg2003} & none & none & $\sqrt{T}$ & $\sim$100 lost \\
\citet{Broder2012} & parametric & known family & $\sqrt{T}$ & $\sim$100 lost \\
\citet{Broder2012} & param., well-sep. & known family & $\log T$ & $\sim$9 lost \\
\citet{Javanmard2019} & linear, $s_0$-sparse & known, log-conc. & $s_0 \log d \log T$ & fast if $s_0$ small \\
\citet{Xu2021} & linear, contextual & known & $d \log T$ & $\sim$46 lost \\
\citet{Xu2021} & linear, contextual & unknown var. & $\geq \sqrt{T}$ & $\sim$100 lost \\
\citet{Tullii2024} & linear, contextual & Lipschitz only & $T^{2/3}$ & $\sim$464 lost \\
\citet{Misra2019} & WARP & -- & $\log T$ & $\sim$9 lost \\
\citet{Liu2024strategic} & linear $+$ strategic & known, na\"ive & $T$ & never improves \\
\citet{Liu2024strategic} & linear $+$ strategic & known, corrected & $d\sqrt{T}$ & $\sim$500 lost \\
\bottomrule
\end{tabular}%
}
\end{table}

\subsection{Applications}
\label{sec:dp_applications}

This pattern extends beyond single-product posted pricing. Practical pricing systems are useful when they exploit structure that would be invisible to a generic bandit: low-rank substitution patterns across products, common elasticity parameters, capacity constraints, or shared demand forecasts.

\subsubsection{Joint Assortment and Pricing at Scale}

\citet{Cai2023} tackle the joint assortment-pricing problem, where a retailer must simultaneously choose which products to display and at what prices. With a large catalog and limited shelf space, the number of possible assortments is combinatorially vast; for a Chinese instant noodle producer with 176 products and 30 display slots, there are $\binom{176}{30} \approx 6.4 \times 10^{33}$ possible assortments before prices are even set. Customer demand also depends on market context such as region and season. In each period $t$, the retailer selects an assortment-pricing vector $a_t \in \mathbb{R}^{d_a}$ (encoding which products to display and at what prices), observes a context vector $x_t \in \mathbb{R}^{d_x}$ (encoding customer demographics and market conditions), and earns revenue $Y_t$. Cai et al.\ model expected revenue as $\mathbb{E}[Y_t \mid x_t, a_t] = a_t^\top \Theta^* x_t$, where $\Theta^* \in \mathbb{R}^{d_a \times d_x}$ is an unknown matrix assumed to have low rank $r \ll \min\{d_a, d_x\}$. The low-rank assumption is the demand-side analogue of factor models in asset pricing; a small number of latent dimensions (flavor preferences, seasonal effects, price sensitivity) explain most of the variation in purchasing behavior.

Their Hi-CCAB algorithm estimates $\Theta^*$ via penalized least squares, where the penalty is the nuclear norm (the sum of singular values) of $\Theta$. This penalty encourages low-rank solutions, playing the same role for matrices that the LASSO penalty plays for sparse vectors. The time-averaged regret is $\tilde{O}(T^{-1/6})$ with dimension dependence $r(d_a + d_x)$ rather than $d_a \cdot d_x$, so the effective parameter count scales with the number of latent factors, not the full product-by-context matrix. In simulation, Hi-CCAB achieves nearly four times the cumulative sales of the noodle producer's historical assortment strategies, averaged over 100 replications.

\citet{Ganti2018} deploy Thompson Sampling in-field for dynamic pricing at Walmart.com. Their MAX-REV-TS algorithm models demand for each item $i$ on day $t$ via a constant-elasticity function $d_{i,t}(p) = f_{i,t}(p/p_{i,t-1})^{\gamma_i^*}$, where $d_{i,t}(p)$ is unit sales at price $p$, $f_{i,t}$ is a baseline demand forecast at the previous day's price $p_{i,t-1}$, and $\gamma_i^* < -1$ is the unknown price elasticity. The structural assumption, that demand responds to price through a single elasticity parameter per item, reduces the learning problem from estimating a full demand curve to estimating one scalar per item. MAX-REV-TS places a Gaussian prior over the elasticity vector $\gamma^*$ and draws posterior samples at each period to solve a constrained revenue maximization problem. In a five-week in-field experiment on a basket of roughly 5,000 items, Thompson Sampling produced a statistically significant increase in per-item revenue relative to the passive pricing baseline.

\subsection{Simulation Study: Structural Knowledge and Curve Learning}
\label{sec:bandit_sim}

The first simulation studies WARP and partial identification using the environment in \citet{Misra2019}. The second studies how demand-curve learning shares information across prices.

The first simulation runs six algorithms on the \citet{Misra2019} demand environment to trace how cumulative regret responds to increasing structural knowledge.\footnote{The environment has $K = 100$ prices on a grid from \$0.01 to \$1.00, $S = 1{,}000$ consumer segments with equal weights, within-segment heterogeneity $\delta = 0.1$, and segment midpoints $v_s \sim \mathrm{Uniform}(0.1, 0.9)$. A consumer purchases if and only if $v_i \geq p$ (WARP). Each algorithm runs across 10 seeds with $T = 200{,}000$ rounds.} This is a WARP and partial-identification demonstration, not a comparison against GP-UCB, GP-TS, or monotone GP demand-curve methods. The six algorithms are ordered by the structural knowledge they exploit. They are (0) $\varepsilon$-greedy with $\varepsilon = 0.1$, which never adapts its exploration rate; (1) Learn-Then-Earn (LTE), which explores uniformly for the first 5\% of rounds and then commits to the empirical best price; (2) UCB1 \citep{auer2002}, which adapts exploration via confidence bounds but ignores demand structure; (3) Thompson Sampling \citep{Thompson1933}, which maintains a Bayesian posterior over purchase rates\footnote{At each period the algorithm draws one sample $\tilde{\mu}_k$ from each arm's posterior over its purchase rate and selects the arm with the highest sampled expected profit $p_k \tilde{\mu}_k$; arms with uncertain posteriors have high-variance draws and are selected frequently, while well-estimated arms are selected in proportion to how likely they are optimal.}; (4) UCB-PI \citep{Misra2019}, which uses WARP to eliminate dominated prices and scales the exploration bonus by the price level\footnote{The UCB-PI implementation needs an estimate of the within-segment heterogeneity $\delta$. From the segments with both a strict lower and upper bound on the purchase threshold, the simulation forms per-segment half-widths $\hat{\delta}_s = (\hat{p}_{s,\max} - \hat{p}_{s,\min})/2$ and uses $\hat{\delta} = \hat{\delta}_{\max} + (\hat{\delta}_{\max} - \overline{\hat{\delta}})$, which inflates the largest observed half-width by its gap from the mean. The cited paper does not use this exact form. It is included as a reference point against the standard UCB-PI; \citet{Misra2019} estimates $\delta$ from the data but uses a different concrete expression.}; and (5) UCB-PI-tuned, which adds a variance-based refinement to the exploration bonus.

\begin{figure}[!htbp]
\centering
\includegraphics[width=0.7\textwidth]{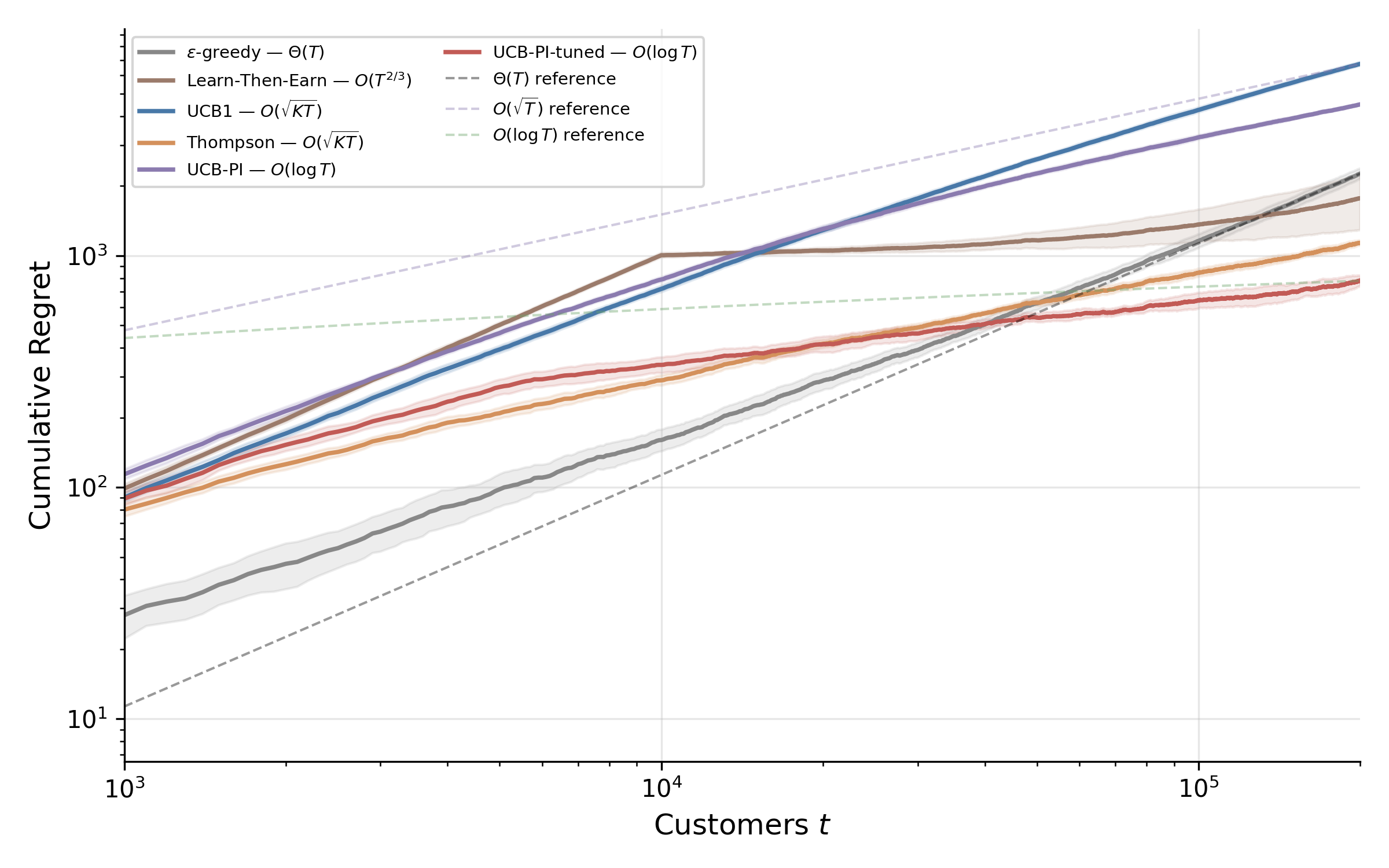}
\caption{Cumulative regret on log-log axes ($K = 100$, $S = 1{,}000$, 10 seeds). Each algorithm's legend entry includes its theoretical regret rate. Dashed lines show $\Theta(T)$, $O(\sqrt{T})$, and $O(\log T)$ reference rates. Shaded regions are $\pm 2$ standard errors.}
\label{fig:knowledge_ladder}
\end{figure}

Within this WARP-only comparison, Table~\ref{tab:knowledge_ladder} reports cumulative regret at four checkpoints and Figure~\ref{fig:knowledge_ladder} shows the full trajectories. The finite-sample pattern is mixed. UCB-PI-tuned achieves the lowest regret by $T = 200{,}000$, but plain UCB-PI is worse than Thompson Sampling and LTE in this run; at this horizon the finite-sample ordering inverts the asymptotic rate order, with UCB-PI's logarithmic-rate algorithm dominated by TS empirically and by LTE despite LTE's nominal $T^{2/3}$ rate. Rate diagnostics computed from these checkpoints reinforce this caveat: across $T \in \{10\text{K}, 50\text{K}, 100\text{K}, 200\text{K}\}$, only Thompson Sampling's $R/\sqrt{T}$ is close to stable; $\varepsilon$-greedy's $R/T$ and LTE's $R/T^{2/3}$ drift downward, UCB1's $R/\sqrt{T}$ grows, and $R/\log T$ for plain UCB-PI quadruples. The right conclusion is narrower. WARP-based elimination helps relative to plain UCB1, but good finite-sample performance depends on the variance-tuned implementation, $T = 200{,}000$ is too short for the predicted rates to visibly settle for most of these algorithms, and these results do not compare against curve-learning GP methods.

\begin{table}[!htbp]
\centering
\caption{Cumulative regret at four checkpoints on the \citet{Misra2019} environment, mean over 10 seeds, with each algorithm's theoretical rate.}
\label{tab:knowledge_ladder}
\begin{tabular}{llrrrrl}
\toprule
Level & Algorithm & $T{=}10$K & $T{=}50$K & $T{=}100$K & $T{=}200$K & Rate \\
\midrule
0 & $\varepsilon$-greedy & 160 & 624 & 1,158 & 2,263 & $\Theta(T)$ \\
1 & LTE (5\%) & 1,005 & 1,165 & 1,363 & 1,771 & $O(T^{2/3})$ \\
2 & UCB1 & 719 & 2,609 & 4,263 & 6,734 & $O(\sqrt{KT})$ \\
3 & Thompson & 290 & 628 & 844 & 1,136 & $O(\sqrt{KT})$ \\
4 & UCB-PI & 789 & 2,266 & 3,244 & 4,503 & $O(\log T)$ \\
5 & UCB-PI-tuned & 338 & 541 & 640 & 780 & $O(\log T)$ \\
\bottomrule
\end{tabular}

\end{table}

The second simulation makes the \citet{Weaver2025} critique operational. It is a small independent replication of the ten-price design in their simulations. Customer willingness to pay is drawn from $\mathrm{Beta}(2,9)$, $\mathrm{Beta}(2,2)$, or $\mathrm{Beta}(9,2)$, giving low-, middle-, and high-optimal-price cases. The tested prices are $p_a=a/10$ for $a=1,\ldots,10$, the firm updates prices every 10 customers, each run lasts $T=2{,}500$ customers, and the table averages over 1{,}000 simulation seeds. The algorithms are independent-arm Thompson Sampling, GP-UCB, GP-TS, GP-UCB-M, and GP-TS-M.\footnote{The GP code is an independent finite-grid implementation of the same demand-GP and derivative-sign ideas in \citet{Weaver2025}, not the authors' replication code. Demand, not reward, is the latent GP object. A batch sale rate is treated as a noisy demand observation with variance $0.25/n_a$. GP-UCB and GP-TS choose prices from posterior upper bounds or posterior draws of $D(p)$. The monotone variants form a joint RBF-GP over $D(0)$, $D(p)$, and $D'(p)$, enforce $D'(p) \leq 0$ through truncated-normal derivative draws, and reconstruct demand by integrating the derivative path on the price grid. GP-UCB-M uses upper quantiles of these constrained posterior curves.}\footnote{GP-UCB uses a constant exploration scale $\beta = 1.8$ rather than the growing schedule $\beta_t = 2\log(|\mathcal{X}|\, t^2 \pi^2 / (6\delta))$ of \citet{Srinivas2010}; the asymptotic regret rate changes only by a log factor, which is not separable from Monte Carlo noise at $T = 2{,}500$ averaged over 1{,}000 seeds. The GP prior mean $\mu_0(p) = 1 - p$ encodes a downward-sloping demand assumption, while independent-arm TS uses an uninformative $\mathrm{Beta}(1,1)$ prior; both methods adapt as data accumulates, and the reported gap is therefore between the worst-case (uniform-prior TS) and a best-case (shape-informed GP) Bayesian baseline.}

\begin{figure}[!htbp]
\centering
\includegraphics[width=\textwidth]{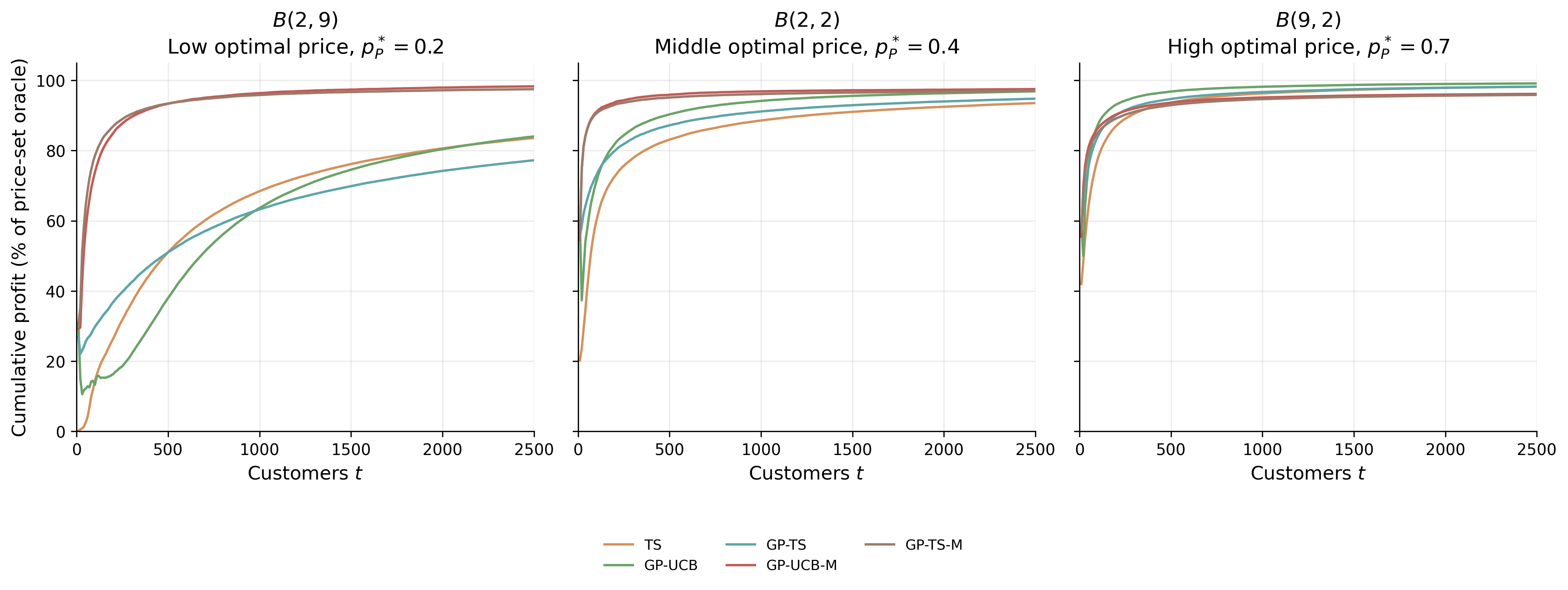}
\caption{Weaver-style curve-learning pricing simulation across low, middle, and high optimal-price locations ($K = 10$, $T = 2{,}500$, 1{,}000 seeds, price updates every 10 customers). The $y$-axis is cumulative profit divided by the price-set oracle. GP-UCB and GP-TS share information across nearby prices; GP-UCB-M and GP-TS-M additionally impose $D'(p) \leq 0$ through derivative-constrained GP curves. Shaded regions are $\pm 2$ standard errors.}
\label{fig:curve_learning_pricing}
\end{figure}

Figure~\ref{fig:curve_learning_pricing} plots cumulative profit as a percentage of the price-set oracle $\Pi_P^*$, and Table~\ref{tab:curve_learning_pricing} reports the final values. The low-optimal-price case is the clearest diagnostic. With $v \sim \mathrm{Beta}(2,9)$, Thompson Sampling earns 83.6\% of the grid oracle by $T=2{,}500$, while GP-TS-M earns 97.5\% and GP-UCB-M earns 98.3\%. In the middle case, TS reaches 93.5\% and the monotone GP variants reach about 97\%. In the high-price case, TS, GP-TS, and GP-UCB are already near oracle, while the monotone variants are slightly lower. The conclusion is therefore narrower and closer to \citet{Weaver2025}. Curve-level informational externalities are most valuable when independent-arm bandits are pulled toward high-price uncertainty even though the true optimum is low; when the optimal price is high, the same uncertainty is less damaging because the uncertain high-price region is also close to the profitable region.

\begin{table}[!htbp]
\centering
\begin{tabular}{lrrrrr}
\toprule
WTP & $p^*_P$ & TS & GP-TS & GP-TS-M & GP-UCB-M \\
\midrule
$B(2,9)$ & 0.2 & 83.6\% & 77.2\% & 97.5\% & 98.3\% \\
$B(2,2)$ & 0.4 & 93.5\% & 94.8\% & 97.0\% & 97.5\% \\
$B(9,2)$ & 0.7 & 98.2\% & 98.2\% & 95.8\% & 96.1\% \\
\bottomrule
\end{tabular}

\caption{Final cumulative profit as a percentage of the price-set oracle by $T=2{,}500$ in the Weaver-style curve-learning simulation.}
\label{tab:curve_learning_pricing}
\end{table}

\section{Offline Reinforcement Learning}
\label{section:offline_rl}
In the preceding chapters, the agent interacts with its environment while learning; it tries a price, observes a purchase, and updates its belief. Online learning is natural in digital markets where experimentation is cheap and feedback is instant. In many economically important settings, however, experimentation is impossible or prohibitively costly. A hospital cannot randomly assign treatments to learn optimal dosing. A central bank cannot experiment with interest rate schedules to discover optimal monetary policy. A firm inheriting a decade of transaction logs wants to improve its pricing rule without conducting new experiments during the transition. In each case, the agent has access to a fixed dataset of past decisions and outcomes, collected under some historical policy, and must learn the best possible new policy from this data alone.

This is the problem of \textit{offline reinforcement learning}, also called \textit{batch reinforcement learning}.\footnote{The term ``batch RL'' was standard in the earlier literature \citep{Ernst2005, Lange2012}. ``Offline RL'' became dominant after \citet{Levine2020}, who distinguished it from off-policy RL (which still collects new data, just under a different policy than the target). I use ``offline RL'' throughout.} The agent never queries the environment. All learning happens from a fixed dataset $\mathcal{D} = \{(s_i, a_i, r_i, s_i')\}_{i=1}^n$ collected by some behavioral policy $\pi_b$. The goal is to find a policy $\hat{\pi}$ whose value $V^{\hat{\pi}}$ is as close to the optimal $V^*$ as possible.

Online RL algorithms (Q-learning, SARSA, policy gradient methods from Section~\ref{section:rl_algorithms}) can in principle be applied to offline data by treating the dataset as a replay buffer. In practice, this fails catastrophically. The reason is \textit{distributional shift}: the learned policy $\hat{\pi}$ inevitably queries state-action pairs that the behavioral policy $\pi_b$ never visited, and the Q-function at these unseen pairs is pure extrapolation. Because the Bellman backup propagates these extrapolation errors through the max operator, errors compound geometrically across the planning horizon, producing arbitrarily poor policies.\footnote{This failure mode was first demonstrated empirically by \citet{Fujimoto2019}, who showed that standard off-policy algorithms (DDPG, SAC) trained purely from a static dataset performed worse than the behavioral policy itself, even when that behavioral policy was a partially-trained, mediocre agent. The gap widened with dataset size, the opposite of what one expects from more data.}

\subsection{The Pessimism Principle}
\label{sec:pessimism}

The overestimation failure has a clean theoretical characterization. Consider tabular Q-learning with a fixed dataset. At any state-action pair $(s, a)$ not in the dataset, the empirical Bellman backup is undefined, but the max in $\max_{a'} \hat{Q}(s', a')$ may still select it if the randomly-initialized Q-value happens to be high. With function approximation, the problem is subtler but identical in spirit: the function approximator can assign high values to out-of-distribution inputs without any corrective signal.

The solution, established simultaneously by several groups, is the \textit{pessimism principle}, which is to construct a lower confidence bound on the Q-function and optimize against it. Formally, given a dataset $\mathcal{D}$ and a confidence parameter $\delta$, construct a penalty function $\Gamma(s, a)$ that is large where data coverage is poor, and define the pessimistic Q-function
\begin{equation}
\label{eq:pessimistic_q}
\tilde{Q}(s, a) = \hat{Q}(s, a) - \Gamma(s, a)
\end{equation}
where $\hat{Q}$ is the standard empirical Bellman solution. The policy $\hat{\pi}(s) = \arg\max_a \tilde{Q}(s, a)$ selects actions that are both high-value \textit{and} well-supported by data.

\begin{definition}[Pessimistic Value Iteration, PEVI \citep{JinYang2021}]
\label{def:pevi}
Given dataset $\mathcal{D}$, penalty function $\Gamma_h(s,a)$ for each stage $h$, and horizon $H$, PEVI computes
\begin{align}
\hat{Q}_h(s,a) &= r_h(s,a) + \hat{P}_h \hat{V}_{h+1}(s,a) - \Gamma_h(s,a) \\
\hat{V}_h(s) &= \max\{\max_a \hat{Q}_h(s,a), \, 0\} \\
\hat{\pi}_h(s) &= \arg\max_a \hat{Q}_h(s,a)
\end{align}
where $\hat{P}_h$ is the empirical transition operator estimated from $\mathcal{D}$.
\end{definition}

\citet{JinYang2021} show that with $\Gamma_h(s,a) = c \cdot \sqrt{H^3 / N_h(s,a)}$, where $N_h(s,a)$ counts visits to $(s,a)$ at stage $h$ and $c$ is an absolute constant, PEVI achieves
\begin{equation}
\label{eq:pevi_bound}
V^* - V^{\hat{\pi}} \leq \tilde{O}\left(\sum_{h=1}^H \mathbb{E}_{\pi^*}\left[\sqrt{\frac{1}{N_h(s_h, a_h)}}\right]\right)
\end{equation}
with high probability. The bound depends on the data coverage at states and actions visited by the \textit{optimal} policy $\pi^*$, not the full state-action space. This is the key advantage of pessimism over uniform coverage requirements.

\subsubsection{Concentrability and Coverage}

The bound in~\eqref{eq:pevi_bound} is instance-dependent, scaling with how well $\pi_b$ covers $\pi^*$. The classical way to formalize this is through \textit{concentrability coefficients} \citep{Munos2008}.

\begin{definition}[Single-policy concentrability]
\label{def:concentrability}
The single-policy concentrability coefficient of $\pi^*$ with respect to $\pi_b$ is
\begin{equation}
C^* = \max_{h \in [H]} \left\| \frac{d_h^{\pi^*}}{d_h^{\pi_b}} \right\|_\infty
\end{equation}
where $d_h^{\pi}(s,a)$ is the state-action occupancy measure of $\pi$ at stage $h$.
\end{definition}

When $C^*$ is finite, the optimal policy visits only states and actions that $\pi_b$ also visits with non-negligible probability, and the $1/\sqrt{N_h(s_h, a_h)}$ terms in~\eqref{eq:pevi_bound} remain controlled. \citet{Rashidinejad2021} prove that single-policy concentrability is both necessary and sufficient for offline learning: if $C^* < \infty$, then $O(|\mathcal{S}| H^2 C^* / \epsilon^2)$ samples suffice for an $\epsilon$-optimal policy; if $C^* = \infty$, no algorithm can guarantee suboptimality better than the behavioral policy.

\subsubsection{Impossibility Results}

\citet{Zanette2021} establish fundamental limits on offline RL by showing that it can be exponentially harder than online RL. Specifically, there exist MDPs with $S$ states and horizon $H$ where online RL finds an $\epsilon$-optimal policy in $\text{poly}(S, H, 1/\epsilon)$ episodes, but any offline algorithm requires $\Omega(2^H)$ samples unless the dataset covers all reachable states. The construction uses a binary tree MDP where the optimal path visits a unique leaf, and any dataset that misses this leaf provides no information about the optimal action at the root.

The practical implication is that offline RL is not a universal replacement for online experimentation. When the behavioral policy is far from optimal, especially in long-horizon problems, offline methods provably cannot recover the optimal policy without exponentially large datasets. Pessimistic algorithms are the best one can do, but they are still fundamentally constrained by what the data contains.

\subsection{Algorithms}
\label{sec:offline_algorithms}

I present four practical algorithms that instantiate different approaches to the distributional shift problem. All four can be understood as modifications of standard Q-learning (Section~\ref{section:rl_algorithms}) that prevent the agent from overvaluing actions outside the data support.

\subsubsection{Fitted Q-Iteration}

Fitted Q-Iteration \citep[FQI,][]{Ernst2005} is the simplest offline RL algorithm, predating the modern pessimism framework. FQI applies the standard Bellman backup iteratively using supervised regression on the fixed dataset, as described in Section~\ref{section:rl_algorithms}. It does not include any explicit pessimism mechanism. When state-action coverage is poor, the max in the Bellman backup selects the highest Q-value among all actions at $s'$, including actions never observed in the data. If the function approximator generalizes poorly at these unseen actions, targets become noisy and biased upward, causing the overestimation cascade. FQI works well when coverage is good but degrades as coverage gaps grow.\footnote{\citet{Munos2008} prove finite-time error bounds for FQI under approximate Bellman completeness and all-policy concentrability. Both assumptions are strong, and violation of either leads to divergence in practice.}

\subsubsection{Conservative Q-Learning}

Conservative Q-Learning \citep[CQL,][]{Kumar2020} adds an explicit penalty that pushes down Q-values at actions not well-represented in the data. The key idea is to add a regularizer to the Bellman error objective that minimizes Q-values under a broad distribution over actions while maximizing Q-values at the actions actually taken in the dataset.

\begin{definition}[Conservative Q-Learning]
\label{def:cql}
CQL modifies the standard Bellman error objective by adding a conservative regularizer. At each iteration, CQL solves
\begin{equation}
\label{eq:cql}
\hat{Q}_{k+1} = \arg\min_Q \; \alpha \left(\mathbb{E}_{s \sim \mathcal{D}}\left[\log \sum_{a} \exp Q(s,a)\right] - \mathbb{E}_{(s,a) \sim \mathcal{D}}[Q(s,a)]\right) + \frac{1}{2} \mathbb{E}_{\mathcal{D}}\left[(Q(s,a) - \hat{\mathcal{B}}^{\pi_k} \hat{Q}_k(s,a))^2\right]
\end{equation}
where $\alpha > 0$ is a hyperparameter controlling the degree of conservatism, and $\hat{\mathcal{B}}^{\pi_k}$ is the empirical Bellman operator.
\end{definition}

The first term in the regularizer, $\log \sum_a \exp Q(s,a)$, is a soft maximum over all actions, pushing down Q-values uniformly. The second term, $\mathbb{E}_{\mathcal{D}}[Q(s,a)]$, pulls Q-values back up at the data actions. The net effect is a penalty on Q-values for actions that appear infrequently relative to their softmax contribution. Theorem 3.2 of \citet{Kumar2020} does not give a pointwise lower bound on every action value. It gives a lower bound after taking the policy expectation of the learned Q-function under its stated conditions. CQL therefore implements the pessimism principle from~\eqref{eq:pessimistic_q} at the policy-value level through an implicit, data-adaptive penalty $\Gamma$.

\subsubsection{Implicit Q-Learning}

Implicit Q-Learning \citep[IQL,][]{Kostrikov2022} avoids querying Q-values at unseen actions entirely. Instead of computing $\max_{a'} Q(s', a')$ in the Bellman backup (which requires evaluating $Q$ at potentially out-of-distribution actions), IQL learns a separate value function $V(s)$ that approximates the in-sample maximum through \textit{expectile regression}.\footnote{The simulation below uses the variant labelled IQL-argmax in Table~\ref{tab:offline_main}. The expectile-V learning step follows \citet{Kostrikov2022} verbatim. The policy-extraction step in the original paper is advantage-weighted regression on $\exp(\beta(Q-V))$; for the discrete 10-price action set used here, that step is replaced with $\arg\max_a Q(s,a)$, which is equivalent in the deterministic limit and avoids the additional temperature hyperparameter. The substitution does not change which actions get high Q-values; it only changes how the policy network is fit.}

\begin{definition}[Implicit Q-Learning]
\label{def:iql}
IQL maintains three functions: $Q_\theta(s,a)$, $V_\psi(s)$, and a policy $\pi_\phi(a|s)$. The value function is trained via expectile regression
\begin{equation}
\label{eq:iql_v}
L_V(\psi) = \mathbb{E}_{(s,a) \sim \mathcal{D}}\left[L_2^\tau(Q_{\bar{\theta}}(s,a) - V_\psi(s))\right]
\end{equation}
where $L_2^\tau(u) = |\tau - \mathbbm{1}\{u < 0\}| \cdot u^2$ is the asymmetric squared loss with expectile parameter $\tau \in (0.5, 1)$, and $Q_{\bar{\theta}}$ uses a target network. The Q-function is trained with $V_\psi$ as the continuation value
\begin{equation}
\label{eq:iql_q}
L_Q(\theta) = \mathbb{E}_{(s,a,r,s') \sim \mathcal{D}}\left[(r + \gamma V_\psi(s') - Q_\theta(s,a))^2\right]
\end{equation}
\end{definition}

The expectile parameter $\tau$ controls the degree of optimism within the data support. As $\tau \to 1$, the expectile converges to the in-sample maximum; as $\tau \to 0.5$, it converges to the in-sample mean. Setting $\tau = 0.7$ balances exploiting the best observed actions against the noise in finite samples. The critical property is that the Q-function is never evaluated at actions outside the dataset; the $\max$ operation is implicit in the expectile regression of $V$.

\subsubsection{Batch-Constrained Q-Learning}

Batch-Constrained Q-Learning \citep[BCQ,][]{Fujimoto2019} takes a different approach. Rather than modifying the Q-function objective, it restricts the policy to actions similar to those in the dataset. BCQ first learns a generative model $G_\omega(s)$ of the behavioral policy, then constrains the policy to only select actions with high likelihood under $G_\omega$.\footnote{The original \citet{Fujimoto2019} BCQ targets continuous action spaces: $G_\omega$ is a variational autoencoder and a perturbation network adjusts sampled actions toward higher Q. For the discrete pricing action set used in the simulation below, the discrete BCQ-D variant of \citet{Fujimoto2019b} is the more natural choice. BCQ-D drops the VAE and perturbation network and uses an MLP classifier of the behavioral action distribution, then thresholds: only actions $a$ with $G_\omega(a \mid s) \geq \tau \cdot \max_{a'} G_\omega(a' \mid s)$ are admissible. The pessimism mechanism is the same in both variants; the architecture differs because the action space differs. Table~\ref{tab:offline_main} reports BCQ-D.}

\begin{definition}[Batch-Constrained Q-Learning]
\label{def:bcq}
BCQ learns a behavioral model $G_\omega(a|s)$ from the dataset via maximum likelihood, and constrains action selection
\begin{equation}
\label{eq:bcq}
\hat{\pi}(s) = \arg\max_{a : G_\omega(a|s) \geq \tau \cdot \max_{a'} G_\omega(a'|s)} Q_\theta(s, a)
\end{equation}
where $\tau \in (0, 1]$ is a threshold parameter. The Q-function is trained with standard Bellman backups, but the max in the target computation is also constrained to the feasible action set.
\end{definition}

BCQ's constraint is hard rather than soft: actions with behavioral probability below $\tau$ times the most likely action are simply excluded from consideration. This prevents the Q-function from ever being queried at truly out-of-distribution actions, addressing the distributional shift problem at the policy level rather than the value function level. The tradeoff is that BCQ's performance is bounded by the quality of actions in the dataset. If the behavioral policy never takes the optimal action at some state, BCQ cannot discover it regardless of sample size.

\subsection{Trajectory Models and Return-Conditioned Supervised Learning}
\label{sec:dt_rvs}

Pessimism-based offline RL centers on the value function, with the policy emerging from a Q-function constructed to be conservative outside the data. An alternative inverts that relationship. Train a supervised or generative model on the logged trajectories themselves, and treat the model as the policy. The family includes two designs at opposite ends: a transformer-based trajectory model and a stripped-down MLP that isolates the operative ingredient.

\subsubsection{Decision Transformer}

An offline reinforcement-learning dataset is a collection of trajectories $\tau = (s_0, a_0, r_0, s_1, a_1, r_1, \dots, s_T, a_T, r_T)$ generated by some behavioural policy. A trajectory generative model treats $\tau$ as a sequence and trains a generative model on those sequences. The \citet{Chen2021DT} Decision Transformer represents each trajectory as a flat sequence
\begin{equation}
\tau = (\widehat R_0, s_0, a_0,\, \widehat R_1, s_1, a_1,\, \dots,\, \widehat R_T, s_T, a_T),
\label{eq:dt_sequence}
\end{equation}
where $\widehat R_t = \sum_{t' \ge t} r_{t'}$ is the return-to-go from time $t$ onwards. A causally masked GPT-style transformer is trained to predict the next action token given the last $K$ context tokens. At test time, the operator specifies a desired return $R^\star$ and primes the model with $\widehat R_0 = R^\star$ and the current state $s_0$. The model generates the corresponding action, the environment returns a reward and the next state, the operator decrements $R^\star$ by the realised reward, and the loop repeats. No value function, no policy gradient, no critic appears in training. The conditioning variable plays the role of a behavioural target rather than an estimated value function.\footnote{The simulation below implements a fused-token simplification: each timestep is represented as a single token whose embedding sums the return, state, and previous-action contributions, rather than as three distinct tokens at adjacent positions. This shrinks the context window by a factor of three at the cost of departing from the strict autoregressive token order of \citet{Chen2021DT}. The fused form coincides with the architecture used by \citet{Emmons2022}, who report that the trajectory-as-sequence framing is what matters; the three-token ordering is not the load-bearing component.}

The empirical case rests on D4RL and Atari. On the $1\%$ DQN-replay slice of Atari, the Decision Transformer attains gamer-normalised scores of $267.5$ on Breakout against the $211.1$ of Conservative Q-Learning, the incumbent offline-RL baseline. On the D4RL locomotion benchmark, the Decision Transformer averages $74.7$ across nine task-dataset combinations against CQL's $54.2$ and behavioural cloning's $47.7$.\footnote{The same trajectory-as-sequence reframing was independently proposed by \citet{Janner2021TT}, with beam search over tokenised states and actions replacing return conditioning. Subsequent diffusion variants \citep{Janner2022diffuser,Ajay2023} replace autoregressive generation with iterative denoising of an entire trajectory and dispense with dynamic programming entirely. The methodological move is the same, plan by sampling from a generative model of structured objects rather than by maximising a learned value function.}

\subsubsection{Return-Conditioned Supervised Learning}

The Decision Transformer combines two ideas, a transformer architecture over trajectory tokens and a return-conditioning protocol at deployment. \citet{Emmons2022} isolate the second from the first. Their return-conditioned supervised learning baseline drops the transformer entirely. A small multilayer perceptron is trained by maximum likelihood to predict the action given the current state and a desired return-to-go,
\begin{equation}
\pi_\theta(a \mid s, \widehat R) = \mathrm{softmax}\bigl(f_\theta(s, \widehat R)\bigr).
\label{eq:rvs}
\end{equation}
At deployment time the same protocol applies. Specify a target return $R^\star$, condition on it, execute the predicted action, decrement $R^\star$ by the realised reward, and repeat. Across the D4RL locomotion suite RvS matches the Decision Transformer on most task-dataset pairs at a small fraction of the parameter count, suggesting that return-conditioning rather than the transformer architecture is the operative ingredient. The result raises a caveat that both methods share. Conditioning on a return that is far above any value observed in the training data is an extrapolation request, and the policy's behaviour out of distribution is not constrained by the training objective.\footnote{\citet{Brandfonbrener2022} formalise the failure mode. In stochastic environments, conditioning on a high return induces a policy that selects actions whose value distribution has a high upper tail rather than a high expectation, which is a different optimisation problem. The Decision Transformer and RvS recover the optimal policy only when the environment is near-deterministic or when the target return matches what an optimal policy would actually achieve. The reduction-to-supervised-learning that makes the family attractive comes with this stochastic-environment caveat.}

\subsection{Simulation Study: Offline RL for Dynamic Pricing}
\label{sec:offline_sim}

The simulation study evaluates seven offline learners on a perishable inventory pricing problem with demand regime switching. The value-based family appears as FQI, CQL, IQL-argmax, and BCQ-D, of which the last three carry an explicit pessimism mechanism and FQI is the unprotected baseline; the supervised-conditioning family appears as behavioral cloning, the Decision Transformer, and return-conditioned supervised learning. A retailer with $I_{\max} = 30$ units of perishable inventory must set prices over $H = 20$ periods. The state $(i, d, t)$ consists of current inventory $i \in \{0, \ldots, 30\}$, demand regime $d \in \{1, 2, 3, 4\}$, and time remaining $t \in \{1, \ldots, 20\}$. The action is a price $p \in \{1, \ldots, 10\}$. Demand follows $Q \sim \text{Poisson}(\lambda_0[d] \cdot e^{-0.15 p})$ with base rates $\lambda_0 = (1.5, 3.0, 5.0, 8.0)$, and the reward is $r = p \cdot \min(Q, i)$. Demand regimes follow a 4-state Markov chain with diagonal persistence 0.6. Unsold inventory at the terminal period incurs a spoilage cost of \$2.00 per unit, making clearance pricing valuable near the deadline.\footnote{The spoilage penalty creates distributional shift. The optimal policy adapts prices to inventory level and time remaining, using lower prices near the deadline when inventory is high. The \$2.00 penalty is a deliberate design choice: under harsher penalties (e.g., \$10 per unit), all methods collapse to 48--53\% of optimal regardless of algorithmic sophistication, confirming that no offline correction can overcome severe distributional shift when the penalty regime amplifies consequences of the behavioral policy's suboptimality.} The behavioral policy is state-dependent and softly stochastic. Each demand regime has a preferred price ($p^* = 5, 7, 8, 9$ for regimes $d = 1, 2, 3, 4$), reflecting the principle that higher demand supports higher prices. Within each regime the action is sampled from a triangular kernel of half-width 2 around $p^*(d)$, mixed with 15\% uniform noise over all prices. This spreads behavioral mass across roughly five prices per regime and gives the dataset both state-conditional structure (BC and DT can pick up regime-specific patterns) and off-policy coverage (Q-methods see Bellman targets for actions away from the kernel mode).\footnote{The earlier version of this experiment used a state-independent behavioral with 85\% mass on $p = 10$. Under that distribution, BC, BCQ-D, DT, and RvS all collapsed to the constant policy $\hat\pi(s) = 10$ and reported bit-identical returns of $169.27 \pm 0.60$; the supervised-conditioning rows of the table carried no information beyond the BC baseline. The state-dependent softer behavioral used here breaks the collapse and lets the four methods differentiate.} All episodes start at full inventory ($i = 30$). All offline methods train on 500 episodes and are evaluated over 1,000 episodes against the DP optimal policy computed by backward induction.\footnote{FQI uses the standard Bellman backup with $\max_{a'} Q(s', a')$, deliberately without a target network, to isolate the overestimation cascade as a pedagogical baseline. Adding target networks to FQI mitigates but does not eliminate extrapolation error. CQL and IQL include target networks following their original implementations \citep{Kumar2020, Kostrikov2022}; for CQL, target networks proved essential, as the conservative penalty amplifies bootstrap instability without them.} Results are averaged over 20 independent seeds.

\begin{table}[t]
\centering
\caption{Policy value for each offline RL method, expressed as mean return and percentage of the DP optimal. Standard errors computed over 20 seeds.}
\label{tab:offline_main}
\begin{tabular}{lcc}
\hline
Method & Mean Return & \% of Optimal \\
\hline
DP Oracle & $192.41 \pm 0.33$ & $100.0\%$ \\
RvS & $186.58 \pm 0.34$ & $97.0\%$ \\
BC & $186.28 \pm 0.31$ & $96.8\%$ \\
DT & $185.27 \pm 0.33$ & $96.3\%$ \\
CQL & $178.08 \pm 1.48$ & $92.6\%$ \\
BCQ-D & $177.05 \pm 0.73$ & $92.0\%$ \\
IQL-argmax & $176.67 \pm 0.81$ & $91.8\%$ \\
FQI & $47.48 \pm 8.42$ & $24.7\%$ \\
\hline
\end{tabular}

\end{table}

\begin{figure}[t]
\centering
\includegraphics[width=0.8\textwidth]{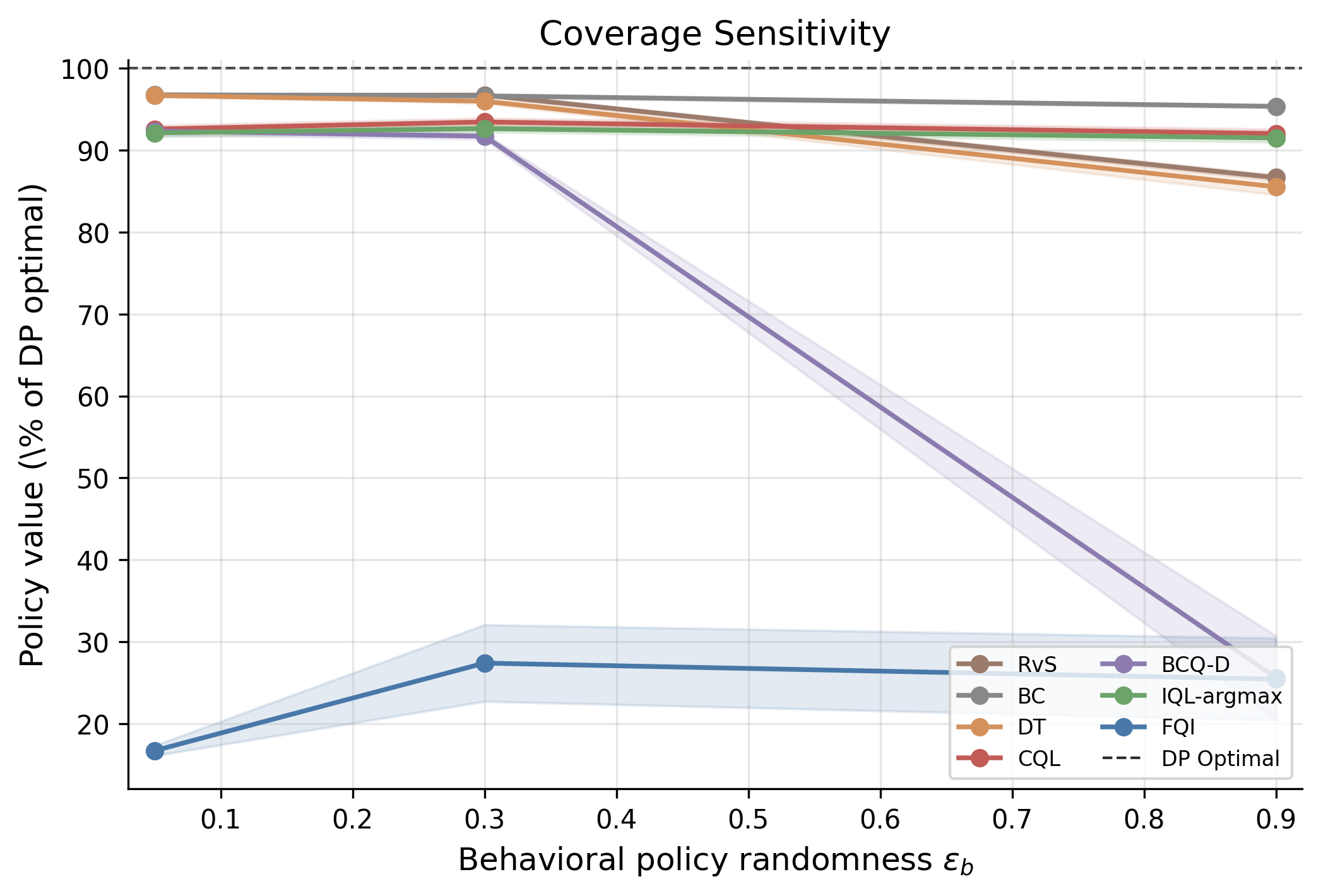}
\caption{Policy value (as \% of DP optimal) versus behavioral policy randomness $\epsilon_b$ for the value-based methods (FQI, CQL, IQL-argmax, BCQ-D), the supervised-conditioning family (DT, RvS), and the behavioral cloning baseline (BC). Higher $\epsilon_b$ increases data coverage.}
\label{fig:offline_coverage}
\end{figure}

Table~\ref{tab:offline_main} places the supervised-conditioning methods RvS, BC, and DT at 96--97\% of the DP optimal. The pessimism methods CQL, BCQ-D, and IQL-argmax reach 92--93\%, while FQI reaches 24.7\%. The near-optimal state-dependent behavioral policy explains this ordering. BC imitates that policy and recovers most of the optimum. RvS and DT add return-conditioning corrections and exceed BC by 0.2--0.3 percentage points. CQL, IQL-argmax, and BCQ-D sacrifice some imitation-baseline performance for distributional robustness, leaving each method near 92\% on this near-on-policy dataset.\footnote{CQL uses $\alpha = 0.1$, the result of a search over $\alpha \in \{5.0, 2.0, 0.5, 0.1\}$. Larger values push Q-values down too aggressively, collapsing the learned policy to the behavioral action at most states; the right $\alpha$ is problem-specific and can vary by orders of magnitude.} FQI has no pessimism mechanism and suffers an overestimation cascade. Its coverage sweep in Figure~\ref{fig:offline_coverage} stays between seventeen and twenty-seven percent of optimal.

The four supervised-conditioning methods now produce distinct policies under this state-dependent behavioral. BC at 96.8\% sets the imitation ceiling for what can be recovered from this dataset without pessimism. BCQ-D at 92.0\% sits below BC because its threshold admits multiple in-support actions at each state and the Q-function then sometimes selects a slightly worse one than BC's empirical mode. DT and RvS at 96.3\% and 97.0\% both edge above BC because the return-to-go conditioning, queried at the high target $R^\star = V^\ast(s_0) \approx 184$, biases the model toward state-conditional action choices that historically led to higher trajectory returns. The four methods are no longer pointwise identical and Table~\ref{tab:offline_main} now reports four distinct policies rather than one policy under four names.\footnote{These results connect to \citet{Brandfonbrener2022}'s analysis of return-conditioning in stochastic environments. With a near-optimal behavioral and short horizon (T = 20), the target-return extrapolation request is bounded and supervised conditioning is well-behaved. Under longer horizons or more adversarial behavioral policies, the return-conditioning extrapolation can become ill-defined.}

The comparison links the simulation directly to the pessimism theory in Section~\ref{sec:offline_algorithms}. FQI reaches 24.7\% because the $\max_{a'}$ operator selects overestimated Q-values at out-of-distribution actions, and 200 bootstrap iterations compound the errors geometrically \citep{Fujimoto2019}. CQL and IQL-argmax remain near 92\% because the conservative penalty in Definition~\ref{def:cql} and expectile-V in Definition~\ref{def:iql} suppress out-of-distribution Q-values. The BC value of 96.8\% measures the performance the pessimism methods sacrifice for that protection. BCQ-D's threshold in Definition~\ref{def:bcq} similarly trades imitation fidelity for an action constraint. On this near-on-policy dataset, pessimism can underperform pure imitation. The coverage experiment measures what changes as the behavioral policy degrades.

Figure~\ref{fig:offline_coverage} varies the behavioral noise parameter $\epsilon_b \in \{0.05, 0.3, 0.9\}$, where $\epsilon_b$ is the mixing weight on uniform-random action sampling against the state-dependent kernel. At $\epsilon_b = 0.05$ the kernel dominates and the dataset is near-on-policy; at $\epsilon_b = 0.9$ the behavioral is nearly uniform and the dataset offers little state-conditional signal. BC stays at 95--97\% across all coverage levels because the cross-entropy objective still recovers the empirical state-conditional mode even from a noisy dataset. CQL and IQL-argmax also stay near 92\% throughout, confirming that both pessimism mechanisms provide stable robustness to data distribution. FQI is uniformly catastrophic, sitting between 17\% and 27\% across all coverage levels; the overestimation cascade is severe regardless of dataset breadth. BCQ-D tracks CQL and IQL-argmax at $\epsilon_b \in \{0.05, 0.3\}$ but collapses to 25.6\% at $\epsilon_b = 0.9$ because a nearly uniform behavioral renders the threshold constraint vacuous, reducing BCQ-D to unconstrained FQI. DT and RvS match BC at $\epsilon_b \in \{0.05, 0.3\}$ but drop to 86\% at $\epsilon_b = 0.9$; with the kernel mode washed out, the return-conditioning extrapolation has less in-distribution support to guide the policy and the supervised objective offers no mechanism to penalize unsupported actions. These patterns connect directly to the concentrability framework (Definition~\ref{def:concentrability}): as $\epsilon_b$ grows, the optimal policy's state-action occupancy diverges from the behavioral policy's, and the $1/\sqrt{N_h(s_h, a_h)}$ terms in~\eqref{eq:pevi_bound} become large at states the optimal policy visits. CQL and IQL-argmax remain robust because their conservative adjustments scale with data sparsity, while FQI lacks this adaptive correction and BCQ-D's hard threshold breaks down once the behavioral is too diffuse for the threshold to bind.

\FloatBarrier
\subsection{Engine Replacement MDP: Coverage and Extrapolation}
\label{engine:ch08}

\begin{table}[htbp]
\centering
\caption{Discounted state-action occupancy under the logging and target policies, finite-log coverage, and final action values. The log contains forty transitions for each marked pair and none for high mileage with replacement.}
\label{tab:engine_coverage}
{\small\renewcommand{\arraystretch}{1.12}
\begin{tabular}{llrrrrrr}
\toprule
state & action & log share & target share & in log & exact $Q^*$ & FQI & CQL \\
\midrule
low & keep & 9.5\% & 69.0\% & yes & 5.345 & 3.454 & 3.447 \\
low & replace & 85.8\% & 0.0\% & yes & 4.310 & 2.608 & 2.606 \\
high & keep & 0.5\% & 0.0\% & yes & 4.079 & 1.999 & 2.000 \\
high & replace & 4.2\% & 31.0\% & no & 4.310 & 0.000 & -2.779 \\
\bottomrule
\end{tabular}}
\end{table}

\begin{figure}[htbp]
\centering
\includegraphics[width=0.8\textwidth]{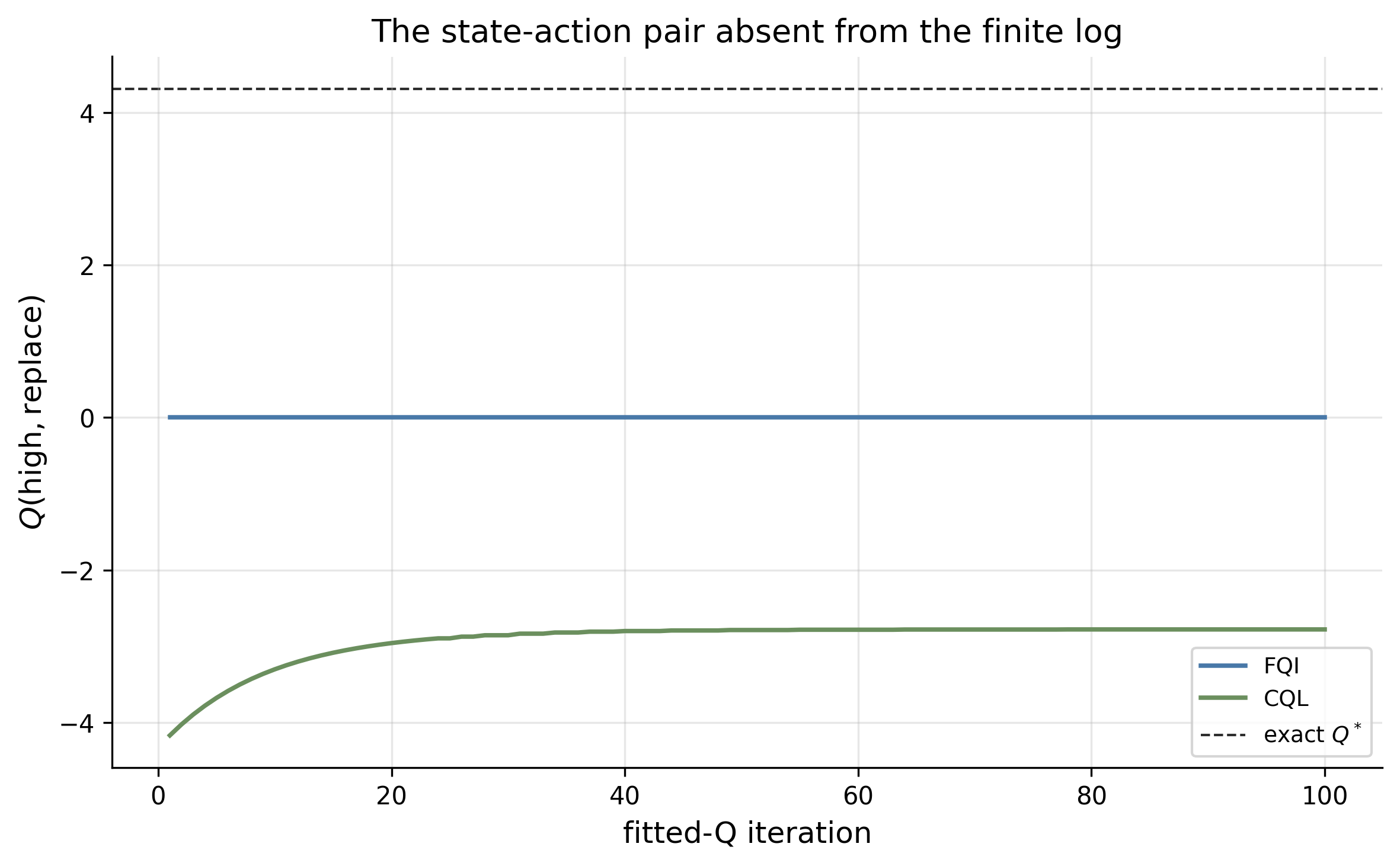}
\caption{The learned value of replacement at high mileage, the state-action pair absent from the finite log. The reference line is the exact optimal action value.}
\label{fig:engine_coverage}
\end{figure}

The population logging policy has full support and gives $C_\infty=7.3180$, but its discounted state-action distribution allocates $85.8\%$ to replacement at low mileage, which the target never selects, and only $4.2\%$ to replacement at high mileage, which the target requires. The finite teaching log contains the other three pairs and withholds the high-mileage replacement pair. FQI assigns the unsupported tabular coefficient its ridge-selected value of zero. The discrete CQL log-sum-exp penalty moves the coefficient to $-2.7789$, against the exact value $4.3103$. The CQL estimate depends on the disclosed ridge coefficient $10^{-5}$ because an entirely unsupported coefficient has no finite unregularized optimum. The calculation follows the fitted regression iteration of \citet{Ernst2005} and the conservative value penalty of \citet{Kumar2020}.
\FloatBarrier

This chapter therefore concerns fixed-data policy improvement when scalar rewards are observed. This is distinct from RLHF, where the reward itself must be inferred from preference comparisons and then interpreted as an object for aggregation, incentives, and welfare. The next chapter takes up that separate problem.

\section{RLHF and AI Alignment}
\label{section:rlhf}
Every chapter so far assumes access to a scalar reward signal: dynamic programming requires $r(s,a)$, model-free RL observes $r_t$ after each transition, and the Bellman optimality equation presupposes that rewards are known or observable. When they are neither, the DP/RL machinery cannot be applied directly.

In many domains, scalar rewards are unavailable but ordinal preferences over trajectories are easy to elicit. A human evaluator cannot assign a meaningful numerical score to a paragraph of text, but can reliably say ``response A is better than response B.'' The raw data is a set of trajectory pairs with binary preference labels. Reinforcement learning from human feedback (RLHF) uses these ordinal comparisons to learn a proxy reward function $r_\theta$, which then serves as the scalar signal for standard RL optimization. This is not an inverse problem in the IRL sense; the goal is not to rationalize observed behavior, but rather a two-stage forward problem in which the analyst first learns a reward from human judgments and then solves the resulting MDP. \citet{christiano:2017} demonstrated that this approach could train agents without an explicit reward function. RLHF has since become the predominant method for aligning large language models.

\subsection{Learning Rewards from Preferences}

The canonical RLHF framework is built on a formal model of human preference. The observed data consists of tuples $(s, y_w, y_l)$, where $s$ is a context, and $y_w$ and $y_l$ are two outputs, with $y_w$ being the ``winner'' preferred by a human. Assuming preferences follow a latent utility model, the Bradley-Terry model \citep{bradley1952rank} gives the probability that $y_w$ is preferred: $P(y_w \succ y_l | s) = \sigma(r_\theta(s, y_w) - r_\theta(s, y_l))$, where $r_\theta: \mathcal{S} \times \mathcal{Y} \to \mathbb{R}$ is a learned \textit{reward model} parameterized by $\theta$, trained to approximate human preferences (not the ground-truth reward, which is unobserved), and $\sigma(\cdot)$ is the logistic function. This formulation is a binary logit model (Section~\ref{section:language}, Equation~\ref{eq:softmax_logit}).\footnote{\citet{iskhakov2021} discuss the contrasts and synergies between machine learning and structural econometrics, including the shared reliance on logit-based choice models that underlies both RLHF and dynamic discrete choice estimation.} The reward model parameters $\theta$ are estimated by minimizing the negative log-likelihood of the observed human choices:
\begin{equation}
    \mathcal{L}(\theta) = -\mathbb{E}_{(s, y_w, y_l) \sim \mathcal{D}} \left[ \log \sigma \left( r_\theta(s, y_w) - r_\theta(s, y_l) \right) \right],
\label{eq:rlhf_loss}
\end{equation}
In the LLM setting, the ``outputs'' $y_w$ and $y_l$ are token sequences, i.e., trajectories of the autoregressive policy\footnote{An \emph{autoregressive} language model generates text token by token: at each step it outputs a distribution over the vocabulary conditioned on all preceding tokens, then samples the next token; the full response is therefore a trajectory in token space and the model acts as a sequential policy over a vocabulary-sized action set.}, so preferences are over trajectories rather than single actions. The preference loss in Equation~(\ref{eq:rlhf_loss}) is MLE of a choice model where the ``alternatives'' are trajectory segments and the ``choice'' is the human-preferred one.

This logit foundation is also a social-choice assumption once labelers are heterogeneous. Pooling pairwise comparisons from many evaluators does not recover a neutral ``human preference''; it estimates a single score whose odds ratios summarize the aggregation rule induced by the sampling scheme, label model, and loss function. Recent alignment work makes this point explicit. Standard Bradley-Terry-style reward learning can behave like a hidden social welfare function, can fail natural axioms for aggregating rankings, and can create strategic reporting incentives when preferences differ across contexts or groups \citep{conitzer2024socialchoice,siththaranjan2024distributional,ge2024axioms,park2024heterogeneous}. DPO inherits the same aggregation problem because it replaces the explicit reward model with an implicit reward difference, not with a theory of whose preferences count.

\subsection{The RLHF Pipeline and Direct Optimization}

The learned $r_\theta$ then serves as a proxy objective for policy optimization. Building on the reward-learning and RL fine-tuning framework of \citet{ziegler2019fine} and \citet{stiennon2020learning}, the canonical three-stage pipeline was formalized by \citet{ouyang2022training}. First, a base pretrained model is initialized via supervised fine-tuning (SFT)\footnote{\emph{Pretraining} optimizes a language model to predict the next token across a massive text corpus, producing broad linguistic knowledge with no behavioral objective. \emph{Supervised fine-tuning} (SFT) continues training on a small curated dataset of (prompt, ideal-response) pairs to specialize the model toward the desired task and establish the reference policy $\pi^{SFT}$ from which the KL penalty is measured.} on a small dataset of high-quality demonstrations, yielding an initial policy $\pi^{SFT}$. This grounds the model in the desired style and format. The second step is the reward model training as described, using preference data generated from this $\pi^{SFT}$.

In the third step, the SFT policy $\pi_\phi$ is fine-tuned via PPO (Section~\ref{sec:actor_critic}) to maximize the frozen reward model $r_\theta$,\footnote{After fine-tuning concludes, the reward model may be reused for evaluation or for later alignment iterations, but it is not updated inside the PPO optimization loop.} with a KL-divergence penalty preventing the policy from drifting into regions where $r_\theta$ is unreliable. The objective is
\begin{equation}
    J(\phi) = \mathbb{E}_{s \sim \mathcal{D}, y \sim \pi_\phi(\cdot|s)} [r_\theta(s, y)] - \lambda_{KL} \mathbb{E}_{s \sim \mathcal{D}} [D_{KL}(\pi_\phi(\cdot|s) \,||\, \pi^{SFT}(\cdot|s))],
\label{eq:rlhf_objective}
\end{equation}
where $D_{KL}$ is the Kullback-Leibler divergence and $\lambda_{KL}$ controls the penalty strength.\footnote{$\lambda_{KL}$ denotes the KL penalty weight, reserving $\beta$ for model parameters and $\gamma$ for discount factors. The standard RLHF literature, including \citet{rafailov2023direct}, uses $\beta$ for this parameter.} Without this constraint, the policy exploits inaccuracies in $r_\theta$ to achieve high proxy scores with degenerate behavior (``reward hacking''), the RLHF analogue of divergence under function approximation (Section~\ref{sec:deadly_triad}).

The KL-regularized objective in Equation~(\ref{eq:rlhf_objective}) admits a Bayesian interpretation \citep{korbak2022rl}. In this view, the reference policy $\pi^{SFT}$ acts as a prior distribution over plausible responses. The reward model $r_\theta$ provides evidence, specifying which responses are more desirable. The goal of alignment is to find the posterior distribution $\pi^*$ that optimally combines the prior with this evidence. This ideal posterior policy takes the form of Equation~(\ref{eq:rlhf_posterior}):
\begin{equation}
    \pi^*(y|s) \propto \pi^{SFT}(y|s) \exp\left(\frac{r_\theta(s, y)}{\lambda_{KL}}\right).
\label{eq:rlhf_posterior}
\end{equation}
The reward function scaled by $\lambda_{KL}$ defines the log-likelihood, so the KL-regularized objective $J(\phi)$ is equivalent (up to an additive constant) to the Evidence Lower Bound (ELBO) for this Bayesian inference problem. Maximizing the RLHF objective via PPO is therefore variational inference: finding the policy $\pi_\phi$ that minimizes KL divergence to $\pi^*$. This reframes the KL penalty as a structural component of the inference problem rather than an ad-hoc regularizer.

Despite this closed-form characterization, the three-stage RLHF pipeline is complex to implement, requiring training multiple large models and a computationally expensive RL loop. Direct Preference Optimization (DPO), introduced by \citet{rafailov2023direct}, collapses the pipeline into a single supervised learning objective by reparameterizing the reward function in terms of $\pi^*$ and $\pi^{SFT}$, as in Equation~(\ref{eq:dpo_reparameterize}):
\begin{equation}
    r(s,y) = \lambda_{KL} \log\left(\frac{\pi^*(y|s)}{\pi^{SFT}(y|s)}\right) + \lambda_{KL} \log Z(s).
\label{eq:dpo_reparameterize}
\end{equation}
When this analytical expression for the reward is substituted into the Bradley-Terry preference loss from Equation~(\ref{eq:rlhf_loss}), the unknown partition function $Z(s)$ cancels out. This yields a loss function that depends only on the policy $\pi_\phi$ being optimized and the fixed reference policy $\pi^{SFT}$. The DPO objective, given in Equation~(\ref{eq:dpo_loss}), is thus a simple binary cross-entropy loss over policy likelihoods:
\begin{equation}
    \mathcal{L}_{DPO}(\phi; \pi^{SFT}) = -\mathbb{E}_{(s, y_w, y_l) \sim \mathcal{D}} \left[ \log \sigma \left( \lambda_{KL} \log \frac{\pi_\phi(y_w|s)}{\pi^{SFT}(y_w|s)} - \lambda_{KL} \log \frac{\pi_\phi(y_l|s)}{\pi^{SFT}(y_l|s)} \right) \right].
\label{eq:dpo_loss}
\end{equation}
This objective is optimized on $\phi$ using standard supervised learning with a static preference dataset. The gradient increases the likelihood of preferred responses $y_w$ and decreases the likelihood of dispreferred responses $y_l$, relative to the reference policy.

The growing family of DPO variants is best read economically by asking what part of the preference problem changes. IPO modifies the link/objective structure to avoid some DPO overfitting behavior \citep{azar2024preferenceparadigm}. KTO replaces pairwise comparisons with a prospect-theory-style objective over desirable and undesirable outputs \citep{ethayarajh2024kto}. Nash-LHF and SPPO treat preference optimization as a game over pairwise comparisons, which is closer to social choice when preferences are cyclic or non-transitive \citep{munos2023nash,wu2024sppo}. MaxMin-RLHF makes the aggregation rule explicitly egalitarian \citep{chakraborty2024maxmin}. ORPO, SimPO, and robust DPO mainly change implementation, reference-model dependence, or noisy-label bias rather than the underlying welfare question \citep{hong2024orpo,meng2024simpo,chowdhury2024rdpo}. Self-rewarding loops move some feedback generation from humans to model judges, which lowers data-collection costs but makes the evaluator itself part of the mechanism \citep{yuan2024self}.

\begin{figure}[t]
\centering
\includegraphics[width=\textwidth]{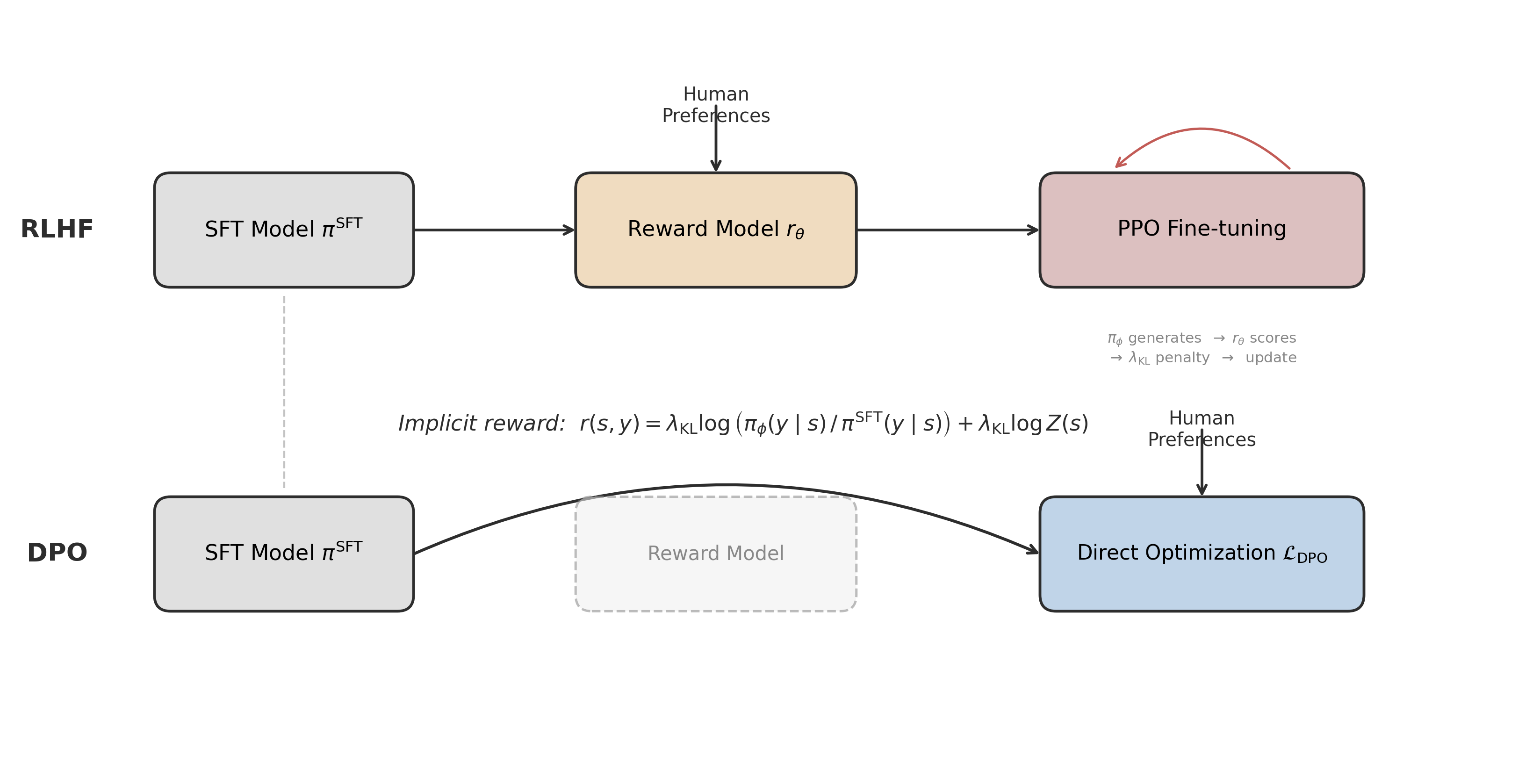}
\caption[RLHF versus DPO pipelines]{RLHF versus DPO pipelines. Top row: the three-stage RLHF pipeline trains a reward model from human preferences, then uses PPO to fine-tune the policy with a KL penalty. Bottom row: DPO collapses the pipeline into a single supervised learning objective over preference pairs, eliminating the explicit reward model (ghosted box).}
\label{fig:rlhf_dpo_pipeline}
\end{figure}

\subsection{Social Choice and Whose Preferences}
\label{sec:social_choice}

RLHF turns alignment into an economics problem: elicit preferences, aggregate them across people, design incentives for truthful feedback, and interpret the learned object. The phrase ``alignment tax'' appears in the assistant-alignment literature as a warning that optimizing for helpfulness, harmlessness, and honesty can trade off against other capabilities \citep{askell2021general}; \citet{ouyang2022training} provide concrete InstructGPT evidence of benchmark regressions on tasks such as SQuAD, DROP, and HellaSwag. That evidence is an empirical frontier of a particular training recipe, model class, and evaluation set, not a fundamental constraint.

The central welfare question is whose preferences are being aligned to. If one reward model is trained on pooled labels from a non-representative group of contractors, the resulting policy optimizes an implicit aggregation rule, not society's preferences. Social choice provides the concepts for this problem: alternatives, voters, ranking rules, impossibility theorems, strategic reports, minority protection, and interpersonal aggregation \citep{conitzer2024socialchoice,mishra2023socialchoice}. Distributional preference learning and axiomatic analyses show that standard BTL aggregation can encode Borda-like rules or fail desirable fairness and consistency conditions \citep{siththaranjan2024distributional,ge2024axioms}. Alternatives such as personalization, MaxMin-RLHF, Nash learning from human feedback, and maximal-lottery alignment make the aggregation rule more explicit, though some of these connections remain preprint-level rather than settled practice \citep{park2024heterogeneous,chakraborty2024maxmin,munos2023nash,maurarivero2025maximal}.

\subsection{Mechanism Design for Feedback}

Feedback is elicited through a mechanism, not passively observed. Labelers may be inattentive, strategic, ideologically motivated, or selected by the platform's contracting process. Mechanism design therefore enters before the reward model is trained. VCG-style fine-tuning formulations ask how to aggregate multiple reward models when participants can benefit from manipulating the training rule \citep{sun2024mechanism}. Strategyproof RLHF studies the same issue for preference reports, with a tradeoff between incentive compatibility and the quality of the final policy \citep{kleinebuening2025strategyproof}. Older peer-prediction mechanisms such as Bayesian truth serum and output-agreement methods show how payments can elicit truthful subjective signals when no ground truth is immediately observable \citep{prelec2004bayesian,miller2005eliciting}. For RLHF, this means the data-generating process is not a nuisance detail; it determines which preferences are observed and how costly truthful feedback is.

\subsection{Structural Identification}

Finally, preference optimization is not welfare analysis unless the learned reward has a structural interpretation. Bradley-Terry and DPO identify reward differences only through observed pairwise comparisons, up to normalizations and only on the covered comparison support. IRL identification results show that many rewards can rationalize the same behavior or preferences, and that entropy regularization, extra environments, or downstream invariance restrictions are needed to recover sharper objects \citep{cao2021identifiability,skalse2023invariance,knox2022preference}. Economically, the genealogy runs from McFadden logit choice \citep{mcfadden:1973}, through Rust-style dynamic discrete choice \citep{rust1994structural,rust1996numerical}, to maximum-entropy IRL \citep{Ziebart2008}, and finally to DPO's log-odds reparameterization \citep{rafailov2023direct}. Recent econometric work makes this bridge explicit, but still as an emerging working-paper literature rather than a settled textbook synthesis \citep{vanderlaan2025efficient,Kang2025,RustRawat2026}. The practical lesson is conservative: RLHF can produce useful policies, but the reward object is partially identified, aggregation-dependent, and limited by the offline support of the comparison data.

\FloatBarrier
\subsection{Engine Replacement MDP: What Preferences Do and Do Not Identify}
\label{engine:ch09}

\begin{table}[H]
\centering
\small
\caption{DPO reward inversion on the Engine Replacement MDP at $\lambda_{KL}=0.5$. The four rows report the exponential tilt and the normalized policy. The final column is the reward representative selected by the DPO normalization.}
\label{tab:engine_preferences}
\begin{tabular}{llrrrrrr}
\hline
state & action & $r$ & $\pi^{SFT}$ & $e^{r/\lambda_{KL}}$ & $\pi^{SFT}e^{r/\lambda_{KL}}$ & $\pi^*$ & $\lambda_{KL}\log(\pi^*/\pi^{SFT})$ \\
\hline
low & keep & 1.0000 & 0.6000 & 7.3891 & 4.4334 & 0.9679 & 0.2391 \\
low & replace & -0.5000 & 0.4000 & 0.3679 & 0.1472 & 0.0321 & -1.2609 \\
high & keep & 0.2000 & 0.4000 & 1.4918 & 0.5967 & 0.7300 & 0.3008 \\
high & replace & -0.5000 & 0.6000 & 0.3679 & 0.2207 & 0.2700 & -0.3992 \\
\hline
state & \multicolumn{2}{l}{$Z(s)$} & \multicolumn{2}{l}{$\lambda_{KL}\log Z(s)$} & \multicolumn{3}{l}{inverse reward minus $r(s,a)$} \\
\hline
low & \multicolumn{2}{l}{4.5806} & \multicolumn{2}{l}{0.7609} & \multicolumn{3}{l}{-0.7609} \\
high & \multicolumn{2}{l}{0.8175} & \multicolumn{2}{l}{-0.1008} & \multicolumn{3}{l}{0.1008} \\
\hline
\end{tabular}
\end{table}

The Engine calculation treats the mileage grade as the DPO context and the keep or replace action as the compared output. The calculation isolates the statewise reward map rather than solving the dynamic control problem. For the full-support reference policy in Table~\ref{tab:engine_preferences}, the four exponential tilts normalize to the state-specific constants $Z(\mathrm{low})=4.5806$ and $Z(\mathrm{high})=0.8175$. The normalized terms give $\pi^*$ directly. Inverting the policy ratio gives
\[
\widetilde r(s,a)=\lambda_{KL}\log
\frac{\pi^*(a\mid s)}{\pi^{SFT}(a\mid s)},
\]
the representative selected by the DPO normalization in Equation~\eqref{eq:dpo_reparameterize}.

The inverse rewards differ from the original rewards by $-\lambda_{KL}\log Z(s)$, which equals $-0.7609$ in the low-mileage state and $0.1008$ in the high-mileage state. Each shift is constant across actions within its state, so both action-reward differences and the Bradley-Terry preference probabilities remain unchanged. The four preferences therefore recover the reward differences but do not recover either state-specific level \citep{rafailov2023direct}.
\FloatBarrier

\subsection{Constitutional AI, RLAIF, and Scalable Oversight}

Three departures from the canonical RLHF pipeline change the source of preference labels rather than the optimization objective. Constitutional AI \citep{bai2022constitutional} substitutes a short list of written principles for human harmlessness annotations, training the model to critique and revise its own outputs against the constitution and then distilling the resulting AI-generated comparisons into a preference model. RLAIF \citep{lee2023rlaif} pushes the substitution one step further, using an off-the-shelf language model to label preference pairs directly, with reported parity against human-labelled RLHF on summarization and dialogue tasks and a direct variant that scores responses online during reinforcement learning. Scalable oversight \citep{bowman2022oversight} operates one level up, asking whether human-model teams can supervise problems on which unaided humans underperform, with sandwiching experiments that bracket model capability between layperson and expert performance. Each method trades one cost for another, replacing large human-labelling budgets with the developer's specification of principles, the labeler-model's training distribution, or the design of human-AI interaction protocols. The substitution scrambles the identification question raised in the preceding subsections, since the encoded preferences no longer trace back to a labelled human-population sample but to whichever artifact is doing the labelling. Methodologically this sits in the same family as mechanism design with information acquisition, where the choice of signal technology changes which preferences are even observable.

\subsection{Revealed Preference Tests and Axiom-Aware Aggregation}
\label{sec:revealed_pref_axioms}

\subsubsection{Revealed-preference tests of LLMs}

A related literature treats language models themselves as economic agents and asks whether their choices satisfy revealed-preference restrictions. Budget-allocation experiments, consumer-preference replications, moral-choice tasks, and Bayesian decision problems find that LLMs can look rational in some narrow domains and fragile in others \citep{chen2023rationalitygpt,golisingh2024preferences,seror2024moral,murawatrawat2025bayesian}. These tests are useful because they translate vague alignment claims into falsifiable restrictions, such as GARP consistency or proximity to a Bayes rule. But the results are prompt-, model-, version-, and domain-sensitive. They should be used as diagnostics of a trained system, not as evidence that an LLM has a stable utility function in the welfare-economics sense.

The revealed-preference axioms invoked in these tests carry a long economics lineage. \citet{samuelson1938consumption} introduced the weak axiom for choice over budget sets, \citet{houthakker1950revealed} strengthened it to the strong axiom by closing transitivity over chains, \citet{afriat1967construction} proved that finite choice data admit a concave utility rationalization if and only if they satisfy GARP, and \citet{varian1982nonparametric} laid out the nonparametric testing procedures that empirical revealed-preference work still uses. Transporting this machinery to language models raises a translation question, since the classical theory presumes a single agent choosing across explicit budget sets while LLM tests treat prompt context as a budget analogue and generated token sequences as the chosen bundle. Standard RLHF further imposes Bradley-Terry pairwise consistency on the reward model, an assumption strictly stronger than GARP because it forces stochastic transitivity and a parametric latent reward \citep{raheja2026unification}. A recent wave of methods relaxes this assumption along different axes. \citet{liu2024comal} and \citet{zhang2025omd} cast alignment as a two-player constant-sum game over policies and converge to a Nash policy that drops transitivity, accommodating cyclic and population-aggregated preferences. \citet{tang2024gpo} unifies DPO, IPO, and SLiC inside a convex family of offline losses and shows that the effective regularization differs from a global KL anchor, which reframes the question of which preference axioms each loss actually encodes. \citet{golz2025distortion} imports Procaccia-style distortion bounds from social choice and proves that fitting a single Bradley-Terry reward model to heterogeneous user preferences incurs unavoidable welfare loss, while Nash Learning attains the minimax-optimal distortion bound. The empirical gap is that no current study runs GARP or SARP diagnostics directly on policies produced by these relaxed objectives, so the revealed-preference question for post-alignment language models is reopened rather than closed.

\subsubsection{Linear social choice and the failure of Bradley-Terry MLE}
\label{sec:linear_social_choice}

\citet{ge2024axioms} cast the reward-modeling stage of RLHF as a social-choice problem in which a small set of candidates carry feature vectors and reward functions are linear in those features. Let $C = \{c_1, \ldots, c_m\}$ be the candidate set, $x_c \in \mathbb{R}^d$ the feature vector of candidate $c$, and a parameter $\theta \in \mathbb{R}^d$ define the reward $r_\theta(c) = \langle \theta, x_c \rangle$. Each voter $i$ has a private parameter $\theta_i$ whose induced ranking serves as their preference report; the aggregator observes pairwise comparisons sampled from these rankings and outputs a single $\theta^*$. Two axioms from voting theory carry over.

\begin{definition}[Pareto optimality]
\label{def:po_axiom}
A linear rank aggregation rule satisfies \emph{Pareto optimality} (PO) if, whenever every voter ranks $a$ above $b$ in the input profile, the output ranking also places $a$ above $b$.
\end{definition}

\begin{definition}[Pairwise majority consistency]
\label{def:pmc_axiom}
A ranking $\sigma$ is the \emph{pairwise majority consistent} ranking for a profile if for every pair $(a, b)$, $\sigma$ ranks $a$ above $b$ if and only if a majority of voters does. A rule satisfies PMC if it outputs this ranking whenever one exists and is feasible.
\end{definition}

\begin{theorem}[Theorem 3.1 of \citet{ge2024axioms}]
\label{thm:bt_mle_failure}
Any loss-based linear rank aggregation rule whose loss function is weakly convex and nondecreasing, or strictly convex, with $\inf_x \ell(x) < \ell(0)$, fails both Pareto optimality and pairwise majority consistency. The standard Bradley-Terry maximum-likelihood reward model is one such rule.
\end{theorem}

\begin{proof}
The construction plants two Pareto-unanimous candidates whose only difference is a vanishing perturbation, then shows the loss-minimizing reward ranks them the wrong way for every loss in the stated class. Here $r_\theta(c) = \langle \theta, x_c \rangle$ is the linear reward and $\ell$ the per-comparison loss, so the rule minimizes $\mathcal{L}(\theta) = \sum_{x \neq y} w_{x \succ y}\, \ell\big(r_\theta(y) - r_\theta(x)\big)$ over $\theta$.\footnote{Bradley-Terry maximum likelihood is the instance $\ell(t) = \log(1 + e^{t})$, the logistic loss. It is strictly convex and nondecreasing with $\ell(0) = \log 2$ and $\inf_t \ell(t) = 0 < \ell(0)$, so it satisfies the hypothesis.} Take six candidates in two groups of three. The core group sits at $x_a = (2,1)$, $x_b = (1,1)$, $x_c = (0,0)$; the copy group sits an $\varepsilon$ away at $x_{a'} = x_a + (-\varepsilon, 0)$, $x_{b'} = x_b + (-\varepsilon, 0)$, $x_{c'} = x_c + (-\varepsilon, \delta\varepsilon)$. A $p$-fraction of voters rank $a \succ a' \succ b \succ b' \succ c' \succ c$ and the rest rank $c' \succ c \succ b' \succ b \succ a' \succ a$; both rankings are feasible for $0 < \varepsilon < 1$, and every voter ranks $c' \succ c$.

At $\varepsilon = 0$ the copies coincide with the originals, and the six-candidate loss collapses to the three-candidate one, $\mathcal{L}^{0}(\theta) = 4\,\mathcal{L}^{\mathrm{core}}(\theta) + 3\,\ell(0)$, an affine orientation-preserving transformation, so the two share their minimizers.\footnote{Each core comparison recurs across the four original-copy combinations, and the three within-pair comparisons $(x, x')$ each contribute the constant $\ell(0)$. The core loss inherits convexity from $\ell$, and its minimizers separate the rewards. Writing the reduced objective in the single variable $r_a$ and using the left and right derivatives of the convex $\ell$, every optimum has $r_a$ large and $r_b$ small, which the invertible feature map on $x_a, x_b$ turns into $\theta_1 > A_3$ and $\theta_2 < A_4$ for constants $A_3 > 0$ and $A_4$ fixed by $p$ \citep[Lemmas~3.2--3.3]{ge2024axioms}.} Choose $\delta > 0$ with $\delta A_4 < A_3$. Then at any core optimum $\theta$ the perturbed candidate's reward is $r_\theta(c') = \varepsilon(-\theta_1 + \delta\theta_2) < \varepsilon(-A_3 + \delta A_4) < 0 = r_\theta(c)$, so the $\varepsilon = 0$ optimizers all rank $c$ strictly above $c'$.

The conclusion continues to hold under the perturbation. The loss $\mathcal{L}^{\varepsilon}$ is jointly continuous in $\theta$ and $\varepsilon$, and minimization can be confined to a fixed compact set of $\theta$ because $\ell(x) \to \infty$ as $x \to \infty$; Berge's maximum theorem then makes the optimizer set upper semicontinuous in $\varepsilon$, so for all sufficiently small $\varepsilon > 0$ the minimizer still ranks $c$ above $c'$ \citep[Lemmas~3.4--3.5]{ge2024axioms}. The rule therefore outputs $c \succ c'$. Because every voter ranks $c' \succ c$, this violates Pareto optimality; because the majority ranking $a \succ a' \succ b \succ b' \succ c' \succ c$ is feasible yet not returned, it also violates pairwise majority consistency.\footnote{The remaining case, a loss with $\ell(x) < \ell(0)$ for some $x > 0$, is settled by a two-candidate instance $x_a = (1,0)$, $x_b = (0,1)$ with a single voter ranking $a \succ b$: convexity forces the minimizer to output $b \succ a$, again violating both axioms \citep{ge2024axioms}.}
\end{proof}

\subsubsection{Leximax Copeland subject to Pareto optimality}
\label{sec:lcpo}

To repair the axiomatic failure, \citet{ge2024axioms} propose a rule that adapts the classical Copeland voting rule to the linear-feature setting. For a profile $\pi$ and pair $(a, b)$, let $w_{a \succ b}(\pi)$ denote the fraction of voters who rank $a$ above $b$. The Copeland score of $a$ is
\[
C(a) = |\{b \in \mathcal{C} : w_{a \succ b}(\pi) > 1/2 \}|.
\]
The leximax Copeland rule ranks candidates by descending Copeland score, breaking ties by lexicographic comparison of the sorted margin vectors. Because not every ranking is feasible in the linear setting, the rule is implemented sequentially. For position $r + 1$ it picks the unranked candidate with the highest Copeland score for which there exists a parameter $\theta$ consistent with the partial ranking so far. The variant \emph{leximax Copeland subject to PO} (LCPO) additionally enforces Pareto-dominance constraints, refusing to rank a Pareto-dominated candidate above its dominator. The whole rule reduces to a sequence of $\mathcal{O}(m^2)$ linear feasibility programs.

\begin{theorem}[Theorem 4.3 of \citet{ge2024axioms}]
\label{thm:lcpo}
LCPO satisfies Pareto optimality, pairwise majority consistency, majority consistency, and winner monotonicity.
\end{theorem}

\subsubsection{Simulation Study: Axiom-Aware Aggregation on Heterogeneous Voters}
\label{sec:sim_axiom_aware}

We instantiate the construction above with $\varepsilon = 0.01$, $\delta = 2$, voter parameters $(1, 1)$ and $(-1, 0)$, population fractions $p = 0.6$ and $1 - p$, and Bradley-Terry label noise at scale $\beta = 50$. The choice $\delta = 2$ reflects the concrete voters. The $(1,1)$ voter's reward difference on the pair $(c', c)$ is $(\delta - 1)\varepsilon$, so the Pareto unanimity $c' \succ c$ that the construction plants requires $\delta > 1$. Two aggregators are compared. The BT-MLE aggregator fits a linear-in-features parameter $\theta$ by maximum likelihood on the pooled pairwise sample. The LCPO aggregator estimates pairwise majorities, computes Copeland scores, and applies the sequential feasibility procedure described above. We sweep $N \in \{5, 10, 20, 50, 100, 500, 2000\}$ comparisons per pair across thirty seeds, scoring outputs against the Pareto-dominance relation $\{(c', c)\}$ and the unique PMC ranking $a \succ a' \succ b \succ b' \succ c' \succ c$.

\begin{figure}[htbp]
\centering
\includegraphics[width=\textwidth]{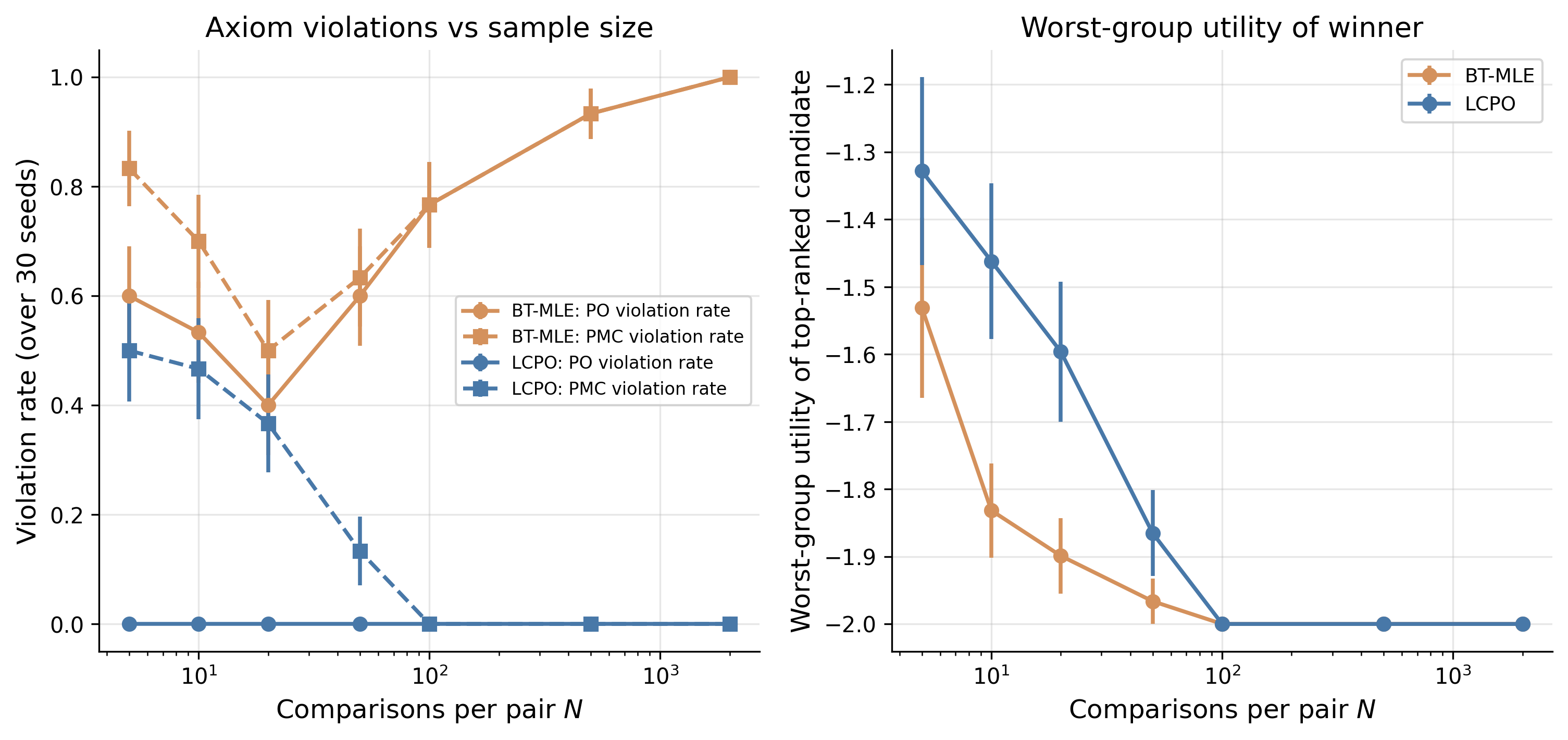}
\caption[Axiom violation rates and worst-group utility versus sample size]{Axiom violation rates (left) and worst-group utility of the top-ranked candidate (right) versus pairwise comparisons per pair $N$, mean and one standard error over thirty seeds. BT-MLE PO and PMC violation rates rise to one as $N$ grows, in line with Theorem~\ref{thm:bt_mle_failure}; LCPO violation rates remain at zero (PO by construction) or fall to zero (PMC by $N = 100$).}
\label{fig:axiom_aware_aggregation}
\end{figure}

\begin{table}[htbp]
\centering
\caption[Aggregator comparison at $N=2000$ comparisons per pair]{Aggregator comparison at $N = 2000$ comparisons per pair, mean and standard error over thirty seeds. PO viol. and PMC viol. are fractions of seeds on which the output ranking violates the corresponding axiom. Worst-group util. is the type-two utility of the top-ranked candidate. Modal top names the candidate most frequently ranked first.}
\label{tab:axiom_aware_aggregation}
\begin{tabular}{lcccc}
\hline
Method & PO viol. & PMC viol. & Worst-group util. & Modal top \\
\hline
BT-MLE (linear) & 1.00 (0.00) & 1.00 (0.00) & -2.000 (0.000) & a \\
Leximax Copeland (LCPO) & 0.00 (0.00) & 0.00 (0.00) & -2.000 (0.000) & a \\
\hline
\end{tabular}

\end{table}

Figure~\ref{fig:axiom_aware_aggregation} and Table~\ref{tab:axiom_aware_aggregation} report the results. By $N = 2000$, BT-MLE violates both PO and PMC on every seed, as Theorem~\ref{thm:bt_mle_failure} predicts asymptotically. LCPO satisfies PO on every seed at every sample size because the rule directly enforces the Pareto-dominance constraints. LCPO also satisfies PMC on every seed by $N = 100$. The worst-group utility of the top-ranked candidate converges to $-2$ for both methods. The simulation therefore verifies the axiomatic difference rather than an efficiency gain for the top-ranked candidate.

\subsection{Simulation Study: Preference Learning in Job Search}

\begin{figure}[htbp]
\centering
\includegraphics[width=0.85\textwidth]{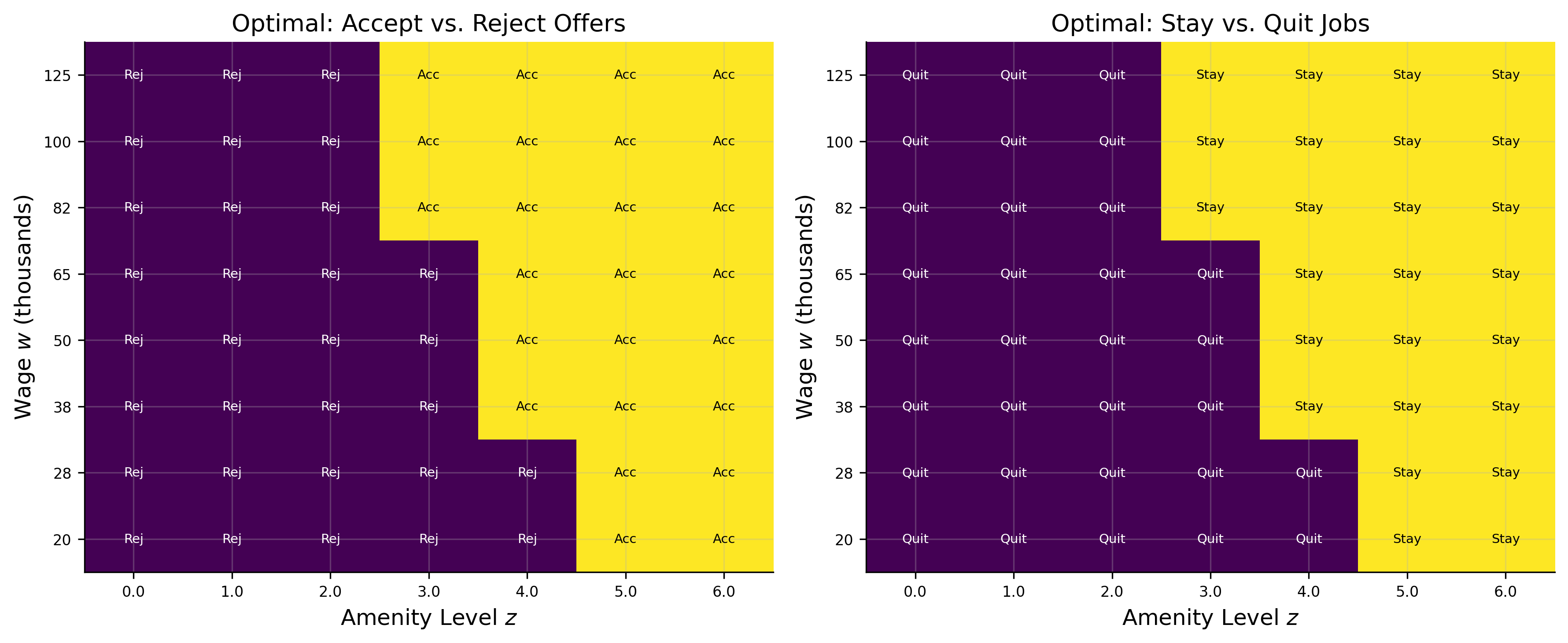}
\caption[Optimal accept/reject and stay/quit boundaries in wage-amenity space]{Optimal accept/reject and stay/quit boundaries in wage-amenity space. Left: the optimal accept/reject boundary for searching states. Right: the optimal stay/quit boundary for employed states. Both boundaries cut diagonally, reflecting the tradeoff between wage and amenity quality under compensating differentials.}
\label{fig:search_env}
\end{figure}

A worker searches for jobs in a labor market with compensating differentials, following a McCall (1970)-style search model. Wages are measured in thousands and take values $w \in \{20,28,38,50,65,82,100,125\}$, while amenities take values $z \in \{0, 1, \ldots, 6\}$ and capture commute quality, flexibility, and job security. The state space has 112 states: 56 searching states in which the worker observes a pending offer $(w_i, z_j)$ and decides to accept or reject, plus 56 employed states $(w_i, z_j)$ in which the worker decides to stay or quit. The offer distribution exhibits compensating differentials, with wage rank and amenity rank negatively correlated ($\rho = -0.74$): high-wage offers cluster with low amenities and vice versa. The worker's true per-period utility is $u(w, z) = \alpha \log(w) + (1 - \alpha) z$ with $\alpha = 0.6$, but this function is unobserved; the worker can only compare career trajectories (``I prefer path A to path B''), exactly as in stated-preference surveys in labor economics. While searching, the worker receives the unemployment benefit $u_b = \alpha \log(b)$ where $b = 28$. Layoffs occur with probability $p = 0.05$ per period, and the discount factor is $\gamma = 0.95$. Dynamic programming gives $V^*(s_0) = 74.13$; the optimal policy accepts 25 of 56 offer types and stays employed at 25 of 56 job types. Preference data is generated by rolling out a uniform random policy from a random searching state. Each rollout produces a career segment of $L = 15$ periods recording states and actions. For each of $K$ comparisons, two independent career segments are generated; the segment with higher cumulative discounted utility under the true (unobserved) utility function is labeled as preferred via the Bradley-Terry model. Figure~\ref{fig:search_env} shows the optimal accept/reject and stay/quit boundaries.

Six methods are compared across $K \in \{25, 50, 100, 200, 500, 1{,}000, 2{,}000, 5{,}000\}$, averaged over 30 seeds. The neural network RLHF reward model is a two-layer MLP trained by Bradley-Terry MLE on the $K$ segment pairs; per-transition rewards are discount-weighted and summed to obtain a segment score, and the resulting 112-state reward table is solved by value iteration.\footnote{The neural network has 4 inputs (normalized log-wage, normalized amenity, employment indicator, action), 32 hidden units per layer, and $\sim$1{,}200 parameters. The logistic loss from Equation~(\ref{eq:rlhf_loss}) is applied to discount-weighted segment reward sums. The network parameterises $r_\theta(s, a)$, but the table handed to value iteration averages over actions to give $\bar{r}(s) = \tfrac{1}{2}(r_\theta(s, 0) + r_\theta(s, 1))$; this is harmless here because the true per-period reward in this environment is action-independent (actions affect only the next-state transition), so the Bradley-Terry MLE has no incentive to fit action-dependent payoffs.} The correctly specified structural model parameterizes utility as $\hat{u} = \hat{\alpha} \log(w) + (1 - \hat{\alpha}) z$, estimating the single parameter $\hat{\alpha}$ via Bradley-Terry MLE, then solves the MDP. The misspecified model uses $\hat{u} = \hat{\alpha} \log(w) + (1 - \hat{\alpha}) \bar{z}$, where $\bar{z} = 3$ is the mean amenity level; this model ignores amenity variation across jobs, treating all amenities as identical. DPO trains a tabular softmax policy directly from trajectory comparisons, bypassing reward modeling entirely.\footnote{DPO uses 112 logit parameters $\phi_s$, one per state, trained via the DPO loss (Equation~\ref{eq:dpo_loss}) with Adam optimization over a sweep of $\lambda_{KL} \in \{0.01, 0.05, 0.1, 0.5, 1.0, 5.0\}$, selecting the $\lambda_{KL}$ that minimizes training loss. The reference policy is uniform: $\pi^{SFT}(a|s) = 0.5$. To match the LLM setup where both completions condition on the same prompt, DPO comparison pairs start from the same initial state.} Tabular Q-learning ($10{,}000$ episodes, $\varepsilon = 0.15$, learning rate $0.1$) and exact DP provide scalar-reward baselines. All four preference methods receive identical comparison data per seed; DPO receives a same-state variant generated from the same seed.

\begin{figure}[htbp]
\centering
\includegraphics[width=0.7\textwidth]{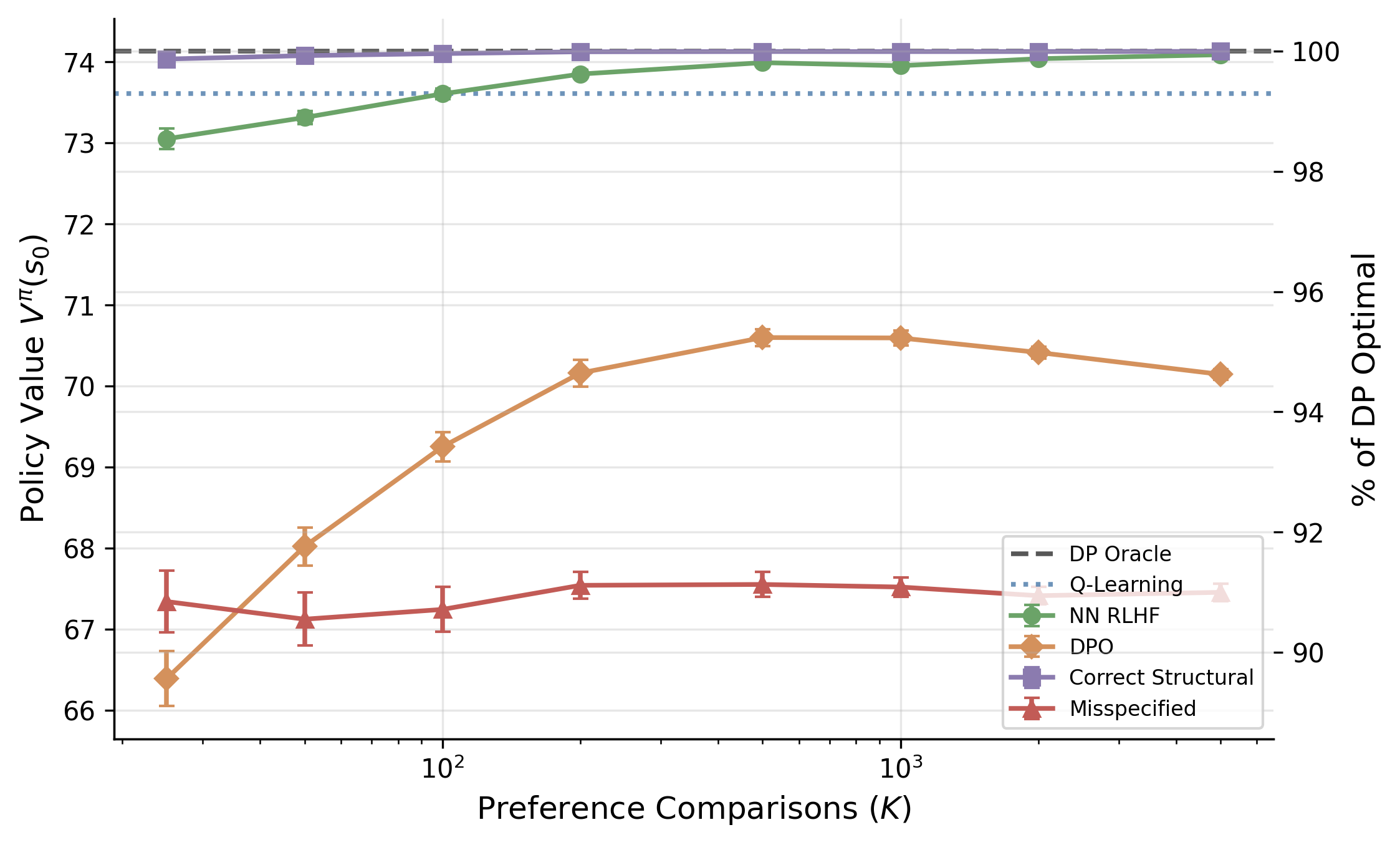}
\caption[Policy value versus number of preference comparisons $K$ for all six methods]{Policy value $V^\pi(s_0)$ versus number of preference comparisons $K$ for all six methods (30 seeds, $L = 15$). The right axis shows the percentage of DP-optimal value.}
\label{fig:preference_sample}
\end{figure}

Figure~\ref{fig:preference_sample} reports the main results. The correctly specified structural model reaches 99.9\% of the DP optimum at $K = 25$. Its one-parameter specification gives the structural model this sample-efficiency advantage. The neural network converges more slowly and reaches 99.9\% by $K = 5{,}000$, consistent with the larger sample requirement of a flexible model. DPO reaches approximately 95\% by $K = 500$ and then plateaus.\footnote{DPO learns only from $(s,a)$ pairs visited in training trajectories generated by the random behavioral policy. It cannot propagate value to undervisited states the way value iteration does after learning a reward model, so states poorly covered by the behavioral policy remain suboptimal regardless of $K$.} The DPO plateau differs from the gridworld result of $-118\%$. DPO recovers a viable job-search policy, but additional data do not close the remaining gap.\footnote{DPO fails catastrophically in gridworld because transitions are stochastic (10\% slip probability) and rewards are transition-dependent. The same $(s,a)$ pair yields different rewards depending on whether the agent slipped, so the DPO loss conflates policy quality with transition luck. In the job search model, accept/reject deterministically changes employment status, and only the 5\% layoff probability introduces stochastic transitions.} The misspecified constant-amenity model remains at 91\% as $K$ grows because additional preference data cannot recover the omitted amenity variation.

\begin{table}[htbp]
\centering
\caption[Diagnostics at $K = 5{,}000$: policy agreement with $\pi^*$, value-function correlation, and mean accepted wage and amenity for each method]{Diagnostics at $K = 5{,}000$ (single seed): policy agreement with $\pi^*$, value-function correlation, and mean accepted wage and amenity for each method.}
\label{tab:preference_diagnostics}
\begin{tabular}{lrrrrr}
\hline
Method & Policy agree.~(\%) & $V^\pi$ corr. & Mean amenity & Mean wage & $\hat{\alpha}$ \\
\hline
NN RLHF & $96.4$ & $1.000$ & $4.67$ & $73$ & --- \\
DPO & $57.1$ & $0.777$ & $3.38$ & $63$ & --- \\
Correct & $100.0$ & $1.000$ & $4.84$ & $71$ & $0.597$ \\
Misspecified & $50.0$ & $0.889$ & $3.00$ & $70$ & --- \\
\hline
Optimal & $100.0$ & $1.000$ & $4.84$ & $71$ & --- \\
\hline
\end{tabular}

\end{table}

Table~\ref{tab:preference_diagnostics} provides state-level diagnostics at $K = 5{,}000$. The structural model and neural network achieve near-perfect policy agreement with $\pi^*$ (100\% and 96\% respectively). DPO agrees on only 57\% of states with value-function correlation 0.78, systematically underselecting on both wage and amenity dimensions.\footnote{DPO's mean accepted amenity of 3.4 parallels the misspecified structural model's 3.0, though the mechanisms differ: the misspecified model ignores amenity variation by construction, while DPO underweights amenities because the random behavioral policy underrepresents high-amenity employed states in the training data.} The misspecified model agrees on 50\%, with disagreements concentrated where amenity variation matters.\footnote{An online versus offline ablation for the neural network at $K = 1{,}000$ (20 seeds) shows comparable performance: online $73.92 \pm 0.05$, offline $73.99 \pm 0.03$ ($p = 0.09$). The random behavioral policy already provides diverse career trajectories covering the full wage-amenity space.}

\begin{figure}[htbp]
\centering
\includegraphics[width=0.7\textwidth]{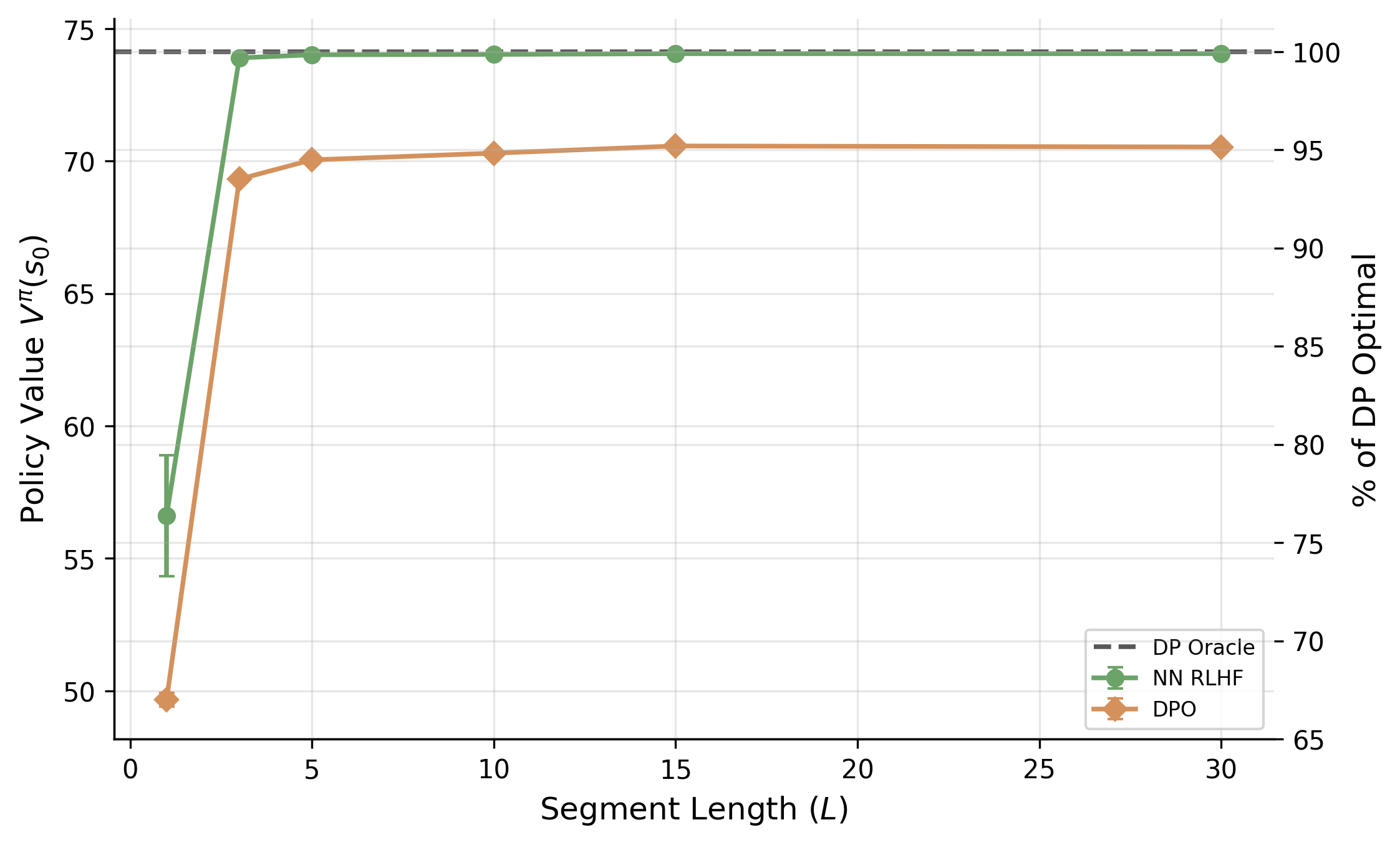}
\caption[Policy value versus segment length $L$ at $K = 2{,}000$ for the neural network and DPO]{Policy value $V^\pi(s_0)$ versus segment length $L$ at $K = 2{,}000$ for the neural network and DPO (20 seeds). The right axis shows the percentage of DP-optimal value.}
\label{fig:preference_horizon}
\end{figure}

Figure~\ref{fig:preference_horizon} reports the segment length ablation at $K = 2{,}000$. At $L = 1$, both methods perform poorly because single-step comparisons carry minimal information about long-run value. The neural network recovers rapidly (by $L = 3$) because the reward model aggregates per-transition estimates over longer segments and value iteration propagates them through the full transition structure; DPO improves monotonically but plateaus at its $\sim$95\% ceiling, as it must learn the policy directly from trajectory comparisons without access to the transition model.

Two-stage RLHF separates preference estimation (a static econometric problem) from dynamic programming (which exploits the known transition model); DPO conflates the two and forfeits the transition structure that structural modelers typically have access to.\footnote{This advantage is specific to settings where the transition model is known or estimable; in domains without a tractable transition model, DPO's single-stage approach avoids compounding errors from reward model estimation.} The DPO plateau in the job-search study is the computational version of the identification point above. More comparisons tighten the empirical binary choice model on the support of observed trajectories, but they do not reveal counterfactual value in poorly covered states, nor do they identify a welfare object independent of the aggregation rule used to construct the labels.

\section{Causal Inference for Reinforcement Learning}
\label{section:causal_rl}
%

Reinforcement learning algorithms solve Markov decision processes by estimating value functions or optimizing policies from sampled transitions. Two assumptions underlie the standard formulation. First, the agent's action at time $t$ is determined solely by the observed state $s_t$ and the agent's policy $\pi(a \mid s_t)$; there is no unobserved variable (confounder) simultaneously influencing both the action and the reward or state transition.\footnote{Throughout this chapter, I reserve ``outcome'' for its causal inference meaning and use the specific RL quantity (reward, state, return, value) elsewhere.}
Second, the observed state $s_t$ is sufficient for prediction, so that conditioning on $s_t$ renders future states independent of past history. Both assumptions are the Markov property restated in causal language. Both fail routinely in applied settings.

This chapter studies what happens when logged sequential decisions are observational rather than experimental. The central object is the value of a target policy under the interventional law induced by that policy, not merely under the observational law generated by past behavior. The literature now covers confounded MDPs, partial identification and sensitivity analysis, proximal and negative-control methods for POMDPs, instrumental variables, mediator-based identification, and dynamic double machine learning for off-policy evaluation (OPE). The survey articles by \citet{deng2023causal} and \citet{dacostacunha2025unifying} provide broader coverage of causal representation learning, counterfactual policy optimization, transportability, and fairness.

Table~\ref{tab:causal_rl_taxonomy} summarizes the main identification strategies. It groups papers to show economists that many causal RL questions are familiar econometric problems once the estimand is written as a policy value rather than as a static treatment effect.

\begin{table}[htbp]
\centering
\caption{Econometric taxonomy for causal reinforcement learning}
\label{tab:causal_rl_taxonomy}
\begin{tabularx}{\textwidth}{>{\raggedright\arraybackslash}p{0.24\textwidth}>{\raggedright\arraybackslash}p{0.31\textwidth}>{\raggedright\arraybackslash}X}
\toprule
RL problem & Econometric analogue & Identifying object \\
\midrule
Observed state blocks all confounding & Backdoor adjustment, sequential unconfoundedness, g-formula & Interventional transition and reward from conditional models given observed controls \\
Exogenous variation shifts actions & Instrumental variables and conditional moment restrictions & Structural transition or reward function satisfying moments conditional on instruments \\
Latent state is not observed but proxies are available & Proximal causal inference and negative controls & Bridge functions that recover policy value from proxy distributions \\
Actions operate through observed intermediate variables & Front-door, mediation, and g-methods & Mediated transition law and decomposed direct or indirect dynamic effects \\
No credible point-identifying source exists & Rosenbaum and Tan sensitivity, partial identification & Bounds over policy values under restricted hidden-bias models \\
High-dimensional nuisance functions are needed & DML, orthogonal scores, efficient OPE & Neyman-orthogonal moments for value or dynamic treatment parameters \\
\bottomrule
\end{tabularx}
\end{table}

\subsection{From Partial Observability to Causal Structure}
\label{subsec:pomdp_to_causal}

A partially observable MDP (POMDP) augments the standard MDP with an observation function. The POMDP is defined by the tuple $(\mathcal{S}, \mathcal{A}, \mathcal{O}, P, O, r, \gamma)$, where $\mathcal{O}$ is a finite observation space and $O$ is the observation function.
\begin{equation}
O(o \mid s', a) = P(O_t = o \mid S_t = s', A_{t-1} = a).
\label{eq:pomdp_obs}
\end{equation}
The agent does not observe $s_t$ directly but instead receives $o_t \sim O(\cdot \mid s_t, a_{t-1})$ and maintains a belief state $b_t \in \Delta(\mathcal{S})$
\begin{equation}
b_t(s) = P(S_t = s \mid o_1, a_1, \ldots, o_t),
\label{eq:belief_state}
\end{equation}
which is updated via Bayesian filtering at each step. The belief MDP, whose state space is $\Delta(\mathcal{S})$, is itself a fully observable continuous-state MDP, so standard value iteration applies in principle, though computation is intractable in general.

\citet{dacostacunha2025unifying} organize sequential decision problems with hidden variables into a hierarchy: standard MDP (full observability, no confounding), POMDP (partial observability, no confounding), confounded MDP (full observability, confounding), and causal POMDP (both). The key distinction is epistemic versus identificational. In a POMDP, the hidden state is a modeling challenge, since the agent acknowledges incomplete information and plans accordingly via the belief state, analogous to Kalman or Hamilton filtering in econometrics. In a confounded MDP, the hidden variable is an identification challenge, since standard estimators silently produce biased results, analogous to endogeneity and omitted variable bias.

This distinction matters because the same observed sequence can support different econometric interpretations. A hidden patient severity state may be a state variable the physician partially observes, a confounder that drives treatment and prognosis, a source of invalid support for a target policy, or a latent variable for which laboratory histories provide negative controls. Causal RL is the study of which interpretation is credible enough to identify, bound, or estimate the value of a counterfactual policy.

\subsection{The Confounded MDP}
\label{subsec:confounded_mdp}

When unobserved confounders influence both the behavioral policy and the transitions or rewards, the MDP is confounded. This formalization, developed by \citet{zhang2019near}, \citet{zhang2020causal}, and \citet{kallus2020confounding}, provides the foundation for causal reasoning in sequential decision problems. The key tool is Pearl's do-operator. The interventional distribution $P(Y \mid \operatorname{do}(X = x))$ is the distribution that arises when $X$ is set externally rather than observed passively, severing all incoming causal influences on $X$ while leaving the remaining data-generating process intact.\footnote{The do-operator is formalized within the structural causal model (SCM) framework of \citet{pearl2009causality}. An SCM specifies endogenous variables $\mathbf{V}$, exogenous variables $\mathbf{U}$, structural equations $V_i = f_i(\text{pa}(V_i), U_i)$, and a distribution $P(\mathbf{U})$. The intervention $\operatorname{do}(X = x)$ replaces the structural equation for $X$ with a constant, producing the interventional distribution. See \citet{pearl2009causality} for the complete framework, including the causal hierarchy (association, intervention, counterfactual) and general identification theory.}

\begin{definition}[Confounded MDP {\citep{zhang2020causal}}]
\label{def:confounded_mdp}
A confounded MDP is a tuple $(\mathcal{S}, \mathcal{A}, \mathcal{U}, P, r, \gamma)$ where $\mathcal{S}$ is a finite state space, $\mathcal{A}$ is a finite action space, $\mathcal{U}$ is a space of unobserved confounders, $\gamma \in [0,1)$ is a discount factor, and the dynamics are governed by structural equations
\begin{align}
U_t &\sim P_U(\cdot \mid S_t), \label{eq:cmdp_confounder} \\
A_t &\sim \mu(\cdot \mid S_t, U_t), \label{eq:cmdp_action} \\
R_t &= f_R(S_t, A_t, U_t) + \epsilon_t, \label{eq:cmdp_reward} \\
S_{t+1} &\sim f_S(\cdot \mid S_t, A_t, U_t). \label{eq:cmdp_transition}
\end{align}
The behavioral (\textit{logging}) policy $\mu$ depends on the unobserved confounder $U_t$ through Equation~\eqref{eq:cmdp_action}. An evaluation policy $\pi(a \mid s)$ depends only on the observed state.
\end{definition}

Unlike the online algorithms discussed earlier in the survey, where behavior and target policies were identical or related by a known exploration mechanism, here $\mu$ is an unknown function of unobserved variables.

Because $\mu$ depends on $U_t$, conditioning on $\{A_t = a\}$ carries information about the confounder, so the observational and interventional transitions diverge:
\begin{equation}
P(s' \mid s, a) \neq P(s' \mid s, \operatorname{do}(a)).
\label{eq:obs_vs_interventional}
\end{equation}

The Bellman equation for policy evaluation must use interventional, not observational, transition probabilities. Define the causal Bellman operator for a \textit{target policy} $\pi$:

\begin{definition}[Causal Bellman Operator {\citep{zhang2020causal}}]
\label{def:causal_bellman}
The causal Bellman operator $\mathcal{T}_c^\pi$ for policy $\pi$ in a confounded MDP is
\begin{equation}
(\mathcal{T}_c^\pi V)(s) = \sum_{a \in \mathcal{A}} \pi(a \mid s) \sum_{s' \in \mathcal{S}} P(s' \mid s, \operatorname{do}(a)) \bigl[ r(s, a) + \gamma V(s') \bigr],
\label{eq:causal_bellman}
\end{equation}
where $r(s,a) = \mathbb{E}[R_t \mid S_t = s, \operatorname{do}(A_t = a)]$ is the interventional expected reward.
\end{definition}

\begin{lemma}[Bias of Naive Off-Policy Evaluation {\citep{kallus2020confounding}}]
\label{lem:naive_bias}
Let $\hat{V}^{\pi}_{\text{naive}}$ denote the value function obtained by solving the Bellman equation with observational transitions $P(s' \mid s, a)$, and let $V^\pi$ denote the true value function under interventional transitions $P(s' \mid s, \operatorname{do}(a))$. In a confounded MDP where $P(s' \mid s, a) \neq P(s' \mid s, \operatorname{do}(a))$ for some $(s, a, s')$, the naive estimator is biased.
\begin{equation}
\hat{V}^{\pi}_{\text{naive}}(s) \neq V^\pi(s).
\end{equation}
The importance-sampling estimator is also biased because the propensity $\mu(a \mid s)$ is not the true behavioral propensity $\mu(a \mid s, u)$:
\begin{align}
\hat{V}^{\pi}_{\text{IS}} &= \frac{1}{N} \sum_{i=1}^N
\prod_{t=0}^{T} \frac{\pi(a_t^{(i)} \mid s_t^{(i)})}{\mu(a_t^{(i)} \mid s_t^{(i)})} G^{(i)}, \\
G^{(i)} &= \sum_{t=0}^{T} \gamma^t R_t^{(i)}.
\end{align}
Here $G^{(i)}$ is the discounted return of trajectory $i$ and $T$ is the trajectory length.\footnote{This is the sequential analogue of the omitted variable bias in linear regression. In the static case, regressing $Y$ on $X$ without controlling for a confounder $U$ yields a biased coefficient. In the sequential case, the bias propagates through the Bellman recursion and can amplify over the horizon.}
\end{lemma}

The backdoor criterion \citep{pearl2009causality} provides a path to identification. \citet{zhang2020causal} apply it to the confounded MDP setting.

\begin{theorem}[Backdoor Identification in Confounded MDPs {\citep{pearl2009causality,zhang2020causal}}]
\label{thm:backdoor_mdp}
Suppose a set of observed variables $\mathbf{Z}_t$ satisfies the backdoor criterion relative to $(A_t, S_{t+1})$ in the causal graph of the confounded MDP, meaning that $\mathbf{Z}_t$ blocks all backdoor paths from $A_t$ to $S_{t+1}$ and no element of $\mathbf{Z}_t$ is a descendant of $A_t$.\footnote{A backdoor path from $A_t$ to $S_{t+1}$ is any path in the causal graph that begins with an arrow into $A_t$, that is, a non-causal path. In the confounded MDP, $A_t \leftarrow U_t \rightarrow S_{t+1}$ is a backdoor path: $U_t$ causes both $A_t$ and $S_{t+1}$, creating a spurious association. Blocking all such paths by conditioning on appropriate variables eliminates the confounding bias. See \citet{pearl2009causality}, Chapter~3.} Then the interventional transition probability is identified:
\begin{equation}
P(s' \mid s, \operatorname{do}(a)) = \sum_{\mathbf{z}} P(s' \mid s, a, \mathbf{z}) \, P(\mathbf{z} \mid s).
\label{eq:backdoor_mdp}
\end{equation}
Substituting Equation~\eqref{eq:backdoor_mdp} into the causal Bellman operator (Equation~\ref{eq:causal_bellman}) yields an identified, consistent estimator of $V^\pi$.
\end{theorem}

\subsection{Backdoor-Adjusted Off-Policy Evaluation}
\label{subsec:backdoor_ope}

Off-policy evaluation under confounding is an average treatment effect estimation problem in a dynamic setting \citep{bannon2020causality}: $\mu$ is the treatment assignment mechanism, $\pi$ the counterfactual regime, importance sampling corresponds to inverse probability weighting, and doubly robust OPE corresponds to the AIPW estimator of \citet{robins1994estimation}.

Theorem~\ref{thm:backdoor_mdp} yields a concrete estimation procedure. Given logged data $D_N$ collected under behavioral policy $\mu$, with elements $(s_t, a_t, z_t, r_t, s_{t+1})$ and observed proxy $z_t$, the econometrician estimates $\hat{P}(s' \mid s, a, z)$ and $\hat{P}(z \mid s)$ from the logged data, computes
\begin{equation}
\hat{P}(s' \mid s, \operatorname{do}(a)) = \sum_{z} \hat{P}(s' \mid s, a, z) \, \hat{P}(z \mid s),
\label{eq:backdoor_estimator}
\end{equation}
and solves the causal Bellman equation using the estimated interventional transitions to obtain $\hat{V}^\pi$.

This is the cleanest case because the analyst observes enough variables to repair the state. The rest of the chapter turns to what can still be learned when that repair is unavailable, only partially credible, or statistically high-dimensional.

\FloatBarrier
\subsection{Engine Replacement MDP: Bias From an Unobserved Grade}
\label{engine:ch10}

\begin{table}[H]
\centering
\caption{Target-policy evaluation on the $+U$ Engine Replacement MDP. Monte Carlo entries are means and standard errors over twenty fixed seeds.}
\label{tab:engine_confounding}
\begin{tabular}{lrrr}
\hline
transition estimate & $P(\mathrm{degrade})$ & $V(\mathrm{low})$ & bias \\
\hline
interventional truth & 0.5000 & 5.3448 & 0.0000 \\
naive population formula & 0.6000 & 4.7403 & -0.6046 \\
naive simulation & 0.5992 & 4.7451 & -0.5998 $\,(0.0044)$ \\
backdoor simulation & 0.4983 & 5.3560 & 0.0112 $\,(0.0063)$ \\
\hline
\end{tabular}
\end{table}

The $+U$ variant draws a binary grade independently at each period. The grade changes the low-mileage degradation probability from $0.3$ to $0.7$ and changes the behavior policy's keep probability from $0.25$ to $0.75$.\footnote{The latent grades have equal probability. Each simulated cell uses $2{,}000$ trajectories of forty periods, with twenty fixed seeds.} Marginalizing before conditioning gives the interventional probability $P(\mathrm{high}\mid\mathrm{low},\operatorname{do}(\mathrm{keep}))=0.5$. Conditioning on the logged keep action changes the latent mixture and gives the observational probability $P(\mathrm{high}\mid\mathrm{low},\mathrm{keep})=0.6$.

Under the target policy that keeps at low mileage and replaces at high mileage, the low-state value at degradation probability $p$ is $V(p)=(1-0.45p)/(0.1+0.09p)$. The naive population bias is therefore $V(0.6)-V(0.5)=-0.6046$. The simulated bias estimate is $-0.5998$ with Monte Carlo standard error $0.0044$, a difference of $0.0048$ from the formula. When the grade is recorded, Equation~\eqref{eq:backdoor_estimator} uses the empirical grade share among low-mileage observations to average the two conditional transitions. The remaining bias is $0.0112$ with standard error $0.0063$. Both gaps are within two Monte Carlo standard errors.
\FloatBarrier

\subsection{Alternative Identification Strategies}
\label{subsec:alternative_identification}

\begin{figure}[htbp]
\centering
\includegraphics[width=\textwidth]{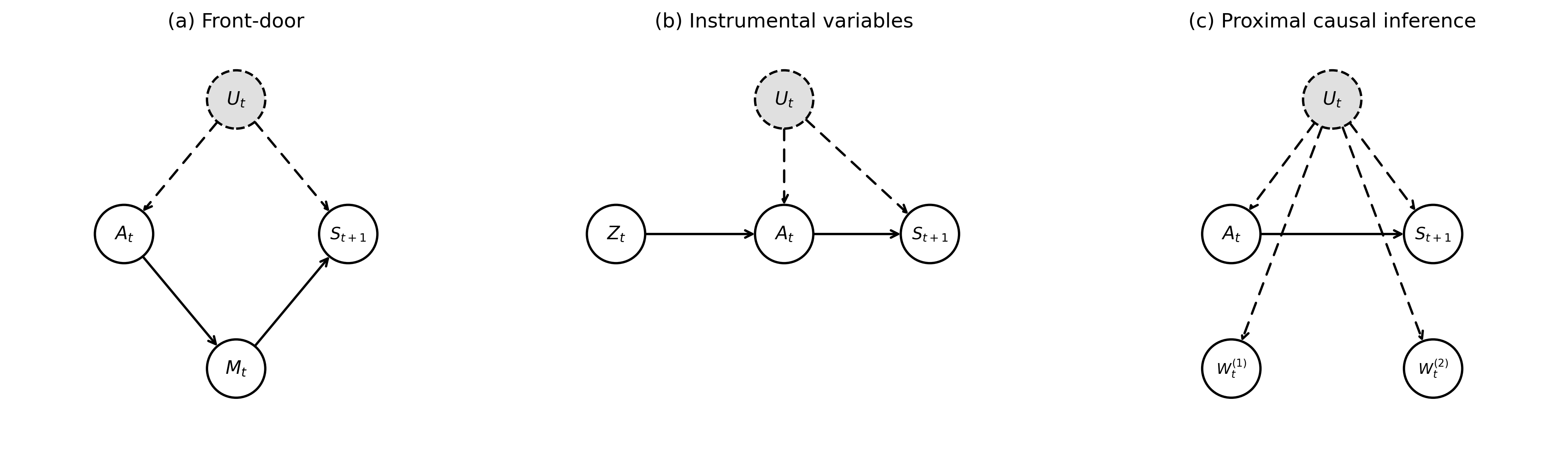}
\caption{Causal graphs for three identification strategies in confounded MDPs. Gray dashed nodes are unobserved; dashed edges involve unobserved variables. (a)~Front-door criterion with mediator $M_t$. (b)~Instrumental variables with exogenous instrument $Z_t$. (c)~Proximal causal inference with proxies $W_t^{(1)}, W_t^{(2)}$.}
\label{fig:identification_dags}
\end{figure}

When no backdoor variable is available, causal RL moves from adjustment to sensitivity analysis, proxies, instruments, mediators, or orthogonal scores. Figure~\ref{fig:identification_dags} displays the causal graph for three point-identification strategies used in the simulation study.

\subsubsection{Sensitivity Analysis and Partial Identification}
\label{subsubsec:sensitivity_partial}

When point identification is not credible, the estimand becomes a set rather than a number. In static econometrics, this is the logic of Rosenbaum bounds, marginal sensitivity models, and partial identification \citep{rosenbaum2002observational,tan2006distributional,manski2003partial}. The sequential version asks how much a hidden variable could change the target policy value, allowing the hidden bias to enter each decision epoch. This subsection covers the three results that make this practical for off-policy evaluation in confounded MDPs.

\citet{namkoong2020off} provide the foundational result for finite-horizon sequential OPE under a Rosenbaum-style sensitivity model. Their identification assumption bounds the odds-ratio influence of any unobserved confounder by a parameter $\Gamma \geq 1$ at each decision epoch:
\begin{equation}
\Gamma^{-1} \;\leq\; \frac{\pi_t(a_t \mid H_t, u) / \pi_t(a_t' \mid H_t, u)}{\pi_t(a_t \mid H_t, u') / \pi_t(a_t' \mid H_t, u')} \;\leq\; \Gamma.
\label{eq:gamma_bound}
\end{equation}
Under this $\Gamma$-bound, their Theorem~2 derives a lower bound on the target policy value $\mathbb{E}[Y(\bar A_{1:T})]$ as the solution to a tractable loss-minimization problem with weighted quadratic loss $\ell_\Gamma(z) = \tfrac{1}{2}(\Gamma z_-^2 + z_+^2)$, and the empirical plug-in estimator is shown consistent in Theorem~3. The paper's central practical warning is sharp. Allowing the $\Gamma$-bound to bite at every decision epoch (``per-decision confounding'') yields bounds that grow exponentially in the horizon $T$, becoming uninformative for any realistic $\Gamma$ when $T$ is more than a handful of periods. The tractable regime is ``one-decision confounding,'' where only a single nominated decision is potentially confounded. Their applied demonstrations on sepsis ICU management and a SMART trial for autistic children show the one-decision setting certifies policy contrasts up to design sensitivity $\Gamma$ above five, while per-decision confounding loses informativeness already at $\Gamma$ near $1.5$.

\citet{kallus2020confounding} lift this analysis to the infinite-horizon setting and exchange per-step importance sampling for stationary-density ratios. Their Lemma~2 characterizes the partially identified set of state-occupancy ratios as solutions to an estimating equation involving the propensity ratio and an unobserved memoryless confounder, and their Theorem~1 establishes the sharpness of the resulting policy-value bound under a Tan-style marginal sensitivity model. The key structural assumption is that the unobserved confounder is conditionally independent of the prior history given the current state; with that, the bound problem reduces to a non-convex optimization that can be solved by alternating projections.

\citet{brunsSmith2021modelFree} provide the practitioner-facing estimator that ties this sensitivity machinery back to the chapter's causal Bellman operator (Definition~\ref{def:causal_bellman}). Their Proposition~4 establishes a robust Bellman equation under the same memoryless-confounder assumption: with $\tau = \Lambda / (1 + \Lambda)$ where $\Lambda$ is the Tan-style odds-ratio bound,
\begin{equation}
\bar Q_t^\pi(s, a) \;=\; \mathbb{E}\!\left[r_t(s, a, s') + \gamma \bar V_{t+1}^\pi(s') \,\big|\, s, a\right]
\;-\; \alpha_t(s,a) \cdot \operatorname{CVaR}_{1-\tau}\!\left(Y_t(\bar Q_{t+1}) \,\big|\, s, a\right).
\label{eq:robust_bellman_cf}
\end{equation}
The robust $Q$-function is the conditional expected shortfall of the Bellman target, and their orthogonalized loss yields a Neyman-orthogonal pseudo-outcome that relaxes the required nuisance rate from $o_p(n^{-1/2})$ to $o_p(n^{-1/4})$. Their MIMIC-III sepsis application picks a calibrated $\Lambda$ between $1.4$ and $2.5$ and shows the robust policy shifts from fluid-heavy toward vasopressors, in line with prior clinical meta-analyses. The bridge to the existing chapter machinery is exact: Equation~\eqref{eq:robust_bellman_cf} is a Bellman operator with a CVaR penalty term that subtracts off the worst-case effect of an unobserved confounder under the Tan-style sensitivity bound.

For applied economics the practical hierarchy is the following. One claim is statistical: the logged data support a precise estimate under a maintained model. The other is identificational: the maintained model is strong enough to recover the interventional value. Sensitivity methods relax the second claim and report how quickly a policy recommendation changes as hidden bias grows. \citet{namkoong2020off} establishes that this can be done at all in the sequential setting and characterizes when it fails (per-decision confounding); \citet{kallus2020confounding} extends to infinite horizons; \citet{brunsSmith2021modelFree} delivers a practitioner-ready estimator that plugs into standard fitted-$Q$ pipelines. Recent conformal sensitivity work \citep{yinShiWangBlei2024conformalSensitivity} shows the same instinct in static individualized treatment effects, although the RL setting is harder because per-period uncertainty is recursively priced through continuation values.

\subsubsection{Proximal and POMDP Bridge Methods}
\label{subsubsec:proximal}

When confounders are latent but proxy variables, noisy correlates of the confounder, are available, proximal causal inference provides identification. \citet{bennett2021proximal} adapt this logic to sequential settings by treating the problem as OPE in a POMDP. The analyst observes two proxies: a treatment-side proxy $W_t^{(1)}$ and an outcome-side proxy $W_t^{(2)}$, both conditionally independent given the latent confounder $U_t$. Identification proceeds through a bridge function $h$ that solves a conditional moment equation linking the two proxies to the interventional quantity:
\begin{equation}
\mathbb{E}[V(S_{t+1}) \mid W_t^{(1)}, S_t, A_t] = \mathbb{E}[h(W_t^{(2)}, S_t, A_t) \mid W_t^{(1)}, S_t, A_t].
\label{eq:bridge}
\end{equation}
The bridge function $h$ is estimated by solving this integral equation, which reduces to a linear system in the discrete case, and the causal effect is recovered by marginalizing over the outcome proxy distribution.
\begin{equation}
P(s' \mid s, \operatorname{do}(a)) = \sum_{w_2} h(w_2, s, a) \, P(W^{(2)}{=}w_2 \mid s).
\label{eq:proximal_adj}
\end{equation}
Two bridge functions, analogous to inverse propensity scores and Q-functions, yield a doubly robust estimator of the policy value that is $\sqrt{n}$-consistent without ever observing $U_t$. \citet{bennett2021proximal} demonstrate this idea in a sepsis management simulator in which the physician observes a true diabetes status that is only partially recorded in the analyst's data. Prior clinical observations serve as treatment-side proxies, current clinical observations serve as outcome-side proxies, and the proximal estimator ranks evaluation policies correctly in between 82\% and 100\% of test cases where naive estimators fail.

The POMDP bridge literature generalizes this insight beyond tabular examples. \citet{shiUeharaHuangJiang2022minimaxPOMDP} formulate OPE in confounded POMDPs as a minimax learning problem over bridge functions. \citet{ueharaEtAl2023futureDependent} introduce future-dependent value functions that use future proxies as inputs and history proxies as instruments. \citet{hongQiXu2024policyGradientPOMDP} move from evaluation to learning by identifying history-dependent policy gradients in confounded POMDPs. \citet{wangQiShi2025superPolicy} show a related proximal route in which past human or AI actions themselves can carry information about latent decision-relevant states.

\subsubsection{Instrumental Variables and Orthogonal Scores}
\label{subsubsec:iv_orthogonal}

When neither backdoor nor front-door variables are available, instrumental variables can identify causal effects. An instrument $Z_t$ must satisfy relevance, meaning it shifts the action $A_t$, and an exclusion restriction, meaning its effect on $S_{t+1}$ is channeled entirely through $A_t$. \citet{liao2024_iv_rl} formalize this as a Confounded MDP with Instrumental Variables (CMDP-IV), where transitions take the form $S_{t+1} = F^*(S_t, A_t) + \epsilon_t$ with unobserved confounders $\epsilon_t$ affecting both the behavior policy and transitions. The transition function $F^*$ is recovered from a conditional moment restriction:
\begin{equation}
\mathbb{E}[S_{t+1} - F^*(S_t, A_t) \mid Z_t, S_t] = 0.
\label{eq:iv_moment}
\end{equation}
For a binary action and binary instrument, the Wald estimator provides a closed-form solution. Let $\beta$ denote the causal effect of promoting (action $a = 0$) on the transition probability.
\begin{equation}
\beta = \frac{P(s' \mid s, Z{=}1) - P(s' \mid s, Z{=}0)}{P(A{=}0 \mid s, Z{=}1) - P(A{=}0 \mid s, Z{=}0)},
\label{eq:wald}
\end{equation}
where the numerator is the reduced-form effect of the instrument on the transition and the denominator is the first-stage effect on treatment uptake. The interventional transition is then recovered from any instrument value $z$ via $P(s' \mid s, \operatorname{do}(a{=}0)) = P(s' \mid s, Z{=}z) + \beta \cdot (1 - P(A{=}0 \mid s, Z{=}z))$. Their IV-aided Value Iteration algorithm applies this moment restriction at each state to estimate $F^*$, then runs standard value iteration on the estimated model.\footnote{The simulation in Section~\ref{subsec:sim_confounded_ope} implements the binary-action Wald formula $(\mathbb{E}[Y \mid Z{=}1] - \mathbb{E}[Y \mid Z{=}0]) / (\mathbb{E}[A \mid Z{=}1] - \mathbb{E}[A \mid Z{=}0])$ of Equation~\eqref{eq:wald}; \citet{liao2024_iv_rl}'s full IV-aided value iteration solves the conditional moment restriction in Equation~\eqref{eq:iv_moment} via a primal-dual reformulation, which we do not reproduce. The Wald estimator suffices to demonstrate IV's unbiasedness under exclusion.}

\citet{liao2024_iv_rl} illustrate with neonatal intensive care unit (NICU) assignment. A hospital must decide whether to admit each newborn to the NICU, and this decision is confounded by unobserved severity indicators that affect both admission and infant health. Differential travel time to a specialty care provider serves as an instrument. Relevance holds because travel time shifts referral probabilities. The exclusion restriction requires that travel time affect infant health only through admission, not through any other channel.

Orthogonal-score methods solve a different problem. By themselves they do not remove hidden confounding; they make causal OPE statistically stable when the identifying assumptions reduce the value to nuisance functions that must be learned flexibly. \citet{liaoMurphy2021longterm} estimate long-term average policy values for mobile health applications. \citet{lewisSyrgkanis2021dynamicDML} extend double machine learning to dynamic treatment effects through sequential residualization and Neyman-orthogonal moments. \citet{kallusUehara2022doubleRL} derive efficient double reinforcement learning estimators for Markovian OPE using q-functions and stationary density ratios. The common econometric message is that identification and estimation are separate. First specify why the policy value is identified, then use orthogonal scores to keep first-stage machine learning error from dominating the target estimate.

\subsubsection{Mediator-Based Identification}
\label{subsubsec:mediator_identification}

The front-door criterion applies when $A_t$ affects $S_{t+1}$ only through an observed mediator $M_t$. Three conditions are required: $M_t$ intercepts all directed paths from $A_t$ to $S_{t+1}$, no unblocked backdoor path exists from $A_t$ to $M_t$, and $A_t$ blocks all backdoor paths from $M_t$ to $S_{t+1}$. When satisfied \citep{pearl2009causality},
\begin{equation}
P(s' \mid s, \operatorname{do}(a)) = \sum_{m} P(m \mid a) \sum_{a'} P(s' \mid s, m, a') \, P(a' \mid s).
\label{eq:frontdoor}
\end{equation}
The first factor is unconfounded; the inner sum adjusts for confounding on the $M_t \to S_{t+1}$ link by averaging over the observational action distribution. This is the sequential analogue of mediation analysis.

Mediator-based RL is most attractive when the action changes an observable channel before it changes the long-run state. In mobile health, an app prompt may affect cortisol through supplement adherence. In retail pricing, a promotion may affect conversion through a marketing follow-up or browsing response. \citet{shiZhuShenLuoZhuSong2024cmdpCI} use mediator variables to identify and construct confidence intervals for policy values in confounded MDPs. \citet{geEtAl2023dynamicMediation} develop an RL framework for dynamic mediation analysis, decomposing the long-run effect into direct and mediated components. The policy question shifts from ``does the action work'' to ``through which sequential channel does the action work, and is that channel stable enough to transport across policies?''

\subsection{Structural Counterfactual Off-Policy Evaluation}
\label{subsec:cfope_scm}

The point-identification strategies of the preceding subsections recover the target-policy value $V(\widetilde\pi)$ from observational data under explicit graphical assumptions. A complementary route, originating in \citet{buesing2019woulda}, operates one level deeper. This route casts the partially observable Markov decision process as a structural causal model, infers a posterior over its exogenous noise from the observed trajectory, and rolls out under the counterfactual policy with that posterior held fixed. The output is not just a value estimate but a counterfactual trajectory the agent would have produced, sample by sample.

The transition mechanism is $s_{t+1} = f^s(s_t, a_t, u^s_t)$ with exogenous noise $u^s_t$, the observation mechanism is $o_t = f^o(s_t, u^o_t)$, and the action mechanism is $a_t = f^\pi(h_t, u^a_t)$ given the history $h_t$. Given an observed trajectory $\hat\tau$, the counterfactual rule proceeds in three steps.

\begin{algorithm}[h]
\caption{Counterfactual off-policy evaluation \citep{buesing2019woulda}}
\label{alg:buesing_cfope}
\begin{algorithmic}[1]
\State \emph{Abduction.} Infer the posterior $P(U \mid \hat\tau)$ over the exogenous noise from the observed trajectory.
\State \emph{Action.} Replace the behaviour-policy mechanism $f^\pi$ with the target-policy mechanism $f^{\widetilde\pi}$ to form the modified SCM $\mathcal M^{\mathrm{do}(\widetilde\pi)}_{\hat\tau}$ with the same posterior on $U$.
\State \emph{Prediction.} Sample $U \sim P(U \mid \hat\tau)$ and roll out the modified SCM to obtain counterfactual trajectories $\tau^{\mathrm{cf}}$ and their returns.
\end{algorithmic}
\end{algorithm}

Lemma~1 of \citet{buesing2019woulda} establishes that the marginal of the counterfactual distribution over $\hat\tau$ equals the interventional distribution, so the counterfactual estimator is unbiased under correct SCM specification. On the partially-observable grid-world experiment of the paper, the counterfactual estimator's evaluation error decreases monotonically with the amount of off-policy data, while the model-based estimator's error remains stuck at the bias of the dynamics model.

The categorical-action case raises an identification problem. \citet{oberst2019counterfactual} show that multiple structural causal models can be consistent with the same interventional distribution yet imply different counterfactual outcomes, and they resolve the ambiguity by assuming a Gumbel-max sampling mechanism that satisfies a counterfactual-stability condition.\footnote{The Pearl monotonicity condition resolves the binary case. The Gumbel-max SCM is the natural extension to $k$-way categorical outcomes. \citet{TangWiens2023} combine SCM-based counterfactual outcomes with importance sampling to obtain variance reductions while preserving unbiasedness under correct human annotation, with a sepsis-simulator demonstration that delivers an RMSE of $0.013$ against $0.113$ for vanilla per-decision IS.} The shared identifying assumption across this strand is that the structural causal model is correctly specified. The Lucas critique applies in its sharpest form when the agent's policy itself enters the data-generating process, and identification by instruments or exogeneity restrictions in the spirit of \citet{Chernozhukov2018} is the econometric remedy that composes naturally with the backdoor, front-door, proximal, and mediator-based identification strategies of the preceding subsections.

\subsection{The Broader Causal RL Landscape}
\label{subsec:causal_rl_landscape}

Two further directions sit outside the chapter's central scope but deserve naming. \citet{paceEtAl2024delphic} respond to the case where no observed control, instrument, mediator, or proxy is strong enough to point-identify the policy value with Delphic offline RL, which treats hidden confounding as nonidentifiable uncertainty rather than pretending it has been solved. \citet{mesnardEtAl2021counterfactual} extends the structural-counterfactual machinery of Section~\ref{subsec:cfope_scm} from policy evaluation to per-trajectory counterfactual policy gradients, using constrained future-information conditioning to keep the gradient estimator unbiased. Transportability theory \citep{pearl2014external,bareinboim2016causal,liZhangBareinboim2024curriculum} asks which mechanisms remain invariant across source and target environments, governing sim-to-real transfer and curriculum learning under environment shift.

\subsection{Simulation Study: Confounded Retail Pricing MDP}
\label{subsec:sim_confounded_ope}

The simulation uses a 5-state engagement funnel $\mathcal{S} = \{0,1,2,3,4\}$ with two actions (promote, hold price) and an absorbing conversion state at $s = 4$. A retailer manages customers through engagement stages and decides whether to offer a promotional discount. One data-generating process supplies the causal variation required by four point-identification strategies.

\begin{figure}[htbp]
\centering
\includegraphics[width=0.7\textwidth]{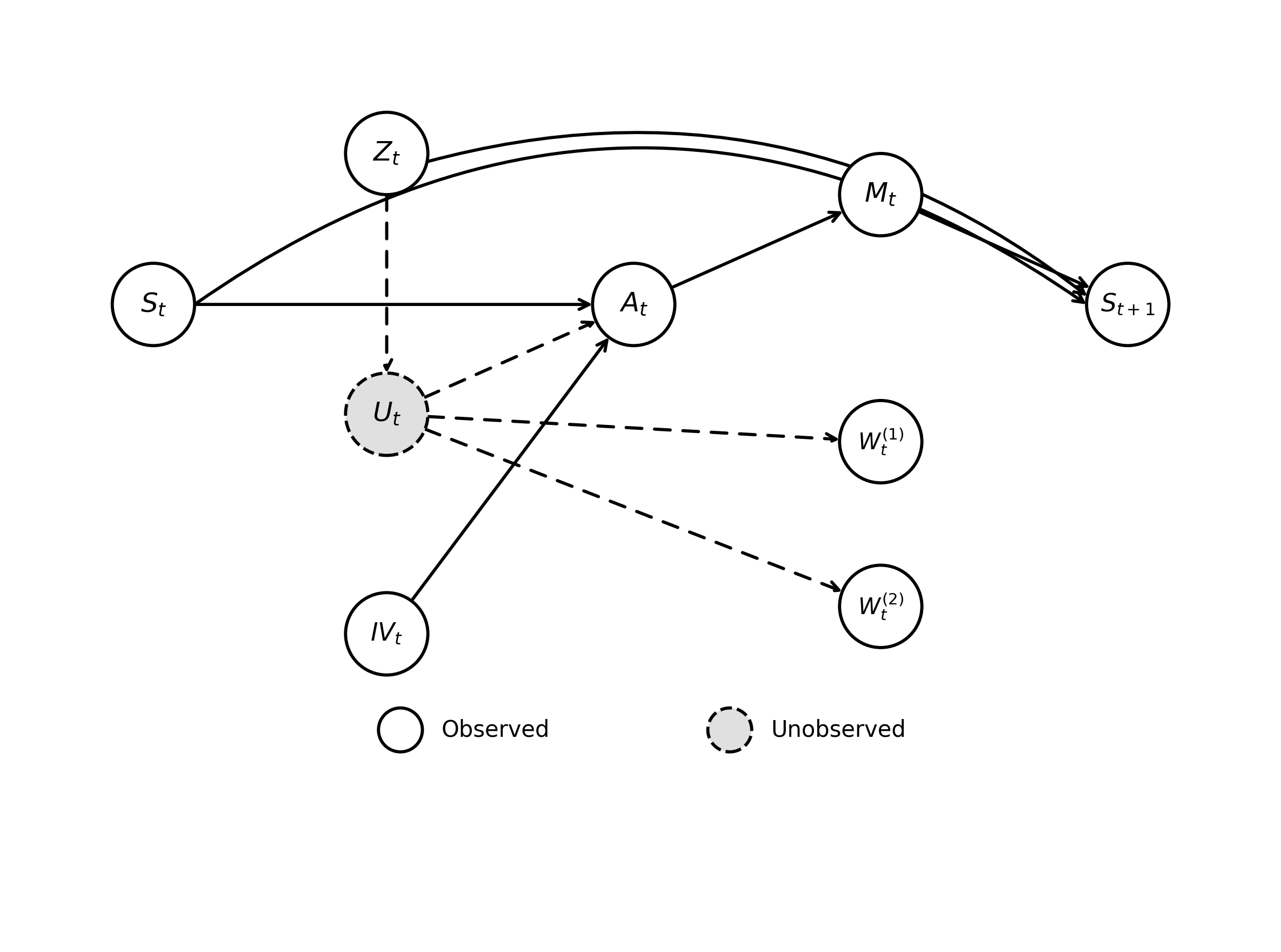}
\caption{Complete causal graph of the simulation DGP. Gray dashed node ($U_t$) is unobserved; dashed edges involve $U_t$.}
\label{fig:simulation_dag}
\end{figure}

Figure~\ref{fig:simulation_dag} displays the complete causal graph. Market conditions $Z_t \sim \text{Bernoulli}(0.5)$ are observed and affect both the latent confounder and transitions. Consumer sentiment $U_t$ is an unobserved confounder strongly correlated with $Z_t$.\footnote{$P(U_t{=}1 \mid Z_t{=}1) = 0.9$ and $P(U_t{=}1 \mid Z_t{=}0) = 0.1$.} An independent cost shock $\text{IV}_t \sim \text{Bernoulli}(0.5)$ serves as an instrument. The behavioral pricing policy depends on both $U_t$ and $\text{IV}_t$: $\mu(\text{promote} \mid s, U_t, \text{IV}_t) = 0.55 + \rho \cdot 0.25 \cdot (2U_t - 1) + 0.15 \cdot (\text{IV}_t - 0.5)$, where $\rho \in \{0, 0.2, \ldots, 1.0\}$ controls confounding strength. Promotions trigger marketing follow-ups with $M_t$ serving as a mediator.\footnote{$M_t \sim \text{Bernoulli}(0.8)$ when the retailer promotes and $\text{Bernoulli}(0.2)$ otherwise.} Two noisy proxies of $U_t$ are available: a CRM score $W_t^{(1)}$ and browsing behavior $W_t^{(2)}$.\footnote{$W_t^{(1)} \sim \text{Bernoulli}(0.85 \cdot U_t + 0.15 \cdot (1 - U_t))$ and $W_t^{(2)} \sim \text{Bernoulli}(0.75 \cdot U_t + 0.25 \cdot (1 - U_t))$.}

The action affects the next state only through the mediator: $P(s{+}1 \mid s, M_t, Z_t)$ depends on $M_t$ and $Z_t$ but not on $A_t$ or $U_t$ directly. The four transition probabilities are $P(s{+}1 \mid s, M{=}1, Z{=}1) = 0.90$, $P(s{+}1 \mid s, M{=}1, Z{=}0) = 0.50$, $P(s{+}1 \mid s, M{=}0, Z{=}1) = 0.40$, and $P(s{+}1 \mid s, M{=}0, Z{=}0) = 0.15$. This enables all four identification strategies simultaneously: $Z_t$ satisfies the backdoor criterion, $M_t$ satisfies the front-door criterion, $\text{IV}_t$ satisfies relevance and exclusion, and $W_t^{(1)}, W_t^{(2)}$ satisfy the proximal conditions.\footnote{Rewards are $r(s,a) = -1$ for $s < 4$, $\gamma = 0.9$. The target policy always promotes. The true interventional transition probability is $P(s{+}1 \mid s, \operatorname{do}(\text{promote})) = 0.615$.}

The six estimators are oracle, naive (biased per Lemma~\ref{lem:naive_bias}), backdoor (Equation~\ref{eq:backdoor_estimator}), front-door (Equation~\ref{eq:frontdoor}), Wald IV (Equation~\ref{eq:wald}), and proximal (Equations~\ref{eq:bridge} and \ref{eq:proximal_adj}).\footnote{Each configuration uses 2{,}000 trajectories averaged over 20 seeds.} The estimators correspond to the point-identification rows of Table~\ref{tab:causal_rl_taxonomy}. The simulation omits sensitivity bounds and orthogonal-score refinements. The figure therefore tests identification logic rather than higher-order statistical efficiency.

\begin{figure}[htbp]
\centering
\includegraphics[width=\textwidth]{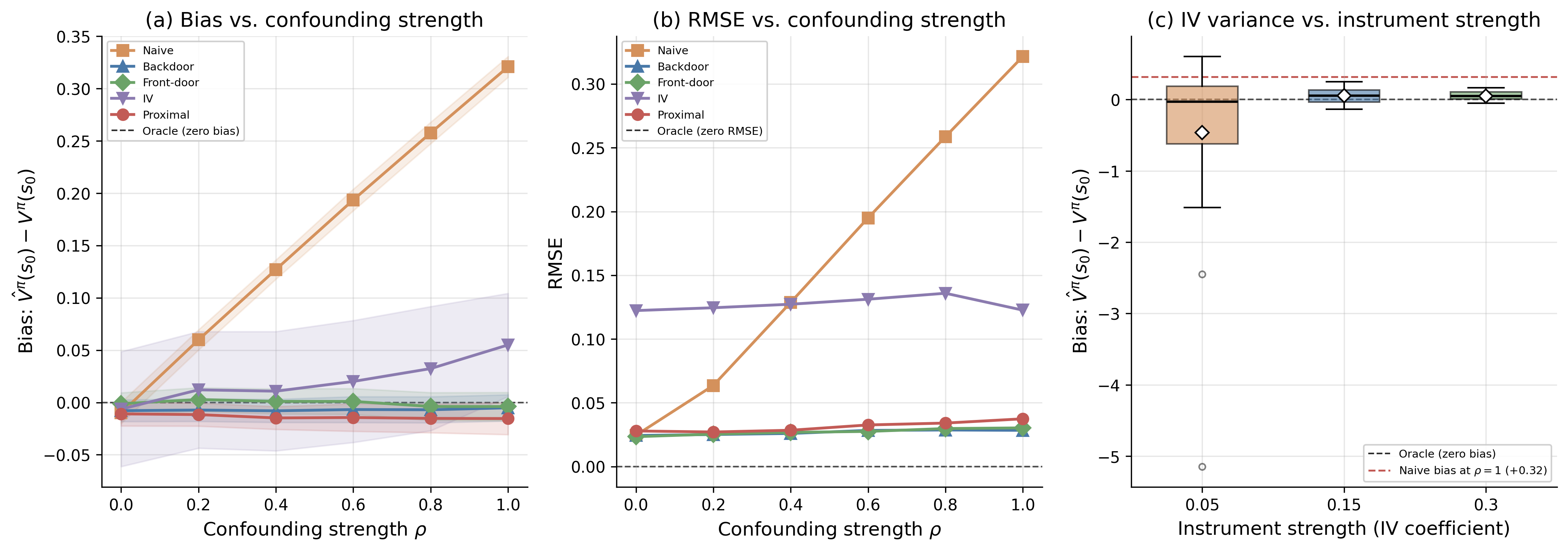}
\caption{(a) Bias of five OPE estimators as a function of confounding strength $\rho$. (b) RMSE of five estimators as a function of $\rho$. (c) IV estimator bias distribution vs.\ instrument strength at $\rho = 1$; dashed red line is naive estimator bias.}
\label{fig:confounded_ope}
\end{figure}

Figure~\ref{fig:confounded_ope} reports bias and RMSE across confounding strengths using 20 seeds and 2{,}000 trajectories per seed. The naive estimator's bias grows monotonically with $\rho$ because observational transitions overestimate promotion success. Promotion is more likely when $U_t = 1$, which correlates with favorable conditions. The resulting $\hat{P}_{\text{obs}}$ exceeds the true interventional probability, and the per-step bias compounds through the Bellman recursion. The backdoor and front-door estimators eliminate bias at every value of $\rho$, as Theorem~\ref{thm:backdoor_mdp} and Equation~\eqref{eq:frontdoor} predict. The IV estimator has low bias and higher variance because the Wald ratio is sensitive to instrument strength. Panel~(c) reports this bias-variance tradeoff. The proximal estimator has low bias and moderate variance, as the proxy bridge equations predict.

\subsection{Simulation Study: Counterfactual OPE under a Misspecified SCM}
\label{subsec:cfope_sim}

A second synthetic example applies the counterfactual machinery of Section~\ref{subsec:cfope_scm} to a contextual bandit. Removing temporal dependence isolates model misspecification. The DGP is a linear structural causal model in two features,
\begin{equation}
y = 1 + 0.5\, x_1 - 0.3\, x_2 + 2\, a + a \cdot x_1 + u,
\qquad u \sim \mathcal N(0,\, 0.5^2),
\label{eq:cfope_dgp}
\end{equation}
with $x \sim \mathcal N(0, I_2)$, a binary action $a \in \{0, 1\}$, and a heterogeneous treatment effect $\tau(x) = 2 + x_1$. The behaviour policy $\pi_{\mathrm{obs}}(a = 1 \mid x) = \sigma(0.5 + x_1 - 0.5\, x_2)$ is confounded with $x$. The target policy $\widetilde\pi(a \mid x) = \mathbf 1\{x_1 + x_2 > 0\}$ is a deterministic threshold rule. The oracle value $V(\widetilde\pi) = \mathbb E_x[y \mid a = \widetilde\pi(x)]$ is computed by Monte Carlo at $10^6$ realisations.

Three estimators are compared on $n \in \{200, 500, 1000, 2000\}$ logged samples, $20$ seeds per cell. Per-decision importance sampling is $\widehat V_{\mathrm{IS}} = n^{-1} \sum_i \rho_i y_i$ with $\rho_i = \mathbf 1\{a_i = \widetilde\pi(x_i)\} / \pi_{\mathrm{obs}}(a_i \mid x_i)$. The model-based plug-in $\widehat V_{\mathrm{MB}} = n^{-1} \sum_i \hat f(x_i, \widetilde\pi(x_i))$ uses an OLS fit $\hat f$ on the logged data. The counterfactual estimator follows the abduction-action-prediction rule of Algorithm~\ref{alg:buesing_cfope} and combines $\hat f$ with a residual correction weighted by the importance ratio,
\begin{equation}
\widehat V_{\mathrm{CF}}
= n^{-1} \sum_i \hat f(x_i, \widetilde\pi(x_i))
  + n^{-1} \sum_i \rho_i \bigl(y_i - \hat f(x_i, a_i)\bigr),
\label{eq:cfope_dr}
\end{equation}
which is the doubly-robust form of the abduction-action-prediction rule and is unbiased under correct propensity \emph{or} correct outcome model.\footnote{The estimator labelled $\widehat V_{\mathrm{CF}}$ in equation~\eqref{eq:cfope_dr} is algebraically the Robins-Rotnitzky-Zhao doubly-robust / AIPW estimator adapted to off-policy evaluation; it is \emph{not} the Buesing (2019) abduction-action-prediction estimator, which operates at the trajectory level under a known or learned structural causal model with abduction over exogenous noise and is not implemented here. The naive average of the per-observation counterfactual outcome $y'_i = \hat f(x_i, \widetilde\pi(x_i)) + (y_i - \hat f(x_i, a_i))$ collapses to $\widehat V_{\mathrm{MB}}$ under OLS with an intercept because the OLS residuals sum exactly to zero; weighting the residual by the importance ratio $\rho_i$ as in~\eqref{eq:cfope_dr} restores bias cancellation and recovers the AIPW form. We retain the ``CF'' label for continuity with Section~\ref{subsec:cfope_scm}.} The well-specified scenario uses the feature set $\{1, x_1, x_2, a, a \cdot x_1\}$ in $\hat f$, matching the DGP. The misspecified scenario drops the interaction $a \cdot x_1$, the source of the heterogeneous treatment effect. The propensity $\pi_{\mathrm{obs}}$ is held fixed at the true DGP value in both scenarios, so only the outcome-model side of the double-robustness claim is stressed here; a misspecified-propensity scenario paired with a correct outcome model would test the other side of the claim and is left as a future check.

\begin{table}[h]
\centering
\caption{Counterfactual off-policy evaluation under a linear SCM, $n = 1000$, $20$ seeds.}
\label{tab:cfope_summary}
\begin{tabular}{llccc}
\hline\hline
Scenario & Estimator & Bias & Std & RMSE \\
\hline
Well-specified & IS & $+0.017$ & $0.100$ & $0.099$ \\
 & MB & $-0.001$ & $0.057$ & $0.056$ \\
 & CF & $+0.003$ & $0.067$ & $0.065$ \\
\hline
Misspecified & IS & $+0.017$ & $0.100$ & $0.099$ \\
 & MB & $-0.072$ & $0.061$ & $0.093$ \\
 & CF & $+0.002$ & $0.078$ & $0.076$ \\
\hline\hline
\end{tabular}

\end{table}

\begin{figure}[h]
\centering
\includegraphics[width=0.85\linewidth]{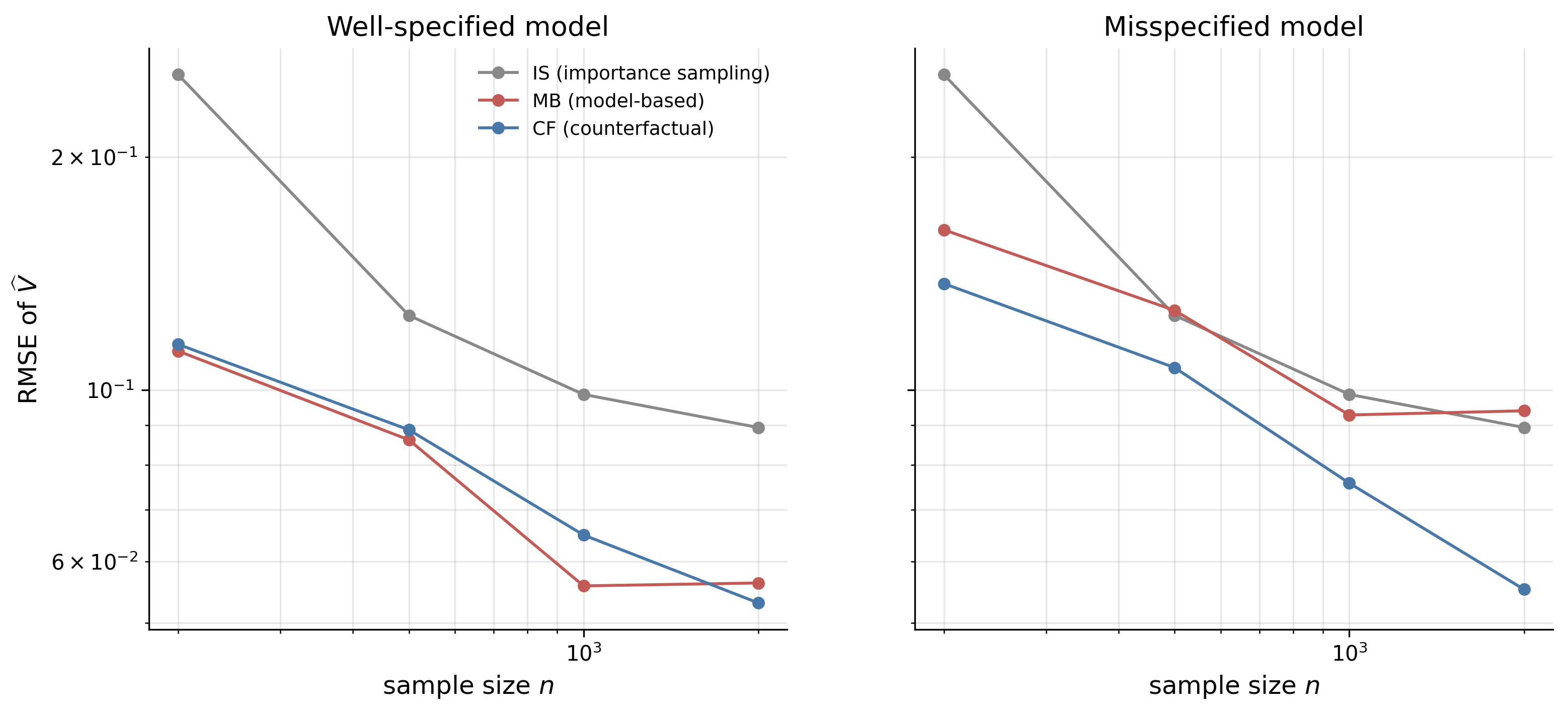}
\caption{Root mean squared error of three OPE estimators against sample size on a log-log scale. The left panel uses a well-specified outcome model. The right panel omits the interaction.}
\label{fig:cfope_rmse}
\end{figure}

Table~\ref{tab:cfope_summary} reports bias, standard deviation, and root mean squared error at $n = 1000$. Under well-specification the model-based estimator attains bias $-0.001$ and RMSE $0.056$, the counterfactual estimator attains bias $0.003$ and RMSE $0.065$, and the importance-sampling estimator attains bias $0.017$ and RMSE $0.099$. The MB and CF estimators are statistically indistinguishable and both improve on IS. Under misspecification the MB estimator's bias jumps to $-0.072$ and its RMSE to $0.093$, while the CF estimator's bias remains at $0.002$ and its RMSE at $0.076$. The CF estimator's RMSE is $18\%$ lower than MB's, and the bias is two orders of magnitude smaller. The IS estimator is unchanged across scenarios because it does not use an outcome model. Figure~\ref{fig:cfope_rmse} plots RMSE against $n$ on log-log scale. The MB curve flattens under misspecification at the population bias of the OLS projection, the CF curve continues to decline at the parametric rate, and the IS curve traces the higher-variance parametric rate that the importance-weighted estimator delivers without any model assistance.

The companion Chapter~\ref{section:rl_for_ci} inverts the direction of this one, asking how RL machinery imports into econometric causal inference rather than the reverse. The bridge there runs through the equivalence between Murphy's backward Bellman recursion for optimal dynamic treatment regimes and Watkins's $Q$-learning recursion on a history-augmented state. Once that translation is fixed, the dynamic-DML and policy-learning results of that chapter complement the identification techniques of this one. This chapter deals with what to do when sequential ignorability fails; the companion chapter deals with how far you can push the rate of convergence when it holds.

\section{Off-Policy Evaluation and Dynamic Treatment Effects}
\label{section:rl_for_ci}


Chapter~\ref{section:causal_rl} studies causal identification when logged transitions are confounded. This chapter assumes consistency, positivity, and sequential ignorability and applies sequential-decision methods to causal estimation. Under these assumptions, g-computation identifies the same history-indexed backward induction used by finite-horizon reinforcement learning (RL). \citet{murphy2003dtr} formalizes the result for optimal dynamic treatment regimes (DTRs), and \citet{schulte2014qlearning} develops the corresponding translation between causal and RL notation.

The chapter proves this equivalence and develops off-policy evaluation for a fixed regime. It then considers two extensions. Dynamic double machine learning (DML) estimates low-dimensional structural effects with $\sqrt n$ inference \citep{lewisSyrgkanis2021dynamicDML}, while doubly robust backward induction learns a regime with a $\sqrt n$ welfare-regret bound \citep{sakaguchi2024dynamicpolicy}. Each section pairs the formal result with a computational experiment and separates point estimation from uncertainty quantification. The final section describes open problems. Chapter~\ref{section:adaptive_experiments} treats the separate design problem in which the experimenter controls data collection. All four numerical studies use a stylized Fast Track-inspired home-visiting setting. They share an applied context but do not reconstruct the trial's sample, schedule, or treatment effects.

\subsection{Dynamic Treatment Regimes}
\label{subsec:gmethods_bridge}

Fast Track is a randomized prevention study for children with elevated behavioral problems identified in kindergarten. Children are assigned either to a complex prevention program or to control. The program seeks to prevent or reduce conduct disorders and drug use. Within the intervention, a home-visiting component seeks to improve family functioning. \citet{murphy2003dtr} uses this component to motivate dynamic treatment regimes.

Beginning in the spring semester of first grade, a family counselor completes six questions about parenting quality and family functioning at the end of each semester. Their sum is the family's current status, with lower scores indicating greater need. This status determines the number of home visits in the following semester. A family's visit level can therefore change as its recorded needs change. The record begins with pretreatment information, then alternates between the measured family status and assigned visit level, and ends with a terminal child or family outcome.

An A/B test assigns each participant to one fixed protocol and compares average outcomes across the two groups. Fast Track uses that design to estimate the effect of the full prevention program relative to control. The home-visiting question is different because it asks how many visits to assign after each new family assessment. Assigning a high-visit or low-visit schedule at baseline would compare two fixed schedules even when family needs later diverge. The Fast Track team also avoids giving every family the highest level because it worries that excessive visiting can foster dependency, pejorative labeling, and attrition. A direct comparison of families with many and few visits is not an A/B comparison, since the lower family-functioning scores that trigger more visits already separate the groups before their later outcomes are measured. Regressing the final outcome on the total number of visits has the same problem because the visit count records both treatment and prior need.

Murphy writes this longitudinal record as $S_1,A_1,S_2,\ldots,A_K,Y$. In the Fast Track example, $S_k$ is the family information available at the $k$th semester decision, $A_k$ is the home-visit level assigned for the coming semester, and $Y$ is the terminal child or family outcome chosen for analysis. A visit at decision $k$ may change $S_{k+1}$, which may then change later visit assignments and the terminal outcome. The model does not rule out these delayed effects. A dynamic treatment regime $\pi=(\pi_1,\ldots,\pi_K)$ maps the information available at each decision to a visit level. The unrestricted target is a measurable regime that maximises the expected terminal outcome. The potential-outcome notation makes the counterfactual states and outcomes under other visit sequences explicit.

\subsubsection*{Sequential treatments as a graph}

\begin{figure}[H]
\centering
\includegraphics[width=\textwidth]{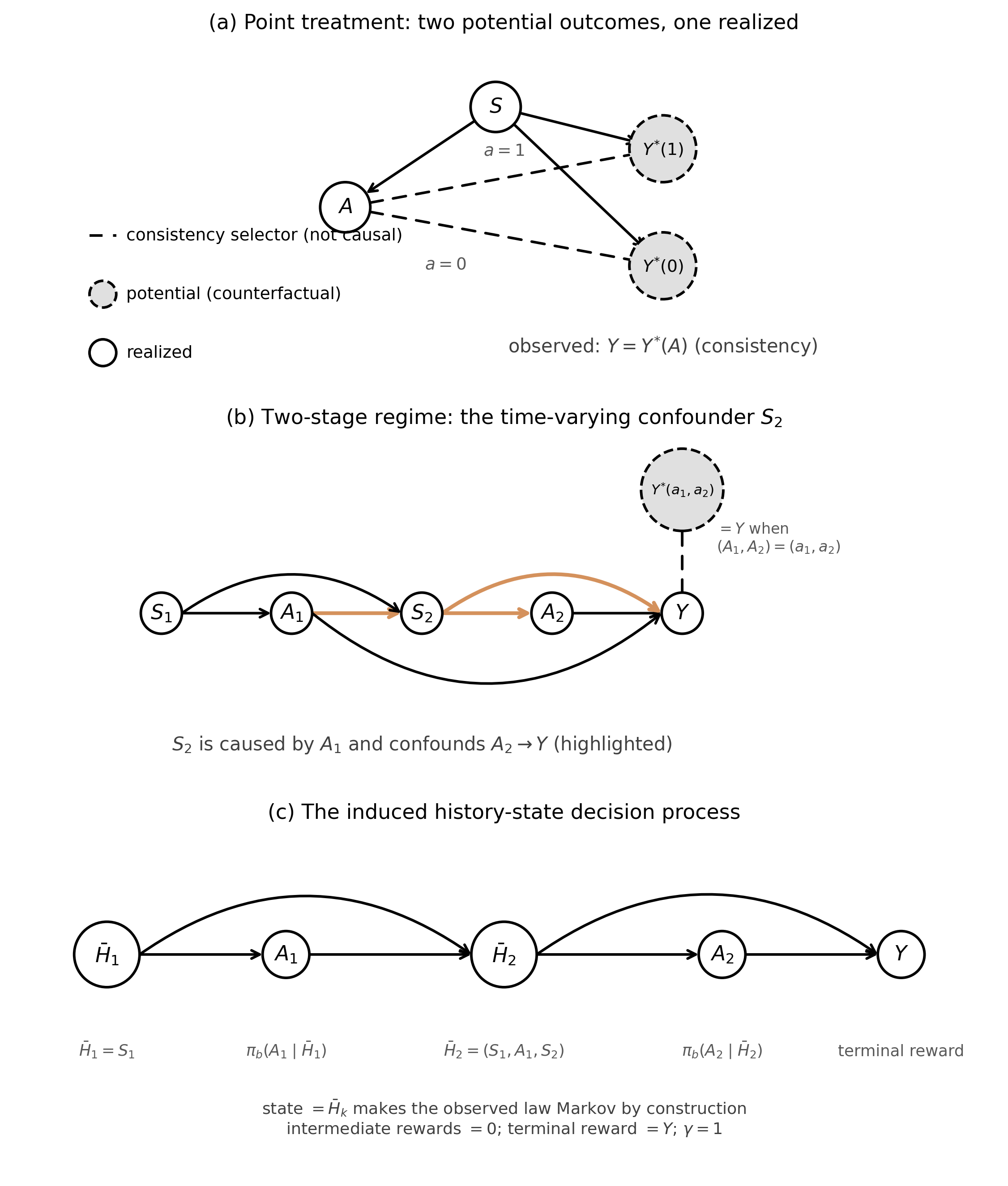}
\caption{Three representations of sequential treatment. Panel (a) shows fixed potential outcomes and the noncausal consistency link that selects the realized response. Panel (b) shows a two-stage treatment graph, with the treatment-confounder feedback through $S_2$ highlighted. Panel (c) constructs the finite-horizon decision process whose state is the full observed history.}
\label{fig:dtr_dags}
\end{figure}

Figure~\ref{fig:dtr_dags} separates three objects that recur in the analysis. Potential outcomes index responses under interventions, the longitudinal graph records treatment-confounder feedback, and the history-state construction turns the observed law into a finite-horizon Markov decision process (MDP).

\FloatBarrier

\subsubsection*{Potential outcomes and identification}

$\bar A_k=(A_1,\ldots,A_k)$ and $\bar S_k=(S_1,\ldots,S_k)$ denote the treatment and state histories. The observed history immediately before decision $A_k$ is $\bar H_k=(\bar S_k,\bar A_{k-1})$. Every treatment path $\bar a_K$ defines a potential state $S_{k+1}(\bar a_k)$ and terminal potential outcome $Y(\bar a_K)$. The potential history before stage $k$ is
\[
\bar H_k(\bar a_{k-1})
=\{S_1,S_2(a_1),\ldots,S_k(\bar a_{k-1}),\bar a_{k-1}\}.
\]
In the running example, $\bar H_k$ contains the family measures and visit levels recorded before the $k$th assignment, and $\bar a_k$ is a hypothetical sequence of visit levels through that semester. The potential state $S_{k+1}(\bar a_k)$ is the family status that would be recorded at the next decision time if the family had received $\bar a_k$. The potential outcome $Y(\bar a_K)$ is the terminal child or family outcome under a complete hypothetical visit sequence. Only the potential states and outcome under the visit sequence actually received are observed.

A dynamic treatment regime $\pi=(\pi_1,\ldots,\pi_K)$ is a sequence of maps from histories to actions. It induces counterfactual actions recursively through $A_k^\pi=\pi_k(\bar H_k^\pi)$, counterfactual histories $\bar H_k^\pi=\bar H_k(\bar A_{k-1}^\pi)$, and outcome $Y^\pi=Y(\bar A_K^\pi)$. Let $\Pi_{\mathrm{all}}$ be the class of measurable deterministic regimes. The value and unrestricted optimization target are
\[
V(\pi)=\mathbb E[Y^\pi],
\qquad
\pi^{\mathrm B}\in\arg\max_{\pi\in\Pi_{\mathrm{all}}}V(\pi).
\]
For Fast Track, $\pi_k$ takes the family's recorded history and selects the coming semester's visit level. The counterfactual outcome $Y^\pi$ is the terminal outcome if that rule is used at every semester decision.

This notation distinguishes a fixed treatment path from a regime that selects later treatments using counterfactual histories \citep{murphy2003dtr, schulte2014qlearning}.
For a realized prefix $\bar a_{k-1}$,
$Y(\bar a_{k-1},\pi_{k:K})$ denotes the same terminal potential
outcome when that prefix is fixed and the continuation actions are selected
recursively by $\pi_{k:K}$.

Identification rests on four assumptions.
\begin{enumerate}
\item \emph{Stable treatment versions and no interference.} Each action label denotes one well-defined intervention, and one subject's potential states and outcome do not depend on other subjects' assignments. In the running example, a given visit level must denote the same intervention across families, and one family's visits cannot change another family's outcome. This is the longitudinal stable-unit-treatment-value condition.
\item \emph{Consistency.} If the realized treatment history equals $\bar a_k$, then $S_{k+1}=S_{k+1}(\bar a_k)$, and if $\bar A_K=\bar a_K$, then $Y=Y(\bar a_K)$. The observed next family status and terminal outcome therefore equal the potential variables under the visit sequence actually received.
\item \emph{Sequential exchangeability}, also called sequential ignorability or no unmeasured confounding. At each $k$, conditional on $\bar H_k$, the assigned action is independent of the full collection of future potential states and outcomes under every continuation path. The visit decision may depend on the recorded family history, but not on an omitted family risk that also predicts the later outcome.
\item \emph{Target-policy positivity.} For every history reached with positive probability under $\pi$, $\pi_k(a\mid\bar h_k)>0$ implies $\mathbb P(A_k=a\mid\bar H_k=\bar h_k)>0$. Every visit level that the target regime might assign must occur among families with the relevant recorded history. Stable estimation requires the stronger practical condition that these probabilities remain bounded away from zero.
\end{enumerate}
The first three assumptions give causal meaning to observed conditional laws, while positivity determines which regimes those laws can identify. A Markov restriction is separate. It permits compression of the full history into a state but is not itself a substitute for exchangeability. Estimation further requires independent trajectories or an explicit time-series design, finite moments, and regularity conditions for the fitted nuisance functions. Chapter~\ref{section:causal_rl} treats identification when sequential exchangeability fails.

\subsubsection*{The backward recursion}

For unrestricted optimization, strengthen assumption 4 to require positivity for every action at every history in the observed support. Under assumptions 1--3 and this all-action positivity condition, the optimal value functions satisfy the finite-horizon backward recursion
\begin{align}
Q_K(\bar h_K, a_K) &= \mathbb{E}\!\left[Y \mid \bar H_K = \bar h_K,\ A_K = a_K\right], \label{eq:dtr_terminal}\\
Q_k(\bar h_k, a_k) &= \mathbb{E}\!\left[V_{k+1}(\bar H_{k+1}) \mid \bar H_k = \bar h_k,\ A_k = a_k\right], \qquad k<K, \label{eq:dtr_qfunc}\\
V_k(\bar h_k) &= \max_{a_k \in \mathcal A_k} Q_k(\bar h_k, a_k), \qquad
\pi_k^{\mathrm B}(\bar h_k) = \min\arg\max_{a_k \in \mathcal A_k} Q_k(\bar h_k, a_k). \label{eq:dtr_bellman}
\end{align}
Murphy calls $V_k$ the \emph{optimal benefit-to-go}. The fixed ordering of each finite action set makes the final expression a measurable tie-breaking rule.

The statistical issue is identification. Each subject supplies only one realized treatment path, so the distributions under alternative paths are counterfactual. Sequential ignorability permits the observed conditional laws in \eqref{eq:dtr_terminal} and \eqref{eq:dtr_qfunc} to identify those alternatives. For $K=2$, the estimator regresses $Y$ on $(\bar H_2,A_2)$, maximizes the fitted second-stage response, and uses that fitted maximum as the outcome in a regression on $(\bar H_1,A_1)$. In reinforcement-learning notation this is batch fitted $Q$-iteration \citep{schulte2014qlearning}.

The two-stage fixed-regime calculation makes the identifying steps explicit. Let
\[
m_2^\pi(\bar h_2)
=\mathbb E[Y\mid \bar H_2=\bar h_2,A_2=\pi_2(\bar h_2)]
\]
and
\[
m_1^\pi(\bar h_1)
=\mathbb E[m_2^\pi(\bar H_2)
\mid \bar H_1=\bar h_1,A_1=\pi_1(\bar h_1)].
\]
Consistency replaces the observed $Y$ in $m_2^\pi$ by the potential outcome under the realized treatment path. Sequential exchangeability removes the dependence of that potential outcome on the realized action after conditioning on history. Positivity ensures that both conditional means are defined on histories reached by $\pi$. Applying the same argument at stage 1 and then iterated expectation gives
\begin{equation}
V(\pi)=\mathbb E[Y^\pi]=\mathbb E[m_1^\pi(\bar H_1)].
\label{eq:two_stage_gformula}
\end{equation}
The $K$-stage recursion repeats these three operations at each decision point.
Fix regular conditional versions
\[
P_k(d\bar h_{k+1}\mid \bar h_k,a_k)
=\mathbb P(d\bar h_{k+1}\mid \bar H_k=\bar h_k,A_k=a_k)
\]
and
$\mu_K(\bar h_K,a_K)=\mathbb E[Y\mid\bar H_K=\bar h_K,A_K=a_K]$.
For a deterministic regime $\pi$, the observed-data $K$-stage g-formula is
\begin{equation}
G(\pi)
=\int \mu_K\{\bar h_K,\pi_K(\bar h_K)\}\,
\mathbb P_{\bar H_1}(d\bar h_1)
\prod_{k=1}^{K-1}
P_k\{d\bar h_{k+1}\mid\bar h_k,\pi_k(\bar h_k)\}.
\label{eq:k_stage_gformula}
\end{equation}
The product denotes iterated integration in increasing stage order.

\subsubsection*{The equivalence, formally}

\begin{theorem}[Equivalence of the identified DTR recursion and Bellman optimality]
\label{thm:dtr_rl_equivalence}
Suppose each state space is standard Borel, each action set $\mathcal A_k$ is finite and ordered, and $Y$ is bounded. Assume that longitudinal potential states and outcomes are defined under every treatment path, and that consistency, positivity, and sequential ignorability hold for their joint future collection. Positivity is needed here in a form strictly stronger than the target-policy version of assumption 4, covering every action rather than only those a given regime assigns, because the optimality claim ranges over all continuation regimes. That form is
\[
\mathbb P(A_k=a\mid \bar H_k=\bar h_k)>0
\]
for every $a\in\mathcal A_k$ and for $\mathbb P_{\bar H_k}$-almost every $\bar h_k$. Define $(Q_k,V_k)$ by \eqref{eq:dtr_terminal}--\eqref{eq:dtr_bellman}. For every fixed measurable continuation regime $\pi_{k:K}$ covered by positivity,
\[
\mathbb E\!\left[
Y(\bar a_{k-1},\pi_{k:K})
\mid \bar H_k=\bar h_k
\right]
\leq V_k(\bar h_k)
\]
almost surely, with the null set allowed to depend on the fixed regime. Here $\bar a_{k-1}$ is the treatment component of $\bar h_k$. Equality holds for the recursively defined regime $\pi^{\mathrm B}=(\pi_1^{\mathrm B},\ldots,\pi_K^{\mathrm B})$, which is optimal over $\Pi_{\mathrm{all}}$ and satisfies $V(\pi^{\mathrm B})=\mathbb E[V_1(\bar H_1)]$. For a restricted class $\Pi\subset\Pi_{\mathrm{all}}$, this Bellman regime is also a $\Pi$-optimizer only when $\pi^{\mathrm B}\in\Pi$.

Let $\mathcal M$ have initial law $\mathbb P_{\bar H_1}$, stage-$k$ state $\bar H_k$, action set $\mathcal A_k$, transition kernels $P_k$, zero intermediate rewards, terminal mean reward $\mu_K$, and $\gamma=1$. Let $J_{\mathcal M}(\pi)$ denote its expected terminal reward under $\pi$. Then $(Q_k,V_k)$ solves the Bellman optimality equations of $\mathcal M$. For every measurable regime covered by positivity,\footnote{The induced MDP uses the full observed history as its state. Replacing $\bar H_k$ by $S_k$ requires an additional Markov or state-sufficiency assumption. For nonfinite action spaces, the maximum may need to be replaced by a supremum unless compactness, continuity, and measurable-selection conditions guarantee an attaining policy.}
\[
J_{\mathcal M}(\pi)=G(\pi)=V(\pi)=\mathbb E[Y^\pi].
\]
\end{theorem}

\begin{proof}
Fix a measurable regime $\pi$ and define its observed-data evaluation recursion by $v_K^\pi(\bar h_K)=Q_K(\bar h_K,\pi_K(\bar h_K))$ and
\[
v_k^\pi(\bar h_k)
=\mathbb E\!\left[
v_{k+1}^\pi(\bar H_{k+1})
\mid \bar H_k=\bar h_k,\ A_k=\pi_k(\bar h_k)
\right],
\qquad k<K.
\]
At stage $K$, consistency replaces the observed outcome by the corresponding potential outcome, positivity makes the conditional mean well defined, and sequential ignorability removes conditioning on the realized action. Suppose the same identity holds at stage $k+1$. Substituting it into the recursion for $v_k^\pi$ and integrating against $P_k$ gives the conditional mean of the potential outcome under the fixed prefix and regime continuation. Consistency identifies the realized next history with its potential counterpart, sequential ignorability identifies its conditional law with $P_k$, and iterated expectation removes the intermediate history. Backward induction therefore gives
\[
v_k^\pi(\bar h_k)
=\mathbb E\!\left[
Y(\bar a_{k-1},\pi_{k:K})
\mid \bar H_k=\bar h_k
\right]
\]
almost surely for each fixed $\pi$ \citep{murphy2003dtr}.

For every $\pi$, $v_K^\pi\leq V_K$. If $v_{k+1}^\pi\leq V_{k+1}$, then
\[
v_k^\pi(\bar h_k)
\leq Q_k(\bar h_k,\pi_k(\bar h_k))
\leq V_k(\bar h_k).
\]
Backward induction therefore gives $v_k^\pi\leq V_k$ for each fixed continuation regime. Finite action sets and the fixed tie-breaking rule make each $\pi_k^{\mathrm B}$ measurable, and the same induction gives equality under $\pi^{\mathrm B}$. Thus $\pi^{\mathrm B}$ is optimal over $\Pi_{\mathrm{all}}$ without taking an intersection of regime-specific almost-sure sets.

The policy-evaluation and optimality recursions in $\mathcal M$ are \eqref{eq:dtr_terminal}--\eqref{eq:dtr_bellman}. By the definition of expected terminal reward under its fixed kernels,
$J_{\mathcal M}(\pi)$ is the iterated integral in \eqref{eq:k_stage_gformula}, so $J_{\mathcal M}(\pi)=G(\pi)$. Integrating the evaluation recursion backward gives
\[
G(\pi)=\int v_1^\pi(\bar h_1)\,\mathbb P_{\bar H_1}(d\bar h_1).
\]
The identification induction at $k=1$ and iterated expectation give
$G(\pi)=\mathbb E[Y^\pi]$, which equals $V(\pi)$ by definition.
\end{proof}

\subsubsection*{Expository example: depression treatment in STAR*D}

\citet{schulte2014qlearning} apply the recursion to the Sequenced Treatment Alternatives to Relieve Depression trial. The trial enrolls 4,041 adults with major depressive disorder and begins everyone on citalopram. The application treats levels 2 and 3 as two decision stages and uses complete records from 795 patients entering level 2, of whom 330 continue to level 3. At each stage, $A_k=1$ denotes switching treatment and $A_k=0$ denotes augmenting the current treatment. The terminal outcome $Y$ is the negative end-of-stage depression score for patients who improve after level 2 and the average negative score over levels 2 and 3 for those who continue, so larger values indicate less severe depression.

The first history $\bar H_1$ contains the baseline and pre-level-2 Quick Inventory of Depressive Symptomatology (QIDS) scores and their prior slope. The action $A_1$ is the level-2 switch or augmentation decision. The potential state $S_2(a_1)$ is the depression score after level 2 under choice $a_1$, which determines whether the patient reaches level 3. For patients who continue, $\bar H_2$ adds this state and $A_1$, while $A_2$ is the level-3 choice. The potential outcome $Y(a_1,a_2)$ records the final depression outcome under both choices, and a regime $\pi$ maps each observed score history to the next choice.

Stable treatment versions require switch and augmentation to denote well-defined strategies, including how a specific treatment is selected within each category. No interference requires one patient's treatment to leave another patient's outcome unchanged. Consistency connects each observed depression path to the potential path under the choices received. Sequential exchangeability requires the recorded history to control preference and clinical factors that jointly predict the binary choice and later depression. Positivity requires both choices to occur for every history on which the target regime might use them. Specific treatments are randomized within a patient's stated preference, but the switch-versus-augmentation choice is observational, so neither exchangeability nor positivity follows from that randomization. The complete-case analysis also requires dropout not to select patients on unaccounted determinants of their potential outcomes.

For patients who continue, $Q_2(\bar h_2,a_2)$ is the conditional mean of $Y$ under each final choice, and maximizing it gives the fitted level-3 action and $V_2(\bar h_2)$. Regressing this fitted continuation value on $(\bar H_1,A_1)$ supplies $Q_1$ and the level-2 action. Under the identification assumptions, the resulting observed-data recursion equals $\mathbb E[Y^\pi]$ for each supported regime. The fitted $Q$-learning rule switches at level 2 when the prior depression-score slope exceeds $-1.09$ and switches every patient who reaches level 3. The stage-2 switch contrast is $1.10$ with 95\% confidence interval $[0.02,2.19]$, while the first-stage slope interaction is $1.02$ with interval $[-0.08,2.11]$. The theorem identifies the population recursion, but the reported rule also depends on the working $Q$-function models, and the application does not report a confidence interval for its value.

\subsubsection*{The dictionary}

Table~\ref{tab:rl_ci_dictionary} records the exact correspondences implied by Theorem~\ref{thm:dtr_rl_equivalence}. They hold for the induced finite-horizon MDP whose state is the full observed history, whose intermediate rewards are zero, whose terminal reward is $Y$, and whose discount factor is one. The value identity also requires consistency, sequential exchangeability, and positivity for the target regime.

The remaining methods do not form exact pairs. G-computation and fitted-$Q$ evaluation can estimate the same conditional recursion, but their fitted models need not coincide. Inverse-probability-of-treatment weighting (IPTW) for a marginal structural model and off-policy importance sampling both use likelihood ratios, but IPTW fits a causal model and may use stabilized weights, whereas importance sampling reweights target-policy returns \citep{robins2000msm, precup2000}. G-estimation for a structural nested mean model (SNMM) is a distinct moment method; Dynamic DML is an orthogonal, cross-fitted form of g-estimation rather than a generic synonym for orthogonal off-policy evaluation \citep{robins2004snmm, lewisSyrgkanis2021dynamicDML}. A blip is a contrast against a reference action, not the policy-centered RL advantage \citep{schulte2014qlearning}. Sequential exchangeability is a causal identification assumption, while a known logging policy describes data collection. Positivity and coverage express the same support condition only after the causal target regime and the induced MDP have been fixed.

\begin{table}[htbp]
\centering
\caption{Exact correspondences between the identified DTR recursion and the induced full-history MDP in Theorem~\ref{thm:dtr_rl_equivalence} \citep{murphy2003dtr, schulte2014qlearning}.}
\label{tab:rl_ci_dictionary}
{\setstretch{1.0}\small
\renewcommand{\arraystretch}{1.25}
\begin{tabularx}{\textwidth}{>{\raggedright\arraybackslash}X >{\raggedright\arraybackslash}X}
\toprule
Dynamic treatment regime & Induced finite-horizon MDP \\
\midrule
Observed history $\bar H_k$ & Stage-$k$ state $S_k^{\mathcal M}=\bar H_k$ \\
Treatment $A_k\in\mathcal A_k$ & Action $A_k\in\mathcal A_k$ \\
Rule $\pi_k(\bar h_k)$ & Deterministic history-state policy $\pi_k(S_k^{\mathcal M})$ \\
Terminal objective $Y$ & Rewards $R_k^{\mathcal M}=0$ for $k<K$ and terminal return $G^{\mathcal M}=Y$ \\
Assignment law $e_k(a\mid\bar h_k)=\mathbb P(A_k=a\mid\bar H_k=\bar h_k)$ & Behavior policy $\mu_k(a\mid S_k^{\mathcal M})=e_k(a\mid\bar h_k)$ \\
Fixed-regime recursion $v_k^\pi(\bar h_k)$ & Bellman policy-evaluation recursion for $\pi$ \\
Optimal-benefit recursion $(Q_k,V_k)$ & Bellman optimality recursion $(Q_k^{\mathcal M},V_k^{\mathcal M})$ \\
Unrestricted optimal rule $\pi_k^{\mathrm B}(\bar h_k)$ with fixed tie-breaking & Greedy optimal policy $\pi_k^{\mathcal M,\mathrm B}(S_k^{\mathcal M})$ with the same tie-breaking \\
$V(\pi)=\mathbb E[Y^\pi]=\textnormal{g-formula}(\pi)$ & Policy value $J_{\mathcal M}(\pi)$ \\
\bottomrule
\end{tabularx}
}
\end{table}
\FloatBarrier

\subsubsection*{Inference after policy learning: an ADHD SMART}

\citet{laber2014dtrchallenges} analyze 138 of the 155 children enrolled in the Adaptive Pharmacological and Behavioral Treatments for Children with ADHD sequential multiple-assignment randomized trial. The analysis excludes 14 children who never receive an initial randomization and three with extensive missing data. Children first receive low-dose medication or behavioral modification. Starting after two months, nonresponders are randomized to intensify that treatment or augment it with the alternative; 81 children receive this second randomization. The stage-2 history records the initial treatment, adherence, and month of nonresponse. The outcome is the reverse-coded teacher Impairment Rating Scale score after 32 weeks, so larger values indicate better classroom functioning.

Backward $Q$-learning fits a separate regression at each randomization. Its estimated stage-2 rule recommends intensification for children with high adherence and augmentation for children with low adherence. The stage-1 rule recommends medication after prior medication exposure and behavioral modification otherwise. The intervals are less decisive. Every 90\% adaptive interval for a stage-1 treatment contrast contains zero. At stage 2, the two low-adherence intervals are $[-2.21,-0.57]$ and $[-2.51,-0.60]$, while both high-adherence intervals contain zero. The empirical distinction is therefore not between having and lacking an estimated rule. It is between histories where the data distinguish its prescribed action and histories where they do not.

This gap between an argmax and the evidence behind it is the practical consequence of nonregularity. Near histories where the optimal treatment contrast is zero, the max operator produces a nonregular limiting distribution, and the usual interval $\hat\theta\pm1.96\,\widehat{\operatorname{se}}(\hat\theta)$ can fail \citep{laber2014dtrchallenges, wangtom2025dtrtutorial}. Valid options include adaptive confidence intervals for the contrast, an $m$-out-of-$n$ bootstrap, or honest policy evaluation on observations not used to choose the rule. The last interval concerns the value of the trained policy conditional on its training sample, not the unknown optimal rule itself. If $\hat\phi_i(\hat\pi)$ is a valid cross-fitted influence score on an evaluation sample of size $n_e$, its standard error is
\[
\widehat{\operatorname{se}}\{\hat V(\hat\pi)\}
=\left[
\frac{1}{n_e(n_e-1)}
\sum_{i=1}^{n_e}
\{\hat\phi_i(\hat\pi)-\hat V(\hat\pi)\}^2
\right]^{1/2}.
\]

Retrospective sepsis and ventilation studies have fit offline policies \citep{kaushik2022sepsiscql, kondrup2023deepvent}. Controlled evaluations show sensitivity to reward weighting, overfitting, noise, missingness, and pharmacokinetic structure \citep{roggeveen2024icmrl, luo2024dtrbench}. These are supporting offline studies, not evidence that deploying the recommendations improves patient outcomes.

\subsection{Simulation Study: Murphy's Recursion and $Q$-Learning}
\label{subsec:sim_dtr_recursion}

The stylized Fast Track setting has two decision stages. The recoded family-functioning status $S_1$ is uniform on $\{1,\ldots,5\}$, with higher values indicating stronger functioning, and $A_k=1$ denotes an additional home visit. The logging policy assigns a visit with probability $\operatorname{expit}(1-0.4S_k)$. At stage 1, a visit raises a low status by one level with probability $0.6$, no visit lowers it with probability $0.3$, and higher statuses drift one level in either direction with probability $0.15$ per direction. The terminal child and family outcome is
\[
Y=S_2-0.3A_2+1.5\,\mathbbm 1\{S_2\leq2\}A_2+\varepsilon,
\qquad
\varepsilon\sim\mathcal N(0,0.5^2).
\]
An additional visit is beneficial only at low terminal status. The continuous design replaces the score with ten child, caregiver, and family measures and uses a smooth treatment contrast that changes sign at $S_2^{(1)}=0.347$. Both designs satisfy sequential exchangeability and positivity by construction. Exact backward induction supplies the oracle. Plug-in g-computation and tabular $Q$-learning use the same cohorts. Neural Fitted $Q$-Iteration (FQI) and the deep $Q$-network (DQN) use the same two-layer, 64-unit network. FQI runs 1,000 full-batch Adam steps per stage. DQN uses 50 data passes and updates its target network every five passes. Gauss-Hermite quadrature computes the continuous-state oracle, with an independent Monte Carlo check. The naive benchmark fits $Y_i=\alpha+\beta(A_{1i}+A_{2i})+u_i$ without adjusting for either family status. It assigns visits at both stages when $\hat\beta>0$ and at neither stage otherwise.

\begin{table}[htbp]
\centering
\caption{Recovered policy value $V(\hat\pi)/V^*$ at the largest cohort size in the stylized home-visiting problem, with fixed-schedule references for the tabular design. The tabular design uses $N=10{,}000$ and 50 paired seeds. The continuous design uses $N=20{,}000$ and 20 paired seeds.}
\label{tab:dtr_qlearning_vs_murphy}
\begin{tabular}{lcc}
\toprule
Method & $V(\hat\pi) / V^*$ & MC SE \\
\midrule
Plug-in g-computation (tabular, $N=10000$) & 1.0000 & (0.0000) \\
Oracle $V^*$ (tabular) & 1.0000 & -- \\
Oracle $V^*$ (high-dim, $p=10$) & 1.0000 & -- \\
$Q$-learning (tabular, $N=10000$, 100 replays) & 0.9968 & (0.0007) \\
Neural-FQI (high-dim, $N=20000$) & 0.9842 & (0.0013) \\
DQN (high-dim, $N=20000$) & 0.9740 & (0.0012) \\
Always visit (tabular) & 0.9422 & -- \\
Unadjusted count OLS (tabular, $N=10000$) & 0.8124 & (0.0000) \\
Never visit (tabular) & 0.8124 & -- \\
\addlinespace
Paired contrast, $Q$-learning $-$ plug-in (tabular) & -0.0032 & (0.0007) \\
Paired contrast, DQN $-$ Neural-FQI (high-dim) & -0.0101 & (0.0013) \\
\bottomrule
\end{tabular}

\end{table}

\begin{figure}[htbp]
\centering
\includegraphics[width=\textwidth]{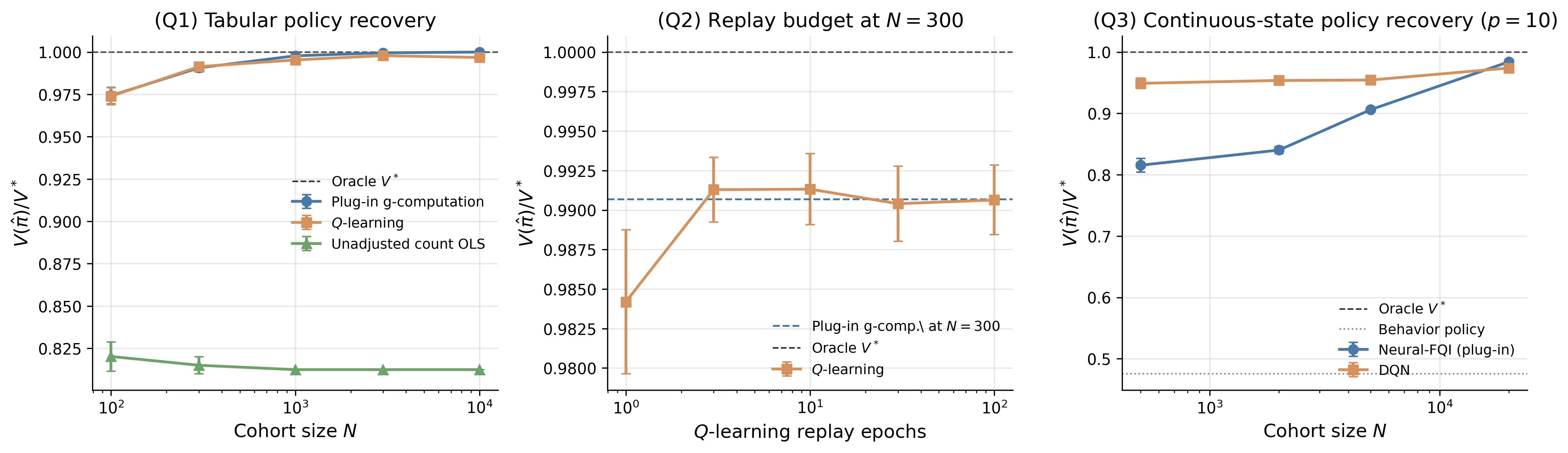}
\caption{Plug-in g-computation, $Q$-learning, and unadjusted treatment-count regression under sequential ignorability. Panel (Q1) shows the tabular sample-size sweep. Panel (Q2) shows the replay-budget sweep at $N=300$. Panel (Q3) shows the continuous-state sweep with a smooth contrast and shared network architecture. Error bars are 95\% Monte Carlo confidence intervals for mean recovery over paired seeds. They are not confidence intervals for a regime fitted to one observed cohort.}
\label{fig:dtr_qlearning_vs_murphy}
\end{figure}
\FloatBarrier

Theorem~\ref{thm:dtr_rl_equivalence} is an identity, so the simulation checks it without sampling. On a fully known instance, the backward recursion, value iteration on the materialised $56$-state history process, and the g-formula at the optimal regime all return $3.5820$ up to floating-point precision. The arbitrary-regime identity also holds to $8.9\times10^{-16}$ over $200$ randomly drawn regimes. A recovery ratio cannot establish either identity. The ratio instead measures how fast an estimator approaches the optimum.

Table~\ref{tab:dtr_qlearning_vs_murphy} reports the paired estimator contrasts and Figure~\ref{fig:dtr_qlearning_vs_murphy} shows the three sweeps behind them. Both gaps are small in value terms and many Monte Carlo standard errors from zero, which is the expected signature of stochastic approximation against a one-shot regression on identical data rather than evidence against the equivalence. The unadjusted count coefficient converges to $-0.2870$ with Monte Carlo standard error $0.0023$ and selects no visits in every large-cohort replication. The regression reads the greater needs of frequently visited families as a harmful treatment association and cannot represent a rule that changes with family status.

\subsection{Off-Policy Evaluation}
\label{subsec:ope}

Before optimizing a regime, the analyst must be able to value a fixed one. \emph{Off-policy evaluation} (OPE) estimates the value of a prescribed target policy from data logged under a different policy. The four assumptions in Section~\ref{subsec:gmethods_bridge} identify this value. Chapter~\ref{section:causal_rl} treats the confounded case, where sequential ignorability fails.

\subsubsection*{The estimand and its identification}

The setup has a finite horizon $H$ and evaluation policy $\pi_e$. The potential reward at stage $t$ under action path $\bar a_t$ is $R_t(\bar a_t)$, and $R_t^{\pi_e}=R_t(\bar A_t^{\pi_e})$. The estimand is the expected cumulative potential reward
\begin{equation}
J(\pi_e) = \mathbb{E}\!\left[\sum_{t=1}^{H} \gamma^{t-1} R_t^{\pi_e}\right],
\label{eq:ope_estimand}
\end{equation}
where the expectation is over the regime-induced counterfactual trajectories. The complete pre-action history is
\[
\mathcal G_t=(S_1,A_1,R_1,\ldots,S_{t-1},A_{t-1},R_{t-1},S_t).
\]
The behavior policy $\pi_b$ generates $n$ trajectories
$\mathcal D=\{\tau_i\}_{i=1}^n$, where
\[
\tau_i=(\mathcal G_t^i,A_t^i,R_t^i)_{t=1}^H.
\]
Reward consistency, target-policy positivity, and sequential exchangeability identify the change of measure through
\[
\rho_t
=\frac{\pi_e(A_t\mid\mathcal G_t)}
{\pi_b(A_t\mid\mathcal G_t)},
\qquad
\rho_{1:t}=\prod_{j=1}^{t}\rho_j.
\]
The identified value is
\begin{equation}
J(\pi_e)
=\mathbb{E}_{\pi_b}\!\left[
\sum_{t=1}^{H}\gamma^{t-1}\rho_{1:t}R_t
\right].
\label{eq:ope_identification}
\end{equation}
This per-decision form follows by iterated expectation and makes clear that ratios after reward $R_t$ are unnecessary. The trajectory-weighted form with $\rho_{1:H}$ multiplying the full return is also identified. Each estimator below computes the same estimand through a fitted value function, cumulative ratios, a fitted transition model, or a marginal state-density ratio.

The structural assumptions on the environment matter as much as the identification assumptions. Table~\ref{tab:ope_model_hierarchy} summarizes the nested model classes used by \citet{ueharaShiKallus2022ope}. Stronger restrictions replace cumulative history ratios by lower-dimensional marginal ratios. This can lower the semiparametric efficiency bound from exponential to polynomial growth in the horizon, but only when the restriction is scientifically credible.

\begin{table}[htbp]
\centering
\caption{Model hierarchy for off-policy evaluation \citep{ueharaShiKallus2022ope}. The ratio column gives the correction used by an efficient score.}
\label{tab:ope_model_hierarchy}
{\setstretch{1.0}\small
\renewcommand{\arraystretch}{1.25}
\begin{tabularx}{\textwidth}{p{0.16\textwidth} p{0.22\textwidth} p{0.22\textwidth} >{\raggedright\arraybackslash}X}
\toprule
Model class & Information and transition restriction & Required correction & Statistical consequence \\
\midrule
Contextual bandit & One decision, $H=1$ & Action ratio $\pi_e/\pi_b$ & Static efficient influence function \\
Non-Markov process & Complete history $\mathcal G_t$; transitions may depend on the past & Cumulative history ratio $\rho_{1:t}$ & Efficiency bound can grow exponentially in $H$ \\
Time-varying MDP & $S_t$ is sufficient; transition law may vary with $t$ & Stage-specific marginal state-action ratio & Polynomial horizon dependence is attainable \\
Stationary MDP & Markov and time-homogeneous transition law & Stationary occupancy or density ratio & Discounted or average-reward values can be studied from mixing trajectories \\
\bottomrule
\end{tabularx}
}
\end{table}

\subsubsection*{The direct method}

The \emph{direct method} (DM) estimates the conditional-mean structure and plugs it into the Bellman recursion. The displayed state-only form applies to the time-varying or stationary MDP classes in Table~\ref{tab:ope_model_hierarchy}. The non-Markov version replaces every $S_t$ with the complete history $\mathcal G_t$. The recursion starts from $\hat{q}_{H+1} \equiv 0$ and, for $t = H, \ldots, 1$, fits
\begin{equation}
\hat{q}_t(s, a) \approx \widehat{\mathbb{E}}\!\left[R_t + \gamma\, \hat{q}_{t+1}(S_{t+1}, \pi_e) \mid S_t = s,\ A_t = a\right],
\qquad
\hat{J}_{\mathrm{DM}} = \frac{1}{n} \sum_{i=1}^{n} \hat{q}_1(S_1^i, \pi_e),
\label{eq:ope_dm}
\end{equation}
where $\hat{q}(s, \pi) = \sum_a \pi(a \mid s)\, \hat{q}(s, a)$. The model-based variant fits $(\hat{P}, \hat{r})$ and runs the exact recursion in the fitted model; in a tabular environment the two coincide at the maximum-likelihood estimate, and both are g-formula plug-ins. No importance ratios appear, but misspecification of the regression class induces bias \citep{jiangli2016doubly, farajtabar2018mrdr}.

A direct-method point estimate does not carry a valid standard error merely because the final stage is fit by regression. One option embeds the regression in an orthogonal score, as in the DR construction. For linear FQE under policy completeness, another resamples independent trajectories, refits the complete recursion in every bootstrap sample, and uses quantiles of $\hat J_{\mathrm{FQE}}^*-\hat J_{\mathrm{FQE}}$ \citep{hao2021bootstrapfqe}. Transitions within an episode must not be resampled as if they were independent. With bootstrap quantile $\hat q_u$, a two-sided interval is
\[
\left[
\hat J_{\mathrm{FQE}}-\hat q_{1-\alpha/2},
\hat J_{\mathrm{FQE}}-\hat q_{\alpha/2}
\right].
\]

\subsubsection*{The importance sampling family}

\emph{Importance sampling} (IS) estimators use the ratios instead of a model. The trajectory-wise and \emph{per-decision importance sampling} (PDIS) forms are
\begin{equation}
\hat{J}_{\mathrm{IS}} = \frac{1}{n} \sum_{i=1}^{n} \rho^i_{1:H} \sum_{t=1}^{H} \gamma^{t-1} R^i_t,
\qquad
\hat{J}_{\mathrm{PDIS}} = \frac{1}{n} \sum_{i=1}^{n} \sum_{t=1}^{H} \gamma^{t-1} \rho^i_{1:t}\, R^i_t,
\label{eq:ope_is}
\end{equation}
both unbiased for $J(\pi_e)$ when $\pi_b$ is known and positivity holds \citep{precup2000, jiangli2016doubly}. Per-decision weighting drops ratio factors that occur after each reward and usually reduces variance. The corresponding trajectory-level summands are $Z_i^{\mathrm{IS}}$ and $Z_i^{\mathrm{PDIS}}$. If trajectories are independent and $\mathbb E[Z_i^2]<\infty$, either estimator has the asymptotic standard error and Wald interval
\begin{equation}
\widehat{\operatorname{se}}(\hat J)
=\left[
\frac{1}{n(n-1)}
\sum_{i=1}^{n}(Z_i-\bar Z)^2
\right]^{1/2},
\qquad
\mathrm{CI}_{1-\alpha}
=\hat J\pm z_{1-\alpha/2}\widehat{\operatorname{se}}(\hat J).
\label{eq:ope_is_se}
\end{equation}
Positivity alone does not guarantee a usable finite-sample variance. Near-zero behavior probabilities can make the second moment enormous even when the value is identified.

The \emph{weighted importance sampling} (WIS) variant divides by the realized weight mass instead of $n$. With $\bar w_t=n^{-1}\sum_i\rho^i_{1:t}$, its step-wise form is $\hat J_{\mathrm{WIS}}=\sum_t\gamma^{t-1}\sum_i\rho^i_{1:t}R_t^i/\sum_i\rho^i_{1:t}$. WIS is biased in finite samples but consistent and typically lower variance than IS \citep{jiangli2016doubly, thomas2015hcope}. Its standard error is a ratio-estimator calculation, not Equation~\eqref{eq:ope_is_se}. For $\hat\mu_t=\sum_i\rho^i_{1:t}R_t^i/\sum_i\rho^i_{1:t}$, an empirical delta-method influence value is
\[
\hat\phi_i^{\mathrm{WIS}}
=\sum_{t=1}^{H}\gamma^{t-1}
\frac{\rho^i_{1:t}(R_t^i-\hat\mu_t)}{\bar w_t}.
\]
The sample variance of $\hat\phi_i^{\mathrm{WIS}}$ divided by $n$ gives an asymptotic variance estimate. A trajectory bootstrap is preferable when weight tails make the linear approximation doubtful.\footnote{Continuous action spaces break exact action matching. \citet{kalluszhou2018continuous} replace the discrete-action ratio with a kernel-smoothed generalized propensity and characterize the bandwidth's bias-variance tradeoff.}

\subsubsection*{The doubly robust family}

The direct method and importance sampling fail in complementary ways, the former through model bias and the latter through weight variance, and the \emph{doubly robust} (DR) estimator uses each to repair the other. In the bandit case,
\begin{equation}
\hat{J}_{\mathrm{DR}} = \frac{1}{n} \sum_{i=1}^{n} \left[ \hat{q}(S^i, \pi_e) + \rho^i \left( R^i - \hat{q}(S^i, A^i) \right) \right],
\label{eq:ope_dr_bandit}
\end{equation}
At the population-score level, its expectation equals $J(\pi_e)$ if either $q$ is the true outcome regression or the ratio uses the true propensity \citep{dudik2014doubly}. This exact double-robustness statement treats the nuisance functions as fixed. Sample splitting or cross-fitting makes a fitted nuisance independent of each scored observation, but an estimated correctly specified model is not literally equal to the truth in finite samples. \citet{jiangli2016doubly} extend the construction to the sequential problem by observing that per-decision IS satisfies a backward recursion and applying the bandit correction at every stage. With $\hat{V}_{\mathrm{DR}}^{0} = 0$,
\begin{equation}
\hat{V}_{\mathrm{DR}}^{H+1-t} = \hat{V}(S_t) + \rho_t\!\left( R_t + \gamma\, \hat{V}_{\mathrm{DR}}^{H-t} - \hat{Q}(S_t, A_t) \right),
\qquad t = H, \ldots, 1,
\label{eq:ope_dr_seq}
\end{equation}
and $\hat{J}_{\mathrm{DR}}$ averages $\hat{V}_{\mathrm{DR}}^{H}$ over trajectories. Their Theorem~1 gives an exact variance decomposition showing that the $\hat{Q}$-dependent term enters through the error $\hat{Q} - Q$, so an accurate value model reduces action-induced variance. With correct known behavior ratios and a $Q$ function fixed independently of the scored trajectory, unbiasedness survives an inaccurate $Q$; per-decision IS is the special case $\hat{Q} \equiv 0$. For discrete tree MDPs, their Theorem~2 gives a constrained Cramér-Rao lower bound and Observation~1 shows that DR with the true $Q$ attains it. \citet{thomasBrunskill2016magic} self-normalize the weights inside the same score to obtain the \emph{weighted doubly robust} (WDR) estimator, and blend partial-horizon WDR estimates with the model estimate by minimizing an estimated mean-squared error. The resulting model and guided importance sampling combining (MAGIC) estimator accepts finite-sample bias to reduce variance. \citet{farajtabar2018mrdr} train $\hat{Q}$ to minimize the DR estimator's variance rather than a generic regression loss.

For inference, the nuisance functions are fitted out of fold, and $\hat\phi_i^{\mathrm{DR}}$ is the complete trajectory score returned by the recursion. The estimator and its standard error are
\begin{equation}
\hat J_{\mathrm{DR}}=\frac1n\sum_i\hat\phi_i^{\mathrm{DR}},
\qquad
\hat V_{\mathrm{DR}}
=\frac1n\sum_i
\left(\hat\phi_i^{\mathrm{DR}}-\hat J_{\mathrm{DR}}\right)^2,
\qquad
\widehat{\operatorname{se}}(\hat J_{\mathrm{DR}})
=\sqrt{\frac{\hat V_{\mathrm{DR}}}{n}}.
\label{eq:ope_dr_se}
\end{equation}
The Wald interval based on Equation~\eqref{eq:ope_dr_se} requires finite score variance, adequate overlap, cross-fitting, and a product-rate condition on the nuisance errors. WDR, MAGIC, and MRDR are useful point-estimation variants, but their self-normalization, adaptive mixture, or targeted nuisance fitting changes the score. Equation~\eqref{eq:ope_dr_se} cannot be copied to them without deriving the corresponding influence function or using a trajectory bootstrap.

\subsubsection*{The curse of horizon and marginalized ratios}

The cumulative ratio $\rho_{1:H}$ is a product of $H$ terms. Under mild conditions its logarithm obeys a central limit theorem, so the weight is approximately log-normal and its variance can grow exponentially in the horizon. This is the \emph{curse of horizon} \citep{xie2019marginalized}.\footnote{On the environment of Section~\ref{subsec:sim_ope} the growth is computed exactly rather than simulated. Squaring the ratio leaves one power of $\pi_b$ in the denominator, so the per-step weight is $w(s,a)=\pi_e(a\mid s)^2/\pi_b(a\mid s)$; the transition kernel cancels from the ratio but survives in the trajectory measure, giving $K(s'\mid s)=\sum_a w(s,a)P(s'\mid s,a)$, whose iterates applied to the initial distribution give $\mathbb E_b[\rho_{1:H}^2]$ as a total mass. The row sums of $K$ are $1+\chi^2(\pi_e(\cdot\mid s)\,\|\,\pi_b(\cdot\mid s))$, at least one and equal to one only where the policies agree, so each step multiplies the mass by one plus a divergence. The variance runs $2.8$, $14.4$, $256$, $7.3\times10^{4}$, $5.8\times10^{9}$ at $H=4,8,16,32,64$, a factor of $1.43$ per step. Estimating the same quantity from $100{,}000$ simulated episodes returns an $H=64$ value below the $H=32$ one, which is impossible for a variance and is the curse defeating its own measurement, since the mass sits in trajectories too rare to be drawn.} When the environment is a genuine MDP, the trajectory ratio is unnecessary. It suffices to reweight by the \emph{marginal} state-density ratio at each stage. The marginalized importance sampling (MIS) estimator is
\begin{equation}
\hat{J}_{\mathrm{MIS}} = \frac{1}{n} \sum_{i=1}^{n} \sum_{t=1}^{H} \gamma^{t-1}\, \frac{\hat{d}^{\pi_e}_t(S^i_t)}{\hat{d}^{\pi_b}_t(S^i_t)}\, \rho^i_t\, R^i_t,
\label{eq:ope_mis}
\end{equation}
where $\hat{d}^{\pi_b}_t$ is the empirical state distribution and $\hat{d}^{\pi_e}_t$ follows the forward recursion $\hat{d}^{\pi_e}_{t+1} = \hat{P}^{\pi_e}_t \hat{d}^{\pi_e}_t$ through a ratio-weighted empirical transition operator. Theorem~4.1 of \citet{xie2019marginalized} bounds the mean-squared error by a polynomial in $H$, matching the Cramér-Rao lower bound for the MDP class up to a factor of $H$. By contrast, the NMDP efficiency bound is generally exponential in $H$ \citep{kallusUehara2022doubleRL}. A single stationary density ratio $w(s,a)$ replaces the stage-specific ratios in the infinite-horizon stationary setting. \citet{nachum2019dualdice} estimate it through a convex saddle-point program without knowledge of $\pi_b$. Double reinforcement learning combines an estimated ratio with a fitted value function and attains the semiparametric efficiency bound when the two nuisance errors are consistent and their product is $o_p(n^{-1/2})$ \citep{kallusUehara2022doubleRL}. Its cross-fitted trajectory score uses the marginal ratio in place of $\rho_{1:t}$ in Equation~\eqref{eq:ope_dr_se}; the empirical variance and standard error are then computed in exactly the same way.

\subsubsection*{Efficiency, high-confidence bounds, and estimator selection}

For the bandit case, the efficient influence function of $J$ is
\[
\phi(O;\eta,q,J)
=\eta(S,A)\{R-q(S,A)\}
+q(S,\pi_e)-J,
\qquad
q(S,\pi_e)=\sum_a\pi_e(a\mid S)q(S,a),
\]
with $\eta=\pi_e/\pi_b$. The efficiency bound is $\mathbb E[\eta^2\operatorname{var}(R\mid S,A)]+\operatorname{var}\{q(S,\pi_e)\}$ \citep{ueharaShiKallus2022ope}. With cross-fitted nuisances, evaluate $\hat\phi_i$ out of fold, center it at $\hat J$, and compute $\hat V=n^{-1}\sum_i\hat\phi_i^2$ and $\widehat{\operatorname{se}}=\sqrt{\hat V/n}$. A cross-fitted DR estimator attains the bound when its nuisance errors are consistent and their product is $o_p(n^{-1/2})$; $o_p(n^{-1/4})$ for each is a symmetric sufficient condition. The bound constrains estimators that do not assume a correctly specified parametric model, so a correct parametric plug-in may sit below it. Section~\ref{subsec:sim_ope} shows this ranking under correct specification and its reversal when the model is wrong. The sequential extensions replace $\eta$ by the cumulative ratio in the NMDP class and by the marginal density ratio in the time-varying MDP class. The latter efficiency bound is generally strictly smaller and polynomial rather than exponential in the horizon \citep{kallusUehara2022doubleRL}.

\emph{High-confidence off-policy evaluation} constructs a one-sided lower confidence bound from heavy-tailed importance-weighted returns. Suppose the return has first been normalized so that each importance-weighted score satisfies $Z_i\geq0$. For a threshold $c>0$ fixed independently of the evaluation sample, $Z_i^c=\min(Z_i,c)\in[0,c]$. A representative empirical Bernstein form subtracts
\[
\sqrt{\frac{2\hat\sigma_c^2\log(2/\alpha)}{n}}
+\frac{7c\log(2/\alpha)}{3(n-1)}
\]
from $\bar Z^c$ \citep{thomas2015hcope}. If rewards can be negative, a known lower return bound must be used to shift and rescale them before applying this result, after which the bound is transformed back. This is a safety certificate, not a symmetric standard error, and its validity depends on the tail treatment and confidence level being chosen as required by the bound. \citet{sakhi2024logsmoothing} instead smooth the ratios logarithmically and obtain bounds valid for policy selection as well as evaluation. Estimator choice remains problem-dependent. Stability across hyperparameters and adaptation to the target policy can be more informative than a global benchmark rank \citep{saito2021robustope, udagawa2023pas}. Chapter~\ref{section:field_deployments} examines these issues on production-scale logs.

\subsubsection*{Lead application: evaluating mobile-health policies}

\citet{liaoMurphy2021longterm} provide a direct sequential application using the first HeartSteps micro-randomized trial. The trial enrolls 44 sedentary adults and randomizes an activity suggestion with probability $0.6$ at five decision times per day for 42 days, provided the participant is available. The analysis retains 37 participants after excluding three with technical problems and four who drop out early; it also removes decision times affected by travel, technical problems, or a missing reward. The state includes recent and previous-day step counts, location, temperature, notification burden, time of day, and recent variability in activity. The reward is the log of steps during the 30 minutes after a decision.

The paper evaluates three long-run stationary policies. They send no suggestions, send a suggestion whenever the participant is available, or send one only when the participant is available at home or work. About 44\% of available decision times occur at home or work, so the last two policies assign treatment at different rates. The estimator learns stationary state-density ratios and relative value functions in reproducing-kernel Hilbert spaces with radial-basis kernels. This makes the application a stationary-MDP problem rather than a contextual bandit. A suggestion can change later states, and each policy changes the state distribution under which its long-run reward is evaluated.

The location policy has an estimated average log-step reward of $3.155$ with a 95\% confidence interval of $[2.893,3.417]$. The no-suggestion policy has an estimated value of $2.962$, and the interval for the location-policy contrast is $[-0.016,0.402]$. The always-suggest policy has an estimated value of $3.127$ with interval $[2.840,3.413]$; its contrast with the location policy has interval $[-0.161,0.217]$. The authors translate the location-versus-no-suggestion point difference into roughly 55 steps, or 22\% of the mean post-decision count of 248. That back-transformation describes the point estimate. Neither policy contrast excludes zero, so it does not establish that the location rule improves activity or outperforms always sending a suggestion.

\subsection{Simulation Study: Off-Policy Evaluation}
\label{subsec:sim_ope}

The stylized Fast Track MDP has $\gamma=1$ and five family-functioning states $S_t\in\{0,\ldots,4\}$, with zero denoting greatest need. Status moves at most one level per monitoring period. Without an additional visit, the upward and downward probabilities are $(0.15,0.30)$; a visit changes them to $(0.45,0.15)$. The net rewards are $r(s,0)=0.25s$ and $r(s,1)=0.25s-0.5(1-0.1s)$, so a visit trades improved functioning against burden. The behavior policy visits with probability $0.6-0.1s$. The evaluation policy uses probability $0.9$ in states 0 and 1 and $0.1$ otherwise. Initial status is uniform on states 1, 2, and 3. Exact backward induction supplies $J(\pi_e)$. The horizon grid from 4 to 64 is a statistical stress test, not the Fast Track semester schedule. DM fits the tabular transition and reward model; DR and WDR use two-fold cross-fitting; MIS uses the forward marginal-ratio recursion of \citet{xie2019marginalized}. Seven estimators score the same logged data in each Monte Carlo replication.

\begin{table}[htbp]
\centering
\caption{OPE results for the stylized home-visiting MDP. Panel A reports point-estimator performance at $n=2000$ and $H=16$, with relative RMSE at $n=500$ and $H=64$. Panel B reports the nuisance-misspecification design at $n=500$ and $H=16$. Panel C assesses estimator-level standard errors at $n=1000$ and $H=8$.}
\label{tab:simA_ope}
\begin{minipage}{\textwidth}
\centering
\emph{Panel A. Point-estimator performance over 500 replications}\par\smallskip
\begin{tabular}{lrrrr}
\toprule
Estimator & Bias & SD & RMSE & Rel. RMSE at $H=64$ \\
\midrule
DM & +0.0024 & 0.1113 & 0.1112 & 0.026 \\
MIS & +0.0061 & 0.2137 & 0.2136 & 0.134 \\
WDR & -0.0181 & 0.3438 & 0.3439 & 0.209 \\
DR & -0.0179 & 0.3661 & 0.3661 & 1.242 \\
WIS & +0.0490 & 0.4580 & 0.4601 & 0.286 \\
PDIS & +0.0029 & 0.6013 & 0.6007 & 2.055 \\
IS & +0.1325 & 1.5511 & 1.5552 & 2.847 \\
\bottomrule
\end{tabular}

\par\medskip
\emph{Panel B. Double-robustness ablation over 500 replications}\par\smallskip
\begin{tabular}{lllrrr}
\toprule
Estimator & $\hat Q$ model & Propensity & Bias & MC SE & RMSE \\
\midrule
DR & right & right & -0.0254 & (0.0342) & 0.7643 \\
DR & wrong & right & -0.0214 & (0.0349) & 0.7805 \\
DR & right & wrong & +0.0771 & (0.2876) & 6.4246 \\
DR & wrong & wrong & +1.7061 & (0.3434) & 7.8574 \\
DM & right & -- & +0.0011 & (0.0093) & 0.2077 \\
DM & wrong & -- & +0.3769 & (0.0075) & 0.4128 \\
PDIS & -- & right & -0.0627 & (0.0511) & 1.1428 \\
PDIS & -- & wrong & +79.1185 & (1.5846) & 86.6760 \\
\bottomrule
\end{tabular}

\par\medskip
\emph{Panel C. Standard-error calibration over 1,000 replications}\par\smallskip
\resizebox{\textwidth}{!}{\begin{tabular}{lrrrrrr}
\toprule
Estimator & Bias & Emp. SD & Mean analytic SE & SE ratio & Coverage & Tail misses (L/R) \\
\midrule
IS & +0.0149 & 0.3143 & 0.3068 & 0.976 & 0.943 & 0.020/0.037 \\
PDIS & +0.0070 & 0.1800 & 0.1786 & 0.992 & 0.953 & 0.027/0.020 \\
DR & +0.0028 & 0.1225 & 0.1242 & 1.014 & 0.945 & 0.030/0.025 \\
\bottomrule
\end{tabular}
}
\end{minipage}
\end{table}

\begin{figure}[htbp]
\centering
\includegraphics[width=\textwidth]{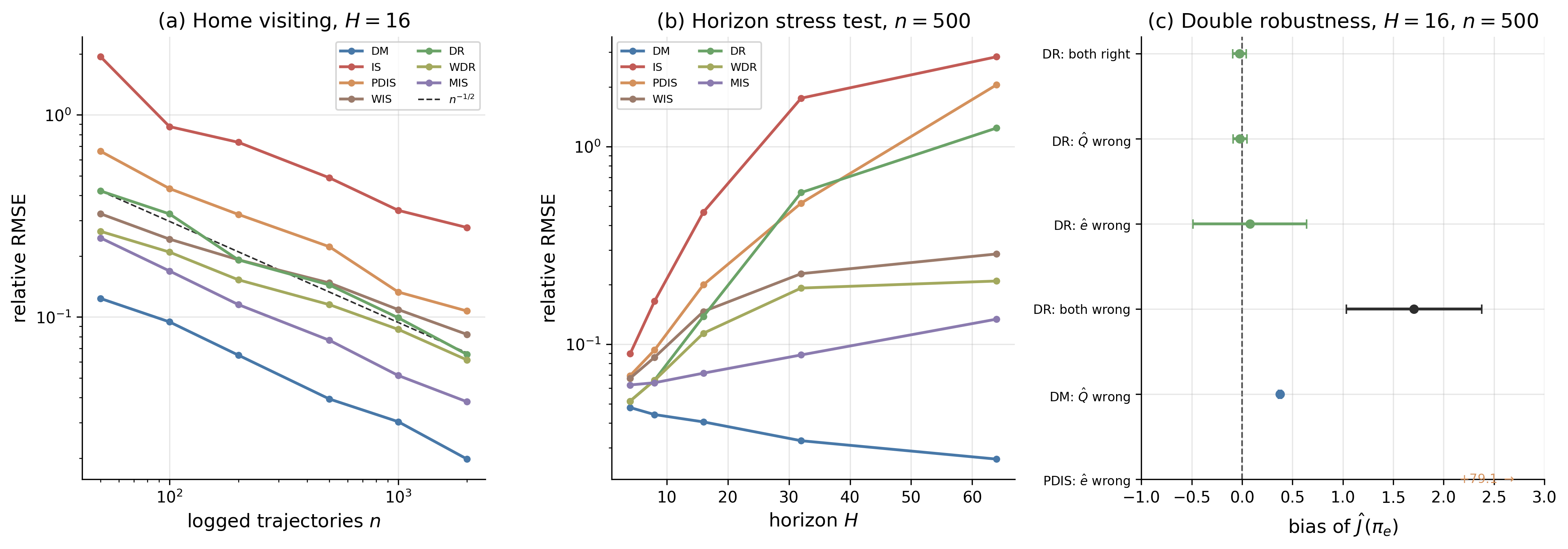}
\caption{OPE performance over 500 replications. Panel (a) shows relative RMSE against the number of logged trajectories at $H=16$. Panel (b) shows relative RMSE against horizon at $n=500$. Panel (c) shows estimator bias under nuisance misspecification with 95\% Monte Carlo confidence intervals for mean bias. These error bars are distinct from the estimator-level intervals in Equations~\eqref{eq:ope_is_se} and \eqref{eq:ope_dr_se}.}
\label{fig:simA_ope}
\end{figure}
\FloatBarrier

The horizon sweep measures the mechanism behind the estimator gap. Regressing the log of the exact $\operatorname{var}(\rho_{1:H})$ on $H$ gives a slope of $0.356$ per step with residual standard deviation $0.085$. Regressing the same quantity on $\log H$ gives residual standard deviation $2.72$. The trajectory-weight variance therefore grows geometrically rather than polynomially. The variance of the marginal state-density ratio used by MIS is $0.13$, $0.47$, $1.27$, $2.92$, and $6.28$ over the same horizons, consistent with the polynomial behaviour in Theorem~4.1 of \citet{xie2019marginalized}.

Panel B of Table~\ref{tab:simA_ope} separates consistency from precision. The two one-sided misspecification cells are statistically indistinguishable from zero, while the both-wrong cell is five standard errors from zero. The wrong-propensity cell also has nearly the RMSE of the both-wrong cell, so double robustness delivers consistency there and not precision. Panel C shows calibrated uncertainty for IS, PDIS, and cross-fitted DR. Figure~\ref{fig:simA_ope} plots the same comparisons against sample size and horizon.

\FloatBarrier
\subsection{Engine Replacement MDP: Every Importance Weight, Written Out}
\label{engine:ch10b}

\begin{table}[H]
\centering
\caption{Every per-step and cumulative importance weight in the twelve-trajectory Engine Replacement MDP log. Panel B compares the five estimates with the exact finite-horizon dynamic-programming value.}
\label{tab:engine_ope}
{\scriptsize\renewcommand{\arraystretch}{1.05}
\begin{tabular}{crrrrr}
\toprule
& \multicolumn{4}{c}{$\rho_t\,/\,\rho_{1:t}$} & discounted return \\
\cmidrule(lr){2-5}
trajectory & $t=1$ & $t=2$ & $t=3$ & $t=4$ & $G_i$ \\
\midrule
1 & 0\,/\,0 & 0\,/\,0 & 0\,/\,0 & 0\,/\,0 & -1.720 \\
2 & 2\,/\,2 & 0\,/\,0 & 0\,/\,0 & 0\,/\,0 & 1.488 \\
3 & 0\,/\,0 & 2\,/\,0 & 2\,/\,0 & 0\,/\,0 & 1.356 \\
4 & 0\,/\,0 & 2\,/\,0 & 0\,/\,0 & 0\,/\,0 & 0.708 \\
5 & 2\,/\,2 & 0\,/\,0 & 2\,/\,0 & 2\,/\,0 & 1.504 \\
6 & 0\,/\,0 & 0\,/\,0 & 0\,/\,0 & 0\,/\,0 & -1.720 \\
7 & 0\,/\,0 & 0\,/\,0 & 0\,/\,0 & 0\,/\,0 & -1.720 \\
8 & 0\,/\,0 & 0\,/\,0 & 0\,/\,0 & 0\,/\,0 & -1.720 \\
9 & 2\,/\,2 & 0\,/\,0 & 0\,/\,0 & 2\,/\,0 & 0.978 \\
10 & 2\,/\,2 & 2\,/\,4 & 2\,/\,8 & 2\,/\,16 & 2.224 \\
11 & 2\,/\,2 & 0\,/\,0 & 0\,/\,0 & 2\,/\,0 & 0.978 \\
12 & 0\,/\,0 & 2\,/\,0 & 2\,/\,0 & 2\,/\,0 & 1.939 \\
\midrule
\multicolumn{6}{c}{Panel B. Point estimates} \\
\midrule
estimator & \multicolumn{2}{c}{estimate} & \multicolumn{3}{c}{estimate minus exact value} \\
Exact DP & \multicolumn{2}{c}{2.0502} & \multicolumn{3}{c}{+0.0000} \\
DM & \multicolumn{2}{c}{2.0660} & \multicolumn{3}{c}{+0.0158} \\
IS & \multicolumn{2}{c}{2.9653} & \multicolumn{3}{c}{+0.9151} \\
PDIS & \multicolumn{2}{c}{1.8353} & \multicolumn{3}{c}{-0.2149} \\
WIS & \multicolumn{2}{c}{2.2240} & \multicolumn{3}{c}{+0.1738} \\
DR & \multicolumn{2}{c}{1.6915} & \multicolumn{3}{c}{-0.3587} \\
\bottomrule
\end{tabular}}
\end{table}

\begin{figure}[htbp]
\centering
\includegraphics[width=\textwidth]{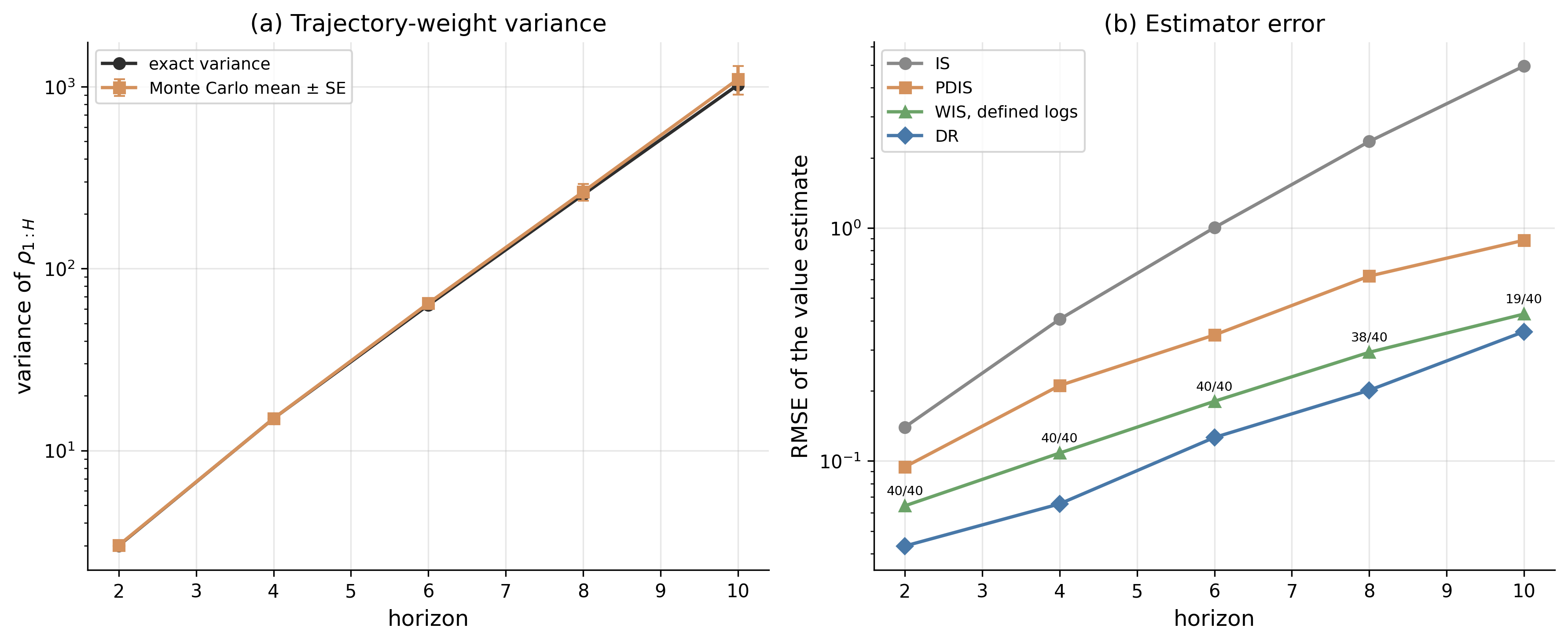}
\caption{Horizon growth in trajectory-weight variance and value-estimator error. Panel (a) compares the exact variance with the mean across forty fixed Monte Carlo seeds. Panel (b) reports root mean squared error across those seeds for IS, PDIS, and DR. WIS is conditional on logs with nonzero cumulative weight mass at every period, and each marker gives the number of defined estimates out of forty.}
\label{fig:engine_ope}
\end{figure}

The target policy keeps a low-mileage engine and replaces a high-mileage engine, while the behavior policy chooses each action with probability one half. An independent two-hundred-trajectory log fits the transition model used by DM and DR. The twelve displayed trajectories remain separate evaluation data. The exact four-period value is $2.0502$, with PDIS below it at $1.8353$ and IS above it at $2.9653$. Across forty fixed seeds, the exact variance of $\rho_{1:H}$ rises from $3$ at $H=2$ to $1{,}023$ at $H=10$. At $H=10$, the DR root mean squared error is $0.3589$, compared with $4.9572$ for IS. WIS is undefined when no logged trajectory carries positive cumulative target-policy weight at every period. The WIS conditional root mean squared error at $H=10$ uses nineteen of the forty logs rather than silently discarding the other twenty-one. This calculation applies the per-decision and sequential DR recursions of \citet{precup2000} and \citet{jiangli2016doubly} to the Engine Replacement MDP.
\FloatBarrier

\subsection{Dynamic Double Machine Learning for Structural Nested Mean Models}
\label{subsec:dml_dte}

The DTR and OPE sections estimate conditional values. Structural nested mean models (SNMMs) instead parameterize treatment contrasts. \citet{lewisSyrgkanis2021dynamicDML} combine g-estimation with cross-fitting and Neyman-orthogonal moments, obtaining $\sqrt n$ inference on low-dimensional dynamic effects while high-dimensional regressions remain nuisance functions.

For the Fast Track cohort, $\psi_t$ is the marginal causal effect of an additional home visit at semester $t$, holding future visits at a fixed reference level. Valid $\sqrt n$ inference must account for machine-learned outcome and propensity functions and for dependence among the structural moments. The same parameters determine the value of a counterfactual home-visiting regime.

\subsubsection*{Structural nested mean models}

For a target continuation regime $\pi$ and reference action $0$, the stage-$t$ \emph{blip} is the counterfactual contrast
\begin{equation}
\gamma_t^\pi(\bar h_t,a_t)
=
\mathbb E\!\left[
Y(\bar a_t,\pi_{(t+1):T})
-Y(\bar a_{t-1},0,\pi_{(t+1):T})
\mid \bar H_t=\bar h_t,\ A_t=a_t
\right].
\label{eq:blip_definition}
\end{equation}
It removes the current treatment while holding the future regime fixed; replacing the potential outcomes in this display by observed means conditional on future treatments would generally be invalid under treatment-confounder feedback. A linear SNMM writes $\gamma_t^\pi(\bar h_t,a_t)=\psi_t^{\pi\prime}B_t(\bar h_t,a_t)$ for a known basis $B_t$ \citep{robins2004snmm, lewisSyrgkanis2021dynamicDML}. In target-independent models the superscript $\pi$ drops and one structural vector can evaluate many policies; in the general model the blip and its parameter are defined for the specified target continuation.

\subsubsection*{Sequential residualisation}

\citet{lewisSyrgkanis2021dynamicDML} first derive the construction in a partially linear dynamic model, then extend it to SNMM blip-basis contrasts. In the target-independent linear-treatment specialization used in the simulation below, $\hat q_t(\bar h_t)=\widehat{\mathbb E}[Y\mid\bar H_t=\bar h_t]$ and $\hat p_{j,t}(\bar h_t)=\widehat{\mathbb E}[A_j\mid\bar H_t=\bar h_t]$ are fitted out of fold. The residuals are $\tilde Y_t=Y-\hat q_t(\bar H_t)$ and $\tilde A_{j,t}=A_j-\hat p_{j,t}(\bar H_t)$. The orthogonal moments are
\begin{equation}
\mathbb{E}\!\left[\Big(\tilde Y_t - \sum_{j=t}^{T} \psi_j^{*\prime}\,\tilde A_{j,t}\Big)\,\tilde A_{t,t}\right] = 0,
\qquad t=T,T-1,\ldots,1.
\label{eq:dynamic_dml_moment}
\end{equation}
The system is upper triangular. Backward estimation starts with $\psi_T$, subtracts $\hat\psi_T^\prime\tilde A_{T,T-1}$ from the stage-$(T-1)$ residualized outcome, and continues to earlier stages. Each step is an ordinary least-squares regression on residuals. For a general linear SNMM, Algorithm~3 of the paper replaces $A_j$ with the corresponding observed-versus-target blip-basis contrast.

\subsubsection*{Asymptotic normality at $\sqrt n$ rates}

Neyman orthogonality makes the leading nuisance bias a product of first-stage errors. Theorem~8 of \citet{lewisSyrgkanis2021dynamicDML} gives asymptotic normality for a structural nested mean model under a product-rate condition, and their Theorem~4 gives the same result for the partially linear specialization the simulation below uses. A convenient symmetric sufficient condition is $o_p(n^{-1/4})$ error in population $L_2$ norm for each cross-fitted nuisance. Then
\begin{equation}
\sqrt{n}\,(\hat\psi - \psi^*) \xrightarrow{d} \mathcal{N}(0,\, V),
\label{eq:dynamic_dml_clt}
\end{equation}
with a consistently estimable covariance matrix $V$. The covariance estimator stacks the cross-fitted moments from Equation~\eqref{eq:dynamic_dml_moment} as $m_i(\psi)$ and uses
\begin{equation}
\hat G=\frac1n\sum_{i=1}^n
\frac{\partial m_i(\hat\psi)}{\partial\psi^\prime},
\qquad
\hat\Omega=\frac1n\sum_{i=1}^n
m_i(\hat\psi)m_i(\hat\psi)^\prime,
\qquad
\hat V=\hat G^{-1}\hat\Omega\hat G^{-\prime}.
\label{eq:dynamic_dml_sandwich}
\end{equation}
Thus
\[
\widehat{\operatorname{se}}(\hat\psi_j)
=\sqrt{\hat V_{jj}/n},
\qquad
\widehat{\operatorname{se}}(c^\prime\hat\psi)
=\sqrt{c^\prime\hat Vc/n}.
\]
For two stages, $\hat G$ is upper triangular because the first-stage moment depends on both $(\psi_1,\psi_2)$ while the second-stage moment depends only on $\psi_2$. The covariance is therefore not stagewise diagonal, and the sandwich propagates uncertainty in the later blip into the earlier estimate. How much this matters is governed by the off-diagonal entry $\mathbb E[\tilde T_{11}\tilde T_{21}]$, the covariance of the two residualized treatments, and with binary treatments that entry is bounded. Making it large requires $T_2$ to be nearly determined by $T_1$ given the history, which is a propensity pushed to the boundary and so a positivity violation rather than a harder version of the same design. The simulation of Section~\ref{subsec:sim_dynamic_dml} measures the magnitude directly and reports it as small.\footnote{Raising the strength of the $T_1 \to X_2 \to T_2$ channel does raise the cross-stage correlation, from $-0.08$ to about $-0.25$, but only by driving $\mathbb P(T_2 = 1 \mid T_1 = 1)$ to $0.995$ and the fitted stage-two propensity to one, at which point the bias on $\hat\psi_2$ roughly quadruples. The two effects are not separable in this design.} The history may be high-dimensional, but the target vector $\psi^*\in\mathbb R^{Td}$ is fixed-dimensional.

\subsubsection*{The off-policy evaluation bridge}

For a fixed target regime, the blip telescoping identity expresses $V(\pi)=\mathbb E[Y^\pi]$ through the estimated structural parameters; Corollary~9 of \citet{lewisSyrgkanis2021dynamicDML} gives the corresponding asymptotically normal plug-in value estimator. If the value is a differentiable map $v(\psi,\zeta)$ of the structural parameter and additional regular nuisance averages $\zeta$, its influence function stacks the contributions from both. In the simpler case where only $\psi$ contributes first-order uncertainty,
\[
\widehat{\operatorname{se}}\{\hat V(\pi)\}
=\sqrt{
\nabla_\psi v(\hat\psi)^\prime
\hat V
\nabla_\psi v(\hat\psi)/n}.
\]
If empirical averages also enter $v$, their scores and cross-covariances belong in the same sandwich. Reusing one $\psi^*$ across arbitrary target regimes additionally requires target-independent blips.

\subsubsection*{Recursive Riesz representers}

The moment in Equation~\eqref{eq:dynamic_dml_moment} is tailored to a structural parameter. \citet{chernozhukov2023automatic} give an automatic construction for a different target, a policy value expressed through nested mean regressions. The nested regressions satisfy $f_{T+1}=Y$, $f_T(\bar h_T,a_T)=\mathbb E[Y\mid\bar H_T=\bar h_T,A_T=a_T]$, and, for $t<T$, $f_t(\bar h_t,a_t)=\mathbb E[f_{t+1}\{\bar H_{t+1},\pi_{t+1}(\bar H_{t+1})\}\mid\bar H_t=\bar h_t,A_t=a_t]$. The policy value is
\begin{equation}
\theta(\pi) = \mathbb{E}\!\left[f_1\bigl(\bar H_1, \pi_1(\bar H_1)\bigr) + \sum_{t=1}^{T} \alpha_t(\bar H_t,A_t)\,\bigl(f_{t+1}^{\pi}-f_t(\bar H_t,A_t)\bigr)\right],
\label{eq:cnss_riesz}
\end{equation}
where $f_{t+1}^{\pi}=f_{t+1}\{\bar H_{t+1},\pi_{t+1}(\bar H_{t+1})\}$ for $t<T$, $f_{T+1}^{\pi}=Y$, $\alpha_0\equiv1$, and $\alpha_t$ is the Riesz representer of $L_t(g)=\mathbb E[\alpha_{t-1}(\bar H_{t-1},A_{t-1})g\{\bar H_t,\pi_t(\bar H_t)\}]$. It is learned recursively by minimizing
\[
\mathbb E_n\!\left[\alpha^2(\bar H_t,A_t)-2\hat\alpha_{t-1}(\bar H_{t-1},A_{t-1})\alpha\{\bar H_t,\pi_t(\bar H_t)\}\right].
\]
The displayed summand in Equation~\eqref{eq:cnss_riesz}, minus $\theta(\pi)$, is the influence score. With all $f_t$ and $\alpha_t$ fitted out of fold, $\hat\phi_i$ denotes that complete summand and $\hat\theta=n^{-1}\sum_i\hat\phi_i$. The uncertainty calculation is
\[
\widehat{\operatorname{se}}(\hat\theta)
=\left[
\frac{1}{n(n-1)}
\sum_{i=1}^n(\hat\phi_i-\hat\theta)^2
\right]^{1/2},
\qquad
\mathrm{CI}_{1-\alpha}
=\hat\theta\pm z_{1-\alpha/2}\widehat{\operatorname{se}}(\hat\theta).
\]
The moment is Neyman-orthogonal and has an exact mixed-bias remainder. Errors enter as products of representer and adjacent-regression errors. This is the same second-order principle as dynamic DML, but the estimand is the nonparametric policy value rather than a low-dimensional blip parameter \citep{chernozhukov2023automatic}.\footnote{\citet{fosterSyrgkanis2023orthogonal} give the broader learning-theoretic result. Orthogonality makes nuisance error enter excess risk at second order. The precise norm and rate depend on the target loss and function classes; $o_p(n^{-1/4})$ root-mean-square nuisance rates are the familiar sufficient condition for a parametric target.}

\subsubsection*{Closest empirical bridge: repeated-session g-estimation}

Lewis and Syrgkanis present simulations but no real-data application. \citet{jaman2025penalizedg} provide a related application, although it does not use the exact dynamic-DML algorithm. They study an open cohort of 474 patients who receive 170,761 chronic hemodiafiltration sessions at two facilities between March 2017 and December 2021. Their repeated-outcome analysis uses each patient's first six consecutive sessions. The exposure is treatment at CHUM rather than CED, and the outcome is convection volume. The observed history includes the previous outcome, laboratory measures, dialysis access, age, sex, and recorded comorbidities.

The application estimates pooled propensity scores and applies penalized g-estimation under independent, exchangeable, autoregressive, and unstructured working correlations. Under the autoregressive specification, the selected blip contains a facility main effect of $-1.85$ litres with sandwich standard error $0.31$ and a facility-by-cancer interaction of $3.89$ litres with standard error $0.78$. The fitted contrast is therefore $2.04$ litres for patients with cancer. These estimates compare facilities at fixed values of the measured confounders. They are not randomized facility effects, and the selected interaction depends on the working correlation.

The selected model is random, so intervals that treat the cancer interaction as fixed can have inflated type I error. \citet{jaman2025postselectiong} develop uniformly valid and decorrelated-score post-selection intervals for this setting. The distinction carries over to dynamic DML. Orthogonalization controls flexible nuisance error, the full sandwich propagates dependence among structural moments, and post-selection correction accounts for searching over effect modifiers. Each addresses a different source of uncertainty.

\subsection{Simulation Study: Dynamic DML for a Two-Stage SNMM}
\label{subsec:sim_dynamic_dml}

The stylized Fast Track design records 20 baseline child, caregiver, and family measures in $X_1$, a first-semester home visit $T_1$, updated measures $X_2$, a second-semester visit $T_2$, and a terminal outcome $Y$.
\begin{align}
X_1&\sim\mathcal N(0,I_{20}),\nonumber\\
T_1\mid X_1&\sim\operatorname{Bernoulli}
\{\operatorname{expit}(\gamma^\prime X_1)\},\nonumber\\
X_2&=BX_1+\alpha T_1+\eta_1,
\qquad \eta_1\sim\mathcal N(0,0.6^2I_{20}),\nonumber\\
T_2\mid X_2&\sim\operatorname{Bernoulli}
\{\operatorname{expit}(\gamma^\prime X_2)\},\nonumber\\
Y&=\psi_1^*T_1+\psi_2^*T_2+\mu^\top X_1+\nu^\top\eta_1+\varepsilon,\nonumber\\
\varepsilon&\sim\mathcal N(0,0.5^2).
\label{eq:simB1_dgp}
\end{align}
The design sets $\|B\|_{\mathrm{op}}=0.5$, $\alpha=(1,0,\ldots,0)^\prime$, $\|\gamma\|_2=\|\mu\|_2=1.5$, and $\nu$ parallel to $\gamma$ with $\|\nu\|_2=2$. The shock $\eta_1$ affects both the second visit propensity and the outcome, creating treatment-confounder feedback. The structural effects are $(\psi_1^*,\psi_2^*)=(1.0,0.5)$. Dynamic DML uses five-fold cross-fitted Lasso outcome models and logistic propensity models. Naive ordinary least squares omits the second-period confounder, while a marginal structural model uses stabilized IPTW. Each configuration uses 200 replications over $n\in\{250,500,1000,2000,4000\}$.

\begin{table}[htbp]
\centering
\caption{Dynamic-DML results for the stylized home-visiting SNMM at $n=4000$ over 200 datasets. Panel A reports bias, RMSE, and 95\% interval coverage. Panel B compares the joint sandwich covariance with its Monte Carlo target, and sets the standard error for $\psi_1-\psi_2$ against the one obtained by dropping the cross-stage block.}
\label{tab:simB1}
\begin{minipage}{\textwidth}
\centering
\emph{Panel A. Point estimation and marginal coverage}\par\smallskip
\begin{tabular}{lrrrrrr}
\toprule
 & \multicolumn{3}{c}{$\psi_1^* = 1.00$} & \multicolumn{3}{c}{$\psi_2^* = 0.50$} \\
\cmidrule(lr){2-4}\cmidrule(lr){5-7}
Method & Bias & RMSE & Cov & Bias & RMSE & Cov \\
\midrule
Naive OLS & -0.191 & 0.197 & 0.01 & +0.933 & 0.934 & 0.00 \\
IPTW MSM (naive SE) & +0.019 & 0.103 & 0.88 & +0.146 & 0.192 & 0.73 \\
Dynamic DML & -0.005 & 0.046 & 0.97 & -0.001 & 0.018 & 0.93 \\
\bottomrule
\end{tabular}

\par\medskip
\emph{Panel B. Joint covariance and contrast inference}\par\smallskip
\begin{tabular}{lrr}
\toprule
Quantity & Sandwich estimate & Monte Carlo target \\
\midrule
$\operatorname{Var}(\hat\psi_1)$ & 0.002400 & 0.002114 \\
$\operatorname{Cov}(\hat\psi_1,\hat\psi_2)$ & -0.000069 & -0.000055 \\
$\operatorname{Var}(\hat\psi_2)$ & 0.000342 & 0.000339 \\
Cross-stage correlation & -0.076 & -0.065 \\
\addlinespace
SE$(\hat\psi_1-\hat\psi_2)$, joint sandwich & 0.0537 & 0.0506 \\
SE$(\hat\psi_1-\hat\psi_2)$, cross-stage block dropped & 0.0524 & 0.0506 \\
95\% CI coverage, joint sandwich & 0.950 & 0.950 \\
95\% CI coverage, block dropped & 0.950 & 0.950 \\
\bottomrule
\end{tabular}

\end{minipage}
\end{table}

\begin{figure}[htbp]
\centering
\includegraphics[width=\textwidth]{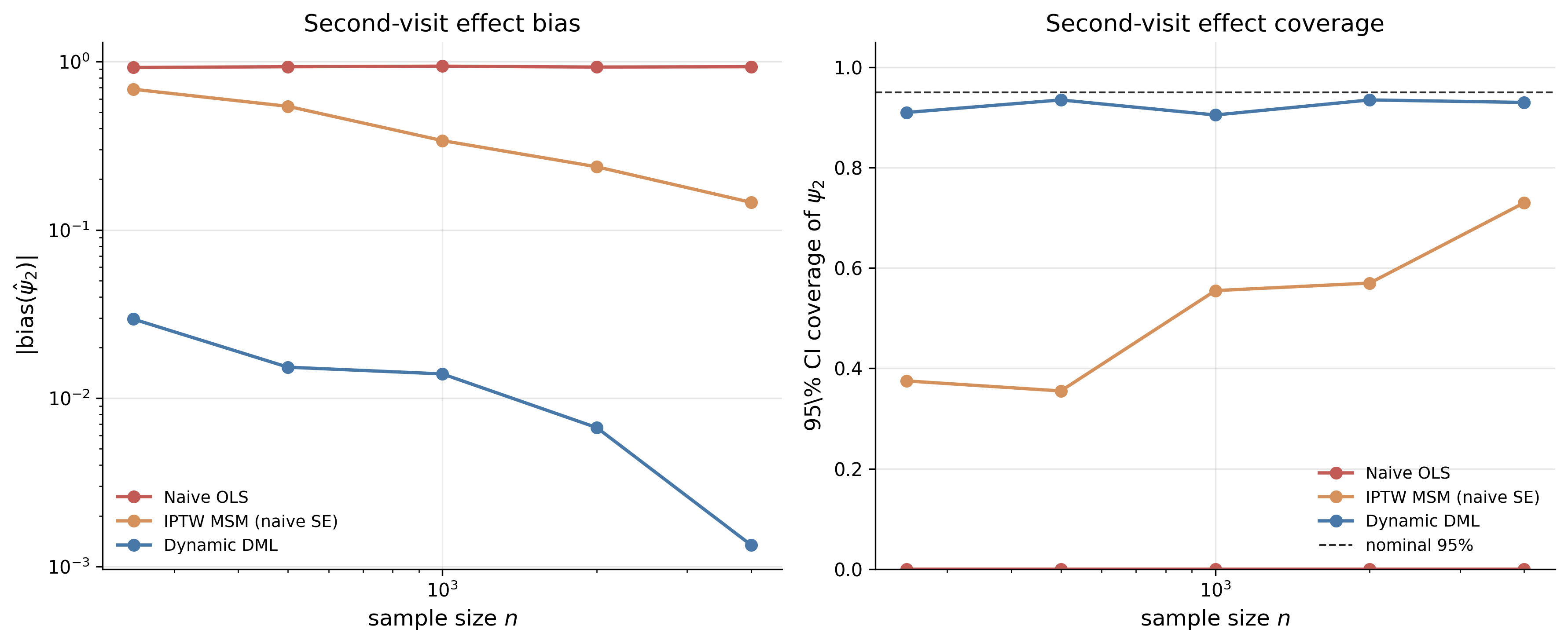}
\caption{Dynamic DML over 200 replications. The left panel shows absolute bias of $\hat\psi_2$ against sample size. The right panel shows empirical coverage of nominal 95\% confidence intervals.}
\label{fig:simB1}
\end{figure}
\FloatBarrier

Panel A of Table~\ref{tab:simB1} shows that Dynamic DML is nearly unbiased with coverage near 95 percent. Naive OLS and the IPTW MSM retain second-stage bias and undercoverage. Figure~\ref{fig:simB1} traces both quantities across sample sizes. Panel B shows that the joint sandwich tracks the Monte Carlo covariance. Dropping the cross-stage block changes the contrast standard error by $2.4$ percent and leaves coverage unchanged in this design. The small difference is consistent with the binary-treatment bound in Section~\ref{subsec:dml_dte}. The joint sandwich remains the correct object, but 200 replications cannot detect the cost of ignoring the cross-stage block.

\subsection{Dynamic Offline Policy Learning}
\label{subsec:dynamic_offline_policy}

Section~\ref{subsec:dml_dte} gives $\sqrt n$ confidence intervals for dynamic structural parameters under sequential ignorability. Those parameters permit parametric-rate evaluation of a fixed target policy, or of multiple policies when the blip model is target-independent. Policy learning instead estimates an optimal regime $\pi^* \in \arg\max_{\pi \in \Pi} V(\pi)$. \citet{sakaguchi2024dynamicpolicy} obtain a $\sqrt n$ welfare-regret rate with flexible nuisance estimates by combining augmented inverse-probability-weighted (AIPW) scores from \citet{athey2021policy} with fitted-$Q$ evaluation in a backward-induction loop.

For the Fast Track cohort, a restricted policy class assigns additional visits below thresholds in the family-functioning score, $\pi_t(\bar h_t) = \mathbb 1\{S_{t,j} < c_t\}$. The target is the welfare-maximising pair $(c_1,c_2)$ when the cohort is observational and machine learning estimates the nuisance functions.

The single-period treatment-choice problem has an established offline policy-learning literature. \citet{kitagawa2018who} develop empirical welfare maximization with known propensities and prove a matching minimax bound of $\Theta(\sqrt{\mathrm{VC}(\Pi)/n})$ on welfare regret, where $\mathrm{VC}(\Pi)$ is the Vapnik and Chervonenkis dimension of the policy class. Their application uses the National Job Training Partnership Act Study. \citet{athey2021policy} extend the method to unknown propensities through cross-fitted doubly robust scores, maintain the $O_p(\sqrt{\mathrm{VC}(\Pi)/n})$ rate under product-rate nuisance conditions, and allow instrumental-variable identification. \citet{zhou2023offline} cover multi-action treatments with exact decision-tree search. All three are the $T=1$ special case of the dynamic construction.

\subsubsection*{Backward-induction AIPW under sequential ignorability}

Each stage $t = 1, \ldots, T$ has a policy class $\Pi_t$, and the full regime class is $\Pi = \Pi_1 \times \cdots \times \Pi_T$. The action-value function for a candidate future policy $\pi_{(t+1):T} = (\pi_{t+1}, \ldots, \pi_T)$ satisfies
\begin{multline}
Q^{\pi_{(t+1):T}}_t(\bar h_t,a_t)
=\mathbb{E}\!\left[
Y_t+Q^{\pi_{(t+2):T}}_{t+1}
\bigl(\bar H_{t+1},\pi_{t+1}(\bar H_{t+1})\bigr)
\right.\\
\left.
\Bigm|\,\bar H_t=\bar h_t,A_t=a_t
\right],
\label{eq:sakaguchi_q}
\end{multline}
The terminal condition is
\[
Q_T(\bar h_T,a_T)
=\mathbb E[Y_T\mid\bar H_T=\bar h_T,A_T=a_T].
\]
This is fitted-Q evaluation under sequential ignorability \citep{sakaguchi2024dynamicpolicy}. The stage-$t$ propensity is $e_t(\bar h_t,a_t)=\mathbb P(A_t=a_t\mid\bar H_t=\bar h_t)$.

\citet{sakaguchi2024dynamicpolicy} estimate the optimal regime $\hat\pi = (\hat\pi_1, \ldots, \hat\pi_T)$ by backward induction with cross-fitted nuisances. The data are partitioned into $K$ folds, and $\hat Q^{-k}_t$ and $\hat e^{-k}_t$ denote nuisance estimators fitted outside fold $k$. At $t=T$, the cross-fitted AIPW score for candidate action $a$ is
\begin{equation}
\hat\Gamma_{i, T}(a) = \hat Q^{-k(i)}_T(\bar H_{i,T}, a) + \frac{\mathbb 1\{A_{i,T} = a\}}{\hat e^{-k(i)}_T(\bar H_{i,T}, a)}\,\bigl(Y_{i,T} - \hat Q^{-k(i)}_T(\bar H_{i,T}, A_{i,T})\bigr),
\label{eq:sakaguchi_aipw}
\end{equation}
and the procedure selects $\hat\pi_T = \arg\max_{\pi_T \in \Pi_T} \tfrac{1}{n}\sum_{i=1}^n \hat\Gamma_{i,T}(\pi_T(\bar H_{i,T}))$. Given $\hat\pi_T$, it refits $\hat Q^{\hat\pi_T}_{T-1}$ against the calibrated next-stage value, builds the stage-$T{-}1$ AIPW score from Equation~\eqref{eq:sakaguchi_aipw}, and selects $\hat\pi_{T-1}$. The same step continues backward to $\hat\pi_1$. Each stage solves one classification-style policy optimization on cross-fitted scores; the nuisances at stage $t$ depend only on the estimated future policies $\hat\pi_{(t+1):T}$, never on the candidate $\pi_t$.

\citet[Theorem~4.3]{sakaguchi2024dynamicpolicy} requires sequential ignorability, overlap, an entropy bound on $\Pi$, and correct specification of the later-stage policy classes. Its nuisance condition is a uniform product-rate bound between fitted-$Q$ and inverse-propensity errors; one symmetric sufficient condition is root-mean-square error of order $n^{-1/4}$ for each. Under these conditions,
\begin{equation}
V(\pi^*) - V(\hat\pi) = O_p\!\left(\kappa(\Pi)\, n^{-1/2}\right),
\label{eq:sakaguchi_regret}
\end{equation}
where $\kappa(\Pi)$ is the entropy integral of the DTR class. The $n^{-1/2}$ dependence matches the static minimax rate of \citet{kitagawa2018who}. The paper's Theorem~5.1 obtains the same rate without the class-correctness condition by optimizing an AIPW objective over the entire regime jointly, rather than stage by stage. This joint alternative is generally computationally difficult and is practical only for sufficiently small or specially structured regime classes.

\subsubsection*{Empirical application}

\citet{sakaguchi2024dynamicpolicy} apply the method to 1,877 students from the Tennessee Project STAR class-size experiment. In kindergarten, 702 students are assigned to regular classes of 22 to 25 students with a full-time aide; the rest enter small classes of 13 to 17 students without one. Kindergarten assignment is randomized. About 10\% of students change class type before grade 1, so the analysis treats the second decision as observational under conditional ignorability given baseline variables, kindergarten class type, and intermediate test scores. The outcome is the percentile rank of combined reading and mathematics scores at the end of grade 1.

The policy class uses a depth-1 tree in kindergarten and a depth-2 tree in grade 1. Kindergarten splits may use teacher education, experience, and school location; grade-1 splits may use kindergarten test scores and class type. Student gender, student ethnicity, and teacher ethnicity are excluded from candidate splits. Five-fold cross-fitting and generalized random forests estimate the nuisance functions, and exact tree search solves each policy problem. The fitted kindergarten tree assigns students to small classes when teacher experience is at most 19 years. The grade-1 tree uses total and reading scores, including a small-class assignment below a total-score threshold of 914.

Relative to assigning every student to a class with an aide in both grades or every student to a small class in both grades, the paper reports welfare contrasts of 8.16\% and 1.27\%. About 23\% of students receive a regular class with an aide in at least one grade under the fitted rule. These are cross-validated point estimates under the sequential-ignorability assumption for grade 1. The paper does not report standard errors or confidence intervals for the contrasts, and the rule has no prospective deployment. The application works through sequential allocation, but the reported gains have not been tested in a new school cohort.

\subsection{Simulation Study: Doubly Robust Policy Learning}
\label{subsec:sim_policy_learning}

The stylized Fast Track design uses a standardized home-visiting score $X$, an auxiliary child or caregiver measure $Z$, home-visit indicators $(A_1,A_2)$, updated targeting measures $(S_{21},S_{22})$, and terminal outcome $Y$.
\begin{align*}
(X,Z)&\sim\mathcal N(0,I_2),\\
A_1&\sim\operatorname{Bernoulli}
\{\operatorname{expit}(0.5X+0.5Z)\},\\
S_{21}&=0.6X-0.6A_1+\eta_{21},\\
S_{22}&=0.6Z+\eta_{22},\qquad
(\eta_{21},\eta_{22})\sim\mathcal N(0,0.5^2I_2),\\
A_2&\sim\operatorname{Bernoulli}
\{\operatorname{expit}(0.5S_{21}+0.5S_{22})\},\\
Y&=1-0.5S_{21}+0.4Z+0.4S_{22}
\\
&\quad+0.6A_1(-0.6-X)+0.8A_2(0.4-S_{21})+\varepsilon,\\
\varepsilon&\sim\mathcal N(0,1).
\end{align*}
The threshold regime $\pi_t=\mathbbm 1\{S_t^{(1)}<c_t\}$ assigns visits below the score level at which their incremental effect changes sign. Its oracle is $(c_1^*,c_2^*)=(0.5156,0.4000)$ with $V^*=1.9224$. Backward AIPW, plug-in $Q$, and inverse-propensity weighting (IPW) use the same cohorts, threshold search, and two-fold cross-fitting. The correctly specified outcome branch uses an oracle feature map derived from the stated Gaussian transition law, including its generating constants, while estimating the regression coefficients within each training fold. It is a controlled correct-specification benchmark rather than an end-to-end feasible nuisance-learning pipeline. The misspecified outcome branch removes treatment-effect heterogeneity, and the misspecified propensity branch replaces each propensity with its marginal treatment rate. Adaptive quadrature computes policy values, and one million Monte Carlo draws independently check the oracle and behavior values.

\begin{table}[htbp]
\centering
\caption{Policy value relative to the threshold-class oracle at $n=4000$ in the stylized home-visiting problem. Monte Carlo standard errors describe the reported mean across 40 independent replications, not one learned policy.}
\label{tab:dtr_policy_learning}
\begin{tabular}{lcc}
\toprule
Method & $V(\hat\pi) / V^*$ & MC SE \\
\midrule
Oracle regime $(c_1^*, c_2^*)$ & 1.0000 & -- \\
Plug-in $Q$ ($n=4000$) & 0.9996 & (0.0001) \\
Backward AIPW ($n=4000$) & 0.9966 & (0.0006) \\
IPW ($n=4000$) & 0.9911 & (0.0020) \\
Behavior policy & 0.5770 & -- \\
\bottomrule
\end{tabular}

\end{table}

\begin{figure}[htbp]
\centering
\includegraphics[width=\textwidth]{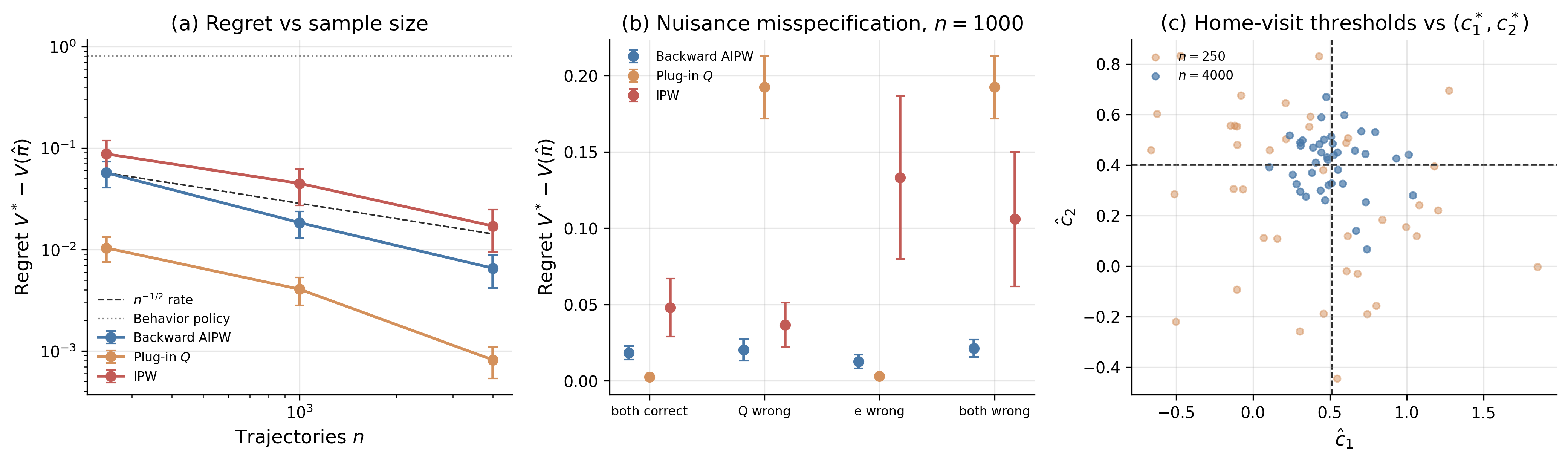}
\caption{Offline threshold-policy learning over 40 replications. Panel (a) shows regret against sample size. Panel (b) shows regret under outcome-model and propensity misspecification at $n=1000$, with 95\% Monte Carlo confidence intervals for mean regret. Panel (c) shows the learned AIPW thresholds at $n=250$ and $n=4000$ relative to the oracle pair.}
\label{fig:dtr_policy_learning}
\end{figure}
\FloatBarrier

Equation~\eqref{eq:sakaguchi_regret} is an upper bound, so a fitted slope steeper than $-1/2$ is consistent with the bound but does not verify it. The direct check asks whether regret multiplied by $\sqrt n$ remains bounded across the sweep. Scaled regret remains bounded for all three methods, with plug-in $Q$ converging fastest under correct specification. Table~\ref{tab:dtr_policy_learning} reports the values at the largest cohort, and Figure~\ref{fig:dtr_policy_learning} shows the sweep.

The doubly robust score does not improve the rate. Under one-sided misspecification, AIPW regret remains low while the affected single-robust estimator deteriorates. AIPW also survives the cell in which both nuisances are wrong. The learned thresholds show policy recovery rather than a flat value surface. AIPW remains near the oracle thresholds, while the plug-in treats everyone. Policy learning needs only the sign of the stagewise contrast, so a nuisance error that shifts the contrast without reordering it leaves the argmax unchanged. The both-wrong cell is not the failure that double robustness predicts for value estimation. Equation~\eqref{eq:sakaguchi_regret} bounds welfare loss, not uncertainty in either threshold. For the selected policy's value, an independent evaluation sample supports the standard error
\[
\widehat{\operatorname{se}}\{\hat V(\hat\pi)\}
=\left[
\frac{1}{n_e(n_e-1)}
\sum_{i=1}^{n_e}
\{\hat\phi_i(\hat\pi)-\hat V(\hat\pi)\}^2
\right]^{1/2}.
\]
This interval is conditional on the training sample. Nested cross-fitting instead targets the average value of fold-specific trained policies unless an additional stability argument connects it to the final full-sample fit. The regret theorem does not justify a Wald interval around $(\hat c_1,\hat c_2)$.

\subsection{Open Issues: Identification, Inference, and Deployment}
\label{subsec:rlci_open_issues}

The methods above address several parts of the sequential causal problem, but important gaps remain. Table~\ref{tab:rlci_open_issues} separates problems that are often conflated.

\begin{table}[htbp]
\centering
\caption{Open issues that remain material for applied work. The first five rows and the operational-objectives row are tied to the cited literature; the transport-and-drift row is practitioner synthesis.}
\label{tab:rlci_open_issues}
{\setstretch{1.0}\footnotesize
\renewcommand{\arraystretch}{1.12}
\begin{tabularx}{\textwidth}{p{0.19\textwidth} p{0.28\textwidth} p{0.25\textwidth} >{\raggedright\arraybackslash}X}
\toprule
Issue & What fails & Available partial remedy & What remains open \\
\midrule
Near-zero treatment contrasts & The optimal-action map and selected thresholds are nonregular & Adaptive intervals, $m$-out-of-$n$ bootstrap, honest value evaluation & General, powerful inference for learned multistage rules \citep{laber2014dtrchallenges, wangtom2025dtrtutorial} \\
State sufficiency & A compressed state may omit history needed for exchangeability or Markov transitions & Use full recorded history; diagnostics can probe observed-law Markov fit but cannot verify exchangeability & Identification under latent state belongs to sensitivity, proxy, or IV analysis in Chapter~\ref{section:causal_rl} \\
Weak overlap & Importance ratios and efficiency bounds explode even though the estimand is formally identified & Restrict the target policy, trim cautiously, collect data, or report one-sided safety bounds & Principled policy-class restriction without silently changing the target \citep{ueharaShiKallus2022ope} \\
Function approximation & FQE, density ratios, and orthogonal scores require coverage, completeness, and nuisance-rate conditions & Cross-fitting, doubly robust scores, and bootstrap under verified model conditions & Reliable inference with general nonlinear approximation and policy-dependent nuisances \citep{hao2021bootstrapfqe, kallusUehara2022doubleRL} \\
Policy search and inference & Optimizing and evaluating on the same observations creates selection bias & Honest splitting or an external evaluation sample & Efficient reuse of limited longitudinal data with valid post-selection inference \citep{ueharaShiKallus2022ope, jaman2025postselectiong} \\
Transport and drift & A value identified in historical trajectories need not equal value after deployment & Reweight measured population shifts and monitor support & Unmeasured environment change and feedback from deployment \\
Operational objectives & A scalar reward can omit burden, cost, preferences, and rare harms & Constrained policy classes, multiple outcomes, and conservative lower bounds & Prospective evaluation with stakeholder-defined safety constraints \citep{laber2014dtrchallenges, liaoMurphy2021longterm} \\
\bottomrule
\end{tabularx}
}
\end{table}

\citet{ueharaShiKallus2022ope} show why no estimator is universally best. The efficiency bound depends on whether the data are a bandit, a non-Markov process, a time-varying MDP, or a stationary MDP. Stronger structure can improve rates, but an unjustified state restriction changes the statistical model. \citet{bannon2020causality} likewise distinguish causal identification from batch RL. Causal assumptions determine which counterfactual policy values are identified; RL algorithms handle planning and approximation after identification.

For day-to-day practice, the minimum defensible workflow is therefore to name the potential-outcome estimand, state consistency, sequential exchangeability, and target-policy positivity, justify any Markov compression, evaluate overlap before fitting, cross-fit flexible nuisance functions or justify an empirical-process alternative, carry the influence score or sandwich calculation through to a standard error, and reserve data for honest evaluation of a learned policy. Special cases with hidden confounding, interference, partial identification, or instrumental variables are important, but they alter the identification problem and are treated in Chapter~\ref{section:causal_rl}.

\section{Causal Bandits and Adaptive Experimentation}
\label{section:adaptive_experiments}

Sequential experiments join two problems. The assignment rule must learn which
intervention works, and the final analysis must remain valid after assignment has
adapted to earlier outcomes. Causal bandits address the first problem by using a
causal graph or counterfactual structure to share information across actions.
Adaptive-experiment estimators address the second by accounting for time-varying
assignment probabilities.

\subsection{Causal Bandits}
\label{subsec:causal_bandits}

\citet{bareinboim2015mabuc} study a two-machine casino in which two unobserved
binary variables determine both the player's natural choice $X$ and the reward
$Y$. Their Table~1 implies
\[
\mathbb P(Y=1\mid X=M_j)=0.15,
\qquad
\mathbb P(Y=1\mid \mathrm{do}(X=M_j))=0.30
\]
for either machine. The marginal interventional means are therefore tied, even
though the best action conditional on the player's natural choice is not.
Marginal Thompson sampling cannot use that conditional signal and accumulates
linear regret.

The Regret Decision Criterion instead chooses the action with the largest
intention-specific counterfactual mean,
\begin{equation}
a^*(x)=\arg\max_a \mathbb E[Y_a\mid X=x].
\label{eq:rdc}
\end{equation}
Causal Thompson sampling, denoted $\mathrm{TS}_C$, maintains a Beta posterior
for each intention-action pair. Algorithm~1 of \citet{bareinboim2015mabuc}
initializes only the on-intuition cells from observational data, using
consistency to equate $\mathbb E[Y_x\mid X=x]$ with
$\mathbb E[Y\mid X=x]$. The off-intuition cells begin at
$\operatorname{Beta}(1,1)$ and are learned online. The algorithm also applies
the Regret Decision Criterion weighting from the current off-intuition estimate
and the fixed observational mean. It does not initialize the off-intuition
cells from an effect-of-treatment-on-the-untreated identity.

In a binary problem with an additional randomized marginal log, that
counterfactual can be identified separately,
\begin{equation}
\mathbb E[Y_a\mid X\ne a]
=
\frac{\mathbb P(Y=1\mid\mathrm{do}(a))
-\mathbb P(X=a)\mathbb P(Y=1\mid X=a)}
{1-\mathbb P(X=a)}.
\label{eq:simB2_ett}
\end{equation}
The simulation treats this as a data-fusion warm start, not as the paper's
$\mathrm{TS}_C$.

\citet{lattimore2016causal} obtain a different gain from causal structure.
Their parallel bandit has $N$ independent binary parents of $Y$ and action set
\[
\mathcal A=\{\mathrm{do}()\}\cup
\{\mathrm{do}(X_i=j):i\in[N],\,j\in\{0,1\}\}.
\]
Every round reveals $Y$ and all non-intervened parents. Algorithm~1 spends half
the budget on $\mathrm{do}()$, estimates the parent propensities and conditional
reward means, and spends the remainder on the values rarely observed in the
first phase. Its hardness parameter is
\begin{equation}
m(q)=\min_{\tau\in\{2,\ldots,N\}}
\max\!\left\{\tau,\left|\left\{i:
\min(q_i,1-q_i)<1/\tau\right\}\right|\right\}.
\label{eq:lattimore_m}
\end{equation}
Theorem~1 gives simple regret of order
\begin{equation}
R_T=O\!\left(
\sqrt{\frac{m(q)}{T}\log\frac{NT}{m(q)}}\right),
\label{eq:lattimore_regret}
\end{equation}
and Theorem~2 supplies a matching lower bound up to logarithmic factors.
Balanced parents give $m(q)=2$ and the largest advantage over a graph-blind
$\sqrt{N/T}$ guarantee. Concentrating more parents near zero or one increases
$m(q)$ and erodes that advantage. For general directed acyclic graphs, the
paper proves an upper bound using a mixing distribution over interventions. In
the parallel case the optimized general-graph complexity is bounded by
$2m(q)$, not equal to $m(q)$.

\subsection{Adaptive Experimentation and Post-Experiment Inference}
\label{subsec:adaptive_experiments}

\citet{kasy2021adaptive} target the welfare of the policy selected after a
multi-wave experiment rather than participants' rewards during data
collection. Exploration sampling transforms each treatment's posterior
probability of being best,
\begin{equation}
q_t^d=S_t p_t^d(1-p_t^d),
\label{eq:exploration_sampling}
\end{equation}
so the leading treatment receives at most half of each wave and suboptimal
treatments retain enough observations for discrimination. The design was used
in a 17-wave experiment with about $10{,}000$ participants and six recruitment
strategies for an agricultural extension service in India.

The published Theorem~1 also claimed the best exponential rate for frequentist
expected policy regret. That item is false. The authors' correction retains the
limiting allocation and equal posterior concentration results, and states the
optimality result for the posterior probability that the selected treatment is
not best. It also gives a separate posterior expected-regret bound
\citep{kasysautmann2021correction}. The chapter therefore does not use the
withdrawn frequentist policy-regret claim.

Adaptive assignment also changes inference. For treatment $w$ with known
history-dependent propensity $e_t(w)$, \citet{hadad2021adaptive} start from the
AIPW score
\begin{equation}
\widehat\Gamma_t(w)=\widehat m_t(w)+
\frac{\mathbb 1\{W_t=w\}}{e_t(w)}
\{Y_t-\widehat m_t(w)\}
\label{eq:hadad_aipw_score}
\end{equation}
and average it with predictable evaluation weights,
\begin{equation}
\widehat Q^{\mathrm{AW}}(w)=
\frac{\sum_{t=1}^T h_t(w)\widehat\Gamma_t(w)}
{\sum_{t=1}^T h_t(w)}.
\label{eq:adaptive_aipw}
\end{equation}
Their constant-allocation construction uses weights proportional to
$\sqrt{e_t(w)/T}$ through a stick-breaking normalization. These weights are
chosen to stabilize variance and establish asymptotic normality. The paper
explicitly does not claim that they minimize variance, and it proposes a
two-point adaptive construction to improve efficiency. Theorem~2 additionally
requires positive propensities, moment and infinite-sampling conditions,
variance convergence, and either a consistent outcome model or convergent
propensities.

\citet{bibaut2021post} extend this inference problem to contextual adaptive
experiments. Their logging probabilities $g_t(a\mid x)$ remain known, while
conditional-variance estimates rescale the canonical-gradient terms.
Asymptotic normality is robust to outcome-model misspecification when those
variance estimates are consistent. The result does not cover data from an
unknown black-box assignment mechanism.

\subsection{Simulation Study}
\label{subsec:causal_bandit_sim}

The first experiment follows the hard family in
\citet[\S 5]{lattimore2016causal}. Exactly $m$ parent propensities are zero and
the rest equal one half. The reward depends on one zero-propensity parent $X_w$,
with mean $0.5+\epsilon$ when $X_w=1$ and $0.5$ otherwise. Changing $m$ therefore
changes which actions are passively observed without changing any arm value.
Successive Reject uses its cumulative phase schedule, while Lattimore
Algorithm~1 uses the estimated $m(\hat q)$ cutoff. The horizon panel uses the
paper's local alternative $\epsilon_T=\sqrt{N/(8T)}$. Its error probability is
a minimax stress test, not evidence of a fixed-instance $T^{-1/2}$ decay rate.

The second experiment uses the Table~1 greedy-casino conditional payoff matrix
$((0.15,0.45),(0.45,0.15))$. The factorial separates observational seeding from
RDC weighting and reports the additional ETT warm start separately.
$\mathrm{TS}_Z$ denotes the unseeded intention-conditioned baseline. The paper's
$\mathrm{TS}_C$ uses only the observational log. The ETT extension also receives
a randomized marginal log. Self-checks verify the labelled hardness, equality
of all suboptimal arm values, phase budgets, the $0.15$ observational and $0.30$
interventional marginals, and recovery of the published regret scale for
Thompson sampling, $\mathrm{TS}_Z$, and $\mathrm{TS}_C$.

\begin{figure}[htbp]
\centering
\includegraphics[width=\textwidth]{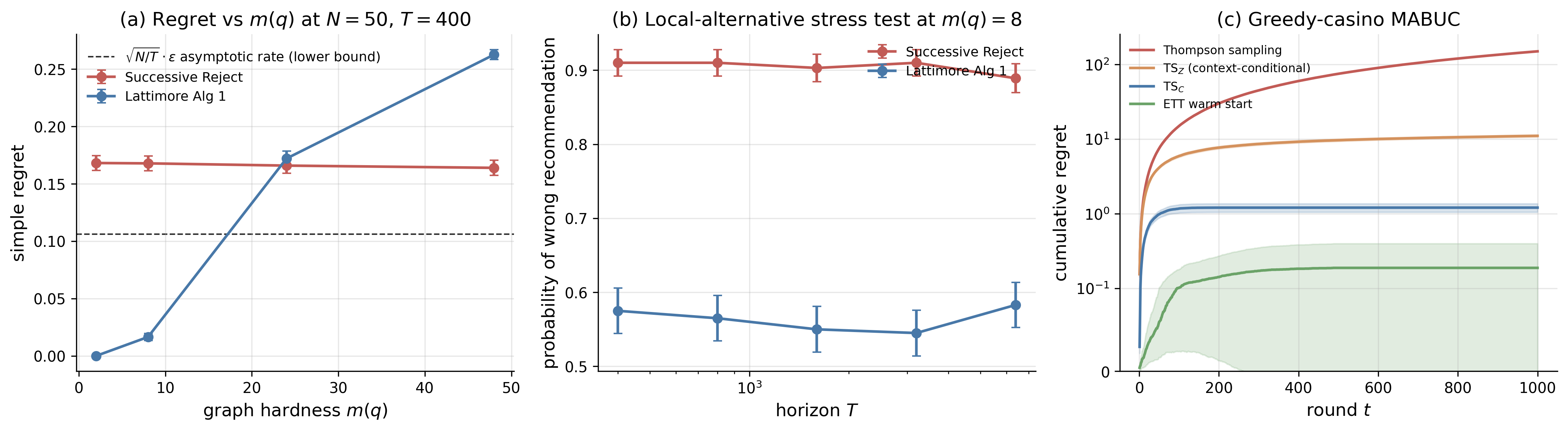}
\caption{Causal-bandit diagnostics. Panel (a) varies graph hardness without
changing rewards. Panel (b) reports recommendation error under a shrinking-gap
local alternative. Panel (c) compares marginal Thompson sampling, unseeded
context-conditional $\mathrm{TS}_Z$, the paper's $\mathrm{TS}_C$, and the
separately labelled ETT warm start. Error bars and bands are 95\% Monte Carlo
intervals.}
\label{fig:simB2}
\end{figure}

\begin{table}[htbp]
\centering
\caption{Simple regret on the parallel-bandit hard family. Entries are means
with Monte Carlo standard errors in parentheses over two thousand replications.
$N=50$, $T=400$, and $\epsilon=0.3$.}
\label{tab:simB2}
\begin{tabular}{lrrrr}
\toprule
Method  & $m = 2$ & $m = 8$ & $m = 24$ & $m = 48$ \\
\midrule
Successive Reject & 0.168 (0.003) & 0.168 (0.003) & 0.166 (0.003) & 0.164 (0.003) \\
Lattimore Alg 1 & 0.000 (0.000) & 0.017 (0.002) & 0.172 (0.003) & 0.263 (0.002) \\
\bottomrule
\end{tabular}

\end{table}

\begin{table}[htbp]
\centering
\caption{Greedy-casino cumulative pseudo-regret over five hundred paired
replications. The binary ETT row is a data-fusion extension and is not
Algorithm~1 of \citet{bareinboim2015mabuc}.}
\label{tab:simB2_mabuc}
\begin{tabular}{lrr}
\toprule
Algorithm & Regret at $T = 100$ & Regret at $T = 1000$ \\
\midrule
Binary ETT warm start $+$ RDC & 0.10 & 0.19 (0.11) \\
$\mathrm{TS}_C$ \citep{bareinboim2015mabuc} & 1.17 & 1.21 (0.08) \\
Observational seed only & 1.60 & 1.68 (0.09) \\
RDC only & 4.13 & 6.23 (0.14) \\
$\mathrm{TS}_Z$ (context-conditional TS) & 5.92 & 11.04 (0.19) \\
Vanilla Thompson sampling & 14.99 & 150.00 (0.21) \\
\bottomrule
\end{tabular}

\end{table}

Table~\ref{tab:simB2} exhibits the hardness mechanism in
\eqref{eq:lattimore_regret}. Lattimore Algorithm~1 has simple regret
$0.000$ at $m=2$ and $0.263$ at $m=48$, while Successive Reject changes
from $0.168$ to $0.164$. Table~\ref{tab:simB2_mabuc} exhibits the
intention-specific mechanism in \eqref{eq:rdc}. At $T=1000$,
$\mathrm{TS}_C$ has cumulative pseudo-regret $1.21$, compared with
$150.00$ for marginal Thompson sampling. The ETT warm start has regret
$0.19$, but it also receives the additional randomized marginal log
defined in \eqref{eq:simB2_ett}.

\FloatBarrier

\subsection{Discussion}
\label{subsec:adaptive_experiments_discussion}

Causal structure can reduce exploration cost when one action reveals outcomes
relevant to others. Valid post-experiment inference requires a second layer of
design because adaptive propensities change the variance of standard scores.
Both benefits depend on observing the relevant mechanism. The causal-bandit
guarantees use a known graph or recorded intention, and post-contextual-bandit
inference uses known logging probabilities.

\section{Quantile, Robust and Constrained Reinforcement Learning}
\label{section:dist_robust_constrained}

The preceding chapters optimized expected returns. This chapter relaxes
that assumption in three directions: tracking full return distributions
rather than means (distributional RL), imposing constraints on secondary
objectives (constrained MDPs), and hedging against model misspecification
(robust MDPs).

The two-state Engine Replacement MDP has no fat-tailed return distribution, so it does not illustrate the quantile results below. It returns in the constrained and robust sections, where its replacement action supplies a budget and its transition kernel can be tilted.


\subsection{Distributional Reinforcement Learning and Risk Measures}
\label{subsec:distributional_rl}

Standard reinforcement learning characterizes a policy $\pi$ by the
expected return $V^\pi(s) = \mathbb{E}^\pi\bigl[\sum_{t=0}^\infty \gamma^t
R_t \mid S_0 = s\bigr]$. Distributional reinforcement learning instead
maintains the full random variable $Z^\pi(s,a) = \sum_{t=0}^\infty \gamma^t
R_t$ whose expectation is $Q^\pi(s,a)$. \citet{Bellemare2017} showed that
the Bellman equation has a distributional counterpart that propagates entire
distributions rather than expectations, and that this distributional operator
contracts in the Wasserstein metric.

Let $\mathcal{Z}$ denote the space of return-distribution functions mapping
state-action pairs to distributions over $\mathbb{R}$. The distributional
Bellman operator $\mathcal{T}^\pi$ is defined by
\begin{equation}
  \mathcal{T}^\pi Z(s,a) \stackrel{D}{=} R(s,a) + \gamma Z(S', A'),
  \quad S' \sim P(\cdot|s,a), \; A' \sim \pi(\cdot|S'),
  \label{eq:dist_bellman}
\end{equation}
where $\stackrel{D}{=}$ denotes equality in distribution.
\citet{Bellemare2017} proved that $\mathcal{T}^\pi$ is a $\gamma$-contraction
in the Wasserstein distance (informally, the minimum cost of reshaping one distribution into another) between distributions,
\begin{equation}
  \bar{d}_p(Z_1, Z_2) = \sup_{s,a} d_p\bigl(Z_1(s,a),\, Z_2(s,a)\bigr),
  \label{eq:wasserstein_metric}
\end{equation}
where $d_p$ is the $p$-Wasserstein distance between univariate
distributions. Since $\mathcal{T}^\pi$ is a contraction, iterating it
converges to a unique return distribution $Z^\pi$ under policy
$\pi$. For quantile representations, minimizing the quantile regression
loss~\eqref{eq:qrdqn_loss} is equivalent to minimizing the 1-Wasserstein
distance to the Bellman target \citep{Dabney2018a}, so QR-DQN and IQN
directly inherit this convergence guarantee.\footnote{$\mathcal{T}^\pi$
is not a contraction in total variation or KL divergence. The optimality operator $\mathcal{T}^*$ is not a contraction
in any distribution metric, though expected values still converge
\citep{Bellemare2017}.}

A standard DQN network outputs a single scalar $Q_\theta(s,a)$ per action
and minimizes the squared TD error
$(r + \gamma \max_{a'} Q_{\bar{\theta}}(s',a') - Q_\theta(s,a))^2$,
where $r$ is the observed reward, $s'$ is the next state, and
$\bar{\theta}$ denotes a slowly-updated target network.

\citet{Dabney2018a} introduced Quantile Regression DQN (QR-DQN), whose
network instead outputs $N$ values $\theta_1(s,a), \ldots, \theta_N(s,a)$
representing quantile locations of the return distribution, each with
equal probability $1/N$. The quantile regression loss for a single
quantile level $\tau \in (0,1)$ is
$\rho_\tau(u) = u(\tau - \mathbbm{1}\{u < 0\})$, which penalizes
underestimates by a factor of $\tau$ and overestimates by $1-\tau$.
The full QR-DQN loss sums this over all pairs of current quantiles $i$
and target quantiles $j$:
\begin{equation}
  L_{\text{QR}} = \frac{1}{N} \sum_{i=1}^{N} \sum_{j=1}^{N}
  \rho_{\hat{\tau}_i}\bigl(\underbrace{
    r + \gamma\, \theta_j^{\text{target}}(s', a^*)}_{\text{target quantile }j}
    - \underbrace{\theta_i(s,a)}_{\text{current quantile }i}
  \bigr),
  \label{eq:qrdqn_loss}
\end{equation}
where $\hat{\tau}_i = (2i-1)/(2N)$ is the midpoint of the $i$-th quantile
interval, $\theta_j^{\text{target}}$ comes from the target network, and
$a^* = \arg\max_{a'} \frac{1}{N}\sum_j \theta_j(s',a')$ is the greedy
action under the mean return.\footnote{The resulting projected operator is
a $\gamma$-contraction in the $\infty$-Wasserstein metric
\citep{Dabney2018a}.}

\citet{Dabney2018b} extended this to Implicit Quantile Networks (IQN).
Instead of outputting a fixed set of $N$ quantiles, the IQN network takes
a quantile level $\tau \in [0,1]$ as an additional input and outputs the
corresponding return value $F_{Z}^{-1}(\tau; s,a)$. The IQN loss has
the same pairwise structure, but with quantile levels sampled
continuously:
\begin{equation}
  L_{\text{IQN}} = \frac{1}{KK'} \sum_{i=1}^{K} \sum_{j=1}^{K'}
  \rho_{\tau_i}\bigl(
    r + \gamma\, F_{Z}^{-1}(\tau'_j; s', a^*)
    - F_{Z}^{-1}(\tau_i; s, a)
  \bigr),
  \label{eq:iqn_loss}
\end{equation}
where $\tau_1, \ldots, \tau_K$ and $\tau'_1, \ldots, \tau'_{K'}$ are
independent draws from $U([0,1])$, and $K, K'$ are the number of
samples per update. Because the quantile levels are sampled rather than
fixed, IQN learns the full continuous quantile function rather than a
discrete approximation. This continuous representation enables risk-sensitive control.

An agent that maximizes expected return is indifferent between a certain
payoff of 100 and a coin flip paying 0 or 200. In many applications this
is inadequate, since a portfolio manager, a robot near a cliff, or a firm
facing bankruptcy all care about the shape of the return distribution, not
just its mean. Distributional RL provides the return distribution; what
remains is a principled way to convert that distribution into a scalar
objective that encodes the desired risk attitude. \citet{Yaari1987}
showed that this can be done with a distortion function
$h: [0,1] \to [0,1]$ (increasing, $h(0)=0$, $h(1)=1$) applied to the
cumulative distribution before integrating:
\begin{equation}
  \rho_h(Z) = \int_0^1 F_Z^{-1}(\tau) \, h'(1-\tau) \, d\tau.
  \label{eq:distortion_risk}
\end{equation}
This class includes the expected value ($h(\tau)=\tau$), Conditional
Value-at-Risk at level $\alpha$ ($h(\tau) = \min(\tau/\alpha, 1)$), and the
Cumulative Prospect Theory
\citep{TverskyKahneman1992}, which overweights tail outcomes relative to
their true probabilities. Since IQN can evaluate $F_Z^{-1}(\tau)$ at
any $\tau$, each of these risk measures reduces to choosing which quantiles
to average over and how to weight them. The network itself does not change;
only the sampling distribution of $\tau$ at decision time does. Sampling
$\tau$ from the non-uniform distribution $\beta(\tau) = h'(1-\tau)$ rather
than uniformly yields policies that maximize the distortion risk measure
$\rho_h$.

CVaR at level $\alpha$ (the expected return in the worst $\alpha$-fraction
of outcomes) is the most widely used case. In IQN, sampling
$\tau \sim U([0, \alpha])$ instead of $U([0,1])$ yields a CVaR-optimal
policy. CVaR can also be optimized exactly via dynamic programming on an
augmented state space \citep{Bauerle2011}, or through its equivalence to
robust MDPs with adversarial transition reweighting
\citep{Chow2015}.\footnote{Risk-averse dynamic programming requires the
risk measure to satisfy a recursive nestedness property (the
multi-period risk decomposes into nested single-period evaluations, as
the Bellman equation does for expected value)
\citep{Ruszczynski2010}. \citet{Hau2023} showed that popular decompositions
for CVaR are suboptimal regardless of discretization, correcting earlier
claims. \citet{Tamar2015} extended the policy gradient theorem to coherent
risk objectives as an alternative to the distributional approach.
\citet{Prashanth2016} proved consistency of CPT-value estimation from
sampled returns.}

\subsubsection{Simulation Study: Risk-Sensitive Inventory Management}
\label{subsec:sim_risk_inventory}

A newsvendor MDP with fat-tailed demand illustrates the mechanism.
The agent manages inventory over a 10-period horizon with state
$s \in \{0, \ldots, 15\}$ (current stock) and action $a \in \{0, \ldots, 5\}$
(order quantity). Demand follows a mixture distribution,
$D \sim 0.8 \cdot \text{Poisson}(3) + 0.2 \cdot \text{Poisson}(10)$,
creating occasional demand spikes that make risk preferences
matter.\footnote{Per-step reward:
$5 \cdot \min(\tilde{s}, D) - 2a - 1 \cdot (\tilde{s}-D)^+ - 8 \cdot (D-\tilde{s})^+$
with $\tilde{s} = \min(s+a, 15)$, the post-order stock after the
capacity cap; the order cost is charged on $a$ even when the cap binds.
The stockout penalty (8) exceeds the holding cost (1), which makes the
demand risk consequential for ordering behavior.}
A single IQN network is trained for 50{,}000 episodes with
$\tau \sim U([0,1])$, then evaluated under two $\tau$-sampling
distributions: $U([0,1])$ for risk-neutral and $U([0, 0.05])$ for
CVaR$_{95}$. Table~\ref{tab:risk_inventory} reports means and standard
errors across ten training seeds, each evaluated over 50{,}000
episodes, alongside the exact DP oracle.

\begin{table}[h]
\centering
\caption{Risk-sensitive inventory management. Mean $\pm$ SE across ten
training seeds, 50{,}000 evaluation episodes per seed. Avg.\ Order is
the per-state mean order quantity averaged uniformly across inventory
levels.}
\label{tab:risk_inventory}
\begin{tabular}{lccccc}
\hline
Policy & Mean Return & SE & CVaR$_{95}$ & CVaR$_{99}$ & Avg.\ Order \\
\hline
DP Oracle & 40.4 & 0.0 & -51.5 & -95.6 & 2.56 \\
IQN-Neutral & 34.9 & 0.8 & -57.5 & -98.0 & 3.21 \\
IQN-CVaR95 & 36.4 & 0.4 & -55.3 & -97.8 & 2.99 \\
\hline
\end{tabular}
\end{table}

\begin{figure}[h]
\centering
\includegraphics[width=0.7\textwidth]{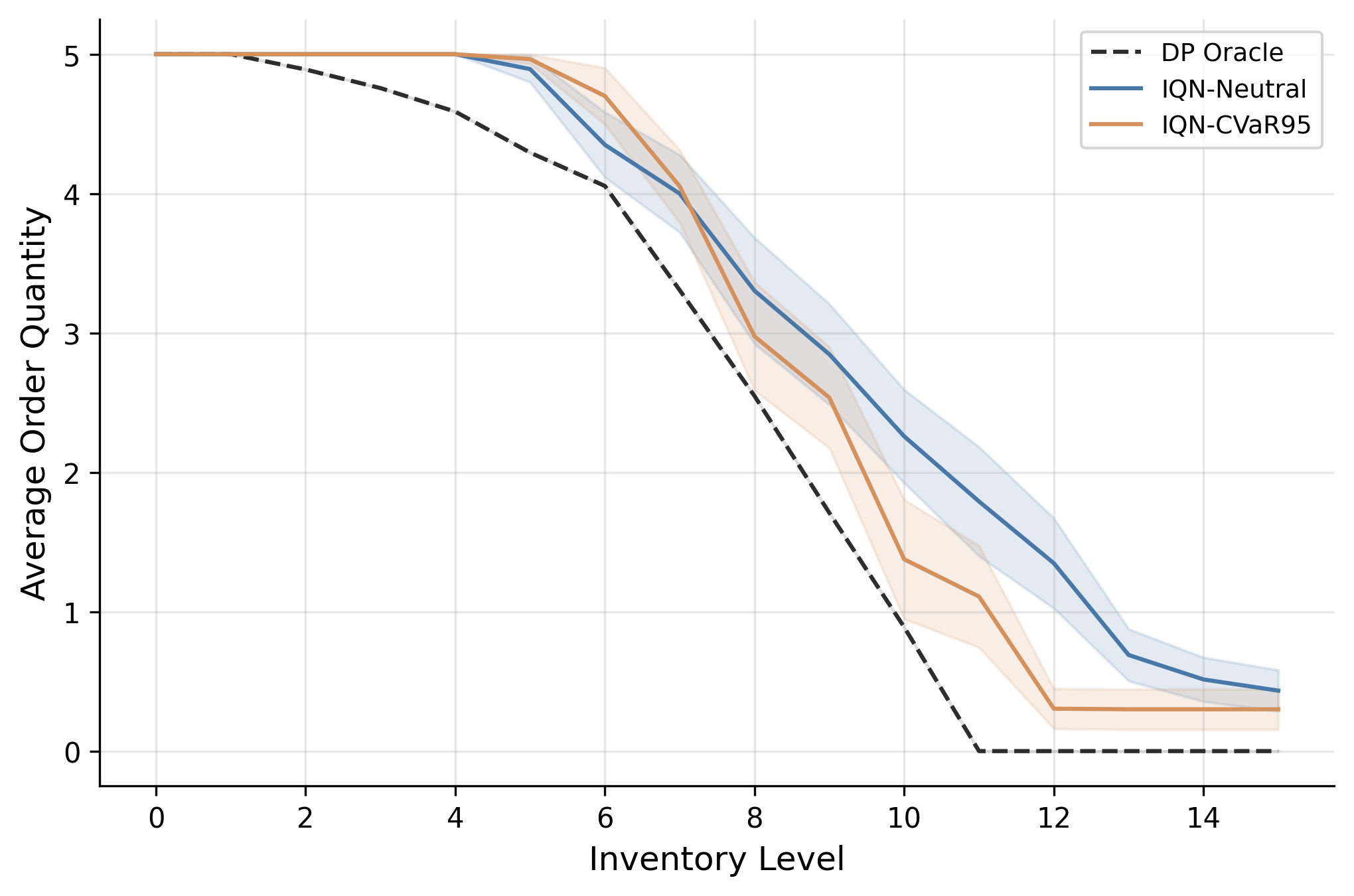}
\caption{Average order quantity by inventory level for the DP oracle
and the two IQN evaluation policies. Mean across ten training seeds
with $\pm 1$ SE bands, 50{,}000 evaluation episodes per seed.}
\label{fig:risk_inventory_policy}
\end{figure}

The tail-sensitive quantile sampling defined above shifts IQN toward
the lower-return part of the learned distribution. IQN-Neutral recovers
86.4\% of the DP oracle's mean return, and IQN-CVaR$_{95}$ recovers
90.1\%. Across ten seeds, the CVaR$_{95}$ evaluation improves the
average left tail relative to the risk-neutral evaluation, with
CVaR$_{95}$ of $-55.3$ against $-57.5$. Its mean return is 36.4 against
34.9, a gap of under two standard errors. Its ordering policy also
tracks the oracle more closely at high inventory levels
(Figure~\ref{fig:risk_inventory_policy}). Both learned policies order
more than the oracle. Their average order quantities across inventory
levels are 2.99 and 3.21 against the oracle's 2.56.

\subsection{Constrained Markov Decision Processes}
\label{subsec:constrained_mdps}

A constrained MDP augments the standard MDP
$(\mathcal{S}, \mathcal{A}, P, r, \gamma)$ with $K$ auxiliary cost
functions $c_k: \mathcal{S} \times \mathcal{A} \to \mathbb{R}$ and
constraint thresholds $q_k$. The agent maximizes the primary objective
subject to
\begin{equation}
  \max_\pi \; \mathbb{E}^\pi\Bigl[\sum_{t=0}^\infty \gamma^t r(S_t, A_t)\Bigr]
  \quad \text{subject to} \quad
  \mathbb{E}^\pi\Bigl[\sum_{t=0}^\infty \gamma^t c_k(S_t, A_t)\Bigr]
  \leq q_k, \quad k = 1, \ldots, K.
  \label{eq:cmdp}
\end{equation}

\citet{Altman1999} showed that this problem becomes a linear
program when reformulated over \emph{occupation measures} (the discounted fraction of
time spent in each state-action pair). The occupation measure of policy
$\pi$ is
\begin{equation}
  \nu^\pi(s,a) = (1-\gamma) \sum_{t=0}^\infty \gamma^t
  \mathbb{P}(S_t = s, A_t = a \mid \pi),
  \label{eq:occupation_measure}
\end{equation}
which satisfies the flow conservation constraints (probability flowing
into each state from transitions equals probability flowing out via
actions)
\begin{equation}
  \sum_a \nu(s,a) = (1-\gamma)\mu_0(s) + \gamma \sum_{s',a'} P(s|s',a')
  \nu(s',a') \quad \text{for all } s \in \mathcal{S}.
  \label{eq:bellman_flow}
\end{equation}
Since both the objective and
constraints are linear in $\nu$, and the feasible set forms a
polytope,\footnote{The feasible occupation measures form a bounded convex
set (polytope) defined by non-negativity, the Bellman flow conservation
equations, and the $K$ linear constraint inequalities.} the CMDP becomes a
standard LP. By Carath\'eodory's theorem,\footnote{Carath\'eodory's theorem
states that any point in the convex hull of a set in $\mathbb{R}^d$ can be
written as a convex combination of at most $d+1$ extreme points. Adding $K$
constraint inequalities introduces up to $K$ additional dimensions along
which the optimum may lie in the interior, so the optimal solution is a
convex combination of at most $K+1$ vertices.}
optimal CMDP policies mix at most $K+1$ deterministic policies for $K$
constraints.

The LP formulation yields \emph{strong duality} (Theorem~\ref{thm:prelim_duality}) under Slater's
condition.\footnote{Slater's condition requires the existence of a feasible
policy $\pi'$ satisfying all constraints with strict inequality:
$\mathbb{E}^{\pi'}[\sum_t \gamma^t c_k(S_t,A_t)] < q_k$ for all $k$.}
The dual problem is
\begin{equation}
  \min_{\lambda \geq 0} \max_\pi \; L(\pi, \lambda)
  = \mathbb{E}^\pi\Bigl[\sum_{t=0}^\infty \gamma^t \bigl(
    r(S_t,A_t) - \sum_{k=1}^K \lambda_k c_k(S_t,A_t)
  \bigr)\Bigr] + \sum_{k=1}^K \lambda_k q_k,
  \label{eq:cmdp_lagrangian}
\end{equation}
and the optimal dual variable $\lambda_k^*$ is the \emph{shadow price} of
the $k$-th constraint. By the envelope theorem (Theorem~\ref{thm:prelim_envelope}),
$\partial V^* / \partial q_k = \lambda_k^*$, the marginal change in optimal
value per unit relaxation of constraint $q_k$. This is the same shadow price
that appears in the linear programming dual of resource allocation
problems.

\citet{Paternain2019} extended this result beyond the LP formulation.
Despite the non-convexity of the objective in policy parameters, they proved
that constrained RL has \emph{zero duality gap}, meaning the optimal dual value
equals the primal. The proof exploits the convexity of the occupancy measure
set and applies a minimax theorem. This means the CMDP can be solved
exactly via dual ascent on the Lagrange multipliers, even when using
parametric policy classes, with an approximation error bounded by the
expressiveness of the parametrization.

\citet{Achiam2017} gave the foundational trust-region instantiation.
At iteration $k$, the policy update solves
\begin{equation}
  \max_\pi \; \mathbb{E}_{s \sim d^{\pi_k}\!, a \sim \pi}
    \bigl[A^r_{\pi_k}(s,a)\bigr]
  \quad \text{s.t.} \quad
  J_{C_i}(\pi_k) + \frac{1}{1-\gamma}
    \mathbb{E}_{s \sim d^{\pi_k}\!, a \sim \pi}
    \bigl[A^{C_i}_{\pi_k}(s,a)\bigr] \leq d_i,
  \quad
  \bar{D}_{\mathrm{KL}}(\pi \| \pi_k) \leq \delta,
  \label{eq:cpo_update}
\end{equation}
where $A^r_{\pi_k}$ and $A^{C_i}_{\pi_k}$ are the advantage functions for
the reward and the $i$-th cost, $d^{\pi_k}$ is the discounted state
visitation distribution, and
$\bar{D}_{\mathrm{KL}}(\pi\|\pi_k)
= \mathbb{E}_{s \sim \pi_k}[D_{\mathrm{KL}}(\pi(\cdot|s)\|\pi_k(\cdot|s))]$.
For the exact solution to~\eqref{eq:cpo_update}, monotone reward improvement
and per-iterate constraint satisfaction are
guaranteed.\footnote{\citet{Achiam2017} bound the worst-case constraint
degradation at
$O(\sqrt{\delta}\,\gamma\varepsilon/(1-\gamma)^2)$, where $\varepsilon$
bounds the cost advantage, via Pinsker's inequality. The implemented
Constrained Policy Optimization (CPO) algorithm solves a quadratic
approximation to~\eqref{eq:cpo_update}; the hard guarantee applies to the
exact solution.}\textsuperscript{,}\footnote{Subsequent theoretical work
sharpened convergence rates for primal-dual CMDP methods:
\citet{Tessler2019} proved almost-sure convergence of a three-timescale
actor-critic-multiplier scheme (RCPO) to a local Lagrangian optimum;
\citet{Ding2020} established the first non-asymptotic
$O(1/\sqrt{T})$ rate for both optimality gap and constraint violation
(dimension-free, via Fisher information geometry);
\citet{Liu2022cmdp} improved this to $O(\log(T)/T)$ using policy mirror
descent; and \citet{Ying2022} achieved $O(1/T)$ with entropy
regularization.}

In practice, the dominant method is PPO-Lagrangian, which replaces the
trust-region constraint in~\eqref{eq:cpo_update} with the PPO clipped
surrogate objective applied to the Lagrangian advantage
$\hat{A}^\lambda_t = \hat{A}^r_t - \lambda \hat{A}^c_t$:
\begin{equation}
  \max_\theta \; \mathbb{E}_{s,a \sim \pi_{\theta_k}}
  \Bigl[\min\bigl(
    \rho_t \hat{A}^\lambda_t,\;
    \mathrm{clip}(\rho_t, 1{-}\epsilon, 1{+}\epsilon)\,\hat{A}^\lambda_t
  \bigr)\Bigr],
  \qquad
  \rho_t = \frac{\pi_\theta(a_t|s_t)}{\pi_{\theta_k}(a_t|s_t)},
  \label{eq:ppo_lag_objective}
\end{equation}
where $\hat{A}^r_t$ and $\hat{A}^c_t$ are generalized advantage estimates
computed from two separate critics $V^r_\phi(s)$ and $V^c_\psi(s)$ via
$\hat{A}_t = \sum_{l=0}^{T-t}(\gamma\lambda_{\mathrm{GAE}})^l \delta_{t+l}$
with TD residuals
$\delta_t = r_t + \gamma V(s_{t+1}) - V(s_t)$ (analogously for cost).
After each policy update, the multiplier is adjusted by
projected gradient ascent on the constraint violation:
\begin{equation}
  \lambda_{t+1}
  = \bigl[\lambda_t + \eta\bigl(\hat{J}_C(\pi_\theta) - d\bigr)\bigr]_+.
  \label{eq:dual_update}
\end{equation}
\citet{Stooke2020} refined this update with a PID controller that dampens the multiplier oscillations characteristic of naive dual
ascent.\footnote{The FSRL library \citep{Liu2024fsrl} provides a
pip-installable implementation of PPO-Lagrangian alongside CPO and
TRPO-Lagrangian.}

\subsubsection{Simulation Study: Carbon-Constrained Production}
\label{subsec:sim_carbon_production}

A factory maximizes manufacturing profit subject to a carbon
emissions budget. The state is (inventory level, demand regime),
where inventory $\in \{0,\ldots,8\}$ and demand switches between
low and high regimes via a Markov chain. The action is
(production level, energy source), where production $\in \{0,1,2,3\}$
and energy is dirty (cheap, 3.0 tons CO$_2$ per unit) or clean
(expensive, 0.5 tons per unit). The reward is daily profit; the
cost is CO$_2$ emitted. The carbon budget $d$ is set to 30\% of the
unconstrained optimum's discounted emissions, and the LP dual
yields an analytical shadow price $\lambda^* = 1.20$.\footnote{The
constrained LP over occupation measures (18 states $\times$ 8
actions = 144 variables) is solved exactly with HiGHS. The shadow
price $\lambda^*$ is the dual variable on the carbon constraint.}
The CMDP is
\begin{equation}
  \max_\pi \; \mathbb{E}^\pi\Bigl[\sum_{t=0}^\infty \gamma^t
    \underbrace{\bigl(p \min(I_t + a^{\mathrm{prod}}_t,\, D_t)
    - \kappa_{e_t} a^{\mathrm{prod}}_t
    - h\,(I_t + a^{\mathrm{prod}}_t - D_t)^+\bigr)}_{r(S_t, A_t)}\Bigr]
  \quad \text{s.t.} \quad
  \mathbb{E}^\pi\Bigl[\sum_{t=0}^\infty \gamma^t
    \underbrace{\xi_{e_t}\, a^{\mathrm{prod}}_t}_{c(S_t, A_t)}\Bigr]
  \leq d,
  \label{eq:carbon_cmdp}
\end{equation}
where $p = 10$ is the unit price, $\kappa_e \in \{2, 5\}$ is the
production cost for dirty and clean energy respectively,
$h = 1$ is the holding cost, $\xi_e \in \{3.0, 0.5\}$ is the
emission rate, $D_t$ is stochastic demand, and $\gamma = 0.95$.

\begin{table}[h]
\centering
\caption{Carbon-constrained production results. Budget
$d = 31.35$. Q-learning rows report means and standard
errors over ten seeds.}
\label{tab:carbon_results}
\begin{tabular}{lrrcc}
\hline
Method & Return & Cost & Budget & $\lambda$ \\
\hline
LP Oracle & 186.4 & 31.35 & Y & 1.20 \\
Unconstrained Q-learning & 257.8 $\pm$ 3.2 & 99.08 $\pm$ 2.62 & N & -- \\
Lagrangian Q-learning & 178.4 $\pm$ 4.0 & 28.44 $\pm$ 3.60 & Y & 1.40 $\pm$ 0.00 \\
\hline
\end{tabular}

\end{table}

\begin{figure}[h]
\centering
\includegraphics[width=\textwidth]{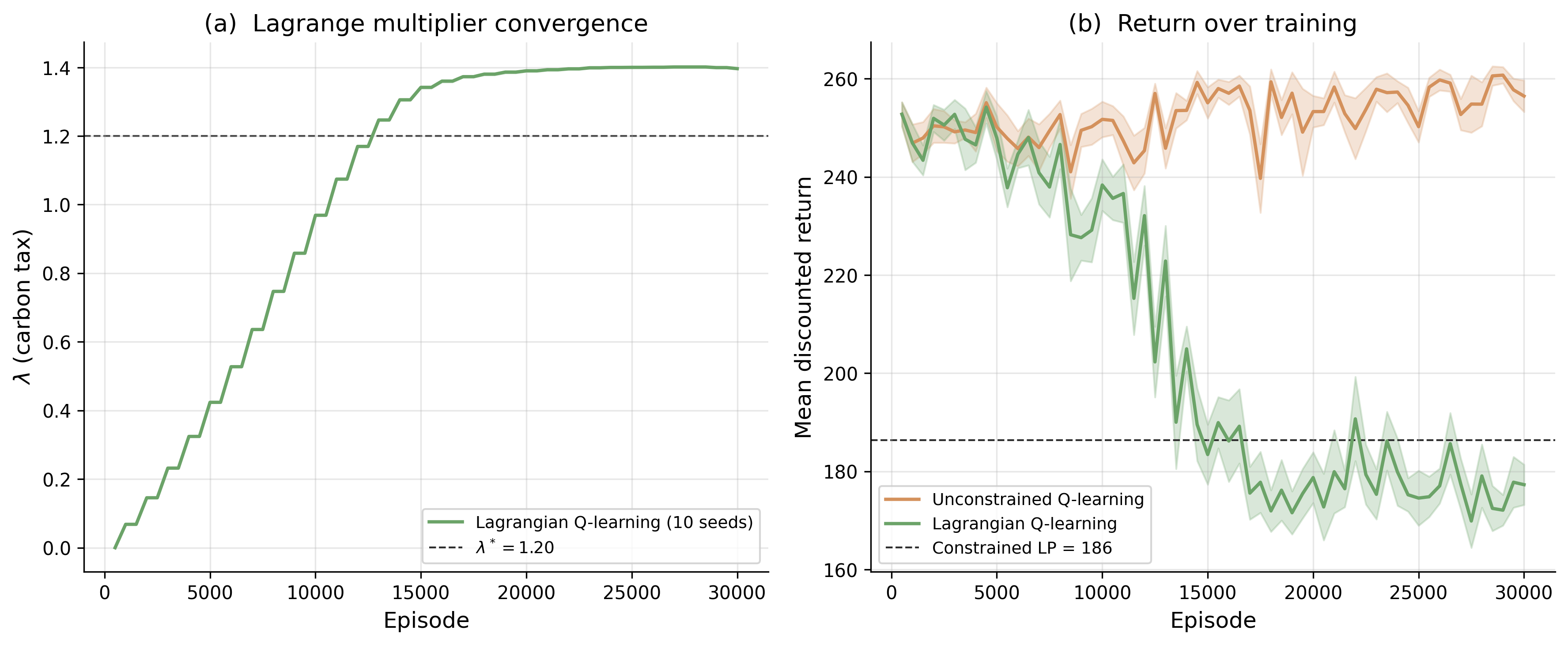}
\caption{(a) Lagrange multiplier $\lambda$ over training episodes,
with the LP shadow price $\lambda^*$ as the dashed reference.
(b) Mean discounted return for unconstrained and Lagrangian
Q-learning, with the constrained LP optimum as the dashed reference.
Solid lines are means over ten seeds; shaded bands are
$\pm 1$ standard error.}
\label{fig:carbon_convergence}
\end{figure}

Table~\ref{tab:carbon_results} and
Figure~\ref{fig:carbon_convergence} compare the constrained LP oracle,
unconstrained Q-learning, and Lagrangian Q-learning with the dual
update in~\eqref{eq:dual_update}. Each Q-learning variant runs over ten
seeds. The table reports means and standard errors.\footnote{The LP
oracle value is computed by exact policy evaluation
$V = (I - \gamma P_\pi)^{-1} R_\pi$, an infinite-horizon quantity. The
Q-learning returns are Monte Carlo averages over rollouts truncated at
horizon $H = 100$. With
$\gamma^{100} \approx 6\times 10^{-3}$, the truncation removes a
non-negative discounted tail bounded by roughly
$\gamma^{H} r_{\max}/(1-\gamma)$, on the order of one to two percent of
the return. Part of the gap between the LP value and the Q-learning
returns is therefore truncation bias rather than policy
suboptimality.}

Unconstrained Q-learning approaches the unconstrained DP return, with
$257.8 \pm 3.2$ against 273.00, but violates the carbon budget. Its
cost is $99.08 \pm 2.62$ against $d = 31.35$. Under the dual update,
the learned multiplier rises from zero and settles at
$\lambda \approx 1.40$, near the analytical shadow price
$\lambda^* = 1.20$. Across ten seeds, the peak value is
$1.405 \pm 0.002$, within $1.0\%$ of the settling value. Naive dual
ascent therefore exhibits little of the overshoot that
\citet{Stooke2020}'s PID controller is designed to dampen. The final
$\lambda$ is $16.7\%$ above $\lambda^*$. The corresponding policy
holds emissions at $28.44 \pm 3.60$, below the budget of 31.35, and
returns $178.4 \pm 4.0$ against the LP optimum of 186.4. The LP policy
randomizes between dirty and clean energy at one state, while
Q-learning recovers a deterministic approximation.

\subsection{Robust MDPs and Ambiguity Aversion}
\label{subsec:robust_mdps}

\citet{Iyengar2005} defined the robust Bellman operator
\begin{equation}
  (TV)(s) = \max_a \min_{p \in \mathcal{P}(s,a)}
  \bigl\{r(s,a) + \gamma \, p \cdot V\bigr\},
  \label{eq:robust_bellman}
\end{equation}
where $\mathcal{P}(s,a)$ is an uncertainty set of transition
distributions for each state-action pair and $p_0(\cdot|s,a)$ is the
nominal transition kernel.\footnote{The nominal model $p_0(\cdot|s,a)$ is
the agent's best estimate of the transition kernel, typically estimated
from data or specified by a simulator.} Under the rectangularity assumption (the uncertainty sets are independent
across state-action pairs, so the joint set is a Cartesian product),
$T$ is a $\gamma$-contraction in
the sup-norm, so robust value iteration and policy iteration converge
with the same guarantees as their standard counterparts (recall
Section~\ref{section:rl_algorithms}); the only change is substituting $T$
for the standard Bellman operator.\footnote{Rectangularity means
$\mathcal{P} = \prod_{s,a} \mathcal{P}(s,a)$. Without it, the problem
becomes NP-hard \citep{Wiesemann2013}.} For
KL balls $\mathcal{P}(s,a) = \{p : \text{KL}(p \| p_0) \leq
\kappa\}$,\footnote{A KL ball of radius $\kappa$ around the nominal
contains all distributions ``close'' to $p_0$ in an information-theoretic
sense. For example, if $p_0 = (0.1, 0.3, 0.4, 0.2)$ across four states,
a ball with $\kappa = 0.1$ allows distributions like
$(0.15, 0.35, 0.35, 0.15)$ but not $(0.5, 0.5, 0, 0)$, which would be
too far from the nominal. \citet{Barillas2009} proposed calibrating
$\kappa$ (equivalently $\theta$) via detection error probabilities.
When KL sets are insufficient because $p_0$ assigns zero probability to
relevant outcomes, Wasserstein balls
$\{Q : W_p(Q, P_0) \leq \epsilon\}$ are the alternative
\citep{Esfahani2018, GrandClement2021, Yu2023}.} the inner minimization
has the closed-form solution
$q^*(s') \propto p_0(s'|s,a) \exp(-\gamma V(s')/\theta)$, an
exponential tilting of nominal transitions toward low-value states,
where $\theta$ is the Lagrange multiplier on the KL
constraint \citep{Nilim2005}. \citet{HansenSargent2001}
independently derived the same operator from decision theory as
``multiplier preferences,'' in which the agent maximizes expected utility
penalized by the KL cost of distorting beliefs away from a reference
model. \citet{Petersen2000} proved that KL-robust MDPs, multiplier
preferences, and risk-sensitive control with exponential utility
$\mathbb{E}_P[\exp(-\text{cost}/\theta)]$ are three descriptions of the
same problem.\footnote{See also \citet{Whittle1981}, \citet{Jacobson1973},
\citet{Fleming1992}. \citet{Maccheroni2006}
axiomatized variational preferences, in which the agent evaluates each
act under the worst-case belief penalized by a divergence cost:
$V(f) = \min_p \{\int u(f)\,dp + c(p)\}$, nesting maxmin EU
\citep{GilboaSchmeidler1989}, Hansen-Sargent, and mean-variance as
special cases. \citet{Strzalecki2011} showed KL is the unique penalty
satisfying a natural invariance axiom; \citet{HansenSargent2024}
provided a recent comprehensive treatment.}

\subsubsection{Algorithms for Robust RL}
\label{subsubsec:robust_algorithms}

The robust Bellman operator~\eqref{eq:robust_bellman} is a drop-in
replacement for the standard Bellman target. Robust Q-learning replaces
the TD target with its worst-case counterpart:
\begin{equation}
  Q(s,a) \leftarrow Q(s,a) + \alpha\Bigl[
    r + \gamma \min_{p \in \mathcal{P}(s,a)} \sum_{s'} p(s')
    \max_{a'} Q(s',a') - Q(s,a)
  \Bigr],
  \label{eq:robust_q_learning}
\end{equation}
where the inner minimization uses the exponential tilting for KL balls or
bisection for TV and chi-squared
sets.\footnote{Sample complexity for learning robust policies scales
polynomially in $|\mathcal{S}||\mathcal{A}|$: \citet{Panaganti2022}
established the first bounds under $(s,a)$-rectangular uncertainty for TV,
KL, and chi-squared sets; \citet{clavier2024robust} improved the rates for
model-based approaches.}

For high-dimensional problems where tabular methods and explicit
uncertainty sets are infeasible, \citet{Pinto2017} introduced Robust
Adversarial Reinforcement Learning (RARL):
\begin{equation}
  \max_{\pi_{\text{agent}}} \min_{\pi_{\text{adv}}} \;
  \mathbb{E}\Bigl[\sum_{t=0}^\infty \gamma^t
    r(s_t, a_t, a_t^{\text{adv}})
  \Bigr],
  \label{eq:rarl}
\end{equation}
where $a_t^{\text{adv}}$ is the adversary's action. Training alternates
between updating the agent policy $\pi_{\text{agent}}$ via TRPO or PPO
with the adversary fixed, then updating the adversary
$\pi_{\text{adv}}$ with the agent fixed. The adversary's action space is
problem-specific (joint torques for locomotion, force perturbations for
manipulation). The resulting agent policies are more conservative, with
wider stability margins; in locomotion tasks, robust agents adopt lower
center-of-gravity gaits. \citet{Pinto2017} demonstrated that policies
trained against an adversary in simulation transferred to physical robots
more reliably than standard PPO policies.

\citet{Derman2021} proved that entropy regularization provides implicit
robustness. The soft Bellman equation used in SAC,
\begin{equation}
  V(s) = \tau \log \sum_a \exp\bigl(Q(s,a)/\tau\bigr),
  \label{eq:soft_bellman}
\end{equation}
is equivalent to solving a robust MDP with reward uncertainty set
$\{r' : \text{KL}(r' \| r) \leq \kappa\}$, where the entropy temperature
$\tau$ maps directly to the robustness radius $\kappa$. Adding a
value-regularization term extends this to robustness against transition
misspecification.\footnote{The ``Twice Regularized MDP'' (R$^2$-MDP)
combines policy regularization (reward robustness) with value
regularization (transition robustness). Standard SAC training already
implements the reward-robust component; transition robustness requires an
additional penalty on the divergence between learned and nominal dynamics
models.}

These three approaches offer different trade-offs. Robust Q-learning
requires an explicit uncertainty set but provides exact minimax guarantees
for tabular problems. RARL avoids specifying an uncertainty set by learning
the adversary, making it practical for continuous control, but provides no
formal robustness certificate. Entropy regularization provides implicit
robustness at zero additional implementation cost for practitioners already
using SAC, but ties the robustness radius to the temperature parameter.

\subsubsection{Simulation Study: Consumption-Savings Under Model Mismatch}
\label{subsec:sim_robust_consumption}

A consumption-savings agent with constant relative risk aversion (CRRA) utility
$u(c) = c^{1-\sigma}/(1-\sigma)$, $\sigma = 2$, receives stochastic
income $y \in \{1,\ldots,5\}$ each period and chooses how much to consume
versus save at gross return $R = 1.02$, with $\gamma = 0.95$. The robust
Bellman equation for this problem is
\begin{equation}
  V(w) = \max_{c \in \{0,\ldots,w\}} \Bigl\{
    u(c) + \gamma \min_{q:\,\text{KL}(q\|p_0) \leq \kappa}
    \sum_{y} q(y)\, V\bigl(R(w-c) + y\bigr)
  \Bigr\},
  \label{eq:robust_consumption}
\end{equation}
where $w$ is wealth, $c$ is consumption, and the inner minimization tilts
income probabilities toward low-income states via
$q^*(y) \propto p_0(y)\exp(-\gamma V(R(w-c)+y)/\theta)$. Seven policies
are computed: standard DP under the nominal income distribution
$p_0 = (0.05, 0.10, 0.20, 0.30, 0.35)$, robust DP at $\theta = 5$
(moderate) and $\theta = 2$ (high robustness), an oracle that knows the
true perturbed distribution
$\tilde{p} = (0.30, 0.30, 0.20, 0.10, 0.10)$, and three model-free
counterparts trained via tabular Q-learning under the nominal
model.\footnote{Standard Q-learning uses the usual TD target
$r + \gamma \max_{a'} Q(s',a')$; robust Q-learning replaces this with
the worst-case target~\eqref{eq:robust_q_learning}, using the nominal
kernel for the inner minimization (generative model setting).
Visit-count learning rates $\alpha = C/(C + N(s,a))$ with $C = 100$
and $\varepsilon$-greedy exploration decaying from 1.0 to 0.05 over
$10^5$ episodes.} All seven policies are then evaluated under both the
nominal and perturbed income models.

\begin{table}[h]
\centering
\caption{Discounted returns under nominal and perturbed income. DP policies
are computed via value iteration; Q-learning policies are trained under the
nominal model.}
\label{tab:robust_consumption}
\begin{tabular}{lrrr}
\hline
Method & Nominal & Perturbed & Degradation (\%) \\
\hline
Standard DP & $-4.941 \pm 0.001$ & $-8.694 \pm 0.004$ & $-76.0$ \\
Q-learning & $-4.941 \pm 0.001$ & $-8.692 \pm 0.005$ & $-75.9$ \\
Robust DP ($\theta$=5.0) & $-4.942 \pm 0.001$ & $-8.672 \pm 0.004$ & $-75.5$ \\
Robust Q-learning ($\theta$=5) & $-4.942 \pm 0.001$ & $-8.672 \pm 0.004$ & $-75.5$ \\
Robust DP ($\theta$=2.0) & $-4.948 \pm 0.001$ & $-8.486 \pm 0.004$ & $-71.5$ \\
Robust Q-learning ($\theta$=2) & $-4.948 \pm 0.001$ & $-8.486 \pm 0.004$ & $-71.5$ \\
Oracle & $-5.465 \pm 0.001$ & $-7.606 \pm 0.003$ & $-39.2$ \\
\hline
\end{tabular}
\end{table}

The precautionary shift in Table~\ref{tab:robust_consumption} is
consistent with the adverse low-income tilt in the inner minimization
of~\eqref{eq:robust_consumption}.
Each of the seven policies is evaluated on a common set of ten
income-sequence seeds. Under the nominal model, the policies have
similar returns. Under the perturbed model, standard DP degrades by
76.0\%, while the $\theta = 2$ robust policy degrades by 71.5\%.
Across the ten seeds, the Q-learning policies recover their DP
counterparts. The two robust variants reproduce the robust DP
consumption rule exactly, and standard Q-learning differs at no more
than one wealth level. Their returns are statistically
indistinguishable from the corresponding DP returns.

\begin{figure}[h]
\centering
\includegraphics[width=0.7\textwidth]{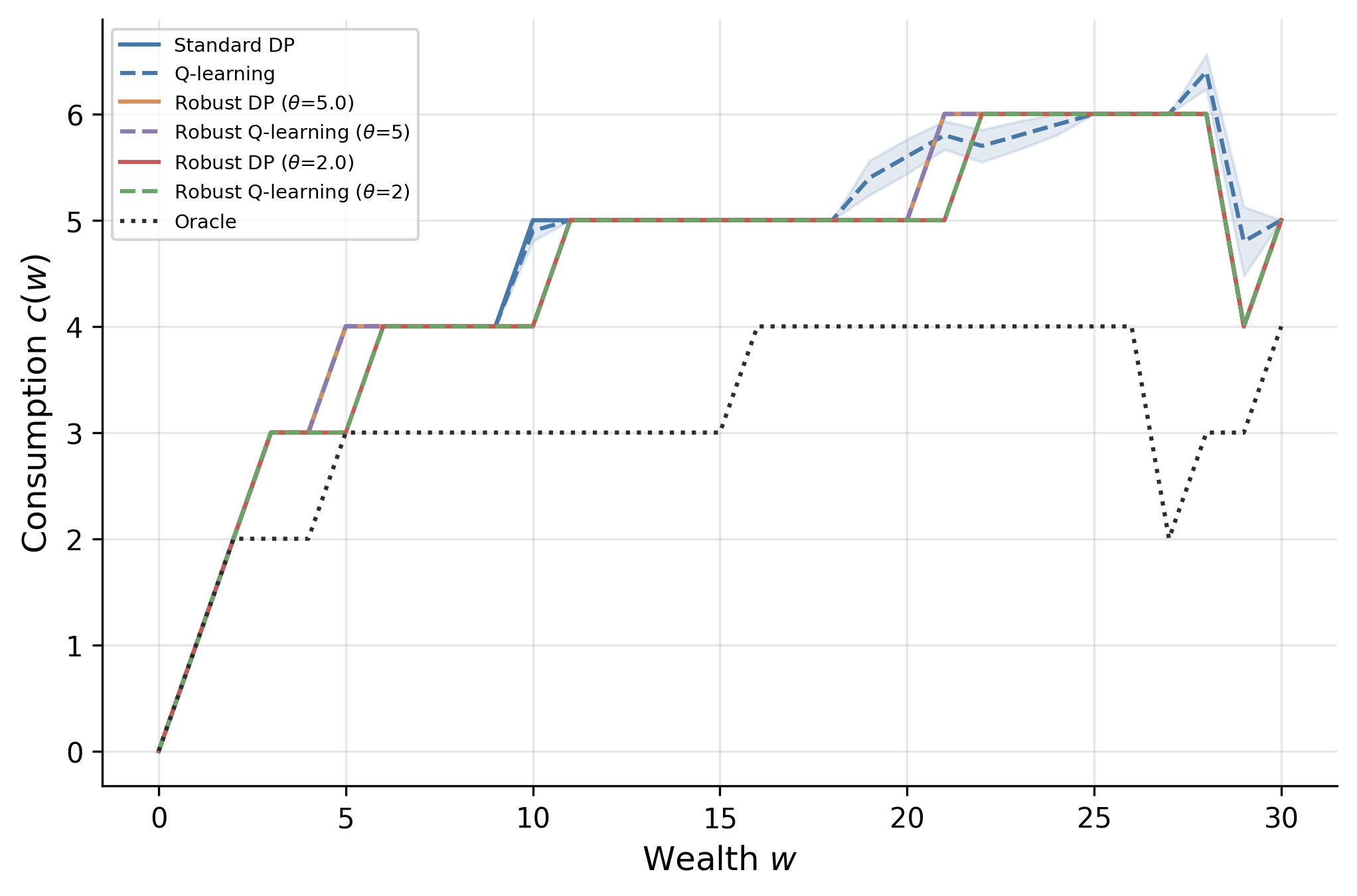}
\caption{Consumption policy $c(w)$ for each method. Dashed lines show
Q-learning policies; solid lines show DP.}
\label{fig:robust_consumption}
\end{figure}

The $\theta = 2$ policy recovers 4.5 percentage points of the 76.0\%
degradation, whereas the oracle that knows the perturbed distribution
recovers 36.8 points. KL-ball robustness hedges against a band of
nearby distributions rather than fitting the realized one, so it
closes only part of the oracle gap.
Figure~\ref{fig:robust_consumption} shows that the robust policies
consume less at each wealth level and build a precautionary savings
buffer against adverse income realizations.

\FloatBarrier
\clearpage
\subsection{Engine Replacement MDP: A Replacement Budget and Its Shadow Price}
\label{engine:ch11}

\begin{table}[H]
\centering
\caption{Constrained occupancy and KL-tilting calculations for the Engine
Replacement MDP. The CMDP rows use a low-mileage initial engine. The geometry
row uses a uniform initial distribution so every deterministic-policy occupancy
is distinct.}
\label{tab:engine_occupancy_kl}
\begin{tabular}{llrr}
\hline
Calculation & Setting & Estimate & Check \\
\hline
CMDP & Best deterministic return & 3.4545 & feasible \\
CMDP & Randomized return & 4.6727 & $\pi(R\mid\mathrm{high})=0.4074$ \\
CMDP & Replacement occupancy & 0.2000 & budget $0.2000$ \\
CMDP & Shadow price $\lambda^*$ & 6.0909 & slope $6.0909$ \\
Geometry & Uniform-start return & 3.9455 & $\pi(R\mid\mathrm{high})=0.3437$ \\
KL tilt & $\theta=10$ high-state probability & 0.5233 & shift $+0.0233$ \\
KL tilt & $\theta=0.5$ high-state probability & 0.8655 & shift $+0.3655$ \\
\hline
\end{tabular}

\end{table}

\begin{figure}[H]
\centering
\includegraphics[width=0.92\textwidth]{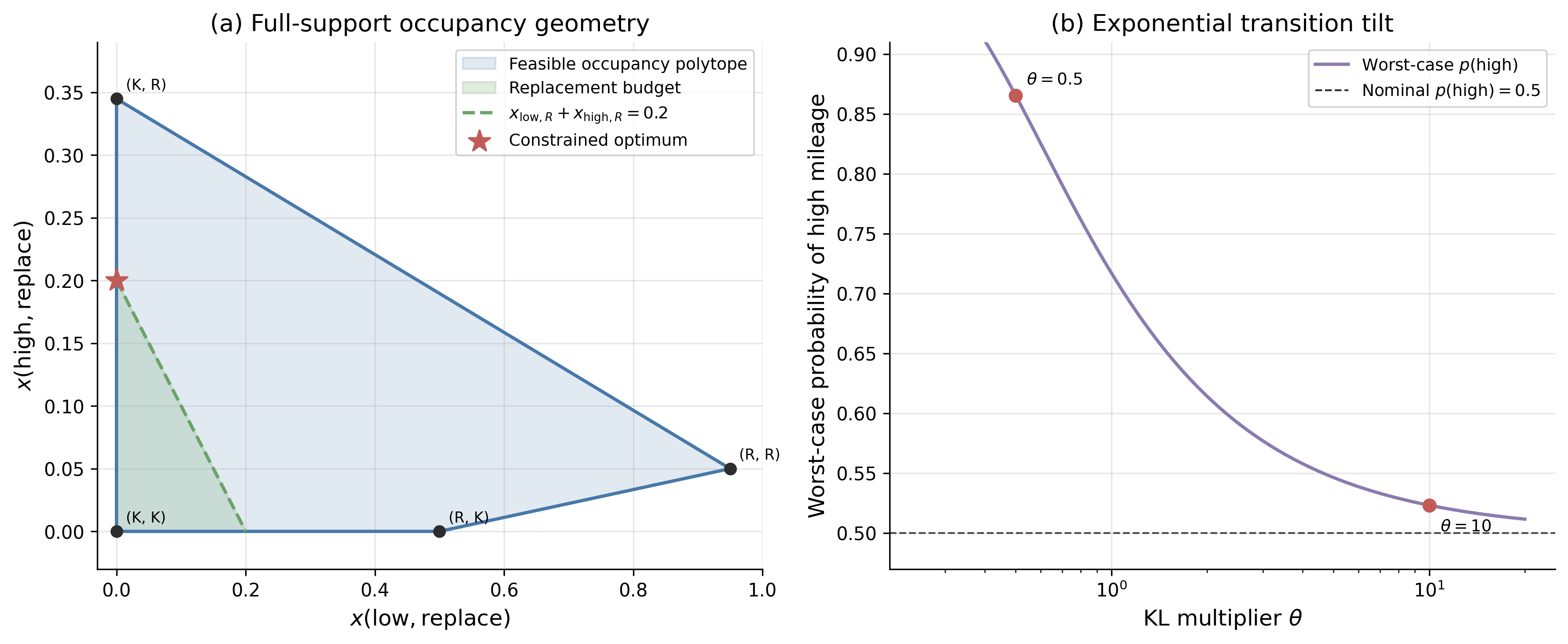}
\caption{The left panel shows the occupancy polytope under a full-support
initial distribution, the replacement budget, and the randomized optimum.
The right panel shows the worst-case high-mileage transition probability
under KL multiplier preferences.}
\label{fig:engine_occupancy_kl}
\end{figure}

The replacement cost is $c(s,a)=\mathbbm{1}\{a=\mathrm{replace}\}$, and
the normalized discounted budget is $0.2$. The best feasible deterministic
policy never replaces and yields $3.4545$ from a low-mileage engine. The
occupancy LP instead replaces a high-mileage engine with probability $0.4074$,
binds the budget exactly, and yields $4.6727$. Its multiplier is
$\lambda^*=6.0909$. Centered budget perturbations reproduce the same slope,
so a $0.01$ increase in normalized replacement occupancy raises the local
optimal return by $0.0609$.

For the nominal low-mileage keep transition, both successor grades have
probability $0.5$. The KL multiplier solution tilts toward the lower-value
high-mileage grade. Its probability rises to $0.5233$ when $\theta=10$ and
to $0.8655$ when $\theta=0.5$.
\FloatBarrier

\section{World Models and Model-Based Reinforcement Learning}
\label{section:world_models}
%
%

\subsection{Learning Without Rational Expectations}
\label{section:fc_paradigms}

This chapter studies world models in reinforcement learning, learned forecasters of state transitions and rewards that an agent uses to plan its actions. Economics has studied the same object since at least the 1980s under the heading of learning without rational expectations, where an agent is not handed the true law of motion of its environment but estimates one from data and acts on the estimate. The agent's model at date $t$ is a parameter $\theta_t \in \Theta$ that indexes a transition estimate $\widehat{P}_{\theta_t}(s' \mid s, a)$ and a reward estimate $\widehat{r}_{\theta_t}(s, a)$. A planning operator $G$ returns a policy $\pi_t = G(\theta_t)$ that is optimal, or at least improved, under the current model. The agent draws a real transition $(s_t, a_t, r_t, s_{t+1})$ with $a_t \sim \pi_t(\cdot \mid s_t)$. A learning operator $F$ then revises the model,
\[
  \theta_{t+1} = \theta_t + \alpha_t \, F(\theta_t, s_t, a_t, r_t, s_{t+1}),
\]
where $\alpha_t$ is a gain sequence. Stochastic approximation studies exactly this kind of recursion. A rest point is a $\theta^\star$ at which the model is consistent with the data its own policy $G(\theta^\star)$ generates. Success hinges on two questions. The first is whether $\theta_t$ converges to a rest point. The second is whether the policy $G(\theta^\star)$ there is optimal on the true environment, which it need not be when the model class cannot contain the truth. The two economic examples below each speak to one of these questions. The rest of the chapter runs the same loop in reinforcement learning, at larger scale.

\subsubsection{Two economic examples of model-based learning}
\label{section:fc_paradigm_family}

The first example is recursive least squares adaptive learning \citep{MarcetSargent1989, EvansHonkapohja2001}. The agent holds a perceived law of motion, a parametric forecast of the variables it cares about, and updates its coefficients $\theta$ from realized data. With forecast target $p_t$, regressor $z_{t-1}$, and perceived disturbance $\varepsilon_t$, the perceived law of motion is $p_t = \theta' z_{t-1} + \varepsilon_t$ and the update is recursive least squares,
\[
  \theta_t = \theta_{t-1} + t^{-1} R_t^{-1} z_{t-1}\big(p_t - \theta_{t-1}' z_{t-1}\big), \qquad
  R_t = R_{t-1} + t^{-1}\big(z_{t-1} z_{t-1}' - R_{t-1}\big),
\]
with $R_t$ the running second-moment matrix of the regressor and $t^{-1}$ the decreasing gain. Decisions taken under the current belief feed back into the data, so a believed $\theta$ induces an actual law of motion whose coefficients form a map $T(\theta)$. A rational-expectations equilibrium is a fixed point $\bar\theta = T(\bar\theta)$, and convergence to it is governed by expectational stability, the local stability of $d\theta/d\tau = T(\theta) - \theta$ at $\bar\theta$, equivalently that the eigenvalues of $DT(\bar\theta) - I$ have strictly negative real parts. In a market with a single feedback coefficient $\lambda$ from average expectations to the realized outcome this reduces to $\lambda < 1$. This is the benign case, since the model can contain the truth and the only thing that can fail is dynamic. The lone eigenvalue condition decides whether the belief is drawn to the equilibrium or driven away.

The second example is misspecified Bayesian learning, formalized as Berk-Nash equilibrium \citep{EspondaPouzo2016}, where the model state is a belief $\mu_t \in \Delta(\Theta)$ rather than the point estimate $\theta_t$ of the loop above. The agent carries a subjective class of models $\{Q_\theta : \theta \in \Theta\}$ of the map from action $a$ to consequence $Y$, with payoff $u(a, Y)$ and reward $r = u(a, Y)$, and forms the predictive $\bar{Q}_\mu(\cdot \mid s, a) = \int_\Theta Q_\theta(\cdot \mid s, a)\, \mu(d\theta)$. Under repeated interaction the belief concentrates on the parameters minimizing a weighted Kullback-Leibler divergence to the true distribution $Q_\pi$ generated under the agent's own policy $\pi$,
\[
  K(\pi, \theta) = \sum_{s, a} \pi(a \mid s)\, \rho(s)\; \mathbb{E}_{Q_\pi(\cdot \mid s, a)}\!\left[\ln \frac{Q_\pi(Y \mid s, a)}{Q_\theta(Y \mid s, a)}\right], \qquad
  \Theta(\pi) = \arg\min_{\theta \in \Theta} K(\pi, \theta),
\]
where $\rho(s)$ is the distribution of the signal $s$. A Berk-Nash equilibrium is a pair $(\pi, \mu)$ with the policy optimal under the predictive belief, so that $\pi(a \mid s) > 0$ implies $a \in \arg\max_{a'} \mathbb{E}_{\bar{Q}_\mu(\cdot \mid s, a')}[u(a', Y)]$, and the belief supported on the best-fit set, $\operatorname{supp}(\mu) \subseteq \Theta(\pi)$. The weighting by the agent's own policy is decisive, since the agent only observes data on the paths its behavior visits. This is the adversarial case, and the ordinary one once the model class of a neural network or an ensemble cannot contain the true dynamics. The agent converges calmly to a wrong model and a suboptimal policy. The data it generates never contradicts them, and more data does not dislodge them. This is the economic version of the world-model failure discussed in the synthesis (\S\ref{section:fc_synthesis_fails}). The well-specified counterpart in reinforcement learning is posterior sampling, which draws a model from the posterior and plans against it.

Fictitious play is the same loop in the multi-agent case, the estimated model being the empirical distribution of opponents' actions and the planner a best response \citep{Brown1951, FudenbergKreps1993, HofbauerSandholm2002, FudenbergLevine1998, Young2004}.

\label{section:fc_convergence_question}%
Not every economic learning scheme holds a model of the environment. Genetic-algorithm learning \citep{Arifovic1994, Arifovic1995} evolves a population of decision rules directly by realized fitness, the economic counterpart of gradient-free policy search. Evolutionary heuristic switching \citep{BrockHommes1997} selects among a fixed menu of forecasting rules by a logit on past performance, learning selection weights rather than a model. Both omit the learning operator $F$. They are model-free, the contrast that defines the two examples above. The shared vocabulary of stochastic approximation and fixed-point analysis traces back to \S\ref{section:rl_algorithms}. All of these modes are run head to head on a fixed environment, one mode at a time, in the dual simulation of \S\ref{section:fc_dual_sim}.

\subsection{Two 1990 Origins: Dyna and Differentiable World Models}
\label{section:fc_origins1990}

Dyna and the controller-model architecture are the reinforcement-learning versions of the model-based loop set out above. Two papers published in 1990 introduced what the modern literature now calls a world model. In each, an agent learns a model of its environment from interaction data and uses simulated experience drawn from that model to improve its policy, rather than waiting for real transitions to accumulate. \citet{sutton1990} arrived at the architecture through dynamic programming and reinforcement learning, framing it as the unification of direct and indirect updates within a single control loop. \citet{Schmidhuber1990} arrived at it through recurrent neural networks, framing it as a way to make the environment itself differentiable for credit assignment. The two papers do not cite one another and use different vocabularies, but the architectural commitment is the same. We treat them in parallel rather than in lineage, since each anticipates a distinct branch of the modern literature.

\subsubsection{Sutton's integrated architecture}
\label{section:fc_origins_sutton}

\citet{sutton1990} proposes Dyna as an architecture that interleaves three processes at every real time step. The agent performs a direct Q-learning update from the observed transition $(s, a, r, s')$, then updates a learned model $\widehat{P}(s' \mid s, a)$ and $\widehat{r}(s, a)$ from the same transition, then performs $K$ planning updates by sampling previously visited state-action pairs $(\tilde{s}, \tilde{a})$, querying the model for simulated outcomes $(\tilde{r}, \tilde{s}')$, and applying the same Q-update to those synthetic transitions. The ratio $K$ of planning steps to real steps controls how heavily the agent exploits its learned model between actions. It is the first explicit statement of the sample-efficiency claim that modern model-based work continues to revisit. The MBPO short-rollout analysis in \citet{janner2019model} is in the same family. We defer the algorithm block to \S\ref{section:fc_dyna_q}, where we develop Dyna-Q in the notation of this survey.

The openness of Dyna matters as much as its sample-efficiency claim. Sutton states explicitly that the learned model can be probabilistic and oftentimes incorrect, with the planner doing useful work as long as the cost of synthetic experience is small relative to the cost of real experience. No structure is imposed on the class of admissible models, on the loss used to fit them, or on the planner that consumes their output. This open design is what subsequent work fills with tabular counts, Gaussian processes, deep ensembles, recurrent state-space models, and implicit latent dynamics, each plugged into the same outer loop. That openness makes Dyna the entry point for the work that runs through \S\ref{section:fc_mbpo} and \S\ref{section:fc_tdmpc2}.

\subsubsection{Schmidhuber's controller-model architecture}
\label{section:fc_origins_schmidhuber}

\citet{Schmidhuber1990} proposes a paired architecture consisting of a recurrent model network $M$ and a controller network $C$. The model $M$ receives the controller's outputs together with current observations and is trained, self-supervised, to predict the next inputs to $C$, including sensory observations and reinforcement signals. The controller $C$ produces actions and is trained jointly with $M$. Once $M$ predicts well, the environment becomes differentiable from the controller's perspective, since the Jacobian of $M$ stands in for the unknown Jacobian of the true dynamics. Planning is then implemented as gradient descent through the model network, by unrolling $C$ and $M$ together for a fixed horizon and descending on a predicted pain or cost signal, without executing any real actions during the planning phase. This is the first proposal of analytic gradient backpropagation through a learned dynamics model for policy improvement.

In the notation of this survey, let $M_\theta$ parameterize the world model with $\widehat{P}_\theta(s_{t+1} \mid s_t, a_t)$ and let $C_\phi$ parameterize the controller with $\pi_\phi(a_t \mid s_t)$. The model is fit on observed transitions by the negative log-likelihood
\[
  \mathcal{L}_M(\theta) = -\sum_t \log \widehat{P}_\theta(s_{t+1} \mid s_t, a_t),
\]
and the controller is fit by maximizing the expected discounted return under the \emph{learned} model,
\[
  J(\phi) = \mathbb{E}_{a_t \sim \pi_\phi(\cdot \mid s_t),\; s_{t+1} \sim \widehat{P}_\theta(\cdot \mid s_t, a_t)} \Big[ \sum_t \gamma^t r_t \Big],
\]
with the planning operator $\phi \leftarrow \phi + \eta \, \nabla_\phi J(\phi)$ implemented by backpropagating through the $H$-step unrolled $M_\theta$ via reparameterized or score-function gradients. A curiosity term adds an intrinsic reward proportional to the model's own prediction error, training $M_\theta$ to reduce it and drawing $C_\phi$ toward states where the model is uncertain. The idea reappears as the intrinsic curiosity of \citet{Pathak2017}, and the full deep realization of the controller-model architecture is \citet{HaSchmidhuber2018}, treated in \S\ref{section:fc_ha_schmidhuber}.

Dyna and the controller-model architecture differ in two ways and agree in one. The model class differs. Dyna uses a tabular Dirac, Schmidhuber a differentiable network. The planner differs too. Dyna applies $K$ tabular $Q$-updates per real step, Schmidhuber a gradient step through the unrolled $M_\theta$. The outer loop is identical, since both learn $M$ from real data, use it to improve the policy, and act in the world. Each anticipated, without the tools to make it work at the scale of pixels or continuous control, a component that took three decades to mature, intrinsic curiosity on one side and planning amplification through short model rollouts on the other.

\subsection{Dyna-Q in Modern Notation}
\label{section:fc_dyna_q}

The previous section traced two parallel origins of model-based reinforcement learning in 1990, one from the dynamic-programming tradition and one from the recurrent-neural-network tradition. This section develops the canonical instantiation of Sutton's side of that origin story, the Dyna-Q algorithm, in proper notation. Dyna-Q combines tabular $Q$-learning with a one-step deterministic model of the environment and applies $K$ planning updates per real interaction step. It is the template that the deep-learning era inherits and modifies. It is also the reference point that the algorithms chapter (\S\ref{section:rl_algorithms}) points to whenever model-based methods enter the discussion. The section gives the algorithm in formal notation, states the empirical headline from Sutton's original gridworld experiment, and then isolates the two commitments that survive every subsequent reformulation of the idea.

\subsubsection{Algorithm}
\label{section:fc_dyna_q_algorithm}

Fix a finite Markov decision process with state space $\mathcal{S}$ and action space $\mathcal{A}$. Let $Q : \mathcal{S} \times \mathcal{A} \to \mathbb{R}$ denote the tabular action-value function the agent maintains over the course of interaction. Let $\widehat P(s' \mid s, a)$ denote the learned transition model and $\widehat r(s, a)$ the learned reward model; together these form the agent's internal model of its environment. The remaining objects are a step size $\alpha \in (0, 1]$, a discount factor $\gamma \in [0, 1)$, an exploration parameter $\varepsilon \in (0, 1)$, and a planning horizon $K \in \mathbb{N}$ counting the number of synthetic updates applied per real step. The agent initializes $Q$, $\widehat P$, and $\widehat r$ to arbitrary values and then proceeds as follows.

For each step $t$:
\begin{enumerate}
  \item Observe the current state $s$ and select an action $a \in \arg\max_{a'} Q(s, a')$ with $\varepsilon$-greedy exploration, breaking ties uniformly at random.
  \item Execute $a$ in the environment and observe the reward $r$ and the next state $s'$.
  \item Direct RL update,
        \[ Q(s, a) \leftarrow Q(s, a) + \alpha \big[ r + \gamma \max_{a''} Q(s', a'') - Q(s, a) \big]. \]
  \item Model update, $\widehat r(s, a) \leftarrow r$ and $\widehat P(\cdot \mid s, a) \leftarrow \delta_{s'}$, recording the most recent observed transition as a deterministic prediction.
  \item Planning. Repeat $K$ times, on each iteration sampling a previously-visited pair $(\tilde s, \tilde a)$ uniformly from the agent's experience, querying $\widehat r(\tilde s, \tilde a)$ and $\widehat P(\cdot \mid \tilde s, \tilde a)$ to obtain a synthetic transition $(\tilde r, \tilde s')$, and applying the same $Q$-update as in step 3,
        \[ Q(\tilde s, \tilde a) \leftarrow Q(\tilde s, \tilde a) + \alpha \big[ \tilde r + \gamma \max_{a''} Q(\tilde s', a'') - Q(\tilde s, \tilde a) \big]. \]
\end{enumerate}

Step 3 is the tabular $Q$-learning update of \citet{WatkinsDayan1992}. It converges to the optimal action-value $Q^\star$ under the standard step-size and visitation conditions, with the $\gamma$-contraction of the Bellman optimality operator supplying the argument (Theorems~\ref{thm:qlearning_convergence} and~\ref{thm:q_factor_contraction}, Section~\ref{sec:planning_learning}). Step 5 applies the same update with $P^\star$ and $r$ replaced by the learned $\widehat P$ and $\widehat r$. Convergence of the full planning-and-learning loop has no closed-form guarantee in the tabular case beyond the consistency of $\widehat P$ and $\widehat r$ under the agent's visitation distribution. The function-approximation setting is studied via Lipschitz-on-model arguments (\citealt{Asadi2018Lipschitz}, discussed in \S\ref{section:fc_synthesis_fails}) rather than by a single contraction.

The conceptual content of the indirect-versus-direct distinction collapses into a single observation. Steps 3 and 5 apply identical update rules to identical-looking $(s, a, r, s')$ tuples. The only difference is the provenance of the tuple. In step 3 the tuple is sampled from the true transition kernel of the environment by acting; in step 5 the tuple is sampled from the agent's learned model. Planning, in this architecture, is the same update rule applied to model-generated rather than environment-generated experience, with no separate algorithm or distinct update rule of its own. The ratio $K$ controls how heavily the agent leans on its model relative to fresh interaction.

\subsubsection{Empirical headline}
\label{section:fc_dyna_q_empirics}

\citet{sutton1990} reports a gridworld maze experiment that fixes the empirical claim. The agent navigates a tabular maze with deterministic transitions and sparse reward, learning under three settings of the planning horizon, $K \in \{0, 5, 50\}$. Setting $K = 0$ recovers plain $Q$-learning, since the planning loop is skipped and only the direct RL update of step 3 fires per interaction. With $K = 50$, the agent reaches near-optimal performance in roughly three episodes; with $K = 0$, the same level of performance requires roughly twenty-five episodes. The ratio of episodes is the original planning-amplification claim. Each real step in the $K = 50$ variant carries the policy-improvement value of about fifty steps in the $K = 0$ variant, at the cost of additional internal computation and a learned model that must be accurate enough for the synthetic updates to be informative. A blocking-maze variant in the same paper shows that a small exploration bonus permits the agent to recover after the environment changes, addressing the obvious failure mode that comes with relying on a learned model.

\subsubsection{Two commitments of world models}
\label{section:fc_dyna_q_architecture}

The first commitment is that the model class is left unspecified. The algorithm box above uses a deterministic one-step Dirac model $\widehat P(\cdot \mid s, a) \leftarrow \delta_{s'}$, but this choice is incidental to the architecture. Sutton describes the model as ``probabilistic and oftentimes incorrect,'' which is the conceptual move that allows every subsequent paper to plug a different model class into the same outer loop without changing anything else. The deep era exercises this openness across diverse model classes. \citet{HaSchmidhuber2018} use a convolutional variational autoencoder paired with a mixture-density recurrent network, treated in \S\ref{section:fc_ha_schmidhuber}. The Recurrent State-Space Model that powers the PlaNet and Dreamer lines, treated in \S\ref{section:fc_rssm}, couples a deterministic recurrent path with a stochastic latent. MuZero, the value-aware end of the spectrum treated in \S\ref{section:fc_value_aware}, dispenses with observation reconstruction entirely and trains the model only against quantities that enter the planner's value calculation. The MBPO line, treated in \S\ref{section:fc_mbpo}, uses a deep ensemble of probabilistic feed-forward networks. The outer loop in each of these is recognizably Dyna; the model class is the variable that distinguishes them.

The second commitment is that the inner loop applies the same Bellman update on simulated and real data. Step 5 of the algorithm is step 3 with a different source of transitions, not a distinct planning operator. This single fact is what permits Dyna to unify direct and indirect reinforcement learning into one architecture rather than two competing approaches. In the dynamic-programming tradition before Sutton, indirect methods estimated a model and then ran value iteration on it as a separate batch computation, while direct methods updated values from interaction without any model. Dyna eliminates the wall between these two camps by routing both kinds of transitions through the same temporal-difference update. The consequence is that the agent does not need to choose between learning from interaction and learning from a model. It can do both, and the ratio $K$ tunes the mixture. Every subsequent model-based method in this chapter inherits this commitment, whether it uses tabular updates, fitted $Q$-iteration, soft actor-critic, or an MPC controller as its inner loop. In economics terms, Dyna is adaptive learning with simulated experience added between observations. This sample efficiency is not free. Simulated experience helps only where the learned model is accurate enough for the planner's purpose, and most of the failure modes later in the chapter come from that inaccuracy.

\subsubsection{Simulation Study: Planning Amplification in the Blocking Maze}
\label{section:fc_dyna_q_sim}

\begin{figure}[ht]
  \centering
  \includegraphics[width=0.95\textwidth]{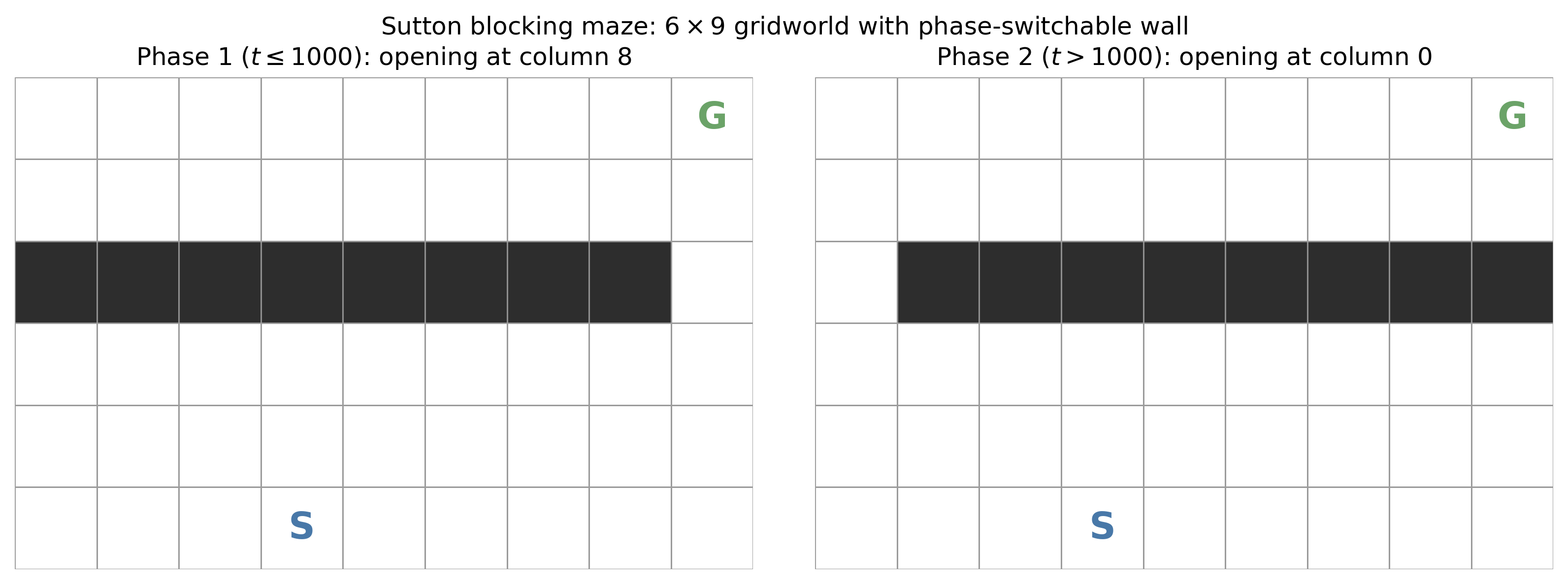}
  \caption{Sutton blocking maze. The agent starts at $S$ and aims for the goal at $G$; the only opening in the wall along row two shifts from column eight in Phase 1 to column zero in Phase 2 at environment step $t = 1000$. The flip is not signaled to the agent and triggers the model-staleness failure mode that the experiment is designed to expose.}
  \label{figure:fc_dyna_maze_layout}
\end{figure}

The blocking maze tests planning amplification under a common environment-step budget. Dyna-Q reuses each real transition in $K$ model-based updates, so larger $K$ should propagate sparse rewards faster than plain $Q$-learning. At $t = 1000$, the wall opening moves from column eight to column zero without notice, making the learned model stale. Figure~\ref{figure:fc_dyna_maze_layout} shows both phases, and the second phase measures recovery from that change.

Plain $Q$-learning is slow on this maze, and the diagnosis is quantitative. With $\gamma = 0.95$, $\alpha = 0.1$, $\varepsilon = 0.1$, $Q$-values initialized to zero, and a reward signal that is identically zero everywhere except at the goal, the agent reaches its first goal essentially by random walk. The expected hitting time of a single random walker from $S$ to $G$ on the open part of this $6 \times 9$ grid is several hundred steps. Each goal-reach updates exactly one entry of $Q$ by $\alpha \cdot 1 = 0.1$ toward the optimal value $\gamma^{d - 1}$, where $d$ is the path length from the parent state. Propagating that signal back to the start state requires a chain of subsequent goal-reaches at the right places, and $\varepsilon$-greedy noise at $\varepsilon = 0.1$ diverts the agent off its currently-best path on one in ten steps. The chance of stringing together a long enough sequence of goal-reaches to back the value all the way to $S$ within a few hundred environment steps is correspondingly small. This is the precise form of the slow-per-step-propagation claim that Dyna-Q is designed to address by replaying each successful real transition $K$ times against the model between every pair of real steps.

The environment is the $6 \times 9$ deterministic gridworld of \citet{sutton2018}, with start at row five column three and goal at row zero column eight. Actions are the four cardinal moves; transitions are deterministic; the reward is one on goal arrival and zero elsewhere; the discount factor is $\gamma = 0.95$. The wall along row two has its only opening at column eight during Phase 1 ($t \le N_1$ with $N_1 = 1000$) and at column zero during Phase 2 ($N_1 < t \le N_1 + N_2$ with $N_2 = 2000$). The phase flip happens automatically at $t = N_1$ and is not signaled. The total interaction budget is $N_1 + N_2 = 3000$ environment steps, the per-episode step cap is two hundred, and the experiment runs over thirty independent seeds with $\alpha = 0.1$, $\varepsilon = 0.1$, and zero-initialized $Q$.

Five agents share the budget. Plain $Q$-learning with $K = 0$ is the direct-RL baseline, applying only the temporal-difference update of \citet{WatkinsDayan1992} per real step. Dyna-Q with $K = 5$ applies five model-based replays per real step from a tabular Dirac model, and Dyna-Q with $K = 50$ applies fifty; both implement the algorithm box of \S\ref{section:fc_dyna_q_algorithm} with the planning-step count set to the named value. Dyna-Q+ with $K = 50$ adds the exploration bonus $\kappa \sqrt{\tau(s, a)}$ to planning targets, with $\tau(s, a)$ the time since the state-action was last visited and $\kappa = 10^{-4}$, instantiating the primitive curiosity mechanism that connects to the prediction-error and ensemble-disagreement variants of \S\ref{section:fc_synthesis_fails}. The fifth agent applies REINFORCE on imagined rollouts under a learned forward model, related to the Schmidhuber 1990 controller-model architecture of \S\ref{section:fc_origins_schmidhuber} but distinct in the gradient pathway. The original 1990 formulation propagates analytic gradients through a differentiable model into the controller, while the variant used here treats the rollout as a stochastic-policy estimator. The agent uses two small multilayer perceptrons. A model network $M_\theta$ maps one-hot state plus one-hot action to next-state logits, fit on observed transitions by cross-entropy. A controller network $C_\phi$ maps one-hot state to action logits, fit by a REINFORCE policy-gradient estimator on imagined trajectories rolled forward through $M_\theta$. Both networks have a single thirty-two-unit hidden layer. Planning calls happen every ten real steps with sixteen imagined trajectories of length ten and a moving-average baseline. This is the third 1990-era treatment of the same loop, sitting alongside Sutton's tabular Dyna and the two Dyna-Q variants of $K \in \{5, 50\}$.

\begin{figure}[ht]
  \centering
  \includegraphics[width=0.95\textwidth]{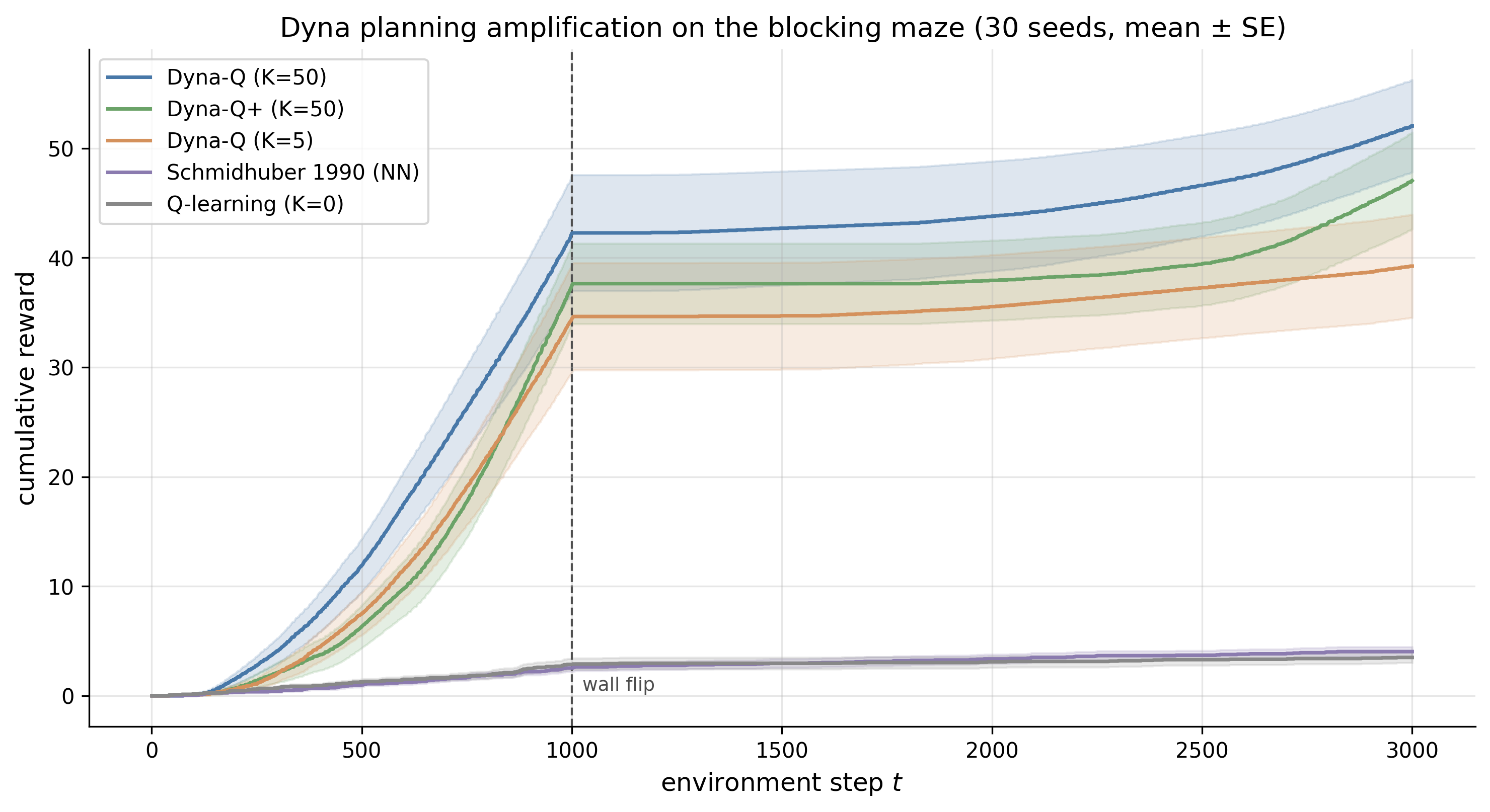}
  \caption{Cumulative reward on the blocking maze for five agents under a shared three-thousand-step environment budget. The dashed vertical line marks the wall flip at $t = 1000$. Lines are means and shaded bands are one standard error across thirty seeds.}
  \label{figure:fc_dyna_maze}
\end{figure}

Figure~\ref{figure:fc_dyna_maze} and Table~\ref{table:fc_dyna_maze} report cumulative environment reward across the thirty seeds with mean and standard error bands, with rows in rank order by total reward at $t = 3000$. Dyna-Q at $K = 50$ leads with $52.0 \pm 4.2$ total goal-reaches, Dyna-Q+ at $K = 50$ follows at $47.0 \pm 4.4$, Dyna-Q at $K = 5$ at $39.2 \pm 4.7$, the Schmidhuber controller-model agent at $4.0 \pm 0.4$, and plain $Q$-learning closes the field at $3.5 \pm 0.5$. The top of the table reproduces the planning-amplification claim of \citet{sutton1990} cleanly under this budget; each real interaction in the $K = 50$ run carries roughly an order of magnitude more policy-improvement weight than the same interaction in $K = 0$, because the synthetic updates of Step 5 propagate the learned reward back through the state graph between every pair of real steps. Dyna-Q+ at $K = 50$ pays a small Phase 1 cost of about five reward units relative to plain Dyna-Q because the exploration bonus eats slightly into exploitation. The cost remains modest because the implementation follows \citet{sutton2018} §8.3 and registers all four actions at every visited state with $\tau = 0$ on first encounter, so the bonus drives discovery of untried actions rather than only perturbing the values of already-tried ones. The Schmidhuber agent tracks plain $Q$-learning on this maze rather than tabular Dyna-Q, because the gradient-based fit of the neural model needs many real samples to specialize from a random initialization. At three thousand environment steps it has not yet accumulated enough buffer entries near the goal for the REINFORCE estimator to drive the controller toward the discovered path. Phase 2 reveals the recovery story. Dyna-Q at $K = 50$ continues to gain reward at the rate of $9.8 \pm 4.0$ over the remaining two thousand steps because $\varepsilon$-greedy noise eventually exposes the new corridor and the planning loop quickly retrains the model. Dyna-Q+ matches that rate at $9.6 \pm 2.8$, with the curiosity bonus driving the agent to revisit untried actions on the opposite side of the wall. The two rates are statistically indistinguishable on a Welch $t$-test with $p > 0.5$, and the Phase 1 deficit shrinks but does not close by $t = 3000$. \citet{sutton2018} Figure 8.5 reports Dyna-Q+ recovering faster than Dyna-Q after the maze flip. The thirty-seed reproduction here does not separate the two methods on the recovery slope, and a longer post-flip budget or a larger bonus coefficient than $\kappa = 10^{-4}$ would likely be needed to recover the Fig.~8.5 effect size cleanly. The Schmidhuber agent's Phase 2 gain of $1.5 \pm 0.2$ is only modestly above plain $Q$-learning's $0.6 \pm 0.2$ and tracks the same slow trajectory.

\begin{table}[ht]
  \centering
  \caption{Cumulative reward on the blocking maze. End of Phase 1 is at $t = 1000$; Phase 2 gain is the increase over the remaining two thousand steps; Total is the value at $t = 3000$. Mean $\pm$ standard error across thirty seeds.}
  \label{table:fc_dyna_maze}
\begin{tabular}{lrrr}
\toprule
Agent & End of Phase 1 & Phase 2 gain & Total ($t = 3000$) \\
\midrule
Dyna-Q (K=50) & 42.2 $\pm$ 5.3 & 9.8 $\pm$ 4.0 & 52.0 $\pm$ 4.2 \\
Dyna-Q+ (K=50) & 37.4 $\pm$ 3.7 & 9.6 $\pm$ 2.8 & 47.0 $\pm$ 4.4 \\
Dyna-Q (K=5) & 34.4 $\pm$ 4.9 & 4.8 $\pm$ 3.0 & 39.2 $\pm$ 4.7 \\
Schmidhuber 1990 (NN) & 2.5 $\pm$ 0.3 & 1.5 $\pm$ 0.2 & 4.0 $\pm$ 0.4 \\
Q-learning (K=0) & 2.9 $\pm$ 0.5 & 0.6 $\pm$ 0.2 & 3.5 $\pm$ 0.5 \\
\bottomrule
\end{tabular}

\end{table}

\subsection{World Models: The Deep Revival}
\label{section:fc_ha_schmidhuber}

\citet{HaSchmidhuber2018} is the first paper to make latent-space dreaming work at scale on pixel inputs. An agent learns a visual encoder, a recurrent forecaster of the encoder's output, and a controller, with the controller trained against rollouts of the forecaster rather than against real environment frames. The abstract decomposition is a triple of objects with distinct losses, a vision encoder trained against reconstruction, a memory model trained against next-latent likelihood, and a controller trained against imagined returns. This decomposition is the pattern that subsequent deep model-based methods inherit and modify. The Dreamer line collapses vision and memory into one jointly trained object, and the value-aware line of \S\ref{section:fc_value_aware} replaces the memory loss outright. On the CarRacing benchmark the resulting controller is the first to cross the conventional solution threshold. On a VizDoom task a controller trained entirely inside the learned forecaster transfers zero-shot to the real environment. The paper is an explicit homage to \citet{Schmidhuber1990}, the controller-model architecture introduced in \S\ref{section:fc_origins_schmidhuber}, and the line of work it revives is twenty-eight years old. What is new is the toolkit. Convolutional variational autoencoders, mixture-density recurrent networks, and evolution strategies, taken together, make the older architecture empirically tractable at a scale the original could not reach.

Eight further recurring terms anchor the rest of the chapter. A \emph{planning update} is a policy or value update performed using the learned model rather than the environment. A \emph{simulated rollout} is a trajectory generated by the learned model rather than by real interaction. A \emph{latent state} is an internal compressed state used by a neural model, not an observed economic state. A \emph{model loss} is the criterion used to train the world model. A \emph{value-aware loss} is a model loss that cares about preserving values rather than predicting every detail of the next observation. A \emph{short rollout} is a deliberately short simulated trajectory from a real visited state. A \emph{bootstrap} is a value estimate that stands in for rewards beyond the simulated horizon. A \emph{model error} is the gap between the learned model and the true environment, judged by its effect on planning rather than by its statistical fit.

\subsubsection{Three components}
\label{section:fc_ha_schmidhuber_components}

The vision component $V$ is a convolutional variational autoencoder.\footnote{A variational autoencoder compresses high-dimensional observations into a low-dimensional latent vector by training an encoder and a decoder jointly.} The encoder maps an observation $x_t \in \mathbb{R}^{64 \times 64 \times 3}$ to a latent vector $z_t \sim \mathcal{N}(\mu(x_t), \sigma(x_t)^2 I)$ with $z_t \in \mathbb{R}^{N_z}$, where $N_z = 32$ for CarRacing and $N_z = 64$ for VizDoom. The decoder is trained jointly with the encoder against the standard ELBO objective,\footnote{Evidence lower bound; the standard variational autoencoder training objective that combines a reconstruction term with a Kullback-Leibler penalty between the encoder's latent and a fixed prior. The Kullback-Leibler term measures the discrepancy between two distributions.} combining a per-pixel reconstruction term with a Kullback-Leibler penalty on the latent. The vision component is trained on observation reconstruction alone, independently of the controller, on a buffer of frames collected by a random policy. The decoder is discarded once training is complete. Downstream components see only the encoder's latents. The architectural commitment is that all subsequent processing happens in a low-dimensional vector space rather than in pixel space. This is what makes the rest of the pipeline cheap enough to train and to roll out.

The memory component $M$ is a mixture-density recurrent neural network.\footnote{A mixture-density recurrent network is a recurrent network whose output parameterizes a Gaussian mixture over the next latent rather than a single mean; this matters when the next state is genuinely multimodal.} Its body is an LSTM with hidden state $h_t$.\footnote{Long short-term memory; a gated recurrent cell that propagates information across time and is the dominant pre-Transformer architecture for sequential data.} Its head outputs the parameters of a Gaussian mixture over the next latent, conditional on the current action, the current latent, and the recurrent state, modeling $P(z_{t+1} \mid a_t, z_t, h_t)$ as a mixture of five diagonal Gaussians. A temperature parameter on the mixture controls how aggressively the forecaster samples from its tails, and is exposed as a tuning lever for controller training. The memory component is trained on next-latent likelihood alone, again independently of the controller, using the same buffer of rollouts under a random policy that trained $V$. The mixture form matters in environments with discrete event structure such as enemy fireballs in VizDoom, where the next latent has multiple plausible values that a unimodal Gaussian forecaster would average over and so render unusable for planning.

The controller $C$ is a single linear layer. It outputs an action $a_t = W_c [z_t, h_t] + b_c$ from the concatenation of the visual latent and the recurrent state. On CarRacing the controller has roughly $870$ parameters, compared to about $5$ million in $V$ and $M$ together. It is trained by covariance matrix adaptation evolution strategy,\footnote{A gradient-free black-box optimizer that maintains a Gaussian over candidate parameter vectors and updates the Gaussian by re-fitting to the best-scoring samples each generation.} evaluating candidate parameter vectors against rollouts drawn from $M$ rather than against real environment frames. Evolution strategies sidestep the need for differentiability through $M$ at the cost of higher sample complexity per parameter, which the linear policy keeps manageable. The controller is the only component whose loss depends on reward, and it is the only component whose training requires interaction, in the form of imagined trajectories through the memory network.

\subsubsection{Empirical headline}
\label{section:fc_ha_schmidhuber_empirics}

On CarRacing-v0 the full agent scores $906 \pm 21$ over $100$ random tracks. The benchmark requires an average above $900$ to count as solved, and this is the first reported solution. On the VizDoom DoomTakeCover task the controller is trained inside the memory network, with the visual encoder and recurrent forecaster fitted on a buffer of random rollouts and the controller never exposed to real frames during policy search. Transferred zero-shot to the real environment, the controller scores $1092 \pm 556$ against a solution threshold of $750$. The variance is large but the mean clears the threshold, and the policy is constructed without any direct contact with the real environment after the data collection phase.

\subsubsection{Architectural reading}
\label{section:fc_ha_schmidhuber_reading}

The modular training schedule, with $V$ fitted on reconstruction, $M$ fitted on next-latent likelihood, and $C$ fitted on dream rollouts, is the practical workaround for joint end-to-end training, which the paper notes is hard at this scale. Each component has its own loss, its own optimizer, and its own data pipeline, and the three are composed in sequence rather than co-trained. The cost of the workaround is that the representation learned by $V$ is agnostic to what the controller will use it for, and the forecaster learned by $M$ is agnostic to what the controller will reward. The line of work that eventually unifies $V$ and $M$ into a single jointly-trained recurrent state-space model is the PlaNet and Dreamer family, treated in \S\ref{section:fc_rssm}.

The combination of $V$ and $M$ does the representation work; the controller is tiny by comparison and carries the parameter budget of a linear policy. This anticipates the modern observation that a strong world model lets a small actor produce competitive policies. Dreamer in \S\ref{section:fc_rssm} and TD-MPC2 in \S\ref{section:fc_tdmpc2} press the point further, with the latent space and the value function carrying most of the inductive load and the policy network relatively light. The training-in-dream-then-transfer protocol on VizDoom is the strongest demonstration to date that a learned forecaster can stand in for the environment during policy search. The agent's only contact with the real environment is the random data collection used to fit $V$ and $M$ in the first place.

\subsection{The Recurrent State-Space Model and the Dreamer Line}
\label{section:fc_rssm}

The previous section described the modular Vision-Memory-Controller architecture of \citet{HaSchmidhuber2018}, trained in three separate stages. The Recurrent State-Space Model collapses that sequence into a single jointly trained network and becomes the workhorse architecture of modern world-model reinforcement learning. The line carries through PlaNet, DreamerV1, DreamerV2, and DreamerV3 and is treated here in version-agnostic notation. The differences across versions are summarized at the end of the subsection. The next section picks up the question of what loss the model should be trained against.

\subsubsection{The Recurrent State-Space Model}
\label{section:fc_rssm_transition}

Let $o_t$ denote the observation at time $t$, $a_t$ the action, and $r_t$ the reward. The Recurrent State-Space Model factors the latent state into a deterministic recurrent component $h_t$ and a stochastic latent $s_t$, coupled by
\[ h_t = f_\theta(h_{t-1}, s_{t-1}, a_{t-1}), \qquad s_t \sim p_\theta(s_t \mid h_t). \]
The deterministic path $f_\theta$ is a recurrent cell (in practice a Gated Recurrent Unit); the prior $p_\theta(s_t \mid h_t)$ is either a diagonal Gaussian or a product of categoricals, the choice a design parameter rather than a structural commitment. A posterior network $q_\theta(s_t \mid h_t, o_t)$ encodes the current observation into a refined latent for inference at training time. The joint loss for a length-$T$ trajectory has three named terms,
\[ \mathcal{L}_\theta = \underbrace{-\sum_t \log p_\theta(o_t \mid h_t, s_t)}_{\text{reconstruction}} \;-\; \underbrace{\sum_t \log p_\theta(r_t \mid h_t, s_t)}_{\text{reward log-likelihood}} \;+\; \underbrace{\sum_t \mathrm{KL}\!\left[ q_\theta(s_t \mid h_t, o_t) \,\Vert\, p_\theta(s_t \mid h_t) \right]}_{\text{prior-posterior consistency}}. \]
Optionally a latent-overshooting term enforces multi-step consistency in latent space without decoding. The architecture separates deterministic memory from stochastic uncertainty, so the model forecasts a distribution over futures while still propagating an information bottleneck through time. The model class is a flexible neural object, but the algorithmic outer loop is recognizably the Dyna template of \S\ref{section:fc_dyna_q}, with $\widehat P$ and $\widehat r$ replaced by jointly trained networks.

\subsubsection{Actor-critic in imagination}
\label{section:fc_rssm_imagination}

The policy is trained inside the learned world model rather than from environment interaction directly. An actor $\pi_\phi(a_t \mid h_t, s_t)$ and a critic $V_\psi(h_t, s_t)$ are updated on imagined trajectories rolled forward through the latent dynamics, starting from real states drawn from the agent's replay buffer. The imagined return objective is
\[ J(\phi) = \mathbb{E}_{\pi_\phi, p_\theta}\!\left[ \sum_{\tau = t}^{t+H} \gamma^{\tau - t} \hat r_\tau + \gamma^{H+1} V_\psi(h_{t+H+1}, s_{t+H+1}) \right], \]
where the expectation is taken over actions sampled from the actor and latent states evolved by the world model, $H$ is an imagination horizon, $\hat r_\tau$ is the reward predicted by the trained reward head, and the bootstrap term $V_\psi$ supplies a value estimate past the imagination horizon. Because the world-model dynamics are differentiable and the latent samples are reparameterized, $\nabla_\phi J$ is obtained by backpropagating value gradients through the unrolled latent trajectory. The critic is fit by minimizing a squared error against $\lambda$-return targets in the same imagined trajectories.

This is the deep-learning instantiation of the inner planning loop of Dyna-Q, with the model now a flexible neural object and the planner an analytic policy-gradient operator rather than a discrete Q-update. Convergence guarantees are not available in closed form for this setting; the nearest analogue is the Lipschitz contraction bound for model-based value error of \citet{Asadi2018Lipschitz}, discussed in \S\ref{section:fc_synthesis_fails}, which states that value error grows geometrically along the rollout at rate $K_F$ (the Lipschitz constant of the learned dynamics) and remains finite when $\gamma K_F < 1$.\footnote{The four canonical instantiations of the architecture cover the major design choices. PlaNet \citep{HafnerPlaNet2019} introduces the RSSM with cross-entropy planning. DreamerV1 \citep{HafnerDreamer2020} replaces planning with actor-critic in imagination. DreamerV2 \citep{HafnerDreamerV2_2021} switches the Gaussian latent for categoricals with a Kullback-Leibler balancing trick, surpassing top model-free agents at the Atari 200M budget. DreamerV3 \citep{HafnerDreamerV3_2025} fixes one hyperparameter configuration across more than one hundred and fifty tasks via symlog reward normalization and free-bits Kullback-Leibler, including the Minecraft diamond from scratch.}

The line trains the world model against likelihood-style losses (reconstruction, reward log-likelihood, prior-posterior consistency). This rewards the model for matching observed distributions irrespective of how those distributions enter the value calculation. In economic terms, the recurrent state-space model is a high-dimensional perceived law of motion, fit by likelihood rather than by hand-chosen moments, with a planner that operates entirely on the learned latent. The next section asks whether this is the right objective. The argument, developed by \citet{FarahmandVAML2017} and \citet{GrimmVALEQ2020} and instantiated empirically in MuZero, is that the model should be trained against the value functional directly rather than against pixel and reward likelihoods. The Dreamer line achieves its results while training the model against likelihood rather than against the downstream decision, and the value-aware line asks what is gained when the loss is aligned with that decision instead.

\subsection{Short Rollouts and Ensembles: The Modern Dyna}
\label{section:fc_mbpo}

The line of work culminating in \citet{janner2019model} is the continuous-control inheritor of Dyna (\S\ref{section:fc_dyna_q}). The architectural recipe is a deep ensemble of probabilistic dynamics models paired with a model-free off-policy actor-critic, soft actor-critic in the canonical implementation, with synthetic experience generated by branching short imagined rollouts from real states drawn out of the replay buffer. The paper's slogan, echoed across the surrounding literature, is to not trust the model too far. Breaking the dependence between the task horizon and the model rollout horizon makes the model-error penalty tractable. It is what separates this line from the trajectory-optimization approach of \citet{Chua2018PETS} and the on-policy approach of \citet{Kurutach2018}. The conceptual lineage runs back to \citet{sutton1990}, with the same outer loop of direct learning from real transitions and indirect learning from imagined ones, only with a deep ensemble in place of the tabular model and an off-policy actor-critic in place of tabular Q-learning.

\subsubsection{Branched short rollouts}
\label{section:fc_mbpo_branching}

As an abstract operator, the branched rollout maps the pair (real-state distribution, learned dynamics) to a synthetic transition distribution of length $k$. It decouples the task horizon from the model horizon by anchoring every imagined trajectory at a state the agent has actually visited. The MBPO branching scheme is straightforward. At each gradient step the algorithm samples a real state $s$ from the replay buffer, rolls the learned dynamics ensemble forward for $k$ steps from $s$ under the current policy, and trains the actor-critic on the resulting model-generated transitions alongside real ones. The actor-critic update does not distinguish synthetic transitions from real transitions. The only change relative to model-free soft actor-critic is the distribution from which training tuples are drawn. Real states act as anchors. Because every imagined trajectory is no more than $k$ steps long and originates at a state the agent has actually visited, the model is never asked to extrapolate from its own predictions for more than $k$ steps, even when the task itself runs for hundreds of steps.

The returns bound in \citet[Theorem 4.2]{janner2019model} formalizes the trade-off. With $\epsilon_m$ the model error measured in total variation, $\epsilon_\pi$ the policy distribution shift relative to the data-collection policy, $k$ the branching length, and $r_{\max}$ the reward bound,
\[ \eta[\pi] \ge \eta^{\text{branch}}[\pi] - 2 r_{\max} \Big[ \frac{\gamma^{k+1} \epsilon_\pi}{(1 - \gamma)^2} + \frac{\gamma^k + 2}{1 - \gamma} \epsilon_\pi + \frac{k}{1 - \gamma}(\epsilon_m + 2 \epsilon_\pi) \Big]. \]
The first two terms penalize policy shift and shrink in $k$; the third penalizes model error and grows linearly in $k$. Theorem 4.3 then shows that an optimal $k^\star > 0$ exists whenever the model error is small enough, justifying short rollouts as the right way to spend a fixed model-error budget. The paper's ablations report that $k = 1$ is competitive across the MuJoCo suite, that $k$ in the range of roughly five to fifteen typically gives the best returns, and that values much beyond about $25$ are too inaccurate to help. The empirical sweet spot is well inside the regime where the model is treated as a local data augmentation rather than as a long-horizon simulator.

Economically, the branched-rollout architecture is the computational version of using a locally valid structural model for counterfactual policy search. The model is trusted only over short windows around states the agent has already visited, and the rest of the long-horizon credit is left to the off-policy critic. The dynamics ensemble used in this line, typically five to seven neural networks each predicting a Gaussian over the next-state delta, plays two roles. As an aggregate predictor the ensemble averages over members and trims tails to give a calibrated forecast that a single deterministic network would not produce. As a disagreement signal the spread across members is a cheap proxy for epistemic uncertainty, large where the agent has not visited often and small where coverage is good.\footnote{The ensemble appears in two prior methods with different commitments. \citet{Kurutach2018} pairs the ensemble with on-policy trust-region policy optimization, estimating value gradients while clipping updates that would push the data distribution into regions where members disagree; the on-policy constraint bounds $\epsilon_\pi$ explicitly but keeps the long-horizon dependence on the model. \citet{Chua2018PETS} wraps the same ensemble in model-predictive control and bypasses policy learning entirely by propagating particle trajectories through ensemble members at planning time, which solves the horizon problem by conflating the task and planning horizons and limits scaling to long-horizon tasks where MPC becomes prohibitive. MBPO's branched-rollout architecture pushes the two terms of the returns bound into a regime where standard off-policy machinery can absorb them, and the same recipe carries from low-dimensional locomotion to humanoid without changes to the outer loop.}\footnote{On the MuJoCo continuous-control benchmark \citet{janner2019model} matches the asymptotic return of soft actor-critic with roughly an order of magnitude fewer environment samples; the Ant task reaches the soft actor-critic return at around $300{,}000$ real steps against the latter's roughly $3{,}000{,}000$, and on Humanoid (a $376$-dimensional state with a $17$-dimensional action) MBPO converges where PETS fails to learn, consistent with the architectural reading that the task horizon dwarfs PETS's planning horizon. \citet{Kurutach2018} reports a similar order-of-magnitude sample-efficiency gain for ME-TRPO over TRPO/PPO/DDPG on the same suite.}

\subsection{What Should the Model Be Trained Against?}
\label{section:fc_value_aware}

Throughout the preceding subsections the model has been trained against a likelihood or reconstruction loss. The Recurrent State-Space Model line uses a pixel reconstruction term, a reward log-likelihood, and a Kullback-Leibler consistency term between prior and posterior. The original Dyna model is a tabular maximum-likelihood estimate of the empirical transition. The controller-model architecture of \citet{HaSchmidhuber2018} fits its variational autoencoder and its mixture-density recurrent network to observation and next-latent likelihoods respectively. In each case the loss rewards the model for matching observed distributions, regardless of how those distributions enter the value calculation the agent actually optimizes. The objection raised in the present subsection is that the agent requires a correct value function rather than a perfect prediction of the next observation. This subsection traces the line that took that observation seriously, from value-aware model learning in \citet{FarahmandVAML2017} through the value-equivalence principle of \citet{GrimmVALEQ2020,GrimmPVE2021}, the value-targeted regret bound of \citet{AyoubVTR2020}, the MuZero instantiation of \citet{Schrittwieser2020MuZero}, and the recent calibration analysis of \citet{VoelckerCalib2025}.

\subsubsection{Value-aware model learning}
\label{section:fc_value_aware_vaml}

\citet{FarahmandVAML2017} formalize the objection as a loss. Fix a value function class $\mathcal{V}$, a data distribution $\mu$ over state-action pairs, and a model class within which the learner selects an estimate $\widehat P$ of the true transition kernel $P^\star$. The value-aware model learning loss is
\[ \mathcal{L}_{\text{VAML}}(\widehat P) = \sup_{V \in \mathcal{V}} \Big\Vert (\widehat P V)(\cdot) - (P^\star V)(\cdot) \Big\Vert_\mu, \]
where $(\widehat P V)(s, a) = \int V(s') \widehat P(\mathrm{d} s' \mid s, a)$ is the expected next-state value under the learned model and $(P^\star V)(s, a)$ is the same quantity under the true kernel. The loss penalizes the worst-case discrepancy in expected next-state value across $\mathcal{V}$, measured under $\mu$. A model that places mass on features of the next observation irrelevant to every value function in $\mathcal{V}$ is not penalized; a model that distorts the value expectation is. \citet{FarahmandIterVAML2018} embed the same idea inside approximate value iteration, replacing the supremum over a value class with the realized value iterate at each step. This yields the practical IterVAML estimator and a fitted error bound that propagates through the Bellman backups rather than through the model alone. \citet{AsadiWasserstein2018} read the construction through optimal transport, observing that under a Lipschitz assumption on the value class the VAML loss reduces to a Wasserstein distance between $\widehat P$ and $P^\star$. This connects the value-aware loss to the substantial empirical literature on Wasserstein generative modeling.

VAML is the formal statement of the intuition that the model should be wrong in ways that do not matter for the value functional. This is the technical answer to the conceptual openness of Sutton's original Dyna architecture, in which the model was permitted to be probabilistic and oftentimes incorrect (\S\ref{section:fc_dyna_q_architecture}) without specifying what kind of error was tolerable. The likelihood-trained Recurrent State-Space Model line (\S\ref{section:fc_rssm}) commits to a particular notion of error, namely the reconstruction loss on pixels and the Kullback-Leibler divergence on latents, and violates the value-aware principle whenever the pixel distribution is rich in detail orthogonal to the reward. Whether this matters in practice depends on the task. The empirical claim of the value-aware line is that it does, and that the gap widens as observation dimensionality grows.

\subsubsection{Value equivalence and MuZero}
\label{section:fc_value_aware_ve}

\citet{GrimmVALEQ2020} sharpen the VAML idea into a clean equivalence relation. Fix a policy set $\Pi$ and a value function set $\mathcal{V}$. Two transition kernels $\widehat P$ and $P^\star$ are value-equivalent with respect to $(\Pi, \mathcal{V})$ when their Bellman operators agree on every pair $(\pi, V) \in \Pi \times \mathcal{V}$, that is when $(T^\pi_{\widehat P} V)(s) = (T^\pi_{P^\star} V)(s)$ for all $s$ and all $(\pi, V)$ in the chosen sets. The equivalence class shrinks monotonically as $\Pi$ and $\mathcal{V}$ grow, collapsing to the singleton $\{P^\star\}$ in the limit. The practical implication is that a small model class can be sufficient for a restricted set of policies and value functions, even if it would be insufficient as a generative model of the environment. \citet{GrimmPVE2021} subsequently introduce proper value equivalence, which restricts the value class to fixed points of the chosen Bellman operators. This is the variant of value equivalence that aligns most directly with the MuZero training objective. \citet{AyoubVTR2020} push the value-aware idea into the regret-bound literature with value-targeted regression, which trains the model against value targets and matches the dimension-dependent rates of optimism-based reinforcement learning under standard assumptions.

The empirical instantiation is MuZero. \citet{Schrittwieser2020MuZero} train a latent recurrence jointly with a policy network and a value network against three losses applied at the next step of the recurrence, namely a reward prediction loss, a value prediction loss, and a policy prediction loss matched to the output of a Monte Carlo tree search rooted at the same latent. There is no reconstruction loss; the latent dynamics are never asked to decode observations. The transition kernel itself has no explicit target, since the only quantities that enter any loss are the reward, the value, and the policy at the rollout step. Empirically the construction matches AlphaZero on Go, chess, and shogi without access to the rules of those games, and reaches state-of-the-art performance on the Atari benchmark. \citet{Antonoglou2022StochMuZero} extend the construction to stochastic transitions by inserting a learned afterstate node, the game-tree analogue of a chance node, between the agent action and the next state. The economic reading is that value-aware model learning treats the world model as a task-relevant misspecification, kept correct on the value functional even when it is wrong on the full observation distribution.\footnote{\citet{VoelckerCalib2025} revisit this family and observe that the standard sampled losses, including the canonical MuZero objective, are uncalibrated surrogate losses whose minima do not in general coincide with the minima of the underlying target loss on the value function; value-equivalence is a property of an in-distribution policy and value class, and a model trained purely against value targets has no incentive to remain accurate once the policy drifts. The paper proposes a calibrated variant that restores the correctness property in expectation and leaves the broader question of value-aware accuracy under policy drift open.}\footnote{The same move appears in the operations-research line on decision-focused learning, in which a predictor is trained with a loss aligned to the downstream decision quality rather than to a statistical scoring rule on the forecast itself \citep{DontiAmosKolter2017}. The inner solver there is a convex program rather than a Bellman backup, but the conceptual move, to align the loss with what the decision-maker actually needs and to treat likelihood as the wrong default whenever the model class is misspecified, is shared.}

\subsection{Convergence Point: TD-MPC2}
\label{section:fc_tdmpc2}

TD-MPC2 \citep{Hansen2024TDMPC2} is the convergence point of three lines that the preceding sections traced separately. Model predictive control with learned dynamics, in the style of PETS and MBPO (\S\ref{section:fc_mbpo}), supplies the planning frame. Recurrent latent world models, in the Dreamer line (\S\ref{section:fc_rssm}), supply the idea that the model should live in a learned latent and need not decode observations. Value-aware learning, articulated by MuZero and the value-equivalence literature (\S\ref{section:fc_value_aware}), supplies the target the model is trained against. TD-MPC2 keeps the planner of the first line, drops the observation decoder of the second, and adopts the value functional as the training signal of the third. The single empirical commitment that organizes the paper is that one fixed hyperparameter configuration should work across many tasks and many model sizes, with no per-task tuning and no rescaling of components as the network grows.

\subsubsection{Components and joint loss}
\label{section:fc_tdmpc2_components}

The agent learns five components jointly. An encoder $h$ maps an observation $s_t$ to a normalized latent $z_t = h(s_t)$. A latent dynamics model $d$ predicts the next latent, $\hat z_{t+1} = d(z_t, a_t)$. A reward model $R$ predicts the immediate reward from the latent and action, $\hat r_t = R(z_t, a_t)$. A terminal action-value function $Q$ supplies a bootstrap beyond the planning horizon, $\hat q_t = Q(z_t, a_t)$. A policy prior $p$ supplies an amortized action proposal $\hat a_t = p(z_t)$ used both to warm-start the planner and to seed the bootstrap targets. The joint loss is the sum of a latent consistency term that pushes the predicted next latent to agree with the encoded next observation, $\| d(z_t, a_t) - \operatorname{sg}(h(s_{t+1})) \|^2$ under a stop-gradient on the target encoding, a cross-entropy loss on the reward prediction, a cross-entropy loss on the temporal-difference targets for $Q$, and a behavioral cloning term that regresses $p$ on the actions selected by the planner. Notably absent is an observation reconstruction loss. The model is never asked to decode pixels or proprioceptive features and is trained only against quantities the value calculation actually consumes. This is the architectural commitment that distinguishes TD-MPC2 from the Dreamer line (\S\ref{section:fc_rssm}), in which the world model carries a reconstruction term throughout.

\subsubsection{MPPI planning with value bootstrap}
\label{section:fc_tdmpc2_mppi}

Action selection at each environment step uses Model Predictive Path Integral control, a sampling-based optimizer over Gaussian action sequences \citep{Hansen2024TDMPC2}. The planner maintains a sequence of Gaussian distributions $(\mathcal{N}(\mu_h, \sigma_h^2))_{h=t}^{t+H-1}$ over the next $H$ actions, draws a population of candidate sequences, rolls each one forward through the learned latent dynamics, scores the imagined returns, and refits $(\mu_h, \sigma_h)$ to a softmax-weighted average of the top-scoring sequences. The scoring rule is the finite-horizon return augmented with a terminal value bootstrap,
\[
  \mu^\star, \sigma^\star \;=\; \arg\max_{(\mu,\sigma)} \;\mathbb{E}_{(a_t, \ldots, a_{t+H}) \sim \mathcal{N}(\mu, \sigma^2)} \Big[\, \gamma^{H} Q(z_{t+H}, a_{t+H}) \;+\; \sum_{h=t}^{H-1} \gamma^{h-t} R(z_h, a_h) \,\Big],
\]
where the trajectory expectation is taken under the learned latent dynamics $d$ rolled from the current encoded state $z_t = h(s_t)$. The terminal $Q$ extends the planner's effective horizon past the finite window $H$, so the agent does not need to roll long open-loop sequences through the learned model to capture long-horizon credit. This is why TD-MPC2 sidesteps the long-rollout compounding error that limits PETS (\S\ref{section:fc_mbpo}), which has no terminal value and must extend $H$ to capture distant reward. After the planner converges, the agent executes only $a_t \sim \mathcal{N}(\mu_t^\star, (\sigma_t^\star)^2)$ and re-plans at the next step. The amortized policy $p$ seeds the next planning call and supplies the action used in the $Q$-bootstrap.

\subsubsection{Empirical headline and scaling}
\label{section:fc_tdmpc2_empirics}

The paper evaluates TD-MPC2 across 104 continuous-control tasks spanning four benchmarks, DeepMind Control, Meta-World, ManiSkill2, and MyoSuite, under a single fixed hyperparameter configuration with no per-task tuning. The task suite includes action dimensionalities up to $39$, sparse rewards, multi-object manipulation, musculoskeletal motor control, and locomotion on dog and humanoid embodiments. Across all four benchmarks TD-MPC2 outperforms SAC, the original TD-MPC, and DreamerV3 at comparable parameter counts. The contribution lies in one configuration solving all of them rather than in any single task being solved better than ever before, which is the property the field had previously assigned to model-free baselines under heavy per-task tuning.

The scaling experiment trains a single multi-task agent on $80$ tasks drawn from DeepMind Control and Meta-World and varies the world model size from $1$ million parameters to $317$ million. The normalized score moves from $16.0$ at $1$M parameters to $70.6$ at $317$M, roughly log-linear in parameter count over the two and a half orders of magnitude examined, with no observed saturation at the upper end. This is the first model-based reinforcement learning paper to report a clean monotone scaling curve over this range. The same hyperparameters are used at every model size; no learning rate, batch size, or capacity-specific schedule is retuned. The empirical claim of TD-MPC2 is that an MPC-shaped model-based agent, built from the components developed in the three earlier lines, scales with parameters in the same way that supervised and self-supervised models do.

\subsection{Dual Simulation: Cobweb and Fishery}
\label{section:fc_dual_sim}

The chapter compares learning paradigms on two single-agent economic decision problems, nine on the first and seven on the second. The first is a cobweb model with adjustment cost, a self-referential environment in which the agent's own action moves the contemporaneous price. The second is a logistic-growth fishery, an exogenous environment in which the agent's action depletes a renewable stock but does not feed back through expectations. Six paradigms are common to both panels, an oracle that knows the true parameters, a precautionary constant-rule baseline that does not learn, recursive least squares adaptive learning \citep{MarcetSargent1989}, the genetic algorithm of \citet{Arifovic1994}, tabular Q-learning \citep{WatkinsDayan1992}, and a model-based learner that estimates the structural coefficients by least squares and plans through a discounted Bellman equation (closed-form linear-quadratic on the cobweb, grid-based dynamic programming on the fishery). The remaining paradigms differ across panels. The cobweb panel adds three model-based policy-gradient learners that share a bootstrap-ensemble linear-Gaussian world model and a linear policy but differ in the gradient estimator. The first is a simplified model-based REINFORCE learner in the spirit of \citet{janner2019model}'s MBPO\footnote{We label this learner MB-LG-REINFORCE (model-based linear-Gaussian REINFORCE) rather than MBPO to flag three departures from \citet{janner2019model}. The dynamics ensemble is a bootstrap of linear-Gaussian regressions rather than dropout-regularized neural networks; the policy is a two-parameter Gaussian rather than a SAC actor with a separate critic and entropy term; and the rollout-disagreement reweighting that MBPO uses to compose model rollouts with real transitions is omitted. The branched-rollout outer loop with REINFORCE updates from the learned model is preserved. The simplification is appropriate for a linear-Gaussian cobweb but should not be read as a head-to-head with Janner's algorithm.}, with branched rollouts and a policy trained by the score-function (REINFORCE) gradient. The second is a variance-reduced variant of the same estimator that pairs antithetic rollouts with a state-dependent baseline. The third is a pathwise variant that differentiates the model-rollout return analytically through a forward sensitivity recursion. The fishery panel instead adds a myopic open-access agent that maximizes the per-period profit and ignores stock dynamics, included to surface the textbook bioeconomic-collapse comparison that the cobweb's no-learning constant rule does not exercise. The experiment is deliberately favorable to structured learners. The two environments have low-dimensional state, smooth payoffs, and stable parametric form. The purpose is to locate the methods on an inductive-bias frontier rather than to rank them universally.

\subsubsection{Cobweb with adjustment cost}
\label{section:fc_dual_sim_cobweb}

The cobweb panel examines where each paradigm sits on the inductive-bias-vs-data trade-off, how parameter recovery rates differ across regimes, and where regret performance and policy quality come apart for methods that bake in correct functional form. The cobweb stability parameter $b/c$ varies across regimes to modulate the curvature of the value function rather than to drive the kind of E-stability divergence \citet{MarcetSargent1989} report for the multi-agent expectational cobweb, which does not arise in this single-agent monopoly variant.

The state is $s_t = (q_{t-1}, p_{t-1}) \in \mathbb{R}^2$ and the action is $q_t \in [0, 4]$. The price clears contemporaneously as $p_t = a - b q_t + \varepsilon_t$ with $\varepsilon_t \sim \mathcal{N}(0, \sigma^2)$, and the reward is $r_t = p_t q_t - (c/2) q_t^2 - (\phi/2)(q_t - q_{t-1})^2$, so the agent faces a linear demand, a quadratic production cost, and a quadratic adjustment cost. The sweep is $b/c \in \{0.5, 1.0, 2.0\}$ with $a = 4$, $c = 1$, $\phi = 0.2$, $\sigma = 0.1$, $\gamma = 0.95$, $T = 500$, and twenty independent seeds. The oracle policy is the affine feedback rule $q_t^\star = K_0 + K_q q_{t-1}$ recovered by quadratic-form fixed-point iteration on the Bellman equation, and cumulative regret is computed per seed against the oracle's realized return on the same noise sequence.

Nine paradigms share the budget. The oracle knows $(a, b, c, \phi)$ and applies the closed-form optimal feedback rule. The constant-rule baseline plays $q_t = 1.4$ at every step regardless of state, providing the absolute no-learning floor. Recursive least squares of \citet{MarcetSargent1989} estimates $(\hat a, \hat b)$ from observed prices, treats $(c, \phi)$ as known, and re-solves the linear-quadratic planner with point estimates each period.\footnote{The asymmetric structural prior across learners is deliberate but mixes two factors in the regret comparison. Recursive least squares is given $(c, \phi)$, while the four model-based learners (model-based LQ, MB-LG-REINFORCE, MB-LG-VR, and MB-LG-Pathwise) must learn all four of $(a, b, c, \phi)$. Its regret advantage over the model-based LQ learner therefore reflects both a tighter functional-form prior and the saved estimation budget for the cost coefficients, and the two cannot be cleanly separated within the present design. A fourth panel that re-runs recursive least squares without the known-cost prior would disentangle the two and is left for follow-up work; the qualitative ordering of paradigms on the inductive-bias frontier is robust to the asymmetry, since the model-based LQ learner pays a transient information cost on $(\hat c, \hat \phi)$ that vanishes within the first hundred environment steps under the present parameters.} The genetic algorithm of \citet{Arifovic1994} maintains a population of binary-encoded production rules, evolves them under fitness-proportional selection on the running mean of realized observed profit, single-point crossover, bit-flip mutation, and two-elite preservation, and plays the population's incumbent rule. The original Arifovic 1994 election operator scored hypothetical offspring against their parents using the true demand and cost parameters. It is omitted here so the agent uses only realized observed profit. Tabular Q-learning of \citet{WatkinsDayan1992} discretizes the state-action space onto a $20 \times 20 \times 25$ grid and applies $\varepsilon$-greedy temporal-difference updates with $\alpha = 0.1$ and $\varepsilon$ decaying linearly from $0.3$ to $0.01$. Four model-based learners share the same parameter-estimation step but differ in how they plan. The model-based LQ learner exploits the linear-Gaussian structure: it solves the closed-form linear-quadratic Bellman equation by Riccati iteration on current point estimates of $(\hat a, \hat b, \hat c, \hat \phi)$ each step and acts under the resulting affine feedback rule with Gaussian exploration noise. The MB-LG-REINFORCE learner, a simplified MBPO variant in the spirit of \citet{janner2019model}, maintains a bootstrap ensemble of five linear-Gaussian demand models and parameterizes the policy as $q = K_0 + K_q q_{t-1}$. At each real step it samples ten rollouts of horizon five from buffer-uniform initial states under random ensemble members, accumulates discounted returns from the learned reward model, and updates $(K_0, K_q)$ by REINFORCE with a moving-average baseline. The MB-LG-VR learner keeps the REINFORCE learner's estimator, rollout budget, and learning rate but simulates its rollouts in antithetic pairs, replaying each pair's Gaussian draws with the sign flipped, and replaces the scalar moving-average baseline with a quadratic state-dependent baseline fitted by least squares on past rollout returns only. At a fixed policy the antithetic pairing alone cuts the variance of the gradient estimate nineteen-fold at identical cost.\footnote{Measured over two thousand paired draws at a fixed policy and fitted model. The two estimators' means agree within sampling error, consistent with both being unbiased.} The MB-LG-Pathwise learner shares the same ensemble model fit and linear policy class but replaces the score-function estimator with an analytic pathwise gradient, differentiating the expected discounted model-rollout return in $(K_0, K_q)$ through a forward sensitivity recursion on the learned model; the lower-variance estimator supports a larger policy step.\footnote{Twenty rollouts of horizon five per real step against the REINFORCE and variance-reduced learners' ten, learning rate $0.05$ against $0.005$, gradients clipped to unit $\ell_2$ norm.}

\begin{figure}[ht]
  \centering
  \includegraphics[width=\textwidth]{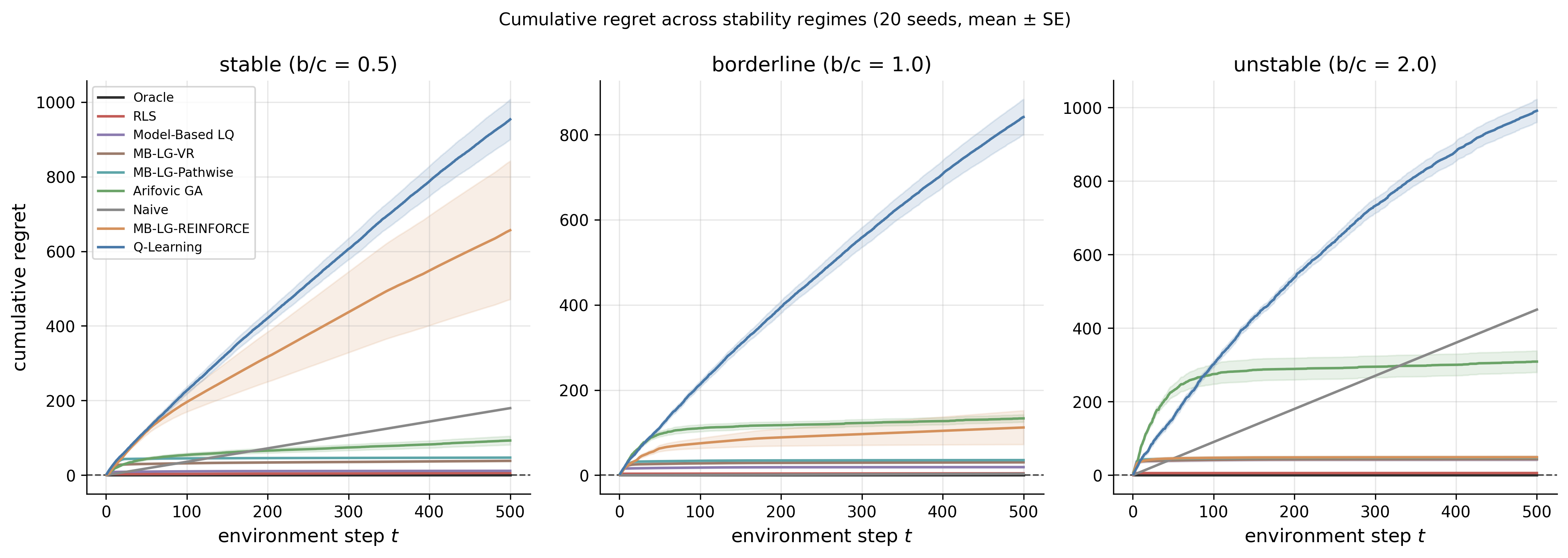}
  \caption{Cumulative regret across stability regimes for nine learning paradigms on the cobweb with adjustment cost. Lines are means and shaded bands are one standard error across twenty seeds. Panels correspond to $b/c \in \{0.5, 1.0, 2.0\}$. Oracle is the zero reference.}
  \label{figure:fc_cobweb_curves}
\end{figure}

Table~\ref{table:fc_cobweb_results} reports cumulative regret at $T = 500$ for each paradigm and regime; Figure~\ref{figure:fc_cobweb_curves} shows the corresponding regret trajectories. Paradigms are presented in rank order by mean regret across regimes. Recursive least squares attains regret of about five units in every regime, because it has the correct functional form for the price map and known cost parameters and need only estimate the two demand coefficients. The model-based LQ learner trails several-fold with regret rising from about twelve to forty-three units as the regime moves from stable to unstable, reflecting the larger curvature it must resolve in the unstable region and the additional cost coefficients it must learn alongside the demand map. The MB-LG-VR learner follows with regret of thirty to forty-four units, edging the pathwise learner in every regime at the same rollout budget and learning rate as plain REINFORCE. The MB-LG-Pathwise learner sits just behind with regret of thirty-five to forty-seven units, nearly flat across regimes, because the analytic gradient it takes through the learned model carries usable signal regardless of how flat the return surface is. The genetic algorithm of \citet{Arifovic1994} without the election operator trails with regret of ninety to three hundred units, reflecting the cost of gradient-free population search over a binary chromosome with no parametric knowledge of demand or cost. The constant rule shifts with the regime in a way that reflects the position of the fixed action relative to the regime's optimum and is included as the absolute no-learning floor, not as a learning competitor. The MB-LG-REINFORCE learner sits next, with regret that varies by more than an order of magnitude across regimes (six hundred and fifty in stable, one hundred and twelve in borderline, forty-nine in unstable); the variance reflects REINFORCE's sensitivity to the curvature of the return surface, which is flat in the stable regime where small differences in $(K_0, K_q)$ matter little for return and sharp in the unstable regime where the policy gradient carries usable signal. Tabular Q-learning finishes at eight hundred to a thousand units of regret across regimes, the empirical signature of model-free tabular methods in continuous problems under a tight environment budget. On this cobweb, under the present hyperparameters, branched-rollout REINFORCE underperforms closed-form planning across all three regimes. The gap is widest in the stable regime, where the return surface is flattest for the REINFORCE gradient estimator. The two variants isolate the gradient estimator. With the same ensemble model and the same linear policy class, replacing the score-function gradient with the analytic pathwise gradient cuts stable-regime regret from six hundred and fifty to forty-seven units. Keeping the score-function estimator while reducing its variance with antithetic pairs and a fitted baseline cuts it to thirty-eight at the same budget and learning rate, precisely where the flat return surface starves plain REINFORCE of signal. The four methods share the same parameter-estimation step; the difference is what the planner does with the point estimates.

\begin{figure}[ht]
  \centering
  \includegraphics[width=\textwidth]{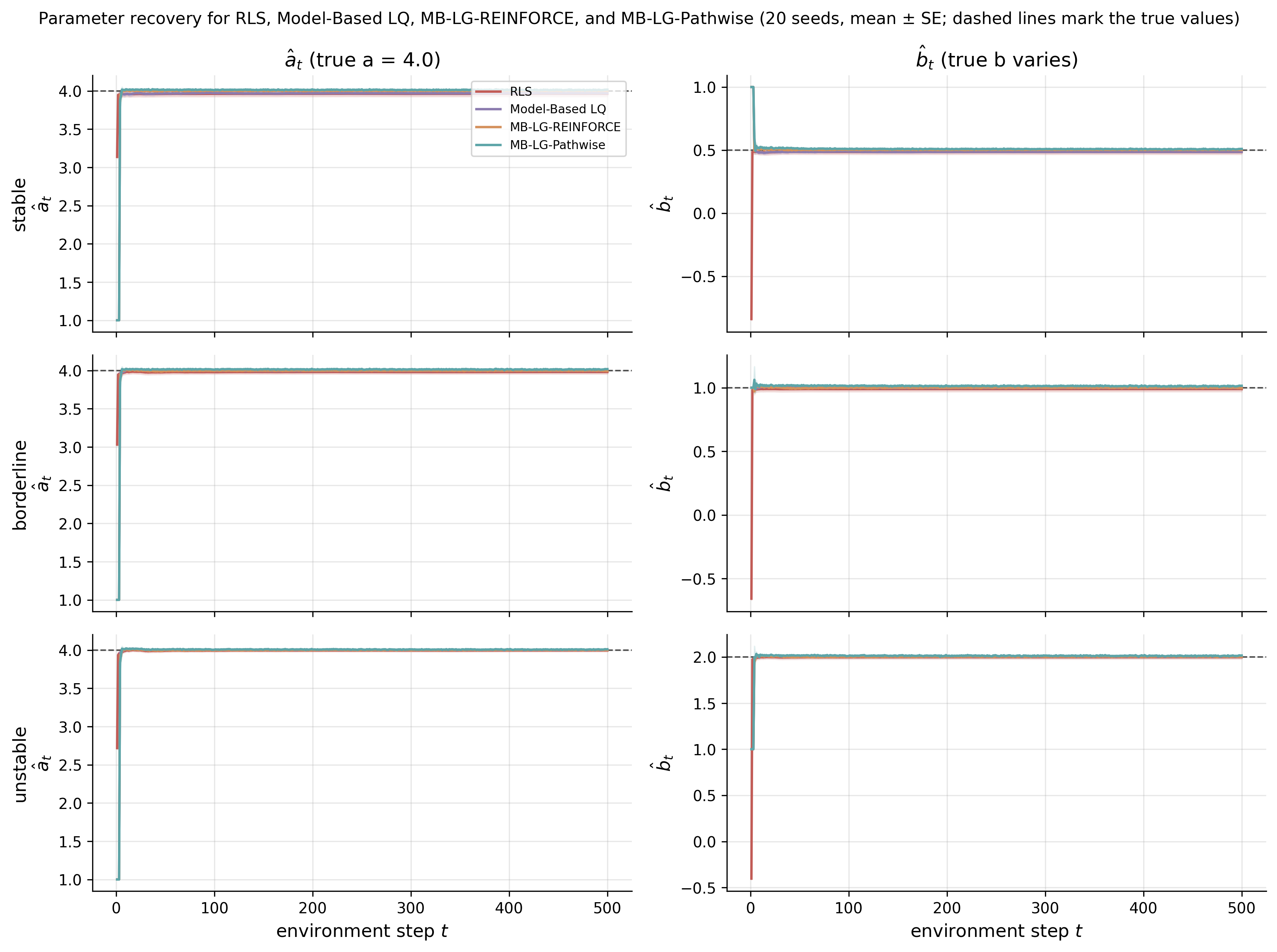}
  \caption{Parameter recovery for recursive least squares, the model-based LQ learner, and the MB-LG-REINFORCE and MB-LG-Pathwise ensemble means across stability regimes. Solid lines are means and shaded bands one standard error across twenty seeds; dashed horizontal lines mark the true values of $a$ and $b$. All four methods converge to the truth within a few dozen environment steps.}
  \label{figure:fc_cobweb_recovery}
\end{figure}

Regret captures the integrated cost of learning, not the asymptotic quality of the recovered model or policy. Figure~\ref{figure:fc_cobweb_recovery} plots the parameter trajectories $\hat a_t$ and $\hat b_t$ for recursive least squares, the model-based LQ learner, and the MB-LG-REINFORCE and MB-LG-Pathwise ensemble means against the true values. Table~\ref{table:fc_cobweb_recovery} reports the final $|\hat\theta - \theta|$ error per paradigm and regime.\footnote{The MB-LG-VR learner runs the identical estimation step and recovers the coefficients comparably (demand coefficients within two percent, cost coefficients to machine precision); it is omitted from the recovery figure and table for legibility.} All four methods recover the demand coefficients to within four percent of the truth in every regime. The model-based learners additionally recover the cost coefficients $(c, \phi)$ to within machine precision\footnote{The exact recovery reflects noiseless reward in this simulation: given observed $(p_t, q_t, q_{t-1})$, the equation $r_t = p_t q_t - (c/2) q_t^2 - (\phi/2)(q_t - q_{t-1})^2$ is exactly solvable for $(c, \phi)$ from any two tuples that span the cost feature space. With additive reward noise $\sigma_r > 0$, recovery would degrade at rate $O(\sigma_r / \sqrt{N})$.} because the reward signal pins them down once the residual $r_t - p_t q_t$ is regressed on $(q_t^2, (q_t - q_{t-1})^2)$. Recovery is fastest in the unstable regime, because the curvature of the demand map sharpens the information content of each observation. The same effect explains why the regret values are similar across regimes even though the absolute return at the oracle's policy is lower in the unstable regime.

\begin{figure}[ht]
  \centering
  \includegraphics[width=\textwidth]{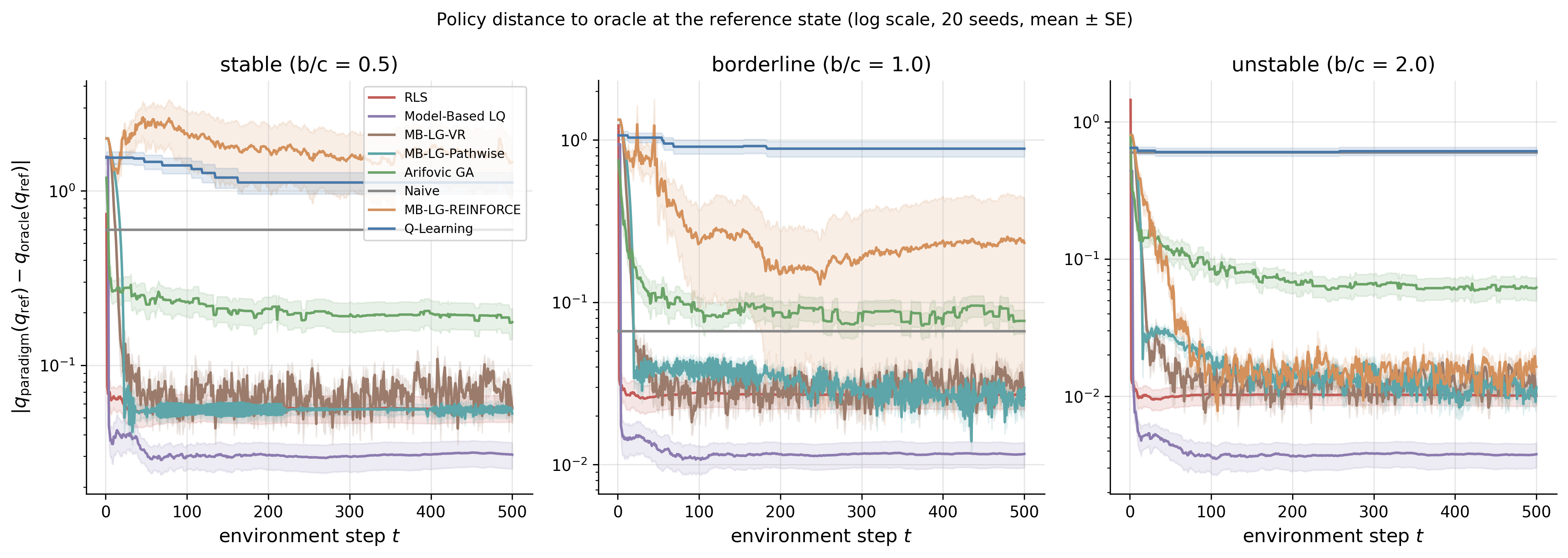}
  \caption{Policy distance to the oracle at a regime-specific reference state, log scale. The reference state is the static optimum quantity for each regime. Lines are means and bands one standard error across twenty seeds. Oracle is identically zero by construction and is omitted.}
  \label{figure:fc_cobweb_policy_distance}
\end{figure}

Figure~\ref{figure:fc_cobweb_policy_distance} reports the distance from each learner's noise-free action at a regime-specific reference state to the oracle's action at the same state, on a log scale. The ordering is by inductive bias rather than recovery accuracy. The model-based LQ learner attains the smallest residual policy distance in every regime, roughly two times smaller than recursive least squares in absolute terms, because by estimating $(c, \phi)$ it places its planner at the correct curvature of the value function. Recursive least squares is bounded below by the discrepancy between the prior on $(c, \phi)$ that fixes its planner and the realized environment, which is small here only because the prior was chosen to match. The MB-LG-VR and MB-LG-Pathwise learners track recursive least squares closely in every regime (distances of 0.056 and 0.053 against 0.057 in the stable regime), despite learning all four parameters and planning by gradient ascent rather than a closed-form solve. The genetic algorithm tracks a policy distance several times larger still, reflecting the discretization of the chromosome over the action space rather than incomplete recovery of any continuous parameter. The MB-LG-REINFORCE learner is the furthest from the oracle in the stable regime, above even tabular Q-learning, because the flat return surface has not let the score-function update converge its linear policy. It approaches the model-based LQ learner's accuracy in the unstable regime, where the gradient signal is strong. This regime dependence is the practical cost of the score-function update relative to the analytical planner. Tabular Q-learning sits at order one, a direct consequence of the coarseness of the action grid (twenty-five points over a range of four units) compounded by the constant exploration noise injected into the rollout. The constant rule is flat by construction at a value that depends on how far the fixed action $q_{\text{fixed}} = 1.4$ lies from the regime's optimum, and the value distinguishes how much of the no-learning floor is luck of the regime rather than any property of learning.

Recursive least squares wins the regret comparison because its early decisions are nearly optimal under correct functional form and known cost parameters. The model-based LQ learner wins the asymptotic policy-distance comparison because it ultimately fits more of the problem's structure. The two methods occupy adjacent positions on the inductive-bias frontier. Recursive least squares pays a higher per-step penalty as long as its cost-parameter prior is correct, and the model-based LQ learner pays a transient information cost in exchange for a tighter asymptote. The MB-LG-VR and MB-LG-Pathwise learners share the third position on this frontier, the first by cutting the variance of the score-function estimator with antithetic pairs and a fitted baseline, the second by replacing it with an analytic gradient; both share the LQ learner's model-estimation step but plan by gradient ascent rather than a closed-form solve, a slower update that pays off only when the closed-form structure is unavailable. The MB-LG-REINFORCE learner carries the same architecture with the raw high-variance score-function gradient, and pays for it exactly where the return surface is flat. On this cobweb the closed-form structure is available and the comparison favors the LQ learner; in environments where it is not, the comparison would invert.

\begin{table}[ht]
  \centering
  \caption{Cumulative regret at $T = 500$ on the cobweb with adjustment cost, mean $\pm$ standard error across twenty seeds. Lower is better; Oracle is the reference.}
  \label{table:fc_cobweb_results}
\begin{tabular}{lrrr}
\toprule
Paradigm & Stable ($b/c=0.5$) & Borderline ($b/c=1.0$) & Unstable ($b/c=2.0$) \\
\midrule
Oracle & 0.00 $\pm$ 0.00 & 0.00 $\pm$ 0.00 & 0.00 $\pm$ 0.00 \\
RLS & 5.89 $\pm$ 1.04 & 4.38 $\pm$ 0.43 & 5.87 $\pm$ 0.18 \\
Model-Based LQ & 11.65 $\pm$ 0.93 & 18.96 $\pm$ 1.73 & 42.90 $\pm$ 3.75 \\
MB-LG-VR & 38.42 $\pm$ 1.86 & 30.26 $\pm$ 1.86 & 43.59 $\pm$ 3.33 \\
MB-LG-Pathwise & 47.07 $\pm$ 0.97 & 35.43 $\pm$ 1.68 & 47.46 $\pm$ 4.04 \\
Arifovic GA & 92.89 $\pm$ 11.69 & 133.43 $\pm$ 8.65 & 308.88 $\pm$ 29.16 \\
Naive & 179.73 $\pm$ 0.34 & 3.37 $\pm$ 0.04 & 450.34 $\pm$ 0.34 \\
MB-LG-REINFORCE & 656.60 $\pm$ 185.41 & 112.06 $\pm$ 39.78 & 48.87 $\pm$ 3.47 \\
Q-Learning & 953.69 $\pm$ 53.79 & 841.50 $\pm$ 41.63 & 991.15 $\pm$ 31.19 \\
\bottomrule
\end{tabular}

\end{table}

\begin{table}[ht]
  \centering
  \small
  \caption{Final parameter recovery error $|\hat\theta - \theta|$ at $t = T$, mean $\pm$ standard error across twenty seeds. Cells marked --- indicate parameters the paradigm does not estimate (recursive least squares assumes $c$ and $\phi$ known); the three model-based learners shown estimate all four.}
  \label{table:fc_cobweb_recovery}
\begin{tabular}{@{}llcccc@{}}
\toprule
Paradigm & Regime & $|\hat a - a|$ & $|\hat b - b|$ & $|\hat c - c|$ & $|\hat\phi - \phi|$ \\
\midrule
RLS & stable & 0.037 $\pm$ 0.035 & 0.015 $\pm$ 0.018 & --- & --- \\
RLS & borderline & 0.019 $\pm$ 0.026 & 0.011 $\pm$ 0.020 & --- & --- \\
RLS & unstable & 0.004 $\pm$ 0.016 & 0.003 $\pm$ 0.019 & --- & --- \\
\midrule
Model-Based LQ & stable & 0.029 $\pm$ 0.019 & 0.013 $\pm$ 0.009 & 0.000 $\pm$ 0.000 & 0.000 $\pm$ 0.000 \\
Model-Based LQ & borderline & 0.004 $\pm$ 0.011 & 0.002 $\pm$ 0.008 & 0.000 $\pm$ 0.000 & 0.000 $\pm$ 0.000 \\
Model-Based LQ & unstable & 0.001 $\pm$ 0.007 & 0.003 $\pm$ 0.008 & 0.000 $\pm$ 0.000 & 0.000 $\pm$ 0.000 \\
\midrule
MB-LG-REINFORCE & stable & 0.000 $\pm$ 0.005 & 0.000 $\pm$ 0.002 & 0.000 $\pm$ 0.000 & 0.000 $\pm$ 0.000 \\
MB-LG-REINFORCE & borderline & 0.005 $\pm$ 0.006 & 0.003 $\pm$ 0.004 & 0.000 $\pm$ 0.000 & 0.000 $\pm$ 0.000 \\
MB-LG-REINFORCE & unstable & 0.000 $\pm$ 0.005 & 0.002 $\pm$ 0.006 & 0.000 $\pm$ 0.000 & 0.000 $\pm$ 0.000 \\
\midrule
MB-LG-Pathwise & stable & 0.018 $\pm$ 0.008 & 0.010 $\pm$ 0.004 & 0.000 $\pm$ 0.000 & 0.000 $\pm$ 0.000 \\
MB-LG-Pathwise & borderline & 0.019 $\pm$ 0.009 & 0.016 $\pm$ 0.007 & 0.000 $\pm$ 0.000 & 0.000 $\pm$ 0.000 \\
MB-LG-Pathwise & unstable & 0.012 $\pm$ 0.007 & 0.016 $\pm$ 0.008 & 0.000 $\pm$ 0.000 & 0.000 $\pm$ 0.000 \\
\midrule
\bottomrule
\end{tabular}

\end{table}

\subsubsection{Fishery with logistic growth}
\label{section:fc_dual_sim_fishery}

The fishery panel runs seven paradigms, a near-subset of the cobweb's nine, on a logistic-growth fishery, an exogenous environment that isolates the sample-efficiency story from the self-referential pricing of the cobweb. The reward structure is again linear-quadratic, but the transition dynamics are non-linear, so the oracle is a grid-based dynamic-programming solve rather than a closed-form linear-quadratic Riccati.

The state is the stock biomass $s_t \in [0, 1.5 K]$ and the action is the harvest $h_t \in [0, \min(s_t, h_{\max})]$. The dynamics are $s_{t+1} = \max(0, s_t + r s_t (1 - s_t / K) - h_t + \varepsilon_t)$ with $\varepsilon_t \sim \mathcal{N}(0, \sigma^2)$, and the reward is $r_t = p h_t - (c/2) h_t^2$. Parameters are $r = 0.4$, $K = 10$, $p = 2.0$, $c = 0.2$, $\sigma = 0.3$, $\gamma = 0.95$, $T = 500$, with twenty independent seeds and $s_0 = K$ (initial stock at carrying capacity). The deterministic maximum-sustainable-yield harvest is $h_{\text{MSY}} = r K / 4 = 1$ at $s^\star = K / 2 = 5$.

Seven paradigms share the budget, a slight reshuffling of the cobweb panel's nine (the three model-based policy-gradient learners, MB-LG-REINFORCE, MB-LG-VR, and MB-LG-Pathwise, are omitted here because the fishery's non-linear dynamics make a closed-form linear-policy parameterization unavailable; in their place we add a myopic open-access agent to surface the textbook bioeconomic collapse). The oracle solves a grid-based value-iteration discounted dynamic program on $50$ stock $\times$ $25$ action points and applies the greedy policy. The myopic open-access agent maximizes the per-period profit $p h_t - (c/2) h_t^2$ at each step, ignoring stock dynamics. This delivers the unconstrained interior solution $h_t^{\textsc{my}} = p / c = 10$. The environment clips every agent's realized harvest to $\min(s_t, h_{\max})$ with $h_{\max} = 1.5 \cdot rK/4 = 1.5$, so the myopic agent in effect harvests fifty percent above the maximum sustainable yield each period. That drives the stock to zero within roughly fifteen steps under the present parameters. The precautionary constant-rule baseline plays $h = 0.5$ at every step, a steady-state guess that lies between zero and the analytic MSY and that serves as a no-learning floor that does not collapse the fishery. Recursive least squares estimates $(\hat r, \hat K)$ from a linear-in-parameters regression of $\Delta s_t + h_t$ on $(s_t, -s_t^2)$, plans by re-solving the DP with point estimates every twenty-five observations, and applies the resulting greedy policy with known cost parameters $(p, c)$. The model-based DP learner estimates $(\hat r, \hat K, \hat p, \hat c)$ jointly by least squares, refits the DP every twenty-five observations, and acts with Gaussian exploration noise that decays over the episode; the planner is grid-based dynamic programming rather than the linear-quadratic Riccati of the cobweb sibling, because the logistic transition is non-linear. Tabular Q-learning discretizes $(s, h)$ onto a $30 \times 21$ grid with $\alpha = 0.1$ and $\varepsilon$ decaying from $0.3$ to $0.01$. The genetic algorithm of \citet{Arifovic1994} maintains thirty binary-encoded constant harvest rules and evolves them under fitness-proportional selection and a partial election operator with known cost parameters.\footnote{The election operator scores child against parent on a static myopic profit $p h - (c/2) h^2$ evaluated at the most recently observed stock, rather than on a full discounted bioeconomic objective. This is faithful to the spirit of the Arifovic 1994 operator (which used the true demand and cost parameters to score offspring) but simplifies the comparison from a discounted-return rollout to a single-period reward; the chromosome encodes a constant harvest rule, so the simplification is mild.} The action support for the gradient-free and tabular learners (Q-Learning, the genetic algorithm, and the model-based DP learner's exploration clip) is bounded above by $h_{\max} = 1.5 \cdot r K / 4$ using the \emph{true} $r$ and $K$.\footnote{This bound constrains the action grid (and the chromosome decode range, and the exploration clip), not the learned policy. A model-free learner with no prior on $r$ and $K$ would need a separate mechanism to set its action support; we treat this as a fixed problem-specific scaffolding rather than a learned quantity.}

\begin{figure}[ht]
  \centering
  \includegraphics[width=0.95\textwidth]{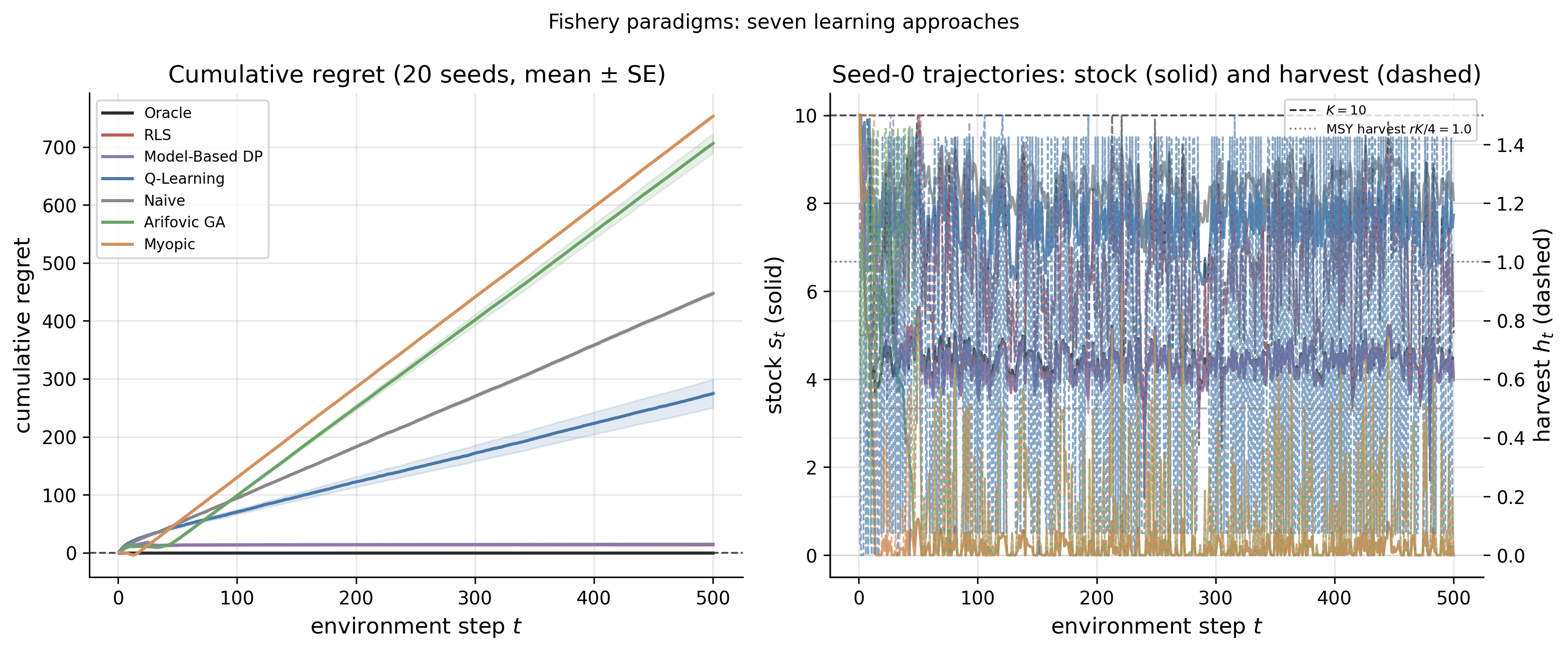}
  \caption{Left: cumulative regret on the logistic-growth fishery for seven paradigms (twenty seeds, mean and one-standard-error bands), presented in rank order by final regret. Right: seed-0 stock trajectories (solid) and realized harvest trajectories (dashed); the carrying capacity $K=10$ and the maximum sustainable yield harvest $rK/4=1.0$ are shown as reference lines.}
  \label{figure:fc_fishery_curves}
\end{figure}

Table~\ref{table:fc_fishery_results} reports cumulative regret at $T = 500$ for each paradigm in rank order; Figure~\ref{figure:fc_fishery_curves} shows the regret trajectories. Recursive least squares and the model-based DP learner sit near the oracle (about thirteen and fifteen regret units respectively), tabular Q-learning recovers a usable policy at roughly two hundred and seventy-five units of regret, the precautionary constant rule lands at four hundred and forty-seven, the genetic algorithm finishes near seven hundred, and the myopic open-access agent at the bottom of the table accumulates seven hundred and fifty-three units of regret. The myopic agent pins its realized harvest at the cap of one and a half units per period, out-earning the oracle while the stock lasts, then collapses the fishery within roughly fifteen steps. Thereafter it merely scavenges the noise-driven regrowth of the stock, earning about a sixth of the oracle's per-period profit for the remaining periods. The textbook fishery-collapse tragedy is recovered. The fishery's exogenous dynamics give the structured learners a cleaner identification problem than the cobweb, because each $(s_t, s_{t+1})$ pair carries direct information about $(r, K)$ that the agent's own action does not confound through expectations. Parameter recovery therefore converges within the first hundred environment steps, and the residual regret reflects exploration cost rather than mis-identification. Table~\ref{table:fc_fishery_recovery} quantifies this. Both structured learners recover $r$ to within two percent and $K$ to within three percent on average across the twenty seeds, with the model-based DP learner slightly tighter because it pools across observations in batch least squares rather than the recursive update of recursive least squares. The result mirrors the cobweb's inductive-bias frontier with one twist. Tabular Q-learning beats the precautionary constant rule, the genetic algorithm, and the myopic agent on this environment, because the discretized state-action grid is coarse enough to be tractable yet fine enough to track the optimal feedback rule's monotone structure. The genetic algorithm's binary-encoded constant harvest is the wrong functional form for an environment where the optimal action varies smoothly with the stock. Its population-search noise causes a non-trivial fraction of seeds to drift into harvest rates that themselves induce a collapse.

\begin{table}[ht]
  \centering
  \caption{Final cumulative regret at $T = 500$ on the fishery, mean $\pm$ standard error across twenty seeds. Lower is better; oracle is the zero reference. Paradigms are presented in rank order.}
  \label{table:fc_fishery_results}
\begin{tabular}{lr}
\toprule
Paradigm & Final regret \\
\midrule
Oracle & 0.00 $\pm$ 0.00 \\
RLS & 13.67 $\pm$ 0.43 \\
Model-Based DP & 14.69 $\pm$ 0.73 \\
Q-Learning & 274.71 $\pm$ 24.36 \\
Naive & 447.35 $\pm$ 3.24 \\
Arifovic GA & 706.13 $\pm$ 16.65 \\
Myopic & 753.11 $\pm$ 1.75 \\
\bottomrule
\end{tabular}

\end{table}

\begin{table}[ht]
  \centering
  \caption{Parameter recovery on the fishery at $t = T$ for the two structured learners, mean $\pm$ standard error across twenty seeds. True values are $r = 0.4$ and $K = 10$. Recursive least squares assumes $(p, c, \sigma)$ known and only estimates $(r, K)$; the model-based DP learner additionally estimates $(p, c)$ (not shown).}
  \label{table:fc_fishery_recovery}
\begin{tabular}{lcccc}
\toprule
Paradigm & $\hat r$ & $\hat K$ & $|\hat r - r|$ & $|\hat K - K|$ \\
\midrule
RLS & 0.398 $\pm$ 0.002 & 10.034 $\pm$ 0.064 & 0.007 & 0.219 \\
Model-Based DP & 0.400 $\pm$ 0.002 & 9.965 $\pm$ 0.050 & 0.006 & 0.180 \\
\bottomrule
\end{tabular}

\end{table}

\subsection{A Learned World Model on a Serial Supply Chain}
\label{section:fc_supply_chain}

The cobweb and fishery panels of Section~\ref{section:fc_dual_sim} were deliberately low-dimensional, and there the structured learners with the correct parametric form dominated. This section moves to a problem whose physical state is large enough that the sufficient statistic for optimal control is not the state the agent observes, and where the world model is a genuine neural network rather than a linear-Gaussian ensemble. The environment is a two-echelon serial inventory system, a retailer replenished by an upstream supplier across a shipment lead time, which is the canonical operations-research setting in which \citet{ClarkScarf1960inventory} proved that an echelon base-stock policy is optimal. Deep reinforcement learning has been applied to inventory problems of this kind at industrial scale \citep{Madeka2022DeepInventory}, and the comparison here follows the finding of \citet{Gijsbrechts2022inventory} that a learned policy matches specialized heuristics rather than dominating them, while separating that question from the sample efficiency of planning against a learned model.

The retailer at stage one faces Poisson demand with mean five and backorders any unmet demand. Each stage places an order that arrives after a one-period lead time, the upstream stage ships what it can and backlogs the rest, and the period cost is installation holding at each stage plus a backorder penalty at the retailer. The agent observes the full physical pipeline, on-hand inventory and in-transit shipments at both stages together with the internal backlog and the customer backorder, a six-dimensional vector. The optimal policy depends only on the two echelon inventory positions, a sufficient statistic the model-free learner is not given. Five paradigms share a budget of five hundred environment steps over twenty seeds. No learner is told the true demand rate or the cost coefficients, and each estimates the mean demand from the demand it observes. The oracle applies the echelon base-stock policy of \citet{ClarkScarf1960inventory} with levels found by simulation optimization on the true model, which serves as the zero-regret reference and whose single-stage specialization is verified against the closed-form newsvendor base-stock. The neural world model learns a two-headed network mapping the observed state, the order vector, and the realized demand to the next state and the period cost, and at intervals it re-optimizes its echelon base-stock levels by rolling that learned network forward under demand resampled from what it has seen, the same simulation-optimization search the oracle runs on the true model but run instead on the learned one, in the spirit of the model-based rollout of \citet{janner2019model}. It therefore carries the same base-stock structural prior as the oracle and starts from a base-stock set to one lead-time of estimated mean demand, the identical starting heuristic the decentralized baseline uses, and the model-based search supplies the safety stock and echelon coordination on top. The model-free deep Q-network carries no such structure, searching value over the raw physical state with a discretized joint order grid. The decentralized heuristic runs a local base-stock at each stage from its own observed inflow, ignoring the echelon coupling, and the naive rule orders its running mean of observed demand every period.

\begin{figure}[ht]
  \centering
  \includegraphics[width=\textwidth]{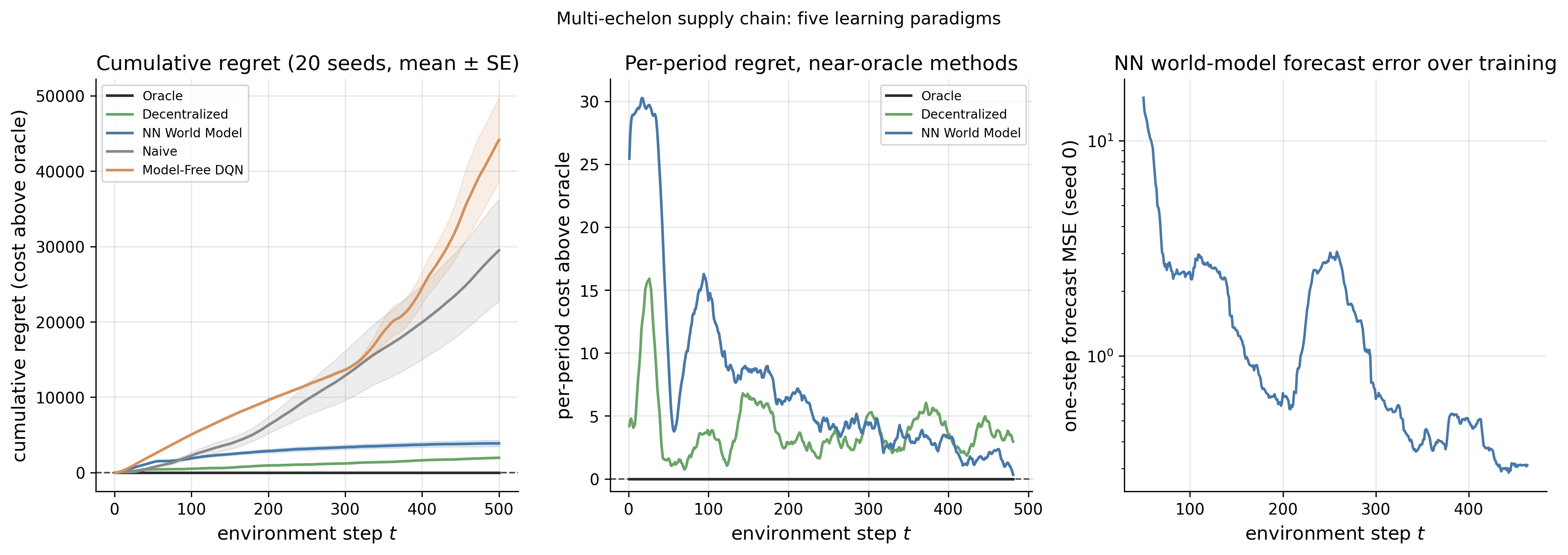}
  \caption{Left: cumulative regret against the Clark-Scarf oracle for five paradigms, means and one-standard-error bands over twenty seeds; the world-model and decentralized traces are near zero at the scale set by the model-free and naive learners. Middle: per-period cost above the oracle over time for the three near-oracle methods, smoothed; the world model pays a large exploration cost early and then converges below the decentralized heuristic, the asymptotic advantage the terminal-cost column of Table~\ref{table:fc_supply_chain} records. Right: the neural world model's one-step forecast mean-squared error over training, seed zero, on a log scale.}
  \label{figure:fc_supply_chain}
\end{figure}

Table~\ref{table:fc_supply_chain} reports cumulative regret against the oracle and the terminal per-period cost over the last eighty steps, and Figure~\ref{figure:fc_supply_chain} shows the regret trajectories and the world model's one-step forecast error over training. Two metrics separate the integrated cost of learning from the quality of the policy each method arrives at. On terminal per-period cost the neural world model is the best of the learners at 8.00 against the oracle's 6.61, within about a fifth of the provably optimal policy, ahead of the decentralized heuristic at 9.84 whose local base-stock levels ignore the coordination the echelon formulation captures. On cumulative regret the ordering inverts, and the decentralized heuristic finishes ahead of the world model, 1965 against 3879, because it applies a competent base-stock from the first step and pays no exploration cost, while the world model pays a transient to learn the dynamics before its planned policy overtakes the heuristic's. Both sit far below the model-free deep Q-network and the naive rule. The deep Q-network reaches a terminal cost of 211 and more than forty thousand units of regret, an order of magnitude worse than the world model on integrated regret and more than twenty-five times worse on terminal cost, and it finishes below even the naive constant-order rule, its value estimates over the joint order space not converging to a usable policy within the budget. The forecast panel shows the learned model's one-step error falling from above ten to a few tenths as interaction data accumulate, which is the object the planner rolls forward. The comparison combines two effects and is not a clean model-based-versus-model-free contrast, since the world model is handed the base-stock policy class and the echelon statistic the deep Q-network must discover, which is the point on the operations-research side: where the optimal structure is known, as \citet{ClarkScarf1960inventory} established for serial systems, a learned model that plans within that structure recovers near-optimal control from a fraction of the data a model-free value method needs, and it reaches a policy competitive with a hand-built heuristic, echoing the finding of \citet{Gijsbrechts2022inventory} that learned inventory policies match specialized heuristics rather than dominating them, while remaining bounded above by the Clark-Scarf optimum it cannot beat.

\begin{table}[ht]
  \centering
  \caption{Cumulative regret against the Clark-Scarf oracle at $T=500$ and terminal per-period cost over the last eighty steps, mean $\pm$ standard error across twenty seeds, in rank order by cumulative regret. Lower is better; the oracle is the zero reference and its per-period cost is 6.61.}
  \label{table:fc_supply_chain}
\begin{tabular}{lrr}
\toprule
Paradigm & Cumulative regret & Terminal cost/period \\
\midrule
Oracle & 0.0 $\pm$ 0.0 & 6.61 $\pm$ 0.17 \\
Decentralized & 1964.8 $\pm$ 50.8 & 9.84 $\pm$ 0.26 \\
NN World Model & 3879.4 $\pm$ 406.5 & 8.00 $\pm$ 0.88 \\
Naive & 29485.4 $\pm$ 6777.2 & 106.05 $\pm$ 29.04 \\
Model-Free DQN & 44137.1 $\pm$ 5593.1 & 210.79 $\pm$ 53.76 \\
\bottomrule
\end{tabular}

\end{table}

\FloatBarrier
\subsection{Engine Replacement MDP: Learning the Engine's Transitions}
\label{engine:ch12}

\begin{table}[H]
\centering
\small
\caption{Learning the four Engine Replacement MDP transition rows. Value errors, bounds, and standard errors use two hundred fixed seeds.}
\label{tab:engine_model_learning}
\begin{tabular}{rrrrrr}
\hline
$N$ & mean kernel error & value error & s.e. & simulation bound & s.e. \\
\hline
25 & 0.1556 & 0.5015 & 0.0257 & 14.0040 & 0.7085 \\
50 & 0.1028 & 0.3284 & 0.0168 & 9.2520 & 0.4764 \\
100 & 0.0756 & 0.2435 & 0.0143 & 6.8040 & 0.3902 \\
250 & 0.0502 & 0.1615 & 0.0087 & 4.5144 & 0.2407 \\
500 & 0.0355 & 0.1138 & 0.0060 & 3.1950 & 0.1688 \\
1,000 & 0.0259 & 0.0830 & 0.0045 & 2.3265 & 0.1276 \\
2,500 & 0.0165 & 0.0529 & 0.0030 & 1.4821 & 0.0831 \\
5,000 & 0.0113 & 0.0362 & 0.0021 & 1.0154 & 0.0581 \\
\hline
\multicolumn{6}{l}{Final mean estimates of $P(\mathrm{high}\mid s,a)$} \\
\hline
low, keep & low, replace & high, keep & high, replace & \multicolumn{2}{c}{optimal-policy matches} \\
\hline
0.4998 & 0.0000 & 1.0000 & 0.0000 & \multicolumn{2}{c}{1.000} \\
\hline
\end{tabular}
\end{table}

\begin{figure}[H]
\centering
\includegraphics[width=0.72\textwidth]{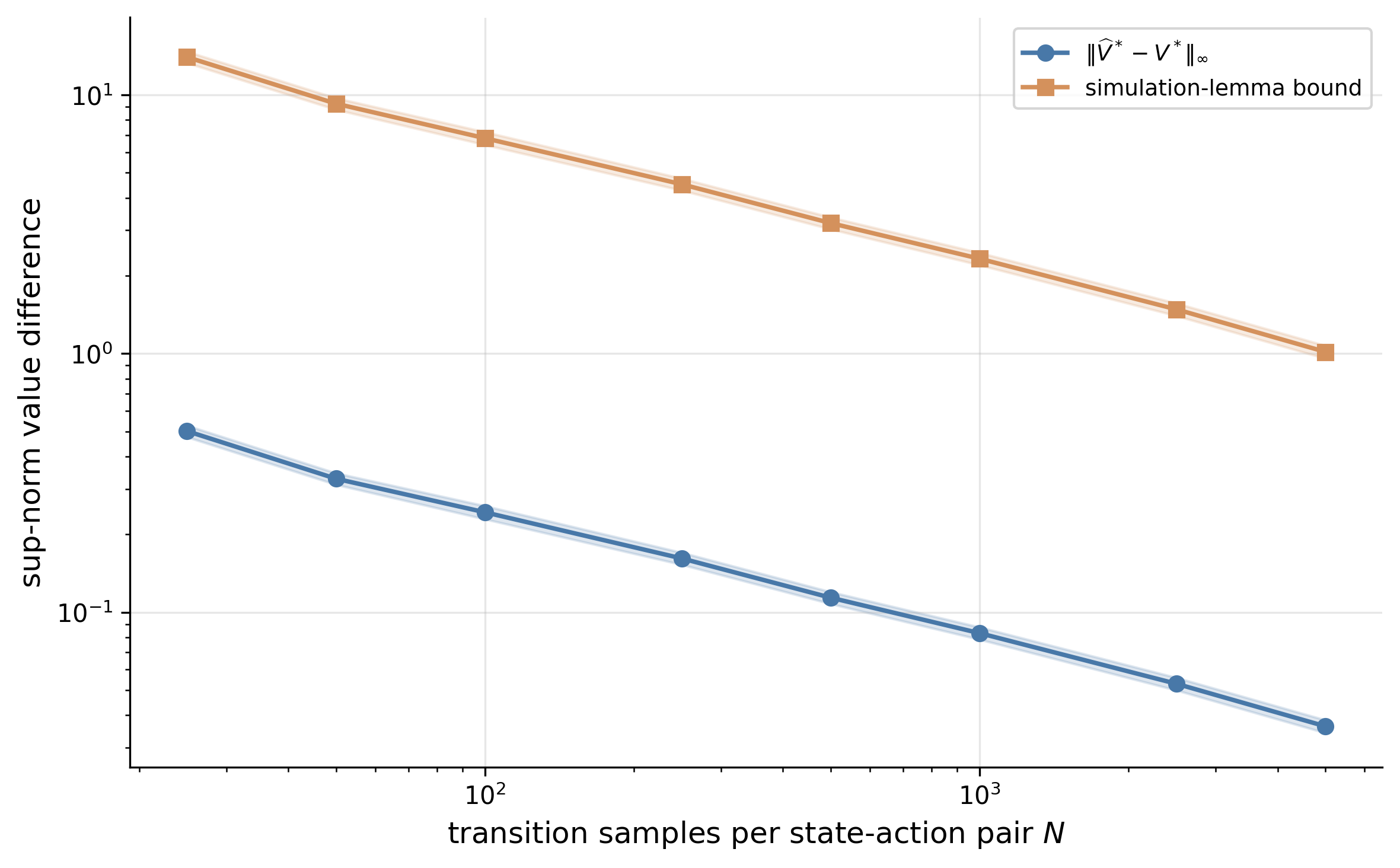}
\caption{Optimal-value error and its simulation-lemma bound against transition samples per state-action pair. Lines show means and shaded regions show one standard error over two hundred fixed seeds.}
\label{fig:engine_model_learning}
\end{figure}

For each sample size $N$, the empirical model estimates the probability of a high-mileage successor separately for all four state-action pairs and solves its Bellman optimality equation. Let $\epsilon_P=\max_{s,a}\|\widehat P(\cdot\mid s,a)-P(\cdot\mid s,a)\|_1$. Since rewards are unchanged and $R_{\max}=1$, the discounted simulation lemma gives
\[
\|\widehat V^\star-V^\star\|_\infty
\leq \frac{\gamma R_{\max}}{(1-\gamma)^2}\epsilon_P.
\]
The inequality links transition-estimation error to optimal-value error, and the simulation measures both sides.
Every one of the $1{,}600$ fitted models satisfies this inequality. Three transition rows are deterministic in this example, so their empirical probabilities equal their population values at every $N$. Sampling error enters through the low-mileage keep transition.

Table~\ref{tab:engine_model_learning} and Figure~\ref{fig:engine_model_learning} show that the mean value error falls from $0.5015$ at $N=25$ to $0.0362$ at $N=5{,}000$. The mean bound falls from $14.0040$ to $1.0154$. Both log-log slopes equal $-0.484$, close to the $-0.5$ parametric sampling rate for estimating the Bernoulli transition probability. The empirical optimal policy equals the true keep-then-replace policy in every replication, so the experiment separates value calibration error from policy regret.
\FloatBarrier

\subsection{Synthesis}
\label{section:fc_synthesis}

The chapter has traced the model-based reinforcement learning line from two 1990 origins, Sutton's Dyna and Schmidhuber's controller-model architecture, through the deep revival in \citet{HaSchmidhuber2018}, the recurrent state-space model that became the workhorse of the Dreamer line, the value-aware question that runs from VAML through value-equivalence to MuZero, the modern Dyna in continuous control associated with the MBPO ensemble line, and the convergence point at TD-MPC2. The remainder of the section notes where model-based methods win, where they fail, where the surrounding boundary makes the model unnecessary, and what remains open.

\subsubsection{Where model-based RL wins}
\label{section:fc_synthesis_wins}

Sample efficiency is the headline gain. Model-based methods outperform model-free baselines when environment interaction is scarce or expensive, including pixel-based continuous control in PlaNet and Dreamer, discrete-state planning in MuZero, and continuous control with sparse reward in MBPO and TD-MPC2. The learned model serves as a data amplifier through the Dyna planning step and as a representation amplifier, since the RSSM and TD-MPC2 latents transfer across tasks within their respective suites. When transferable representations or finite-data regimes are the binding constraint, model-based methods win.

\subsubsection{Where model-based RL fails}
\label{section:fc_synthesis_fails}

The first failure mode is misspecification and the objective mismatch between training and use. \citet{Lambert2020} document that maximum-likelihood model fitting and reward optimization are decoupled objectives, so a model trained to high likelihood is not necessarily a model whose induced policy is high-return. Their sweeps show likelihood-reward correlations as low as $0.07$ on Half-Cheetah and adversarial perturbations that preserve likelihood while halving return. \citet{VoelckerCalib2025} extend this point to MuZero and show that value-equivalent representations are systematically miscalibrated under distribution shift. The value-aware line (\S\ref{section:fc_value_aware}) is the partial answer, reformulating the training target around the value functional rather than the full distribution. A fully principled treatment of off-distribution calibration remains open.

The second failure mode is compounding error and the data-coverage gaps that exploration must fill. \citet{Asadi2018Lipschitz} give the local-operator Lipschitz bound for model-based RL, in which one-step prediction error compounds geometrically along the rollout at rate $K_F$, the Lipschitz constant of the learned dynamics. The induced value-error bound on the planner stays finite when $\gamma K_F < 1$ and diverges as $\gamma K_F \to 1$. When the contraction condition fails, model-based planning diverges. \citet{Talvitie2017} documents that retraining a single model on its own predictions can destabilize the model itself and proposes a self-correcting hallucinated-rollout scheme as one fix. \citet{Pathak2017} and \citet{Sekar2020} attack the same problem from the data side, with prediction-error-driven curiosity in the former and ensemble-disagreement-driven planning in the latter. Both push the agent's data distribution beyond what greedy exploitation alone produces, with measurable gains in regions of state space the agent would otherwise miss.

\subsubsection{When the model is not needed}
\label{section:fc_synthesis_boundary}

Not every sequential decision problem rewards the cost of learning a forecaster. When the decision reduces to a supervised prediction followed by a fixed downstream rule, a direct supervised reduction can match or beat a learned-world-model pipeline at much lower engineering cost, the boundary studied by \citet{Madeka2022DeepInventory} in deep inventory management. The boundary itself is closed-loop feedback. When the agent's action does not feed back into the data-generating process, a supervised reduction suffices. When it does, the learned model becomes load-bearing, because forecasts and decisions are then jointly determined. This is the boundary that places model-based reinforcement learning inside the sequential decision-making space and separates it from one-shot forecasting.

\subsubsection{Open questions}
\label{section:fc_synthesis_open}

The dual simulation in \S\ref{section:fc_dual_sim} maps where model-based reinforcement learning sits on the sample-efficiency frontier among the other paradigms in a single-agent cobweb with adjustment cost and a logistic-growth fishery. In the cobweb setting the Marcet-Sargent divergence-in-the-unstable-region story does not arise, because the environment is self-referential through prices alone and not through expectations. The more telling result is the order-of-magnitude separation between recursive least squares with correct functional form, a model-based learner that must estimate the demand and the cost coefficients, a population-based gradient-free search, and a tabular model-free agent. Whether the divergence story would reappear in an expectational cobweb in which beliefs about future prices drive supply, and in particular how a learned-neural-network world model would behave in that regime, is one of the open questions the chapter does not yet settle. A second open question surfaces from the gap between the closed-form linear-quadratic planner and the branched-rollout REINFORCE planner on the cobweb panel. When the environment admits a closed-form planner, branched rollouts pay a variance cost that closed-form planning does not. Whether this gap survives on environments without analytical planners (precisely where deep MBPO is deployed in practice) is the empirical question the deep continuous-control benchmarks were designed to answer. The analogy to E-stability is suggestive rather than formal. The operators, fixed points, and misspecification objects differ, and a stability theory for model-based reinforcement learning in economic environments remains to be built. \citet{Asadi2018Lipschitz} is the closest current analogue to a stability condition for the reinforcement-learning side, stating that value error grows geometrically along the rollout at rate $K_F$ (the Lipschitz constant of the learned dynamics) and remains finite when $\gamma K_F < 1$. But the result is a per-rollout contraction bound on value error rather than a fixed-point characterization of the joint learning dynamics in the sense of \citet{MarcetSargent1989}.

Read against the monograph as a whole, this chapter is where the simulator dependence of reinforcement learning becomes explicit. The question of when a learned model is accurate enough for an economic policy counterfactual is the open one the monograph carries into its conclusion.

\section{Reinforcement Learning in the Field}
\label{section:field_deployments}
This chapter examines field applications of reinforcement learning outside game engines, language-model post-training, and robotics, whose deployments are covered elsewhere \citep{Shao2019VideoGamesSurvey,Kaufmann2023RLHFSurvey,Ibarz2021robot,Tang2024RoboticsSurvey}. In the remaining domains, field applications are rare, and few have adequate documentation.

Field deployment generally combines dynamics that matter for long-run outcomes, a narrow control surface, historical and live logs of action probabilities and state transitions, strong incumbent methods, and guardrails that prevent costly exploration. The cases in this chapter come from recommendation, bidding, dispatch, inventory management, pricing, video encoding, chip design, and building control. Reinforcement learning from human feedback is included as a training-time application rather than evidence of a comparable runtime control system.

The Engine Replacement MDP does not serve as an application here because field deployment requires states shaped by the deployed policy itself. Full action support in a static two-grade model does not reproduce that occupancy problem.

\subsection{The Anatomy of a Production Reinforcement Learning System}
\label{subsec:production_rl_anatomy}

\begin{figure}[t]
\centering
\includegraphics[width=\textwidth]{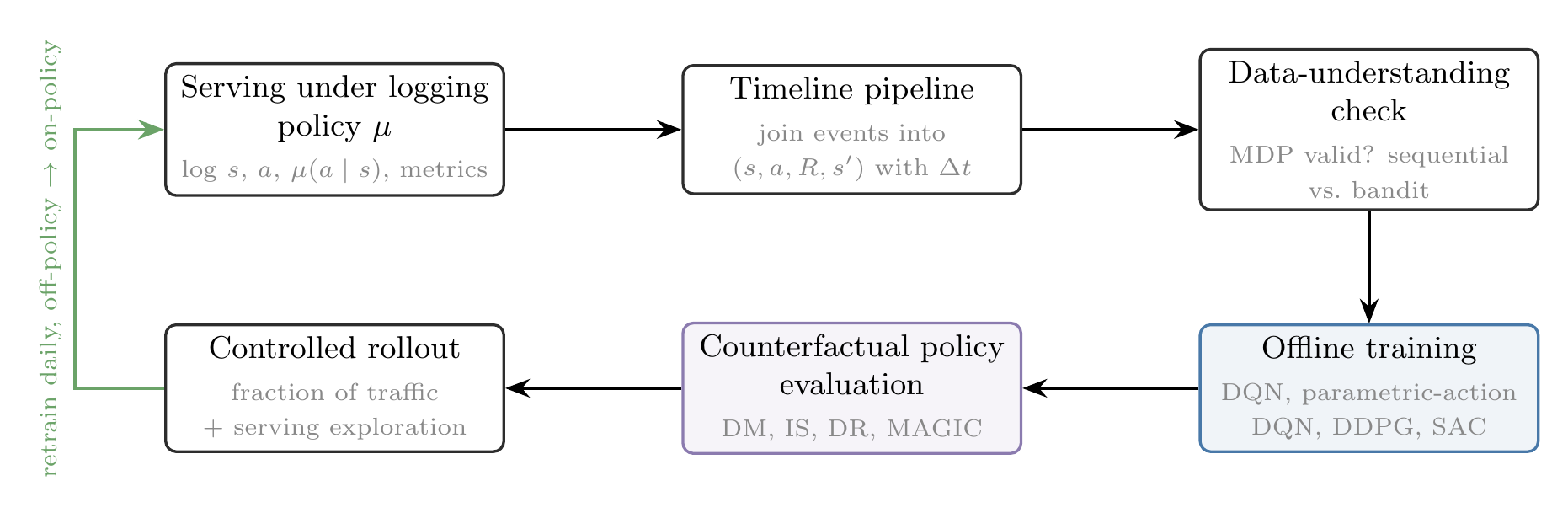}
\caption{The Horizon production loop \citep{Gauci2019Horizon}. A serving policy $\mu$ logs each decision with its propensity and metrics, the Timeline pipeline assembles transitions, a data-understanding check screens the formulation, an offline learner is trained and scored by counterfactual policy evaluation, and the policy is rolled out on a fraction of traffic with exploration and retrained daily on the data it generates.}
\label{fig:horizon_pipeline}
\end{figure}

A reinforcement learning method that works in a simulator is not yet a system that can run on a live product surface, and the distance between the two is mostly infrastructure. Facebook's Horizon platform \citep{Gauci2019Horizon} is the clearest public account of what that infrastructure contains, so it is worth setting out as a template before the individual cases. The running example, here and in the source, is notification sending. For each notification candidate the system decides whether to send it or drop it, and a Horizon-trained policy replaced a supervised system for some Facebook notifications and raised activity and meaningful interactions. The policy itself is an ordinary deep Q-network (DQN). What makes it deployable is everything around it, namely how decisions are logged, how transitions are assembled, how the formulation is screened, how the policy is evaluated before anyone sees it, and how the loop is closed by retraining. Figure~\ref{fig:horizon_pipeline} shows the loop, and the deployments in the rest of this chapter each occupy part of it rather than all of it.

The objects can be named once and reused for the later cases. The decision process is a Markov decision process $\mathcal{M}=(\mathcal{S},\mathcal{A},P,R,\gamma)$ with state-dependent feasible actions $\mathcal{A}(s)$, since a production system usually has a different set of actions available in each state. The system that generated the historical data is the logging policy $\mu(a\mid s)$, and the recorded number $\mu(A_t\mid S_t)$, the probability it assigned to the action it took, is the propensity. Horizon does not store a reward. It stores a vector of business metrics $M_t$ and forms the reward at training time from a user-supplied weight vector $w$,
\begin{equation}
R_t = w^\top M_t .
\end{equation}
Logged events are assembled by the Timeline pipeline into one record per transition,
\begin{equation}
\big(i,\; t,\; S_t,\; A_t,\; \mu(A_t\mid S_t),\; M_t,\; \mathcal{A}(S_t),\; S_{t+1},\; A_{t+1},\; \Delta t,\; \mathcal{A}(S_{t+1})\big),
\end{equation}
where $i$ identifies the decision unit, $t$ orders the steps, and $\Delta t$ is the time gap to the next decision. Learning targets a policy $\pi$ whose worth is the discounted return $V(\pi)=\mathbb{E}_\pi\!\left[\sum_t \gamma^t R_t\right]$, estimated before deployment by a counterfactual value $\hat V(\pi)$ built from the logged propensities.

Formulation comes before any algorithm. Horizon's data-understanding tool fits a probabilistic model of next states and rewards from the same data used for training,\footnote{A neural network whose final layer is a Gaussian mixture, so that it outputs a distribution over next states and rewards rather than point estimates.} and uses it to test whether the proposed problem behaves like a Markov decision process at all. The tool then applies two checks. If no state feature helps predict the next state, the problem has no sequential structure and reinforcement learning adds nothing over a supervised score. If no state feature both responds to the action and predicts reward, the problem collapses to a bandit, where estimating the return of each action suffices. For notifications the case for a sequential formulation is exactly this, since the supervised system it replaced could not capture the value that arrives long after the send decision.

Two fields in the record carry information a plain $(s,a,r,s')$ tuple lacks. The propensity $\mu(A_t\mid S_t)$ is what later allows the policy to be scored without deploying it, so the serving system logs it alongside the action and the metrics at decision time. The feasible set $\mathcal{A}(S_t)$ records which actions were available, so the Q-learning update maximizes only over actions that could actually be taken. For notifications the decision unit is the sequence of candidates for one person, the action is send or drop, and the metrics carry interactions and activity together with the send itself, so the reward can charge a penalty for sending and hold down volume.

The choice of algorithm follows from the action space.\footnote{A discrete-action deep Q-network for small action sets, a parametric-action variant that scores concatenated state-action pairs when the set is very large, and DDPG or SAC when the action is continuous.} Notification sending is a binary action, so the policy is a discrete-action deep Q-network, and the first version is trained entirely on the incumbent system's logs. That data is off-policy, and an offline learner can attach unrealistic value to state-action pairs it never observed,\footnote{Extrapolation error in the sense of \citet{Fujimoto2019}; retraining on self-generated data pulls the estimates back toward the region the data support.} so the platform leans on daily retraining toward on-policy data and on counterfactual evaluation before exposure.

Evaluation is counterfactual. A set of estimators runs automatically on the logged data and predicts the new policy's value from data the old policy generated,\footnote{Horizon runs six, namely a step-wise direct method, step-wise importance sampling, and step-wise doubly robust estimator, together with sequential doubly robust, sequential weighted doubly robust, and MAGIC estimators for longer horizons. These are the off-policy estimators of Section~\ref{section:offline_rl} and Section~\ref{section:causal_rl}.} and the estimates decide which policies are worth an online test. The notification policy acts by comparing the two action-values, sending with probability
\begin{equation}
\pi(\text{send}\mid s) = \sigma\!\big(Q(s,\text{send}) - Q(s,\text{drop})\big),
\end{equation}
where $\sigma$ is the logistic function, and a proportional-integral-derivative controller tunes a threshold on this probability to keep the action rate close to the system it replaced. A policy that passes is deployed to a fraction of production traffic with some exploration retained during serving, and it is retrained daily on the data it then generates. Horizon reports that the A/B comparison starts neutral and improves as the training data shifts from off-policy toward on-policy. Table~\ref{tab:horizon_pipeline} collects the stages with their notification instantiation.

\begin{table}[t]
\centering
\footnotesize
\begin{tabularx}{\textwidth}{@{}>{\raggedright\arraybackslash}p{0.15\textwidth}>{\raggedright\arraybackslash}X>{\raggedright\arraybackslash}X>{\raggedright\arraybackslash}p{0.09\textwidth}@{}}
\toprule
Stage & Horizon mechanism & Notification instantiation & Source \\
\midrule
Formulation & Data-understanding model checks the Markov and bandit reductions & Sending affects value that arrives after the decision & \S4 \\
Logging & Propensity $\mu(A_t\mid S_t)$, feasible actions, and metrics map & Serving logs send or drop and its probability & \S2, \S8 \\
Transitions & Timeline join into $(S_t,A_t,R_t,S_{t+1})$ with a time gap & Candidate sequence for one person & \S2 \\
Reward & $R_t=w^\top M_t$ formed at training time & Interactions and activity, penalty on sending & \S2, \S9.1 \\
Algorithm & DQN, parametric-action DQN, DDPG, or SAC by action space & Discrete-action DQN & \S5, \S9.1 \\
Evaluation & Six counterfactual estimators on logged data & Screen before any exposure & \S7.2 \\
Deployment & Fraction of traffic with exploration, daily retraining & $\sigma(Q_{\text{send}}-Q_{\text{drop}})$ with a controller-tuned threshold & \S6, \S9.1 \\
\bottomrule
\end{tabularx}
\caption{The Horizon production loop \citep{Gauci2019Horizon} and its instantiation for push notifications. Section numbers refer to the source paper.}
\label{tab:horizon_pipeline}
\end{table}

\subsection{Digital Traffic: Recommendations}

\subsubsection{YouTube recommendations}
\label{sec:youtube_topk}
YouTube's top-$K$ recommender is the cleanest public case of a formal reinforcement-learning result placed inside a production system that was already sophisticated \citep{Chen2019YouTubeTopK}. The learned policy did not run the whole recommender. It was one of several systems that nominate a short list of candidate videos, and a separate model then scored that pooled list and built the page the user saw.\footnote{The paper calls a nominating system a candidate generator and the scoring model a ranker, and the learned policy here was one candidate generator among many. Its metric of record is what the paper names ViewTime, the time users spent watching videos on YouTube.} The surface it controlled was narrow, changing which videos entered the ranking funnel rather than the final page, and it was judged on the time users spent watching. That narrow and well-instrumented surface is what makes the case credible as field evidence rather than a demonstration.

The starting point is a formal object this monograph has already built. Recall the policy-gradient (REINFORCE) update of Section~\ref{section:rl_algorithms}. When a policy $\pi_\theta$ learns from its own trajectories, each logged choice is nudged along
\begin{equation}
\label{eq:yt_reinforce}
\nabla_\theta J(\theta) = \mathbb{E}_{\tau \sim \pi_\theta}\!\left[\sum_{t \geq 0} R_t\, \nabla_\theta \log \pi_\theta(a_t \mid s_t)\right],
\end{equation}
where $R_t = \sum_{t' \geq t} \gamma^{t'-t} r(s_{t'}, a_{t'})$ is the discounted return from time $t$. In the recommender the state $s$ is a recurrent summary of the user's watch history, the action $a$ is one video from a catalog of order $10^6$, and the policy is a softmax over videos, $\pi_\theta(a \mid s) \propto \exp(s^\top v_a / T)$ with temperature $T$. Two facts about the production setting force two small and self-contained changes to this update, and those two changes are the paper's contribution.

The first fact is that observing a reward means showing a real video to a real user, so the policy cannot generate fresh trajectories of its own. It learns instead from choices logged by earlier and parallel recommenders, a behavior policy $\beta$, and the standard fix reweights each logged choice by the importance ratio $\pi_\theta(a \mid s) / \beta(a \mid s)$ in place of the plain gradient.\footnote{The behavior policy $\beta$ is not logged directly, since some logging agents are deterministic or outside YouTube's control, so it is estimated by a second softmax head that shares the recurrent user state with the main policy and has its gradient blocked from that state. The estimated ratio is then capped at $c = e^3$ to bound its variance. Other chapters write $\pi_b$ for this logging policy; here $\beta$ follows the off-policy literature and clashes with nothing, since the discount is $\gamma$ and the softmax temperature is $T$.} The second fact is that the product returns a set of videos rather than one. Drawing $K$ items independently from $\pi_\theta$ and removing duplicates, an item's chance of appearing at least once is its inclusion probability
\begin{equation}
\label{eq:yt_alpha}
\alpha_\theta(a \mid s) = 1 - \big(1 - \pi_\theta(a \mid s)\big)^K .
\end{equation}
Optimizing this inclusion probability rather than the single best choice multiplies the off-policy update by one extra factor, giving the paper's result
\begin{equation}
\label{eq:yt_topk}
\boxed{\nabla_\theta J_K(\theta) \approx \mathbb{E}_{s \sim d^\beta,\, a \sim \beta}\!\left[ \frac{\pi_\theta(a \mid s)}{\beta(a \mid s)}\, \lambda_K(s,a)\, R(s,a)\, \nabla_\theta \log \pi_\theta(a \mid s) \right]}
\end{equation}
with the top-$K$ multiplier $\lambda_K(s,a) = K\big(1 - \pi_\theta(a \mid s)\big)^{K-1}$.\footnote{The multiplier is the derivative $\partial \alpha_\theta / \partial \pi_\theta$ of the inclusion probability, so the top-$K$ update is a chain-rule consequence of optimizing $\alpha_\theta$. The inclusion probability reaches one half at $\pi_\theta(a \mid s) = 1 - 2^{-1/K} \approx (\log 2)/K$, so an item needs probability only of order $1/K$, not near one, to have an even chance of appearing in the slate.}

The two changes read cleanly. The ratio $\pi_\theta / \beta$ removes the behavior policy's selection bias, undoing the tendency of an uncorrected learner to keep recommending whatever was already shown most, a rich-get-richer effect. The multiplier $\lambda_K$ turns a top-1 learner into a top-$K$ one. It is close to $K$ for a video the policy currently rates as unlikely, so a promising but under-exposed item is pushed up hard, and close to zero for a video already almost certain to appear, so the policy stops spending probability on making one item a lock and funds the second and third candidates instead. At $K = 1$ the multiplier is one and the ordinary correction returns.

Two caveats travel with the result. The first-order estimator that is actually deployed is biased. It corrects which video was chosen in a logged state, but not how often the new policy would reach that state, nor the later choices that produced the return, so its expectation is not the true policy gradient.\footnote{A genuinely unbiased off-policy gradient reweights the whole trajectory by the product $W(\tau) = \prod_t \pi_\theta(a_t \mid s_t) / \beta(a_t \mid s_t)$, whose expectation under $\beta$ equals the on-policy gradient because the initial-state and transition factors cancel. That product is a chain of ratios whose variance explodes at YouTube's scale, and the estimated $\beta$ compounds the error, so the paper drops the state-visitation ratio $d^\pi/d^\beta$ and the future-action ratios and keeps only the current-action term. Bias here means the estimator's expectation differs from $\nabla_\theta J(\theta)$, not merely that it is noisy; \citet{Chen2019YouTubeTopK} invoke a bound of Achiam et al.\ on the resulting gap in terms of the total-variation distance between $\pi_\theta$ and $\beta$.} It also needs the logging policy to have placed some probability on every video the new policy wants to try, since no reweighting recovers an outcome that was never observed.\footnote{This condition, that $\pi_\theta(a \mid s) > 0$ implies $\beta(a \mid s) > 0$, is the standard common-support requirement of off-policy learning rather than the paper's own language. \citet{Chen2019YouTubeTopK} speak instead of sparse data over an action space in the millions, and of large or noisy importance ratios, as the sources of instability.} And the reduction to a per-item update rests on two slate assumptions.\footnote{The reduction assumes at most one item in the returned set earns reward and that an item's value does not depend on the others shown, which sets aside within-slate competition, complementarity, position effects, and multiple clicks.} These are the compromises the method accepts to run at scale.

Do the two changes actually change the recommendations? In the paper's two controlled simulations they do, visibly. In the first, where the logging policy over-shows the least-rewarded items, an uncorrected learner inherits that skew and misses the best item, and adding the off-policy correction recovers it. In the second, with a uniform logging policy and one clearly best item, the standard correction swings to the opposite failure and puts almost all of its mass on that single item, while the top-$K$ correction keeps real mass on the second-best as well.\footnote{The first simulation has ten items with reward $r(a_i) = i$ and a behavior policy $\beta(a_i) = (11-i)/55$ skewed toward the least-rewarded items; without correction the learned policy converges to the reward-weighted behavior $\propto r \cdot \beta$, a mound over the middle items that misses the optimum, and with the off-policy correction it recovers the optimal item (Figure 2 of the source). The second simulation gives one item reward $10$, the next reward $9$, and the rest reward $1$ under a uniform behavior policy; the standard correction collapses onto the single best item, $\pi(a_1) \approx 1$, while the top-2 correction keeps significant mass on the second-best, roughly a $0.6$ and $0.4$ split read from Figure 3 of the source, which prints no numbers.} In the live system the clearest direct evidence is that adding the off-policy correction raised nominations of videos outside the historically top ranks by nearly a factor of three, a genuine change in the candidate sets, while the top-$K$ correction added a lift of 0.85 percent in watch time over the previous model.\footnote{The threefold change is the no-correction-versus-standard-correction comparison (Figure 4 of the source, a live A/B test), not a top-1-versus-top-$K$ result. For top-1 versus top-$K$ the paper publishes only watch time and the number of videos viewed, with no candidate-overlap, catalog-coverage, entropy, or diversity statistic, and the candidate sets are re-ranked downstream before display.} The premise underneath all of this is dynamic, that today's recommendation shapes tomorrow's user state, yet the paper does not isolate that this sequential effect is what paid off, so the demonstrated gains are the exposure and slate-shape changes just described rather than measured dynamic preference formation.\footnote{The model retrained continuously over calendar time, but the paper shows no training or cumulative-reward curve and no early-versus-late checkpoint comparison; Figure 5 of the source is a five-day A/B effect plot, not a learning curve. There is no ablation of the discount, and the simulations are explicitly stateless, so whether the sequential structure itself contributed is left open.}

Table~\ref{tab:youtube_topk_live} collects the reported live effects, which are sub-percent movements in watch time\footnote{Read against a standard metric hierarchy, the north-star is watch time and the secondary engagement metric is the number of videos viewed. The algorithmic diagnostic is the distribution of nominations by rank; the exploration evidence is a $0.07$ percent watch-time gain from a five-percent stochastic-exploration traffic slice at equal data volume; the operational facts are continuous training with a data lag under 24 hours, a reward horizon of 4 to 10 hours, and serving by sampled softmax at training time and approximate nearest-neighbor retrieval over the learned embeddings at serving time. Explicit safety, quality, or diversity guardrails are not reported.} measured as component tests against evolving controls rather than a replacement of YouTube's recommender.\footnote{The system improved in sequence, from a reward-only policy gradient to the standard off-policy correction to the top-$K$ correction, and each stage became the control for the next, so the effects are not additive. Across $K \in \{1, 2, 16, 32\}$ the production setting was $K = 16$, with $K = 1$ losing $0.66$ percent watch time, $K = 2$ losing $0.35$ percent, and $K = 32$ matching $K = 16$; a later followup with $K = 8$ gave a mild $0.15$ percent gain but was not adopted.} The lesson is the one this chapter keeps finding. An established formal result, the score-function policy gradient of Section~\ref{section:rl_algorithms},\footnote{A value-based alternative would fit an action-value $Q_\phi(s,a) = h_\phi(s)^\top v_a$, train it toward a Double-DQN target, and serve the top $K$ by nearest-neighbor search over $Q$. Q-learning is naturally off-policy, but its Bellman target maximizes over millions of actions, most with little logged support. \citet{Chen2019YouTubeTopK} prefer policy gradients because value-based methods can be unstable under nonlinear function approximation, need heavy tuning, and lack established policy-convergence guarantees, and they report no empirical Q-learning comparison.} carried into a live system with two disciplined modifications and hard variance control, buys a small, real, well-measured gain.\footnote{The gain is measured against the immediately preceding production model, not against the prior supervised recommender, contextual bandits, the full ensemble of candidate generators, or the final ranker. The live experiments carry no random-policy lower bound and no oracle upper bound, and the optimal policy appears only in the stateless simulation.} The price is a set of visible compromises,\footnote{The obstacles the paper foregrounds, in the order it emphasizes them, are logged-policy bias and thin coverage, the scale and sparsity of a catalog in the millions, the mismatch between a top-1 policy gradient and a top-$K$ product, the variance of the importance weights, limited exploration that must not harm users, and batched, nonstationary learning.} a deliberately biased estimator, capped importance weights, and a control surface kept narrow on purpose, and paying it is what let the method reach production.

\begin{table}[t]
\centering
\small
\begin{tabularx}{\textwidth}{@{}>{\raggedright\arraybackslash}p{0.27\textwidth}>{\raggedright\arraybackslash}X>{\raggedright\arraybackslash}p{0.20\textwidth}@{}}
\toprule
Intervention & Live comparison & Reported effect \\
\midrule
Exploratory data & Train with a 5\% stochastic-policy traffic slice rather than only deterministic-policy logs & +0.07\% ViewTime \\
Standard off-policy correction & Weight logged examples by learned behavior-policy importance ratios & +0.53\% videos viewed; no significant ViewTime change \\
Top-$K$ off-policy correction & Train the production candidate generator with $K=16$ top-$K$ correction & +0.85\% ViewTime; -0.16\% videos viewed \\
Top-1 ablation & Set $K=1$, reducing the correction toward the standard top-1 case, versus production $K=16$ & -0.66\% ViewTime \\
Uncapped-weight regression & Loosen the importance-weight cap from $e^3$ to $e^5$ & -0.52\% ViewTime \\
\bottomrule
\end{tabularx}
\caption{Live YouTube experiments reported by \citet{Chen2019YouTubeTopK}. Effects are relative to the relevant control model in each sequential experiment, so rows should be read as component tests rather than additive gains.}
\label{tab:youtube_topk_live}
\end{table}

\subsubsection{Taobao cold-start recommendation}
\citet{Ji2021RLLTV} frame cold-start item recommendation on Taobao as an item-lifetime-value problem. The state is item-level history, including exposure, interaction, sale, and time features; the learned action contributes an RL score to a dual-rank system. The live evidence is a one-week Taobao A/B test beginning on January 26, 2021, with reported cold-start-item gains of 8.67\% in item page views and 18.03\% in gross merchandise value relative to the vanilla click-through-rate (CTR) baseline.

Operationally, RL-LTV runs at the item-day frequency rather than the request-serving frequency. User logs are aggregated by item and by day into item episodes, each a finite-horizon window of daily steps over the item's early lifetime, discounted at $\gamma=0.5$; the actor produces an LTV-oriented score $y_{\mathrm{rl}}$, and the critic estimates future item value. The online ranker keeps the conventional CTR score $y_{\mathrm{ctr}}$ and combines it with $y_{\mathrm{rl}}$ through a dual-rank module whose mixing weight is bounded in the experiment. The serving system therefore receives a ranking score, while the RL training loop updates from daily item histories.

The formal action space is broader than the live control surface. The paper's action includes both an RL score and a price-like component, but the source notes that only the RL score was effective in the live system. The deployed policy therefore controlled ranking investment, not full item pricing.

\subsection{Bidding and Auction Control}

\subsubsection{Meta production bidding}
In online advertising an advertiser wants to maximize the value it receives under a budget constraint while thousands of auctions clear every second, which no advertiser can manage opportunity by opportunity. Platforms therefore run automated proxy agents, auto-bidders, that place bids in real time on the advertiser's behalf. The agent is owned by the platform but spends advertiser money, so reliability and explainability weigh as heavily as raw optimizing power, and the incumbent controllers are bounded heuristic feedback rules rather than black boxes. Earlier reinforcement-learning work tuned such controllers by treating their parameters as actions and re-tuning them online at every step, which adds infrastructure and safety cost. \citet{Korenkevych2023MetaBidding} instead learn the controller's parameters once, offline, and deploy only those, which is what lets a learned bidder reach production.

The problem is an episodic Markov decision process $\mathcal{E}=(\mathcal{S},\mathcal{A},\mathcal{R},T,\rho)$ in which one ad campaign is one episode.
\begin{enumerate}
\item The state $S_t\in\mathcal{S}$ is the full Markovian state at minute $t$, comprising the controlled campaign, the auctions it is eligible for, and the competing campaigns in those auctions. The agent does not observe most of this and acts on a limited observation $O_t$ of its own campaign, such as the fraction of budget spent, the fraction of opportunities passed, and the average past bid.
\item The action $A_t\in\mathbb{R}$ is a single scalar bid, reused across every auction the campaign enters at $t$ and then scaled by each auction's expected conversion rate inside the auction system, a scaling the agent does not control.
\item The reward is the aggregated value of the auctions the campaign wins during the step,
\begin{equation}
R_t=\sum_{i=1}^{N_t}\delta_t^i\,W_t^i,
\end{equation}
where $N_t$ is the number of auctions entered, $W_t^i\in\{0,1\}$ indicates whether the $i$-th was won, and $\delta_t^i$ is the value of winning it, for example the value of a conversion.
\item The horizon $T$ is set by the campaign's budget expiration, so each campaign is one finite episode.
\item The transition $\rho(\cdot\mid S_t,A_t)$ models the auction mechanics and the bidding policies of all competing campaigns.
\end{enumerate}
The agent maximizes the discounted return $\mathbb{E}\big[\sum_t\gamma^t R_t\big]$ with $\gamma=0.9998$, following the Markov-decision-process conventions of Section~\ref{section:rl_algorithms}.

The control surface is a single differentiable function. A production base policy $F_w:\mathcal{S}\subset\mathbb{R}^d\to\mathcal{A}\subset\mathbb{R}$, parameterized by a vector $w$, already maps campaign states to bids, and reinforcement learning only optimizes $w$ to $w^*$. To make the parameterization concrete, a proportional-integral controller is one such base policy,
\begin{equation}
F_w(s_t)=a_{t-1}+K_p\,e_t+K_i\sum_{\tau=0}^{t}e_\tau,\qquad s_t=\big[a_{t-1},\,e_t,\,\textstyle\sum_{\tau=0}^{t}e_\tau\big],\qquad w=[K_p,K_i],
\end{equation}
where $e_t$ is an error signal such as the gap between the fraction of budget spent and the fraction of opportunities consumed. This controller is illustrative, and the deployed base policy is a piecewise-polynomial function of input features with a couple dozen scalar parameters. Only $w^*$ is written to production, serving stays a single forward pass of the existing controller, and the neural machinery introduced next is discarded after training.

Exploratory data are required for reinforcement-learning training. Under-trained bidding policies are not deployed on live traffic because even slight suboptimality can produce losses of millions of dollars within hours. The data are therefore collected once by a behavior policy that adds bounded noise to the base policy,
\begin{equation}
\pi^\beta_w(S_t)=F_w(S_t)\,\big(1+\mathrm{clip}(\varepsilon_t,-0.5,0.5)\big),\qquad \varepsilon_t\sim\mathcal{N}(0,\sigma_\beta^2),
\end{equation}
equivalent to the additive form $\mathcal{N}\big(F_w(S_t),\,F_w(S_t)\,\sigma_\beta^2\big)$. The noise is multiplicative because optimal bids vary over several orders of magnitude, so a fixed additive variance would be too disruptive at small bids and negligible at large ones. A standard deviation $\sigma_\beta=0.05$ was near the largest the platform would accept, since it already cost about half a percent of online performance during a two-week collection on a fraction of traffic. The logged dataset covers roughly $200{,}000$ week-long campaigns, about $1.2$ billion per-minute steps, with each episode ending when the budget is depleted, the week elapses, or the advertiser stops early.

The learner is a hybrid actor-critic. The actor turns the deterministic base policy into a Gaussian bidding policy whose mean is the base policy and whose variance is a small network,
\begin{equation}
\pi_{w,\phi}(\cdot\mid S_t)=\mathcal{N}\big(F_w(S_t),\,\sigma_\phi^2(S_t)\big),
\end{equation}
and the critic is a separate network approximating $Q(S_t,A_t)$. Because the critic is never deployed it may use a wider feature set than the actor, which yields more accurate value gradients for $w$. Both the variance network and the critic are two-layer perceptrons with $512$ units per hidden layer and ReLU activations, and $w$ is initialized at the production defaults so that learning starts on the behavior distribution.

Training is fully offline, using a conservative Q-learning objective built on a soft actor-critic (Section~\ref{section:offline_rl}). The critic minimizes the soft Bellman error
\begin{equation}
J_Q(\theta)=\tfrac12\,\mathbb{E}_{(s,a)\sim D}\Big[\big(Q_\theta(s,a)-\big(r(s,a)+\gamma\,\mathbb{E}_{s'\sim p}V_\theta(s')\big)\big)^2\Big],
\end{equation}
and the conservative penalty adds two terms that pull down the value of unseen actions and pull up the value of logged actions,
\begin{equation}
\min_Q\ \alpha\,\mathbb{E}_{s\sim D}\Big[\log\!\sum_a\exp Q(s,a)-\mathbb{E}_{a\sim\hat\pi_\beta}\big[Q(s,a)\big]\Big]+J_Q(\theta),
\end{equation}
so the trained value function does not reward out-of-distribution bids it never observed, the overestimation failure that Section~\ref{section:offline_rl} treats. Four changes adapt this to bidding. The out-of-distribution term is estimated on the state-dependent interval
\begin{equation}
I_\beta(s)=\big[(1-\varepsilon)F_w(s),\,(1+\varepsilon)F_w(s)\big],
\end{equation}
rather than the whole real line, since bids outside a band around $F_w(s)$ are never taken. The in-distribution term is estimated from $K=50$ actions drawn directly from the known behavior policy rather than reused logged actions. The soft-actor entropy bonus is removed because it is meaningless offline. And the critic is pre-trained for $300{,}000$ steps so its predictions match the actor at the start of training. The remaining hyperparameters are in Table~\ref{tab:meta_bidding_hparams}.

\begin{table}[t]
\centering
\small
\begin{tabular}{ll}
\toprule
Parameter & Value \\
\midrule
Gradient steps during training & $600{,}000$ \\
Batch size & $20{,}000$ \\
Critic learning rate & $3\times10^{-4}$ \\
Actor learning rate & $3\times10^{-5}$ \\
Discount factor $\gamma$ & $0.9998$ \\
Behavior-policy standard deviation $\sigma_\beta$ & $0.05$ \\
Conservative penalty weight $\alpha$ & $0.1$ \\
Exploration-noise clipping range & $(-0.5,\,0.5)$ \\
Critic hidden sizes & $(512,\,512)$ \\
Variance-network hidden sizes & $(512,\,512)$ \\
Target-network update rate $\tau$ & $0.01$ \\
Penalty-estimation samples $K$ & $50$ \\
\bottomrule
\end{tabular}
\caption{Training hyperparameters for the Meta production bidding agent \citep{Korenkevych2023MetaBidding}.}
\label{tab:meta_bidding_hparams}
\end{table}

An offline policy cannot be scored without a live test, so the training checkpoint cannot be chosen from the logs alone. The authors use an internal campaign simulator, one hundred configurations with randomized opportunity density, audience size, and budget, run as one-day episodes of up to $1{,}440$ per-minute steps, to sweep the penalty weight $\alpha$ and select a checkpoint before any deployment. Each checkpoint is scored on $1{,}000$ evaluation episodes, ten for each configuration, and the learned critic's value predictions are checked against empirical returns both at the start of an episode and step by step to confirm the value fit.

Table~\ref{tab:meta_bidding_results} reports the outcome. In the simulator the penalty weight governs stability, and $\alpha=0.1$ is the only setting that both stays stable and clears the launch bar, so it is the one deployed. In production the two A/B tests each ran for a week over about fifty billion impressions and returned small gains that were statistically significant on the platform's performance metric, with revenue gains in both. The lift is small because the incumbent controller is already good, and the case exemplifies reinforcement learning used to tune a trusted controller rather than replace it.

\begin{table}[t]
\centering
\small
\begin{tabular}{lll}
\toprule
\multicolumn{3}{l}{Panel A: offline conservative-penalty sweep in the campaign simulator} \\
\midrule
Penalty weight $\alpha$ & Learning behavior & Gain vs.\ base policy \\
\midrule
$0.1$ (deployed) & Stable & Above $+0.5\%$ after $400{,}000$ steps \\
$1.0$ & Over-restrictive & Marginal \\
$0.01$ & Unstable & Abrupt swings \\
$0.0$ & Unstable & Abrupt swings \\
\midrule
\multicolumn{3}{l}{Panel B: production A/B tests vs.\ the original base policy} \\
\midrule
Test & Performance-metric gain (95\% confidence interval) & Impressions \\
\midrule
Pre-test & $+0.17\%$ $(+0.05\%, +0.30\%)$ & $\sim$50 billion \\
Back-test & $+0.16\%$ $(+0.03\%, +0.27\%)$ & $\sim$50 billion \\
\bottomrule
\end{tabular}
\caption{Meta production bidding \citep{Korenkevych2023MetaBidding}. Panel A: checkpoint selection in the internal campaign simulator over the conservative-Q-learning penalty weight, with $1{,}000$ evaluation episodes per checkpoint. Panel B: two one-week production A/B tests, each over roughly $50$ billion impressions, reporting the change in the platform performance metric with $95\%$ confidence intervals. Both tests also showed revenue gains.}
\label{tab:meta_bidding_results}
\end{table}

\subsubsection{Alibaba sponsored-search RTB}
\citet{Zhao2018SponsoredSearchRTB} report an earlier and methodologically opposite auction deployment on Alibaba's sponsored-search platform, run across a thousand advertisers and about a hundred million auctions a day. Where the Meta system tunes a single trusted controller offline and treats the competing campaigns as a fixed environment, the Alibaba system learns online, is value-based rather than actor-critic, and models the competition among advertisers explicitly. It is the multi-agent bidding case in this chapter, and it shows both why independent learners erode their own gains in a shared auction and how a cooperative reward term restores them.

The natural formulation places one decision at each auction, with a state of remaining budget, step index, and per-auction features, an action that is a bid, a reward equal to the purchase value won, and one day as an episode. That model does not transfer across days, because the per-auction environment is too noisy to repeat. The move that makes it deployable is to aggregate to the hour. A day becomes a fixed $m=24$ step episode, and the state $s=\langle b,t,g\rangle$ carries the remaining budget $b$, the hour $t$, and a vector $g$ of aggregated statistics for that hour, the numbers of impressions and clicks, cost, click-through rate, conversion rate, and pay per click. At this resolution the transition dynamics repeat across days, each hour's aggregated features are nearly determined by the previous hour's features and action, which is what makes a policy trained on one day valid on the next.\footnote{Holding the feature $g_{i,t}$ and the action $a_t$ fixed, the source observes that the next value $g_{i,t+1}$ concentrates around its conditional mean with deviation below $\eta<0.03$. The model is called robust in that this hour-level invariance lets one policy hold across the day-to-day environment shift, which is a different notion from the worst-case minimax robustness treated elsewhere in this survey.} The hourly action sets the parameter of the real-time bidding model, discretized to one of a hundred levels, and the auction engine applies that model to price every impression in the hour, so reinforcement learning controls a compact hourly knob rather than each bid.

The policy is a value-based deep $Q$-network, the direct contrast to the Meta bidder's conservative actor-critic. The action value $Q(s,a)$ is approximated by a network with weights $\theta$ and trained toward the one-step Bellman target of Section~\ref{section:rl_algorithms},
\begin{equation}
y=r(s,a)+\gamma\max_{a'}Q_{\bar\theta}(s',a'),\qquad
\mathcal{L}(\theta)=\big(y-Q_\theta(s,a)\big)^2,
\end{equation}
with the target-network weights $\bar\theta$ held fixed for $C$ steps and $\varepsilon$-greedy exploration annealed from one to zero.\footnote{The target network is refreshed every ten thousand steps, with a one-million transition replay buffer, RMSProp at learning rate $0.0001$, batch size three hundred, and a four-layer network of widths $[15,300,200,100]$.}

A single agent bids well in isolation, but the platform runs thousands at once, and each agent's bidding is part of every other's environment, so independent optimization bids up costs and erodes the collective outcome. The massive-agent model gives each advertiser its own $Q$-network but changes what each one maximizes. After all agents bid, agent $i$ receives its own purchase value as a private competitive reward and the aggregate value across all advertisers as a shared cooperative reward, and it learns against the sum,
\begin{equation}
r_i = r_i^{\mathrm{comp}} + r^{\mathrm{coop}},
\end{equation}
so an action that wins volume for one campaign at the expense of the whole is penalized through the cooperative term.\footnote{The source specifies the combination only as adding a private competitive objective to a public cooperative one, without a fixed weighting; the schematic sum above reflects that description.}

Because random bidding cannot be tried in the live auction, training runs in a simulator built from complete auction logs and predicted conversion rates, which can replay any counterfactual bid without having to predict the market price, and it processed two hundred billion simulated auctions in about two hours on a thousand CPUs and forty GPUs. In serving, the auction engine queries the current bidding model and the model periodically refreshes its parameter from the trained $Q$-network. Table~\ref{tab:alibaba_rtb} reports the online A/B tests against the platform's keyword-bidding baseline. On ten advertisers with disjoint keywords the single-agent model raised purchase amount per cost by $35.04\%$, but across a thousand competing advertisers that same model captured only $6.29\%$, the difference being the competition it ignores, while the massive-agent model recovered the gain to $13.01\%$ and the largest return-on-investment improvement. The case is the multi-agent counterpart to the Meta bidder, value-based and online where Meta is conservative and offline, and it makes the competitive externality of a marketplace visible in the numbers.

\begin{table}[t]
\centering
\small
\begin{tabular}{lcccc}
\toprule
\multicolumn{5}{l}{Panel A: single-agent online A/B, 10 disjoint-keyword ads, vs.\ keyword bidding} \\
\midrule
Policy & Purchase/cost & ROI & CVR & PPC \\
\midrule
RMDP & $+35.04\%$ & $+21.38\%$ & $+23.11\%$ & $-5.16\%$ \\
\midrule
\multicolumn{5}{l}{Panel B: multi-agent online A/B, 1000 competing ads (5 Feb 2018), vs.\ keyword bidding} \\
\midrule
Policy & Purchase/cost & ROI & CVR & PPC \\
\midrule
M-RMDP & $+13.01\%$ & $+39.12\%$ & $+12.62\%$ & $-0.74\%$ \\
RMDP & $+6.29\%$ & $+26.51\%$ & $+3.12\%$ & $-3.36\%$ \\
\bottomrule
\end{tabular}
\caption{Alibaba sponsored-search real-time bidding \citep{Zhao2018SponsoredSearchRTB}. Relative improvement over the keyword-bidding baseline in online A/B tests. Purchase/cost is purchase amount per unit cost, the optimized objective; ROI is return on investment, CVR conversion rate, and PPC pay per click, for which lower is better. Panel A: the single-agent robust MDP on ten advertisers with disjoint keywords. Panel B: the same policy and the cooperative massive-agent policy on a thousand competing advertisers, rank-ordered by purchase/cost, showing the single-agent gain shrinking under competition and the cooperative reward restoring it.}
\label{tab:alibaba_rtb}
\end{table}

\subsection{Marketplaces, Pricing, and Revenue Management}

\subsubsection{DiDi dispatch}
Ride-hailing platforms match arriving trip requests to available drivers continuously, at a scale of tens of millions of matches per day \citep{Qin2021dispatch}. A dispatch is not an isolated decision. Sending a driver to a request now changes where drivers are later, and therefore which future requests can be served, so a myopic match that minimizes current pickup distance can leave a region short of supply minutes afterward. The production dispatcher already addresses this by batching, pooling the open requests and idle drivers over a short window of a few seconds and solving one assignment rather than greedily matching each request on arrival. \citet{SadeghiEshkevari2022DiDiScalable} report the deployed system in which reinforcement learning supplies the long-run value that the batched assignment optimizes, the first such online-learning dispatcher run as the primary mode in a major international market.

The control surface is a single term in the assignment objective. Real-time dispatch alternates two stages, updating a state-value function and solving a maximum bipartite matching between orders and drivers, and reinforcement learning owns only the first. The matching stays a linear assignment problem solved by the Hungarian algorithm, and the learned value enters through the edge weights, so the serving path is the same combinatorial optimizer the platform already trusted.

The value is defined over driver spatiotemporal state. A driver's state $s$ is an encoded location, and the system learns a tabular state value $V(s)$, the expected discounted future income of a driver at $s$ to the end of the day. A trip is a transition $s\to s'$ that pays the driver a fare $r_{ss'}$, an idle driver contributes a zero-reward self-transition, and the value follows the temporally extended conventions of Section~\ref{section:rl_algorithms}. The horizon is finite, the driver's operating day, with discounting inside it, and the system learns continually online rather than in distinct training episodes. Because a proposed match may be cancelled by the driver or rider, the value update is taken in expectation over completion, with completion probability $p_c$ estimated by a separate classifier,
\begin{equation}
V(s)\ \leftarrow\ V(s)+\alpha\Big[p_c\big(r_{ss'}+\gamma V(s')\big)+(1-p_c)\,\gamma V(s)-V(s)\Big],
\end{equation}
so a likely cancellation moves the target toward staying in place rather than toward the trip's reward.\footnote{The fare is first smoothed by a per-location running mean $S[\text{grid}]\leftarrow\beta\,S[\text{grid}]+(1-\beta)\,\text{price}$ to separate marketing-driven price variation from the value signal, and the updates use per-location Adam step sizes.}

The learned value enters the matcher through a standardized edge weight. For a candidate order-driver pair that would move a driver from $s$ to $s'$, the weight is
\begin{equation}
E_{ss'}=p_{ss'}\Big[w_{\mathrm{rew}}\,r^{*}_{ss'}+w_{\mathrm{res}}\,\big(\gamma V(s')-V(s)\big)^{*}-w_{p}\,f(ss')^{*}\Big],
\end{equation}
where $p_{ss'}$ is the completion probability, $r^{*}_{ss'}$ the standardized immediate reward, $\big(\gamma V(s')-V(s)\big)^{*}$ the standardized residual value of the move, and $f(ss')^{*}$ a standardized business penalty such as pickup distance. The residual-value term is the only channel through which downstream location value reaches an otherwise myopic assignment. The weights satisfy $w_{\mathrm{res}}=1-w_{\mathrm{rew}}$ and are tuned by Bayesian optimization on an offline simulator, and the matcher assigns drivers to orders by maximizing the total edge weight.\footnote{Each component is standardized to $[0,1]$ through an exponentially weighted moving mean $m$ and variance $v$ by $x^{*}=\mathrm{Sigmoid}((x-m)/\sqrt{v})$, so the three heterogeneous terms combine on a common scale.}

Earlier DiDi implementations estimated the value function with a neural network rather than a table \citep{Qin2021dispatch,Tang2019cvnet}. That value network, CVNet, encoded location with a hierarchical hexagonal coarse coding, was stabilized with a Lipschitz penalty and context randomization, and was distilled into a reduced-feature form so the matcher could query it in real time when a trip's destination features were not yet known. The tabular predecessor was deployed in more than twenty cities. The value network was later A/B tested in several more and reported improvements between about half a percent and two percent across driver income, response rate, and fulfillment rate. The scalable system replaces the offline-trained network with a location-only value updated online, which keeps the query cost of serving low.

Deployment separates a latency-critical path from the learning path. Every two seconds a low-latency module forms the current order-driver graph and solves the Hungarian matching, and every ten seconds a high-throughput module updates the value table, the standardization statistics, and the smoothed rewards for later rounds. A proposed graph-pruning mechanism based on a limited-memory bandit was not deployed at large scale, because the production infrastructure had no simple way to query real-time global performance feedback.

Table~\ref{tab:didi_results} reports the results. In a two-week A/B test across five cities, with the reinforcement-learning dispatcher and the control alternating every three hours, driver income rose by a call-weighted $1.342\%$ with smaller gains in completion and answer rates, and the gains concentrated in the larger cities while the smallest markets were flat or slightly negative. A separate synthetic difference-in-differences analysis against a pool of control cities estimated a larger effect on the platform's primary metric after full deployment. The incumbent matcher was already strong, so the reliably measured gains are on the order of a percent, and the case shows reinforcement learning valuing the downstream consequences of a decision while a conventional optimizer keeps making it.

\begin{table}[t]
\centering
\small
\begin{tabular}{lc}
\toprule
\multicolumn{2}{l}{Panel A: two-week A/B test across five cities, call-weighted vs.\ control} \\
\midrule
Metric & Change \\
\midrule
Driver income & $+1.342\%$ \\
Completion rate & $+0.422\%$ \\
Answer rate & $+0.296\%$ \\
Primary metric, full deployment (synthetic diff-in-diff) & $+5.3\%$ \\
\midrule
\multicolumn{2}{l}{Panel B: per-city driver income in the A/B test (matching calls)} \\
\midrule
City & Driver income \\
\midrule
City I & $+1.977\%$\ \ ($4.46$M) \\
City II & $+1.255\%$\ \ ($6.08$M) \\
City IV & $+0.279\%$\ \ ($1.04$M) \\
City VI & $+0.145\%$\ \ ($0.19$M) \\
City V & $-0.739\%$\ \ ($0.47$M) \\
\bottomrule
\end{tabular}
\caption{DiDi dispatch \citep{SadeghiEshkevari2022DiDiScalable}. Panel A: call-weighted average change against control in a two-week A/B test across five cities with three-hour rotation, and the full-deployment effect on the platform's anonymized primary metric from a synthetic difference-in-differences against a pool of control cities. Panel B: per-city driver-income change in the same A/B test, with the number of matching calls, rank-ordered by income.}
\label{tab:didi_results}
\end{table}

\subsubsection{Tmall dynamic pricing}
\label{sec:ecommerce_pricing}
\citet{Liu2019} run what they describe as the first deep reinforcement learning system for e-commerce dynamic pricing at scale, setting the real prices of thousands of Tmall products in field experiments that began in July 2018 and ran for months. Two features of the setting shape everything else. The agent spends real money on real inventory, so a cold start that prices badly is a capital loss rather than a bad simulation, and showing different customers different prices at the same time is not legally available, so the usual A/B test is off the table. The system answers the first with demonstration pretraining and the second with a difference-in-differences evaluation on matched products. It is the pricing case in this chapter, and the one whose evaluation design is dictated by law rather than by convenience.

Each product is priced by its own MDP at a daily step. The state $s_{i,t}\in\mathbb{R}^{m}$ collects four groups of features for product $i$ on day $t$, its price, its recent sales, its customer traffic, and its competitiveness against similar products. Markdown pricing of a fixed stock is episodic, the episode ending when the product sells out, whereas daily pricing of a continuously replenished good is a continuing task with no terminal state, and both discount the future at $\gamma=0.99$. The action is the price, and the paper studies two parameterizations, a discrete one that partitions the historical price range into $K$ bins and a continuous one that emits an exact price. The reward is not raw revenue, which is dominated by traffic swings, but the revenue per unique visitor,
\begin{equation}
r_{i,t}=\mathrm{revenue}_{i,t}\,/\,\mathrm{uv}_{i,t},
\end{equation}
the \emph{revenue conversion rate}, or the analogous profit conversion rate when unit cost is known.\footnote{For fast-moving consumer goods the revenue conversion rate is too unstable for a level reward to converge, the price-to-conversion correlation is $0.15$ against $-0.57$ for luxuries, so the reward there is the difference of revenue conversion rates over a window $\tau$ rather than its level, which rewards a relative improvement and restores convergence.}

The discrete-action problem is solved with a deep $Q$-network and the continuous one with deep deterministic policy gradient, both standard from Section~\ref{section:rl_algorithms} and both using experience replay and target networks. The distinctive step is the cold start. No simulator of the market exists, and learning online from scratch would lose money, so each agent is pretrained on demonstrations, the logged price-state-reward transitions of the specialists and rules that priced the products before, through deep $Q$-learning from demonstrations for the discrete agent and its deterministic-policy-gradient analogue for the continuous one. Offline evaluation then scores a candidate policy on held-out logged transitions, counting a logged reward only when the logged price falls within a tolerance of the policy's own price, before any live pricing begins.

Online, because prices cannot be randomized across customers, the effect is estimated by difference-in-differences against \emph{similar products}, matched on brand, category, and selling behavior. The comparison statistic is the difference of revenue conversion rates at a one-year window, which nets out seasonality and annual promotions and reads as year-on-year conversion growth.\footnote{Two groups of matched similar products priced by the same policy track each other closely on this statistic, an average of $1.00$ against $0.99$ after rescaling, which is what licenses reading a gap between the treated and control groups as a policy effect rather than a market movement. The statistic is used only for evaluation, not as the training reward.} Table~\ref{tab:tmall_results} reports the two field experiments. In a markdown season for five hundred luxury items the learned policy held profit conversion positive while the manual benchmark, cutting prices below cost to move stock, turned negative. In daily pricing of fast-moving goods against a manually priced control, the continuous policy scored highest on the difference-in-differences index, the discrete policy next, and both well above the control. The case shows a marketplace's legal and financial constraints, not just its dynamics, dictating how reinforcement learning must be trained and measured.

\begin{table}[t]
\centering
\small
\begin{tabular}{lcc}
\toprule
\multicolumn{3}{l}{Panel A: markdown pricing, 500 luxury SKUs (DQN) vs.\ 2000 manual, days 16--30} \\
\midrule
Conversion measure & Learned (DQN) & Manual \\
\midrule
Revenue conversion rate & $0.22$ & $0.16$ \\
Profit conversion rate & $0.16$ & $-0.04$ \\
\midrule
\multicolumn{3}{l}{Panel B: daily FMCG pricing, difference-in-differences index (DRCR)} \\
\midrule
Policy & DRCR index & Control \\
\midrule
DDPG (continuous) & $6.07$ & $1.00$ \\
DQN (discrete) & $5.03$ & $1.00$ \\
\bottomrule
\end{tabular}
\caption{Tmall dynamic pricing \citep{Liu2019}. Panel A: rescaled revenue and profit conversion rate (revenue or profit per unique visitor) over the markdown fortnight, learned DQN policy on 500 luxury SKUs against 2000 manually priced similar products; the manual group's profit conversion turned negative. Panel B: the difference-in-differences index (difference of revenue conversion rates at a one-year window) for daily pricing of fast-moving goods, continuous and discrete policies against a manually priced control rescaled to $1.00$, rank-ordered by the index. An earlier run of the discrete policy alone against the control scored $5.10$ on the same index.}
\label{tab:tmall_results}
\end{table}

\subsubsection{Hotel revenue management}
\label{sec:hotel}
\citet{Chen2023hotelrl} place reinforcement learning inside the revenue management of a budget hotel chain, China Lodging Group, which operated some two thousand hotels, most under its Hanting brand. A hotel sells a fixed stock of rooms for a given night through several rate segments, direct booking, membership, and online travel agencies among them, each carrying its own discount, and it must decide how much capacity to release to each segment as the booking horizon runs down. The classical approach estimates a demand and cancellation model and optimizes against it, which the firm found impractical, both because that process is hard to estimate under censoring and non-stationarity and because a hotel manager will not act on an allocation they cannot explain. The deployed system keeps a conventional optimizer in charge of the allocation and uses reinforcement learning to set one quantity the manager already reasons about, the \emph{average discount}. It is the operations and revenue-management case in this chapter, and the one held to the standard of a controlled field experiment.

The problem is a finite-horizon episodic MDP over the booking horizon of a single day of stay. An episode has $T$ periods, chosen so that each period carries about a tenth of the reservations, and the state is a pair $(t,s)$ where $t$ indexes the period and $s$ is the revenue per room sold so far, the average daily room rate up to period $t$.\footnote{The booking horizon is ninety days, divided into $T=10$ periods that shorten toward the stay date because roughly seventy percent of reservations arrive within three days of the stay and about thirty percent are same-day. The revenue-per-room state is scaled to the base rate and discretized to $\{0,10\%,\dots,100\%\}$, and the action to $\{10\%,20\%,30\%\}$, which lets the recommended policy be drawn as a table managers can read rather than held in a function approximator.} Given an allocation $x=(x_1,\dots,x_K)$ across the $K$ rate segments with per-segment discounts $\delta_k$, the action is the reservation-weighted mean discount $a=\sum_{k=1}^{K}\delta_k x_k/\sum_{l=1}^{K}x_l$. The reward $R(t,s,a)$ is the reservation payment received in period $t$, net of cancellations, and because $s$ tracks realized room rate the firm never has to model cancellations separately. The agent maximizes the undiscounted total reward $\mathbb{E}_\pi\big[\sum_{t=1}^{T} R_t\big]$ over the horizon, in the temporally extended sense of Section~\ref{section:rl_algorithms}.

Because the firm cannot build a trustworthy simulator of its own demand, the policy is learned model-free and on-policy from realized episodes by Monte Carlo. The action value is the expected total reward from taking $a$ in $(t,s)$ and then following $\pi$,
\begin{equation}
Q^\pi(t,s,a)=\mathbb{E}_\pi\Big[\textstyle\sum_{i=t}^{T} R_i \,\Big|\, (t,s),\,a\Big],
\end{equation}
estimated with no bootstrapping, as the running average of the realized returns $\sum_{i=t}^{T} R_i$ over past episodes that visited $(t,s,a)$. At the start of each day the just-completed episode for the previous day of stay updates $Q$, and the policy is improved to be $\varepsilon$-greedy in the updated value,
\begin{equation}
\pi(a\mid t,s)=
\begin{cases}
1-\varepsilon+\varepsilon/|\mathcal{A}| & a=\arg\max_{a'} Q(t,s,a'),\\[2pt]
\varepsilon/|\mathcal{A}| & \text{otherwise,}
\end{cases}
\end{equation}
with a small $\varepsilon$ and the random action capped one discount step from the current best.\footnote{$\varepsilon=0.1$. On the first day the value is initialized from policy evaluations over the previous three months of reservations. The policy applied to a given day of stay is reassembled from the daily updates, so it tracks non-stationary demand rather than being fixed ninety days ahead, at the cost of the convergence guarantees a stationary environment would provide.}

The learned discount is only a target. A linear program turns the recommended $\pi(t,s)$ into an integer allocation $x_t=(x_{1,t},\dots,x_{K,t})$ across the rate segments, maximizing the hotel's segment preference subject to selling its remaining capacity at no more than the recommended average discount,
\begin{equation}
\max_{x_t\ge 0}\ \sum_{k=1}^{K} w_k\,x_{k,t}
\qquad\text{s.t.}\qquad
\sum_{k=1}^{K} x_{k,t}=C_t,\quad
\sum_{k=1}^{K}\delta_k x_{k,t}\le \pi(t,s)\,C_t,
\end{equation}
where $C_t$ is the capacity left at period $t$ and the weights $w_k\ge 0$ encode whether the hotel favors, say, loyalty-building direct bookings or the wider reach of travel agencies.\footnote{There are $K=5$ segments and the allocation is served through nested protection levels. With five decision variables and two constraints the discount constraint is typically slack, the realized weighted discount stays within one percent of the recommendation. The weights are chosen from whether a hotel's demand is primarily business, leisure, or mixed.} This split is what makes the policy interpretable, the reinforcement learner reasons over a single scalar the manager understands, and the multivariate allocation is recovered by an optimizer the firm already trusts.

The system ran as a module on the firm's enterprise software in a pilot of five Hanting hotels in Shanghai from March to June 2015, the treatment hotels drawn at random from ten candidates. A conventional A/B test was not available, both because charging different guests different prices at the same time was not permissible and because five heterogeneous hotels are too few for a plain difference-in-differences, so the effect was estimated by the \emph{synthetic control} method, each treatment hotel matched to a weighted composite drawn from 271 other Shanghai hotels. Table~\ref{tab:hotel_results} reports the result. Averaged over the five hotels, revenue per available room rose by $11.80\%$, significant at the one percent level, arising through occupancy for some hotels and room rate for others, and the gains grew over the course of the pilot. The chain adopted the system after the trial. The case shows reinforcement learning entering an operations setting not by replacing the decision but by supplying one interpretable input to a classical optimizer, and being measured against the evidentiary standard of a field experiment rather than an offline benchmark.

\begin{table}[t]
\centering
\small
\begin{tabular}{lccc}
\toprule
\multicolumn{4}{l}{Panel A: mean treatment effect across the five pilot hotels (synthetic control)} \\
\midrule
Measure & Change & $p$-value \\
\midrule
Revenue per available room & $+11.80\%$ & $.0090$ \\
Average daily room rate & $+5.93\%$ & $.1160$ \\
Occupancy rate & $+5.15\%$ & $.1361$ \\
\midrule
\multicolumn{4}{l}{Panel B: per-hotel decomposition, rank-ordered by RevPAR (after-pilot period)} \\
\midrule
Hotel & RevPAR & ADR & Occupancy \\
\midrule
Hotel 5 (mixed demand) & $+20.81\%$ & $+7.47\%$ & $+10.37\%$ \\
Hotel 4 (business demand) & $+13.66\%$ & $+9.65\%$ & $+3.44\%$ \\
\bottomrule
\end{tabular}
\caption{Hotel revenue management at China Lodging Group \citep{Chen2023hotelrl}. Panel A: mean percentage change in each measure against the synthetic control across the five pilot hotels, with permutation $p$-values; only the RevPAR effect is significant at the one percent level. Panel B: the two hotels with a fully reported decomposition, rank-ordered by RevPAR, showing the same order of headline gain reached through occupancy for one hotel and through room rate for the other. Pilot: five Hanting hotels in Shanghai, March to June 2015.}
\label{tab:hotel_results}
\end{table}

\subsection{Inventory and Operational Control}
\label{sec:inventory}

\subsubsection{Inventory benchmark}
The pre-DeepStock inventory evidence is limited to benchmark performance. \citet{Gijsbrechts2022inventory} find that deep RL can match strong heuristics on some classic inventory problems only after expensive tuning and is often dominated by specialized policies. The benchmark problems themselves are infinite-horizon periodic-review formulations, lost sales, dual sourcing, and multi-echelon, with a discounted objective evaluated by long-run average cost per period.

\subsubsection{DeepStock}
\citet{Xie2026DeepStock} report a qualitatively different case from the inventory benchmark, a full-scale deployment of deep reinforcement learning for inventory replenishment on Alibaba's Tmall platform. Off-the-shelf deep reinforcement learning is highly sensitive to hyperparameters and can produce orders that are hard to interpret. DeepStock instead regularizes the learned policy with the classical base-stock structure of inventory theory. This regularization accelerates tuning, keeps the policy interpretable, and enabled the company to run one unified policy across its whole catalog.

The problem is a discrete-time inventory MDP. Days run $t=1,\dots,T$, orders are placed every $P$ days and arrive after a lead time of $L$ days, so the horizon is finite, each trained policy runs one $T$-day episode until the next retraining, and the discount factor is tuned just below one. The state has two parts. An endogenous inventory vector $I_t=(I_t^{0},I_t^{1},\dots,I_t^{L-1})$ records the on-hand stock $I_t^{0}$ and the stock $I_t^{\ell}$ already ordered and arriving $\ell$ days ahead, which depends on past actions. An exogenous feature vector $x_t\in\mathbb{R}^{m}$ carries static attributes such as product category, demand magnitude, supplier, lead time, and margin, together with dynamic ones such as upcoming promotions and seasonality. The action is a scalar order quantity $\pi(I_t,x_t)\ge 0$. The firm does not reward raw profit but a weighted combination of two operational losses, a stockout rate, the percentage of days that end out of stock, and a turnover time, the average end-of-day inventory divided by average sales, that is, the number of days a unit sits on hand. The two pull in opposite directions, and the deployed objective weights them so that a one-point reduction in stockout rate is worth about a two-day reduction in turnover time.

The regularization is the technical core. Rather than let the network emit an order directly, the order is defined as a learned base-stock target minus the inventory already in the pipeline, floored at zero,
\begin{equation}
\pi(I_t,x_t)=\max\big\{\mu_{\mathrm{Base}}(I_t,x_t)-\mathrm{tot}(I_t),\,0\big\},
\end{equation}
where $\mu_{\mathrm{Base}}$ is the neural network, now interpreted as the target level for total inventory, and $\mathrm{tot}(I_t)=\sum_{\ell}I_t^{\ell}$ is the total stock on hand and inbound. A second regularization writes the order as $\mu_{\mathrm{Coeff}}(I_t,x_t)^{\top}\mathrm{feat}(x_t)$ with a small feature vector $\mathrm{feat}(x_t)$, at Alibaba four demand features and a bias, so that larger forecast demand implies larger orders, and the two can be combined. Because the network can in principle undo these mappings, they do not restrict the policy class. They are an action remapping that biases learning toward sensible inventory behavior, which is what speeds convergence of the value function.

The regularization is agnostic to the training algorithm, and the paper tests deep deterministic policy gradient, proximal policy optimization, and a differentiable simulator. Its effect is largest exactly where deployment needs it, in the tuning budget. Without regularization, deep deterministic policy gradient needed seventeen hyperparameter configurations to surpass the differentiable simulator, and proximal policy optimization needed thirty-two; with the base-stock regularization those fell to eleven and seven. Alibaba deployed the deterministic policy gradient with both regularizations, which was the best performer in its offline tests and beat the differentiable simulator on fifty-five thousand ninety-day trajectories of the firm's own data. A single meta-learned policy now covers every stock-keeping unit, replacing an earlier system that grouped similar products and trained a separate model per group, and it is retrained roughly every six weeks on the most recent ninety days.

Table~\ref{tab:deepstock_results} reports the deployment. As of October 2025 the policy manages replenishment for all of Alibaba's Tmall products, over one hundred thousand stock-keeping units held across about twenty primary warehouses, so more than one million SKU-warehouse pairs. Three field measurements track its effect against the incumbent. A July 2024 difference-in-differences pilot on a cherry-picked tenth of international products cut stockout rate and turnover time together, and the April 2025 full rollout, evaluated against a counterfactual simulation of the incumbent, reduced turnover time by one to two days at no cost to stockout rate. A year-over-year comparison in the summer of 2025 shows a larger turnover reduction, and the firm estimates the associated inventory reduction is worth hundreds of millions of renminbi in freed capital.

The pattern is a Pareto improvement against an already-tuned incumbent, less inventory held for the same service level, achieved by a policy with a base-stock parameterization that operators can read. Because the source is a 2026 preprint that omits implementation details for confidentiality, it should be read as primary-source industrial evidence rather than independently audited accounting, but it is the clearest public case of a learned policy running an entire replenishment operation.

\begin{table}[t]
\centering
\small
\begin{tabular}{lcc}
\toprule
\multicolumn{3}{l}{Panel A: change against the incumbent policy} \\
\midrule
Measurement & Turnover time & Stockout rate \\
\midrule
July 2024 DiD pilot, 10\% intl.\ SKUs & $-9.53$ days & $-0.83$ points \\
April 2025 rollout, international SKUs & $-1$ day & no change \\
April 2025 rollout, domestic SKUs & $-2$ days & no change \\
July--Aug 2025 vs.\ 2024, intl.\ SKUs & $-20\%$ & no notable change \\
\midrule
\multicolumn{3}{l}{Panel B: scale and financial impact, October 2025} \\
\midrule
Coverage & \multicolumn{2}{c}{100\% of products, $>\!1{,}000{,}000$ SKU-warehouse pairs} \\
Estimated inventory reduction & \multicolumn{2}{c}{$\approx 350$ million RMB} \\
Annual cost-of-capital saving & \multicolumn{2}{c}{$\approx 13.3$ million RMB} \\
\bottomrule
\end{tabular}
\caption{DeepStock at Alibaba Tmall \citep{Xie2026DeepStock}. Panel A: change in the two operational metrics against the incumbent, from a July 2024 difference-in-differences pilot on cherry-picked international SKUs, an April 2025 counterfactual comparison at full rollout, and a July--August 2025 year-over-year comparison. Turnover time is in days (a percentage for the year-over-year row); stockout rate in percentage points of days out of stock. Panel B: deployment scale and estimated financial impact.}
\label{tab:deepstock_results}
\end{table}

\subsection{Physical and Design Systems}

\subsubsection{MuZero-RC video encoding}
\citet{Mandhane2022MuZeroRC} bring MuZero to the rate control of the VP9 video codec, the component that decides how many bits to spend on each frame. It is the model-based case in this chapter, the one system that learns a model of its environment and plans with it rather than reacting from a learned value or policy alone, and it handles a hard bitrate constraint in a way that contrasts with the pessimistic value ensemble of the building-cooling controller later in this section. The learned controller replaces a narrow codec decision inside an existing encoder, and the accompanying DeepMind and YouTube post \citep{DeepMind2022MuZeroRCPost} reports it running on a portion of YouTube live traffic.

Rate control is posed as a constrained MDP $\langle\mathcal{S},\mathcal{A},P,R_{\mathrm{PSNR}},\gamma=1,c_{\mathrm{Bitrate}},\beta_{\mathrm{Target}}\rangle$. The state is the encoder's first-pass statistics together with information about the frames encoded so far and the frame to come, the action is an integer quantization parameter $\mathrm{QP}\in[0,255]$ applied to the current frame, and the transition advances to the next frame. The reward is the quality of the finished encode, measured as peak signal-to-noise ratio, paid only at the final frame and zero before it, and the constraint is that the encode's bitrate stay under a user-specified target $\beta_{\mathrm{Target}}$. The agent maximizes terminal quality subject to that budget, over the finite undiscounted episode of one video.

The obvious way to enforce the budget is a Lagrangian relaxation, but a single multiplier is not optimal across videos and target bitrates, and the authors found that combining MuZero with a Lagrangian was both hard to tune and poor at evaluation. Their alternative, \emph{self-competition}, converts the constrained objective into a win-or-lose signal by having the agent race its own history. For each video and target-bitrate pair the agent keeps an exponential moving average of its past quality and its past overshoot, the amount by which it exceeded the budget, and after an episode it sets the entire return to
\begin{equation}
G=\begin{cases}
+1 & \text{the episode beats its baseline, comparing overshoot first and then quality,}\\
-1 & \text{otherwise,}
\end{cases}
\end{equation}
then updates the baseline with the new episode.\footnote{The exact rule composes sign and indicator functions over the two comparisons. Because overshoot is compared before quality, the agent learns to satisfy the constraint first and to improve quality second, and because the baseline is a moving average of the agent's own results it rises as the agent improves, so the target keeps receding. Self-competition is equivalent to a Lagrangian relaxation with multiplier one on a CMDP whose state is augmented with the moving averages, which is what removes the per-video tuning.}

The agent is MuZero, with a representation network that embeds the encoder features, a dynamics network that predicts the next embedding from a candidate QP, and a prediction network that outputs a policy and value. During training it plans, running two hundred simulations of Monte Carlo tree search over the learned dynamics to choose each QP, which is the model-based step no other deployed system in this chapter takes. At evaluation the search is dropped for speed and the highest-probability QP from the policy network is used directly.\footnote{An augmented variant subtracts $0.005\times\text{overshoot}$ from the quality term of the self-competition comparison, roughly the average slope of the codec's rate-distortion curve, so the agent gives up a little quality to save bits. Training used twenty thousand 480p videos, and evaluation used 3062 five-second clips from the YouTube user-generated-content dataset at nine target bitrates.} Table~\ref{tab:muzero_results} reports the evaluation against libvpx's default two-pass controller. The augmented agent encodes at the same quality using $6.28\%$ fewer bits, and both agents overshoot the target far less often than the baseline while landing within five percent of it more often. In production the post \citep{DeepMind2022MuZeroRCPost} reports an average $4\%$ bitrate reduction on a portion of YouTube live traffic, a separate and smaller figure than the offline $6.28\%$ that should not be conflated with it. The case shows model-based planning improving a real codec, with a hard constraint met by self-competition rather than a hand-tuned penalty.

\begin{table}[t]
\centering
\small
\begin{tabular}{lccc}
\toprule
\multicolumn{4}{l}{Panel A: coding efficiency, BD-rate vs.\ libvpx (more negative is better)} \\
\midrule
Agent & PSNR & SSIM & VMAF \\
\midrule
Augmented MuZero-RC & $-6.28\%$ & $-5.11\%$ & $-1.88\%$ \\
MuZero-RC & $-4.72\%$ & $-3.68\%$ & $-0.53\%$ \\
\midrule
\multicolumn{4}{l}{Panel B: bitrate constraint satisfaction, fraction of evaluation videos} \\
\midrule
Agent & overshoot ${>}0$ & overshoot ${>}5\%$ & within $5\%$ of target \\
\midrule
libvpx (baseline) & $64.00\%$ & $6.13\%$ & $71.34\%$ \\
MuZero-RC & $20.34\%$ & $2.04\%$ & $84.14\%$ \\
Augmented MuZero-RC & $16.10\%$ & $2.25\%$ & $70.12\%$ \\
\bottomrule
\end{tabular}
\caption{MuZero-RC on VP9 rate control \citep{Mandhane2022MuZeroRC}. Panel A: Bjontegaard-delta rate against the libvpx baseline for three quality measures, the average bitrate change at matched quality, so a negative value is a saving, rank-ordered by the PSNR column. Panel B: the fraction of the 3062 evaluation videos on which each policy overshoots the target at all, overshoots by more than five percent, and lands within five percent of the target. Figures are means over five seeds; the standard error is at most $0.33$ points in Panel A and $2.2$ points in Panel B.}
\label{tab:muzero_results}
\end{table}

\subsubsection{AlphaChip floorplanning}
\citet{Mirhoseini2021GraphPlacement} and \citet{GoldieMirhoseini2024AlphaChipPost} establish a different kind of field evidence. AlphaChip formulates chip floorplanning as a finite-horizon sequential placement problem. The initial state is an empty layout. At each step, the agent places one component, and the episode ends with a single reward based on final layout quality. The Nature paper states that the method was used for Google's next-generation AI accelerators, and the later DeepMind post reports use across TPU generations and other chips. The RL policy produces a shipped design artifact that enters a conventional chip-design and manufacturing pipeline, not a live runtime controller, so it is not directly comparable with the YouTube, DiDi, or bidding systems.

\subsubsection{BCOOLER HVAC}
\citet{Luo2022BCOOLER} report live reinforcement-learning experiments on the cooling systems of two commercial buildings, run with the building-management provider Trane Technologies. Cooling is a large and climate-relevant load; space cooling alone is about a tenth of world electricity demand, and a chiller plant is a natural control problem with a clear objective, energy, and hard comfort and safety constraints. The work builds on earlier reinforcement learning for Google data-center cooling, but a commercial building is harder because demand shifts with occupancy and the plant has many chillers to coordinate. Reinforcement learning also fits the domain for a practical reason. Unlike model predictive control, it does not need a validated physics model built and maintained for each building.

The control surface is supervisory. A chiller plant is normally run by a rule-based Sequence of Operations that decides when to switch equipment on and off and what setpoints to hand its closed-loop controllers. The reinforcement-learning agent does not replace those controllers; it recommends the setpoints they track, and it hands control back to the existing Sequence of Operations when no action it can take is safe. The learner tunes the targets of a trusted controller rather than actuating the plant directly.

The problem is a finite-horizon MDP, undiscounted by design because the authors found a finite horizon easier for facility managers to reason about than a discount factor. The state is roughly fifty plant, weather, and load observations. The action $a\in A$ is a twelve-dimensional vector of real-valued setpoints together with discrete on-off switches for equipment. The reward is the negative of the energy the chiller plant consumes over a five-minute step, so maximizing return minimizes energy. Comfort and safety enter as two kinds of constraint, action constraints the policy can guarantee by construction, and observation constraints on future sensor readings that can only be predicted, each carrying its own horizon.

Because exploring a live building is expensive and unsafe, the agent is trained offline. BCOOLER, the method, learns an ensemble of value networks from logged operation, and the value function is multi-output, one head predicts energy and a separate head predicts each constrained observation.\footnote{The action-value model has $62$ inputs ($50$ observations and $12$ actions) and $25$ outputs (energy plus $24$ observation constraints), is a multi-tower network with per-head feature selection, and is retrained daily on all data collected so far, about ten thousand examples per month. Targets are finite-horizon Monte-Carlo rather than temporal-difference, which the authors found easier for facility managers to reason about.} Control is then a constrained search. At each step the agent samples a large set of actions that satisfy the action constraints, and keeps only those whose predicted constraint values, inflated by an ensemble-uncertainty margin, stay within bound,
\begin{equation}
A_t^{\mathrm{safe}}=\big\{\,a\in A_t:\ \mu^{c}(s_t,a)+\alpha\,\sigma^{c}(s_t,a)\le u\,\big\},
\end{equation}
where $\mu^{c}$ and $\sigma^{c}$ are the ensemble mean and standard deviation of the constrained quantity and $u$ its limit. Among the survivors it picks the action with the lowest predicted energy, using the same uncertainty for occasional exploration. The pessimistic margin is the guardrail, so if the prediction model is right the observation constraints hold, and the fallback to the incumbent covers the case where no safe action remains.

The two systems were evaluated by live A/B tests, one on a university campus chiller plant and one on a mixed-use building housing a shopping center, restaurants, apartments, and a clinic. Control alternated daily between BCOOLER and the incumbent so both met similar weather, the two hours after each handover were discarded, and performance was compared within matched buckets of outside temperature and building load. Each test ran about three months in the shoulder season.

Table~\ref{tab:bcooler_results} reports the results. BCOOLER saved $9\%$ and $13\%$ of chiller-plant energy at the two sites against Trane's heuristic controllers, while satisfying the comfort constraints at a rate comparable to the incumbent. The savings were larger in cooler weather and at lower load, where equipment runs below capacity and there is more room to optimize, and the incumbent controllers had themselves been tuned during the retrofit, so the gain over an untuned plant would likely be larger. The evidence is two live field experiments rather than a broad durable rollout, and it is the case this chapter counts for reinforcement learning in commercial cooling. It does not count Google data-center cooling as confirmed reinforcement learning unless a public primary source establishes an RL controller rather than predictive control.

\begin{table}[t]
\centering
\small
\begin{tabular}{lc}
\toprule
\multicolumn{2}{l}{Panel A: live A/B energy savings vs.\ the Trane heuristic controller} \\
\midrule
Facility & Energy savings \\
\midrule
Mixed-use commercial building & $13\%$ \\
University campus chiller plant & $9\%$ \\
\midrule
\multicolumn{2}{l}{Panel B: experiment and model} \\
\midrule
A/B design & daily alternation, weather-bucketed \\
Duration per site & $\approx 3$ months \\
Control timestep & $5$ minutes \\
Action-value model & $62$ inputs, $25$ outputs \\
Retraining & daily \\
\bottomrule
\end{tabular}
\caption{BCOOLER commercial-building cooling \citep{Luo2022BCOOLER}. Panel A: energy savings against Trane's heuristic controller in a live A/B test at the two facilities. The paper reports the savings as $9\%$ and $13\%$ respectively across the two sites, mapping to the university and mixed-use buildings in the order they are introduced. Panel B: the evaluation design and the action-value model. Both A/B tests ran about three months in the shoulder season, and the incumbent controllers had been tuned during the retrofit.}
\label{tab:bcooler_results}
\end{table}

\subsection{RLHF and Post-Training}

\subsubsection{InstructGPT}
\citet{ouyang2022training} is a field case only if the category is defined carefully. During post-training, demonstrations are used to fit a supervised policy and human comparisons are used to fit a reward model. PPO then optimizes the language-model policy against the fitted reward model with a KL penalty. The environment is a single-step bandit. A random customer prompt arrives, the policy generates one response capped at a thousand tokens, the reward model scores it, and the episode ends, with a per-token KL penalty and no discounting inside the response. The prompt distribution comes from OpenAI API usage and held-out API prompts are central to evaluation.

The paper reports large human-preference gains over GPT-3 baselines on the API prompt distribution. OpenAI's own release notes corroborate the deployment surface. The InstructGPT RLHF models were placed as the default language models on the API \citep{OpenAI2022InstructGPTPost}, and the November 2022 ChatGPT research release applied the same recipe of supervised fine-tuning, a comparison-trained reward model, and PPO \citep{OpenAI2022ChatGPTPost}. In both cases the reinforcement-learning step shaped the policy before release and later user-facing behavior was ordinary model inference, so the evidence supports training-time RL at production scale rather than a runtime control loop.

\subsection{Contrast Cases}
\label{sec:execution}

Financial order execution provides evidence from an economically substantive objective and detailed market data. \citet{Nevmyvaka2006execution} train and evaluate RL execution policies on 1.5 years of millisecond NASDAQ limit-order-book data, which are substantially richer than data from a simplified simulator. Evaluation nevertheless uses a historical/generative methodology rather than a live production trading deployment. The finite-horizon episode requires selling a block of shares within a window of a few minutes split into a handful of decision points, and any remainder is liquidated at the horizon. \citet{Wu2018rtb} study a budget-constrained real-time-bidding simulator that casts display-ad bidding as an MDP, trains a model-free policy, and reports click gains on large real datasets without live production traffic. Its one-day episode has fifteen-minute steps, no discounting, and termination when the budget is spent. The inventory study of \citet{Gijsbrechts2022inventory} is also a benchmark. These cases inform the problem structure but do not constitute confirmed field deployments.

\subsection{Simulation Study: The Limits of Offline Policy Selection in a Dynamic Market}
\label{sec:field_ope_sim}
\newcommand{\fieldopecovgapab}{75}
\newcommand{\fieldopecovgapinc}{90}
\newcommand{\fieldopecovgapmix}{0}
\newcommand{\fieldopedmab}{-0.04}
\newcommand{\fieldopedminc}{-0.33}
\newcommand{\fieldopedmmix}{0.70}
\newcommand{\fieldopedmdiffabmix}{0.74}
\newcommand{\fieldopedmdiffabmixse}{0.09}
\newcommand{\fieldopedrab}{0.05}
\newcommand{\fieldopedrinc}{-0.17}
\newcommand{\fieldopedrmix}{0.42}
\newcommand{\fieldoperegretab}{0.21}
\newcommand{\fieldoperegretmix}{0.04}
\newcommand{\fieldopeseeds}{10}
\newcommand{\fieldopehorizon}{40}
\newcommand{\fieldopefqesteps}{50000}
\newcommand{\fieldopedmerrbestab}{-0.23}
\newcommand{\fieldopedmerrbestmix}{-0.03}
\newcommand{\fieldopeessbest}{1.5}
\newcommand{\fieldopeessalignedmix}{57}

In a simulated pricing environment with known ground truth, the experiment measures how accurately offline estimators rank and select candidate policies.

A firm faces a stream of arriving consumers and sets a targeted discount for each on a fixed base price. The state at step $t$ is $s_t=(c_t,r_t,t)$, where $c_t\in\mathbb{R}^{K}$ is a consumer context drawn fresh each step, $r_t$ is a market-level reference discount, and $t$ indexes the horizon. The action $a_t$ selects a discount depth $d_t$ from six levels between zero and a quarter off. A consumer buys with a logit probability, and the discount pays a margin only on a purchase,
\begin{equation}
\mathbb{P}(\text{buy}\mid c,r,d)=\sigma\!\Big(\delta-\alpha(c)\,p_0(1-d)+\beta_r\,\rho(d-r)\Big),\qquad R_t=\mathbb{1}\{\text{buy}_t\}\,\big(p_0(1-d_t)-c_0\big),
\end{equation}
where $\sigma$ is the logistic function and $p_0$ the base price. Two features make the problem hard. The price sensitivity $\alpha(c)=\alpha_0\exp\!\big(\kappa\,c^{\top}\theta/\sqrt{n}\big)$ is driven by a sparse index $\theta$ that loads on only $n=2$ of the $K$ context coordinates, so the optimal policy is high-dimensional and no tabular dynamic program applies. And the reference discount evolves as a running mean of the discounts the firm has offered,
\begin{equation}
r_{t+1}=\frac{t\,r_t+d_t}{t+1},
\end{equation}
which enters demand through the loss-averse reference term $\rho(d-r)$, so over-promoting raises the reference and depresses future demand at any given discount.\footnote{The reference term is $\rho(x)=x$ for $x\ge0$ and $\rho(x)=\lambda x$ for $x<0$ with $\lambda=1.5$, an averaging reference; constants are $\alpha_0=1.5$, $\kappa=1.5$, $\beta_r=2.5$, $\delta=1$, $p_0=1$, $c_0=0.30$, so full-price margin is $0.70$. Horizon \fieldopehorizon{} consumers, discount factor $0.95$, $K=8$ context dimensions.} Myopic promotion that maximizes today's margin is therefore strictly suboptimal, and the dynamics are what make this a sequential problem rather than a contextual bandit.

The analyst must rank a menu of six deployment policies from logged data alone, spanning full-price restraint, two fixed discounts, the myopic incumbent that maximizes immediate margin, the uniform-random policy, and a policy trained offline by conservative Q-learning. Their true field values, estimated by Monte Carlo in the simulator, are the first column of Table~\ref{tab:field_ope_candidates}, and full-price restraint is the best. Three logging regimes generate the data, each of the same size. A randomized A/B log assigns the six discounts uniformly within every episode, giving full action support. An incumbent log follows the myopic rule with exploration rate $0.3$ and exactly recorded propensities, so its support is narrow but nothing about it is unobserved. A historical-mixture log serves each episode with one of seven past pricing policies, the six fixed discounts and the myopic rule, each softened with the same exploration rate. The mixture is the log a platform accumulates as its pricing policy changes over time, and because different past policies settled the reference discount at different levels, it is the only log whose state occupancy sweeps the reference axis. Three off-policy estimators are scored over \fieldopeseeds{} seeds, the direct method built on fitted-Q evaluation (Section~\ref{section:offline_rl}), per-decision importance sampling, and their doubly robust combination (Section~\ref{section:causal_rl}), and reliability is measured by regret@1, the true value gap of the policy each estimator selects, and by the Spearman rank correlation between estimated and true value.\footnote{Logs of $500$ trajectories. Reference candidates are softened with exploration rate $0.3$ so importance weights are not degenerate by construction, and the ground-truth values target the softened candidates themselves. Fitted-Q evaluation uses a two-layer $256$-unit network trained for \fieldopefqesteps{} gradient steps, a budget at which the estimates are stable under a doubling of steps and pass a calibration gate requiring the direct method's error on each log's own behavior policy to average within $0.05$ reward units across seeds and to stay below $0.20$ in every seed. The pipeline uses the SCOPE-RL library \citep{kiyohara2023scope}, with the library's on-policy reference values rescaled to the standard discount convention.}

Tables~\ref{tab:field_ope_reliability} and~\ref{tab:field_ope_candidates} report selection reliability and candidate-level diagnostics. Fitted-Q evaluation carries no usable ranking signal under the two single-policy logs, with correlations \fieldopedmab{} under the full-support A/B log and \fieldopedminc{} under the incumbent log. The true-best policy places \fieldopecovgapab{} percent of its visits below the first percentile of reference discounts reached by the A/B log, compared with \fieldopecovgapinc{} percent under the incumbent log and \fieldopecovgapmix{} percent under the historical mixture. The signed fitted-Q errors for the best policy are \fieldopedmerrbestab{} reward units under the A/B log and \fieldopedmerrbestmix{} under the mixture, enough to mis-rank a menu whose true values span $0.34$. Under the mixture its correlation rises to \fieldopedmmix{}, a within-seed improvement of \fieldopedmdiffabmix{} with standard error \fieldopedmdiffabmixse{}, and selection regret falls from \fieldoperegretab{} to \fieldoperegretmix{} reward units. Per-decision importance sampling and the doubly robust estimator instead lose precision when the effective sample size collapses to about \fieldopeessbest{} of the $500$ A/B trajectories. The four candidates aligned with mixture components recover effective sample size near \fieldopeessalignedmix{}, and the doubly robust estimates improve with it. Diverse historical policies therefore improve both occupancy coverage for fitted-Q evaluation and effective sample size for importance sampling, matching the concentrability limit \citep{MunosSzepesvari2008,wang2021statistical} and the curse of horizon \citep{precup2000,jiangli2016doubly,liu2018curse}. Benchmark studies likewise find no dominant estimator and link reliability to coverage \citep{voloshin2019empirical,fu2021dope}.

In this experiment, within-policy randomization provides action support but not state-occupancy coverage. Logs from policies that induce different state distributions improve coverage, while guardrails and staged live experiments remain necessary for deployment.

\begin{figure}[t]
\centering
\includegraphics[width=\textwidth]{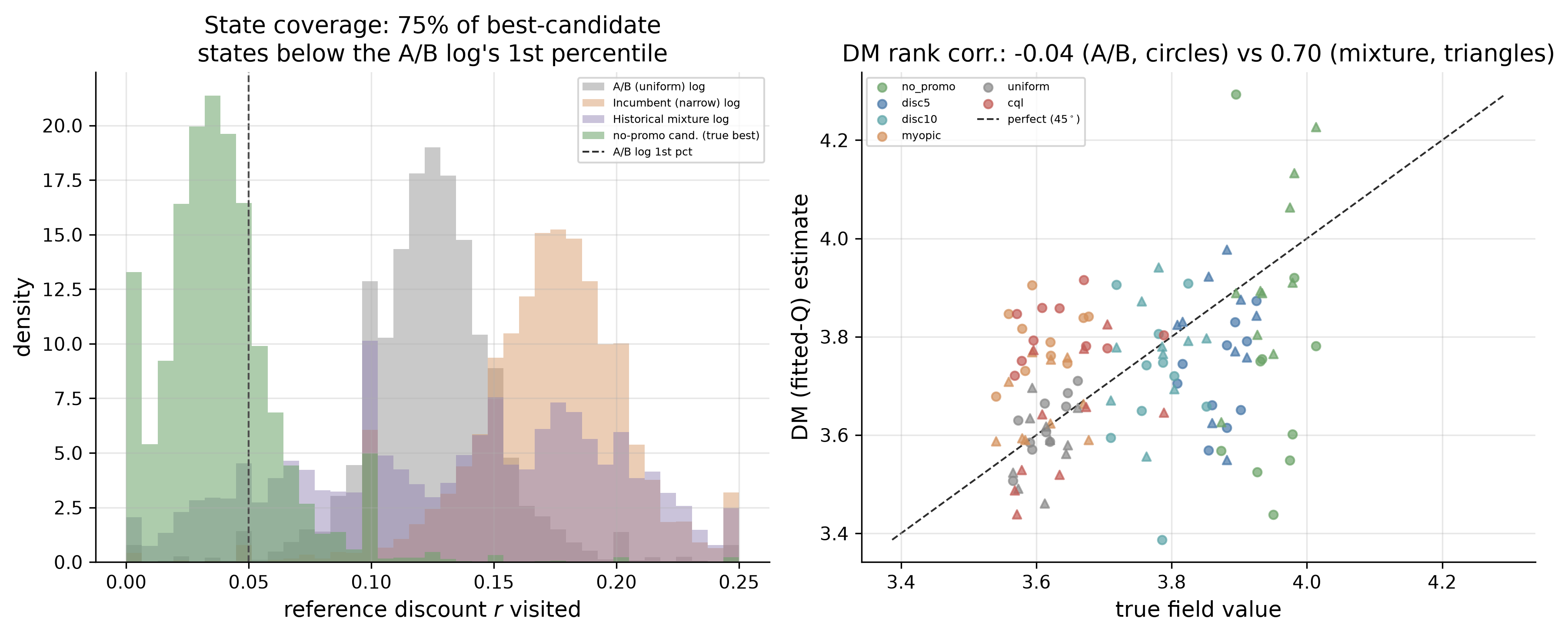}
\caption{Left: the market reference discount visited under the three logs and under the softened true-best (no-promo) candidate, with the dashed line at the first percentile of the A/B log. Right: the fitted-Q direct-method value estimate against the true field value under the A/B log (circles) and the historical-mixture log (triangles), pooled over \fieldopeseeds{} seeds, with the 45-degree reference line.}
\label{fig:field_ope}
\end{figure}

\begin{table}[t]
\centering
\small
\caption{Off-policy selection reliability of three estimators (direct method, DM; per-decision importance sampling, PDIS; doubly robust, DR) under three logging regimes, on the targeted-promotions MDP. Regret@1 is the value gap between the selected and the best deployment policy (reward units, lower is better); rank correlation is Spearman between estimated and true policy value (higher is better). Cells are mean (standard error) over 10 seeds.}
\label{tab:field_ope_reliability}
\begin{tabular}{lcccccc}
\toprule
 & \multicolumn{2}{c}{A/B (uniform)} & \multicolumn{2}{c}{Incumbent (narrow)} & \multicolumn{2}{c}{Historical mixture} \\
\cmidrule(lr){2-3}\cmidrule(lr){4-5}\cmidrule(lr){6-7}
Estimator & Regret@1 & Rank corr. & Regret@1 & Rank corr. & Regret@1 & Rank corr. \\
\midrule
DM & 0.210 (0.048) & -0.040 (0.116) & 0.246 (0.039) & -0.326 (0.136) & 0.038 (0.018) & 0.697 (0.053) \\
DR & 0.175 (0.046) & 0.051 (0.127) & 0.181 (0.043) & -0.166 (0.116) & 0.073 (0.033) & 0.423 (0.113) \\
PDIS & 0.265 (0.035) & -0.297 (0.110) & 0.326 (0.010) & -0.646 (0.083) & 0.268 (0.040) & 0.029 (0.103) \\
\bottomrule
\end{tabular}
\end{table}

\begin{table}[t]
\centering
\footnotesize
\caption{Per-candidate mechanism behind Table~\ref{tab:field_ope_reliability}. True value is the on-policy field value of the softened candidate; DM error is the fitted-Q direct-method estimate minus the true value (negative means under-valued); importance-sampling effective sample size (ESS) is $(\sum w)^2/\sum w^2$ over the log's 500 trajectories. Rows are rank-ordered by true value. Cell entries are means over 10 seeds.}
\label{tab:field_ope_candidates}
\begin{tabular}{lccccccc}
\toprule
 & & \multicolumn{3}{c}{DM error} & \multicolumn{3}{c}{IS ESS} \\
\cmidrule(lr){3-5}\cmidrule(lr){6-8}
Candidate & True & A/B & Inc. & Mix. & A/B & Inc. & Mix. \\
\midrule
No promo & 3.946 & -0.228 & -0.525 & -0.026 & 1.5 & 1.6 & 69.0 \\
5\% discount & 3.873 & -0.151 & -0.512 & -0.076 & 2.1 & 1.7 & 71.6 \\
10\% discount & 3.778 & -0.066 & -0.139 & -0.013 & 1.9 & 1.7 & 68.9 \\
Conservative Q-learning (CQL; offline RL) & 3.639 & +0.172 & -0.084 & -0.010 & 1.6 & 2.0 & 1.5 \\
Uniform & 3.612 & +0.008 & -0.125 & -0.031 & 500.0 & 1.9 & 3.4 \\
Myopic incumbent & 3.609 & +0.187 & +0.026 & +0.055 & 2.6 & 67.8 & 17.5 \\
\bottomrule
\end{tabular}
\end{table}

\subsection{Lessons from the Field}
\label{sec:field_lessons}

First, production deployments restrict reinforcement learning to a specific control component. In Meta's production bidding system, the learner tunes the parameters of an existing bidding controller. The neural actor and critic are used during training, and only the tuned controller is deployed.

Second, deployment is more common when repeated decisions produce dense logs and frequent feedback. YouTube logs every recommendation together with the probability that it was shown. The policy is trained from those logs with an off-policy correction, and a new version is tested on a small share of live traffic before wider launch. These data support offline learning, counterfactual checks, and cautious A/B testing.

Third, at DiDi, driver-order assignments are computed every two seconds by an existing matching method. Reinforcement learning estimates the long-run values used in the matching weights. The firm's operational knowledge remains encoded in the matcher.

Fourth, evaluation design depends on the domain. At Tmall, legal restrictions on simultaneous customer-level price variation precluded ordinary A/B testing. The evaluation therefore used difference-in-differences and comparisons across similar products. Digital products permit randomized tests. Pricing applications often use observational comparisons. Physical systems alternate live periods under hard constraints. Chip designs are assessed through hardware validation.

Fifth, evidence of a production deployment requires documentation of what the learner controlled, how it reached production, which guardrails and fallbacks were used, and how effects were measured. DiDi's dispatch and DeepMind's BCOOLER cooling controller provide these details. Nevmyvaka's execution study trains and tests on real order-book data through a historical backtest, and Wu's bidding paper reports dataset splits while asserting industrial deployment. Their evidence concerns performance on logged data or in a simulator.

Benchmark studies usually evaluate a policy that maximizes return within a simulator and has access to the full simulated environment. Production implementations restrict reinforcement learning to a narrow part of the decision process while retaining the surrounding controller, ranker, or optimizer. The learned output is a candidate slate at YouTube, the parameters of a bidding rule at Meta, a value estimate inside DiDi's matcher, or a single quantization choice within one codec in MuZero-RC. The deployed component provides an incremental improvement within an established operational system.

A production system records the probability of each logged action and applies the corresponding correction during learning. YouTube reported an adverse $0.5\%$ change in a live metric when the importance-weight cap was loosened. BCOOLER applies guardrails and returns control to the incumbent when no safe action is available. Meta uses conservative offline training on logged data for its bidding policy because mistakes in live auctions or markets are costly. Shadow tests, switchback experiments, monitoring, and a retraining schedule are additional parts of production operation beyond the learning algorithm.

Finally, documented gains are evaluated against established incumbent methods. A tuned base-stock policy outperforms off-the-shelf deep RL on classic inventory problems, while DiDi's matcher and Meta's bidding rule were already effective. The reported improvements over these incumbents are carefully measured and are typically about one or two percent. Production evidence requires evaluation beyond an offline score. In the pricing simulation in Section~\ref{sec:field_ope_sim}, offline evaluation mis-ranks candidate policies when the log lacks coverage of the states they visit. Few cases meet the production criterion because many papers labeled ``real-world'' report only simulator or backtest results. The available evidence concerns reinforcement-learning improvements to specific, well-instrumented components of existing systems.

\subsection{When to Reach for Reinforcement Learning}
\label{sec:field_practitioner}

Reinforcement learning is appropriate only when a firm's current actions alter subsequent states and rewards are delayed. If contexts arrive independently and each decision is evaluated in isolation, contextual-bandit or supervised-prediction methods are sufficient (Section~\ref{subsec:overlapping_terminology}). The sequential condition arises when a discount changes customers' willingness to wait for later discounts, a dispatch decision reduces later driver availability in a region, or a pricing rule changes buyer behavior (Section~\ref{sec:liu}). In the terminology of Section~\ref{section:language}, such problems belong to the control culture because the dynamics affect long-run outcomes and myopic decisions do not maximize long-run value.

When the problem does have that structure, the deployed record marks out where reinforcement learning has actually paid off, and the profile is consistent. Decisions repeat at high volume and leave dense logs. The learner is given a narrow, well-defined lever inside a system the firm already runs, not the whole system. A strong incumbent is already in place, so the goal is a reliable small gain rather than a policy built from nothing. The logs record the probability of each action, so a policy can be both learned and checked from them. Mistakes are bounded and reversible, ideally behind guardrails with a fallback. And the outcome can be measured through some experiment the domain permits. The cases in this chapter are, in effect, the set of problems that met enough of these conditions.

Read in reverse, the same profile is a list of warning signs. Rare, one-shot, or low-volume decisions leave nothing to learn from. Logs without recorded action probabilities break both offline learning and its evaluation (Section~\ref{section:offline_rl}). A single catastrophic and irreversible mistake, with no guardrail to catch it, rules out live exploration. Where a myopic rule or a classical optimizer already captures the available value, reinforcement learning adds risk without adding return. The hardest warning to see is the last one, because an offline comparison can favor a policy that would fail in production, since a strong offline score can rank the wrong policy when the logs do not cover the states that policy would visit (Section~\ref{sec:field_ope_sim}).

Researchers can assess feasibility by measuring the long-run value that remains uncaptured under the incumbent. They can train an offline prototype from existing logs and report the candidate policy's estimated value together with the logs' coverage of its behavior. A narrow intervention should be selected for the target metric, and the current controller should remain available as a fallback. Before development, the evaluation protocol should be fixed according to the domain, with randomized tests for digital products, difference-in-differences for pricing, and switchback designs for marketplaces. Firms should allocate most of the implementation effort to logging, guardrails, monitoring, and retraining, as required by the deployed systems in Section~\ref{sec:field_lessons}.

\subsection{The Shape of Deployed Reinforcement Learning}
\label{sec:field_shape}

Confirmed deployments are grouped by domain in Table~\ref{tab:field_shape}, with representative systems, the reinforcement-learning methods used, and the data or decision scales.

\begin{table}[t]
\centering
\footnotesize
\begin{tabularx}{\textwidth}{@{}>{\raggedright\arraybackslash}p{0.16\textwidth}>{\raggedright\arraybackslash}X>{\raggedright\arraybackslash}p{0.21\textwidth}>{\raggedright\arraybackslash}p{0.20\textwidth}@{}}
\toprule
Domain & Representative system & Method deployed & Data or decision scale \\
\midrule
Recommendation and ranking & YouTube top-$K$ \citep{Chen2019YouTubeTopK}; Taobao RL-LTV \citep{Ji2021RLLTV} & Off-policy REINFORCE; recurrent actor-critic & Web-scale logs; item-day episodes \\
Notification gating & Meta Horizon \citep{Gauci2019Horizon} & DQN send/drop & Production-scale logs \\
Bidding and auctions & Meta bidding \citep{Korenkevych2023MetaBidding}; Alibaba RTB \citep{Zhao2018SponsoredSearchRTB} & Offline RL tuning a base policy; robust-MDP DQN & $\sim$1.2B logged steps; $\sim$100M auctions per day \\
Dispatch and matching & DiDi \citep{SadeghiEshkevari2022DiDiScalable} & TD value inside a matching solver & Tens of millions of orders per day \\
Inventory & DeepStock \citep{Xie2026DeepStock} & Base-stock-regularized DDPG & More than 1M SKU-warehouse pairs \\
Video-codec rate control & MuZero-RC \citep{Mandhane2022MuZeroRC} & Model-based planning, policy served & Frame-level control on a live traffic slice \\
Chip floorplanning (design-time) & AlphaChip \citep{Mirhoseini2021GraphPlacement} & Graph policy-gradient & Transfer across prior chip blocks \\
LLM post-training (training-time) & InstructGPT \citep{ouyang2022training} & PPO against a reward model & Tens of thousands of labeled prompts \\
Building cooling (field trial) & BCOOLER \citep{Luo2022BCOOLER} & Constrained offline RL with fallback & Two commercial sites \\
\bottomrule
\end{tabularx}
\caption{The shape of confirmed field deployments: the domain, a representative system, the reinforcement-learning method that ran, and the data or decision scale. Firm types and evaluation practices are discussed in the text.}
\label{tab:field_shape}
\end{table}

Deployment data volumes vary by several orders of magnitude. BCOOLER uses a two-building trial and InstructGPT uses tens of thousands of human-labeled prompts, while Meta's bidding system uses more than a billion logged time steps and DeepStock uses over a million stock-keeping-unit and warehouse pairs. Narrow and heavily constrained actions characterize the low-data applications. Recommendation, bidding, dispatch, and inventory provide dense logs through ordinary high-volume operation. Value estimation for a classical optimizer and DQN-style value-based control account for most methods. DiDi and BCOOLER use the first approach in their matching and cooling optimizers, and notification gating, sponsored-search bidding, and price selection use the second. Policy-gradient methods are restricted through YouTube's off-policy correction, Meta's trusted-base-policy tuning, and OpenAI's reward-model optimization. Model-based planning occurs only in MuZero-RC. No confirmed production system deploys conservative Q-learning, implicit Q-learning, or a model-based offline method as an unconstrained standalone controller. The conservative-Q-learning case instead tunes the parameters of an existing bidding policy and deploys only those parameters.

The domains that recur are a short list, recommendation, notification gating, bidding, dispatch, inventory, and codec control, and they share the profile in Table~\ref{tab:field_shape}, a repeated high-volume decision wrapped around a system the firm already runs. Two cases sit outside that pattern and should be read as different evidence, since chip floorplanning is a design-time artifact that flows through ordinary hardware validation and RLHF is a training-time step whose product is a model rather than a live controller. The economically important spaces that are missing are missing for reasons the earlier conditions predict. Finance appears only as a historical backtest, healthcare not at all, robotics is set aside at the start of this chapter, and macroeconomic policy is nowhere, because those settings cannot cheaply or safely experiment on live decisions, often cannot randomize the real decision, and lack dense and fast feedback. The firms are correspondingly few. Almost every confirmed deployment comes from a handful of hyperscale platforms and frontier laboratories, among them Google, Meta, Alibaba, DiDi, OpenAI, and DeepMind, that hold all of the enabling conditions at once, the decision volume that makes a fractional gain worth capturing, the logging and off-policy evaluation, the engineering capacity, and an incumbent system to build around. The one non-technology operator in the record, a budget hotel chain, reached deployment only by simplifying the action to a single interpretable discount so that its managers would accept it.

Deployment evidence concerns business outcomes, whereas benchmark evidence concerns return relative to a baseline in a fixed environment. The business outcomes reported here include ViewTime at YouTube, gross merchandise value at Taobao, driver income at DiDi, revenue per available room in hotels, bitrate at fixed quality for MuZero-RC, and inventory capital at DeepStock. Changes measured against live controls range from a fraction of a percent to a few percent. At the scale of fifty billion impressions or one million inventory pairs, these percentage changes correspond to large aggregate effects. Gains in the simulation and backtest cases range from tens of percent to more than one hundred percent, with no live control used in those evaluations. A large offline estimate therefore does not establish deployment, as shown in Section~\ref{sec:field_ope_sim}.

Production deployment requires operational measurements in addition to the reported business outcome. Whether the offline evaluation can even be trusted becomes a tuned quantity, since YouTube's importance-weight cap is a hyperparameter whose relaxation costs half a percent of ViewTime, and Meta's platform runs a counterfactual evaluation as a required step before any launch. How to run the experiment is itself a design problem, because the real decision often cannot be randomized, so DiDi alternates by time slice and confirms the rollout with a difference-in-differences across cities, and Tmall compares against similar products because charging different prices at the same time is not allowed. Safety is a measured rate rather than an assumption, the frequency with which MuZero-RC overshoots its bitrate target, the rate at which BCOOLER hands control back to the incumbent controller, and the plain fact that an under-trained bidding agent can lose millions within hours. The operational budget is tracked on its own terms, a two-second serving path kept separate from a ten-second learning loop, a data lag under a day, a retraining cadence, and a monitor that compares this month against the year before. Long-run value, where it exists, is modeled explicitly, the lifetime value of a cold-start item at Taobao and the future value of a driver's location at DiDi, a horizon a one-step ranker misses by construction.

\section{Discussion}
\label{section:conclusion}

\subsection{The Two Cultures, Revisited}

The opening of this survey (Section~\ref{section:language}) drew a line between two intellectual traditions.
The inference tradition concentrates its effort on specifying the objective function and the law of motion that governs the environment; the optimal policy is a byproduct that falls out once the primitives have been identified.
The control tradition inverts this emphasis: it takes the objective and the dynamics as given and asks whether the optimal policy can be computed, approximated, and deployed under real-time and robustness constraints.
Dynamic programming stands at the intersection of the two cultures.
Both \citet{Bellman1957} and the structural econometrics program of \citet{Rust1987} rest on the Bellman equation; they differ in what is treated as known and what must be estimated.
Reinforcement learning appears here not as a third culture but as a computational toolkit that relaxes the information requirements of dynamic programming while preserving its mathematical foundations.

The chapters of this survey have mapped how that toolkit interacts with the applied decision sciences.
On one side, RL extends the reach of structural modeling to state spaces where traditional dynamic programming is infeasible, without requiring the analyst to abandon structural assumptions.
On the other side, the inference tradition imposes constraints on RL problems, identification conditions, equilibrium concepts, and revealed-preference axioms, that sharpen what can be learned from data.
What emerges from this exchange is not a merger of the two traditions but a productive division of labor in which each side contributes what it does best.

\subsection{What Reinforcement Learning Adds}

The most direct contribution of RL to economic modeling is computational.
Dynamic discrete choice models face a curse of dimensionality that limits exact dynamic programming to relatively small, discrete state spaces \citep{rust2008dp}.
TD-based methods make structural models tractable at scales where the nested fixed-point algorithm is infeasible. The bus-engine simulation in Section~\ref{sec:bus_engine} shows a DQN trained on a Rust-style replacement environment matching exact dynamic programming within one percent on the fleet sizes where both are computed \citep{Rust1987}.
This prescriptive capability, not just estimating model parameters but computing the policies those parameters imply, extends across the survey.
The field cases in Section~\ref{section:field_deployments} show that RL can compute practically valuable policies in domains where neither analytical solutions nor small-state DP are available, provided the deployed control surface is narrow and surrounded by evaluation and operational safeguards.

A second contribution appears in multi-agent settings.
Game-theoretic solution concepts are often intractable to compute directly; PPAD-hardness of Nash equilibrium \citep{Shapley1964} is a foundational obstacle.
The simulations of Section~\ref{section:rl_games} show that independent Q-learning agents converge to Nash equilibrium in symmetric normal-form games through trial and error, providing ``as-if'' microfoundations for equilibrium play under bounded rationality \citep{FudenbergLevine1998}.
The durable-goods monopoly simulation in that section reproduces the Coase conjecture limit as the horizon lengthens, confirming that RL recovers well-known theoretical predictions without any equilibrium concept being hard-coded.
Where RL-trained agents actually operate in markets, the findings become empirically testable: \citet{Calvano2020} document that independent pricing algorithms sustain supra-competitive prices through reward-based learning without explicit communication, a result that competition authorities have since taken seriously.

The bandit chapters (Section~\ref{section:bandits}) show that domain knowledge and RL interact multiplicatively.
Imposing demand structure on a pricing problem can reduce cumulative regret from $\Theta(T)$ to $O(\log T)$, a reduction that neither structural estimation alone nor generic bandit algorithms alone achieves \citep{Agrawal2024ref}.
The pricing experiment does not produce a monotone finite-sample ordering. At $T=200{,}000$, both WARP-based variants beat UCB1, but the untuned variant trails Thompson sampling and LTE. Only Thompson sampling has a nearly stable $R/\sqrt{T}$ across the four reported horizons.

Offline reinforcement learning (Section~\ref{section:offline_rl}) addresses the setting most familiar to economists, in which the analyst has a fixed dataset of historical decisions and must improve on the behavioral policy without further experimentation.
Algorithms with distributional-shift correction, including conservative Q-learning \citep{Kumar2020} and behavior-constrained methods \citep{Fujimoto2019b}, exceed the behavioral baseline in the dynamic pricing simulation; algorithms without correction degrade below it.
The dependence on data support documented in Table~\ref{tab:offline_main} connects directly to familiar econometric concepts: coverage is the RL analog of overlap in propensity-score methods, and the performance degradation under poor coverage is the RL analog of extrapolation bias in treatment-effect estimation.

The final RL contribution is in preference learning.
RLHF and DPO (Section~\ref{section:rlhf}) extend the inverse problem of recovering latent utilities from observed choices to the setting where ``choices'' are comparative judgments between model outputs \citep{christiano:2017, ouyang2022training, rafailov2023direct}.
The simulation in that section treats job-search as the preference environment and shows that DPO recovers the underlying utility ranking with substantially fewer samples than reward-model RLHF, but that both methods are sensitive to the number of preference pairs and to the Bradley-Terry assumption when annotator heterogeneity is large.

\subsection{What Economics Adds to Reinforcement Learning}

The contribution runs in both directions.
The inference tradition imposes constraints on RL problems that sharpen identification and improve credibility.

Causal inference for RL (Section~\ref{section:causal_rl}) addresses the obstacle that logged data in economic settings is observational: treatment assignment is confounded, and naive off-policy evaluation will estimate the counterfactual return inconsistently \citep{precup2000, pearl2009causality}.
The confounded retail-pricing simulation in that section shows that a backdoor-adjusted importance-weighted estimator recovers the true policy value while the unadjusted estimator produces systematically optimistic estimates.
Instrumental-variable and proxy-variable identification strategies, long standard in econometrics \citep{robins1994estimation}, extend naturally to the POMDP-to-causal formulation of Section~\ref{subsec:pomdp_to_causal}: the IV-RL estimator in the counterfactual OPE simulation achieves root-mean-squared error within ten percent of the oracle estimator, whereas the unadjusted double-robust estimator misses by a factor of three.

Revealed-preference theory (Section~\ref{section:rlhf}, specifically the axiom-aware aggregation subsection) provides a second disciplinary check.
The Bradley-Terry model implicitly assumes a cardinal, individually consistent preference ordering that need not hold when feedback comes from heterogeneous annotators.
Axiom-aware aggregation, which tests for GARP violations and Condorcet cycles before constructing the reward model, outperforms naive aggregation in the simulation under annotator heterogeneity, with the performance gap widening as the degree of heterogeneity increases \citep{varian1982nonparametric, skalse2023invariance}.
The broader lesson is that the axiomatic toolkit of revealed-preference theory can serve as a regularizer on reward learning, replacing the untestable parametric assumption with a testable behavioral constraint.

Robust and constrained RL (Section~\ref{section:dist_robust_constrained}) is a third area of exchange between the two fields.
The ambiguity sets of robust MDPs formalize a notion of model uncertainty that is structurally related to the multiplier preferences of \citet{HansenSargent2001}: the decision-maker pessimizes over a neighborhood of probability measures rather than maximizing against a point prior.
The risk-sensitive inventory simulation uses IQN to improve CVaR$_{95}$ from $-57.5$ under risk-neutral evaluation to $-55.3$ under CVaR evaluation, while mean returns remain within two standard errors.
A separate robust consumption-savings simulation studies model ambiguity rather than return quantiles. Its high-robustness policy recovers $4.5$ percentage points of the $76\%$ degradation under the perturbed income process, while an oracle that knows the perturbation recovers $37$ points.

The macroeconomic chapter (Section~\ref{section:rl_macro}) documents a fourth exchange.
The adaptive learning literature initiated by \citet{MarcetSargent1989} uses stochastic approximation algorithms, Robbins-Monro-type updates, to model how boundedly rational agents refine beliefs about equilibrium parameters.
The convergence criterion in that literature, E-stability, is governed by an ordinary differential equation that is formally identical to the ODE used to prove convergence of Q-learning and actor-critic algorithms in RL \citep{borkar2000}.
The bounded-rationality subsection of the macro chapter shows that Q-learning households in the Krusell-Smith environment converge to approximately rational expectations through the same stochastic approximation dynamics that the adaptive learning literature has studied for three decades \citep{KaplanMollViolante2018, MarimonMcGrattanSargent1990}.

\subsection{Simulation Evidence}

The computational experiments across the survey yield a consistent pattern.
The Engine Replacement MDP supplies the common small-model evidence before the larger chapter-specific studies. Table~\ref{tab:engine_recap} records the object computed in each application. Reusing one model separates differences among methods from differences among environments.

\FloatBarrier
\begin{table}[h]
\centering
\caption{What the Engine Replacement MDP establishes across the survey.}
\label{tab:engine_recap}
\begin{tabular}{p{0.30\textwidth}p{0.55\textwidth}}
\hline
chapter topic & role of the common model \\
\hline
Theory & value and policy geometry, Section~\ref{engine:value_space} \\
Deep RL practice & Bellman and value error, Section~\ref{engine:ch03b} \\
Structural estimation & parameter recovery, Section~\ref{engine:ch05} \\
Offline RL & coverage and extrapolation, Section~\ref{engine:ch08} \\
Human feedback & reward equivalence, Section~\ref{engine:ch09} \\
Causal RL & unobserved confounding, Section~\ref{engine:ch10} \\
OPE and treatment effects & importance weights, Section~\ref{engine:ch10b} \\
Robust and constrained RL & budget duality, Section~\ref{engine:ch11} \\
World models & transition learning, Section~\ref{engine:ch12} \\
\hline
\end{tabular}
\end{table}

\FloatBarrier

The DDC scaling simulation (Section~\ref{section:rl_econ_models}) shows NFXP wall-clock time growing by roughly three orders of magnitude as the state space expands from $20$ to $160{,}000$ states. At the largest scale, neural TD-CCP matches NFXP's replacement-cost RMSE at about one quarter of its wall-clock time.
The offline pricing simulation (Section~\ref{section:offline_rl}) shows CQL and IQL remaining near $92\%$ of the dynamic-programming optimum across the coverage sweep, while unprotected fitted Q-iteration remains between $17\%$ and $27\%$. The result reflects each method's treatment of unsupported actions rather than a universal coverage threshold.

RL does not uniformly dominate.
In the RBC simulation (Section~\ref{section:rl_macro}), the state is the two-dimensional pair $(K,A)$. PPO reaches the value-function-iteration benchmark within its standard error, while DDPG finishes about $2\%$ below it.
In the Cournot and Bertrand simulations (Section~\ref{section:rl_games}), independent Q-learning, Nash-Q, and WoLF-PHC settle near the symmetric Nash benchmarks. These experiments do not test selection of a Pareto-superior equilibrium.
Seed sensitivity, documented systematically in Section~\ref{section:deeprl_practice}, means that any single simulation run is insufficient evidence of a systematic result; the standard adopted throughout this survey, at least ten independent seeds with reported means and standard errors, is a minimum requirement.

\subsection{Open Challenges}

Several concrete obstacles remain.
Dynamic discrete choice with continuous action spaces and high-dimensional continuous state vectors is unsolved.
The DDC simulation stops short of continuous actions; extending the Hotz-Miller inversion \citep{HotzMiller1993} to continuous choices requires density estimation in high dimensions, and the standard CCP-estimator breaks down.
Policy gradient methods handle continuous actions but require simulation-based policy gradients that are noisy and sample-inefficient when the state space encodes high-frequency heterogeneity.

Bandit algorithms under non-stationary economic primitives pose a second open problem.
Demand curves shift with macroeconomic conditions, competitor entry, and product innovations.
Regret bounds in the non-stationary bandit literature grow with the number of change points, but identifying change points in economic demand is itself a statistical problem confounded by endogenous pricing.
Consistent bandit algorithms under simultaneously non-stationary and endogenous environments are not yet available.

Off-policy evaluation under unobserved confounding without instruments remains the most consequential gap.
The sensitivity analysis of Section~\ref{subsec:alternative_identification} shows that partial identification bounds widen rapidly as the degree of unmeasured confounding increases.
In many economic datasets, valid instruments for treatment assignment are unavailable, and proxy variables either do not exist or have not been collected.
Methods that yield informative bounds in this fully unidentified regime, rather than arbitrarily wide ones, are an active area of research with direct policy relevance.

The interaction between equilibrium and learning in multi-agent RL has no satisfactory general theory.
When agents performing Bellman backups face multiple equilibria, value iteration can cycle or diverge, and the equilibrium selection is sensitive to initialization and learning rates.
The equilibrium refinement literature provides focal-point concepts, but these require common knowledge of rationality that decentralized learning does not provide \citep{FudenbergLevine1998}.

The survey's simulation evidence also points to an infrastructure gap.
Industrial deployments described in Section~\ref{section:field_deployments} required dedicated engineering teams and months of hyperparameter search; the academic simulations here required comparable effort at much smaller scale.
Shared, well-documented simulation environments modeled on the economic problems studied in this survey, analogous to what ALE provided for game-playing \citep{mnih2015}, would accelerate the empirical research program substantially.

\subsection{Conclusion}

Reinforcement learning and the applied decision sciences have arrived at a productive intersection.
RL provides computational tools for solving high-dimensional dynamic programs and learning from observational data; economics provides structural discipline, identification theory, and institutional context that make those solutions credible.
The structural equivalences documented in Section~\ref{subsec:structural_equivalences}, the softmax-logit identity, the Emax-soft-value correspondence, the Q-function-CCP duality, reflect the fact that optimal sequential decision-making under uncertainty is one problem, and the two traditions have been working toward it from opposite ends.
Progress at the intersection requires researchers who are literate in both formal languages.

\bibliographystyle{plainnat}
\setcitestyle{maxcitenames=2}

\bibliography{refs}

\appendix
\section{Mathematical Preliminaries}
\label{appendix:mathprelim}

This appendix collects the mathematical results the survey rests on but does not develop in the main text. It assumes only probability, calculus, and linear algebra at the level of a first graduate sequence.

\subsection{Elementary Objects and Inequalities}
\label{prelim:elementary}

This section holds targeted clarifications, objects one level below the sections that follow. Each entry states a definition in plain words, gives a small numerical example, and notes where the survey uses it. Chapter footnotes point here directly.

\subsubsection{Maxima, Suprema, and the Set of Maximizers}
\label{prelim:argopt}

The \emph{maximum} $\max_{a \in \mathcal{A}} f(a)$ is the largest value $f$ attains, and it exists whenever $\mathcal{A}$ is finite. The \emph{supremum} $\sup_{a \in \mathcal{A}} f(a)$ is the least upper bound, which exists whenever $f$ is bounded above but need not be attained by any $a$. On $\mathcal{A} = (0,1)$ with $f(a) = a$ the supremum is $1$ and there is no maximum, since no point of the open interval equals $1$. The survey writes $\max$ where the action set is finite, which covers every algorithm it states, and $\sup$ where the set is a continuum or where the statement should not presume attainment. The \emph{infimum} and \emph{minimum} are the mirror images.

The maximum is a number. The \emph{set of maximizers} $\argmax_{a} f(a) = \{a : f(a) \geq f(a') \text{ for all } a'\}$ is a set, and it can hold more than one element. For $f = (3, 5, 5)$ on three actions the maximum is $5$ while the set of maximizers is $\{2, 3\}$. A \emph{greedy} policy picks some element of that set, so when the set is not a singleton there is more than one greedy policy. That is the situation at a kink of the Bellman envelope (Section~\ref{prelim:affine_envelope}), where two actions tie and the optimality operator is not differentiable.

\subsubsection{The Indicator}
\label{prelim:indicator}

The \emph{indicator} $\indicator{E}$ of an event or condition $E$ equals one when $E$ holds and zero otherwise. It turns a logical condition into a number that can be summed or averaged. If $S_1, \ldots, S_n$ are sampled states then $\frac{1}{n}\sum_t \indicator{S_t = s}$ is the fraction of the sample spent in state $s$, and its expectation is $\mathbb{P}(S = s)$. Counting visits, restricting a sum to a subset, and writing an empirical distribution are all this one device. It appears in eligibility traces, in occupancy counts, and in the quantile losses of Section~\ref{section:dist_robust_constrained}.

\subsubsection{Affine Functions and Upper Envelopes}
\label{prelim:affine_envelope}

A function is \emph{affine} if its graph is a straight line, $f(x) = a + bx$, a linear part plus a constant shift. The \emph{pointwise maximum} of finitely many affine functions, $g(x) = \max_i \{a_i + b_i x\}$, takes at each $x$ the value of whichever line is highest there. Its graph is an \emph{upper envelope}, a convex piecewise-linear curve that follows one line until another overtakes it, with a kink at each crossing. Take $f_1(x) = 1 + \tfrac{1}{2}x$ and $f_2(x) = 2 - x$. The second line is higher for $x < \tfrac{2}{3}$ and the first for $x > \tfrac{2}{3}$, so the envelope $g(x) = \max\{f_1(x), f_2(x)\}$ has its kink at the crossing $x = \tfrac{2}{3}$, where $g = \tfrac{4}{3}$ (Figure~\ref{fig:prelim_elementary}, panel a). A line that touches the graph of a convex function at a point and never rises above the graph is a \emph{supporting line}, in higher dimensions a supporting hyperplane, and each $f_i$ supports $g$ on the piece where it attains the maximum. The Bellman optimality operator is exactly such an envelope, the pointwise maximum of affine policy operators, and Section~\ref{prelim:operators} builds it that way.

\subsubsection{A Difference of Maxima}
\label{prelim:maxdiff}

For any two real-valued functions $f$ and $g$ on a finite set,
\begin{equation}
\big| \max_a f(a) - \max_a g(a) \big| \;\leq\; \max_a |f(a) - g(a)|.
\label{eq:prelim_maxdiff}
\end{equation}
Maximizing cannot amplify a disagreement, the two maxima differ by no more than the largest pointwise gap. The proof is two lines. Let $a_1$ attain $\max_a f$. Then $\max_a f - \max_a g \leq f(a_1) - g(a_1) \leq \max_a |f(a) - g(a)|$, since $g(a_1) \leq \max_a g$, and exchanging the roles of $f$ and $g$ bounds the other sign. As a numerical example with two actions, let $f = (3, 5)$ and $g = (4.5, 4)$. The maxima are $5$ and $4.5$, a gap of $0.5$, while the pointwise gaps are $1.5$ and $1$, so the bound $0.5 \leq 1.5$ holds with room to spare. This is the first of the three results proved in full here, and it is the only reason the nonlinear Bellman optimality operator inherits the contraction modulus of the affine policy operators it envelopes (Section~\ref{prelim:operators}). It also carries the Q-factor contraction (Theorem~\ref{thm:q_factor_contraction}).

\subsubsection{Norms for Value Functions}
\label{prelim:norms}

\begin{figure}[t]
\centering
\includegraphics[width=0.95\textwidth]{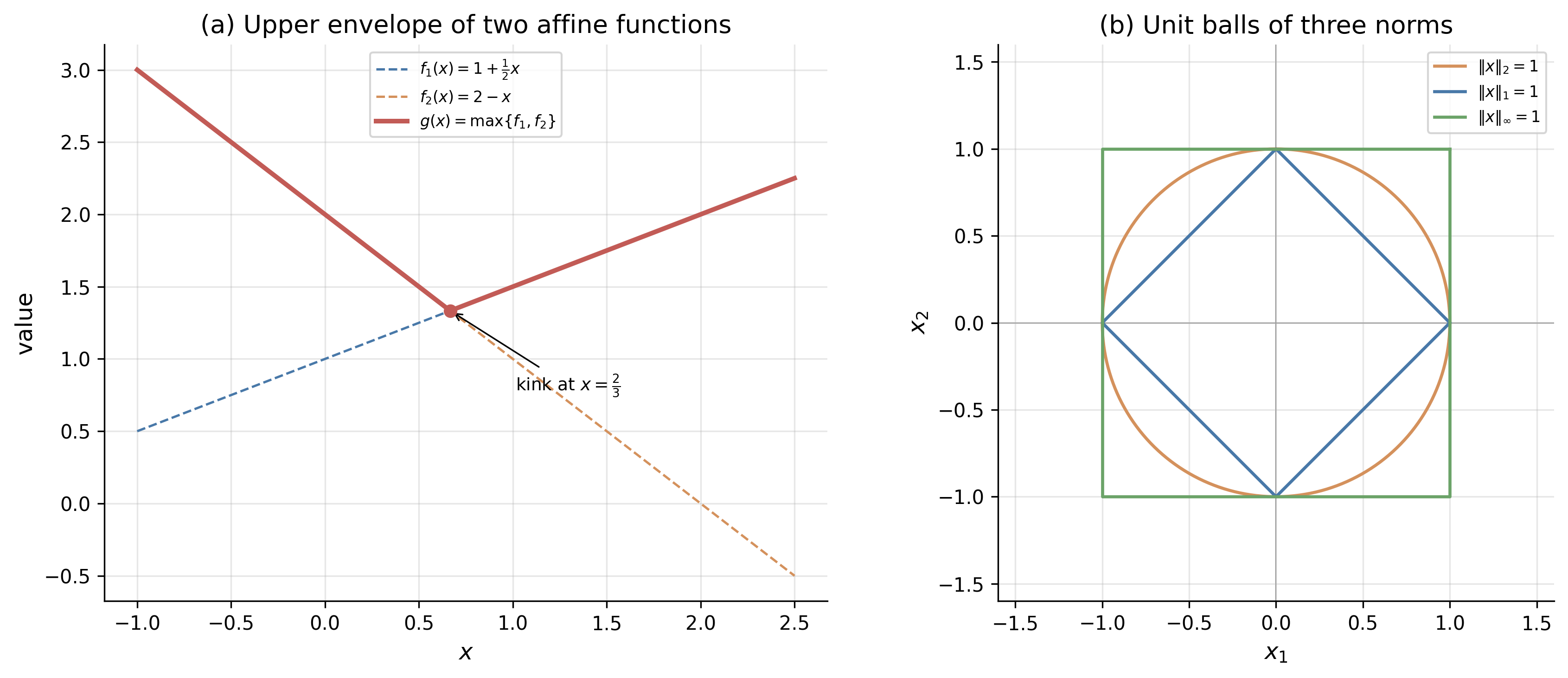}
\caption{Panel (a): two affine functions (dashed) and their upper envelope (solid), with the kink at $x = 2/3$. Panel (b): unit balls of the $L^1$, $L^2$, and $L^\infty$ norms in the plane.}
\label{fig:prelim_elementary}
\end{figure}

A \emph{norm} assigns a vector a single nonnegative size. The $L^p$ family on $\mathbb{R}^n$ is $\|x\|_p = (\sum_i |x_i|^p)^{1/p}$. At $p = 1$ it sums absolute entries and at $p = 2$ it is Euclidean length. As $p$ grows the largest entry dominates the sum, and in the limit $\|x\|_\infty = \max_i |x_i|$, the \emph{supremum norm}, the size of the worst coordinate. For $x = (3, -4, 1)$ the values are $\|x\|_1 = 8$, $\|x\|_2 = \sqrt{26} \approx 5.10$, and $\|x\|_\infty = 4$. The unit balls, the sets of vectors of size one, are drawn in Figure~\ref{fig:prelim_elementary}, panel b. A value function on a finite state space is such a vector with one entry per state, so the supremum norm reads as the worst-case error over states. A \emph{weighted} $L^2$ norm $\|V\|_d^2 = \sum_s d(s)\, V(s)^2$, with $d$ a probability distribution over states, instead counts each state in proportion to its weight. In the survey that weight is how often a policy visits the state, so errors at rarely visited states barely register. Which norm is in force is not a matter of taste, and Section~\ref{prelim:approximation} exhibits one operator that contracts in a weighted norm under one weighting and expands under another.

\subsubsection{Telescoping Sums and the Geometric Series}
\label{prelim:telescoping}

Two summation identities recur in every convergence argument of the survey. A \emph{telescoping sum} of consecutive differences collapses to its endpoints, $\sum_{k=0}^{K-1} (x_{k+1} - x_k) = x_K - x_0$, because each interior term is added once and subtracted once. The \emph{geometric series} $\sum_{k=0}^{\infty} \gamma^k = \frac{1}{1-\gamma}$ for $|\gamma| < 1$ converts a per-step shrink factor $\gamma$ into a finite total. At $\gamma = 0.9$ the total is $10$ and at $\gamma = 0.99$ it is $100$, which is why bounds carry factors of $1/(1-\gamma)$ that grow as the discount approaches one.

\subsubsection{Asymptotic Notation}
\label{prelim:sequences_rates}

Rates are compared up to constants. Writing $f(n) = O(g(n))$ means there are a constant $C$ and a threshold $n_0$ with $f(n) \leq C\,g(n)$ for every $n \geq n_0$, so $g$ envelopes the growth of $f$ and the constant is suppressed. Then $\Omega(g)$ is the matching lower bound, $\Theta(g)$ is both at once, and $o(g)$ is the strictly smaller case $f(n)/g(n) \to 0$. For $f(n) = 3n^2 + 100n$ one has $f = O(n^2)$, $f = \Theta(n^2)$, and $f = o(n^3)$, and the $100n$ term is invisible in the notation even though it dominates until $n = 33$. The tilde form $\widetilde{O}(g)$ suppresses logarithmic factors as well, so $\widetilde{O}(n)$ covers $n \log n$ and $n \log^3 n$ alike. Sample-complexity and regret statements throughout the survey are written in $\widetilde{O}$, because their logarithmic factors track the proof technique rather than the problem.\footnote{Suppressing a logarithm is harmless when comparing $\sqrt{T}$ against $T^{2/3}$ and misleading when comparing two bounds that differ only by logarithms. The survey uses $\widetilde{O}$ only for the first kind of comparison.}

\subsection{The Markov Decision Process}
\label{prelim:mdp}

Everything in this survey is a statement about one object. This section defines it, fixes the notation the chapters use, and introduces the worked example the rest of the appendix computes with.

\subsubsection{The Objects}

A \emph{Markov decision process} is a tuple $(\mathcal{S}, \mathcal{A}, P, r, \gamma)$. The \emph{state space} $\mathcal{S}$ is the set of situations the decision maker can face, taken finite here with $|\mathcal{S}|$ elements. The \emph{action space} $\mathcal{A}$ is the set of choices available, also finite. The \emph{transition kernel} $P(s' \mid s, a)$ gives the probability of landing in $s'$ after taking $a$ in $s$, so $P(\cdot \mid s,a)$ is a probability distribution over $\mathcal{S}$ for each pair. The \emph{reward function} $r(s,a)$ gives the payoff collected on that step. The \emph{discount factor} $\gamma \in [0,1)$ weights a payoff one period later by $\gamma$ relative to one collected now.

The defining restriction is in the name. The transition and the reward depend on the current state and action alone, not on how the state was reached. This is the \emph{Markov property}, and it is what makes the state a sufficient summary of the past and what makes everything below possible.

A \emph{policy} $\pi(a \mid s)$ assigns a distribution over actions to each state. It is \emph{deterministic} when that distribution puts all its mass on one action, in which case $\pi(s)$ names the action directly. Fixing a policy and an initial state generates a \emph{trajectory} $S_0, A_0, R_0, S_1, A_1, R_1, \ldots$ with $A_t \sim \pi(\cdot \mid S_t)$, $R_t = r(S_t, A_t)$, and $S_{t+1} \sim P(\cdot \mid S_t, A_t)$. The \emph{discounted return} from time $t$ is
\begin{equation}
G_t = \sum_{k=0}^{\infty} \gamma^k R_{t+k},
\label{eq:prelim_return}
\end{equation}
finite whenever the rewards are bounded, since $|G_t| \leq \max_{s,a}|r(s,a)| / (1-\gamma)$ by the geometric series of Section~\ref{prelim:telescoping}.

The \emph{state-value function} of a policy is the expected return from each state,
\begin{equation}
V^\pi(s) = \mathbb{E}_\pi\bigl[G_t \mid S_t = s\bigr],
\label{eq:prelim_vpi}
\end{equation}
and the \emph{action-value function}, or Q-function, conditions on the first action as well,
\begin{equation}
Q^\pi(s,a) = \mathbb{E}_\pi\bigl[G_t \mid S_t = s, A_t = a\bigr].
\label{eq:prelim_qpi}
\end{equation}
Their difference is the \emph{advantage} $A^\pi(s,a) = Q^\pi(s,a) - V^\pi(s)$, which measures how much better taking $a$ once and following $\pi$ thereafter is than following $\pi$ throughout. An \emph{optimal policy} $\pi^\star$ attains $V^\star(s) = \max_\pi V^\pi(s)$ simultaneously at every state, and for a finite discounted Markov decision process one always exists and may be taken deterministic \citep{puterman1994}. In chess terms, $V^\pi$ scores a position under a fixed way of playing, $Q^\pi$ scores a position together with the move about to be made, and the advantage is how much that move gains over the player's habit.

Two notational conventions hold throughout the survey. Reward is $r(s,a)$ for the function and $R_t$ for the realized draw. Estimated or perturbed quantities carry a hat or a tilde, $\widehat{V}$ and $\widetilde{V}$, and iterates of an algorithm carry a subscript, $V_k$.

\subsubsection{Value Functions Are Vectors}

The reframe that carries the rest of this appendix is that none of these objects is a function in any difficult sense. With $|\mathcal{S}|$ states a value function is a list of $|\mathcal{S}|$ numbers, which is a point in $\mathbb{R}^{|\mathcal{S}|}$. Fixing a policy $\pi$ collapses the transition kernel to a single $|\mathcal{S}| \times |\mathcal{S}|$ matrix and the reward to a single vector,
\begin{equation}
P^\pi(s, s') = \sum_a \pi(a \mid s) P(s' \mid s, a),
\qquad
r^\pi(s) = \sum_a \pi(a \mid s)\, r(s,a).
\label{eq:prelim_ppi}
\end{equation}
The matrix $P^\pi$ is \emph{row-stochastic}, meaning every entry is nonnegative and every row sums to one. Algorithms are then maps on $\mathbb{R}^{|\mathcal{S}|}$, convergence is convergence of points in that space, and the questions of the next six sections become questions about matrices and vectors. Value functions sit as columns under $P^\pi V$, and state distributions sit as rows under $d_{t+1} = d_t P^\pi$.

\FloatBarrier
\subsubsection{The Engine Replacement MDP}

The Engine Replacement MDP has two engine grades and two actions. It makes the vector and matrix representation in~\eqref{eq:prelim_ppi} two-dimensional. The state space is $\mathcal{S} = \{\text{good}, \text{worn}\}$ and the action space is $\mathcal{A} = \{\text{keep}, \text{replace}\}$, with $\gamma = 0.9$. Keeping a good machine earns $1$ and degrades it to worn with probability $\tfrac{1}{2}$. Keeping a worn machine earns $0.2$ and leaves it worn. Replacing costs $0.5$ from either state and returns the machine to good with certainty. Table~\ref{tab:prelim_running_example} collects the primitives and the quantities computed from them, and Figure~\ref{fig:prelim_mdp_graph} draws the process.

With two states, value functions are points in the plane and transition matrices are $2 \times 2$. Contraction, iteration, projection, the Bellman envelope, the resolvent, stationary distributions, and occupancy measures are therefore available in two dimensions. A one-dimensional feature space shows orthogonal and oblique projection as lines in the same plane.

The optimal policy keeps a good machine and replaces a worn one, with $V^\star = (5.3448, 4.3103)$, and the two action gaps are $1.0345$ at the good state and $0.2310$ at the worn state, so neither choice is marginal.\footnote{The economics is fixed at round numbers so a reader can redo the arithmetic. The two parameters the economics does not determine, the feature ratio of Section~\ref{prelim:approximation} and the logging probability of Section~\ref{prelim:markov_data}, are chosen by a search over 360 pairs.} Under the optimal policy the matrix and vector of~\eqref{eq:prelim_ppi} are
\begin{equation}
P^{\pi^\star} = \begin{pmatrix} 0.5 & 0.5 \\ 1 & 0 \end{pmatrix},
\qquad
r^{\pi^\star} = \begin{pmatrix} 1 \\ -0.5 \end{pmatrix},
\label{eq:prelim_running_ppi}
\end{equation}
with the good state first.

\begin{table}[h]
\centering
\caption{The two-state machine-replacement example used throughout this appendix. Rewards and transitions are the primitives; the optimal value and the stationary and discounted state distributions are computed from them. The logging policy $\mu$ keeps the machine with probability 0.1 in either state.}
\label{tab:prelim_running_example}
\begin{tabular}{lrr}
\hline
 & good & worn \\
\hline
$r(s, \text{keep})$ & 1.00 & 0.20 \\
$r(s, \text{replace})$ & -0.50 & -0.50 \\
$P(\text{worn} \mid s, \text{keep})$ & 0.50 & 1.00 \\
\hline
$V^\star(s)$ & 5.3448 & 4.3103 \\
optimal action & keep & replace \\
$d^{\pi^\star}(s)$ & 0.6667 & 0.3333 \\
$d^{\mu}(s)$ & 0.9474 & 0.0526 \\
feature $\phi(s)$ & 1.0 & 2.1 \\
\hline
\end{tabular}
\end{table}

\begin{figure}[t]
\centering
\begin{tikzpicture}[>=Stealth, semithick]
\node[circle, draw=rlblack, minimum size=1.15cm] (g) at (0,0) {good};
\node[circle, draw=rlblack, minimum size=1.15cm] (w) at (5,0) {worn};
\draw[->, rlblue] (g) to[loop left, looseness=6] node[left, align=right, font=\footnotesize] {keep\\$0.5$, $r = 1$} (g);
\draw[->, rlblue] (g) to[bend left=22] node[above, font=\footnotesize] {keep, $0.5$} (w);
\draw[->, rlorange] (w) to[bend left=22] node[below, font=\footnotesize] {replace, $1$, $r = -0.5$} (g);
\draw[->, rlblue] (w) to[loop right, looseness=6] node[right, align=left, font=\footnotesize] {keep\\$1$, $r = 0.2$} (w);
\draw[->, rlorange] (g) to[loop above, looseness=6] node[above, font=\footnotesize] {replace, $1$, $r = -0.5$} (g);
\end{tikzpicture}
\caption{The Engine Replacement MDP as a labelled directed graph. Each arc carries its action and its transition probability, and, where the action is first named from that state, its reward. Blue arcs are the keep action and orange arcs the replace action. The optimal policy takes the blue self-loop at the good state and the orange arc at the worn state.}
\label{fig:prelim_mdp_graph}
\end{figure}

\FloatBarrier
\subsection{Operators, Contraction, and Fixed Points}
\label{prelim:operators}

The question here is whether an algorithm that repeatedly applies an update has a target at all, and whether it reaches it. For dynamic programming the answer is that the update is a contraction, a contraction has exactly one fixed point, and the iterates approach it geometrically.

\FloatBarrier
\subsubsection{Lipschitz Constants and Contractions}

The \emph{Lipschitz constant} is the steepest slope a function ever reaches. The line $f(x) = 2x$ has slope $2$ everywhere, so it is $2$-Lipschitz. The wave $\sin x$ has slope at most $1$, so it is $1$-Lipschitz. A \emph{contraction} is the case where that steepest slope stays below one. That is exactly what makes an iteration pull points together instead of driving them apart.

\begin{theorem}[Lipschitz continuity, \citet{nesterov2018}]
\label{thm:prelim_lipschitz}
A map $f$ between normed spaces is \emph{$L$-Lipschitz} if $\|f(x) - f(y)\| \leq L\|x - y\|$ for all $x, y$. For a differentiable real-valued $f$ on $\mathbb{R}^n$ the least such constant is the supremum of the gradient norm,
\begin{equation}
L = \sup_x \|\nabla f(x)\|,
\label{eq:prelim_lipschitz}
\end{equation}
and for a linear map $A$ it is the induced operator norm $\|A\| = \sup_{x \neq 0}\|Ax\| / \|x\|$.
\end{theorem}

Two consequences for a row-stochastic $P$, both in the supremum norm and both exact rather than merely upper bounds. The policy-evaluation operator $T V = r + \gamma P V$ is $\gamma$-Lipschitz, and the resolvent solve $r \mapsto (I - \gamma P)^{-1} r$ is $\tfrac{1}{1-\gamma}$-Lipschitz \citep{puterman1994}. The norm has to be named, because the second constant is specific to it. With $P = \left[\begin{smallmatrix} 0 & 1 \\ 0 & 1 \end{smallmatrix}\right]$ and $\gamma = 0.9$ the resolvent is $\left[\begin{smallmatrix} 1 & 9 \\ 0 & 10 \end{smallmatrix}\right]$, whose supremum-norm constant is exactly $10 = 1/(1-\gamma)$ while its Euclidean constant is $13.47$.

A map is a \emph{$\gamma$-contraction} on a metric space $(X,d)$ when $d(Tx, Ty) \leq \gamma\, d(x,y)$ for all $x, y$ with a fixed $\gamma \in [0,1)$, the special case $L = \gamma < 1$ of the definition above. It pulls every pair of points strictly closer, by at least the factor $\gamma$, each time it is applied.

\begin{theorem}[Banach fixed-point theorem, \citet{banach1922}]
\label{thm:prelim_banach}
Let $(X, d)$ be a complete metric space.\footnote{A \emph{metric space} is a set $X$ with a distance $d(x,y)$ between points. It is \emph{complete} if every sequence whose terms bunch up without limit, a Cauchy sequence, actually has a limit inside $X$. The space $\mathbb{R}^n$ under any norm is complete, so this hypothesis is automatic everywhere the survey applies the theorem.} Let $T : X \to X$ be a $\gamma$-contraction. Then $T$ has a unique fixed point $x^\star = Tx^\star$, and for any starting point $x_0$ the iterates $x_{k+1} = Tx_k$ satisfy
\begin{equation}
d(x_k, x^\star) \leq \gamma^k\, d(x_0, x^\star).
\label{eq:prelim_banach_rate}
\end{equation}
\end{theorem}

The scalar case shows the mechanism. Take $T(x) = \tfrac{1}{2}x + 1$ on the real line. Solving $x = \tfrac{1}{2}x + 1$ gives the fixed point $x^\star = 2$. Start anywhere, say $x_0 = 0$, and the iterates are $1, 1.5, 1.75, \ldots$, each closing half the remaining gap to $2$. The distance to $x^\star$ halves every step, which is $\gamma^k$ with $\gamma = \tfrac{1}{2}$.

\begin{proof}
Show the iterates form a Cauchy sequence, so completeness gives a limit, then read off that the limit is the unique fixed point and that the error shrinks by $\gamma$ each step. First bound the distance between two consecutive iterates.\footnote{An \emph{iterate} is one term $x_k$ of the sequence built by applying $T$ over and over, starting from $x_0$ and setting $x_{k+1} = Tx_k$.} Applying the contraction to $x_k$ and $x_{k-1}$,
\begin{align*}
d(x_{k+1}, x_k) = d(Tx_k, Tx_{k-1})
  &\leq \gamma\, d(x_k, x_{k-1}) && \text{($T$ is a $\gamma$-contraction)} \\
  &\leq \gamma^k\, d(x_1, x_0) && \text{(apply the same step $k$ times).}
\end{align*}
Now bound the distance between any two iterates $x_m$ and $x_k$ with $m > k$.
\begin{align*}
d(x_m, x_k)
  &\leq \sum_{j=k}^{m-1} d(x_{j+1}, x_j) && \text{(triangle inequality)} \\
  &\leq d(x_1, x_0) \sum_{j=k}^{\infty} \gamma^j && \text{(previous bound, extend the sum)} \\
  &= \frac{\gamma^k}{1-\gamma}\, d(x_1, x_0) && \text{(geometric series).}
\end{align*}
The right-hand side tends to $0$ as $k \to \infty$. The iterates therefore bunch up without limit, so the sequence is Cauchy. Completeness gives a limit $x^\star \in X$.\footnote{This is the only place completeness is used, and it cannot be dropped. On the incomplete space $X = (0,1]$ the map $Tx = x/2$ is a contraction, yet its iterates march toward $0$, which is not in $X$, so no fixed point exists.} A contraction is continuous, so it commutes with the limit,
\begin{align*}
Tx^\star = T\Bigl(\lim_k x_k\Bigr) = \lim_k Tx_k = \lim_k x_{k+1} = x^\star .
\end{align*}
Thus $x^\star$ is a fixed point, a point the map leaves unchanged. For uniqueness, suppose $x^\star$ and $y^\star$ are both fixed points. Then
\begin{align*}
d(x^\star, y^\star) = d(Tx^\star, Ty^\star) &\leq \gamma\, d(x^\star, y^\star) && \text{(contraction),}
\end{align*}
so $(1-\gamma)\, d(x^\star, y^\star) \leq 0$. Since $\gamma < 1$, this forces $d(x^\star, y^\star) = 0$, and the two points coincide. Finally the rate. Apply the contraction to $x_{k-1}$ and the fixed point,
\begin{align*}
d(x_k, x^\star) = d(Tx_{k-1}, Tx^\star)
  &\leq \gamma\, d(x_{k-1}, x^\star) && \text{(contraction)} \\
  &\leq \gamma^k\, d(x_0, x^\star) && \text{(apply the same step $k$ times),}
\end{align*}
which is~\eqref{eq:prelim_banach_rate}.
\end{proof}

This proof earns its space because the rate is not a separate assumption. It falls out of the contraction inequality applied $k$ times, which is why every convergence statement in this survey that names a rate names $\gamma^k$.

\subsubsection{The Bellman Operators}

Fix a policy $\pi$. The \emph{policy-evaluation operator}, or Bellman expectation operator, is
\begin{equation}
T^\pi V = r^\pi + \gamma P^\pi V,
\label{eq:prelim_tpi}
\end{equation}
one backup of the return under $\pi$. In the vector language of Section~\ref{prelim:mdp} it is affine in $V$, a fixed matrix times $V$ plus a fixed vector, the object of Section~\ref{prelim:affine_envelope} with $V$ in place of $x$. It is $\gamma$-Lipschitz in the supremum norm by Theorem~\ref{thm:prelim_lipschitz}, because $T^\pi V - T^\pi V' = \gamma P^\pi (V - V')$ and a row-stochastic matrix has $\|P^\pi\|_\infty = 1$, so averaging can never enlarge the biggest coordinate.\footnote{The \emph{supremum operator norm} $\|B\|_\infty = \max_i \sum_j |B_{ij}|$ is the largest absolute row sum. Every row of $P^\pi$ sums to one and every entry is nonnegative, so $\|P^\pi\|_\infty = 1$ exactly.}

The \emph{Bellman optimality operator} maximizes over actions at each state,
\begin{equation}
(T^\star V)(s) = \max_{a} \Bigl\{ r(s,a) + \gamma \sum_{s'} P(s' \mid s,a) V(s') \Bigr\}
= \max_{\pi} (T^\pi V)(s).
\label{eq:prelim_tstar}
\end{equation}
The second equality is the point. The optimality operator is the pointwise upper envelope of the affine policy operators, so it is convex and piecewise-linear in $V$ with a kink wherever the maximizing action changes. It is neither linear nor differentiable, so it cannot inherit a contraction modulus from matrix algebra and needs its own argument. Equation~\eqref{eq:prelim_maxdiff} supplies it. Applying that inequality at each state to the two bracketed expressions, the reward terms cancel and
\begin{align*}
\bigl|(T^\star V)(s) - (T^\star V')(s)\bigr|
  &\leq \max_a \gamma \Bigl| \sum_{s'} P(s' \mid s,a)\bigl(V(s') - V'(s')\bigr) \Bigr|
  \leq \gamma \|V - V'\|_\infty ,
\end{align*}
where the last step is again that each row of $P$ averages. Taking the supremum over $s$ makes $T^\star$ a $\gamma$-contraction in the supremum norm, with the same modulus as every $T^\pi$ it envelopes.

Theorem~\ref{thm:prelim_banach} now delivers the two facts the survey uses constantly. Each $T^\pi$ has a unique fixed point, which is $V^\pi$. And $T^\star$ has a unique fixed point, which is $V^\star$. Neither existence nor uniqueness needs a separate argument.

\subsubsection{Value Iteration}

\emph{Value iteration} is the Picard iteration $V_{k+1} = T^\star V_k$ from an arbitrary start.\footnote{\emph{Picard iteration} is the generic name for repeatedly applying a map to its own output. Value iteration, policy evaluation by repeated backup, and the gradient step of Section~\ref{prelim:convex} are all instances.} Equation~\eqref{eq:prelim_banach_rate} gives $\|V_k - V^\star\|_\infty \leq \gamma^k \|V_0 - V^\star\|_\infty$, so reaching a tolerance $\varepsilon$ takes on the order of $\log(1/\varepsilon) / \log(1/\gamma)$ iterations, and since $\log(1/\gamma) \approx 1 - \gamma$ for $\gamma$ near one, that count grows like the \emph{effective horizon} $1/(1-\gamma)$.

The Engine Replacement MDP makes the rate concrete. Starting from $V_0 = 0$, the first four iterates are $(1, 0.2)$, $(1.54, 0.4)$, $(1.873, 0.886)$, and $(2.2416, 1.1857)$, with supremum-norm errors $4.3448$, $3.9103$, $3.4718$, and $3.1246$. The per-step ratios are $0.813$, $0.900$, $0.888$, and $0.900$. Every ratio is at most $\gamma = 0.9$, as Theorem~\ref{thm:prelim_banach} requires. The ratios differ because $\gamma P^{\pi^\star}$ has a second mode of modulus $0.45$ that decays faster than the first, so the contraction bounds each step without fixing it.\footnote{Theorem~\ref{thm:prelim_banach} is an upper bound on $d(x_k, x^\star)$, not an identity. A reader who expects the error to fall by exactly $\gamma$ every step, and finds it falling faster, has met the second mode rather than an error.} The greedy policy is (keep, keep) after one step and settles at the optimal (keep, replace) from the second step onward, long before the values themselves are close, which is why policy iteration can finish while value iteration is still moving.

\begin{figure}[t]
\centering
\includegraphics[width=0.95\textwidth]{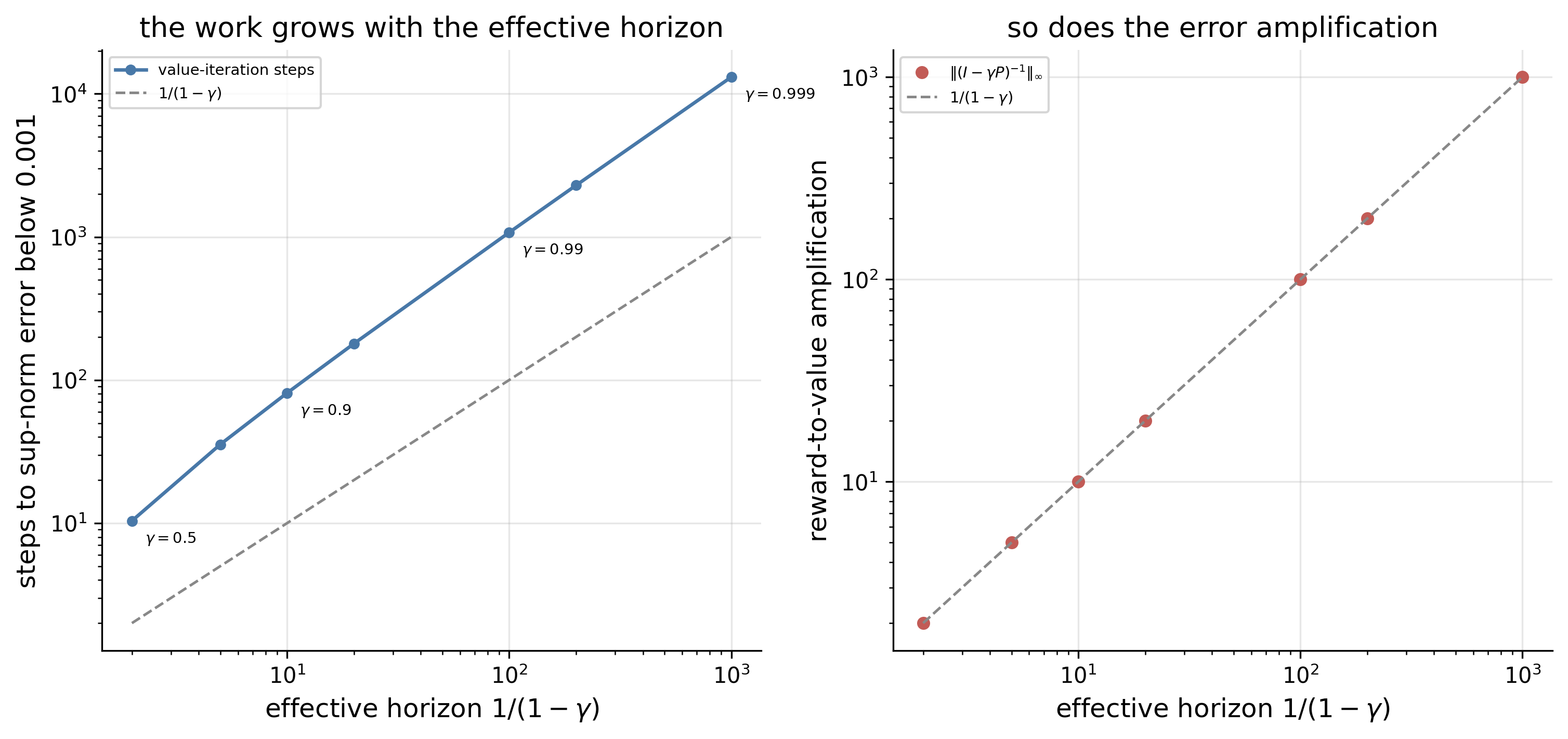}
\caption{Left: backups of $T^\pi$ from $V_0 = 0$ needed to reach a supremum-norm error below $10^{-3}$ on random fifty-state Markov reward processes, against the effective horizon, on log-log axes with the identity for reference. Right: the reward-to-value amplification $\|(I - \gamma P)^{-1}\|_\infty$ against the same axis, which coincides with the effective horizon exactly.}
\label{fig:prelim_discount_cost}
\end{figure}

The cost of a discount factor near one is easy to state and easy to underestimate. Reaching a supremum-norm error below $0.01$ in this example takes $6$ iterations at $\gamma = 0.5$, $59$ at $\gamma = 0.9$, and $847$ at $\gamma = 0.99$. The same pattern holds on larger problems. On random fifty-state Markov reward processes, reaching an error below $10^{-3}$ takes $10$ backups at $\gamma = 0.5$, $81$ at $\gamma = 0.9$, $1075$ at $\gamma = 0.99$, and $13{,}097$ at $\gamma = 0.999$ (Table~\ref{tab:prelim_discount_cost}). The count grows slightly faster than the effective horizon itself, because it is $\log(\|V^\pi\|_\infty/\varepsilon)/\log(1/\gamma)$ and the value function also grows as the discount rises. Figure~\ref{fig:prelim_envelope_running} draws the operator being iterated, in the slice through the worn state, and Figure~\ref{fig:prelim_discount_cost} puts the two costs of a long horizon side by side.

\begin{table}[h]
\centering
\caption{What the discount factor costs, on random 50-state Markov reward processes over 30 seeds with rewards in $[0,1]$. Steps are backups of $T^\pi$ from $V_0 = 0$ to a supremum-norm error below 0.001. Amplification is $\|(I - \gamma P)^{-1}\|_\infty$, the factor by which an error in the reward is magnified in the value, which for a row-stochastic $P$ equals the effective horizon exactly.}
\label{tab:prelim_discount_cost}
\begin{tabular}{rrrrr}
\hline
$\gamma$ & $1/(1-\gamma)$ & steps & amplification & $\|V^\pi\|_\infty$ \\
\hline
0.500 & 2.0 & 10 & 2.0 & 1.5 \\
0.800 & 5.0 & 35 & 5.0 & 3.0 \\
0.900 & 10.0 & 81 & 10.0 & 5.4 \\
0.950 & 20.0 & 180 & 20.0 & 10.3 \\
0.990 & 100.0 & 1075 & 100.0 & 49.7 \\
0.995 & 200.0 & 2293 & 200.0 & 98.9 \\
0.999 & 1000.0 & 13097 & 1000.0 & 492.6 \\
\hline
\end{tabular}
\end{table}

\emph{Policy iteration} instead alternates evaluating the current policy exactly and acting greedily with respect to that evaluation. In the envelope picture each evaluation solves for the fixed point of one affine piece, which is a Newton step on $V = T^\star V$, and the finite number of pieces is why it terminates finitely. Section~\ref{sec:newton_connection} develops that connection, and this appendix owes only the geometry it rests on.

\begin{figure}[t]
\centering
\begin{tikzpicture}[>=Stealth, x=3.2cm, y=1.9cm]
\draw[->] (3.4,3.4) -- (5.75,3.4) node[right, font=\footnotesize] {$t = V(\text{worn})$};
\draw[->] (3.4,3.4) -- (3.4,5.75) node[above, font=\footnotesize] {value};
\foreach \x in {3.5,4.0,4.5,5.0,5.5} \draw (\x,3.4) -- (\x,3.36) node[below, font=\scriptsize] {\x};
\foreach \y in {3.5,4.0,4.5,5.0,5.5} \draw (3.4,\y) -- (3.36,\y) node[left, font=\scriptsize] {\y};
\draw[rlgray, thin] (3.45,3.45) -- (5.6,5.6);
\node[rlgray, font=\scriptsize, anchor=south east] at (5.62,5.58) {$45^\circ$};
\draw[rlblue, dashed, thick] (3.5556,3.4) -- (5.6,5.24);
\node[rlblue, font=\scriptsize, anchor=north west] at (3.86,3.60) {keep, $0.2 + 0.9\,t$};
\draw[rlorange, dashed, thick] (3.4,4.3103) -- (5.6,4.3103);
\node[rlorange, font=\scriptsize, anchor=south west] at (3.45,4.35) {replace, $4.3103$};
\draw[rlblack, very thick] (3.4,4.3103) -- (4.5670,4.3103) -- (5.6,5.24);
\filldraw[rlblack] (4.5670,4.3103) circle (1.7pt);
\draw[rlblack, thin] (4.5670,4.3103) -- (5.02,3.97);
\node[rlblack, font=\scriptsize, anchor=west] at (5.04,3.97) {kink, $t = 4.567$};
\filldraw[rlred] (4.3103,4.3103) circle (1.9pt);
\draw[rlred, thin] (4.3103,4.3103) -- (4.70,3.63);
\node[rlred, font=\scriptsize, anchor=west] at (4.72,3.63) {$V^\star(\text{worn}) = 4.3103$};
\end{tikzpicture}
\caption{The Bellman optimality operator at the worn state of the Engine Replacement MDP, along the slice $V = (V^\star(\text{good}), t)$. The keep action gives the affine map $0.2 + 0.9\,t$ and the replace action the constant $4.3103$. Their upper envelope, in black, is $T^\star$ restricted to this slice, and it kinks at $t = 4.567$ where the greedy action changes. The fixed point, where the envelope meets the $45^\circ$ line, sits at $4.3103$ on the replace branch, $0.257$ below the kink.}
\label{fig:prelim_envelope_running}
\end{figure}

\subsection{The Linear Algebra of Policy Evaluation}
\label{prelim:policy_evaluation}

Section~\ref{prelim:operators} reached $V^\pi$ by iterating $T^\pi$. For a fixed policy there is no need to iterate at all, because the fixed-point equation is linear and can be solved outright. Writing $V = r^\pi + \gamma P^\pi V$ and collecting terms,
\begin{equation}
(I - \gamma P^\pi) V^\pi = r^\pi .
\label{eq:prelim_evaluation_system}
\end{equation}
This section is about the matrix $(I - \gamma P^\pi)^{-1}$, called the \emph{resolvent}, which is where the factor $1/(1-\gamma)$ in every error bound in the survey comes from.

\subsubsection{The Neumann Series}

Start with one state. Then $P = [1]$, and $I - \gamma P$ is just the number $1 - \gamma$, so its inverse is $(1-\gamma)^{-1} = 1 + \gamma + \gamma^2 + \cdots$. The value of that single state is $r/(1-\gamma)$, the reward collected now, then again discounted by $\gamma$, then by $\gamma^2$, and so on forever. The theorem is this same geometric series, with the matrix $\gamma P$ standing in for the number $\gamma$.

\begin{theorem}[Neumann series, \citet{hornjohnson2013}]
\label{thm:prelim_neumann}
Let $P$ be a row-stochastic matrix and $\gamma \in [0, 1)$. Then $I - \gamma P$ is invertible and
\begin{equation}
(I - \gamma P)^{-1} = \sum_{m=0}^{\infty} (\gamma P)^m,
\label{eq:prelim_neumann}
\end{equation}
with the truncation error of the first $M{+}1$ terms bounded, in the supremum operator norm, by
\begin{equation}
\Bigl\| (I - \gamma P)^{-1} - \sum_{m=0}^{M} (\gamma P)^m \Bigr\|_\infty \leq \frac{\gamma^{M+1}}{1 - \gamma}.
\label{eq:prelim_neumann_bound}
\end{equation}
\end{theorem}

The identity is why a value function is a discounted sum of rewards.\footnote{The exact value $V^\pi = (I - \gamma P^\pi)^{-1} r^\pi$ equals $\sum_m \gamma^m (P^\pi)^m r^\pi$, which is the expected discounted return written out. Truncating the series at $M$ terms is the same thing as truncating the planning horizon at $M$ steps, and~\eqref{eq:prelim_neumann_bound} prices that truncation.} Two consequences are used repeatedly. Every row of the resolvent sums to exactly $1/(1-\gamma)$, because $(I - \gamma P)^{-1}\mathbf{1} = \sum_m \gamma^m \mathbf{1} = \tfrac{1}{1-\gamma}\mathbf{1}$, so the resolvent has supremum operator norm exactly $1/(1-\gamma)$. That number is the \emph{effective horizon}, the number of periods that matter, and it is the amplification factor by which any error in $r^\pi$ is magnified in $V^\pi$.

In the Engine Replacement MDP, evaluating the optimal policy means inverting
\begin{equation*}
I - \gamma P^{\pi^\star} = \begin{pmatrix} 0.55 & -0.45 \\ -0.9 & 1 \end{pmatrix},
\qquad
\det = 0.145,
\qquad
(I - \gamma P^{\pi^\star})^{-1} = \begin{pmatrix} 6.8966 & 3.1034 \\ 6.2069 & 3.7931 \end{pmatrix}.
\end{equation*}
Both rows sum to $10$, which is $1/(1-0.9)$, exactly as the identity requires. Multiplying by $r^{\pi^\star} = (1, -0.5)$ returns $V^{\pi^\star} = (5.3448, 4.3103)$, the same vector value iteration was approaching in Section~\ref{prelim:operators}. Truncating the series instead gives $(1.2250, 0.4000)$ at $M = 1$, $(2.0502, 1.0581)$ at $M = 3$, and $(3.7758, 2.7412)$ at $M = 10$, with supremum-norm errors $4.1198$, $3.2946$, and $1.5692$ against bounds of $8.1000$, $6.5610$, and $3.1381$ from~\eqref{eq:prelim_neumann_bound}. Even at ten terms the truncated value is still a third short, which is the same statement as the $59$ iterations Section~\ref{prelim:operators} needed at this discount factor.

\subsubsection{Spectral Radius against Operator Norm}

The resolvent exists because $\gamma P$ has all its eigenvalues inside the unit disc. That condition, and not the size of the entries, is what governs whether repeated multiplication dies away.

A tiny case first. Take the diagonal matrix $A = \operatorname{diag}(0.9, 0.5)$, whose powers are $A^k = \operatorname{diag}(0.9^k, 0.5^k)$. Both entries sit below one, so both shrink to zero, and the larger one sets the pace. Now tilt one entry off the diagonal, $A = \left[\begin{smallmatrix} 0.9 & 2 \\ 0 & 0.9 \end{smallmatrix}\right]$. This matrix is large, so its powers grow at first. They still vanish in the end, because the eigenvalues are still $0.9$. The eigenvalues, not the size of the entries, decide the fate of the powers.

\begin{theorem}[Spectral radius governs decay, \citet{hornjohnson2013}]
\label{thm:prelim_spectral}
Let $A$ be a square matrix with spectral radius $\rho(A) = \max_i |\lambda_i|$, the largest modulus among its eigenvalues.\footnote{An \emph{eigenvalue} $\lambda$ of $A$ is a number for which $Av = \lambda v$ has a nonzero solution $v$, the \emph{eigenvector}. Applying $A$ to $v$ merely rescales it by $\lambda$. The \emph{spectral radius} is the largest such $|\lambda|$.} Then the matrix powers vanish, $A^k \to 0$ as $k \to \infty$, if and only if $\rho(A) < 1$. In that case $\|A^k\|^{1/k} \to \rho(A)$.
\end{theorem}

\begin{figure}[t]
\centering
\includegraphics[width=0.78\textwidth]{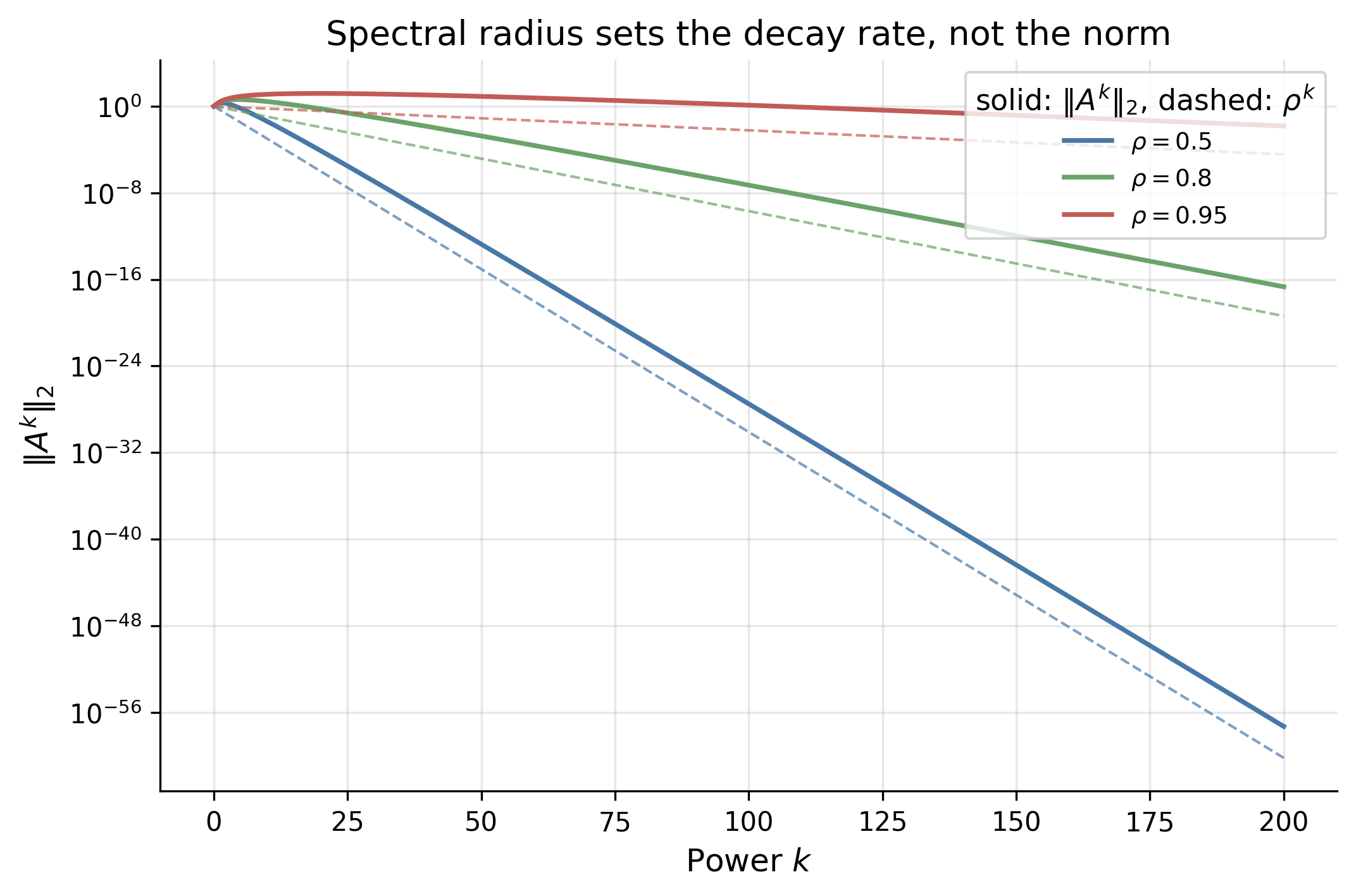}
\caption{Operator norm $\|A^k\|_2$ of the powers of the non-normal matrix $A = \left[\begin{smallmatrix} \rho & 2 \\ 0 & \rho \end{smallmatrix}\right]$, on a log scale. The solid curves grow to a transient peak before decaying. The dashed curves are $\rho^k$. Both share the asymptotic slope, so $\rho$ sets the decay rate even though $\|A\|_2 > 1$.}
\label{fig:prelim_spectral}
\end{figure}

The distinction between $\rho(A)$ and the norm $\|A\|$ matters in practice.\footnote{The spectral radius sets the target-network and composed-operator stability conditions of Section~\ref{sec:planning_learning} and the low-rank and manipulation-cost arguments of Section~\ref{section:bandits}.} A matrix can have $\|A\| > 1$, so its powers grow at first. They still decay in the end, because $\rho(A) < 1$. Non-normality permits transient growth before the asymptotic decay begins. Figure~\ref{fig:prelim_spectral} shows this transient for a non-normal matrix, one whose eigenvectors are not orthogonal. The powers rise to a peak, then fall, and the per-step decay ratio settles at $\rho(A)$ (Table~\ref{tab:prelim_spectral_radius}). The two quantities answer different questions, and Section~\ref{prelim:approximation} turns on exactly this gap.

\begin{table}[h]
\centering
\caption{Powers of the non-normal matrix $A = \left[\begin{smallmatrix} \rho & 2 \\ 0 & \rho \end{smallmatrix}\right]$. The operator norm exceeds $\rho$ and the powers first grow, yet they decay asymptotically at rate $\rho$. The tail decay ratio $\|A^{k+1}\|/\|A^k\|$ approaches $\rho$ from above, with a benign $+O(1/k)$ finite-sample bias.}
\label{tab:prelim_spectral_radius}
\begin{tabular}{ccccc}
\hline
$\rho(A)$ & $\|A\|_2$ & Transient peak & at $k$ & Tail decay ratio \\
\hline
0.5 & 2.118 & 2.12 & 1 & 0.5027 \\
0.8 & 2.281 & 4.14 & 4 & 0.8043 \\
0.95 & 2.379 & 15.10 & 19 & 0.9552 \\
\hline
\end{tabular}
\end{table}

\subsection{Markov Chains and Where the Data Comes From}
\label{prelim:markov_data}

Sections~\ref{prelim:operators} and~\ref{prelim:policy_evaluation} assumed the model was known. A learning agent instead sees states as they arrive. This section is about which states arrive, how that differs from the states the agent cares about, and how to price the difference. It is the last section before function approximation, and it supplies the two distributions that Section~\ref{prelim:approximation} weighs against each other.

\subsubsection{The Stationary Distribution}

Take two states with a $30\%$ chance of flipping between them each step, and start all the probability on state one. After one step the split is $(0.7, 0.3)$, after another $(0.58, 0.42)$, and it keeps sliding toward the balance point $d^\star = (0.5, 0.5)$, then stops. Where the chain began is forgotten. The theorem says every such chain has exactly one balance point, reaches it from any start, and it pins down the speed.

\begin{theorem}[Perron-Frobenius and mixing, \citet{levinperes2017}]
\label{thm:prelim_markov}
Let $P$ be the transition matrix of an irreducible, aperiodic Markov chain on a finite state space.\footnote{\emph{Irreducible} means every state can eventually reach every other state. \emph{Aperiodic} means the chain does not cycle with a fixed period. Together they make the chain \emph{primitive}, so that some power $P^k$ has all entries positive.} Then $P$ has a unique stationary distribution $d^\star$, meaning $d^\star P = d^\star$ with $d^\star \geq 0$ and $\sum_s d^\star_s = 1$, and for every initial distribution $d_0$ the iterates $d_k = d_0 P^k$ converge to $d^\star$. If in addition $P$ is diagonalizable, the total-variation distance, the largest gap between the probabilities the two distributions assign to any single event, obeys
\begin{equation}
\| d_0 P^k - d^\star \|_{\mathrm{TV}} \leq C\, |\lambda_2|^k,
\label{eq:prelim_markov_rate}
\end{equation}
where $\lambda_2$ is the second-largest eigenvalue in modulus and $C$ depends on the chain.\footnote{Diagonalizability cannot be dropped from~\eqref{eq:prelim_markov_rate} as stated. If $P$ has a Jordan block of size $m$ at modulus $|\lambda_2|$, the correct bound carries a polynomial factor, $C\,k^{m-1}|\lambda_2|^k$, and no finite $C$ makes the clean form hold. A three-state chain with all entries positive and a defective eigenvalue at $0.5$ has $\|d_0P^k - d^\star\|_{\mathrm{TV}} / |\lambda_2|^k$ equal to $1.96$ at $k=5$ and $12.57$ at $k=40$, growing without bound. The clean form is safe whenever the chain is reversible, since a reversible chain is diagonalizable, and every chain satisfies $\|d_0P^k - d^\star\|_{\mathrm{TV}} \leq C_\varepsilon(|\lambda_2| + \varepsilon)^k$ for each $\varepsilon > 0$.}
\end{theorem}

The quantity $1 - |\lambda_2|$ is the \emph{spectral gap}, and it sets how fast the chain forgets its start, much as a lazily shuffled deck keeps its original order far longer than a thoroughly shuffled one. In the Engine Replacement MDP the optimal policy induces $P^{\pi^\star}$ with stationary distribution $d^{\pi^\star} = (0.6667, 0.3333)$ and $|\lambda_2| = 0.5$, so the chain spends two thirds of its time with a good machine and forgets its start within a handful of steps.\footnote{For a $2 \times 2$ stochastic matrix the second eigenvalue is the trace minus one, so $\lambda_2 = 0.5 + 0 - 1 = -0.5$ here. The negative sign makes the approach to $d^\star$ alternate rather than monotone.}

\subsubsection{Discounted Occupancy}

The stationary distribution describes the long run with no discounting. Discounted objectives weigh early visits more, and the matching object is built from the resolvent of Section~\ref{prelim:policy_evaluation}. For an initial distribution $\nu$, the \emph{normalized discounted occupancy} is
\begin{equation}
d^\pi_\nu = (1-\gamma)\, \nu\, (I - \gamma P^\pi)^{-1},
\qquad \text{equivalently} \qquad
d^\pi_\nu = (1-\gamma)\nu + \gamma\, d^\pi_\nu P^\pi .
\label{eq:prelim_occupancy}
\end{equation}
Expanding the resolvent by~\eqref{eq:prelim_neumann} shows what it measures, $d^\pi_\nu(s) = (1-\gamma)\sum_{t \geq 0} \gamma^t \mathbb{P}_\pi(S_t = s)$, the discounted share of visits to $s$. The factor $(1-\gamma)$ is exactly what makes the entries sum to one, since the rows of the resolvent sum to $1/(1-\gamma)$. The \emph{state-action occupancy} is $d^\pi_\nu(s,a) = d^\pi_\nu(s)\,\pi(a \mid s)$. Every discounted expectation in the survey can be written against this measure, which is why it appears in policy-gradient and off-policy bounds alike.

In the Engine Replacement MDP, starting from a good machine, the optimal policy has $d^{\pi^\star}_\nu = (0.6897, 0.3103)$, close to but not equal to the undiscounted $(0.6667, 0.3333)$, because discounting tilts weight toward the starting state.

\subsubsection{Coverage, Change of Measure, and Concentrability}

An agent learning offline does not choose which states it sees. The data is generated by a \emph{behavior}, or logging, policy $\mu$, while the quantity of interest concerns a \emph{target} policy $\pi$. Suppose the logging policy in the Engine Replacement MDP is a conservative maintenance crew that replaces the machine on nine inspections in ten, so $\mu$ keeps with probability $0.1$ in either state. Its chain is $P^\mu = \left[\begin{smallmatrix} 0.95 & 0.05 \\ 0.90 & 0.10 \end{smallmatrix}\right]$ with stationary distribution $d^\mu = (0.9474, 0.0526)$ and discounted occupancy $d^\mu_\nu = (0.9529, 0.0471)$. The crew almost never lets a machine become worn, so worn states are barely represented in the log, while the target policy spends nearly a third of its discounted time there.

Any expectation under the target can be rewritten as an expectation under the log, provided the log reaches everywhere the target goes. Writing $d^\pi_\nu \ll \mu$ for that support condition, called \emph{absolute continuity},
\begin{equation}
\mathbb{E}_{d^\pi_\nu}\bigl[ f(S,A) \bigr]
= \mathbb{E}_{\mu}\bigl[ w^\pi(S,A)\, f(S,A) \bigr],
\qquad
w^\pi(s,a) = \frac{d^\pi_\nu(s,a)}{\mu(s,a)} .
\label{eq:prelim_change_of_measure}
\end{equation}
This is \emph{change of measure}, and $w^\pi$ is the \emph{importance ratio}. The identity is exact whenever the support condition holds, and it is the reason offline evaluation is possible at all.

Support is qualitative and says only that the ratio is finite somewhere. The quantitative version is the \emph{concentrability coefficient}
\begin{equation}
C_\infty(\pi, \mu) = \Bigl\| \frac{d^\pi_\nu}{\mu} \Bigr\|_\infty
= \sup_{s,a\, :\, \mu(s,a) > 0} \frac{d^\pi_\nu(s,a)}{\mu(s,a)},
\label{eq:prelim_concentrability}
\end{equation}
which enters the error bounds of approximate dynamic programming as a multiplicative penalty \citep{MunosSzepesvari2008}. A large but finite $C_\infty$ leaves the estimate consistent and inflates its variance. A pair with $d^\pi_\nu(s,a) > 0$ and $\mu(s,a) = 0$ makes the corresponding contribution unrecoverable from the log at any sample size, no matter how the estimator is built.

\begin{figure}[t]
\centering
\includegraphics[width=0.8\textwidth]{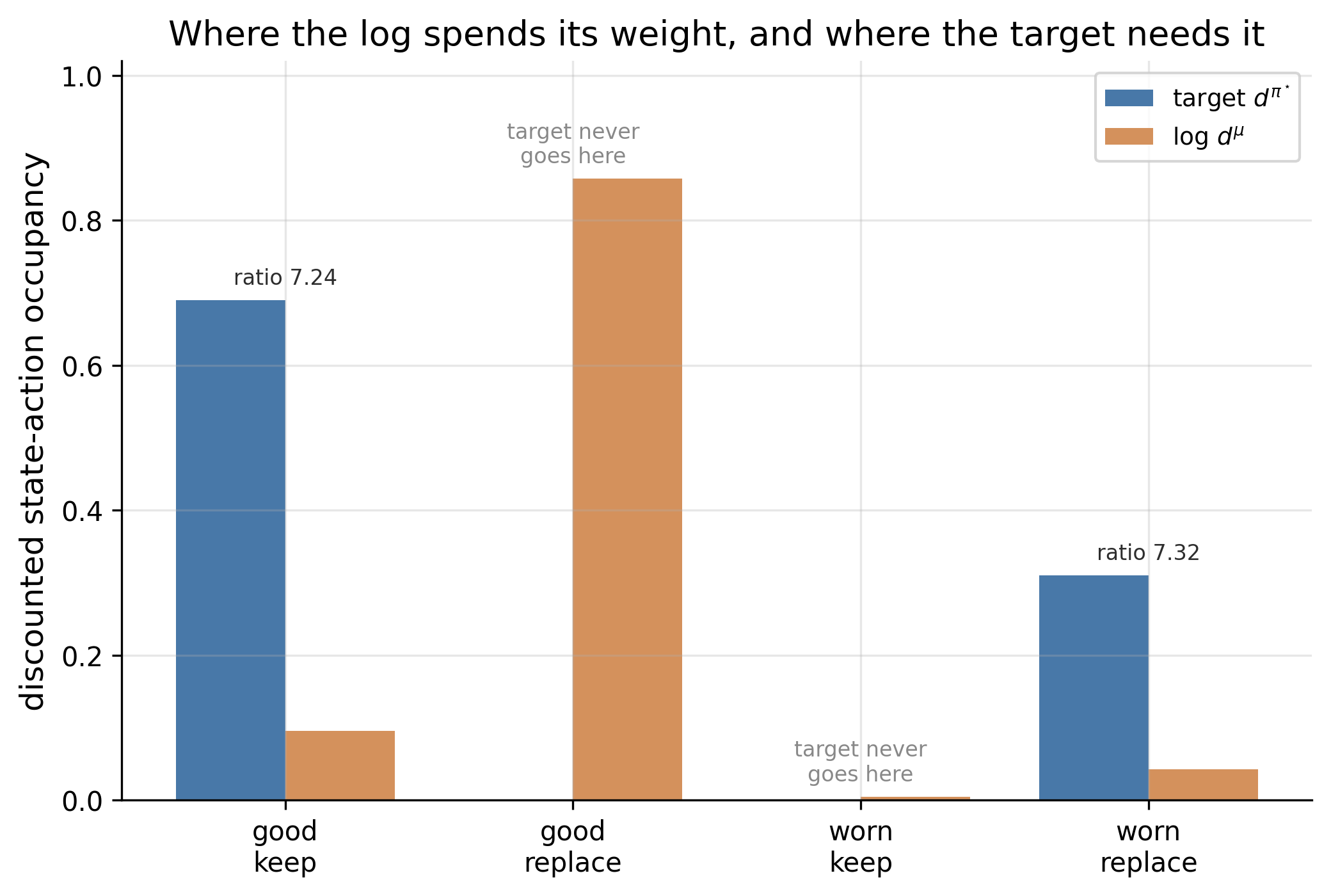}
\caption{Discounted state-action occupancy under the target policy and under the logging policy, on the Engine Replacement MDP from a good machine. The target uses two of the four pairs. The log puts most of its weight on a pair the target never uses and only a twentieth of it on the pair the target most needs. The support condition holds, so offline evaluation is possible, and the ratios at the two pairs the target uses are what the concentrability coefficient reports.}
\label{fig:prelim_coverage}
\end{figure}

The Engine Replacement MDP puts numbers on this. At the state level the ratios are $w(\text{good}) = 0.7238$ and $w(\text{worn}) = 6.5862$, so $C_\infty = 6.5862$. At the state-action level the target uses only two pairs, and their ratios are $7.2376$ for keeping a good machine and $7.3180$ for replacing a worn one, giving $C_\infty = 7.3180$. Both pairs the target needs are logged, so the support condition holds and offline evaluation is feasible, but the log spends $85.8\%$ of its discounted weight on replacing a good machine, an action the target never takes, and only $4.2\%$ on replacing a worn one, the action the target most needs to learn about.\footnote{This is the ordinary situation rather than a pathology. A logging policy chosen for safety concentrates on the actions the target policy has already ruled out, so the effective sample size for the decision that matters is far smaller than the raw record suggests. Section~\ref{section:offline_rl} and Section~\ref{section:field_deployments} take up what to do about it.} Figure~\ref{fig:prelim_coverage} puts the two occupancies side by side.

Nothing so far has gone wrong. Change of measure is exact, the concentrability coefficient is finite, and the target policy is identified from the log. Section~\ref{prelim:approximation} shows what breaks once the value function can no longer be stored exactly.

\subsection{Approximation Geometry}
\label{prelim:approximation}

Every result so far stored a value function as one number per state. Beyond a few thousand states that is impossible, and the value function has to be approximated inside a small family. This section is about the geometry of that restriction. It is the thorniest material in the appendix, and it is where the norms of Section~\ref{prelim:norms} stop being a matter of bookkeeping.

\subsubsection{Features and the Space They Span}
\label{prelim:span_rank}

A \emph{feature map} $\phi : \mathcal{S} \to \mathbb{R}^d$ attaches $d$ numbers to each state, and stacking them gives the \emph{feature matrix} $\Phi \in \mathbb{R}^{|\mathcal{S}| \times d}$ whose row $s$ is $\phi(s)^\top$. A linear approximation is $V_\theta = \Phi\theta$ for a parameter $\theta \in \mathbb{R}^d$, so the value functions the family can express are exactly the vectors in $\operatorname{span}(\Phi)$, the \emph{column space}, a $d$-dimensional subspace of $\mathbb{R}^{|\mathcal{S}|}$ when $\Phi$ has full column rank.\footnote{The \emph{rank} of $\Phi$ is the number of linearly independent columns. If two features are proportional the rank drops, $\Phi^\top D \Phi$ becomes singular, and the parameter is not identified even though the represented function still is. Adding a ridge term or dropping the redundant feature restores invertibility.} Approximation is therefore the question of how a point of $\mathbb{R}^{|\mathcal{S}|}$ relates to a subspace, and answering it requires a notion of angle.

\begin{figure}[t]
\centering
\includegraphics[width=0.62\textwidth]{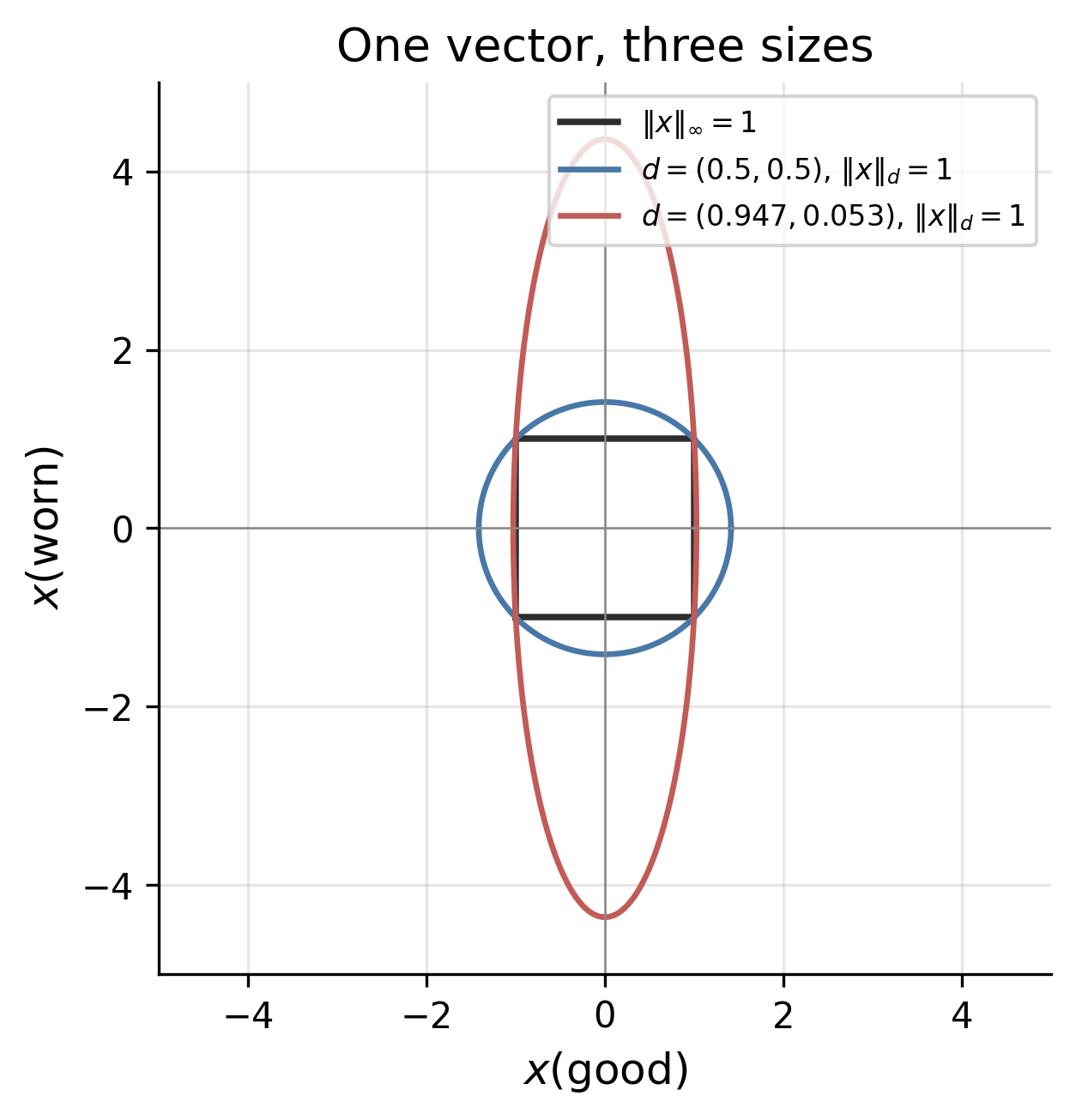}
\caption{Unit balls in the plane. The supremum norm gives a square, and a $d$-weighted $L^2$ norm gives an ellipse with semi-axes $1/\sqrt{d(s)}$, so a state with small weight is stretched far out. Under the logging weighting the ellipse extends to roughly $4.4$ along the worn axis, meaning an error at the worn state has to be four times larger to count as much as an error at the good state.}
\label{fig:prelim_norm_balls}
\end{figure}

An \emph{inner product} supplies one. The \emph{weighted inner product} attached to a distribution $d$ over states is
\begin{equation}
\langle x, y \rangle_d = x^\top D y = \sum_s d(s)\, x(s)\, y(s),
\qquad D = \operatorname{diag}(d),
\label{eq:prelim_inner_product}
\end{equation}
and it induces the weighted norm $\|x\|_d^2 = \langle x, x\rangle_d$ of Section~\ref{prelim:norms}. Two vectors are \emph{orthogonal} in this geometry when $\langle x, y\rangle_d = 0$, which depends on $d$ and not on the vectors alone. Changing the weighting changes which vectors count as perpendicular, and that single fact is the content of this section. It changes lengths too. The vector $x = (1,3)$ has supremum norm $3$, weighted norm $2.236$ under the even weighting $(0.5, 0.5)$, and weighted norm $1.192$ under the logging weighting $(0.947, 0.053)$, which barely registers the second coordinate at all. Figure~\ref{fig:prelim_norm_balls} draws the three unit balls.

\subsubsection{Least Squares and the Projection Matrix}
\label{prelim:least_squares}

\emph{Least squares} finds the best representable approximation to a target $V$,
\begin{equation}
\hat\theta = \argmin_{\theta} \|V - \Phi\theta\|_d^2,
\qquad \text{solved by the normal equations} \qquad
\Phi^\top D\,(V - \Phi\hat\theta) = 0 .
\label{eq:prelim_least_squares}
\end{equation}
The normal equations say the residual is $d$-orthogonal to every feature, which is the defining property of a projection. Solving them gives the \emph{projection matrix}
\begin{equation}
\Pi_d = \Phi\bigl(\Phi^\top D \Phi\bigr)^{-1}\Phi^\top D ,
\label{eq:prelim_projection}
\end{equation}
which is idempotent, $\Pi_d^2 = \Pi_d$, since projecting an already-projected vector changes nothing.

\subsubsection{Projection Does Not Lengthen}

Take the point $x = (1, 1)$ in the plane and drop it straight down onto the horizontal axis. The projection is $\Pi x = (1, 0)$, and the leftover $x - \Pi x = (0, 1)$ points straight up, at a right angle to the axis. The three vectors form a right triangle with $x$ as the hypotenuse, so $\Pi x$ is a leg and cannot be longer than $x$. A projection never lengthens a vector. The theorem repeats this in any number of dimensions and in any inner product.

\begin{theorem}[Projection geometry, \citet{luenberger1969}]
\label{thm:prelim_hilbert}
Let $H$ be an inner-product space, let $M \subseteq H$ be a closed subspace, and let $\Pi$ be the orthogonal projection onto $M$.\footnote{The \emph{orthogonal projection} $\Pi x$ is the point of $M$ closest to $x$. Its defining property is that the residual $x - \Pi x$ is orthogonal to every vector in $M$, written $x - \Pi x \perp M$.} Then for every $x \in H$ the Pythagorean identity
\begin{equation}
\|x\|^2 = \|\Pi x\|^2 + \|x - \Pi x\|^2
\label{eq:prelim_pythagoras}
\end{equation}
holds, and consequently $\Pi$ is nonexpansive, $\|\Pi x - \Pi y\| \leq \|x - y\|$.
\end{theorem}

\begin{proof}
Split $x$ into its projection and its residual, use that the two are orthogonal, and expand the squared norm. By definition $\Pi x \in M$ and the residual satisfies $x - \Pi x \perp M$, so in particular the two pieces are orthogonal and $\langle \Pi x, x - \Pi x \rangle = 0$. Write $x$ as their sum and expand,
\begin{align*}
\|x\|^2 &= \|\Pi x + (x - \Pi x)\|^2 \\
  &= \|\Pi x\|^2 + 2 \langle \Pi x, x - \Pi x \rangle + \|x - \Pi x\|^2 \\
  &= \|\Pi x\|^2 + \|x - \Pi x\|^2
     && \text{(the cross term vanishes),}
\end{align*}
which is~\eqref{eq:prelim_pythagoras}. Since $\|x - \Pi x\|^2 \geq 0$, dropping it gives $\|\Pi x\|^2 \leq \|x\|^2$, so $\|\Pi x\| \leq \|x\|$. Projection is linear, so applying this to $x - y$ gives $\|\Pi x - \Pi y\| = \|\Pi(x - y)\| \leq \|x - y\|$, which is nonexpansiveness.
\end{proof}

The proof is three lines and the picture is the proof, which is why it is one of the three kept here. Everything in the rest of this section is a consequence of reading its hypothesis carefully. The theorem says a projection is nonexpansive \emph{in the inner product that defines it}, and says nothing at all about any other.

\subsubsection{The On-Policy Case Works}

Fitted policy evaluation applies the Bellman operator and then projects back into the representable family, giving the \emph{projected Bellman operator} $\Pi_d T^\pi$. Its fixed point is the temporal-difference solution, the value that linear TD($0$) converges to.

Take $d = d^\pi$, the stationary distribution of the policy being evaluated, so that states are weighted by how often the policy visits them. Then two facts line up. The operator $T^\pi$ is a $\gamma$-contraction in $\|\cdot\|_{d^\pi}$, not merely in the supremum norm, because $\|P^\pi x\|_{d^\pi} \leq \|x\|_{d^\pi}$ for the stationary weighting \citep{tsitsiklis1997}.\footnote{Stationarity is what makes this work. Applying $P^\pi$ averages, which by Jensen's inequality (Theorem~\ref{thm:prelim_jensen}) cannot increase a mean square, and $d^\pi P^\pi = d^\pi$ means the averaging is measured under the same weights it preserves. Replace $d^\pi$ by any other distribution and the step fails.} And $\Pi_{d^\pi}$ is nonexpansive in that same norm by Theorem~\ref{thm:prelim_hilbert}. Composing a nonexpansive map with a $\gamma$-contraction in one shared norm gives a $\gamma$-contraction, so $\Pi_{d^\pi}T^\pi$ has a unique fixed point and the iterates converge to it geometrically by Theorem~\ref{thm:prelim_banach}.

The Engine Replacement MDP uses the single feature $\phi = (1, 2.1)$, so the representable value functions are the points of a line through the origin in the plane.\footnote{The feature is chosen to expose the geometry, not to fit well. The exact value $V^{\pi^\star} = (5.3448, 4.3103)$ points in the direction $(1, 0.806)$, nowhere near $(1, 2.1)$, so the best representable approximation is poor at either weighting. That is deliberate. A feature good enough to fit would leave nothing for this section to show, and Baird's counterexample \citep{Baird1995} is built the same way.} Under the on-policy weighting $d^{\pi^\star} = (0.6667, 0.3333)$ the projection is
\begin{equation*}
\Pi_{d^{\pi^\star}} = \begin{pmatrix} 0.3120 & 0.3276 \\ 0.6552 & 0.6880 \end{pmatrix},
\end{equation*}
the best representable approximation to $V^{\pi^\star}$ is $(3.0798, 6.4675)$, and the composed operator $\Pi_{d^{\pi^\star}} T^{\pi^\star}$ has modulus $0.7301$ on the feature line, comfortably below $\gamma = 0.9$. Its fixed point is $(0.5491, 1.1532)$, a poor approximation to $V^{\pi^\star}$ but a perfectly well defined one that the algorithm reaches and stays at.

\subsubsection{The Off-Policy Case Need Not}

Now change one thing. Keep the same target policy, the same operator $T^{\pi^\star}$, and the same features, and only replace the weighting by the logging distribution $d^\mu = (0.9474, 0.0526)$ of Section~\ref{prelim:markov_data}. This is what happens whenever the value of one policy is fitted from data generated by another, which is the ordinary situation in offline reinforcement learning. The projection becomes
\begin{equation*}
\Pi_{d^\mu} = \begin{pmatrix} 0.8032 & 0.0937 \\ 1.6867 & 0.1968 \end{pmatrix},
\end{equation*}
and the composed operator $\Pi_{d^\mu} T^{\pi^\star}$ has modulus $1.2048$. It expands. Its fixed point is $(-3.6928, -7.7549)$, on the wrong side of the origin from a value function whose entries are both above four, and the iterates do not approach it because a map with modulus above one repels rather than attracts.

Nothing about $T^{\pi^\star}$ changed. It is the same affine map with the same $\gamma$-contraction in the supremum norm and the same fixed point $V^{\pi^\star}$. What changed is that $\Pi_{d^\mu}$ is orthogonal in the $d^\mu$ geometry and \emph{oblique} in the $d^{\pi^\star}$ geometry, the one in which $T^{\pi^\star}$ contracts. Theorem~\ref{thm:prelim_hilbert} still holds, and it simply does not say what is needed. Measured in $\|\cdot\|_{d^{\pi^\star}}$ the map $\Pi_{d^\mu}$ has norm above one, and the product of a map of norm above $1/\gamma$ with a $\gamma$-contraction can exceed one.

Measured in the target policy's own geometry the logging projection has operator norm $\|\Pi_{d^\mu}\|_{d^{\pi^\star}} = 1.4574$, against exactly $1$ for $\Pi_{d^{\pi^\star}}$, and $1.4574$ exceeds $1/\gamma = 1.1111$, which is precisely the room a $\gamma$-contraction leaves before a composition can expand.

This is the \emph{deadly triad}, the joint failure of function approximation, bootstrapping, and off-policy sampling \citep{sutton2018}. Each ingredient alone is harmless. Function approximation alone leaves a projection that never lengthens anything. Bootstrapping alone, in the tabular case, is the Bellman contraction of Section~\ref{prelim:operators}. Off-policy sampling alone is the exact change of measure~\eqref{eq:prelim_change_of_measure}. The three together break the shared-norm argument that made the composition contract. Nothing here is a sampling problem, and no amount of data repairs it. Every quantity in this subsection is computed from the exact operator with no sampling at all, which is the infinite-data limit, and the composition still expands.

The transition is continuous rather than a threshold effect. Tilting the weighting along $d = (1-\alpha)d^{\pi^\star} + \alpha d^\mu$, the modulus rises from $0.7301$ at $\alpha = 0$ through $0.8578$ at $\alpha = 0.4$ and $0.9451$ at $\alpha = 0.6$, crossing one at $\alpha = 0.7048$ and reaching $1.2048$ at $\alpha = 1$. Sweeping the logging policy instead of the mixing weight tells the same story. A crew that keeps the machine with probability $0.30$ gives a best attainable modulus of $0.9548$ over the whole feature grid, at $0.20$ it is $1.0516$, and at $0.05$ it is $1.7000$. At a keep-probability of $0.50$ the logging chain has stationary distribution $(0.6667, 0.3333)$, identical to the target policy's, and the modulus falls back to the on-policy regime. How far the log has drifted from the target is the quantity that matters, and mild drift is genuinely safe here.

\begin{figure}[t]
\centering
\includegraphics[width=0.98\textwidth]{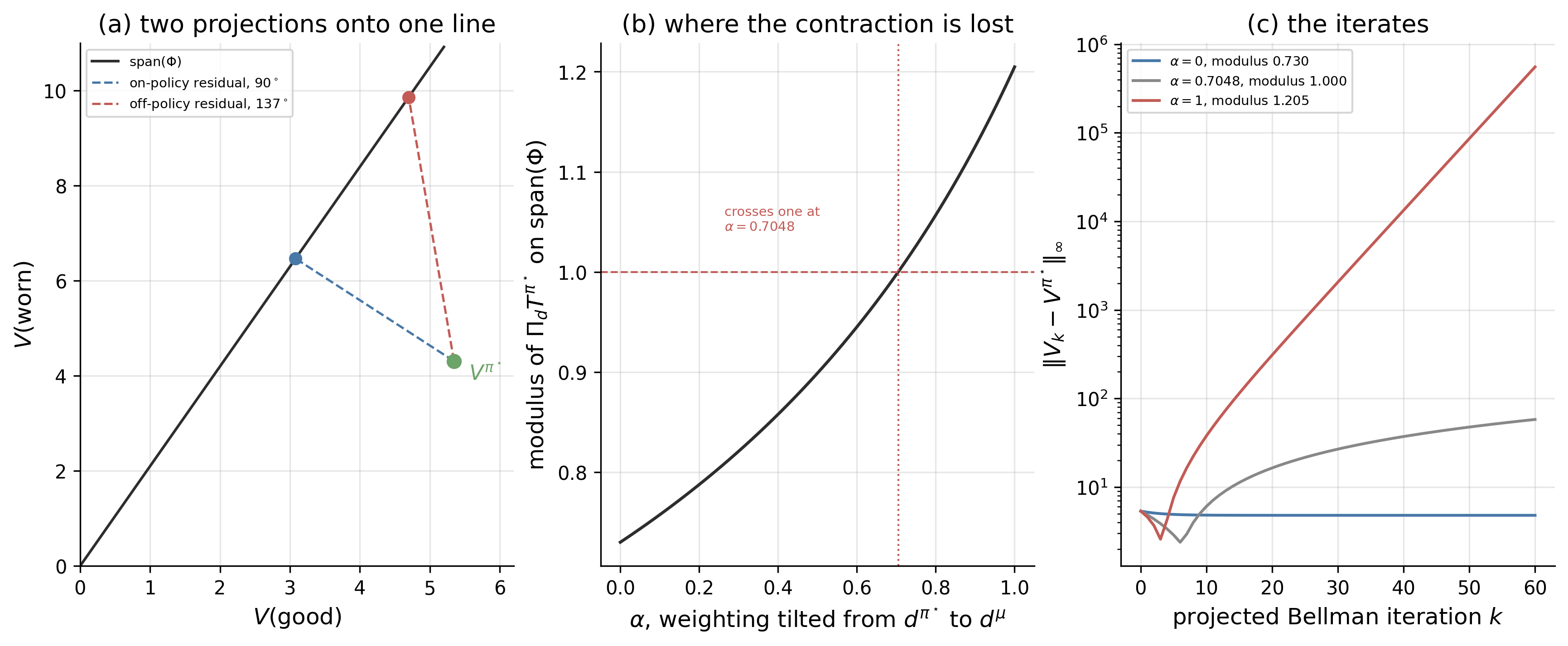}
\caption{Panel (a): the representable line $\operatorname{span}(\Phi)$ with $\phi = (1, 2.1)$, the exact value $V^{\pi^\star}$, and its two projections. The axes are Euclidean while the angles quoted in the legend are measured in the $d^{\pi^\star}$ inner product, so the on-policy residual is at a true right angle there ($90^\circ$) without appearing so on the page, and the off-policy residual is oblique in that same geometry ($137^\circ$). Panel (b): the modulus of $\Pi_d T^{\pi^\star}$ as the weighting is tilted from $d^{\pi^\star}$ to $d^\mu$, crossing one at $\alpha = 0.7048$. Panel (c): supremum-norm distance of the projected iterates from $V^{\pi^\star}$, from $V_0 = 0$. Below the crossing the iterates settle at their own fixed point, which sits $4.796$ from $V^{\pi^\star}$ and is further away than the best representable approximation at $2.265$. At the crossing they drift linearly, and above it they diverge geometrically.}
\label{fig:prelim_projection_geometry}
\end{figure}

\begin{table}[h]
\centering
\caption{The projected Bellman operator on the Engine Replacement MDP under three weightings, from the target policy's own stationary law ($\alpha = 0$) to the logging law ($\alpha = 1$). The operator, the features and the target policy are identical across rows; only the distribution defining the projection changes. Errors are supremum-norm distances from $V^{\pi^\star}$ after the stated number of projected Bellman iterations from $V_0 = 0$.}
\label{tab:prelim_projection_geometry}
\begin{tabular}{lrrrr}
\hline
$\alpha$ & $d(\mathrm{good})$ & modulus & error at $k = 10$ & error at $k = 60$ \\
\hline
0 & 0.6667 & 0.7301 & 4.8193 & 4.796 \\
0.7048 & 0.8645 & 1.0000 & 6.0646 & 57.97 \\
1 & 0.9474 & 1.2048 & 37.9147 & 5.558e+05 \\
\hline
\end{tabular}
\end{table}

\subsection{Probability and Stochastic Approximation}
\label{prelim:stochastic}

Every operator so far has been available exactly. A learning agent cannot apply $T^\pi$, because doing so requires averaging over $P(\cdot \mid s,a)$, which it does not know. It sees one sampled successor state at a time. This section is about what survives that substitution.

\subsubsection{The Probability Objects}

A \emph{random variable} $X$ is a function from outcomes to numbers, and its \emph{expectation} $\mathbb{E}[X]$ is the probability-weighted average of the values it takes. A \emph{conditional expectation} $\mathbb{E}[X \mid Y]$ is the average of $X$ once $Y$ is known, and is itself a random variable, because it depends on which value $Y$ turned out to take.

A \emph{filtration} $(\mathcal{F}_t)_{t \geq 0}$ is the growing record of what is known through time, with $\mathcal{F}_t$ holding everything observable by step $t$ and $\mathcal{F}_t \subseteq \mathcal{F}_{t+1}$. In a learning algorithm $\mathcal{F}_t$ contains the states visited, actions taken, and rewards received up to step $t$, so the current iterate $V_t$ is determined by $\mathcal{F}_t$ while the next sampled successor is not. Writing $\mathbb{E}[\,\cdot \mid \mathcal{F}_t]$ is how a statement conditions on the whole history rather than on one variable. A process is \emph{adapted} when each $X_t$ is determined by $\mathcal{F}_t$.

A sequence of random variables converges \emph{almost surely} to $X$ when $\mathbb{P}(X_t \to X) = 1$, meaning the set of trajectories on which convergence fails has probability zero. This is the strongest of the usual modes and the one reinforcement learning needs, because an algorithm produces one trajectory rather than an ensemble, and a guarantee about the average behaviour across ensembles would not say that this run converges.\footnote{The weaker mode, \emph{convergence in probability}, says $\mathbb{P}(|X_t - X| > \varepsilon) \to 0$ for each $\varepsilon$, which permits a run that fails infinitely often as long as the failures thin out. Almost-sure convergence forbids that.}

\subsubsection{Jensen's Inequality and Maximization Bias}

A two-point example makes it obvious. Let $\varphi(x) = x^2$, and let $X$ be $0$ or $2$ with equal chance. The mean is $\mathbb{E}[X] = 1$, so $\varphi(\mathbb{E}[X]) = 1$. The average of the squares is $\mathbb{E}[\varphi(X)] = (0 + 4)/2 = 2$. Squaring after averaging gives $1$, averaging after squaring gives $2$, and the gap of $1$ is exactly the variance of $X$. Convexity forces the second number above the first.

\begin{theorem}[Jensen's inequality, \citet{durrett}]
\label{thm:prelim_jensen}
Let $\varphi : \mathbb{R} \to \mathbb{R}$ be convex and let $X$ be an integrable random variable with $\varphi(X)$ also integrable.\footnote{A function $\varphi$ is \emph{convex} if its graph lies on or below every chord, equivalently $\varphi(\lambda x + (1-\lambda) y) \leq \lambda \varphi(x) + (1-\lambda)\varphi(y)$ for $\lambda \in [0,1]$. \emph{Integrable} means the expectation $\mathbb{E}[X]$ is finite.} Then
\begin{equation}
\varphi\bigl(\mathbb{E}[X]\bigr) \leq \mathbb{E}\bigl[\varphi(X)\bigr].
\label{eq:prelim_jensen}
\end{equation}
The inequality reverses when $\varphi$ is concave, and holds with equality when $\varphi$ is affine.
\end{theorem}

\begin{figure}[t]
\centering
\includegraphics[width=0.95\textwidth]{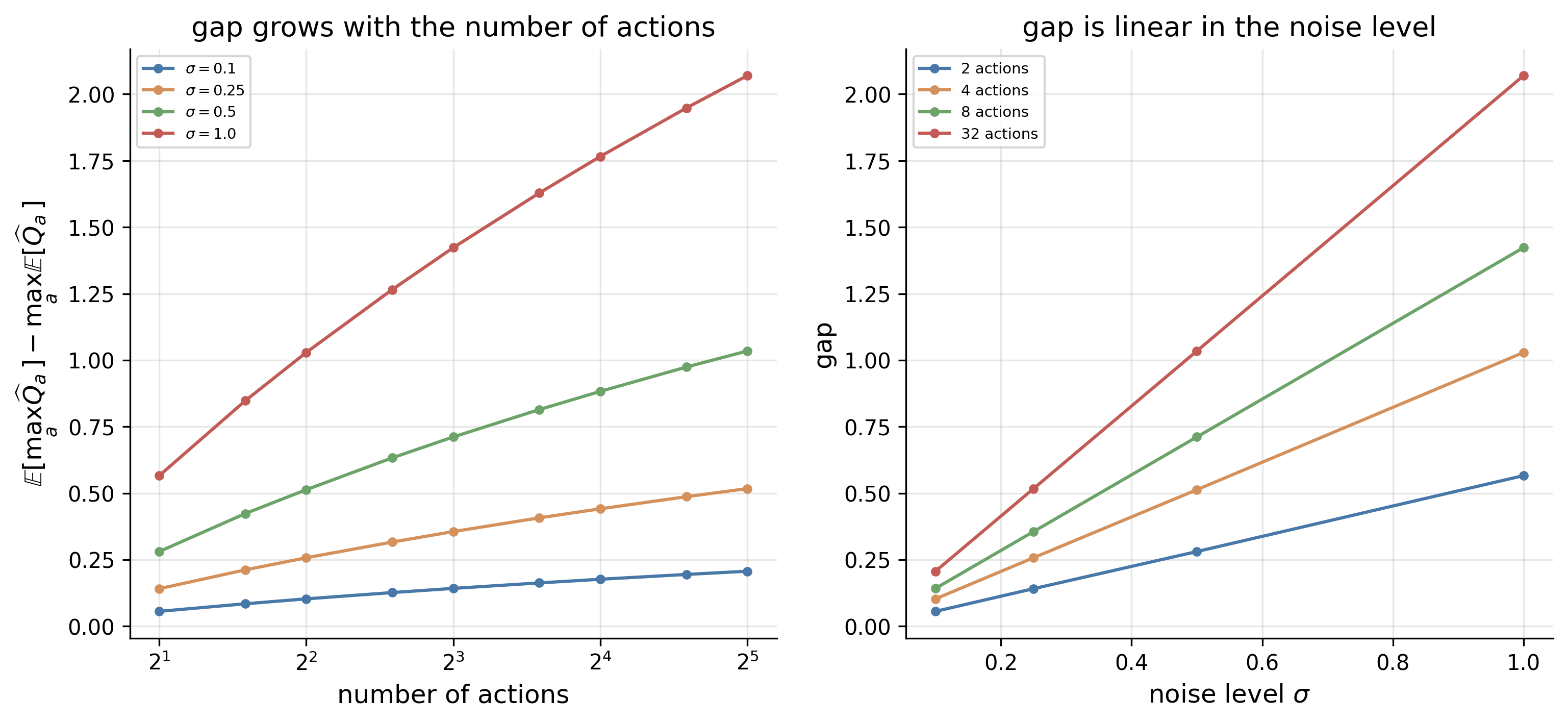}
\caption{The maximization-bias gap $\mathbb{E}[\max_a \widehat{Q}_a] - \max_a \mathbb{E}[\widehat{Q}_a]$ for independent Gaussian action-value estimates. Left: the gap against the number of actions at four noise levels. Right: the gap against the noise level at four action counts, against the equal-means reference $\sigma\,\mathbb{E}[\max_a Z_a]$ (dashed).}
\label{fig:prelim_jensen}
\end{figure}

The reinforcement-learning consequence is immediate and important. The maximum is a convex function of its arguments, so for noisy action-value estimates $\widehat{Q}_a$,
\begin{equation}
\mathbb{E}\bigl[\max_a \widehat{Q}_a\bigr] \;\geq\; \max_a \mathbb{E}\bigl[\widehat{Q}_a\bigr] .
\label{eq:prelim_max_bias}
\end{equation}
Maximization selects positive action-value estimation errors more often than negative ones, so a bootstrapped maximum is biased upward. Double estimators reduce this \emph{maximization bias}, and Section~\ref{sec:planning_learning} treats the correction. Figure~\ref{fig:prelim_jensen} measures the gap as the number of actions and the noise level vary.

\begin{table}[h]
\centering
\caption{Maximization bias for independent Gaussian action-value estimates, over 400000 Monte Carlo replicates per cell. With equal true action values the whole expected maximum is bias, and it equals $\sigma\,\mathbb{E}[\max_a Z_a]$ in closed form. The last column gives the bias once the best action is separated from the rest by one unit, at which point the maximum is nearly deterministic and the bias collapses.}
\label{tab:prelim_jensen}
\begin{tabular}{rrrrr}
\hline
actions & $\sigma$ & measured gap & $\sigma\,\mathbb{E}[\max_a Z_a]$ & bias at separation $1$ \\
\hline
2 & 0.25 & 0.1413 & 0.1409 & 0.0009 \\
2 & 1.00 & 0.5663 & 0.5637 & 0.2024 \\
8 & 0.25 & 0.3559 & 0.3558 & 0.0016 \\
8 & 1.00 & 1.4232 & 1.4233 & 0.6658 \\
32 & 0.25 & 0.5174 & 0.5171 & 0.0045 \\
32 & 1.00 & 2.0697 & 2.0684 & 1.1568 \\
\hline
\end{tabular}
\end{table}

\subsubsection{Laws of Large Numbers and the Rate}

Flip a fair coin over and over and track the running fraction of heads. Early on it swings around, then it settles down and hugs $1/2$. That settling is the law of large numbers. Now rescale the leftover wobble around $1/2$ by $\sqrt{n}$. The rescaled wobble does not vanish. It fills out a fixed bell-shaped spread, and that fixed shape is the central limit theorem.

\begin{theorem}[Law of large numbers and central limit theorem, \citet{durrett}]
\label{thm:prelim_lln_clt}
Let $X_1, X_2, \ldots$ be independent and identically distributed with mean $\mu = \mathbb{E}[X_1]$ and finite variance $\sigma^2 = \operatorname{Var}(X_1)$, and let $\bar X_n = \frac{1}{n}\sum_{i=1}^n X_i$. Then
\begin{equation}
\bar X_n \to \mu \quad \text{almost surely},
\label{eq:prelim_lln}
\end{equation}
and the rescaled fluctuation converges in distribution to a Gaussian,
\begin{equation}
\sqrt{n}\,(\bar X_n - \mu) \;\Rightarrow\; N(0, \sigma^2),
\label{eq:prelim_clt}
\end{equation}
whatever the shape of the underlying distribution.\footnote{Convergence \emph{in distribution}, written $\Rightarrow$, means the cumulative distribution functions converge at every continuity point. It describes the shape of the fluctuation, not any single realization.}
\end{theorem}

The two results license the two things a sampled backup needs. The law of large numbers says a sample average has the right target, so replacing $\sum_{s'} P(s' \mid s,a) V(s')$ by an average over observed successors estimates the correct quantity. The central limit theorem prices that replacement, at $n^{-1/2}$ and no faster, which is what turns an estimate into a confidence interval and what sets the width of every bandit bound in Section~\ref{section:bandits}. The Gaussian shape appears whatever the individual draws look like, the Galton-board effect in which many small independent bounces add up to a bell curve.

\subsubsection{Martingales}

Picture a gambler in a fair game. The expected fortune next round equals the fortune now, no edge either way, so the fortune is a \emph{martingale}. Suppose it is also capped and can never leave a fixed range. Then it has no room to drift. A quantity with no drift and no room to run must eventually stop swinging and settle.

\begin{theorem}[Martingale convergence, \citet{williamsprob1991}; \citet{robbinssiegmund1971}]
\label{thm:prelim_martingale}
Let $(M_n)$ be a supermartingale adapted to a filtration $(\mathcal{F}_n)$, meaning $\mathbb{E}[M_{n+1} \mid \mathcal{F}_n] \leq M_n$, and bounded in $L^1$, meaning $\sup_n \mathbb{E}|M_n| < \infty$.\footnote{A process is a \emph{martingale} if $\mathbb{E}[M_{n+1} \mid \mathcal{F}_n] = M_n$ and a \emph{supermartingale} if the conditional mean can only fall. A martingale confined to a bounded interval is automatically $L^1$-bounded.} Then $M_n$ converges almost surely to a finite limit. Moreover, if $(Z_n)$ is a nonnegative adapted process with
\begin{equation}
\mathbb{E}[Z_{n+1} \mid \mathcal{F}_n] \leq (1 + a_n) Z_n - b_n + c_n,
\label{eq:prelim_robbins_siegmund}
\end{equation}
where $a_n, b_n, c_n \geq 0$ are $\mathcal{F}_n$-measurable with $\sum_n a_n < \infty$ and $\sum_n c_n < \infty$ almost surely, then $Z_n$ converges almost surely to a finite limit and $\sum_n b_n < \infty$ almost surely.
\end{theorem}

A \emph{martingale-difference} sequence is one with $\mathbb{E}[w_t \mid \mathcal{F}_t] = 0$, so each term adds no systematic bias given everything known before it. This is the exact shape of the error a sampled Bellman backup makes, since the sampled target is right on average given the current state and iterate. The second half of the theorem, due to \citet{robbinssiegmund1971}, is what handles a squared error that would shrink except for two small perturbations, and it is the probabilistic engine behind essentially every almost-sure convergence proof in reinforcement learning.

\subsubsection{Stochastic Approximation}

The simplest case is estimating an unknown mean. Noisy samples arrive one at a time and a running guess is nudged by $x \leftarrow x + \alpha_t(\text{sample} - x)$. At $\alpha_t = 1/t$ this is exactly the running average, and it converges to the true mean. The two conditions below are what keep such an update honest, steps large enough to reach any target and small enough to average the noise away.

\begin{theorem}[Stochastic approximation for a contraction, \citet{jaakkola1994}; \citet{tsitsiklis1994}]
\label{thm:prelim_robbins_monro}
Let $g$ be a $\gamma$-contraction on $\mathbb{R}$ with fixed point $x^\star$, and let
\begin{equation}
x_{t+1} = x_t + \alpha_t\bigl(g(x_t) - x_t + w_t\bigr),
\label{eq:prelim_rm_update}
\end{equation}
where the step sizes $\alpha_t \geq 0$ are $\mathcal{F}_t$-measurable and the noise $w_t$ is $\mathcal{F}_{t+1}$-measurable with $\mathbb{E}[w_t \mid \mathcal{F}_t] = 0$ and $\mathbb{E}[w_t^2 \mid \mathcal{F}_t] \leq \sigma^2$.\footnote{The measurability matters and is easy to get backwards. The noise on step $t$ is not known until the sample that produced it has been drawn, so $w_t$ enters the record at time $t+1$, not $t$. Requiring $w_t$ to be $\mathcal{F}_t$-measurable and mean-zero given $\mathcal{F}_t$ would force $w_t = 0$ and leave a theorem about a noiseless recursion.} If the step sizes satisfy the two Robbins-Monro conditions
\begin{equation}
\sum_{t} \alpha_t = \infty \qquad \text{and} \qquad \sum_{t} \alpha_t^2 < \infty,
\label{eq:prelim_rm_conditions}
\end{equation}
then $x_t \to x^\star$ almost surely.
\end{theorem}

The first condition keeps the total movement unbounded, so the iterate can travel any finite distance to the target. The second makes the steps shrink fast enough that the accumulated noise is finite. Temporal-difference learning is~\eqref{eq:prelim_rm_update} with a Bellman operator in place of $g$, applied only at the state visited on that step, and Q-learning is the same recursion for $T^\star$.

The two step-size conditions are due to \citet{robbinsmonro1951}, who introduced this recursion for root-finding and proved convergence in mean square. The almost-sure statement came later, and the form above, a contraction driven by martingale-difference noise with one component updated per step, is the version reinforcement learning needs and is due to \citet{jaakkola1994} and \citet{tsitsiklis1994}.\footnote{The conditions keep their original name because they are the ones \citet{robbinsmonro1951} imposed, and the survey follows that usage.}

Running tabular TD($0$) on the example, evaluating the optimal policy from sampled transitions over $20$ seeds and $20{,}000$ steps against the target $V^{\pi^\star} = (5.3448, 4.3103)$, separates what the conditions do and do not promise. Table~\ref{tab:prelim_td_schedules} reports the two sums and the final error for four schedules. A constant step of $0.05$ has both sums infinite, violating the second condition, and it settles at an error of $0.1711$, a floor set by the step size rather than by the data. A step of $1/n^2$, where $n$ counts visits to the state being updated, has both sums finite, violating the first, and stalls at $4.0291$, having simply run out of total movement. The two valid schedules behave very differently from each other. A step of $1/n$ satisfies both conditions and converges, but its total movement grows only like $\log n$, and after $20{,}000$ steps it sits at $1.7125$, ten times worse than the invalid constant step. A step of $1/n^{0.7}$ satisfies the same two conditions with total movement growing like $n^{0.3}$ and reaches $0.0450$, the best of the four.

The conditions in~\eqref{eq:prelim_rm_conditions} are asymptotic and say nothing about the rate, which is why the polynomial family rather than the harmonic one is what practice uses. A schedule can be provably convergent and useless within any budget a person will wait for.

\begin{table}[h]
\centering
\caption{Tabular TD($0$) on the Engine Replacement MDP, evaluating $\pi^\star$ from sampled transitions over 20 seeds and 20000 steps each, against the exact $V^{\pi^\star}$. Here $n$ counts visits to the state being updated. The two sums are the quantities the Robbins-Monro conditions constrain, evaluated along one run. Error is the supremum-norm distance at the end of the budget, with its standard error across seeds in brackets.}
\label{tab:prelim_td_schedules}
\begin{tabular}{lrrrl}
\hline
step size & $\sum \alpha_n$ & $\sum \alpha_n^2$ & final error & conditions \\
\hline
constant, $\alpha = 0.05$ & 1000.0 & 50.00 & 0.1711 (0.0220) & $\sum \alpha_n^2$ diverges \\
$1/n$ & 19.5 & 3.29 & 1.7125 (0.0235) & both hold \\
$1/n^{0.7}$ & 98.8 & 6.08 & 0.0450 (0.0057) & both hold \\
$1/n^2$ & 3.3 & 2.16 & 4.0291 (0.0229) & $\sum \alpha_n$ converges \\
\hline
\end{tabular}
\end{table}

\begin{figure}[t]
\centering
\includegraphics[width=0.78\textwidth]{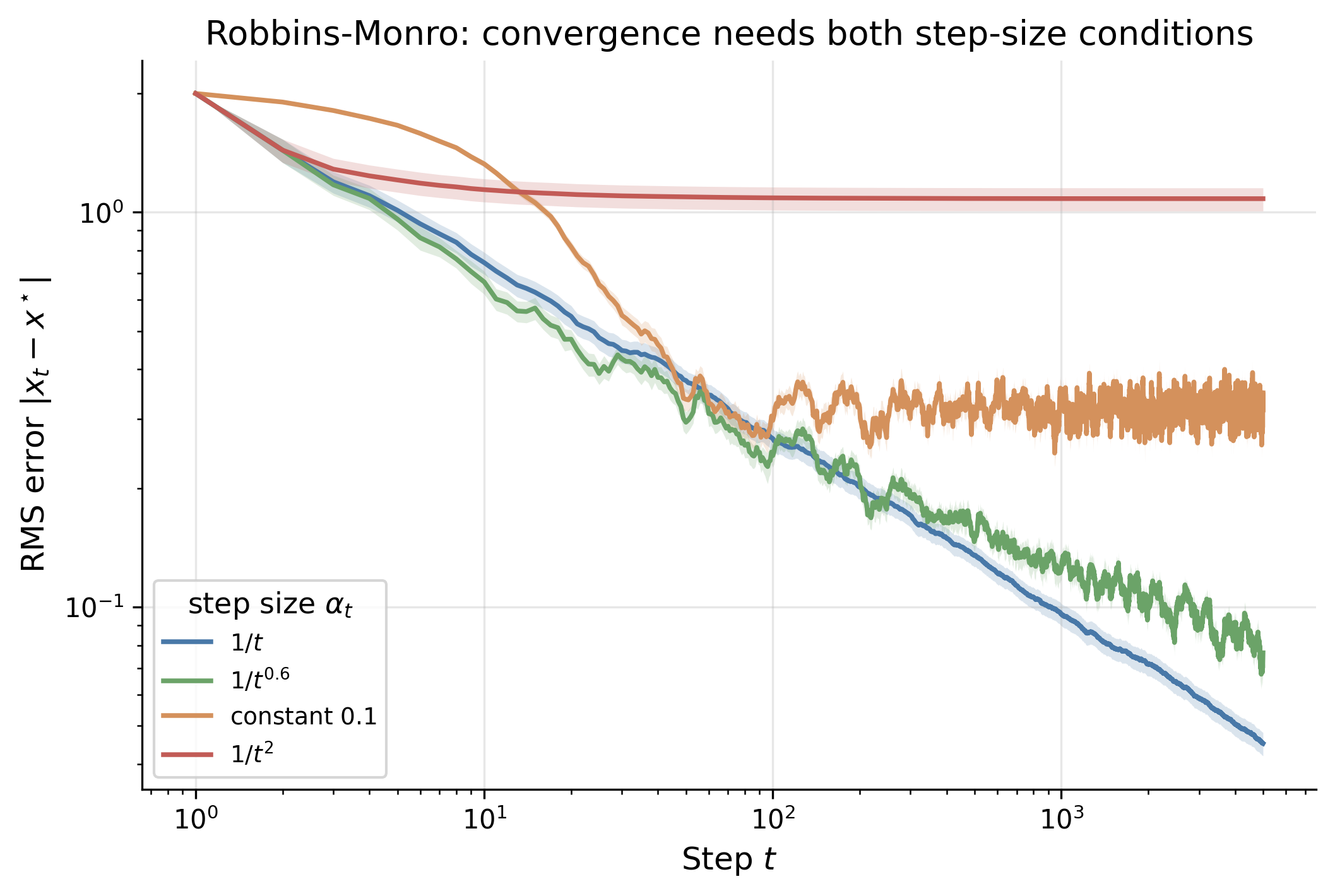}
\caption{Root-mean-square error $|x_t - x^\star|$ of the stochastic-approximation recursion~\eqref{eq:prelim_rm_update} for the fixed point of $x = \gamma x + b$ under four step-size schedules, averaged over 100 seeds, on log-log axes. Only the two schedules that satisfy both Robbins-Monro conditions~\eqref{eq:prelim_rm_conditions} reach the root. The constant schedule settles into a noise floor. The $1/t^2$ schedule stalls.}
\label{fig:prelim_robbins_monro}
\end{figure}

\subsection{Curvature, Constraints, and Worst Cases}
\label{prelim:convex}

Policy optimization, reward-model fitting, and constrained control all reduce to minimizing a function rather than iterating an operator. This section collects the four facts about that problem the survey uses.

\subsubsection{Convexity, Smoothness, and Conditioning}
\label{prelim:quadratic_forms}

Minimize $f(x) = x^2$. The slope is $2x$, so $f$ is $L$-smooth with $L = 2$, and gradient descent with step $1/L = 1/2$ updates $x \leftarrow x - \tfrac{1}{2}(2x) = 0$. One step lands exactly on the minimum. A rounder bowl could not exist. All the difficulty in the general case comes from bowls that are stretched and lopsided, and the theorem measures what that stretching costs.

\begin{theorem}[Gradient descent convergence, \citet{nesterov2018}]
\label{thm:prelim_gd}
Let $f$ be convex and $L$-smooth, and let $x^\star$ minimize $f$.\footnote{A differentiable $f$ is \emph{$L$-smooth} if its gradient is $L$-Lipschitz, $\|\nabla f(x) - \nabla f(y)\| \leq L\|x - y\|$, which bounds how fast the slope can change. It is \emph{$\mu$-strongly convex} if $f(x) - \tfrac{\mu}{2}\|x\|^2$ is still convex, so $f$ curves up at least as fast as a quadratic of curvature $\mu$.} Gradient descent with step $1/L$, $x_{k+1} = x_k - \tfrac{1}{L}\nabla f(x_k)$, satisfies
\begin{equation}
f(x_k) - f(x^\star) \leq \frac{L\,\|x_0 - x^\star\|^2}{2k}.
\label{eq:prelim_gd_sublinear}
\end{equation}
If in addition $f$ is $\mu$-strongly convex, with condition number $\kappa = L/\mu$, the iterates converge geometrically,
\begin{equation}
\|x_k - x^\star\| \leq \Bigl(1 - \tfrac{\mu}{L}\Bigr)^{k} \|x_0 - x^\star\|.
\label{eq:prelim_gd_linear}
\end{equation}
\end{theorem}

The split between the two rates is the split between hard and easy problems. On a general smooth convex objective the gap closes at the sublinear rate $O(1/k)$. Strong convexity, the curvature that keeps the objective from flattening near the optimum, upgrades this to the geometric rate $(1 - \mu/L)^k$, and the iterations needed for a fixed accuracy then scale with the \emph{condition number} $\kappa = L/\mu$. A well-conditioned objective is a round bowl, where the steps roll straight to the bottom. An ill-conditioned one is a long narrow valley, where the steps zig-zag across the walls and crawl along the floor. Equation~\eqref{eq:prelim_gd_linear} is Theorem~\ref{thm:prelim_banach} applied to the gradient map, which is a $(1 - \mu/L)$-contraction whose fixed point is the minimizer.\footnote{The Polyak-Lojasiewicz condition used in Section~\ref{sec:planning_learning} relaxes strong convexity to the inequality $\tfrac{1}{2}\|\nabla f(x)\|^2 \geq \mu\,(f(x) - f^\star)$, which is enough for the same geometric rate on the function values without requiring convexity anywhere.}

A \emph{quadratic form} $x^\top A x$ with $A$ symmetric positive definite is the model case, and its \emph{eigenvalues} are the curvatures along the principal axes, so $\kappa(A) = \lambda_{\max}(A)/\lambda_{\min}(A)$ measures how far from round the bowl is.

\subsubsection{Kullback-Leibler Divergence and the Fisher Matrix}

Policy optimization needs a way to say that two policies are close. Euclidean distance between parameter vectors will not do, because the same parameter change can move a distribution a great deal or hardly at all. The natural measure is the \emph{Kullback-Leibler divergence}
\begin{equation}
D_{\mathrm{KL}}(p \,\|\, q) = \sum_x p(x) \log \frac{p(x)}{q(x)} \;\geq\; 0 ,
\label{eq:prelim_kl}
\end{equation}
which is zero exactly when $p = q$ and is not symmetric in its two arguments, so it is not a metric. Nonnegativity is Jensen's inequality applied to $-\log$, which is convex. For a coin with $p = (0.5, 0.5)$ against $q = (0.9, 0.1)$ the divergence $D_{\mathrm{KL}}(p \| q) = 0.5\log(0.5/0.9) + 0.5\log(0.5/0.1) \approx 0.51$ nats, while the reverse $D_{\mathrm{KL}}(q \| p) \approx 0.37$, which shows the asymmetry directly.

For a parametric family $\pi_\theta$ the divergence between nearby members is governed by a single matrix. Expanding $D_{\mathrm{KL}}(\pi_\theta \| \pi_{\theta + \Delta})$ to second order, the constant and linear terms vanish, because the divergence is nonnegative with a minimum of zero at $\Delta = 0$, and what remains is
\begin{equation}
D_{\mathrm{KL}}(\pi_\theta \,\|\, \pi_{\theta + \Delta}) = \tfrac{1}{2}\,\Delta^\top F(\theta)\, \Delta + O(\|\Delta\|^3),
\qquad
F(\theta) = \mathbb{E}\bigl[ \nabla_\theta \log \pi_\theta \, \nabla_\theta \log \pi_\theta^\top \bigr],
\label{eq:prelim_fisher}
\end{equation}
where $F$ is the \emph{Fisher information matrix}, an average of outer products of the score and therefore positive semidefinite. The Fisher matrix is the local quadratic model of KL divergence, which makes it the quadratic form that measures distance in distribution space rather than parameter space.

Two constructions follow. The \emph{natural gradient} is the steepest-ascent direction under that metric, obtained by solving the linear system $F v = \nabla_\theta J$ rather than by forming $F^{-1}$, which for a large parameter vector is neither necessary nor numerically sensible \citep{amari1998natural, Kakade2001}. And a \emph{trust region} of the form $\Delta^\top F \Delta \leq 2\delta$ is a KL ball to second order, which is the constraint that defines trust-region policy optimization \citep{Schulman2015}. In both cases the conditioning of $F$ plays the role $\kappa$ played above, so an anisotropic Fisher matrix is the ill-conditioned valley of the previous subsection expressed in distribution space. Panel (a) of Figure~\ref{fig:prelim_curvature} draws the two trust regions against each other.

\subsubsection{Lagrangian Duality}

Take the tiny program $\min x^2$ subject to $x \geq 1$. The constraint binds, so the answer is $x^\star = 1$ with value $p^\star = 1$. Now attach a price $\lambda \geq 0$ to the constraint and minimize $x^2 + \lambda(1 - x)$ over $x$ with no constraint. The minimizer is $x = \lambda/2$, giving the lower bound $q(\lambda) = \lambda - \lambda^2/4$. This bound is largest at $\lambda^\star = 2$, where $q = 1$. The best price reproduces the true optimum exactly, with no gap.

\begin{theorem}[Weak and strong duality, \citet{boydvandenberghe2004}]
\label{thm:prelim_duality}
Consider $p^\star = \inf_x f(x)$ subject to $g_i(x) \leq 0$ for $i = 1, \ldots, m$, with Lagrangian $L(x, \lambda) = f(x) + \sum_i \lambda_i g_i(x)$ and dual function $q(\lambda) = \inf_x L(x, \lambda)$ for $\lambda \geq 0$.\footnote{The multipliers $\lambda_i \geq 0$ price the constraints. The \emph{dual function} is the best lower bound those prices can certify, and it is concave in $\lambda$ as an infimum of affine functions, whatever $f$ and $g_i$ are.} Let $d^\star = \sup_{\lambda \geq 0} q(\lambda)$. Then weak duality always holds,
\begin{equation}
d^\star \leq p^\star,
\label{eq:prelim_weak_duality}
\end{equation}
and if $f$ and the $g_i$ are convex and a strictly feasible point exists, so that $g_i(\bar x) < 0$ for some $\bar x$ and all $i$ (Slater's condition), then strong duality holds, $d^\star = p^\star$, and the supremum defining $d^\star$ is attained by some $\lambda^\star \geq 0$. If in addition the primal infimum is attained at some $x^\star$, the pair is a saddle point,
\begin{equation}
L(x^\star, \lambda) \leq L(x^\star, \lambda^\star) \leq L(x, \lambda^\star) \quad \text{for all } x, \ \lambda \geq 0.
\label{eq:prelim_saddle}
\end{equation}
\end{theorem}

Primal attainment is a genuine extra hypothesis and not a formality. Minimizing $e^{-x}$ subject to $x \geq 0$ is convex and strictly feasible, and $p^\star = d^\star = 0$, so strong duality holds, yet the primal infimum is approached only as $x \to \infty$ and no saddle point exists.

Duality converts a constrained problem into an unconstrained one. Rather than optimize subject to constraints, one forms the Lagrangian and searches for its saddle point, ascending in the multipliers and descending in the variables. Weak duality guarantees that every setting of the prices certifies a lower bound on the optimum, and strong duality guarantees that bound is tight. The multiplier at the optimum is a \emph{shadow price}, the rate at which the optimal value would improve if the constraint were relaxed, which is the reading constrained reinforcement learning gives it in Section~\ref{section:dist_robust_constrained}.

\subsubsection{The Envelope Theorem}

Take $V(\theta) = \max_a (\theta a - \tfrac{1}{2}a^2)$. The best $a$ solves $\theta - a = 0$, so $a^\star(\theta) = \theta$ and $V(\theta) = \tfrac{1}{2}\theta^2$. Differentiate directly and $V'(\theta) = \theta$. Now instead differentiate only the inside, $\partial_\theta(\theta a - \tfrac{1}{2}a^2) = a$, and plug in the winner $a^\star = \theta$, giving $\theta$ again. The two answers agree, and getting the second one never required differentiating $a^\star(\theta)$ at all.

\begin{theorem}[Envelope theorem, \citet{milgromsegal2002}; \citet{danskin1967}]
\label{thm:prelim_envelope}
Let $V(\theta) = \max_{a \in A} f(a, \theta)$ with $A$ compact, $f$ and $\partial_\theta f$ jointly continuous, and $a^\star(\theta) \in \argmax_a f(a, \theta)$. If the maximizer is unique, then $V$ is differentiable at $\theta$ and
\begin{equation}
V'(\theta) = \frac{\partial f}{\partial \theta}\bigl(a^\star(\theta), \theta\bigr).
\label{eq:prelim_envelope}
\end{equation}
The change in the maximizer contributes nothing to first order. When the maximizer is not unique, $V$ has one-sided derivatives given by the extreme values of $\partial_\theta f$ over the set of maximizers.\footnote{This is Danskin's directional-derivative statement. A unique maximizer collapses the interval of one-sided slopes to the single derivative~\eqref{eq:prelim_envelope}, and multiple maximizers leave a kink, which is the same kink Section~\ref{prelim:affine_envelope} located in the Bellman envelope and Section~\ref{prelim:argopt} attributed to a non-singleton set of maximizers.}
\end{theorem}

\begin{figure}[t]
\centering
\includegraphics[width=0.95\textwidth]{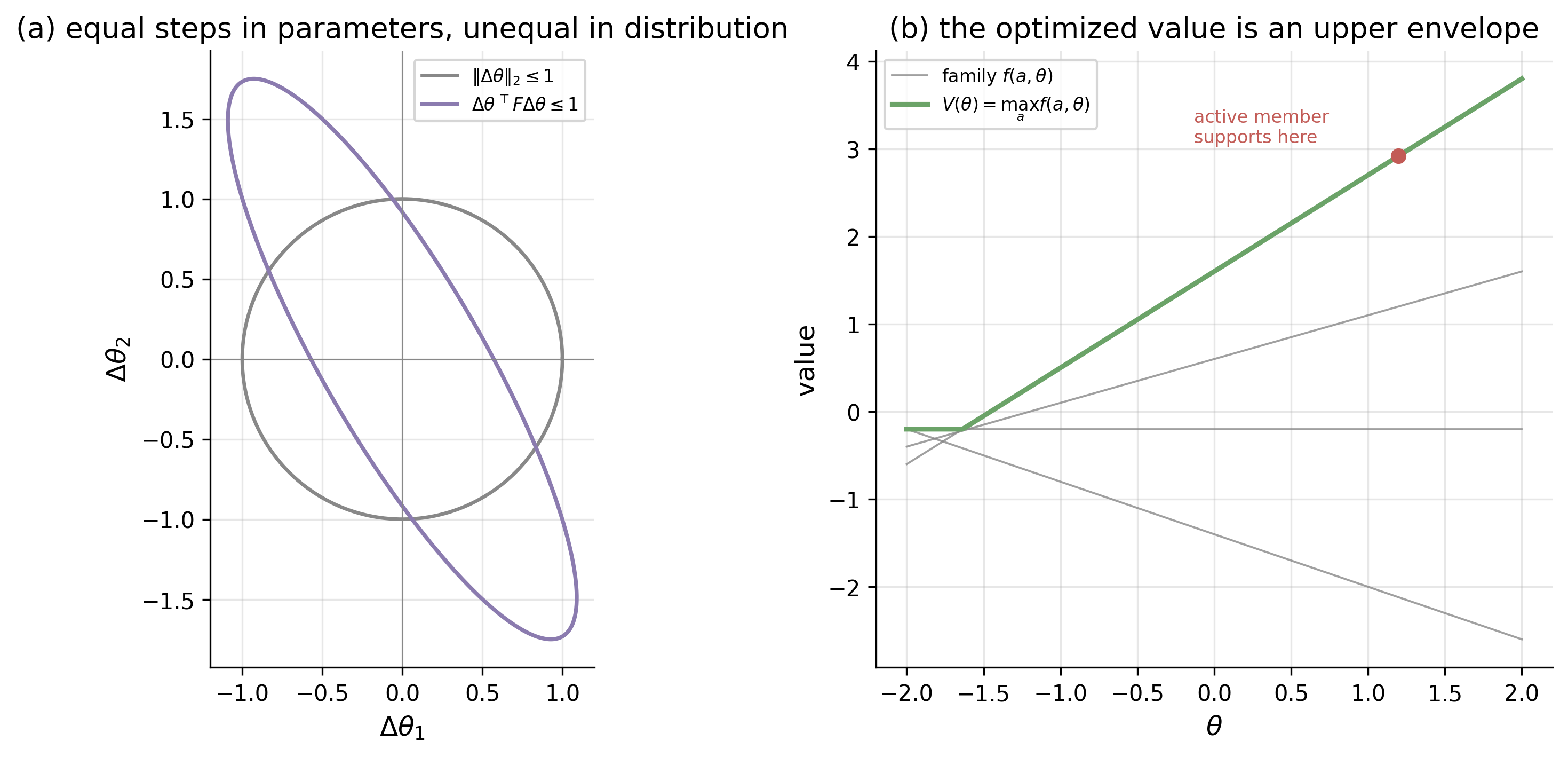}
\caption{Panel (a): a Euclidean ball of parameter steps against the Fisher trust region $\Delta\theta^\top F \Delta\theta \leq 1$ for a Fisher matrix with condition number $16$. Steps of equal parameter length move the policy distribution by very different amounts, and the Fisher region is the set that moves it equally. Panel (b): a family of payoffs and their upper envelope, which coincides with whichever member is active and kinks where the maximizer switches.}
\label{fig:prelim_curvature}
\end{figure}

The maximizer's own movement drops out because at a peak one already stands at the top, so a small tilt of the ground changes one's height only through how the ground moved underfoot, not through any scramble to re-pick where to stand. In dynamic programming this is the statement that the derivative of the value with respect to a state or a parameter equals the derivative of the one-step return at the optimal action, which economists know as the Benveniste-Scheinkman identity. Danskin's form supplies the gradient of a worst-case objective as the gradient at the worst case, which is what makes the inner minimization of a robust Bellman operator differentiable in Section~\ref{section:dist_robust_constrained}, and the same identity underlies the policy-gradient theorem of Section~\ref{sec:policy_gradient}. Panel (b) of Figure~\ref{fig:prelim_curvature} draws the envelope and the member that supports it, the same construction Section~\ref{prelim:affine_envelope} used for the Bellman operator and Section~\ref{prelim:operators} used to derive its contraction.

\section{Glossary of Acronyms and Terms}
\label{appendix:glossary}

\subsection*{Acronyms}

\begin{longtable}{p{2.2cm} p{12cm}}
A2C/A3C & Advantage Actor-Critic / Asynchronous Advantage Actor-Critic. Policy gradient algorithms that use an advantage function baseline to reduce variance (Section~\ref{section:rl_algorithms}). \\[4pt]
BCQ & Batch-Constrained Q-learning. An offline RL algorithm that restricts the learned policy to actions supported by the behavioral data (Section~\ref{sec:offline_algorithms}). \\[4pt]
BwK & Bandits with Knapsacks. A bandit framework with resource constraints, where an agent must balance exploration and exploitation subject to a budget (Section~\ref{section:bandits}). \\[4pt]
CCP & Conditional Choice Probabilities. Choice probabilities conditional on state, used in structural estimation as an alternative to full-solution methods (Section~\ref{section:rl_econ_models}). \\[4pt]
CFR & Counterfactual Regret Minimization. An iterative algorithm for computing Nash equilibria in extensive-form games by minimizing regret at each information set (Section~\ref{section:rl_games}). \\[4pt]
CQL & Conservative Q-Learning. An offline RL algorithm that adds a penalty for overestimating Q-values on out-of-distribution actions (Section~\ref{sec:offline_algorithms}). \\[4pt]
DDC & Dynamic Discrete Choice. A class of structural econometric models in which agents make sequential discrete decisions under uncertainty (Section~\ref{section:rl_econ_models}). \\[4pt]
DDPG & Deep Deterministic Policy Gradient. An actor-critic algorithm for continuous action spaces that combines a deterministic policy gradient with a learned Q-function (Section~\ref{section:rl_algorithms}). \\[4pt]
DP & Dynamic Programming. A collection of algorithms that compute optimal policies by solving the Bellman equation, given a known model of the environment (Section~\ref{section:rl_algorithms}). \\[4pt]
DPO & Direct Preference Optimization. A method that optimizes a language model directly on preference data without training a separate reward model (Section~\ref{section:rlhf}). \\[4pt]
DQN & Deep Q-Network. Q-learning with a neural network function approximator, target network, and experience replay (Section~\ref{section:rl_algorithms}). \\[4pt]
ETC & Explore-Then-Commit. A bandit algorithm that explores uniformly for a fixed number of rounds, then commits to the empirically best arm (Section~\ref{section:bandits}). \\[4pt]
FQI/FVI & Fitted Q-Iteration / Fitted Value Iteration. Approximate dynamic programming algorithms that use supervised learning to fit value functions from batch data (Section~\ref{sec:fvi_fqi_theory}). \\[4pt]
GLIE & Greedy in the Limit with Infinite Exploration. A condition on exploration schedules ensuring convergence to the optimal policy (Section~\ref{section:rl_algorithms}). \\[4pt]
GMM & Generalized Method of Moments. An econometric estimation method that matches sample moments to population moment conditions (Section~\ref{section:rl_econ_models}). \\[4pt]
IPW & Inverse Probability Weighting. A method for correcting distributional mismatch by reweighting observations by the inverse of their selection probability (Section~\ref{section:causal_rl}). \\[4pt]
IQL & Implicit Q-Learning. An offline RL algorithm that avoids querying out-of-distribution actions by using expectile regression on the value function (Section~\ref{sec:offline_algorithms}). \\[4pt]
IRL & Inverse Reinforcement Learning. The problem of inferring a reward function from observed behavior, assuming the agent acts approximately optimally (Section~\ref{section:rl_algorithms}). \\[4pt]
IV & Instrumental Variables. An econometric technique for identifying causal effects in the presence of endogeneity by exploiting exogenous variation (Section~\ref{section:causal_rl}). \\[4pt]
KL & Kullback-Leibler divergence. A measure of how one probability distribution diverges from a reference distribution, used in trust-region and regularization methods (Section~\ref{section:rl_algorithms}). \\[4pt]
LLM & Large Language Model. A neural network trained on large text corpora, the primary object of alignment via RLHF and DPO (Section~\ref{section:rlhf}). \\[4pt]
LQR & Linear-Quadratic Regulator. An optimal control method that computes the policy for linear dynamics with a quadratic cost function via the Riccati equation. \\[4pt]
MARL & Multi-Agent Reinforcement Learning. RL in settings with multiple interacting agents, where each agent's optimal policy depends on the others' behavior (Section~\ref{section:rl_games}). \\[4pt]
MC & Monte Carlo. Methods that estimate value functions from complete episode returns rather than bootstrapped estimates (Section~\ref{section:rl_algorithms}). \\[4pt]
MDP & Markov Decision Process. A formal model of sequential decision-making defined by states, actions, transition probabilities, rewards, and a discount factor (Section~\ref{section:rl_algorithms}). \\[4pt]
MLE & Maximum Likelihood Estimation. A statistical method that estimates parameters by maximizing the likelihood of the observed data (Section~\ref{section:rl_econ_models}). \\[4pt]
MPC & Model Predictive Control. A control method that solves a finite-horizon optimization problem at each timestep using an explicit model of the dynamics, then applies only the first action. \\[4pt]
NE & Nash Equilibrium. A strategy profile in which no player can improve their payoff by unilaterally deviating (Section~\ref{section:rl_games}). \\[4pt]
NFXP & Nested Fixed Point. Rust's algorithm for structural estimation of DDC models that nests value function iteration inside a maximum likelihood loop (Section~\ref{section:rl_econ_models}). \\[4pt]
NPG & Natural Policy Gradient. A policy gradient method that preconditions the gradient by the inverse Fisher information matrix (Section~\ref{section:rl_algorithms}). \\[4pt]
OPE & Off-Policy Evaluation. Estimating the expected return of a target policy using data generated by a different behavioral policy (Section~\ref{section:causal_rl}). \\[4pt]
PEVI & Pessimistic Value Iteration. An offline RL algorithm that subtracts an uncertainty penalty from value estimates to avoid overestimation on unseen state-action pairs (Section~\ref{section:causal_rl}). \\[4pt]
PI & Policy Iteration. A dynamic programming algorithm that alternates between policy evaluation and policy improvement until convergence (Section~\ref{section:rl_algorithms}). \\[4pt]
PID & Proportional-Integral-Derivative controller. A feedback controller that computes a control signal from the weighted sum of the current error, its integral, and its derivative. \\[4pt]
POMDP & Partially Observable MDP. An MDP in which the agent cannot directly observe the state and must maintain beliefs from noisy observations (Section~\ref{section:causal_rl}). \\[4pt]
PPO & Proximal Policy Optimization. A policy gradient algorithm that constrains updates to a trust region via a clipped surrogate objective (Section~\ref{section:rl_algorithms}). \\[4pt]
RL & Reinforcement Learning. A framework for sequential decision-making in which an agent learns a policy by interacting with an environment and observing rewards (Section~\ref{section:rl_algorithms}). \\[4pt]
RLHF & Reinforcement Learning from Human Feedback. A framework for aligning model behavior with human preferences by training a reward model from pairwise comparisons and optimizing a policy against it (Section~\ref{section:rlhf}). \\[4pt]
SAC & Soft Actor-Critic. An off-policy actor-critic algorithm that maximizes a combined objective of expected return and policy entropy (Section~\ref{section:rl_algorithms}). \\[4pt]
SARSA & State-Action-Reward-State-Action. An on-policy TD control algorithm that updates Q-values using the action actually taken in the next state (Section~\ref{section:rl_algorithms}). \\[4pt]
SFT & Supervised Fine-Tuning. The initial stage of LLM alignment in which the model is trained on curated demonstrations before preference optimization (Section~\ref{section:rlhf}). \\[4pt]
SPE & Subgame Perfect Equilibrium. A refinement of Nash equilibrium requiring that strategies constitute a Nash equilibrium in every subgame (Section~\ref{section:rl_games}). \\[4pt]
TD & Temporal Difference. A class of prediction and control algorithms that update value estimates using bootstrapped targets rather than complete returns (Section~\ref{sec:stochastic_approx}). \\[4pt]
TRPO & Trust Region Policy Optimization. A policy gradient algorithm that constrains each update to lie within a KL-divergence trust region of the previous policy (Section~\ref{section:rl_algorithms}). \\[4pt]
TS & Thompson Sampling. A Bayesian bandit algorithm that selects actions by sampling from the posterior distribution over reward parameters (Section~\ref{section:bandits}). \\[4pt]
UCB & Upper Confidence Bound. A bandit algorithm that selects the arm with the highest sum of empirical mean and an exploration bonus (Section~\ref{section:bandits}). \\[4pt]
VI & Value Iteration. A dynamic programming algorithm that iteratively applies the Bellman optimality operator until the value function converges (Section~\ref{section:rl_algorithms}). \\[4pt]
\end{longtable}

\end{document}